\documentclass[a4paper,12pt,twoside]{StyleThese}

\usepackage{amsmath,amssymb,mathabx}             % AMS Math
\usepackage[french]{babel}
\usepackage[latin1]{inputenc}
\usepackage[T1]{fontenc}
\usepackage{booktabs}
\usepackage{threeparttable} %[para] pour alligner les footnote
\usepackage{float}
\usepackage[left=1.1in,right=1.in,top=.5in,bottom=.5in,includefoot,includehead,headheight=13.6pt]{geometry}

\usepackage[nottoc, notlof, notlot]{tocbibind}
\usepackage[french]{minitoc}
\usepackage{aecompl}

\usepackage[intoc]{nomencl}

\makenomenclature

\usepackage{ifpdf}

\ifpdf
  \usepackage[pdftex]{graphicx}
  \DeclareGraphicsExtensions{.jpg}
  \usepackage[a4paper,pagebackref,hyperindex=true]{hyperref}
\else
  \usepackage{graphicx}
  \DeclareGraphicsExtensions{.ps,.eps}
  \usepackage[a4paper,dvipdfm,pagebackref,hyperindex=true]{hyperref}
\fi

\graphicspath{{.}{images/}}

\renewcommand*{\backref}[1]{}
\renewcommand*{\backrefalt}[4]{%
\ifcase #1 %
(Non cité.)%
\or
(Cité en page~#2.)%
\else
(Cité en pages~#2.)%
\fi}

\usepackage{color}
\definecolor{linkcol}{rgb}{0,0,0.4} 
\definecolor{citecol}{rgb}{0.5,0,0} 

\hypersetup
{
bookmarksopen=true,
pdftitle="Imagerie à très haute résolution spatiale et dynamique photométrique pour l'étude de l'environnement des systèmes stellaires",
pdfauthor="Massinissa Hadjara", %auteur du document
pdfsubject="Interférométrie \& coronographie des environnements stellaires", %sujet du document
pdfmenubar=true, %barre de menu visible
pdfhighlight=/O, %effet d'un clic sur un lien hypertexte
colorlinks=true, %couleurs sur les liens hypertextes
pdfpagemode=None, %aucun mode de page
pdfpagelayout=SinglePage, %ouverture en simple page
pdffitwindow=true, %pages ouvertes entierement dans toute la fenetre
linkcolor=linkcol, %couleur des liens hypertextes internes
citecolor=citecol, %couleur des liens pour les citations
urlcolor=linkcol %couleur des liens pour les url
}

\usepackage{rotating}                    % Sideways of figures & tables
\usepackage{graphicx}
\usepackage{epstopdf}
\usepackage{txfonts}
\usepackage{natbib}
\bibpunct{(}{)}{;}{a}{}{,} % to follow the A&A style
\usepackage{fancyhdr}                    % Fancy Header and Footer

\let\headruleORIG\headrule
\renewcommand{\headrule}{\color{black} \headruleORIG}

\usepackage{colortbl}
\arrayrulecolor{black}

\fancypagestyle{plain}{
  \fancyhead{}
  \fancyfoot{}
  
}

\usepackage{MyAlgorithm}
\usepackage[noend]{MyAlgorithmic}

\makeatletter

\def\cleardoublepage{\clearpage\if@twoside \ifodd\c@page\else%
  \hbox{}%
  \thispagestyle{empty}%              % Empty header styles
  \newpage%
  \if@twocolumn\hbox{}\newpage\fi\fi\fi}

\makeatother
 
{%

\hrulefill
\vspace*{0.5cm}%
\end{minipage}
}

\let\minitocORIG\minitoc
\renewcommand{\minitoc}{\minitocORIG \vspace{1.5em}}

\usepackage{multirow}
\usepackage{slashbox}

{ \begin{list}%
	{$\bullet$}%
	{\setlength{\labelwidth}{25pt}%
	 \setlength{\leftmargin}{30pt}%
	 \setlength{\itemsep}{\parsep}}}%
{ \end{list} }

\renewcommand{\epsilon}{\varepsilon}

\newenvironment{vcenterpage}
{\newpage\vspace*{\fill}\thispagestyle{empty}}
{\vspace*{\fill}}

\usepackage[final]{pdfpages} 

\begin{document}

%\begin{samepage}
\begin{titlepage}

\begin{center}
\noindent \Large \textbf{T H È S E} \\
\vspace*{0.1cm}
\noindent \normalsize {Présentée afin d'obtenir le titre de} \\
\vspace*{0.1cm}
\noindent {\large \textbf{DOCTEUR DE L'UNIVERSITE NICE-SOPHIA ANTIPOLIS - UFR SCIENCE}} \\
\vspace*{0.1cm}
\noindent {\textbf{École Doctorale des Sciences Fondamentales et Appliquées}} \\
\vspace*{0.5cm}
\noindent \textbf{Specialité : \textsc{SCIENCES DE L'UNIVERS}}\\
\vspace*{0.4cm}
\noindent \normalsize {Soutenue par\\}
\noindent \Large \textbf{Massinissa \textsc{Hadjara}} \\
\noindent \normalsize {et préparée au sein du:} \\
\noindent \normalsize {Laboratoire J.-L. Lagrange (UMR 7293)} \\
\noindent \normalsize {\& de l'Observatoire d'Alger -Algérie- (CRAAG)} \\
\noindent \normalsize {sur le sujet :} %\\
%\vspace*{0.3cm}
\begin{center}
\begin{tabular}{c}
\hline
\hline
%\vspace*{0.2cm}
%\noindent {\Huge \textbf{Imagerie à très haute résolution }} \\
%\noindent {\Huge \textbf{spatiale et dynamique photométrique }} \\
%\noindent {\Huge \textbf{pour l'étude de l'environnement}} \\
%\noindent {\Huge \textbf{des systèmes stellaires}} \\
\noindent {\Huge \textbf{Observations et modélisations}} \\
\noindent {\Huge \textbf{spectro-interférométriques}} \\
\noindent {\Huge \textbf{longue base des étoiles et de leur}} \\
\noindent {\Huge \textbf{environnement proche}} \\
%\vspace*{0.2cm}
%\framebox{\begin{figure}
%\includegraphics{stellar}
 %\raisebox{-\totalheight}{\includegraphics[width=0.6\textwidth, height=60mm]{Titlepage/stellar3}}\\
 \raisebox{-\totalheight}{\includegraphics[width=1.\textwidth, height=60mm]{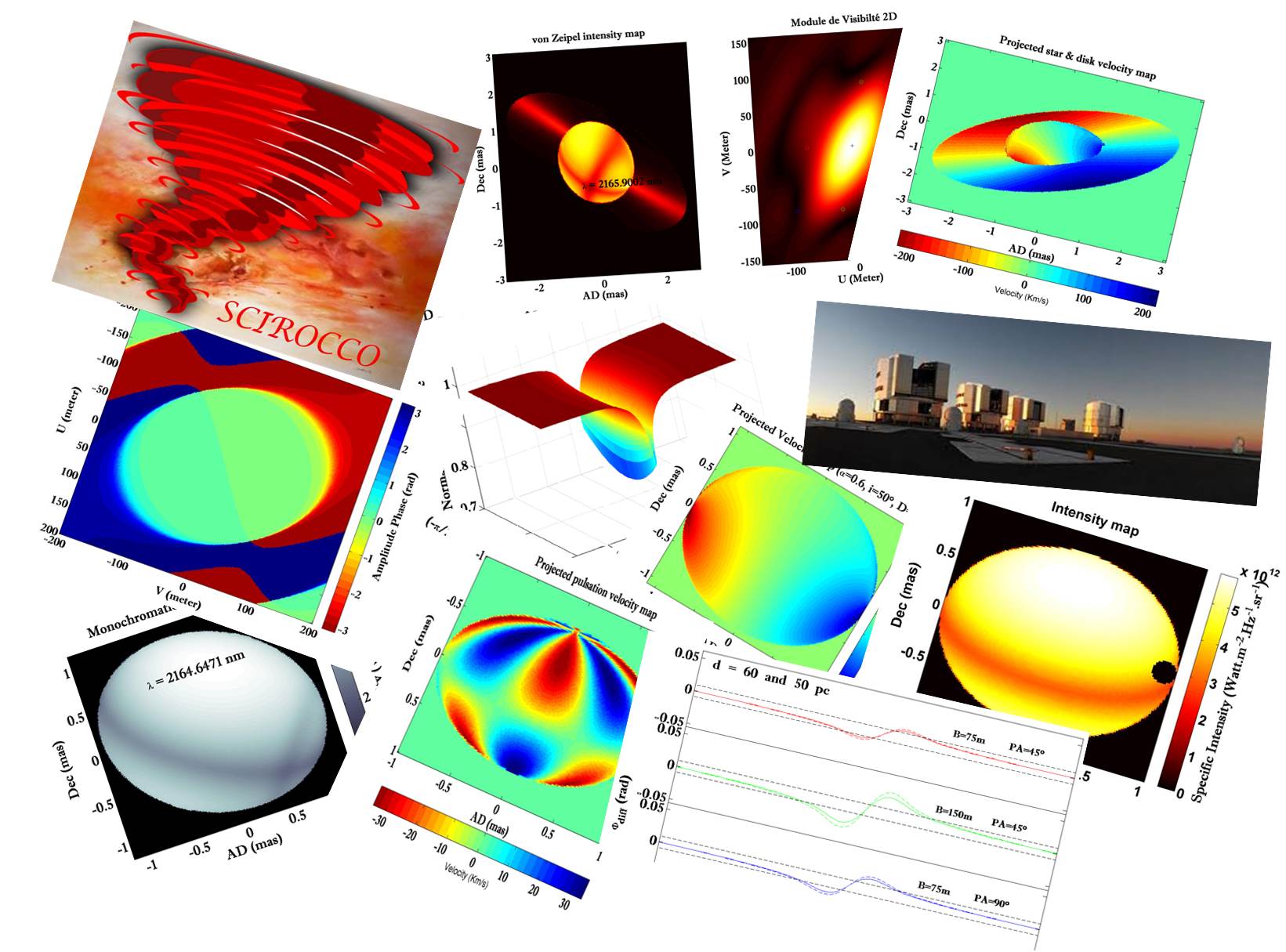}}\\
%  \end{figure}}
\hline
\hline
%\vspace*{0.2cm}
\end{tabular}
\end{center}
%\vspace*{0.2cm}
\end{center}

\vspace*{0.2cm}

\noindent {\normalsize Soutenue le 31 mars 2015 devant un jury composé de :}  \\
\begin{center}
\noindent \normalsize
\begin{tabular}{clccc}
		{}	& Philippe  \textsc{Stee}	& - & \textit{(Président)} & {} \\ [0.5ex]
		
      \multirow{2}{*}{\parbox[c]{5em}{\includegraphics[width=0.12\textwidth, height=14mm]{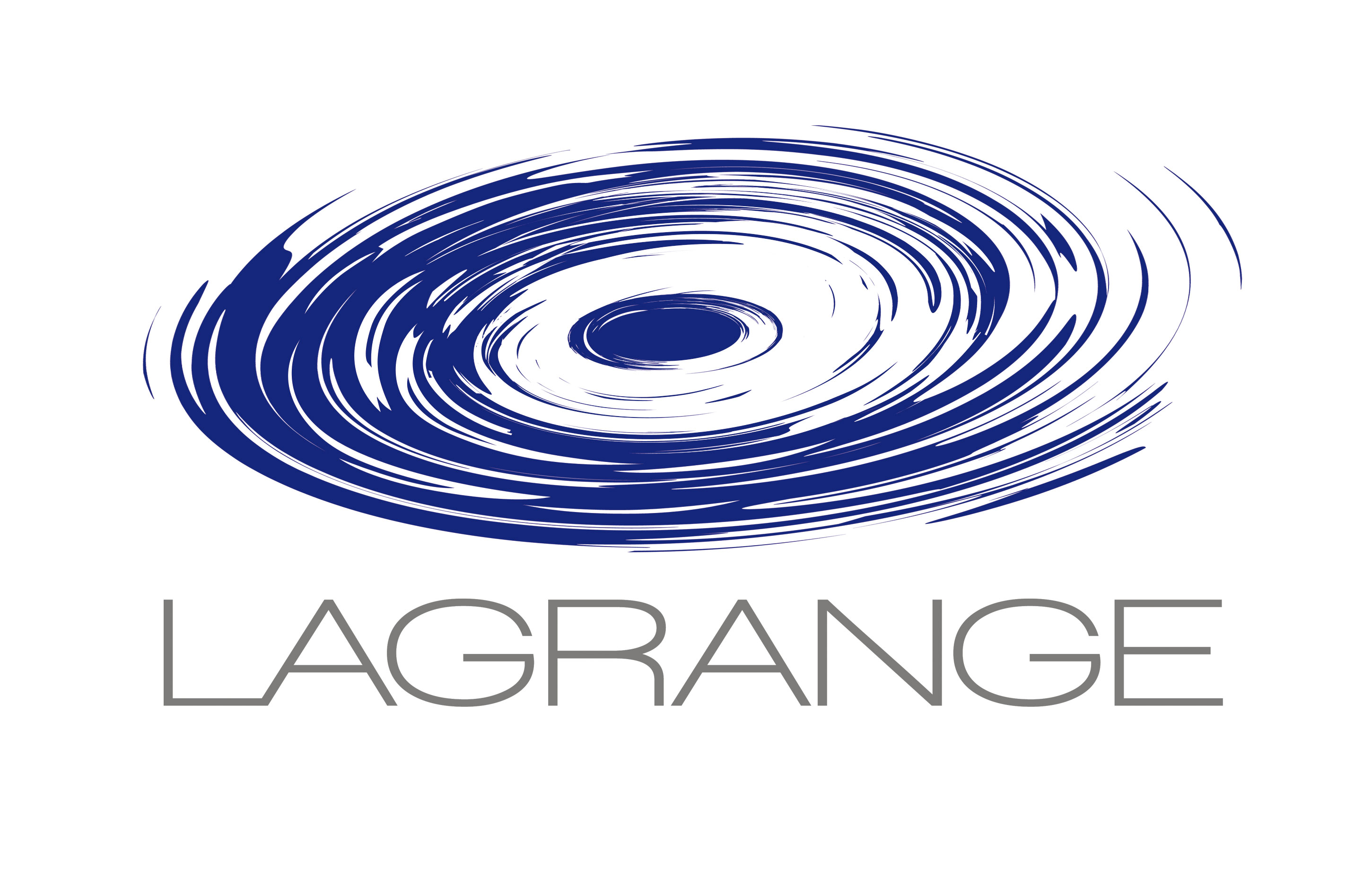}}} &  Pierre \textsc{Kervella}		& - & \textit{(Rapporteur)} & \parbox[c]{4em}{\includegraphics[width=0.15\textwidth, height=12mm]{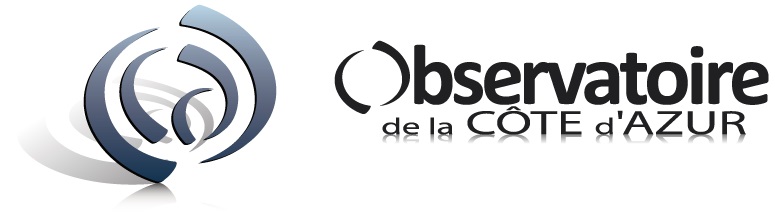}}\\ [0.5ex]
      
      {}	& Coralie  \textsc{Neiner}	& - & \textit{(Rapporteur)} & {} \\ [0.5ex]
      
     \multirow{2}{*}{\parbox[c]{5em}{\includegraphics[width=0.15\textwidth, height=10mm]{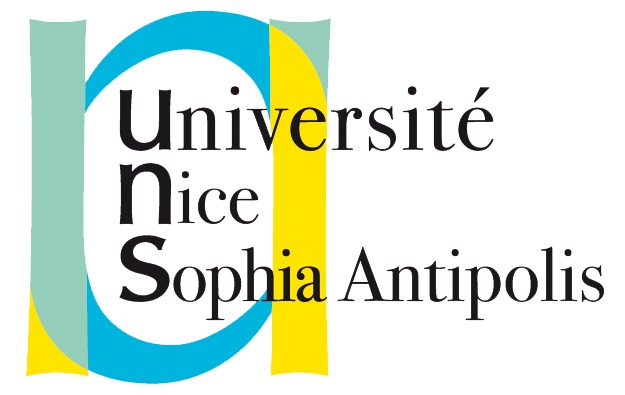}}} & Jean-Baptiste \textsc{Le Bouquin}	 & - & \textit{(Examinateur)}& \parbox[c]{1em}{\includegraphics[width=0.09\textwidth, height=10mm]{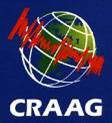}}\\ [0.5ex]
     
     {}	& Juan  \textsc{Zorec}	& - & \textit{(Invité)} & {} \\ [0.5ex]
     
     {} & Farrokh \textsc{Vakili}	& - & \textit{(Directeur de thèse)} & {}\\ [0.5ex]
     
     {} & Armando \textsc{Domiciano de Souza}	& - & \textit{(Co-directeur de thèse)} & {}
%\begin{tabular}{cllccc}
%		{}	& Pr	& Xxxxx \textsc{Yyyy}		& - & \textit{(Président)} & {} \\
%      \parbox[c]{5em}{\includegraphics[width=0.12\textwidth, height=14mm]{Titlepage/Logo_Lagrange}} & Dr	& Xxxxx \textsc{Yyyy}		& - & \textit{(Rapporteur)} & \parbox[c]{4em}{\includegraphics[width=0.15\textwidth, height=12mm]{Titlepage/Logo_OCA}}\\
%     \parbox[c]{5em}{\includegraphics[width=0.15\textwidth, height=10mm]{Titlepage/Logo_Unice}} {} & DR	& Xxxxx \textsc{Yyyy}			& - & \textit{(Directeurde thèse)}& \parbox[c]{1em}{\includegraphics[width=0.09\textwidth, height=10mm]{Titlepage/Logo_CRAAG}}\\
%     {} & DR	& Xxxxx \textsc{Yyyy}			& - & \textit{(Invité)} & {}
\end{tabular}
\end{center}

\end{titlepage}
\sloppy

\titlepage
%\end{samepage}

\dominitoc

\pagenumbering{roman}

 \cleardoublepage

\section*{Remerciements}

\noindent {\small Je tiens tout d'abord à exprimer ma profonde reconnaissance à mes directeurs de thèse, Farrokh Vakili et Armando Domiciano de Souza, pour m'avoir ouvert les portes du monde merveilleux et fascinant de l'interférométrie stellaire et des rotateurs rapides.
Je voudrais remercier les rapporteurs de ma thèse, Pierre Kervella et Coralie Neiner pour avoir consacré de leurs temps pour lire et corriger mon manuscrit. Leurs pertinentes remarques et suggestions m'ont permis de fortement l'améliorer. J'aimerais tout particulièrement remercier Philippe Stee pour avoir accepté de présider le jury de ma soutenance. Ainsi que les autres membres de mon jury : Jean-Baptiste Le Bouquin et Juan Zorec.
\\
Mes remerciements les plus chaleureux s'adressent à l'ensemble de mes collègues et amis du PES ; Bruno le magnifique, l'étincelant Florentin  et le brillant Anthony (avec qui j'ai eu beaucoup de plaisir à travailler),  Nicolas esprit vif, le père Éric Fossat (et ces randonnées  thérapeutiques), l'aimable Martin, l'honorable Yan, le joyeux Sébastien Flament, le sympathique Sébastien Ottogali, la gentille Carolyn, le minutieux Yves, Magic Christophe, le paisible Jean-Baptiste Daban, le vaillant Pierre Antonelli, Michel Faguet le patron, Alain Spang la force tranquille, les M\&Ms Frantz le Jedi \& Patrice l'athlétique, l'agréable Philippe Berio, Pierre Cruzalebes le Grand, la talentueuse Roxane, Eric Lagadec le taquin, Djamel l'aventurier, l'admirable Andrea Chiavassa et l'adorable Christine (qui sont au CION plutôt), la pétillante Aurèlie, et le regretté Olivier. Sans oublier le bienveillant FX, la divine Sylvie et le seigneur de tous les seigneurs Stéphane. Merci les amis.
\\
C'est tout naturellement, que  je dois remercier aussi Romain Petrov et Slobodan Jankov pour leurs aides, conseils et soutien ainsi que pour leurs amitiés (ce fut pour moi à la fois un honneur et un plaisir de travailler avec eux).
\\
Je me dois de remercier aussi mes collègues de l'université : David Mary, Lyu et tout particulièrement Marcel Carbillet. Mes collègues de Calern : Yves Rabbia et  tout particulièrement Jean-Pierre Rivet pour sa grande gentillesse et son soutien. Les copains d'Artemis \& mes amis doctorants : Gillaume, Karrelle, Mamadou, Gaetan, Suvendu, Florent et Zienab. Que ceux que j'ai oublié m'excusent.
\\
Je remercie l'ensemble de tout le personnel de l'OCA, pour leur sympathie et bonne humeur à toute épreuve ; l'Accueil : Isabelle, Nathalie Christian  et Aziz. Les secrétaires : Isabelle G., Murielle, Jocelyne, Delphine, Sylevie et Gérard. Le restaurant : Khaled, Giselin, Nadia et Karima. L'atelier mécanique (Serge et Thierry). Ainsi que tous les informaticiens (Marie-Laure, Jean-Philippe, Serge et Daniel en particulier), Sans oublier David, Jean-Marie, Mohamed et Robert. Que ceux que j'ai oublié m'excusent.
\\
Je tiens également à remercier, les amis que je me suis fait en dehors de l'OCA : Les Alem, les Susini, les Rekkas, mes amis d'Astrorama et d'Auchan Trinité, ainsi que Stéphanie de Recherche et Avenir. Mes amis de longue date : François Impens, Lionel Bigot, Dario Vincenzi, Marco Delbo et Wassila.
\\
Mes amis algériens de l'Observatoire d'Alger : Fodil, Rabah, Yacine Athmani et tout particulièrement Zouleikha (ma grande s\oe{}ur). Sans oublier Noureddine Moussaoui de l'Université d'Alger (USTHB), Toufik Moutefaoui (de l'Université de Béjaïa) et Samir et son Association Sirius Béjaïa.
\\
Mes remerciements les plus profonds s'adressent tout naturellement à mes parents, ma s\oe{}ur et mon frère, ainsi qu'à mon épouse et sa famille, pour leurs tendresses, leurs encouragements et leurs soutiens sans faille. Sans oublier toute ma famille à Béjaïa (Particulièrement mon cousin Nadhir et toute la famille Lalaoui).}
\\

\begin{figure}[ht]
\centering
\includegraphics[width=0.16\hsize,draft=false]{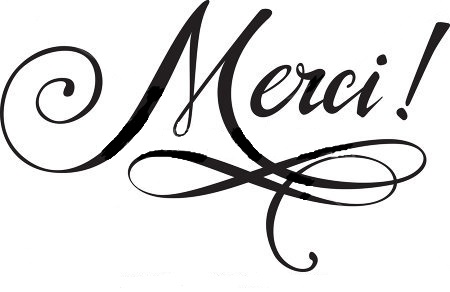}
\end{figure}

\newpage
\vspace*{0.1cm}
\emph{"Vos enfants ne sont pas vos enfants. Ils sont fils et filles du désir de vie en lui-même. Ils viennent par vous mais non de vous, et bien qu'ils soient avec vous, ce n'est pas à vous qu'ils appartiennent. Vous pouvez leur donner votre amour mais non vos pensées, car ils ont leurs propres pensées. Vous pouvez loger leurs corps mais non leurs âmes, car leurs âmes habitent la demeure de demain, que vous ne pouvez vous efforcer de leur ressembler, mais n'essayez pas qu'ils vous ressemble. Car la vie ne retourne pas en arrière ni s'attarde à hier. Vous êtes les arcs qui projettent vos enfants telles des flèches vivantes. L'archer voit la cible sur le chemin de l'infini, et il vous courbe avec toute sa force pour ses flèches aillent vite et loin. Que cette courbure, dans les mains de l'archer, tende à la joie; car comme il aime la flèche qui vole, il aime aussi l'arc qui est stable."}\\
\noindent {\footnotesize \textbf{Khalil Gibran (1883-1931) : Extrait du recueil Le Prophète 1923.}}\\
\\ \\ \\ \\ \\
\begin{figure}[ht]
\centering
\includegraphics[width=1.0\hsize,draft=false]{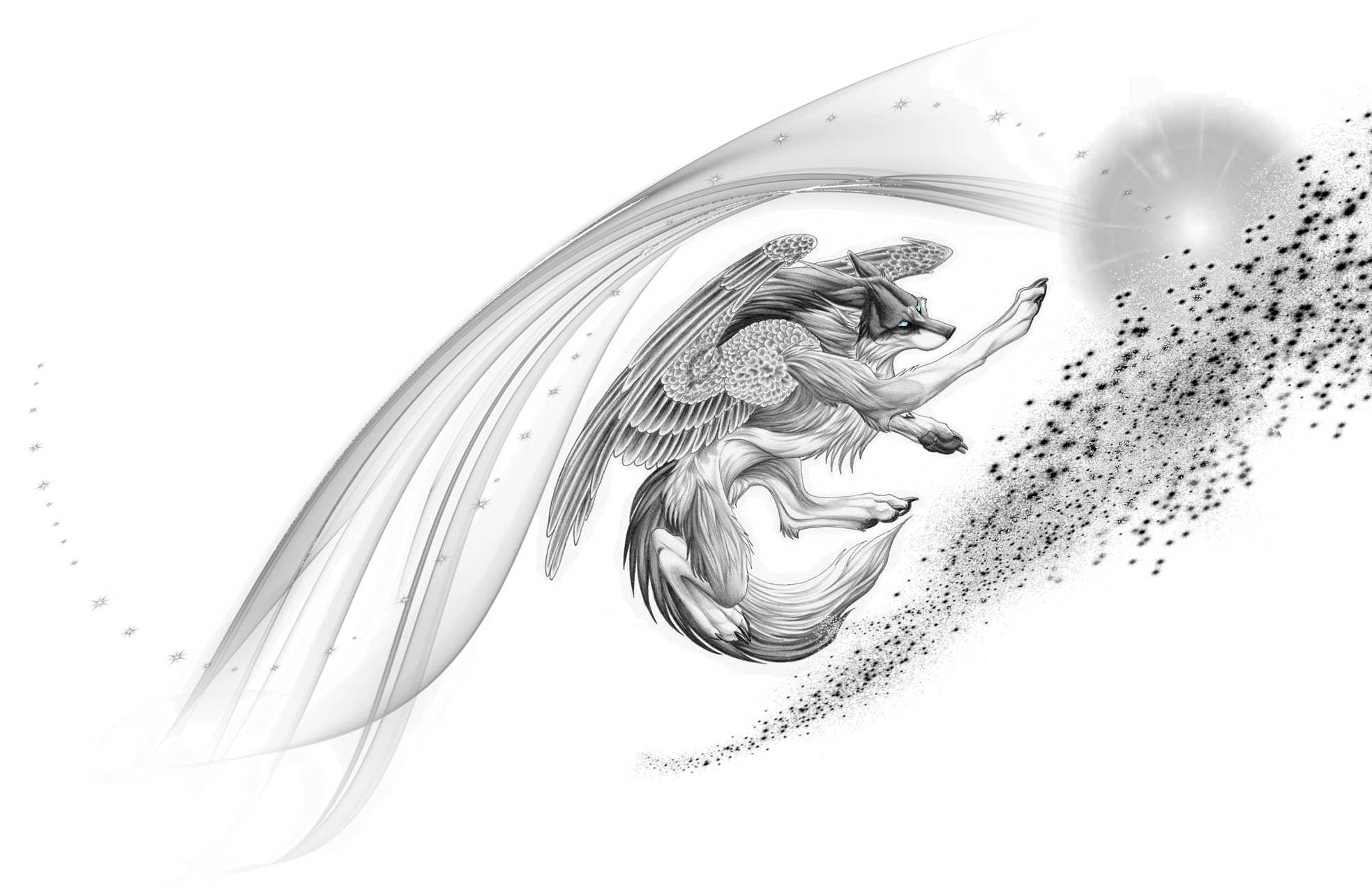}
\end{figure}
\\ \\ \\ \\ \\
\emph{"Si tu donnes un poisson à un homme, il mangera un jour. Si tu lui apprends à pêcher, il mangera toujours."}\\
\noindent {\footnotesize \textbf{Lao Tseu (Ve - IVe siècle av. J.-C.).}}\\

\tableofcontents
\listoftables
\listoffigures

\mainmatter

\chapter{Introduction}
%\label{chap:Introduction}
%\addcontentsline{toc}{chapter}{Introduction}
\begin{figure}[h!]
\centering
\includegraphics[height=0.5\hsize,draft=false]{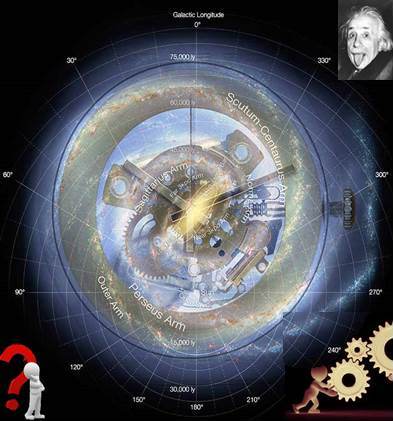}
\end{figure}
\label{chap:intro}
\minitoc

%%%%%%%%%%%%%%%%%%%%%%%%%%%%%%%%%%%%%%%%
\def\vsini{v_\mathrm{eq} \sin i} 
\def\kms{\mathrm{km.s}^{-1}}
\def\phidiff{\phi_\mathrm{diff}}
\def\Rsun{\mathrm{R}_{\odot}}
\def\Lsun{\mathrm{L}_{\odot}}
\def\Msun{\mathrm{M}_{\odot}}
\def\Tmean{\overline{T}_\mathrm{eff}}
\def\diameq{\diameter_\mathrm{eq}}
\def\chir{\chi_\mathrm{r}}
\def\chimin{\chi_\mathrm{min}}
\def\chirmin{\chi_\mathrm{min,r}}
%%%%%%%%%%%%%%%%%%%%%%%%%%%%%%%%%%%%%%%%

\section{Notations et abréviations utilisées dans ce manuscrit}

\begin{table*}[htbp]
\centering
%\begin{sideways}
\caption{Principales abréviations utilisées dans ce manuscrit.}\label{tab_abrevia}
\centering
\begin{tabular}{||c||c|c||}
  \hline\hline
  & \multicolumn{2}{c||}{ABRÉVIATIONS}\\
  \hline
  & En Français & En Anglais\\
  \hline\hline\hline
  %ALI & Lambda-Itération Accélérée & Accelerated Lambda Iteration\\
  %\hline
  AMBER & \multicolumn{2}{c||}{Astronomical Multi-BEam combineR}\\
  \hline
  AT & \multicolumn{2}{c||}{Auxiliary Telescope}\\
  \hline
  CDS & \multicolumn{2}{c||}{Centre de Données astronomiques de Strasbourg}\\
  \hline
  CL & Linéarisation Complète & Complete Linearisation\\
  \hline
  CSE & Environnement circumstellaire & CircumStellar Environment\\
  \hline
  ddm & Différence de Marche Optique & optical path difference (OPD)\\
  \hline
  DFE & Éléments Finis Discrets & Discret Finite Elements\\
  \hline
  DI & Interférométrie Différentielle & Differential interferometry\\
  \hline
  E-ELT & \multicolumn{2}{c||}{European Extremely Large Telescope}\\
  \hline
  ESO & Observatoire Européen Austral & European Southern Observatory\\
  \hline
  ETL & Équilibre Thermodynamique Local & Local Thermodynamic Equilibrium\\
  \hline
  ETR & Équation de Transfert Radiatif & Radiative Transfer Equation\\
  \hline
  FFT & Transformation de Fourier Rapide & Fast Fourier Transform\\
  \hline
  GI2T & Grand Interféromètre à 2 Télescopes & Large Interferometer of 2 Telescope\\
  \hline
  HR & \multicolumn{2}{c||}{Hertzsprung-Russell}\\
  \hline
  HRA & Haute Résolution Angulaire & High Angular Resolution\\
  \hline
  IDI & Imagerie Doppler Interférométrique & Interferometric Doppler Imaging\\
  \hline
  IDL & \multicolumn{2}{c||}{Iterative Data Language}\\
  \hline
  JMMC & Centre Jean-Marie Mariotti & Jean-Marie Mariotti Center\\
  \hline
  mas & milli second d'angle & milli arc second\\
  \hline
  LBV & \multicolumn{2}{c||}{Luminous Blue Variable stars}\\
  \hline
  MATISSE & \multicolumn{2}{c||}{Multi AperTure mid-Infrared SpectroScopic Experiment}\\
  \hline
  Matlab & \multicolumn{2}{c||}{MATtrix LABoratory}\\
  \hline
  MIDI & \multicolumn{2}{c||}{MID-infrared Interferometric instrument}\\
  \hline
  OLBI & Interférométrie Optique à Longue Base & Optical Long Baseline Interferometry\\
  \hline
  PNR & Pulsations Non-Radiales & Non-Radial Pulsations(NRP)\\
  \hline
  PR & Pulsations Radiales & Radial Pulsations\\
  \hline
  PRIMA & \multicolumn{2}{c||}{Phase-Referenced Imaging and Micro-arcsecond Astrometry}\\
  \hline
  SED & distribution spectrale d'énergie  & Spectral Energy Distribution\\
  \hline
  SPB & Étoiles B avec pulsation lente & Slowly Pulsating B stars\\
  \hline
  SP & Séquence Principale & Main Sequence\\
  \hline
  TF & Transformation de Fourier & Fourier Transform\\
  \hline
  UAI & Union Astronomique Internatinale & International Astronomical Union\\
  \hline
  UT & \multicolumn{2}{c||}{Unit Telescope}\\
  \hline
  VLTI & \multicolumn{2}{c||}{Very Large Telescope Interferometer}\\
  \hline
  WR & \multicolumn{2}{c||}{Wolf-Rayet}\\
\hline\hline
\end{tabular}
\end{table*}

\begin{table*}[htbp]
\centering
%\begin{sideways}
\caption{Principales notations utilisées dans ce manuscrit.}\label{tab_nota}
\centering
\begin{tabular}{||c||c||}
  \hline\hline
  \multicolumn{2}{||c||}{NOTATIONS DES PRINCIPAUX PARAMÈTRES PHYSIQUES}\\
  \hline
  Notation & Description\\
  \hline\hline\hline
  $\alpha$ & Coefficient de rotation différentielle\\
  \hline
  $\beta$ & Coefficient d'assombrissement gravitationnel\\
  \hline
  $i$ & Angle d'inclinaison de l'étoile par rapport à la ligne de visée\\
  \hline
  $\vsini$ & Vitesse radiale de rotation équatoriale  le long de l'axe de visée\\
  \hline
  $\phidiff$ & Phase différentielle \\
  \hline
  $E_c$ & Estimateur de rotation critique\\
  \hline
  $\eta$ & Estimateur de rotation critique $N^\circ 2$ \\
  \hline
  $\Rsun$ & Rayon solaire \\
  \hline
  $R_{eq}$ & Rayon équatorial\\
  \hline
  $R_{eq,crit}$ & Rayon équatorial critique\\
  \hline
  $R_{pol}$ & Rayon pôlaire\\
  \hline
  $V_{proj}$ & Vitesse projetée\\
  \hline
  $v_{eq}$, $v_{eq,rot}$ & Vitesse radiale de rotation équatoriale\\
  \hline
  $v_{eq,crit}$ & Vitesse radiale de rotation équatoriale critique \\
  \hline
  $\Omega_{eq}$ & Vitesse angulaire de rotation équatoriale \\
  \hline
  $\Omega_{crit}$ & Vitesse angulaire de rotation équatoriale critique \\
  \hline
  $\Lsun$ & Luminosité solaire \\
  \hline
  $\Msun$ & Masse solaire\\
  \hline
  $M_*$ & Masse stellaire\\
  \hline
  $G$ & Constante gravitationnelle (constante de Newton)\\
  \hline
  $g$ & La gravité\\
  \hline
  $\Tmean$ & Température effective moyenne\\
  \hline
  $T_{eq}$ & Température effective à l'équateur\\
  \hline
  $T_{pol}$ & Température effective aux pôles\\
  \hline
  $\diameq$ & diamètre angulaire équatorial \\
  \hline
  $d$ & Distance terre-étoile\\
  \hline
  $I$ & Intensité lumineuse \\
  \hline
  $I_{cont}$ & Intensité lumineuse dans le continuum\\
  \hline
  $PA_{rot}$ & Angle de position de l'axe de rotation\\
  \hline
  $H$ & Profil de raie\\
  \hline
  $\lambda$ & Longueur d'onde\\
  \hline
  $c$ & Vitesse de la lumière\\
  \hline
  $\theta, \theta'$ & Latitude, co-Latitude\\
  \hline
  $\phi$ & Longitude\\
  \hline
  $S$, $F$ & Spectre, Flux\\
  \hline
  $P$ & Photo-centre\\
  \hline
  $\Phi$ & Phase\\
  \hline
  $\Psi$ & Phase de Clôture\\
  \hline
  $\epsilon_{\lambda}$ & Coefficient d'assombrissement centre-bord à la longueur d'onde $\lambda$\\
  \hline
  $\mu$ & Cosinus de l'angle entre la normale à la surface d'un point considéré\\
   & et de la direction d'observation\\
  \hline
  $h$ & Constante de Planck\\
  \hline
  $\sigma_{SB}$ & Constante de Stefan-Boltzmann\\
  \hline
  $\sigma_{B}$ & Constante de Boltzmann\\
  \hline
  $\in$ & Paramètre de Rieutord\\
  %$\chir$ & \\
  %\hline
  %$\chimin$ & \\
  %\hline
  %$\chirmin$ & \\
  \hline\hline
\end{tabular}
\end{table*}

%\newpage\newpage
\cleardoublepage

\section{Introduction générale}

\textit{"C'est en réalité tout notre système de conjectures qui doit être prouvé ou réfuté par l'expérience. Aucune de ces suppositions ne peut être isolée pour être examinée séparément. . . Les concepts physiques sont des créations libres de l'esprit humain et ne sont pas, comme on pourrait le croire, uniquement déterminés par le monde extérieur. Dans l'effort que nous faisons pour comprendre le monde, nous ressemblons quelque peu à l'homme qui essaie de comprendre le mécanisme d'une montre fermée. Il voit le cadran et les aiguilles en mouvement, il entend le tic-tac, mais il n'a aucun moyen d'ouvrir le boîtier. S'il est ingénieux il pourra se former quelque image du mécanisme, qu'il rendra responsable de tout ce qu'il observe, mais il ne sera jamais sûr que son image soit la seule capable d'expliquer ses observations. Il ne sera jamais en état de comparer son image avec le mécanisme réel, et il ne peut même pas se représenter la possibilité ou la signification d'une telle comparaison. Mais le chercheur croit certainement qu'à mesure que ses connaissances s'accroîtront, son image de la réalité deviendra de plus en plus simple et expliquera des domaines de plus en plus étendus de ses impressions sensibles. Il pourra aussi croire à l'existence d'une limite idéale de la connaissance que l'esprit humain peut atteindre. Il pourra appeler cette limite idéale la vérité objective."}\\

C'est en ces termes qu'Albert Einstein \footnote {Albert Einstein et Léopold Infeld; L'évolution des idées en physique, 1936.} s'exprima au sujet de l'approche scientifique, qui selon lui consiste à concilier des modèles théoriques avec les observations empiriques.\\

Le rouage que j'ai entrepris de comprendre durant ma thèse se situe dans l'étude du phénomène d'évolution des étoiles et de leur environnement proche. La technique d'observation produisant les données utilisées est la spectro-interférométrie.\\

Habituellement en Astronomie on utilise un télescope pour scruter le ciel. Plus le diamètre du télescope est grand, plus la résolution spatiale (la finesse des détails observés) s'améliore. Hélas la construction d'un télescope ayant un miroir monolithique supérieur à 10m est difficilement envisageable de nos jours pour des raisons technologiques. La résolution maximale $\alpha= 1.22\frac{\lambda}{D}$ (où $\lambda$ est la longueur d'onde de la lumière observée et $D$ le diamètre du télescope) pour un télescope de $10m$ de diamètre est de $14 mas$, dans le visible. L'interférométrie s'affranchit de cette limite, en recombinant simultanément au moins 2 télescopes séparés, l'un de l'autre d'une distance $B$ qu'on appelle base interférométrique. Ainsi et dans le cas d'une base $B$ de $100m$ la résolution maximale $\alpha= \frac{\lambda}{B}$, sera dans le visible, de $1.0mas$ à $500nm$ \citep{1972ailt.conf..389L}.\\

L'observation sur une gamme de différentes longueurs d'onde (ou observation spectroscopique) nous permet d'avoir une idée sur la composition chimique de l'objet ainsi que sur sa cinématique grâce à l'effet Fizeau-Doppler. Cela permet d'étudier la rotation stellaire par exemple. En effet, toutes les étoiles tournent sur elles-mêmes à des vitesses différentes. Nées au début de leur vie au sein d'un disque d'accrétion, de gaz et de poussière, elles tournent à très grande vitesse sous l'effet de la gravité. Certaines étoiles gardent cette caractéristique tout le long de leur vie par inertie et tournent même tellement vite jusqu'à affecter leurs caractéristiques géométriques, lumineuses et chimiques, (voir chapitre \ref{chap:rota}). On les appelle des rotateurs stellaire rapides. Par la suite seules les étoiles riches en métaux, ayant entretenus un champ magnétique fossile en formant une magnétosphère (pour les étoiles chaudes), ou ayant une zone convective, et générant un fort champ magnétique par effet dynamo (pour les étoiles froides), perdent de leur moment cinétique (ce fut le cas du Soleil -notre étoile froide-). Sous l'effet de la force centrifuge, plus l'étoile tourne vite, plus elle s'élargit à l'équateur et plus elle sera de forme aplatie le long de son axe de rotation, engendrant à l'équateur, une baisse de la température, de la gravité ainsi que de l'intensité lumineuse. C'est l'effet appelé von Zeipel \citep{1924MNRAS..84..665V}.\\

C'est donc à l'étude de ces objets passionnants, que je compare souvent à une patineuse artistique, que fut consacré l'essentiel de ma thèse. Les étoiles en rotation étudiées, et dont les données étaient accessibles à notre équipe, sont Achernar, Altair, $\delta$ Aquilae et Fomalhaut.\\

Pour modéliser ce phénomène (mon rouage) et en prenant en compte plusieurs concepts physique importants, j'ai développé un code au nom sec et poussiéreux : SCIROCCO (Simulation Code for Interferometric-observation of ROtators and CirCumsteller Objects). Ce code m'a permis par comparaison de mesures interférométriques à partir des données réelles de retrouver les paramètres fondamentaux de ces étoiles: degré d'aplatissement, vitesse de rotation, rayon équatorial, masse, températures (aux pôles, à l'équateur et à chaque latitude) et gravités de surface.\\

La mesure, caractérisée par sa précision et son exactitude, constitue une notion fondamentale en astrophysique observationnelle. Et si l'astrométrie arpente la position des astres et sa variation perceptible, la photométrie et la spectroscopie ont longtemps dominé le terrain des techniques de mesure en astrophysique. Ce n'est que très récemment que l'interférométrie stellaire optique est devenue une méthode couramment utilisée pour fournir des contraintes "directes" sur la morphologie (presque toujours dépendante de la couleur) des objets célestes qu'elle permet de scruter.\\

L'interférométrie et l'observation des objets stellaires ont été intimement liées dès les premières tentatives d'obtenir des franges d'interférences suggérées par A.H. Fizeau \citep{1868PB....66..932F} et mises en \oe{}uvre par Stéphan \citep{1874PB....78..1008S} pour mesurer le diamètre apparent des étoiles les plus brillantes du ciel, et ensuite pleinement exploitée par A.A. Michelson et F.G. Pease et son équipe par la suite, dans la limite des technologies disponibles au début du 20\up{ième} siècle \citep{1921PASP...33..171P}. L'interférométrie d'intensité a permis de fournir le premier catalogue de paramètres fondamentaux stellaires par Hanbury Brown et Twiss dans les années 1970 \citep{1974MNRAS.167..121H}. L'interférométrie d'amplitude, reprise depuis 1974 par A. Labeyrie à Nice, puis sur le plateau de Calern \citep{1975ApJ...196L..71L}, s'approche de la technique utilisée par Michelson. Elle fut appliquée aux interféromètres prototypes I2T et GI2T du plateau de Calern \citep{1988ESOC...29..695K, 1988ESOC...29..729M}.
Avec l'avènement du VLTI et de ses différents instruments focaux : MIDI, AMBER, PIONIER et bientôt Gravity et MATISSE \citep{2003Ap&SS.286...73L, 2007A&A...464....1P, 2010SPIE.7734E..35B}, ainsi que CHARA et ses instruments recombinateurs comme VEGA \citep{2000SPIE.4006..465M}, l'interférométrie stellaire, combinée à la synthèse d'ouverture par rotation terrestre (expliquée dans la sous-section \ref{OLBI}), est devenue une technique à découvertes et en même temps de suivi des étoiles (ex. \citet{2008poii.conf.....R}) mais de manière beaucoup plus restreinte, des objets extra-galactiques (ex. \citet{2012ApJ...750...33L}). Elle est de plus en plus employée et considérée par les astronomes en général (ex. la reconstruction d'image d'Altair via \citet{2007Sci...317..342M}).\\

Le champ d'application de l'imagerie interférométrique optique permet aujourd'hui, au-delà des contraintes qu'elle apporte sur les paramètres stellaires (le rayon angulaire, la vitesse angulaire équatoriale, l'inclinaison et l'angle de projection de l'axe de rotation; \citet{2012A&A...545A.130D}), la détection de binaires très serrées, d'environnements stellaires asymétriques \citep{1987JAF....29...20V} ou des effets quantifiés par les paramètres fondamentaux tels que la rotation, la variabilité, la perte de masse sous l'effet du vent radiatif, et éventuellement la détection directe des oscillations radiales ou non des étoiles pulsantes (ex. sur les céphéïdes; \citet{2011A&A...525A..67N}).\\

Le travail décrit dans ce manuscrit se focalise principalement sur un aspect particulier de la physique stellaire qui consiste à détecter et à mesurer le degré d'aplatissement d'une étoile possédant une rotation importante par l'effet dit de von Zeipel \citep{1924MNRAS..84..665V}. Via la technique d'interférométrie optique qui en s'attachant à exploiter la possibilité remarquable de l'interférométrie différentielle \citep{1982AcOpt..29..361B, 1989dli..conf..249P} permet d'analyser les propriétés des franges d'interférence en fonction de la longueur d'onde. Ces méthodes sont offertes en particulier par le spectro-interféromètre AMBER sur le VLTI.  Bien sûr, cette description n'a pas la prétention réductrice de résumer les différentes facettes de la rotation stellaire au seul effet dit de von Zeipel. Elle a pour objet de tenter de démontrer qu'à partir d'un modèle analytique simple, mais suffisamment complet on peut exploiter la technique, somme toute novatrice d'interférométrie spectrale différentielle, comme un outil de diagnostic pour comprendre la rotation rapide et ses effets à travers le diagramme HR. Ce faisant notre approche se veut aussi d'ouvrir la voie pour tester des théories sous-jacentes de formation et d'évolution stellaire par une approche observationnelle couplant l'interférométrie et la modélisation analytique qui pourrait, en ligne de perspective, diagnostiquer l'impact d'autres mécanismes et processus physiques dans le champ vaste de la physique stellaire.\\

Mon manuscrit  de thèse s'organise en 6 chapitres :\\

\textbf{Le premier chapitre} présente le contexte général du travail ainsi que les notations et abréviations utilisées dans ce manuscrit.\\

\textbf{Le deuxième chapitre} se consacre à l'introduction du phénomène de la rotation stellaire, où après un bref historique et une succincte présentation de l'intérêt scientifique de l'étude de ce phénomène, j'expose le cas des rotateurs rapides ainsi que les forces qui animent une étoile à devenir et à en rester une. J'énumère les différents types de rotateurs, où je consacre tout un paragraphe aux étoiles dites Be, le groupe qui possède le taux de rotation le plus elevé, qui sont capables d'éjecter de la matière sous effet de la force centrifuge et de former un disque circumstellaire autour d'elles. Après un bref paragraphe sur l'éjection de la matière et la formation des environnements circumstellaires, je clos ce chapitre en parlant aussi d'autres types d'activités susceptibles d'engendrer des environnements circumstellaires.\\

\textbf{Le troisième chapitre} introduit les concepts et l'historique (en bref) de l'interférométrie, de la spectroscopie, ainsi que la fructueuse combinaison des deux techniques en astronomie ; la spectro-interférométrie. Quelques instruments faisant appel à cette technique d'observation sont brièvement cités, en particulier l'instrument AMBER/VLTI grâce auquel nous avons pu observer, réduire/traiter et interpréter les données observées. Dans un paragraphe de ce chapitre, je décris surtout des spécificités d'AMBER,  des observations, des observables interférométriques où je m'attarde un peu plus longuement sur la phase différentielle  $\phidiff$, la réduction des données ainsi que des traitements de certains biais qui s'y attachent. A la fin de ce chapitre un poster résume les différents traitements de biais réalisés sur Achernar observée en 2009. Ces traitements nous ont permis  de restreindre, avec CHARRON (CODE FOR HIGH ANGULAR RESOLUTION OF ROTATING OBJECTS IN NATURE), les paramètres fondamentaux de cette étoile et dont les résultats sont résumés dans un papier A\&A dont je suis co-auteur \citep{2012A&A...545A.130D}, à la fin de ce chapitre, juste après le poster cité ci-haut.\\

\textbf{Le quatrième chapitre} présente ma principale contribution à la communauté restreinte des "interférométristes" ;   un code numérique que j'ai élaboré, dédié à l'étude des rotateurs et leur environnement proche en interférométrie différentielle (DI) ;  SCIROCCO (Simulation Code of Interferometric-observations for ROtators and CirCumstellar Objects). Ce chapitre décrit l'approche théorique physique et de modélisation numérique que j'ai adoptée pour interpréter les mesures interférométriques avec AMBER/VLTI, qui a nécessité de mettre en place un ensemble d'outils et de techniques de réduction de données adaptés à mon modèle pour exploiter les mesures observées.\\

\textbf{Le cinquième chapitre} détaille les principaux résultats de mon code, à travers  un article A\&A, montrant l'étude menée et les résultats obtenus sur 4 rotateurs, où \citet{2014A&A...569A..45H} y est inclus. Une étude comparative de l'influence de certains paramètres physiques sur la phase differentielle $\phidiff$ y est aussi consacrée.\\

\textbf{Le sixième chapitre} est dédié à toutes les possibilités de simulation de SCIROCCO, à savoir inclure les pulsations non-radiales, un disque circumstellaire, une tache et/ou une exoplanète autour de l'étoile en rotation. A la fin de ce chapitre j'inclus deux papiers (proceedings), qui détaillent une étude sur les rotateurs rapides avec et sans l'effet de la pulsation non radiale (SCIROCCO ; \citet{2012sf2a.conf..533H} \& SCIROCCO+ ; \citet{2013EAS....59..131H}). Enfin j'y inclus un poster qui résume toute l'étendue d'application du code SCIROCCO.\\

\textbf{Le septième chapitre} clos mon manuscrit de thèse par une discussion, des conclusions et perspectives.\\

\textbf{Une annexe} est également rattachée à ce manuscrit, où je présente brièvement d'autres travaux scientifiques parallèles que j'ai dû mener ; à savoir une étude sur un instrument d'imagerie haute dynamique de détection directe d'Exoplanète (voir papier SPIE où mon nom y est en second auteur ; \citet{2012SPIE.8446E..7EA}), et un travail de reconstruction d'image sur Achernar (nième auteur dans un papier A\&A \citet{2014A&A...569A..10D}).

\chapter{La rotation au c\oe{}ur de l'activité stellaire}
\begin{figure}[h!]
\centering
 \includegraphics[height=0.5\hsize,draft=false]{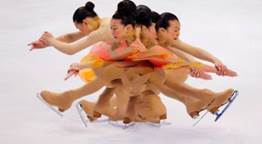}
\end{figure}
\label{chap:rota}
\minitoc

%%%%%%%%%%%%%%%%%%%%%%%%%%%%%%%%%%%%%%%%
\def\vsini{v_\mathrm{eq} \sin i} 
\def\kms{\mathrm{km.s}^{-1}}
\def\phidiff{\phi_\mathrm{diff}}
\def\Rsun{\mathrm{R}_{\odot}}
\def\Lsun{\mathrm{L}_{\odot}}
\def\Msun{\mathrm{M}_{\odot}}
\def\Tmean{\overline{T}_\mathrm{eff}}
\def\diameq{\diameter_\mathrm{eq}}
\def\chir{\chi_\mathrm{r}}
\def\chimin{\chi_\mathrm{min}}
\def\chirmin{\chi_\mathrm{min,r}}
%%%%%%%%%%%%%%%%%%%%%%%%%%%%%%%%%%%%%%%%

\section{La rotation}
\subsection{Rappel historique}

Nées d'un nuage de gaz et de poussière tournoyant à très haute vélocité, sous forme de disque d'accrétion, toutes les étoiles se forment au début de leur vie avec un fort moment cinétique. Par la suite seules les étoiles à zone convective, engendrant un fort champ magnétique par effet dynamo, ou ayant entretenus un champ magnétique fossile \citep{2015arXiv150200226N} en formant une magnétosphère (comme les Ap), se retrouvent ralenties et finissent par perdre leur impressionnante vitesse de rotation. Ce fut le cas de notre Soleil par exemple (qui a une vitesse de rotation à l'équateur proche de $2\kms$).
Toutes les autres, i.e. celles qui ont un champ magnétique faible où  inexistant, gardent leurs caractéristiques innées de grande vitesse de rotation. Ces dernières sont appelées : étoiles toupies ou bien rotateurs rapides.\\

\begin{figure}[h!]
\centering
 	\includegraphics[height=0.5\hsize,width=0.5\hsize,draft=false]{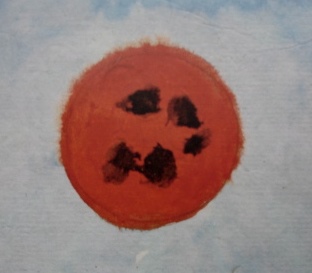}
	\caption[Taches solaires observées au 4\up{ième} siècle av. J.-C par les Chinois]{Illustration de taches solaires provenant du manuscrit des prévisions astronomique et météorologique de l'empereur Zhu Gaoji (Xuanzong) des Ming, datant de 1425, et dont on pense qu'elle a été exécutée par la main de l'empereur lui-même. (Bibliothèque de l'Université de Cambridge).}\label{taches_sol_chine}
\end{figure}

Les premières tentatives d'observation de la rotation stellaire, furent réalisées sur notre étoile ; le Soleil. En effet, c'est dans les Manuscrits de prévisions astronomiques et météorologiques de l'empereur Zhu Gaoji des Mings, de l'état de Qi en 1425 qu'on retrouve la première illustration des taches solaires (Fig.\ref{taches_sol_chine}). Bien que ce soit au 4\up{ième} siècle avant JC que les 3 astronomes chinois ;  Shi Shen, Wu Xian \& Gan De, élaborèrent le premier grand catalogue d'étoiles connu de l'humanité (catalogue antérieur à celui du Grec Hipparque de 200 ans). C'est là que furent mentionnées des observations de taches sur le Soleil par l'astronome chinois Gan De qui fut ainsi le premier à reconnaître les taches solaires comme phénomène purement solaire (non lié à l'atmosphère, à des obstacles naturels ou autres), reliant de ce fait le mouvement apparent des taches de la surface solaire à la rotation du Soleil \citep{Temple86,2012A&ARv..20...51V}.
Tandis qu'à l'occident du 9\up{ième} siècle, on considérait encore les taches solaires comme étant causées par des d'objets extérieurs au Soleil, tel qu'en témoigna le moine bénédictin Adelmus dans ses transcriptions d'observations des taches solaires, du 17 au 24 Mars 807 \citep{1917PA.....25...88W,2008ASSL..352.....M}.\\

Plus tard, les taches solaires furent scientifiquement et rigoureusement observées par Thomas Harriot (1560-1621), puis par Johann Fabricius (1587-1615) qui fut, en 1611, le premier à publier une explication correcte du phénomène. Peu après ce fut Christophe Scheiner (1575-1650) qui publia ses "Lettres sur les taches solaires".\\

C'est en 1613 qu'une démonstration claire et rigoureuse fut apportée par Galilée (1564-1642) dans un texte considéré par les épistémologistes comme étant un texte exemplaire de travail scientifique, et dont voici un court extrait :
\textit{" Quand on n'ignore pas totalement la perspective, du changement apparent des figures et des vitesses du mouvement, il faut conclure que les taches sont contiguës au corps solaire et que, touchant sa surface, elles se meuvent avec lui ou sur lui (...). À preuve, leur mouvement : il paraît très lent au bord du disque solaire et plus rapide vers le centre ; autre preuve encore, la forme des taches : au bord de la circonférence elles paraissent beaucoup plus étroites qu'au centre ; c'est qu'au centre on les voit en majesté, telles qu'elles sont vraiment, alors que près de la circonférence, quand se dérobe la surface du globe, on les voit en raccourci ". }(Galilée," Dialogue sur les deux grands systèmes du monde")\\

C'est donc en ces termes de grande éloquence scientifique que Galilée, tout en s'affirmant en faveur de la théorie de Copernic et contre la vision de l'immuabilité aristotélicienne, qu'il apporta pour la toute première fois une preuve irréfutable de la rotation d'une étoile sur elle-même.\\

\begin{figure}[h!]
\centering
 	\includegraphics[height=0.5\hsize,width=0.5\hsize,draft=false]{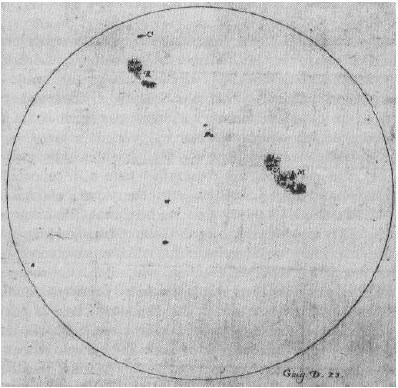}
	\caption[Taches solaires observées par Galilée en 1613]{Un dessin de taches solaires exécuté par la main de Galilée, paru en 1613, observées grâce la lunette qu'il avait inventée quelques années plus tôt et projetées sur un écran.}\label{taches_sol_Galiee}
\end{figure}

Fabricius observa aussi une variabilité de luminosité de l'astre Omicron Ceti  (une étoile binaire de la constellation de la Baleine) sans apporter d'explication, étoile que l'astronome polonais Johannes Hevelius nomma Mira ("merveilleuse" ou "étonnante" en latin) plus tard au 17ième siècle. Ce fut Ismail Bouillaud (1605 - 1694) qui observa par photométrie une régularité dans la variation lumineuse de cette étoile avec une périodicité de 333 jours. A l'époque ce phénomène fut interprété comme étant la rotation intrinsèque de l'étoile. Les travaux furent repris par Cassini, Fontenelle, et Miraldi \citep{Brunet31} dans la même optique avant que l'explication de la rotation intrinsèque soit abandonnée au 19ième siècle, et que fut adoptée, pour Mira A, celle des variations de la transparence des couches extérieures de son atmosphère\footnote{Au minimum de l'activité lumineuse de l'étoile, des molécules de gaz et de poussières absorbent le rayonnement visible et rayonnent dans l'infrarouge. Dans la phase de dilatation, une onde de choc dissocie les molécules et les poussières de l'atmosphère, qui redevient transparent au rayonnement visible. Lorsque l'atmosphère se refroidit, les molécules se recombinent et les poussières se condensent à nouveau, bloquant ainsi le rayonnement visible.}.\\

200 ans après Galileo,  en 1877, William de Wiveleslie Abney (1843-1920)  proposa une hypothèse selon laquelle l'élargissement des raies spectrales observées sur l'étoile serait causé par  la rotation axiale de celles-ci.  Hermann Carl Vogel (1841-1907) qui fut le premier à déterminer la période de rotation du Soleil à l'aide de l'effet Doppler, contesta l'hypothèse d'Abney la même année, argumentant que dans le spectre observé par Abney, seule la raie d'Hydrogène était élargie en comparaison avec les autres raies qui semblaient rester relativement étroites. De nos jours on sait que pour la plupart des étoiles toutes les raies sont élargies par rotation.\\ 

En 1893, JR Holt proposa une méthode de mesure de la rotation d'étoiles basée sur la vitesse radiale. Il imagina ainsi que dans un système d'étoiles binaires, quand l'étoile secondaire éclipse sa primaire en rotation (dans son sens de rotation), le spectre observé de la primaire se décalera d'abord vers le rouge lorsque la première moitié sera éclipsée, puis le spectre se décalera vers le bleu à la deuxième moitié, ce qui provoquera un changement apparent de la vitesse radiale en plus de l'apport du mouvement orbital de l'étoile éclipsée \citep{2013A&A...549A..18T}.\\

C'est dans l'astrophysique observationelle en 1910 qu'une mesure réelle de l'effet de rotation sur raies spectrales fut réalisée, d'abord par Schlesinger (1871-1943) sur le système binaire $\lambda$ Tauri - $\delta$ Librae. Ce fut la première mesure de la vitesse radiale apparente sur un rotateur.\\
 
\begin{figure}[h!]
\centering
 	\includegraphics[height=0.4\hsize,draft=false]{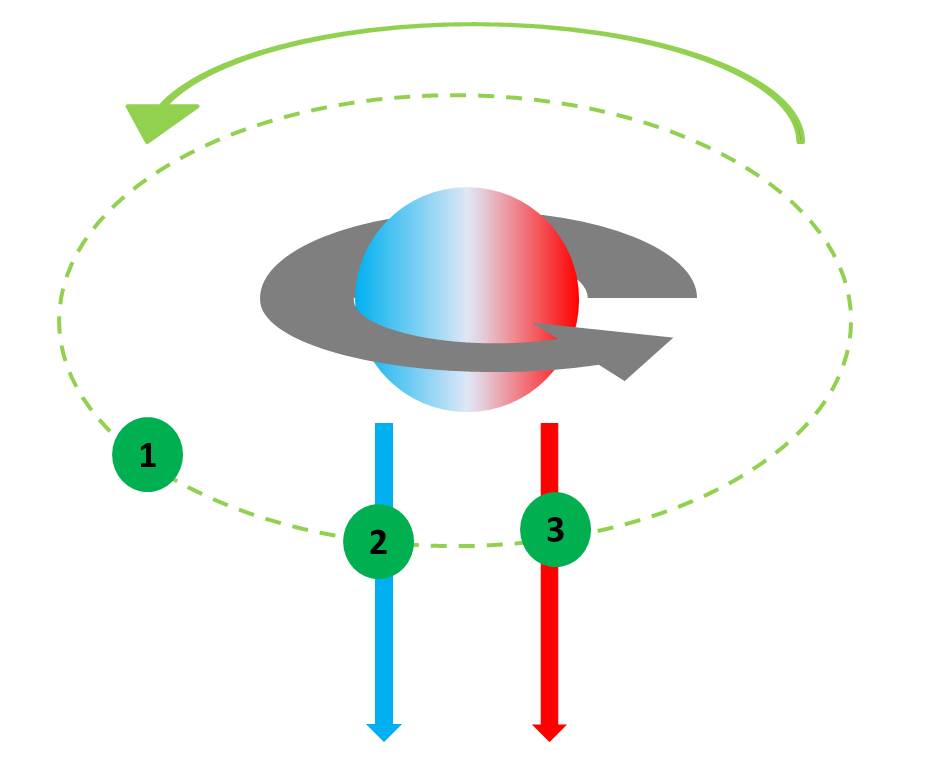}
	\caption[Effect Rossiter-McLaughlin]{Figure descriptive de l'effet Rossiter-McLaughlin. Système d'étoiles binaires, où l'étoile primaire est en rotation et autour de laquelle révolutionne une l'étoile secondaire (dans le sens de rotation de la primaire). Quand la secondaire éclipse la première moitié de sa primaire, le spectre observé se décalera d'abord vers le rouge (position 2), puis  vers le bleu à la deuxième moitié (position 3).}\label{effect_RM}
\end{figure}

Plus tard, en 1924 Rossiter et McLaughlin firent les mêmes observations du phénomène qui porte désormais leurs noms " effet Rossiter-McLaughlin " ou phénomène RM (Fig.\ref{effect_RM}); couramment observé par transit exoplanétaire pour sonder l'alignement du plan orbital par rapport à l'axe de rotation de l'étoile (voir, par exemple, \citet{2005ApJ...631.1215W}).\\

En 1929 Shajn et Struve prédirent la forme des raies spécifiques attendue par élargissement Doppler rotationnel \citep{1929MNRAS..89..222S}. Bien que là aussi l'observation de cet effet fut attendu pour les binaires, il fut une bonne prédiction pour la rotation d'étoiles solitaires, ce qui permit à Elvey en  1930 d'observer via cet effet et de publier une première liste de vitesses de rotation, ce qui marqua cette décennie-là de rapides progrès observationnels \citep{1930ApJ....71..221E}.\\

A la même époque, \citet{1924MNRAS..84..665V, 1924MNRAS..84..684V} démontra, avec l'hypothèse d'une rotation d'un corps rigide, que la luminosité de surface locale est proportionnelle à la gravité locale effective en tout point sur une étoile, ce qui se traduit pour une étoile en rotation, par une température plus élevée aux pôles qu'à l'équateur.\\

En 1931 Struve et Elvey \citep{1931MNRAS..91..663S} réussirent à formuler un lien entre le taux de rotation d'une étoile et son type spectral, constatant ainsi que les étoiles de type A sont les plus susceptibles d'être des rotateurs rapides; \citet{1933ApJ....77..141W, 1933ApJ....78...46W, 1934ApJ....79..357W} publia un vaste catalogue d'observation pour des centaines d'étoiles de types spectraux de O à F.\\

Par la suite, dans une douzaine d'articles entre 1949 et 1956, Slettebak avait découvert que les rotateurs les plus rapides se trouvaient parmi les étoiles Be, et établi une relation entre la rotation et la masse. En 1949, Slettebak s'appuyant sur les prédictions de "l'effet von Zeipel" déduit les premières modifications de forme des raies spectrales de rotateurs rapides brillants.\footnote{Implications qui ont été développées en détail dans \citet{1963ApJ...138.1134C, 1965ApJ...142..265C} pour l'émission dans le continuum, et \citet{1966AJ.....71R.381C}, qui incorporent une distorsion de la forme, des effets d'aspect, de la gravité, de l'assombrissement centre-bord, et de la variation latitudinale dans le calcul des profils $H\beta$.}.\\

\citet{1968ApJ...151.1051H} caractérisèrent la polarisation intrinsèque des rotateurs rapides de type "précoce", et \citet{1977ApJS...34...41C} démontrèrent que les rotateurs rapides étaient 2 à 3 fois plus présents que prévu dans la séquence principale du diagramme HR.\\ 

Dans l'ensemble, la position apparente des étoiles sur le diagramme HR est nettement affectée (par exemple les sous-types B2-3) en comparant les étoiles en rotation rapide de leur  homologues non-rotatives (voir \citet{1980ApJ...242..171S}; \citet{1985MNRAS.213..519C}, et références qui y sont). Slettebak (1985) publia l'histoire détaillée du profil de raie avec analyse et traitement approffondi du sujet.\\
 
La gyrochronologie \citep{2003ApJ...586..464B} relient l'âge d'une étoile (type solaire i.e. de faible masse) à sa rotation, est aussi un élément important de la rotation stellaire. Ainsi, pour les étoiles froides  de la séquence principale du diagramme HR, une relation connue sous le nom de loi de Skumanich \citep{Tassoul00} prédit que la vitesse de rotation équatoriale $\Omega_{eq}$ perçue d'une étoile est inversement proportionnelle à la racine carrée de son âge $t$ \citep{1972ApJ...171..565S}.\\

La rotation stellaire fut mesurée pour la première fois en interférométrie, à partir du déplacement du photo-centre (le terme au premier ordre de la phase différentielle selon le développement de Mac Lauren \citep{2001A&A...377..721J}), par Lagarde \& Petrov en 1994 \citep{sl94}, sur l'étoile Aldébaran ($\alpha$ Tau), rotateur lent de la constellation du Taureau, observé en 1988 à l'OHP sur un télescope de 152 cm via une méthode d'observation appelée Interférométrie Différentielle des Tavelures (IDT).\\

\begin{figure}[h!]
\centering
 	\includegraphics[height=0.4\hsize,draft=false]{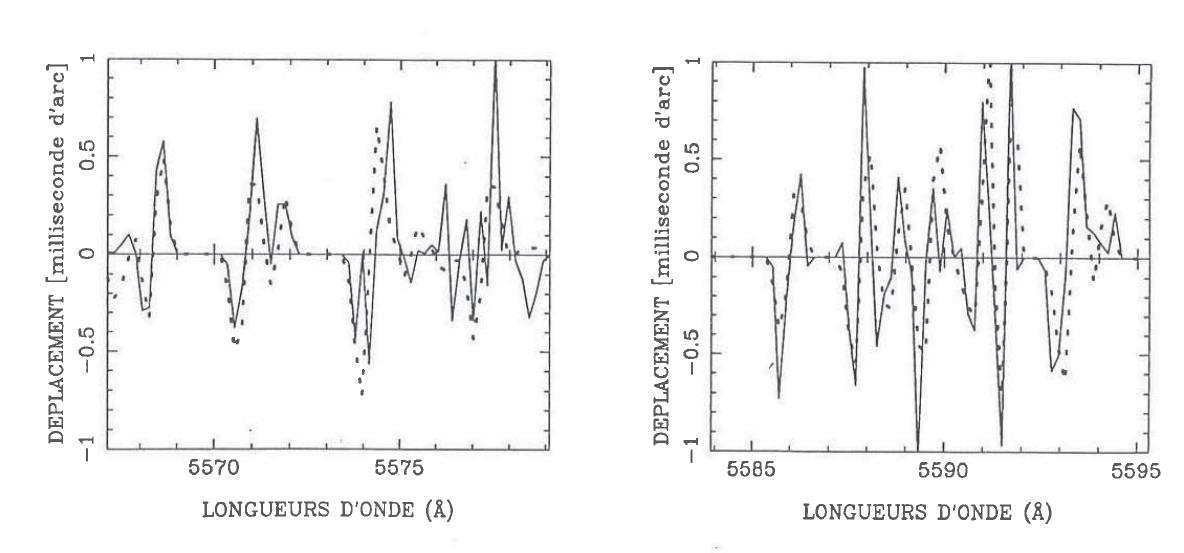}
	\caption[Déplacement de photo-centre simulé et mesuré sur Aldébaran]{Déplacement photo-centre simulé (en pointillés) et mesuré (en ligne continue) sur le rotateur lent Aldébaran par IDT. $\vsini$ déduit: $5.8\pm1$ $\kms$ \citep{sl94}}\label{Lagarde_1994}
\end{figure}

Depuis cette date, plusieurs observations et mesures interférométriques furent menées sur les rotateurs rapides, que le Tab. \ref{tab_vanbelle2012} ci-dessous récapitule de manière succincte. Un historique détaillé sur la rotation stellaire a été produit par \citet{Tassoul00}.\\

\begin{table*}[htbp]
\centering
\begin{sideways}
\centering
\resizebox{1.7\textwidth}{!}{\begin{threeparttable}
\centering

%\begin{table*}[htbp]
%\centering
%\begin{sideways}
\caption[Résumé de résultats d'observations sur des rotateurs rapides]{Résumé de résultats d'observations sur les rotateurs rapides à ce jour (inspiré de van Belle (2012)).}\label{tab_vanbelle2012}
%\centering
%\resizebox{\textwidth}{!}{\begin{threeparttable}
%\begin{threeparttable}
%\centering
\begin{tabular}{||c||c|c|c|c|c|c|c|c|c|c|c|c|c||}
  \hline\hline
  Étoiles & Type & Vitesse $v$ & Inclinaison $i$ & $v/v_{crit}$ & $\Omega/\Omega_{crit}$ & Orientation & Assombrissement & $T_{pol}$ & $T_{eq}$ & $R_{pol}$ & $R_{eq}$ & Aplatissement & Ref\\
    & spectral & $\kms$ & ($^\circ$) & & & $PA_{rot}$ ($^\circ$) & gravitationnel $\beta$ & ($K$) & ($K$) & ($\Rsun$) & ($\Rsun$) & $R_{eq}/R_{pol}$ & \\
  \hline\hline
   Achernar & B3Vpe & 225\tnote{a} & >50 & 0.79-0.96 &  & 39$\pm$1 & 0.25 (fixé) & 20000 (fixé) & 9500-14800 & 8.3-9.5 & 12.0$\pm$0.4 & 1.56$\pm$0.05 & (5)\\ 
   ($\alpha$ Eri) &  & 298$\pm$9 & 101.5$\pm$5.2 & 0.96$\pm$0.03 &  & 34.9$\pm$1.6 & 0.20 (fixé) & $18013^{+141}_{-171}$ & $9955^{+1115}_{-2339}$ & 8.0$\pm$0.4 & 11.6$\pm$0.3 & 1.45$\pm$0.04 & (12)\\
   \hline
   Regulus & B8IVn & 317$\pm$3 & $90^{+0}_{-15}$ & 0.86$\pm$0.03 & 0.974$\pm$0.043 & 85.5$\pm$2.8 & 0.25$\pm$0.11 & 15400$\pm$1000 & 10300$\pm$1000 & 3.14$\pm$0.06 & 4.16$\pm$0.08 & 1.325$\pm$0.036 & (8)\\
   ($\alpha$ Leo) & & $336^{+16}_{-24}$ & $86.3^{+1.0}_{-1.6}$ & 0.839$\pm$0.030 & $0.962^{+0.014}_{-0.026}$ & $258^{+2}_{-1}$ & $0.188^{+0.012}_{-0.029}\tnote{c}$ & $14520^{+550}_{-690}$ & $11010^{+420}_{-520}$ & $3.22^{+0.05}_{-0.04}$ & $4.21^{+0.07}_{-0.06}$ & 1.307$\pm$0.030 & (11)\\
   \hline
   Vega & A0V & 270$\pm$15 & 4.7$\pm$0.3 & 0.746$\pm$0.034 & 0.91$\pm$0.03 & Non cité & 0.25 (fixé) & 10500$\pm$100 & $8250^{+415}_{-315}$ & 2.26$\pm$0.07 & 2.78$\pm$0.02 & 1.230$\pm$0.039 & (6)\\
   ($\alpha$ Lyr) &  & 274$\pm$14 & 4.54$\pm$0.33 & 0.769$\pm$0.021 & 0.921$\pm$0.021 & 8.6$\pm$2.7 & 0.25 (?, fixé) & 9988$\pm$61 & 7557$\pm$261 & 2.306$\pm$0.031 & 2.873$\pm$0.026 & 1.246$\pm$0.020 & (7)\\
   \hline
   Rasalhague & A5IV & 237 &  87.70$\pm$0.43 & 0.709$\pm$0.011& 0.885$\pm$0.011& -53.88$\pm$1.23 & 0.25 (fixé) & 9300$\pm$150 & 7460$\pm$100 & 2.390$\pm$0.014 & 2.871$\pm$0.020 & 1.201$\pm$0.011 & (9)\\
   ($\alpha$  Oph) &  &  &  &  & &  &  &  &  &  &  &  & \\
   \hline
   Altair & A7IV-V & \multicolumn{2}{|c|}{$\vsini=210\pm13$, i > 30} &  &  & -68.4$\pm$6.2\tnote{b} & Non appliqué &7680$\pm$90 &  & 1.8868$\pm$0.0066 & & & (1)\\
   ($\alpha$ Aql)& & & & & & 35$\pm$18 & Voir\tnote{d} & 7750 (fixé) & & & & & (2)\\
   &  & 273$\pm$13 & $\sim$64 & 0.729$\pm$0.019 & 0.90$\pm$0.02 & 123.2$\pm$2.8 & 0.25 (fixé) & 8740$\pm$140 & 6890$\pm$60 & 1.636$\pm$0.022 & 1.988$\pm$0.009 & 1.215$\pm$0.017 & (3)\\
   & & 285$\pm$10 & 57.2$\pm$1.9 & 0.764$\pm$0.008 & 0.923$\pm$0.006 & -61.8$\pm$0.8 & 0.19$\pm$0.012\tnote{c} & 8450$\pm$140 & 6860$\pm$150 & 1.634$\pm$0.011 & 2.029$\pm$0.007 & 1.242$\pm$0.009 & (4)\\
   \hline
   Alderamin & A7IV-V & 283$\pm$19 & $88.2^{+1.8}_{-13.3}$ & $0.8287^{+0.0482}_{-0.0232}$ & 0.958$\pm$0.068 & 3$\pm$10 & $0.084^{+0.026}_{-0.049}$ & 8440+430-700 & $\sim$7600 & 2.175$\pm$0.046 & 2.82$\pm$0.10 & 1.297$\pm$0.054 & (10)\\
   ($\alpha$ Cep) &  & 225 & 55.70$\pm$6.23 & 0.795$\pm$0.025 & 0.941$\pm$0.020 & -178.84$\pm$4.28 & 0.216$\pm$ 0.021\tnote{c} & 8588$\pm$300 & 6574$\pm$200 & 2.162$\pm$0.036 & 2.74$\pm$0.044 & 1.267$\pm$0.029 & (9)\\
   \hline
   Caph & F2III-IV & $72.4^{+1.5}_{-3.5}$ & $19.9^{+1.9}_{-1.9}$ & 0.760$\pm$0.040 & $0.920^{+0.024}_{-0.034}$ & $-7.09^{+2.24}_{-2.40}$ & $0.146^{+0.013}_{-0.007}\tnote{c}$ & $7208^{+42}_{-24}$ & $6167^{+36}_{-21}$ & $3.06^{+0.08}_{-0.07}$ & $3.79^{+0.10}_{-0.09}$ & 1.239$\pm$0.046 & (11)\\
   ($\beta$ Cas) &  &  &  &  & &  &  &  &  &  &  &  & \\
   \hline\hline
   %&  &  &  &  & &  &  &  &  &  &  &  & \\
\end{tabular}

\begin{tablenotes}
		\footnotesize
		  \item[a] Fixé par \citet{1982ApJS...50...55S}
   		\item[b] Reflète une correction des coordonnées {u,v}, où $PA_{rot}$ prend donc la valeur de $-21.6\pm6.2$; la physique ne se trouve nullement affectée ici
   		\item[c] Deuxième solution avec $\beta = 0.25$ (fixé) ont également été présenté dans ce papier
   		\item[d] Les auteurs ont appliqué une répartition asymétrique de la luminosité à hauteur de $\sim5\%$ pour simuler l'effet de l'assombrissement gravitationnel

(1) \citet{2001ApJ...559.1155V}, (2) \citet{2004ApJ...612..463O}, (3) \citet{2006ApJ...636.1087P}, (4) \citet{2007Sci...317..342M}, (5) \citet{2003A&A...407L..47D}, (6) \citet{2006ApJ...645..664A}, (7) \citet{2006Natur.440..896P},(8) \citet{2005ApJ...628..439M}, (9) \citet{2009ApJ...701..209Z}, (10) \citet{2006ApJ...637..494V}, (11) \citet{2011ApJ...732...68C}, (12) \citet{2012A&A...545A.130D}.
\end{tablenotes}

\end{threeparttable}
}
\end{sideways}
\end{table*}

\subsection{Intérêt scientifique}
Le paramètre du moment angulaire est de plus en plus intégré dans la modélisation stellaire en même temps que les autres paramètres tels que la masse, les abondances chimiques et le champ magnétique par exemple. Ceci entraîne une plus grande précision dans le classement des étoiles dans le diagramme HR basé sur leurs abondances chimiques, la métallicité plus précisément, ainsi que l'hélium et l'azote (He \& N) pour les étoiles type géantes/super-géantes.\\

La rotation peut directement entraîner d'importantes conséquences sur la physique des étoiles, telles que :

\begin{itemize} 
\item Une altération du moment angulaire interne ainsi que la distribution des éléments chimiques, provoquée par une amplification du mouvement de flux de matière, de type turbulence et circulation méridionale, impactant aussi la physique de l'atmosphère stellaire comme la distribution de température et de gravité, ce qui peut entraîner une possible perte de masse et du moment angulaire.

\item Une modification des caractéristiques hydrostatiques ainsi qu'une diminution de la pression interne de l'étoile, causée par une force centrifuge importante, entraînant une variation du type spectral apparent et un rallongement du temps évolutif de l'étoile.\\
\end{itemize} 

Tel qu'on a pu le voir dans le bref historique cité plus haut, seules deux méthodes ont été principalement utilisées au XX\up{ième} siècle pour la mesure de la rotation stellaire : la mesure temporelle photométrique et/ou spectroscopique à l'aide d'irrégularités surfaciques de l'étoile en rotation (taches par exemple), et la mesure de l'élargissement Doppler de raies spectrales issues de la rotation. L'élargissement (en supposant une rotation rigide) qui nous renseigne sur la quantité $\vsini$ (vitesse linéaire équatoriale projetée le long de la ligne de visée (Fig.\ref{reference})) est une conséquence directe de la rotation, car ce phénomène est très important pour l'observation astronomique, étant donné que la vitesse linéaire est maximale à l'équateur et nulle aux pôles. Par exemple et tout récemment \citet{2006Natur.440..896P} et \citet{2006ApJ...645..664A} ont découvert que Vega, une étoile référence de notre voute céleste, est en fait un rotateur rapide vu par le pôle.\\

Des études statistiques spectroscopiques ont maintes fois été menées sur la rotation stellaire \citep{1970IAUCo...4.....S}, montrant par exemple aucun lien entre l'inclinaison des étoiles et leur position dans la galaxie. La quantité $\vsini$ peut dépasser $100$ $\kms$ pour certaines étoiles de type spectral O, B et A. Tandis que les étoiles plus tardives (au delà du type A5V) perdent leur moment angulaire par freinage magnétique , engendré par la zone convective ou par la magnétosphère, et provoquant une perte de particules ionisées véhiculées par un vent stellaire qui suit des lignes de champ peuvant atteindre plusieurs rayons stellaires et qui accompagnent le mouvement rotatif du rotateur \citep{1962AnAp...25...18S}.\\

\begin{figure}[h!]
\centering
 	\includegraphics[height=0.5\hsize,width=0.5\hsize,draft=false]{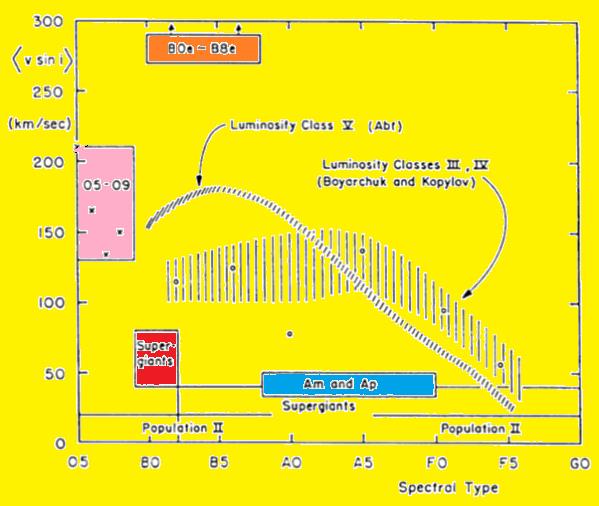}
 	\includegraphics[height=0.5\hsize,width=0.5\hsize,draft=false]{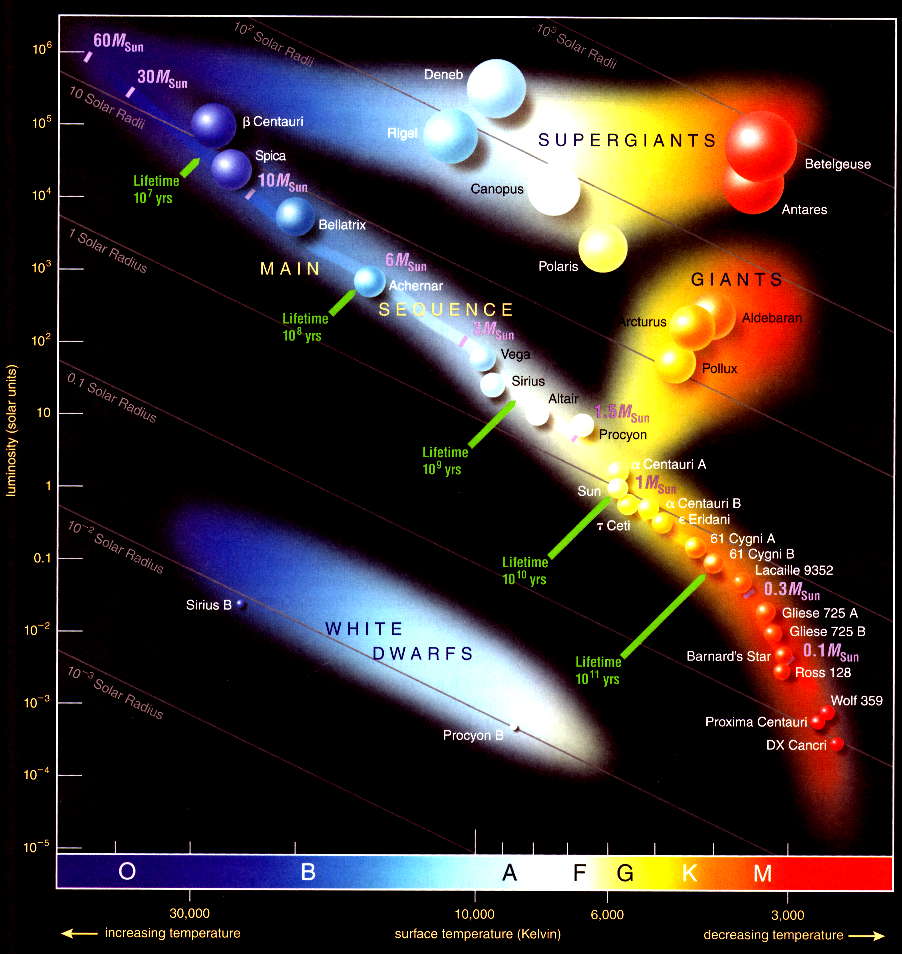}
	\caption[La rotation stellaire selon le type spectral \& le diagramme HR]{\textbf{En haut:} Une figure qui montre la répartition des vitesse des rotateurs stellaires selon leur type spectral, et \textbf{En bas:} Un diagramme HR.}\label{Slettebak_HR}
\end{figure}

D'une manière générale, plus la métallicité est élevée dans une étoile, plus celle-ci subit le freinage magnétique. C'est ce qui est illustré dans la Fig. \ref{Slettebak_HR} (en compagnie du diagramme HR) avec les étoiles A, B et F ayant respectivement des raies métalliques (Am, Ap, Bp \& Fm ), à l'inverse des étoiles Be qui peuvent atteindre des vitesses de rotation avoisinant les $500$ $\kms$. Plusieurs catalogues dédiés à la rotation stellaire peuvent être consultés en ligne sur le site internet du CDS (Centre de Données astronomique de Strasbourg)\footnote{ http://cdsweb.u-strasbg.fr/}.\\

Aussi, avec l'avancée constante des techniques d'observation, on accroit notre compréhension de la rotation stellaire et ses conséquences sur l'évolution stellaire, à la fois quantitativement et qualitativement. Du point de vue interférométrique, on a pu récemment faire des reconstructions d'images d'étoiles en rotation faisant apparaître de manière directe, leur forme aplatie, le gradient de température entre les pôles et l'équateur (l'effet von Zeipel), ainsi que  la complexité de leur atmosphère (e.g.  \citet{2007Sci...317..342M}). Un récapitulatif de l'historique ainsi que des récentes et importantes découvertes concernant les rotateurs rapides et la rotation stellaire fut publié par \citet{2012A&ARv..20...51V}.\\

Mon travail de thèse se concentre essentiellement sur la rotation stellaire et sur l'observation interférométrique longue base. Dans le chapitre \ref{chap:scirocco}, je décris le code que j'ai developpé afin d'étudier l'effet de la variation de certains paramètres fondamentaux stellaires sur des observables interférométriques, plus particulièrement la phase différentielle.\\

J'ai pu valider cette étude sur des données AMBER/VLTI de plusieures étoiles: Achernar, Altair, $\delta$ Aquilae et Fomlahaut.\\

\subsection{Les rotateurs stellaires rapides}

\begin{figure}[h!]
\centering
 	\includegraphics[height=0.5\hsize,width=0.6\hsize,draft=false]{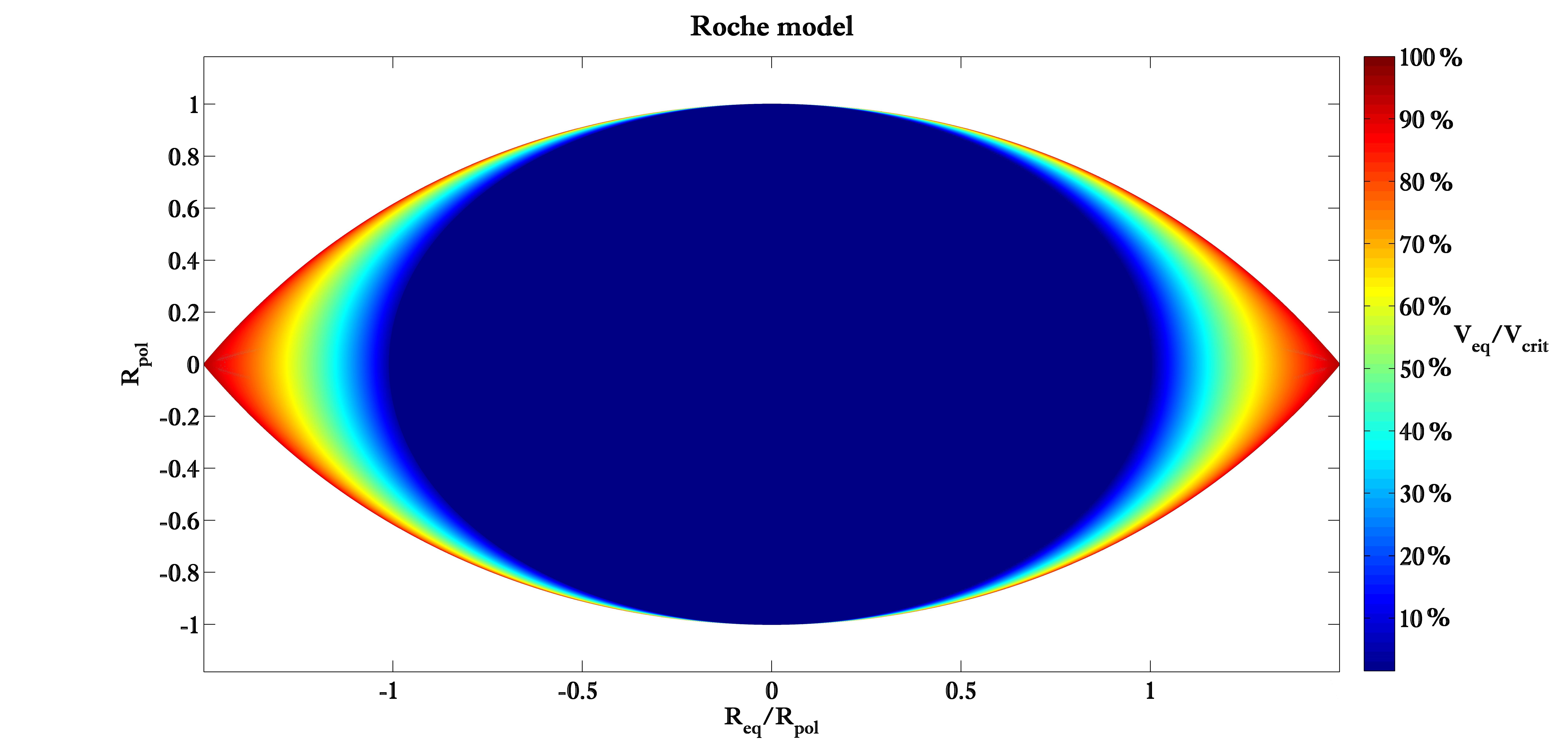}
	\caption[Modèle de Roche]{Aplatissement d'une étoile pour différentes vitesses interprété par le modèle de Roche. Au centre (en bleu foncé) l'étoile qui ne tourne pas ($v_{eq}=0$) est parfaitement sphérique  (son rayon polaire et équatorial sont identique ($R_{eq}=R_{pol}$). Plus l'étoile tourne vite sur elle-même plus son rayon équatorial augmente la rendant de forme aplatie ($R_{eq}>R_{pol}$), jusqu'à se rapprocher ainsi de sa vitesse critique (Eq.\eqref{v_crit}) -en rouge foncé- et au-delà l'étoile perd de la matière.}\label{roche_model}
\end{figure}

C'est la force centrifuge qui caractérise le plus tout rotateur rapide. En effet toute étoile en équilibre est soumise à deux forces principales qui se stabilisent mutuellement; la force de pression (radiative et hydrodynamique) qui tend à faire exploser l'étoile, et la force de gravité qui tend à la faire imploser. Dans le cas des rotateurs rapides, la force centrifuge s'impose donc d'elle même comme troisième force au côté des deux précédentes. De plus, cette force induit d'importantes transformations dans l'évolution et la structure du rotateur rapide. C'est donc à cet effet que fut introduite une notion quantitative importante, qui sert à déterminer à partir de quel seuil la force centrifuge d'une étoile est considérée assez importante pour que cette dernière soit classée comme un rotateur rapide. Cette quantité est la vitesse critique à l'équateur stellaire :

\begin{equation}\label{v_crit}
v_{crit}=\sqrt{\frac{GM_{*}}{R_{eq, crit}}}
\end{equation}

où $G$ est la constante gravitationnelle, $M_{*}$  la masse de l'étoile et $R_{eq, crit}$ le rayon équatorial à la vitesse critique. 
Comme exemple concret prenons les étoiles Be (les rotateurs non dégénérés les plus rapides de l'univers). Celles-ci peuvent atteindre une vitesse de rotation équatoriale $v_{eq, rot}$  de l'ordre de $90\%$ à $95\%$ de leur vitesse critique $v_{eq, crit}$ \citep{1968MNRAS.140..141S, 2005A&A...440..305F}. Cependant il faut noter que certaines études \citep{1982ApJS...50...55S, 1996MNRAS.280L..31P, 2001A&A...368..912Y, 2004MNRAS.350..189T} contredisent ces chiffres et estiment un $v_{eq, rot}$ autour de $50\%$ à $80\%$ de $v_{eq, crit}$. Le plus souvent, la vitesse de rotation à la surface d'une étoile est différentielle, i.e. qu'elle varie selon la latitude. Elle est maximale à l'équateur et nulle aux pôles, sans oublier que dans le cas d'une vitesse de rotation non nulle la photosphère de l'étoile est déformée et n'est plus sphérique mais plutôt ellipsoïdale (nommée ellipsoïde de McLaurin, voir le chapitre "stellar rotation" de \citet{Tassoul00}), le grand axe étant situé le long de l'équateur et le petit axe le long de l'axe de rotation de l'étoile (traversant l'étoile d'un pôle à un autre). Le modèle d'une forme d'ellipsoïde de révolution s'applique assez bien dans le cas d'une force centrifuge inferieure à la force gravitationnelle de l'étoile. Quand la force centrifuge s'approche de la force gravitationnelle, i.e. les deux forces se compensent, la forme de l'étoile est plutôt sous forme de modèle de Roche (voir Fig.\ref{roche_model} ci-dessous). Au-delà, et quand la force centrifuge est supérieure à la force gravitationnelle, l'étoile commence à perdre de la matière.\\

La question qui nous vient automatiquement à l'esprit  à ce stade est :"à partir de quelle vitesse de rotation une étoile peut-elle être classée comme un rotateur rapide?". Pour y répondre nous faisons appel à  l'estimateur de rotation critique le plus souvent utilisé $E_c$, faisant appel aux vitesses angulaires avec:

\begin{equation}\label{Ec}
E_c=\frac{\Omega_{eq}}{\Omega_{crit}}
\end{equation}

où $\Omega_{eq}$ est la vitesse angulaire équatoriale de l'étoile et $\Omega_{crit}$ la vitesse angulaire équatoriale critique au-delà de laquelle l'étoile perd sa cohésion gravitationnelle. Pour $E_c=0$ l'étoile ne tourne pas et pour $E_c=1$ la vitesse de rotation de l'étoile a atteint le stade critique ($v_{eq}= v_{eq, crit}$). D'un autre côté, on peut aussi utiliser un autre estimateur de la rotation critique, faisant appel celui-là aux forces ; L'estimateur $\eta$, qu'on définit comme suit :

\begin{equation}\label{eta}
\eta=\frac{F_{centrifuge}}{F_{gravitationnelle}}
\end{equation}

où $F_{centrifuge}=m\Omega^2_{eq}R_{eq}$ est la force centrifuge causée par la rotation de l'étoile et $F_{gravitationnelle}=\frac{G m M_*}{R_{eq}^2}$ la force gravitationnelle engendrée par la masse $m$ à volume élémentaire localisée à l'équateur de l'étoile. $\Omega_{eq}$ est la vitesse angulaire $\Omega_{eq}=\frac{v_{eq}}{R_{eq}}$, $v_{eq}$ et $R_{eq}$  sont respectivement la vitesse de rotation équatoriale et le  rayon équatorial de l'étoile. Ainsi l'étoile a atteint une rotation critique pour $\eta=1$, et est au repos pour $\eta=0$. En remplaçant les forces centrifuge et gravitationnelle par leurs expressions respectives, $\eta$ s'écrit sous la forme \citep{2011A&A...526A..87Z}:

\begin{equation}\label{eta2}
\eta=\frac{\Omega_{eq}^2 R_e^3}{GM_*}
\end{equation}

Avec $M_*$ la masse de l'étoile et G la constante de gravitation. Ainsi après quelques petites manipulations mathématiques  et sachant que $\Omega_{eq,crit}=\frac{v_{eq,crit}}{R_{eq,crit}}$, on peut écrire :

\begin{eqnarray}\label{v_gamma}
\frac{\Omega_{eq}}{\Omega_{crit}}=\sqrt{\eta \frac{R^3_{crit}}{R^3_{eq}}}\nonumber \\
% \& \nonumber \\
\frac{v_{eq}}{v_{crit}}=\sqrt{\eta \frac{R_{crit}}{R _{eq}}}
\end{eqnarray}

\citet{2011A&A...529A..87D} démontra via les équations de \citet{2011A&A...526A..87Z} qu'aucune relation linéaire n'existait entre $\frac{v_{eq}}{v_{crit}}$  (ou $\frac{\Omega_{eq}}{\Omega _{crit}}$) et $\eta$. Lorsque la vitesse de rotation est  $0.9$ de la vitesse critique, i.e. le rapport $E_c=\frac{\Omega_{eq}}{\Omega _{crit}}=90\%$ (ce qui est élevé), la force centrifuge elle n'est, en fait qu'à $50\%$ de la force gravitationnelle  (i.e. $\eta$ lui n'est qu'autour des $50\%$), ce qui nous permet d'affirmer que le rapport $E_c$ n'est pas un bon estimateur de rotateur en état critique ou pas.\\

Ainsi généralement, on commence à considérer une étoile comme un rotateur rapide à partir du moment où la rotation de celle-ci est assez importante pour que :

\begin{enumerate}
\item	L'étoile devienne aplatie géométriquement.
\item	Un gradient de gravité et de température effective commence à apparaître entre l'équateur et les pôles. On appelle cet effet l'assombrissement gravitationnel (e.g. \citet{1924MNRAS..84..665V, 1974MNRAS.167..199C, 1977A&A....58..267K, 1999A&A...347..185M}).
\end{enumerate}

Par conséquent toute étoile ayant un $E_c>20\%$ peut être considérée comme un rotateur rapide. Pour $E_c>80\%$, l'étoile, en rotation rapide, est soumise à d'importantes modifications physiques dont certaines sont directement mesurables par observation, comme par exemple :

\begin{itemize}
\item L'étoile qui s'élargit selon son équateur et donnant l'impression de s'aplatir fortement le long de son axe de rotation (la symétrie ici n'est plus sphérique mais azimutale par rapport à l'axe de rotation).
\item Un gradient de gravité et de température effective apparaît entre l'équateur et les pôles. C'est l'effet de l'assombrissement gravitationnel (e.g. \citet{1924MNRAS..84..665V, 1974MNRAS.167..199C, 1977A&A....58..267K, 1999A&A...347..185M}). Cet effet entraîne une réduction de la largeur à mi-hauteur des raies dans le domaine de l'ultraviolet (une différence par rapport au visible qui peut atteindre $200$ $\kms$ \citep{1977PASP...89...19H}), ce qui peut être un bon moyen d'estimation de l'axe de rotation et de l'inclinaison de l'étoile \citep{1979PASP...91..313H}.
\item L'assombrissement gravitationnel provoque un surplus infrarouge dans le flux mesuré \citep{1977ApJS...34...41C}.
\item Un classement erroné de l'étoile, du fait de sa rotation rapide, et de son inclinaison qui influencent énormément le flux observé  (e.g. \citet{1972A&A....21..279M}), entraînant une mauvaise interprétation du type spectral de l'étoile observée, dans un intervalle de la séquence principale du diagramme HR, pouvant atteindre la famille des étoiles géantes \citep{1966ApJ...146..152C}.
\item En plus d'un flux  polarisé, les raies spectrales différemment dépendantes de la température et de la gravité se distribuent de manière inhomogène sur l'étoile, rendant tout classement spectral relatif à ses raies totalement erroné \citep{1974ApJ...191..157C}.
\item Sans oublier les circulations méridionales et autres turbulences dues à des mouvements de masses, provoquant des taux de mélanges chimiques de plus en plus importants au fur et à mesure que le rotateur est de plus en plus rapide. La circulation méridionale, qui peut aussi enrichir le c\oe{}ur de l'étoile en Hydrogène, augmentant de ce fait la durée de sa vie, comme elle alimente aussi les couches extérieuress de celle-ci en métaux, transférant le moment angulaire du c\oe{}ur (là où il était à l'origine plus important qu'à la surface) vers la photosphère, et inversant le gradient radial de rotation initiale \citep{2000A&A...361..101M, 2008A&A...478..467E}.\\
\end{itemize}

Il est tout à fait primordial de rappeler que l'axe de rotation d'un rotateur rapide peut être quelconque par rapport à l'observateur (il est même rare qu'il soit aligné avec l'axe de visée). De ce fait, toute mesure indirectement déduite et/ou directement observée, telle que celles des rayons (équatorial ou polaire), des températures effectives, de la gravité de surface et/ou de la luminosité, doit être interprétée avec la plus grande des prudences (pour plus de détails voir les chapitres \ref{chap:scirocco} \& \ref{chap:appli}).\\

Tel que nous avons pu le voir plus haut, les étoiles du type Be sont considérées comme étant les étoiles tournant le plus rapidement. Cependant il existe un type d'étoile appelé "étoiles dégénérées" qui, après que leur noyau ait terminé de consommer leurs combustibles par fusion thermonucléaire, évolue en une forme condensée de matière dégénérée (faisant intervenir le principe d'exclusion de Pauli à l'échelle macroscopique) pouvant acquérir un très fort moment cinétique. Ce processus intervient sur des étoiles en fin de vie, à l'instar d'une patineuse artistique qui replie ses bras. Plus l'étoile s'effondre sur elle-même de manière importante, plus elle subit une augmentation drastique de sa rotation.
Parmi ces étoiles dégénérées, on peut citer ; les naines blanches, les étoiles à neutrons et les trous noirs.\\   

Pour clore ce chapitre important, il serait tout fait inapproprié de ne pas parler de manière plus étendue du phénomène Be. C'est donc à cela que sera consacré le sous chapitre suivant.\\

\subsection{Les étoiles toupies; les Be}

\textit{"... Mais parmi le nombre très considérable des étoiles examinées, je trouve une exception bien singulière. L'étoile $\gamma$ Cassiopée est parfaitement complémentaire de ce type, et au lieu d'avoir une raie obscure à la place F, elle a une bande lumineuse d'une longueur sensible. Il est facile de s'en convaincre, et en regardant $\beta$ Cassiopée qui est du premier type ordinaire et en portant ensuite l'instrument sur $\gamma$ Cassiopée : on voit qu'à la place de la raie noire de la première on a une raie brillante dans la seconde. Après avoir beaucoup cherché si cette exception se présentait pour d'autres étoiles, je viens d'en trouver une autre, c'est $\beta$ Lyre ; mais sa raie est très fine et très difficile à voir. Ces exceptions si peu nombreuses méritent toute l'attention des théoriciens. Car s'il est vrai que les raies noires sont dues à une absorption par une certaine substance (l'hydrogène dans le cas actuel), ici nous trouvons la lumière directe émanée de cette substance ; cela prouverait ce que nous avons avancé ailleurs, que toutes les raies ne sont pas produites par simple absorption..."}\\

Père Angelo Secchi, Rome, 8 septembre 1866.\\

C'est en ces termes que le Père Angelo Secchi avait posé sans le savoir le problème du phénomène Be, le 8 septembre 1866 à Rome, dans la revue Astronomische Nachrichten, en observant pour la première fois de mystérieuses émissions, inexpliquées à l'époque, de la raie F de Fraunhofer ($H\beta$  correspondant à la transition entre les niveaux 4 et 2 de l'atome d'hydrogène), sur les deux étoiles $\gamma$ Cassiopée et $\beta$ Lyre.\\
 
\begin{figure}[h!]
\centering
 	\includegraphics[width=0.5\hsize,draft=false]{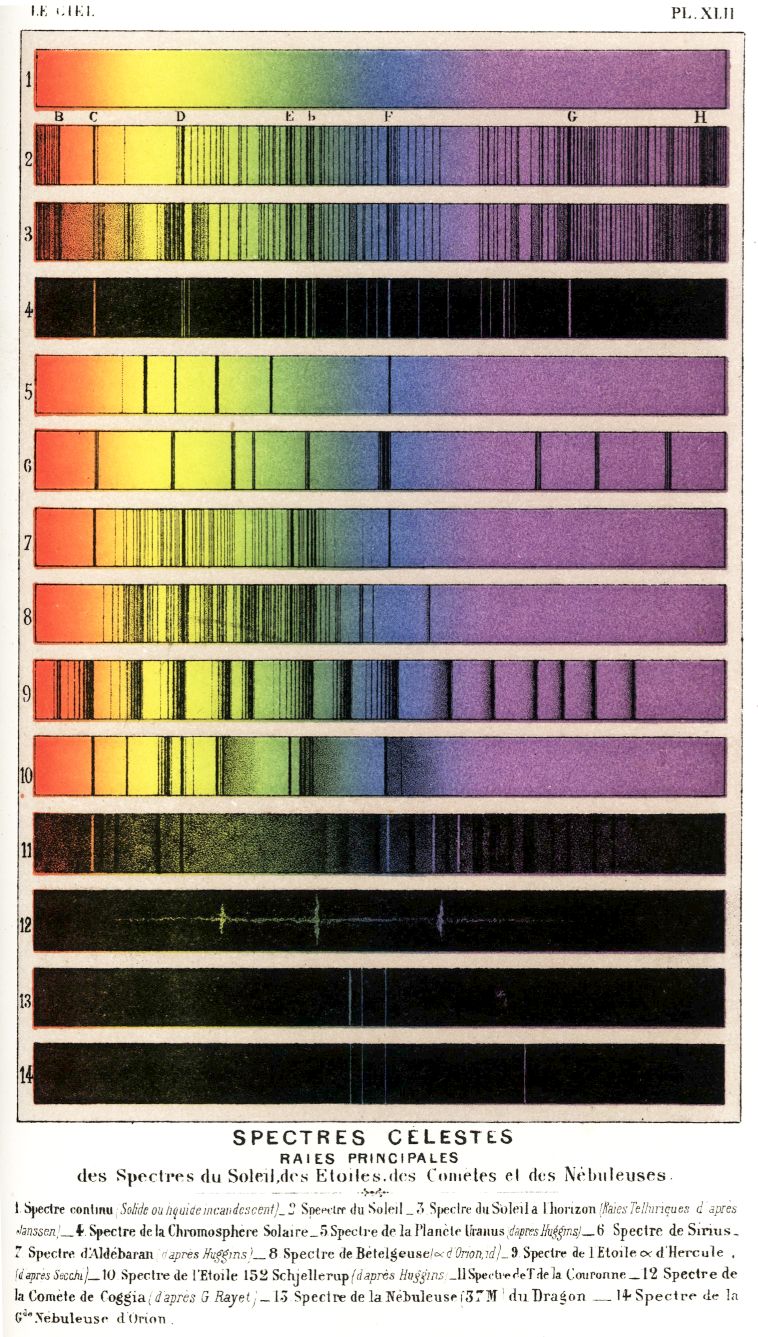}
	\caption[Spectres stellaires observés par Secchi au XIXe siècle]{Différents spectres stellaires observés par l'astronome italien Secchi au XIXe siècles. Crédit : Bibliothèque de l'Observatoire de Paris.}\label{Spectres_Secchi}
\end{figure}

Mise à part la raie $H\alpha$, le père Secchi observa par spectroscopie l'effet de l'émission sur toute la série de Balmer (transition, dans l'atome d'hydrogène, des niveaux supérieurs vers le niveau 2).\\

Une année plus tard, en 1867, MM. Wolf et Rayet s'exprimèrent ainsi au sujet d'une observation sur l'étoile P-Cygni, presque semblable à celle de $\gamma$ Cas : 
\textit{" Parmi les nombreuses étoiles dont la lumière a été étudiée à l'aide d'un prisme, on n'en connaît qu'une seule, Gamma de Cassiopée, dont le spectre offre constamment des lignes brillantes. Nous avons l'honneur de signaler à l'Academie l'existence de semblables lignes dans trois étoiles de la constellation du Cygne... Leur spectre se compose d'un fond éclairé dont les couleurs sont à peine visibles. Tous trois présentent une série de lignes brillantes. L'identification des lignes lumineuses de ces étoiles avec celles des spectres des gaz incandescents nous a été impossible..."}
(Comptes rendus de l'Académie des sciences, 1867, vol 65, p. 292)\\

Bien que présentant des raies à forte brillance, le spectre de P-Cygni  (ainsi que d'autres étoiles de ce type découvertes ultérieurement) était suffisamment différent des Be, pour que ce type d'étoiles, baptisées ultérieurement étoiles Wolf-Rayet en hommage à ceux qui les ont découvertes en premier, soit classé à part.\\

Toujours en 1867, Huggins, qui échangeait souvent avec Secchi, reprit l'observation de $\gamma$ Cassiopée et releva que $H\alpha$ était en émission également. Par la suite plusieurs autres observations firent état de ce phénomène de raies brillantes sur certaines étoiles, où elles furent recensées dans les revues spécialisées comme "étoiles à raie d'émission" (emission line stars), avant que la communauté scientifique adopte la classification stellaire d'Harvard (OBAFGKM) développée  par Henry Draper au 19\up{ième} siècle.
En 1926, Svein Rosseland , présenta un essai de 4 scenarios possibles et assez complexe pour interpréter la physique d'excitation susceptible d'engendrer ces fameuses raies en émission ; Un rayonnement occasionné par des températures et pressions excessives, l'introduction d'un rayonnement d'origine externe dans l'atmosphère de l'étoile, une volumineuse atmosphère en équilibre thermique, et l'existence de régions extrêmement chaudes dans une atmosphère en  déséquilibre hydrostatique.\\

Mais ce n'est qu'en 1931 qu'Otto Struve \citep{1931ApJ....73...94S} présenta une hypothèse simple et limpide pour expliquer le phénomène Be et leurs raies en émission. Il attribua de facto une enveloppe gazeuse (ou une nébuleuse) autour de l'étoile, où il compara de manière imagée l'apparence d'une étoile Be à la planète Saturne, mais aplatie par une rotation rapide (s'inspirant sans doute du travail de Sir \citet{1928Natur.121..279J} qui démontra  qu'un corps gazeux en rotation rapide pouvait éjecter le long de son équateur). Il décrivit ainsi un rotateur critique de forme altéré par une force centrifuge élevée et projetant de la matière à hauteur de son plan équatorial, qui forme autour de l'étoile un anneau de gaz en mouvement Képlérien.\\

Le modèle d'une enveloppe de gaz entourant une étoile critique centrale, ou de disque de décretion (décretion car émanant de l'étoile par analogie au terme disque d'accrétion qui lui peut produire une étoile sous certaines conditions ou indiquer de la matière tombant sur l'étoile), tel que proposé par Struve, fut assez vite accepté et repris par les astronomes et les astrophysiciens, à l'opposé de l'explication traitant de la forme du disque circumstellaire qui suscita de nombreux débats passionnants au sein de la communauté scientifique. En effet, après le développement des méthodes d'observation (dans les années 50) autres que spectroscopique, on constata une variabilité dans l'intensité des ailes rouge et bleu des raies Balmer en émission dans ces étoiles là, directement liée à la forme du disque circumstellaire entourant l'étoile ainsi qu'à l'interaction de ces derniers (étoile-disque).\\

Plusieurs modèles plus ou moins compliqués ont été élaborés et proposés, en  passant par des modèles d'anneaux (tel que proposé par Struve) mais elliptiques (e.g. \citet{1961JRASC..55...13M,1961JRASC..55...73M} et \citet{1973ApJ...183..541H}) ou bien des modèles d'enveloppe sphérique dont le plus fameux sont ceux de \citet{1976IAUS...70..335M} et \citet{1982IAUS...98..453P} qui ont depuis été largement adoptés. Ce n'est que dans les années 80 et avec l'avènement de l'interférométrie que l'aplatissement de l'enveloppe a pu être avéré et mesuré.\\

\begin{figure}[h!]
\centering
 	\includegraphics[width=0.5\hsize,draft=false]{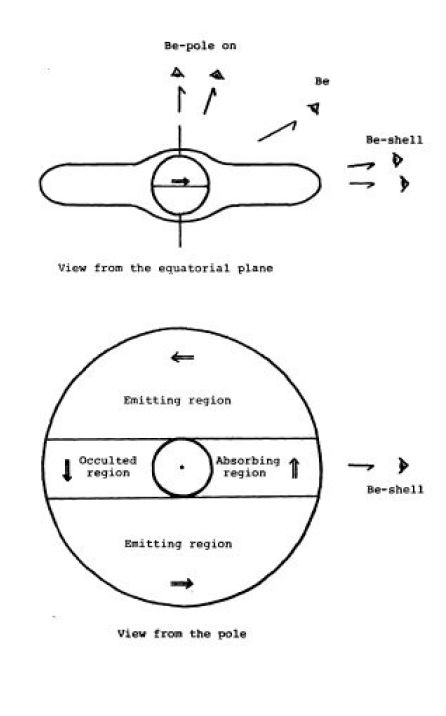}
 	\caption[Les différentes types d'étoiles Be]{Différents types d'étoiles Be, classées selon l'angle de visée.}\label{classes_Be}
\end{figure}

En 1982 Tomakazu Kogure et Ryuko Hirata classèrent les étoiles Be selon l'angle de vue en trois catégories bien distinctes: les étoiles "Be-shell" quand la ligne de visée est équatoriale; les étoiles "Be-pole on" vues du pôle ; et les "Be" simplement classées entre les deux (voir le schéma ci-dessous Fig. \ref{classes_Be}) \citep{1988PASP..100..770S}.\\

Dans les années 70 des étoiles Be possédant des raies interdites à faible émission furent observées avec un excès infrarouge. Conti suggéra de noter cette nouvelle classe de Be aux raies interdites par notation B[e], lors du colloque IAU " Be and shell stars " de 1976. Cette classe est répartie dans le diagramme HR entre les Be dites "classique" et les supergéantes enveloppées de gaz et de poussière.\\

\section{Effet de la rotation sur la formation des enveloppes circumstellaires}

Les étoiles B[e] font partie de la famille des étoiles actives chaudes.  Si le facteur de perte de masse par l'équateur  est considéré comme insignifiant pour les B[e] supergéantes qui perdent une grande portion de leur moment cinétique lors du cycle d'expansion des couches externes de leur atmosphère, ce n'est surement pas  le cas pour les Be classiques, et à un niveau moins important pour les Ae/Be d'Herbig (étoiles PMS -Pre Main Sequence- avec un disque d'accrétion), là où la rotation est  plus importante. Mais dans tous les cas la proportion de perte de masse est toujours plus importante à l'équateur qu'aux pôles \citep{2000A&A...361..159M}.\\

Avant les années 2000 l'hypothèse de Struve fut assez critiquée, car toutes les observations menées sur les étoiles Be pour prouver que leur vitesse de rotation était suffisamment puissante pour que la force centrifuge résultante puisse contrer la force de gravité et provoquer une perte de masse équatoriale par rotation ; \citet{1996MNRAS.280L..31P} mesura des vitesses proches de $70\%$ de la vitesse critique, ce qui  implique une gravité réduite d'un facteur 2.\\

Il n'y a qu'une décennie que  \citet{2004MNRAS.350..189T} démontrèrent que l'effet de l'assombrissement gravitationnel pouvait provoquer une saturation dans l'élargissement des profils de raies photosphériques et entraîner ainsi une sous-estimation de l'ordre de $12$ à $33\%$ sur la vitesse de rotation d'étoiles critiques (étoiles proches de leur vitesse critique).
Ce qui renforce l'hypothèse de \citet{1931ApJ....73...94S} que les étoiles Be laissent échapper de la matière par leur régions équatoriales, à cause d'une forte vitesse de rotation, beaucoup plus proche de la vitesse critique que ce qui été envisagé avant. Cependant une récente théorie incluant le champ magnétique fossile est de plus en plus admise pour expliquer la présence et le maintien d'un environnement circumstellaire. En effet, les particules éjectées par un vent stellaire, contraint par les lignes de champ magnétique fossile et formant une magnétosphère pouvant emprisonner de la matière circumstellaire ou pas, selon la vitesse du vent, la force du champ magnétique et la vitesse de rotation de l'étoile. Vitesse qui maintient la matière de l'environnement circumstellaire dans le plan équatorial magnétique sous l'effet de la force centrifuge \citep{1997A&A...325..195B}.\\

Une observation avec l'instrument VINCI/VLTI sur l'étoile Be Achernar ($\alpha$ Eri) par \citet{2003A&A...407L..47D} révéla un rapport du rayon équatorial sur le rayon polaire de l'ordre de $1.5$ (voir Figure). Rapport qui ne peut être expliqué que par une vitesse très proche de la vitesse critique de l'ordre de $95\%$. Résultats confirmés par la suite dans \citet{2012A&A...545A.130D} ($R_{eq}/R_{pol}=1.45$), \citet{2014A&A...569A..45H} ($R_{eq}/R_{pol}=1.42$) et \citet{2014A&A...569A..10D} ($R_{eq}/R_{pol}=1.35$).
Bien que l'enveloppe circumstellaire d'Achernar connait plusieurs stades de variation cyclique de densité, et qui est en une étroite corrélation avec le cycle de révolution d'un compagnon détecté par \citet{2008A&A...484L..13K}.\\

\begin{figure}[h!]
\centering
 	\includegraphics[width=0.5\hsize,draft=false]{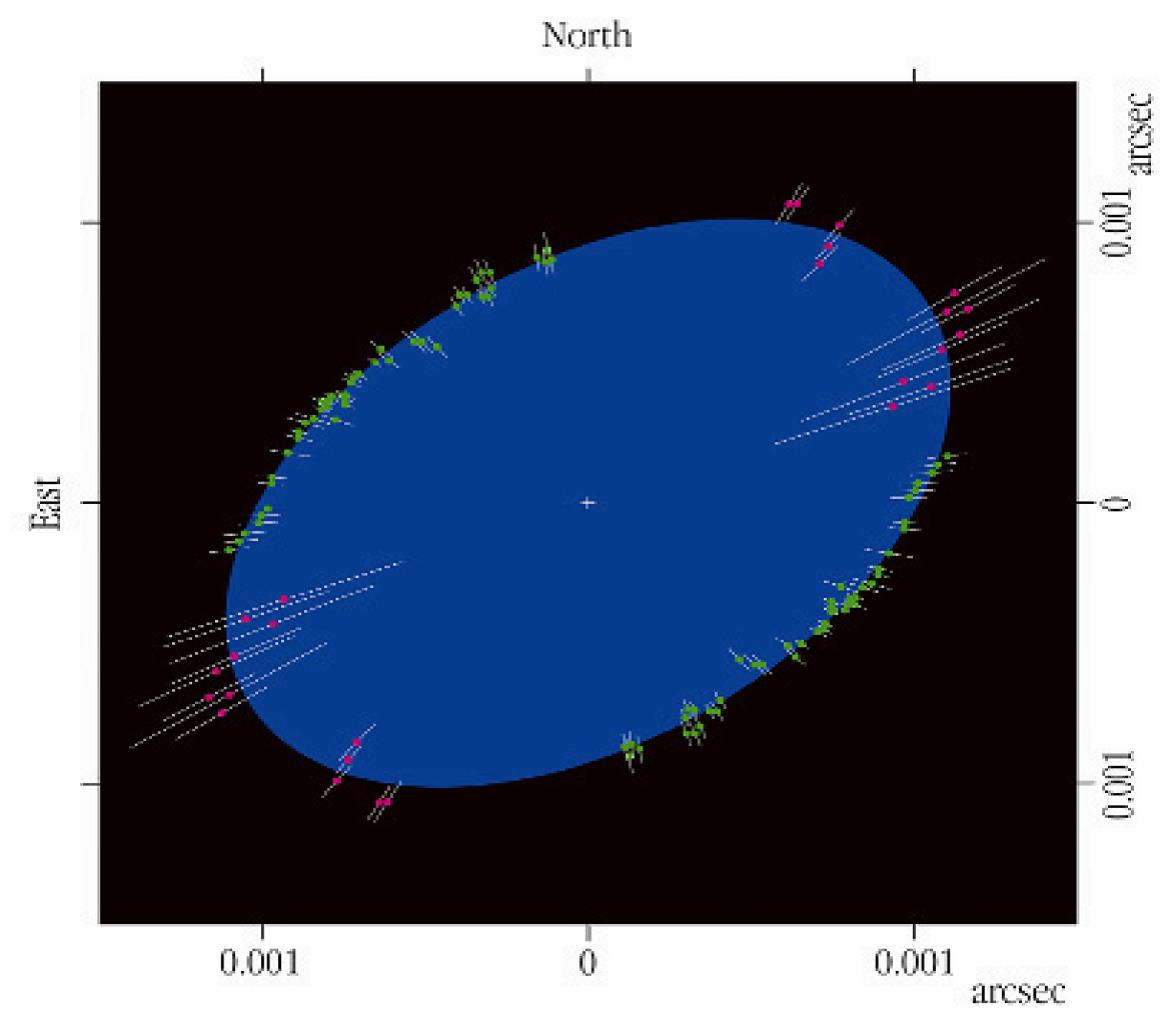}
 	\caption[Déduction de l'aplatissement d'Achernar par les mesures interférométriques]{Déduction de l'aplatissement d'Achernar via des mesures interférométriques Source : \citet{2003A&A...407L..47D}.}\label{Domiciano03}
\end{figure}

Enfin \citet{2004IAUS..215...23F}, dans une étude minutieuse de l'impact de l'assombrissement gravitationnel sur le $\vsini$, déduisirent que la vitesse moyenne de rotation des étoiles Be est de l'ordre de $88\%$ de la vitesse critique, pour une gravité qui se restreint d'un $1/4$.\\

%*** Parler de la Morphologie et cinématique des enveloppes, ainsi que des raies circumstellaires.

\section{Autres types d'activités susceptibles d'engendrer des enveloppes circumstellaires}

Le phénomène de perte de masse des étoiles chaudes est un aspect important de la vie stellaire évolutive.  Les O, B et A chaudes peuvent ainsi perdre jusqu'à $10^{-9}$ à $10^{-5}$ masse solaire par an ($\Msun/an$) et $10^{-3}$ pour les LBV (Luminous Blue Variable stars), comme $\eta$ Car (l'une des étoiles les plus massives de notre galaxie), et les WR (Wolf-Rayet) comme P-Cygni.\\

Ainsi, les plus massives des étoiles sont capables de se défaire de $90\%$ de leur masse, nourrissant ainsi le milieu interstellaire en hydrogène, hélium et autres métaux légers. Une telle perte de masse, qui a aussi un impact sur le moment cinétique de l'étoile, ne peut qu'influer de manière significative sur l'évolution de cette dernière au cours du temps.\\

Néanmoins le phénomène des enveloppes circumstellaires ne peut être simplement expliqué par une perte de masse propre à l'étoile. Il peut aussi provenir de sources externes, telles que les Ae/Be de Herbig qui doivent leur enveloppe à un reste de nébuleuse proto-stellaire, en plus de  générer également un vent stellaire lié à un mécanisme propre à l'étoile centrale, et telles que les Be binaires en interaction. Outre la rotation rapide (citée plus haut), parmi les autres phénomènes susceptibles d'engendrer les enveloppes circumstellaires, on peut compter les vents radiatifs, les pulsations non-radiales ainsi que le  magnétisme. Ce dernier est très présent dans le Soleil et il a une grande responsabilité dans la réorganisation de la matière autour de notre étoile (e.g. les éjections de masse coronale). Ces 4 phénomènes sont explicités ci-dessous.\\
 
\subsection{Magnétisme}

Le champ magnétique peut influencer de manière impressionnante l'éjection de la matière d'une étoile, et pour preuve notre Soleil, très actif magnétiquement, tel que le révèlent les taches solaires visibles sur sa photosphère depuis la terre (voir plus haut, Figs. \ref{taches_sol_chine} \& \ref{taches_sol_Galiee}), les éruptions solaires (capable d'atteindre le volume de plusieurs terres) et les éjections coronales (responsables des belles aurores boréales qu'on peut admirer près de nos régions polaires)...etc. De fortes éruptions solaires peuvent même provoquer d'importants dégâts sur nos installations électriques, et ce à plus de 150 millions de kilomètres (e.g., effondrement du réseau électrique d'Hydro-Québec, le 13 mars 1989).\\

\begin{figure}[h!]
\centering
 	\includegraphics[width=0.5\hsize,draft=false]{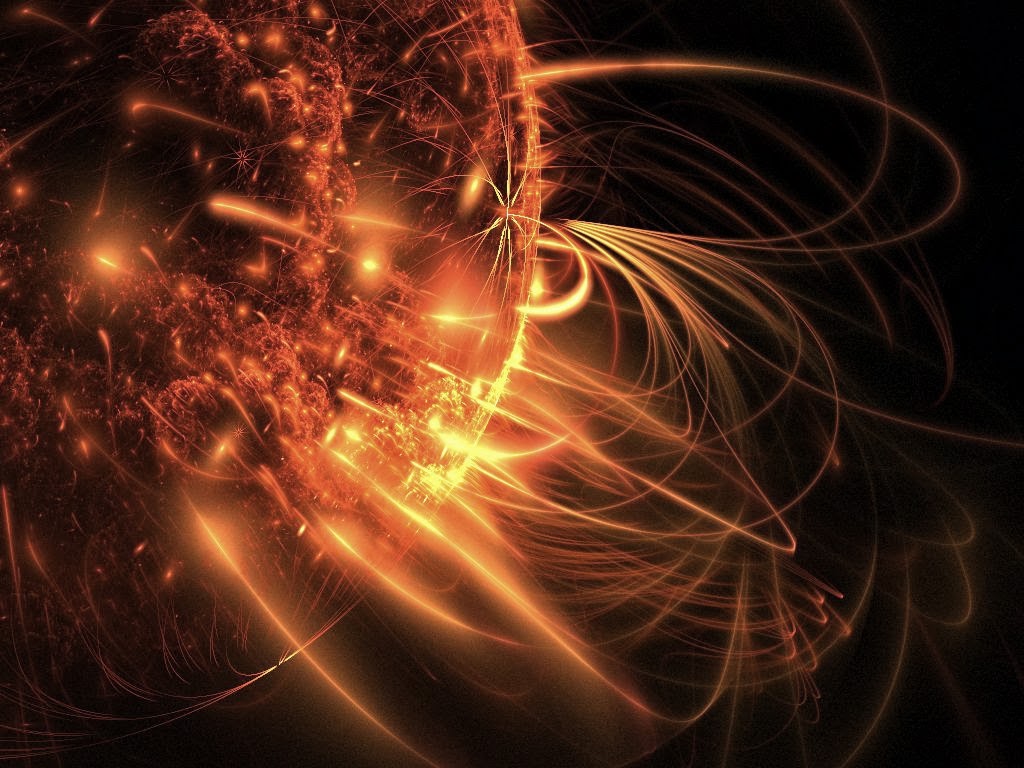}
 	\includegraphics[width=0.5\hsize,draft=false]{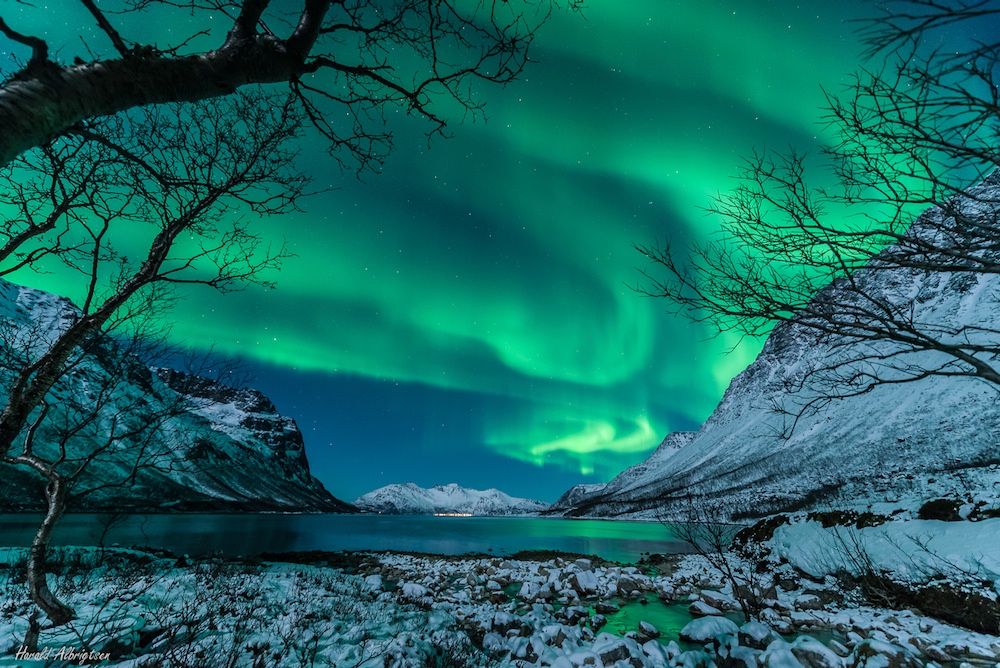}
 	\caption[Champ magnétique solaire et aurores boréales]{\textbf{En haut:} Vue d'artiste du champ magnétique solaire. \textbf{En bas:} Aurore boréale au-dessus de la Norvège dans la nuit du 10 janvier 2014 (source : Harald Albrigsten).}\label{Mag_Sun}
\end{figure}

Cependant et d'un point de vue purement théorique, la structuration d'enveloppes circumstellaires autour d'étoiles chaudes actives semble guère envisageable. Effectivement, ce type d'étoiles dites précoces, ne possède pas une zone convective assez consistante pour générer par effet dynamo un champ magnétique continu et important. Aussi, il faut compter en moyenne une intensité du champ magnétique de l'ordre d'une centaine de Gauss pour espérer un changement sensible des vents stellaires.\\

Néanmoins, certains ont émis d'intéressantes hypothèses sur le rôle du champ magnétique dans le fonctionnement des étoiles actives chaudes. Notamment, pour le champ magnétique fossile (décrit ci-haut) qui entretien un environnement circumstellaire dans le plan équatorial magnétique de l'étoile entre le rayon d'Alfven (limite du champ magnétique stellaire) et le rayon de corotation Képlérienne (où la matière du disque reste synchrone avec la rotation propre de l'étoile) \citep{2013MNRAS.429..398P}. En effet, \citet{2002ApJ...578..951C} augmente la densité du disque en rajoutant un champ magnétique à ces modèles afin d'être en adéquation avec les observations. \citet{2001MNRAS.326.1265D} décrit l'importance des lignes de champ sur les $\beta$ Cephei pour expliquer l'apport de matière dans l'enveloppe circumstellaire provenant de la photosphère (bien qu'il a bien été établie plus tard que $\beta$ Cep soit bien une étoile magnétique mais pas une Be, et que c'est son compagnon qui en est une Be). \citet{2003A&A...409..275N} fournit quelques preuves de l'existence d'un champ magnétique de l'ordre de $530\pm230$ Gauss autour de la Be classique $\omega$ Orionis.\\

A cause de leur rotation rapide, les éventuelles lignes de champ des Be, ont de forte chance de se recombiner dans le plan équatorial de l'étoile, imposant de ce fait, si celle-ci est en rotation rigide, le même type de rotation qu'aux régions photosphériques adjacentes à l'enveloppe circumstellaire. \citet{2006A&A...460..821A} a observé ce genre de phénomène en spectro-interférométrie dans la raie du Fe II.\\

A l'aide du programme MiMeS (Magnetism in Massive Stars), destiné à la prospection de caractéristiques encore inconnues du magnétisme des étoiles massives, \citet{2009IAUS..259..333W} ont suspecté la présence de champ magnétique autour de certaines Be, mais les récents résultats de \citet{2014arXiv1411.6165W} ne font état d'aucune Be magnétique détectée. La seule Be dont on a indirectement observé le champ magnétique par MIMeS reste $\omega$ Ori par \citet{2012MNRAS.426.2738N}.\\

\subsection{Vent radiatif}\label{Vent_radiatif}
Contrairement aux étoiles de type solaire dont le vent stellaire est généré par la présence du champ magnétique et la pression thermique du gaz (c'est le cas du Soleil qui possède une zone convective), les étoiles massives chaudes créent elles leurs vents par effet de pression radiative. Ce phénomène est très important dans les Be. En effet  la luminosité de l'étoile est proportionnelle à la température photosphérique de l'étoile puissance 4. Plus l'étoile est lumineuse, plus l'énergie des photons sera considérable et plus important sera le mouvement de la matière photosphérique \citep{1975ApJ...195..157C}. Dans un milieu optiquement épais et au rayonnement continu, les raies observées n'en seront que plus remarquables.\\ 

Il y a deux types de  raies susceptibles d'engendrer un vent radiatif ; les raies faibles (optiquement minces) relatives à une transition d'un niveau excité à un autre et les raies fortes (optiquement épaisses) qui sont imputées à une transition entre un niveau excité et un niveau fondamental, ce qui est le cas dans un milieu dit standard où les atomes sont en principe dans un état fondamental.\\ 

Tout comme dans le cas de la rotation où la force centrifuge doit compenser la force de gravité pour qu'il y ait une perte de masse stellaire vers l'enveloppe (voir plus haut), là aussi il faut que la force de pression radiative soit plus importante que la force gravitationnelle pour  enclencher un vent radiatif. Ainsi \citet{1979IAUS...83..237A} établit que ce phénomène n'était observé que chez les Be de type précoce. Ce qui n'est pas le cas des étoiles de types plus tardifs où au-delà du type spectral B8V la force de pression radiative à elle seule est insuffisante pour maintenir un vent radiatif.\\

Le premier modèle de vent radiatif comprenant des milliers de raies dans la modélisation de la pression radiative, fut développé en 1975 par Castor, Abbott et  Klein pour les étoiles du type Of \citep{1975ApJ...195..157C}. Ce modèle nommé CAK inspira par la suite plusieurs autres modèles de vents radiatifs d'étoiles chaudes actives de différents types dans les années 80 \& 90 (y compris le code SIMECA - SIMulation d'Etoiles Chaudes Actives- de Philppe Stee à l'OCA, pour plus de détail voir le chapitre 4.1 de la thèse de \citet{am07}). Dans les cas des étoiles actives chaudes à haut moment cinétique, comme les Be par exemple, l'effet von Zeipel (expliqué brièvement plus haut) y est prédominant. De ce fait, la pression radiative est plus importante aux pôles qu'à l'équateur, ce qui peut entraîner des vents radiatifs avec des vitesses d'environ $1000$ $\kms$ aux pôles alors qu'au niveau de l'équateur ces vitesses-là ne dépassent guère les quelques dizaines de $\kms$. Ainsi les vents stellaires dans la population d'étoiles chaudes actives sont hautement anisotrope.\\

Des modèles, tels que "Wind Compressed Disk" (WCD) (e.g. \citet{1993ApJ...409..429B}) ont été avancés pour expliquer la présence de disque équatorial relativement dense dans l'environnement des Be à partir d'un vent quasi-sphérique. Ce modèle suppose des lignes d'écoulement du vent tordues depuis les deux hémisphères, se rejoignant vers l'équateur sous forme de disque mince (voir figure ci-dessous) et émettant dans la bande X (ce que confirment les observations).\\

\begin{figure}[h!]
\centering
 	\includegraphics[width=0.5\hsize,draft=false]{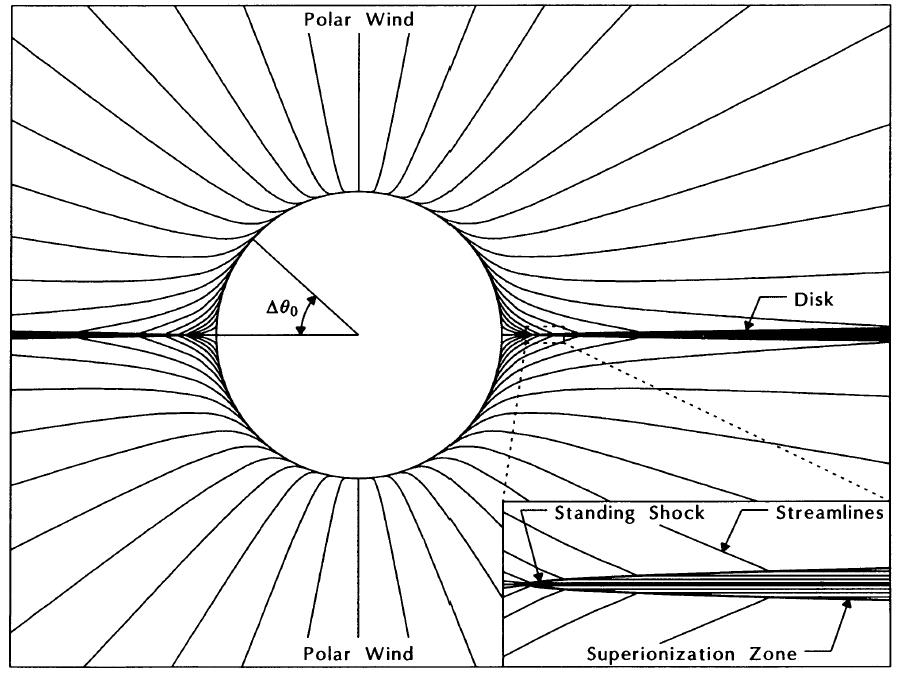}
	\caption[Simulation WCD de lignes d'écoulement d'un vent radiatif-lignes de flux et choc équatorial dans le cas du Wind compressed Disk model (WCD)]{Lignes d'écoulement d'un vent radiatif qui s'entrechoque  dans le plan équatorial de l'étoile, simulation effectuées par le code  Wind Compressed Disk model (WCD). Source \citet{1993ApJ...409..429B}}\label{Bjorkman&Cassinelli_1993}
\end{figure}

D'autre part, les calculs approfondis d'\citet{1996ApJ...472L.115O} ont clairement démontré que toute force radiative non radiale rendait impossible toute formation d'un disque stellaire par ce processus. Pour ce faire, il faut donc envisager la présence de forces/effets supplémentaires, tel que le champ magnétique, (abordé dans le sous-chapitre précédent). Et dans ce cas-là on parle plus de WCD mais de Magnetically Wind Compressed Disk (MWCD). D'autre part le disque modèlisé par WCD, et qui n'a qu'une durée de vie de quelques jours, n'arrive pas à bien expliquer les longues variations temporelles du disque circumstellaire. Pour ce faire, des modèles de disque dits visqueux ont été imaginés par \citet{1991MNRAS.250..432L}, \citet{1996MNRAS.280L..31P} et \citet{2001PASJ...53..119O} et inspirés de \citet{1981ARA&A..19..137P}.\\

\subsection{Binarité}
Malgré le fait que la proportion de systèmes multiples soit entre 60\% et 75\% parmi les étoiles tout type confondu, le phénomène de la binarité à moins d'une séparation faible, ne peut à lui seul expliquer la présence d'une enveloppe circumstellaire/circumbinaire. Un bon nombre de Be binaires sont énumérées par \citet{1987pbes.coll..339H} et \citet{2000ASPC..214..668G}.\\

\begin{figure}[h!]
\centering
 	\includegraphics[width=0.5\hsize,draft=false]{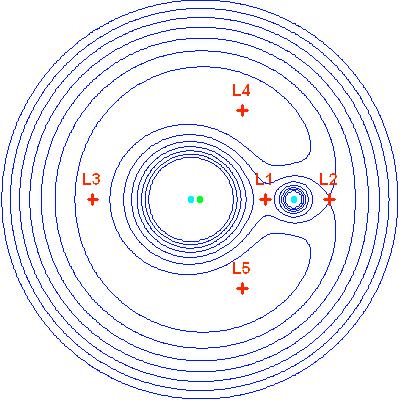}
 	\includegraphics[width=0.5\hsize,draft=false]{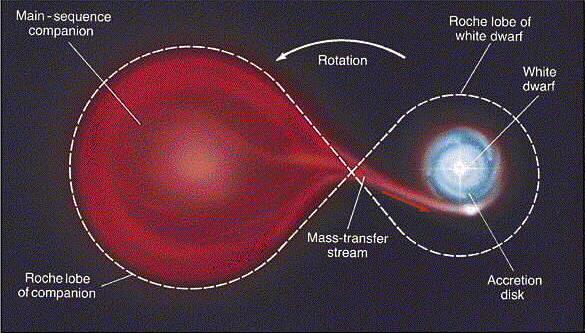}
	\caption[Systèmes binaires et Points de Lagrange]{\textbf{En haut:} Les 5 points de Lagrange, extrema du potentiel gravitationnel d'un système à 2 corps (source:Astrophysique sur Mesure). \textbf{En bas:} Une naine blanche dans un système binaire assez proche de son compagnon que son champ gravitationnel peut arracher de la matière de la surface de son compagnon par le point Lagrange L1. }\label{systeme_binaire}
\end{figure}

Pour les systèmes binaires, plusieurs scénarios sont possibles pour expliquer  la présence d'enveloppes. Tout dépend de la séparation des deux étoiles (distance étoile principale - compagnon), des limites du lobe de Roche (masse et densité des étoiles) et dans quelle configuration se trouve le système binaire (à quel point de  Lagrange se trouve positionné le compagnon par rapport à l'étoile principale). Ainsi dans le cas d'une petite séparation, d'un compagnon assez avancé dans son évolution (moins compact donc) et occupant le lobe de Roche de l'étoile principale, la matière du compagnon enrichit l'environnement circumstellaire de la primaire en empruntant le point de Lagrange L1 \citep{1987pbes.coll..339H}, (voir Fig.\ref{systeme_binaire}).\\

Un autre scénario possible a été  avancé par \citet{1975BAICz..26...65K} pour interpréter certaines fluctuations spectro-photométriques observées: la matière de la secondaire enrichit l'enveloppe de la primaire en passant par le point L1, le tout en éjectant de la matière par les points de Lagrange L2 et L3 et en formant une spirale (ou deux) sous l'effet de la rotation du système binaire.\\

Dans le cas de séparations plus importantes il ne peut y avoir aucun transfert de masse entre les deux étoiles de la binaire. Néanmoins, par effet de la gravité et pour une orbite elliptique, le compagnon, lors de son passage au périastre, peut  brièvement  perturber l'étoile principale et provoquer chez elle une éjection de masse qui enrichit son enveloppe circumstellaire (surtout que l'étoile primaire est en rotation critique \citep{2002A&A...396..937H}), ainsi qu'une éventuelle stimulation de mode de pulsations non-radiales (PNR) à ce moment-là. Ceci peut expliquer, par exemple, l'apparition de certaines raies d'hydrogène en émission dans le spectre de l'étoile $\delta$ Scorpii, lors de l'approche de son compagnon du périastre \citep{2001A&A...377..485M}. Ce phénomène est fortement suspecté sur l'évolution de l'enveloppe Achernar avec la révolution de son compagnon (détecté par \citet{2008A&A...484L..13K}). La présence d'un compagnon dans l'environnement circumstellaire peut aussi influencer la structure et la densité de ce dernier, ainsi que l'ont révélé \citet{2007ApJ...654..527G} sur l'étude des quatre étoiles Be ; $\gamma$ Cassiopeiae, $\phi$ Persei, $\zeta$ Tauri et $\kappa$ Draconis. Aussi, pour une séparation de quelques dizaines de rayons stellaires, le compagnon peut avoir une influence gravitationnelle sur l'enveloppe de l'étoile principale, où cette dernière peut s'étendre, avec toutes les répercutions ressenties et observées du point du vue spectroscopique par effet Doppler (e.g. \citet{2005A&A...435..275C} sur $\alpha$ Ara et \citet{2004AJ....127.1194T} pour $\xi$ Tau).\\  

\begin{figure}[h!]
\centering
 	\includegraphics[width=0.5\hsize,draft=false]{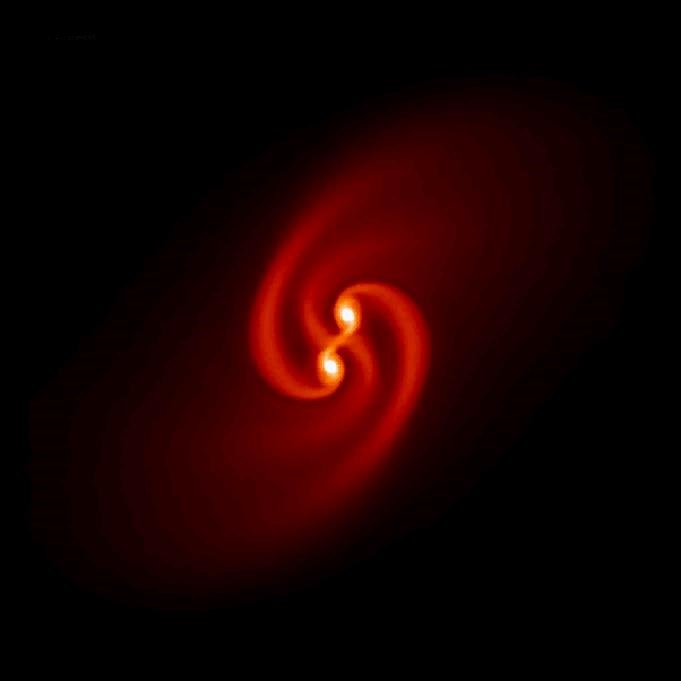}
	\caption[Un bel exemple de formation binaire en rotation]{Un bel exemple de formation binaire en rotation, où on voit de la matière s'éjecter par les points de Lagrange L2 et L3 en forme de 2 spirales (simulation de \citet{2007MNRAS.377...77P}, avec un champ magnétique relativement faible perpendiculaire à l'axe de rotation).}\label{binary_systeme2}
\end{figure}

\subsection{Pulsations}
Bien que la pulsation elle seule soit insuffisante pour arracher de la matière photosphérique pour enrichir le disque circumstellaire, elle peut néanmoins amoindrir de manière assez sensible la gravité surfacique effective de l'étoile pour permettre à d'autres phénomènes précédemment cités tels qu'une rotation critique \citep{2003PASP..115.1153P}, un vent radiatif et/ou un champ magnétique puissant, d'extraire de la matière de la photosphère vers l'enveloppe circumstellaire. On peut donc classer ce phénomène comme un éventuel catalyseur qui peut favoriser la création et/ou le maintien d'une enveloppe.\\

Les étoiles sont généralement gouvernées par deux forces majeures (la force de pression qui tend à les faire exploser et la force de gravité qui tend à les faire imploser). En conséquence elles peuvent être le siège d'oscillations, interprétées comme étant la superposition d'ondes se propageant à l'intérieur de l'étoile, conduisant à la formation par interférence d'ondes stationnaires, identifiables par le mouvement cohérent de la surface, et appelées modes propres de vibration ; où chacune est définie par une fréquence caractéristique. Chaque onde stationnaire (mode propre) est caractérisée essentiellement par 3 nombres entiers (les nombres harmoniques sphériques) décrivant la position des points et lignes de n\oe{}uds dans notre étoile:

\begin{enumerate}
\item \textbf{n} : L'ordre radial du mode, nombre de n\oe{}uds le long du rayon de l'étoile.
\item \textbf{l} : L'ordre du mode, nombre total de lignes de n\oe{}uds.
\item \textbf{m }: L'ordre azimutal du mode, nombre de lignes de n\oe{}uds qui passent par les pôles de vibrations.
\end{enumerate}

\begin{figure}[h!]
\centering
 	\includegraphics[width=0.5\hsize,draft=false]{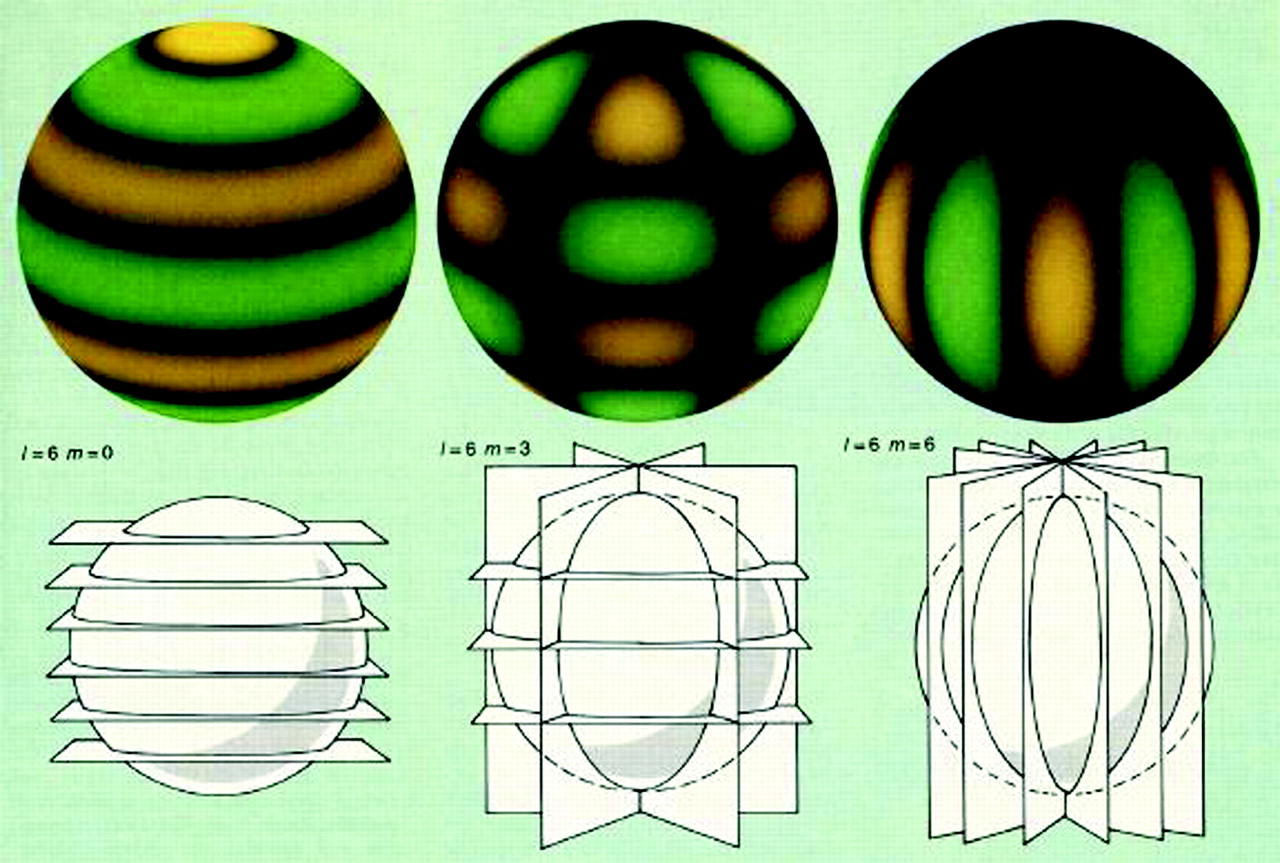}
	\caption[Exemples de pulsations représentées par les harmoniques sphériques]{Trois exemples sur les nombres d'harmoniques sphériques : (m,l) = (6,0), (6,3), (6,6) à partir de la gauche (source : http ://gong.nso.edu/).}\label{puls1}
\end{figure}

Comme pour une corde vibrante, qui compte des points fixes (n\oe{}uds), une surface vibrante (un tambour par exemple) compte aussi des lignes fixes appelées lignes de n\oe{}uds, 'l' et 'm' traitant la partie tangentielle de la vibration 3D, tandis que 'n' traite la partie radiale.\\

On compte en général 2 types de modes:
\begin{enumerate}
\item \textbf{Les modes p} : modes de pression (ou mode acoustiques), de fréquences plus élevées (de l'ordre de quelques heures) et d'ordre radial positif (n > 0). Ils sont générés près de la surface. La force de rappel est la pression du gaz. Ce type de mode a été observé chez les $\beta$ Cephei et les étoiles Be/Bn.
\item \textbf{Les modes g }: modes de gravité, de fréquences relativement basses (de l'ordre du jour) et d'ordre radial négatif (n < 0). Ils sont créés par la force d'Archimède. Prédits théoriquement pour le Soleil, mais jamais observés avec certitude car essentiellement générés dans les zones les plus denses de l'étoile (le c\oe{}ur plus une partie de la zone radiative), ils sont noyés et annihilés à  la surface par les modes p, de fréquences et d'intensité plus élevées. On peut par contre les observer dans les SPB  ("slowly pulsating B-stars") et les étoiles Be.
\end{enumerate}
Un autre mode de fréquences intermédiaires et d'ordre radial nul (n = 0) existe. Appelé \textbf{mode f} (mode fondamental ou également connu sous le nom de mode de gravité de surface), il peut être classifié comme étant un troisième type de mode. Ces derniers sont très présent dans le Soleil.\\

Ce sont les pulsations non-radiales qui favorisent le plus encore l'éjection de matière en induisant une turbulence photosphérique avec une énergie cinétique concentrée sur de petites zones oscillantes. C'est ce qu'ont révélées certaines observations CoRoT (Convection, Rotation and planetary Transits) \citep{2009A&A...506...95H}.\\

C'est grâce à la variation de certains profil de la raie photosphérique tel que le MgII ($4481 \AA$) qu'on a pu observer des pulsations sur les Be \citep{2003A&A...411..181M}, avec une périodicité allant  de 2.5 heures à 3 jours.\\

Quelques modes d'oscillations ont été observés chez les Be. En effet à cause de l'aplatissement qui est dû à la rotation rapide de l'étoile, un certain type de mode peut être privilégié autour de l'équateur.  Ainsi, \citet{1996A&A...310..849F} détectèrent un mode $l = |m| = 8$, \citet{1997AGAb...13...36R}, les modes : $l = -m = 3$, $l = -m = 2$ pour certaines étoiles chaudes actives, et \citet{2003A&A...411..229R} observèrent un mode $l = m = 2$ sur une vingtaine d'étoiles Be. Sans oublier, qu'en plus des modes p et g, il existerait de modes de pulsations stochastiques dont \citet{2013ASPC..479..319N, 2014IAUS..301..465N} ont révélées l'importance dans l'augmentation du moment angulaire, jusqu'à sa valeur critique, à la surface des Be.\\

Tel que nous l'avons vu dans ce chapitre, qui est dédié au sujet principal traité dans ma thèse, et tel que nous allons le voir dans chapitre suivant, qui lui traite du moyen de mesure utilisé pour mener à bien mon étude (l'interférométrie), les différents chercheurs de différentes époques et de différentes civilisations (nationalités) se sont  appuyés les uns sur les autres, se sont passé le relais, générations après générations tout en apportant de remarquables améliorations et découvertes à chaque fois. Sans ces remarquables qualités intellectuelles et surtout humaines, on en serait surement toujours à l'âge de pierre encore de nos jours. Heureusement pour nous, que la science est universelle et qu'elle est basée sur la générosité, l'échange et le partage (la citation ci-dessous, que je partage volontiers avec vous, résume assez bien ma pensée à ce sujet).\\

\textit{" Nous sommes comme des nains assis sur des épaules de géants. Si nous voyons plus de choses et plus lointaines qu'eux, ce n'est pas à cause de la perspicacité de notre vue, ni de notre grandeur, c'est parce que nous sommes élevés par eux. "}

\textit{Bernard de Chartres, dans le livre III Metalogicon de Jean de Salisbury (1159).}

\chapter{Combiner la haute résolution spatiale et spectrale}
\begin{figure}[h!]
\centering
\includegraphics[height=0.5\hsize,draft=false]{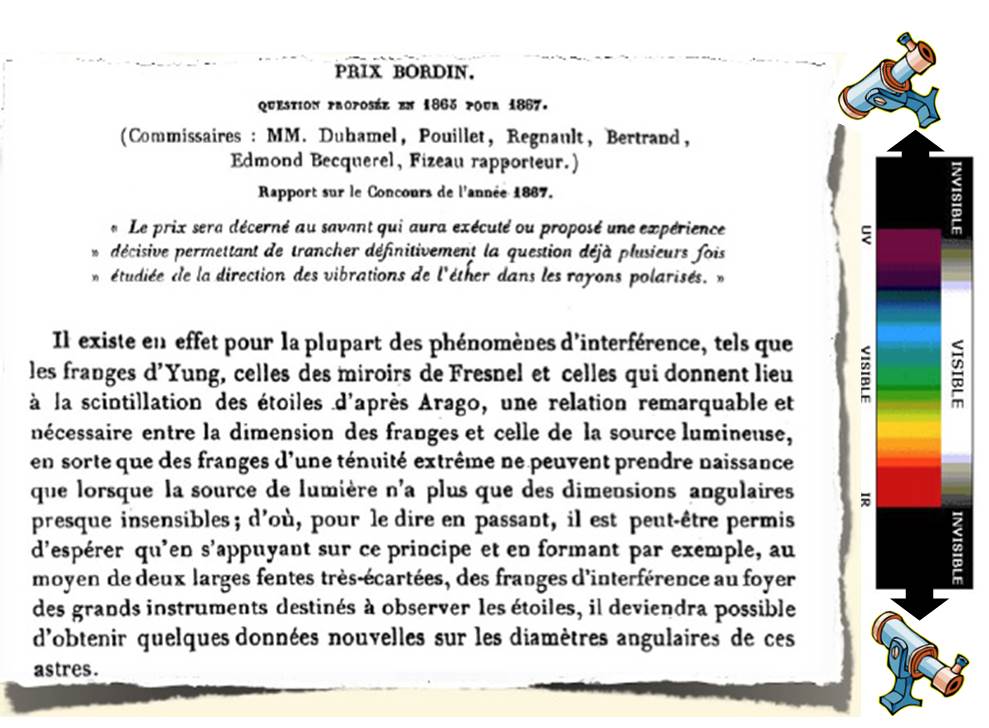}
\end{figure}
\label{chap:spec-interfero}
\minitoc

\section{Les outils de mesures astronomiques et leurs limites}

Le principal messager qu'utilise l'astronome/astrophysicien dans sa quête pour comprendre le ciel est la lumière qui nous parvient des astres et des lointaines galaxies, ainsi il est plus qu'impératif pour tout astronome ou astrophysicien de bien maitriser et comprendre les équations de base de l'électromagnétisme, de l'optique géometrique et ondulatoire, ainsi que d'importantes notion de physique contemporaine, relativiste et même quantique.\\

\textbf{Onde électromagnétique \& intensité lumineuse:} Une onde électromagnétique se compose de deux champs perpendiculaires l'une par rapport à l'autre ; d'un champ électrique ($\vec{E}$) et d'un champ magnétique ($\vec{B}$) qui oscillent à la même fréquence avec une pulsation $\omega$, dans le temps $t$ et qui se propagent de manière transversale (Fresnel 1821\footnote{Dans un ouvrage intitulé "Mémoire sur la double réfraction", regroupant trois Mémoires présentés le 26 novembre 1821, le 22 janvier 1822 et le 26 avril 1822, à l'Académie des sciences de l'Institut de France (T 7, 45-176)}) selon une direction orthogonale ($\vec{r}$) dans un milieu donné. Dans le vide, la vitesse de propagation est égale à la vitesse constant, celle de la vitesse de la lumière $c \sim 3.10^8$ $m.s^{-1}$. Une telle onde se doit de satisfaire l'équation d'Alembert contrainte par la jauge de Lorentz et est décrite (dans le vide) par Maxwell via les équations suivantes :

\begin{gather}
 \vec{E}( \vec{r},t)= \vec{E_0}( \vec{r})e^{-i\omega t} \\
 \label{3.1}
 \vec{B}( \vec{r},t)= \vec{B_0}( \vec{r})e^{-i\omega t}
\end{gather}

Ainsi et dans un milieu isotrope, le vecteur de propagation de l'onde électromagnétique qui transporte l'énergie de celle-ci n'est autre que le vecteur de "Poynting" $\vec{S}$ et qui est connu sous la forme :

\begin{equation}
 \vec{S}(\vec{r},t)=\frac{\vec{E}(\vec{r},t) \wedge \vec{B}(\vec{r},t)}{\mu_0}
\end{equation}

Où $\mu_0$ est la perméabilité dans le vide. Le vecteur de Poynting ayant une pulsation temporelle d'environ $10^{-14}s$ dans le visible (l'\oe{}il humain lui a un temps d'intégration d'environ 40 ms alors qu'une caméra rapide de 1 ms), il ne peut donc  pas être mesuré de manière instantané, ainsi  on n'aperçoit qu'une valeur moyenne de ce dernier. Cette valeur est appelée existence, irradiance ou bien intensité lumineuse et est définit comme suit :

\begin{equation}
 I(\vec{s})=\left\langle \vec{S}(\vec{s},t) \right\rangle_t= \frac{1}{2\mu_0} \Re(\vec{E}(\vec{s},t)  \wedge \vec{B}^* (\vec{s},t))= \left\langle \left\| \vec{E}(\vec{s},t) \right\|^2 \right\rangle_t
\end{equation}

\begin{figure}[h!]
\centering
\includegraphics[height=0.35\hsize,draft=false]{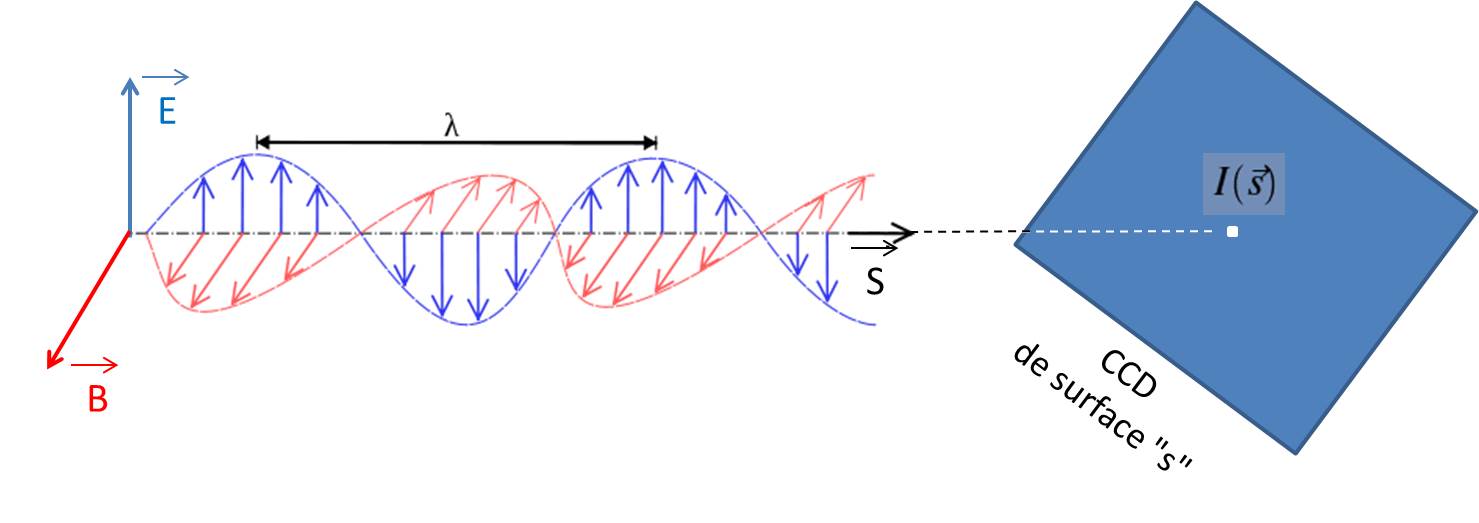}
\caption[Onde lumineuse collectée par une caméra CCD]{Onde lumineuse collectée par une CDD.}\label{Onde_lumineuse}
\end{figure}

En pratique nos cameras CCD (Couple Charge Device en anglais ou Appareil à Transfert de Charges en français) sont sensibles à la moyenne du carré de la norme du champ électrique en un point $\vec{s}$ à la surface de ladite camera (voir Fig.\ref{Onde_lumineuse}).\\

\begin{figure}[h!]
\centering
\includegraphics[height=0.5\hsize,draft=false]{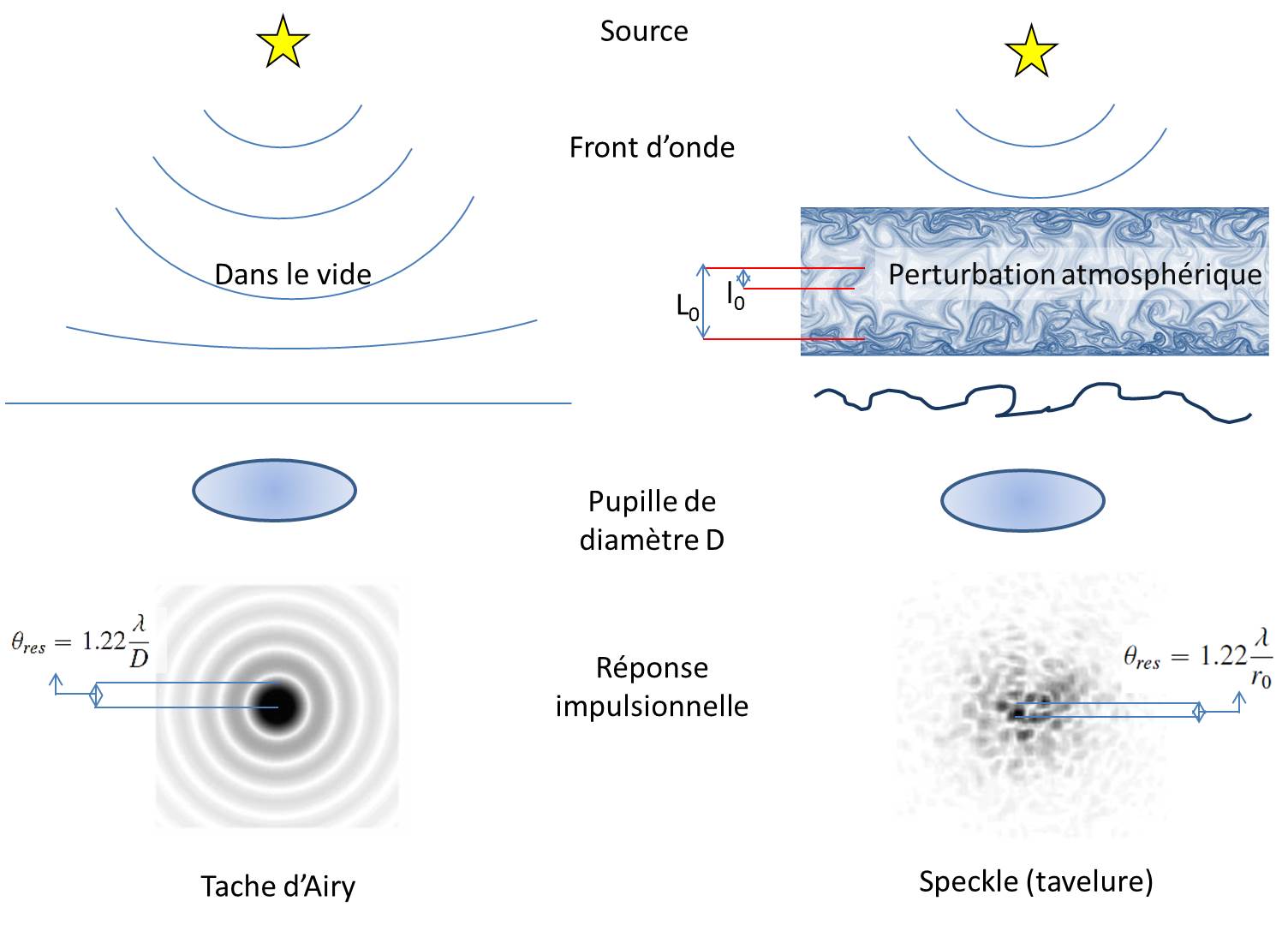}
\caption[L'effet de la perturbation atmosphérique sur le pouvoir de résolution angulaire d'un télescope]{L'effet de la perturbation atmosphérique sur le pouvoir de résolution angulaire d'un télescope. Notre perturbation atmosphérique est représenté ici par une échelle interne $l_0$ et une échelle externe $L_0$ ; Selon le modèle de Kolmogorov l'énergie cinétique associée au mouvement de l'air dans les couches atmosphériques turbulentes peut créer des structures de l'ordre d'une dizaines de mètres (l'échelle externe), énergie qu'elles transfèrent à leurs tours à des structures de tailles de plus en plus réduites jusqu'à atteindre la longueur $l_0$ (l'échelle interne). Enfin ces structures de cellules de turbulence, qui affectent le front d'onde lumineux qui les traverse, sont supposées garder leurs formes pendant une certaine durée. Ce temps est nommé temps de cohérence de la perturbation atmosphérique et est régie par l'équation suivante : $\tau_0 = 0.31\frac{r_0}{V_h}$, où $V_h$ est la vitesse horizontale des cellules de turbulence selon le modèle de Taylor qui décrit la dynamique d'atmosphérique.}\label{Atmos_peturb}
\end{figure}

L'instrument de mesure et la forme et taille de la pupille du télescope impacte significativement la distribution d'intensité de l'objet observé. En effet chaque instrument à ses propres caractéristiques qui peuvent être quantifiés par l'autocorrélation de sa pupille, exprimée par la fonction porte $\Pi$ qui désigne la forme et la taille géométrique de l'ouverture de l'instrument optique ; c'est la réponse impulsionnelle $RI(\vec{s}, D,\lambda)=\Pi\otimes\Pi(\vec{s})$. Ainsi, pour un télescope de diamètre $D$ observant dans une longueur d'onde $\lambda$ dans le vide, sa réponse impulsionnelle sera une tache d'Airy qui est représentée dans la Fig. \ref{Atmos_peturb} et dont l'expression est en puissance 2 de la fonction de Bessel (et dont la formulation est similaire à l'Eq.\eqref{3.27}).

En pratique, l'intensité $I(\vec{s},\lambda)$ mesurée à la longueur d'onde $\lambda$ est assimilée au produit de convolution (noté ici $\otimes$) de la distribution spatiale d'intensité $I_{obj}(\vec{s}, \lambda)$ de l'objet par la réponse impulsionnelle d'un télescope de diamètre $D$, $RI(\vec{s}, D,\lambda)$ ; $ I(\vec{s},\lambda)=I_{obj}(\vec{s}, \lambda)\otimes RI(\vec{s}, D,\lambda)$. Ceci dans le plan image, mais grâce aux propriétés de la transformation de Fourier noté ici par ($\sim$), on peut écrire dans le plan de Fourier: $\tilde{I}(\vec{s},\lambda)= \tilde{I}_{obj}(\vec{s}, \lambda) \tilde{RI}(\vec{s}, D,\lambda)$, ce qui simplifie grandement les calculs.\\

\textbf{Propriétés de la lumière:} En plus des propriétés de la lumière, tel que la réflexion et la réfraction régit par les lois de Snell-Descartes, le principe de Fermat qui énonce la trajectoire à durée minimale,  et la diffraction de Grimaldi et Newton, toutes énoncées au début du 17\up{ième}, et que je ne vais pas énumérer ici en détail, je vais surtout introduire le principe de Huyghens-Fresnel qui nous permettra de mieux adopter le phénomène de l'interférence (aborder dans le chapitre suivant), tel qu'il a été décrit par Huyghens en 1678, où il suppose que chaque partie d'une surface d'une onde agit comme une source secondaire qui émet à son tour une quantité de lumière proportionnelle à celle reçue. Théorie complétée par Fresnel en 1818 avec la notion d'addition cohérente des amplitudes des ondes émises par chaque source secondaire. Le nom de "principe de Huyghens-Fresnel" a vu le jour en 1818, et la première démonstration mathématique a été réalisée par Kirchhoff  en 1882. C'est ce phénomène qui permet d'expliquer l'interférence (ex. expérience de Young).\\

\begin{figure}[h!]
\centering
\includegraphics[height=0.5\hsize,draft=false]{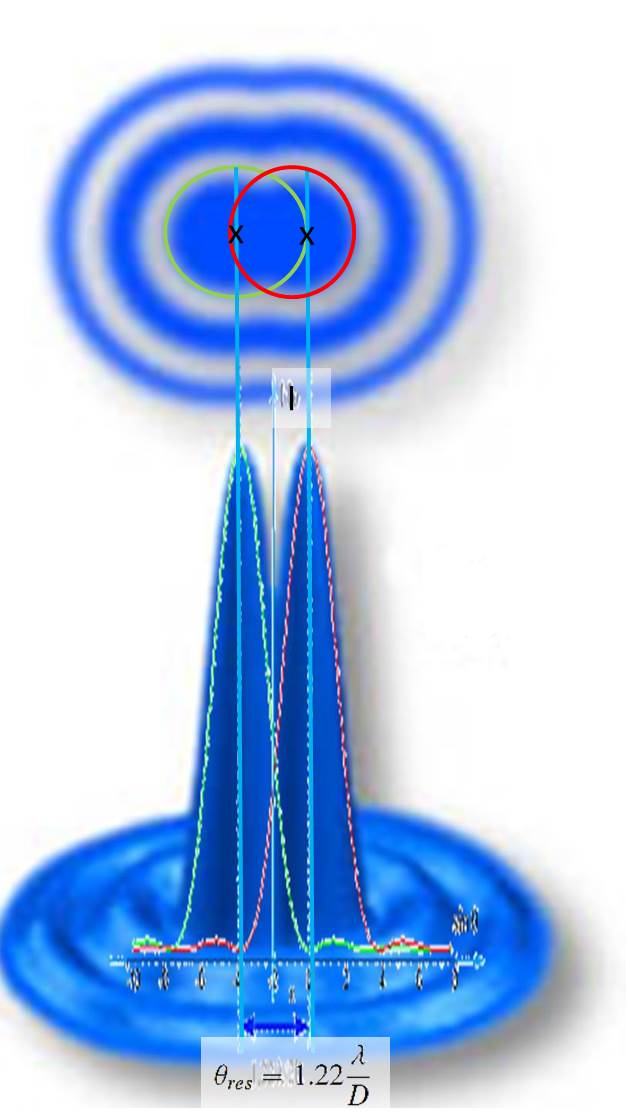}
\caption[Critère de Rayleigh]{Pouvoir de résolution angulaire décrit par le critère de Rayleigh.}\label{Rayleigh_Criterion}
\end{figure}

\textbf{Pouvoir de résolution angulaire \& perturbation atmosphérique:} Ce sont les télescopes qui se chargent de collecter la lumière issue d'un astre donné. Ces derniers sont caractérisés par un pouvoir de résolution angulaire "optimal" (i.e. sans la présence de perturbation atmosphérique -dans l'espace par exemple-) qui est soumis à un critère dit de Rayleigh, qui est déterminé par la séparation angulaire de deux sources considérées ponctuelles, à partir de la superposition de leurs taches d'Airy respectives au plan focal, où le maxima de l'une coïncide avec le minima de la seconde (voir Fig.\ref{Rayleigh_Criterion}). Ainsi, la résolution angulaire $\theta_{res}$ (la séparation angulaire minimale) pour un télescope de diamètre $D$ observant dans une longueur d'onde $\lambda$ est définit comme suit: 

\begin{equation}
\theta_{res}=1.22\frac{\lambda}{D}
\end{equation}

La finesse du pouvoir de résolution est proportionnelle au diamètre et inversement proportionnelle à la longueur d'onde. Néanmoins notre technologie actuelle ne nous permet guère  de construire un miroir d'un télescope monolithique excédant 10 m.\\

Ceci est la définition empirique de la résolution angulaire, mais l'origine physique de cette limite de résolution est due au phénomène diffraction, qui fait en sorte que dans un système optique, un point objet génère toujours une tache au lieu d'un point image (principe de Huygens-Fresnel où chaque point ébranlé par une onde est considéré comme étant une nouvelle source secondaire).\\

En présence de perturbation atmosphérique, ce pouvoir de résolution ne sera que plus diminué. Ce qui est le cas de tout télescope terrestre. En effet, un front d'onde, provenant d'une étoile, et  qui serait localement plat se retrouvera froissé dès son contact avec l'atmosphère terrestre. De ce fait on ne parle plus de tache d'Airy au plan focale de l'instrument mais de "Speckle" (tavelure en français). La perturbation atmosphérique est caractérisée par un bon nombre de paramètres, tels que l'échelle interne, l'échelle externe, le temps de cohérence (voir Fig.\ref{Atmos_peturb}) et le paramètre de Fried ($r_0$) qui peut directement être mesuré sur le Speckle, qui est défini comme la distance angulaire pour laquelle l'écart-type sur la phase du front d'onde atmosphérique n'excède pas 1 radian, il peut être aussi assimilé au diamètre d'un télescope équivalent non astreint par la perturbation atmosphérique (dans l'espace) et qui ne peut être qu'inférieur. De ce fait le pouvoir de résolution angulaire d'un télescope peut être formulé comme suit :

\begin{equation}
\theta_{res}=1.22\frac{\lambda}{r_0}
\end{equation}

Bien que les perturbations de la turbulence atmosphérique peuvent être atténuées par l'optique adaptative, que je ne vais pas aborder dans ce manuscrit, cette technique dite corrective reste néanmoins complexes et couteuses. Avec toutes ces contraintes technologiques, naturelles et instrumentales, l'arrivée de l'interférométrie stellaire a permis d'avoir un gain considérable en résolution spatiale en contrepartie d'une perte de flux, tout en permettant de dépasser la limite de la résolution théorique. Et malgré tous les défis techniques importants qu'il a fallu relever, elle a apporté de conséquentes contributions à l'astrophysique moderne. Le chapitre ci-dessous résume bien l'historique, les bases et l'apport de cette technique.\\

\section{L'Interférométrie}
%\subsection{Un peu d'histoire}
%\subsubsection{La genèse}
\subsection{La genèse}
  C'est en totale contradiction avec la théorie corpusculaire de la lumière d'Isaac Newton (1642-1727) que Thomas Young (1773-1829) proposa une nature ondulatoire de la lumière dans les années 1800, via  sa fameuse expérience des deux fentes dites d'Young réalisée en 1801 (la théorie de l'optique ondulatoire fut établit par Augustin Fresnel (1788-1827) un peu plus tard).
Inversement et dans la même philosophie, Louis de Broglie (1892-1987) prédit, dans  sa théorie de la mécanique ondulatoire, que les particules matérielles devaient elles aussi se comporter comme des ondes dans leur propagation. Ce n'est qu'au 20\up{ième} siècle, et avec l'avènement de l'optique quantique que fut réconciliées et/ou réuniées les deux approches (corpusculaire et ondulatoire de la lumière).
Ainsi, l'expérience de Young fut aussi utilisée dans les années 1970 pour démontrer la nature ondulatoire des électrons et même de corpuscules plus grands. C'est l'expérience de Young qui a permis à la base d'ouvrir la porte à une nouvelle discipline: "L'Interférométrie".\\

Ce n'est qu'après une étude sur les interférences en lumière dispersée en 1845 par A. Foucault (1819-1868) qu'en 1868 A. H. L. Fizeau (1819-1896) suggéra l'utilisation de l'interférométrie pour mesurer le diamètre angulaire des étoiles à partir des franges d'interférence, lors de la remise du Prix Bordin de l'Académie des Sciences (\citet{1868PB....66..932F}; voir sa citation ci-haut, dans la figure introductive du chapitre \ref{chap:spec-interfero}), où il déclara  qu'une possibilité de mesure d'un diamètre stellaire via cette méthode dépendait essentiellement de deux facteurs, conjointement du diamètre angulaire de la source et de la distance entre les deux ouvertures interférométriques.\\

\begin{figure}[h!]
\centering
\includegraphics[height=0.5\hsize,draft=false]{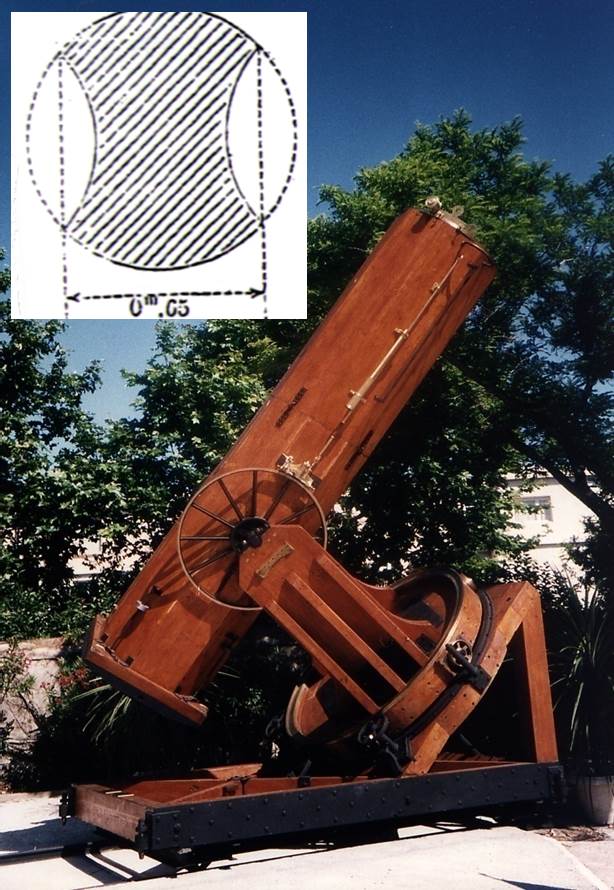}
\caption[L'interféromètre de Fizeau/Stéphan]{Le télescope monolithique de 80 cm transformé en interféromètre via un masque à deux ouvertures espacées de 65 cm, utilisé par Stéphan en 1973, exposé de nos jours à l'OHP.}\label{Stephan_interfero}
\end{figure}

L'expérience est réalisée cinq ans après (1873) par le directeur de l'Observatoire de Marseille Edouard  Stéphan à Marseille via l'utilisation d'un masque pupillaire sur un télescope monolithique de 80 cm (Fig.\ref{Stephan_interfero}), qui arrêta ses conclusions à des diamètres d'étoiles très inférieurs à 0.158 seconde d'arc \citep{1874PB....78..1008S}.\\

\begin{figure}[h!]
\centering
\includegraphics[height=0.5\hsize,draft=false]{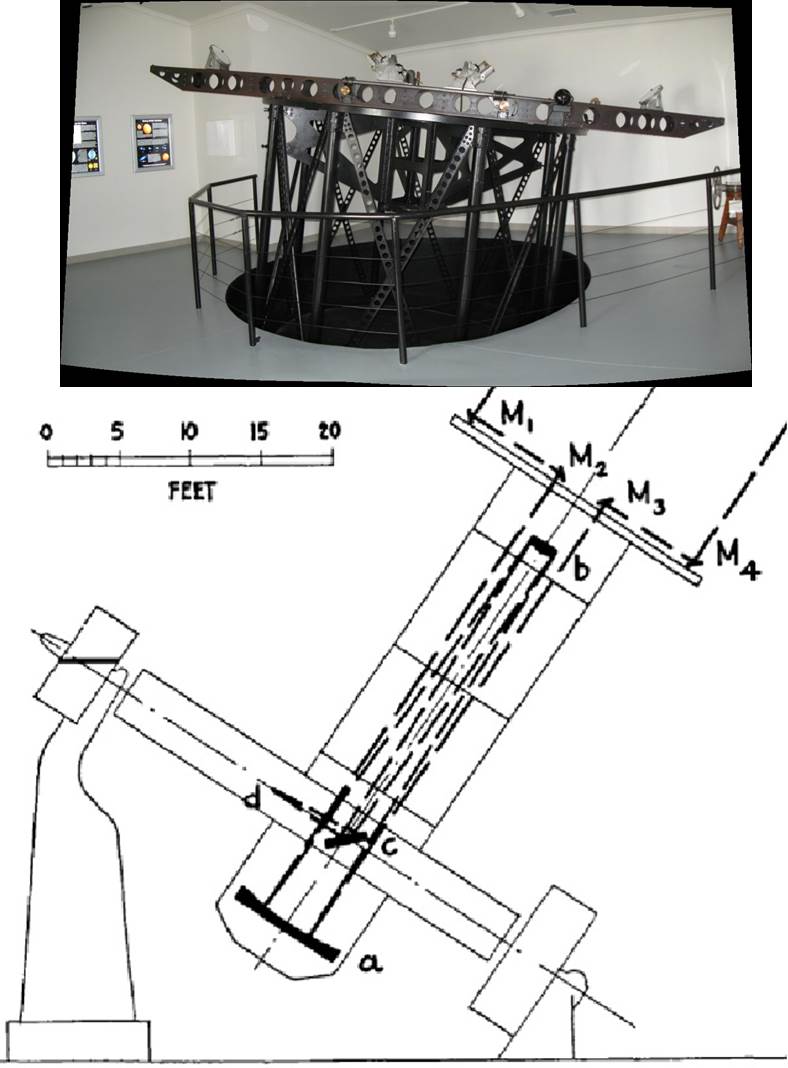}
\includegraphics[height=0.5\hsize,draft=false]{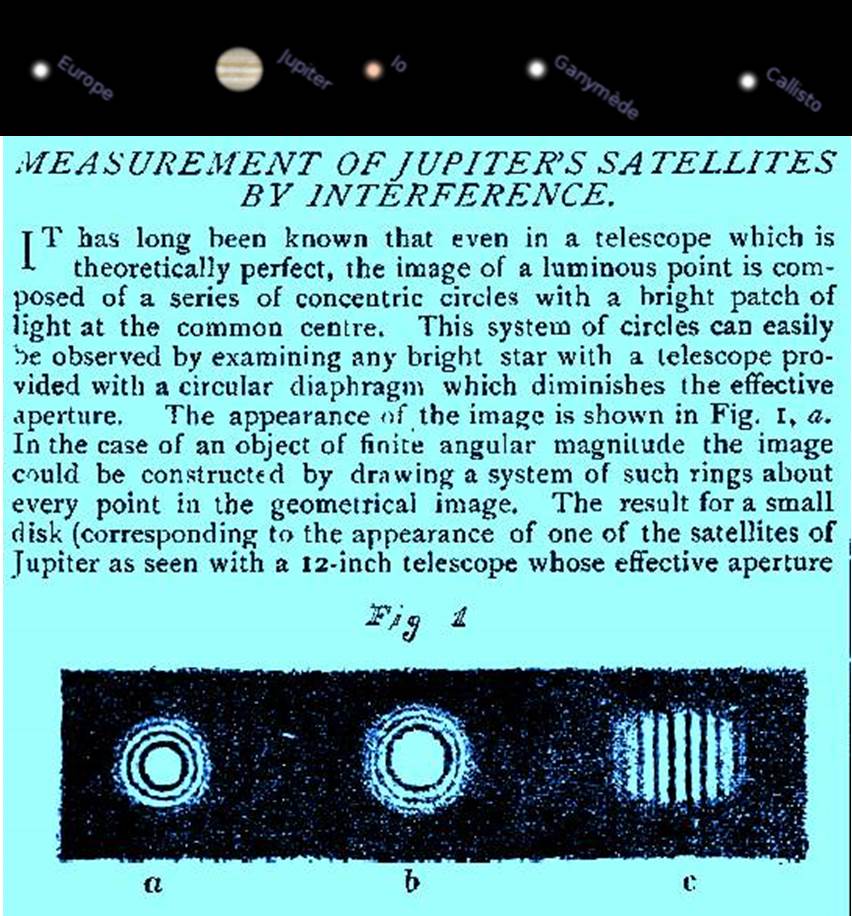}
\caption[Interféromètre de Michelson et ses premiers résultats sur les satellites de Jupiter]{L'interféromètre (schéma tiré de \citep{1921ApJ....53..249M} et photo de http://the-great-silence.blogspot.fr/), (extrait du papier Nature de \citet{1891Natur..45..160M})
.}
\label{Michelson_schema_interfero}
\end{figure}

Plus tard au début du 20\up{ième} siècle A. A. Michelson et F. G. Pease ont réussi à construire et utiliser le premier interféromètre stellaire avec une base plus grande que l'ouverture d'un télescope monolithique (télescope de Hooker du mont Wilson de 2,5 m à l'époque), pour mesurer d'abord (en 1891) les diamètres des 4 satellites galiléens de Jupiter (Io, Europe, Ganymède et Callisto), puis déterminer pour la première fois, en 1921, le diamètre angulaire d'autres étoiles que le Soleil, en particulier celui de Betelgeuse, supergéante rouge de la constellation d'Orion ($\alpha$ $Orionis$, d'environ $ 0.047''\pm0.005''$, en utilisant à l'entrée dudit télescope, quatre miroirs positionnés aux extrémités d'une poutre à séparation maximale d'environ 6 m \citep{1921ApJ....53..249M} (Fig.\ref{Michelson_schema_interfero}). Entre temps, Karl Schwarzschild, qui eut connaissance de la méthode de Michelson et de ses mesures des rayons des satellites joviens (en 1891), s'en inspira en 1896 pour mesurer la séparation de 13 étoiles doubles \citep{1896AN....139..353S}.\\

La méthode de mesure de rayon apparent d'un astre considéré comme étant un disque uniforme par Michelson \& Pease était essentiellement visuelle, où via une relation assez triviale ils liaient la distance de séparation des miroirs relative à la disparition des franges d'interférence (la résolution de l'objet) avec les rayons angulaires des objets observés (plus amples explications dans la Sec.\ref{Source_disk}). Pease continua le travail en mesurant le diamètre de quelques étoiles géantes proches tel que $\alpha$ Scorpii \citep{1921PASP...33..171P, 1921PASP...33..204P}.\\

En 1920, J. A. Anderson décrivit une méthode qui permit de mesurer le mouvement orbital apparent des binaires spectroscopiques qu'il appliqua à Capella. Cette méthode fut utilisée plus tard par \citet{1922CMWCI.240....1M}, \citet{1925PNAS...11..356P, 1927PASP...39..313P} sur $\kappa$ Ursae Majoris, v2 Bootis et Mizar.\\

Les observations étant toujours effectuées dans le visible, la différence de chemin optique entre les deux faisceaux ne devait guère excéder quelques micromètres (la longueur de cohérence dans ce domaine), ce qui était un exploit à l'époque en raison d'une grande instabilité de la structure interférométrique; une poutre de 6 m, l'alignement des deux faisceaux lumineux qui devait être réfléchies par 3 miroirs chacun.  Malgré toutes ces difficultés techniques et le manque de moyens d'asservissement optique, Pease entreprit, en 1931, la réalisation d'un second interféromètre plus volumineux, avec une base interférométrique de plus de 15 m (50 pieds). Projet qu'il n'a hélas pas pu mener à terme à cause d'une forte instabilité et des vibrations de l'imposante structure d'une part et du début de la seconde guerre mondiale d'autre part.\\

\begin{figure}[h!]
\centering
\includegraphics[height=0.5\hsize,draft=false]{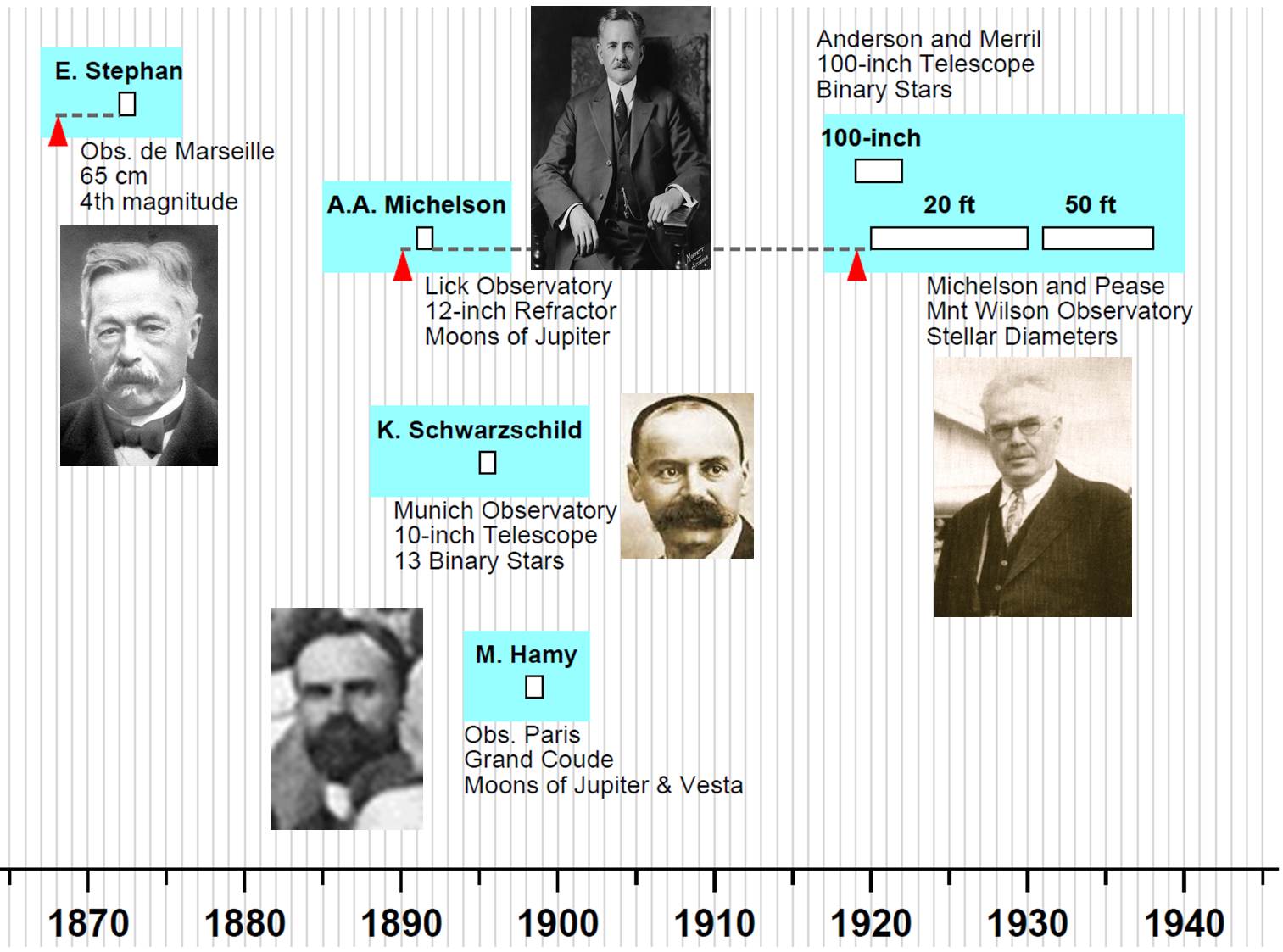}
\caption[L'interférométrie stellaire de 1868 à 1940.]{Les débuts de l'interférométrie stellaire de 1868 à 1940 (inspiré du diagramme de \citet{1999AAS...195.2301L}).}\label{premises_interfero}
\end{figure}

\begin{table*}[htbp]
\centering
\caption[Principaux résultats au début de l'interférométrie stellaire]{Principaux résultats au début de l'interférométrie stellaire (inspiré de \citet{1999AAS...195.2301L}).}\label{Lawsan_1999_1}
\centering
\resizebox{\textwidth}{!}{\begin{tabular}{ccc}
\hline\hline
Année & Événements & Auteurs \& référence\\
\hline
1868 & Suggestion de l'interférométrie stellaire & H. Fizeau, CR Acad. Sci. 66, 932 (1868)\\
1872-73 & Diamètres stellaires $<< 0,158$ seconde d'arc & E. Stéphan, CR Acad. Sci. 78, 1008 (1874)\\
1890 & Fondement de la théorie mathématique de l'interférométrie stellaire & AA Michelson, Phil. Mag. 30, 1 (1890)\\
1891 & Mesure des satellites de Jupiter & A.A. Michelson, Nature 45, 160 (1891)\\
1896 & Mesures d'étoiles binaires  & K. Schwarzschild, Astron. Nachr. 139 3335 (1896)\\
1920 & Mesure de l'orbite de Capella & J.A. Anderson, Astrophys. J. 51, 263 (1920)\\
1921-31  & Mesurée du premier diamètre stellaire  & A.A. Michelson, F.G. Pease, Astrophys. J. 53, 249 (1921)\\
1931-38 & Interféromètre de 50 pieds & F.G. Pease, Erg. Exakt. Natur. 10, 84 (1931)\\
\hline\hline
\end{tabular}}
\end{table*}

L'instrumentation de l'interférométrie stellaire consistait, ainsi, jusqu'aux années quarante, dans l'utilisation d'un télescope monolithique agrémenté d'artifices optiques (masque, poutre à miroirs,...etc.). Ce n'est qu'au début des années cinquante, et avec l'avènement de la radioastronomie  \citep{1947Obs....67...15R}, que Ryle suggéra en 1952 la combinaison cohérente de deux antennes radio \citep{1952RSPSA.211..351R}, un autre fut réalisé la même année par \citet{1952MNRAS.112..497S}. Depuis la radioastronomie interférométrique a connue un essor rapide et une perpétuelle évolution jusqu'à la réalisation de radio-interféromètres capables d'atteindre une résolution de quelques milli-arcsecondes dans le domaine des fréquences radio, tel que  le VLA (Very Large Array, \citet{1967Sci...158...75H, 2006IAUSS...1E..18B} et le VLBA (Very Long Baselines Array, \citet{1975ApJ...201..249C}). Pour le domaine des ondes à haute fréquences (visible, IR, ...etc.), la technique qui consiste à combiner la lumière issue de deux télescopes (ou plus) n'a pu être réalisée qu'à partir des années 60. Cette technique communément connue sous le nom de : "Interférométrie Optique à Longue Base" (OLBI : "Optical Long Baseline Interferometry" en anglais) est expliquée en détail dans le sous chapitre suivant.\\

\subsection{Les équations de base en interférométrie}
Ce sous-chapitre est inspiré à la fois des cours de Jean Surdej (VLTI School 2010 et 2013), des lectures notes of Michelson School de Peter \citet{1999AAS...195.2301L} et des cours d'Eric Aristidi en optique ondulatoire (http://webs.unice.fr/site/aristidi/optique/).\\

\begin{figure}[h!]
\centering
\includegraphics[height=0.5\hsize,draft=false]{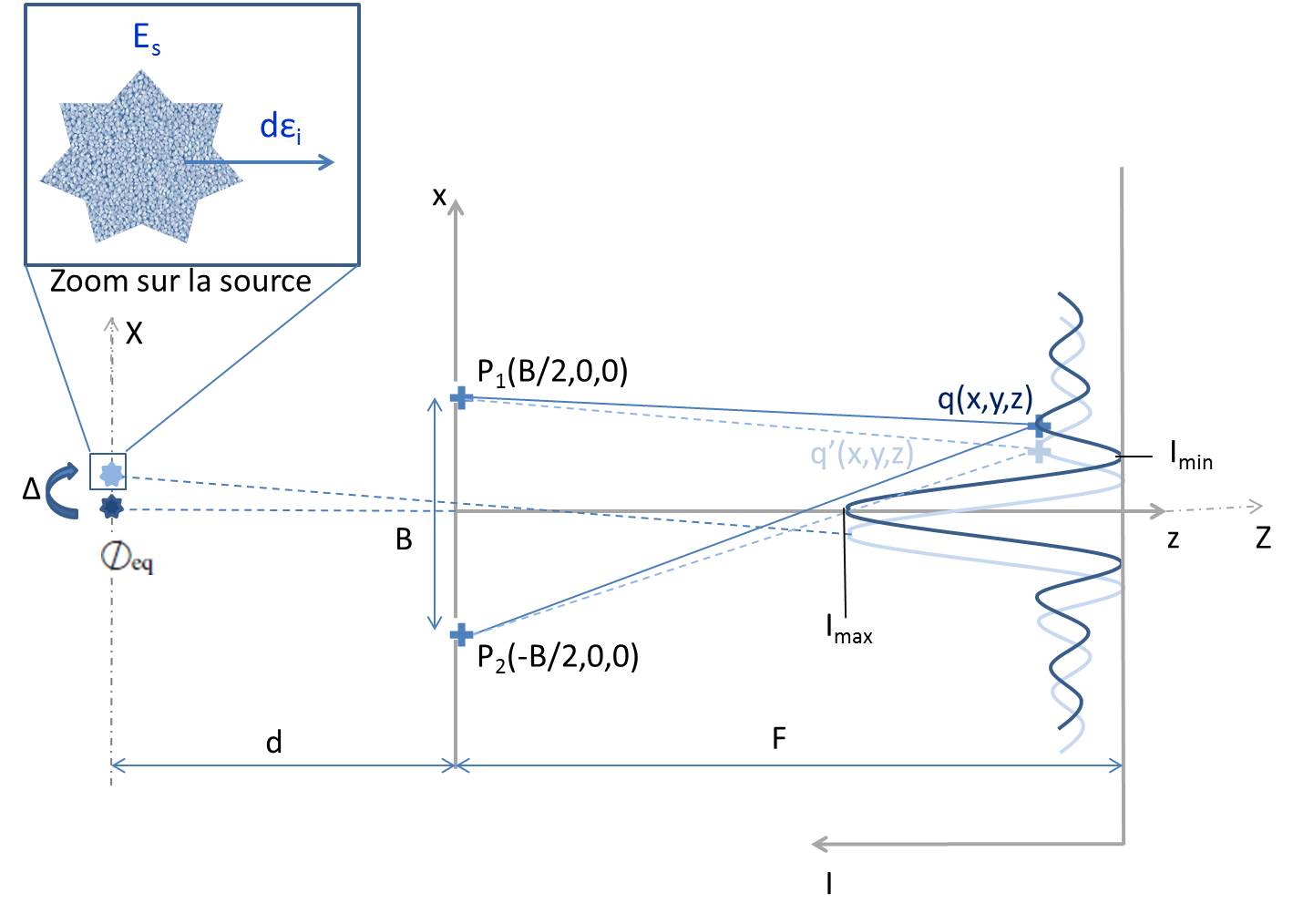}
\caption[Schéma illustrant la formation de franges d'interférences dans l'expérience des fentes de Young.]{Schéma illustrant la formation de franges d'interférences avec l'expérience des fentes de Young.}\label{Equa_Interfero}
\end{figure}

Afin d'illustrer les équations et principe de base de l'interférométrie de manière succincte et efficace, il n'y a rien de mieux que d'utiliser l'expérience des fentes de Young (Fig. \ref{Equa_Interfero}), qui consiste à faire interférer sur un écran deux faisceaux lumineux supposés monochromatiques et provenant d'une même source (non résolue) à travers deux fentes $p1$ \& $p2$ (dites fentes de Young) espacées d'une distance $B$. La distance entre le plan des fentes et l'écran est noté ici par $F$.\\

\textbf{La différence de marche:} La différence des parcours optique entre chaque fente et l'écran, qui peut être assimilée ici à la différence entre les distances $ \left|p_1q\right|$ \& $\left|p_2q\right|$ est appelée différence de marche (ddm) (ou bien OPD pour Optical Path Difference en anglais). Elle est proportionnelle au déphasage $\phi$ dans le cas d'une onde monochromatique, fait apparaitre tantôt une frange brillante (dans le cas où les deux ondes lumineuses sont en phase $\Phi=0$, ce qu'on appelle aussi une interférence constructive), et tantôt une frange sombre (dans le cas où les deux ondes lumineuses sont en opposition de phase $\Phi=\pm\pi$, interférence destructive). Ainsi on peut écrire :

\begin{equation}
\Phi= \left|p_1q\right|- \left|p_2q\right|=n\lambda
\end{equation}

Où $\lambda$ est la longueur d'onde et $n$ un nombre entier. Cette équation peut être réécrite, dans le cas où au point $q$; $\left|x\right|, \left|y\right|<<z$, comme suit :

\begin{equation}
\Phi=z \left[\frac{1 + ((x + \frac{B}{2})^2 + y2) }{2z^2}\right] -z\left[\frac{1 + ((x - \frac{B}{2})^2 + y2) }{2z^2}\right] = n\lambda
\label{3.8}
\end{equation}

La séparation angulaire entre deux franges brillantes successives (ou bien l'interfranges/résolution $\delta x$ mesurée sur l'écran à une distance $F$ de l'objet diffractant) est donc régit par l'équation suivante :

\begin{equation}
\Phi=\frac{x}{z} = \frac{\lambda}{B} \Rightarrow \delta x = \frac{F\lambda}{B}
\label{3.9}
\end{equation}

Par exemple, $\Phi$ serait égal à $113"$ (la limite de notre résolution visuelle) pour une longueur d'onde $\lambda = 5500$ $\text{\AA{}}$ et une séparation des fentes $B = 1$ $mm$.\\

\textbf{La visibilité:} Le brouillage des franges qu'on observe sur l'écran et qui se compose donc de franges brillantes à intensité maximale $ I_{max}$ et de franges sombres à intensité minimale $I_{min}$, peut être mesuré de manière objective et quantitative via la "visibilité" $V$ (appelé aussi le contraste), qui est déterminée par l'expression suivante:

\begin{equation}
V = \frac{ I_{max}-I_{min}} {I_{max}+I_{min}}
\label{3.10}
\end{equation}

Pour un objet non résolu, l'intensité en tout point est maximale ; $I=I_{max}$ \& $I_{min}=0$ et la visibilité prend alors la valeur 1 ($V=1$). Par contre pour un objet parfaitement résolu, l'intensité en tout point est minimale ; $I=I_{max}=I_{min}$ et la visibilité est nulle ($V=0$).\\

Le brouillage de frange ne peut exister que si le phénomène d'interférence est réalisé. Et pour ce, il faut qu'il y ait cohérence temporelle et spatiale entre les faisceaux lumineux combinés. Ces cohérences-ci sont explicitées ci-dessous :\\

\textbf{La cohérence temporelle:} C'est la largeur de bande spectrale d'une source qui détermine s'il y a cohérence temporelle ou pas. Dans la pratique, il n'existe pas d'onde monochromatique (ou mono-fréquence) au sens propre et pour qui la  longueur et le temps de cohérence seraient infinis. Toutes les ondes lumineuses dans la nature sont polychromatiques, i.e. centrées autour d'une fréquence centrale $\nu_0$ et couvrant une bande spectrale de largeur $\delta\nu$. De ce fait le temps d'oscillation associé est de l'ordre de $\frac{1}{\delta\nu}$. Cette durée est toujours très courte devant le plus petit temps d'intégration qu'on sait réaliser dans le domaine du visible et de l'infrarouge mais pas de le domaine radio. Ainsi le temps de cohérence $\tau_c$ qui définit la durée utile du train d'onde s'écrit comme suit :

\begin{equation}
\tau_c=\frac{1}{\delta\nu}
\end{equation}

La longueur de cohérence temporelle associée est donc : $L_c = \tau_c v=\frac{\lambda^2}{\delta\lambda}$, où $v$ est la vitesse de propagation de l'onde dans un milieu donné ($v=c$, la vitesse de la lumière dans le vide). $L_c$ détermine la longueur du train de l'onde, qui est fini pour une onde polychromatique et infini pour une onde monochromatique. Ainsi, les franges disparaissent en cas d'incohérence temporelle, quand la $ddm > L_c$, i.e. que le retard temporelle entre les deux fronts d'onde $\tau=ddm/c$ est grand devant $\tau_c$. Le brouillage des franges n'est visible qu'en cas de cohérence temporelle, quand $\tau \leq \tau_c $ (ou $ddm \leq L_c$).\\

\textbf{La cohérence spatiale:} Du moment où sur une largeur donnée d'un front d'onde, tous les points gardent la capacité d'interférer entre eux, on parle alors de cohérence spatiale, de largeur de cohérence (notée ici $\Lambda_c$) et on dit que l'onde (de largeur $\Lambda_c$) est cohérente. De ce fait et pour qu'il y ait interférence dans l'expérience de Young, la distance de séparation des deux fentes doit impérativement être inferieur ou égale à la distance maximale entre deux points d'un front d'onde pour lesquels les battements restent cohérents (i.e. $B \leq \Lambda_c$), où :

\begin{equation}
\Lambda_c=\frac{\lambda}{\diameter}
\end{equation}

Avec $\lambda$ la longueur d'onde de la source et $\diameter$ son diamètre angulaire (ici la source étant supposée parfaitement sphérique, le diamètre angulaire est uniformément égal à $\diameter_{eq}$ ; le diamètre angulaire équatoriale). Pour le Soleil par exemple, dont le diamètre angulaire équatoriale $\diameter_{eq}=0.5^\circ$, il nous serait impossible d'obtenir des franges d'interférence sans faire passer sa lumière à travers une première fente d'une largeur minimale de $57\mu m$ (la largeur de cohérence solaire), sans compter le polychromatisme de la lumière blanche solaire qui peut causer de fortes incohérences temporelles (tel que vu précédemment), contrairement au Laser qui offre une bonne cohérence spatiotemporelle et qui rend l'expérience de Young aisément réalisable. Le fait de jouer avec la distance de séparation des fentes ($B$), en dessous de la largeur de cohérence se mesure directement sur le brouillage des franges ; plus $B$ est grand ($B>\Lambda_c$), meilleure est la résolution spatiale, et plus faible est le contraste. Enfin, en cas d'incohérence spatiale les intensités lumineuses s'additionnent au lieu d'interférer.\\

\textbf{Degré complexe de cohérence mutuelle:} Le calcul de l'intensité lumineuse au point $q$ résultant des ondes aux sources $p_1$ \& $p_2$  (voir la fig.\ref {Equa_Interfero}), nous permet d'écrire : 

\begin{gather}
 I(\vec{q})= \left\langle \left\| \vec{E}(\vec{q},t) \right\|^2 \right\rangle_t\\
           =\left\langle \left\| \vec{E}(\vec{p_1},t) +\vec{E}(\vec{p_2},t-\tau) \right\|^2 \right\rangle_t,
\label{3.14}
\end{gather}

où $\tau=t_{p_2q}-t_{p_1q}$ est la différence temporelle des deux faisceaux lumineux $p_1q$ et $p_2q$. En supposant que l'intensité lumineuse aux deux fentes  est strictement la même -les deux fentes ont la même ouverture- ($I_0=\left\langle \left\| \vec{E}(\vec{p_1},t) \right\|^2 \right\rangle_t=\left\langle \left\| \vec{E}(\vec{p_2},t-\tau) \right\|^2 \right\rangle_t$), l'Eq.\eqref{3.14} devient:

\begin{gather}
 I(\vec{q})=I_0+I_0+\left\langle \vec{E}(\vec{p_1},t)\vec{E}^*(\vec{p_2},t-\tau)\right\rangle_t+\left\langle \vec{E}^*(\vec{p_1},t)\vec{E}(\vec{p_2},t-\tau) \right\rangle_t\\
= 2I_0+2I_0\Re(\gamma_{12}(\tau)),
\label{3.16}
\end{gather}

avec $\gamma_{12}(\tau)= \left\langle \vec{E}^*(\vec{p_1},t)\vec{E}(\vec{p_2},t-\tau) \right\rangle_t/I_0$ qu'on désigne sous le nom de \textbf{degré complexe de cohérence mutuelle}. En utilisant l'expression de l'Eq.\eqref{3.1} et en supposant que la différence entre les temps de parcours des deux faisceaux est inférieure à la période de battement (i.e. $\tau<<\frac{1}{\delta\nu}$),  $\gamma_{12}(\tau)$ devient:

\begin{equation}
\gamma_{12}(\tau)= \left|\gamma_{12}(0) \right| e^{i\Phi_{12}-i\omega\tau}/I_0,
\end{equation}

$\Phi_{12}$ étant le déphasage entre les deux faisceaux lumineux émergeants. De ce fait l'Eq.\eqref{3.16}, avec l'utilisation de la loi d'Euler,  peut être réécrite comme suit:

\begin{equation}
I(\vec{q})= 2I_0 \left[1+\left|\gamma_{12}(0)\right|\cos(i\Phi_{12}-i\omega\tau) \right]
\end{equation}

Cette équation qui décrit l'intensité est connue sous le nom d'interférogramme, et représente l'intensité totale $2I_0=I(\vec{p_1})+I(\vec{p_2})$ fluctuée par une fonction cosinusoïdale. Dans le vide, l'interférogramme est donc constitué d'une tache d'Airy frangée, alternativement en claires et sombres, que le degré de cohérence mutuelle $\left|\gamma_{12}(0)\right|$ décrit parfaitement. En effet, on remarque clairement que ce dernier n'est autre que le contraste ($\left|\gamma_{12}(0)\right|=V$, voir Eq.\eqref{3.10}). En effet, sur une frange brillante centrale; $I(\vec{q)}=I_{max}=I_{min}$ quand $\Phi_{12}=0$ et $\left|\gamma_{12}(0)\right|=0$ et sur une frange sombre minimale; $I(\vec{q)}=I_{min}=0$ quand $\Phi_{12}=\pm\pi$ et $\left|\gamma_{12}(0)\right|=1$.\\

\textbf{Théorème de Van Cittert-Zernike:}
Dans le cas d'une source non ponctuelle, la surface de la source $E_s$ serait donc constituée d'éléments $d\epsilon_i$ quasi-monochromatiques et incohérents, où $E_s=\sum_{i=1}^{n_{\epsilon}} d\epsilon_i$. Le centre de la source $E$ est situé à une distance $d$ du milieu de l'objet diffractant (les fentes de Young). Notons $\eta$, $\xi$ les coordonnées d'un point de la source à partir d'un autre repère dont l'origine est confondue avec le centre de la source et  $X_{1,2}$,  $Y_{1,2}$ les coordonnées des deux fentes $P_1$ et $P_2$.\\

A partir de l'Eq.\eqref{3.16}, on avait déduit que $\gamma_{12}(\tau)= \left\langle \vec{E}^*(\vec{p_1},t)\vec{E}(\vec{p_2},t-\tau) \right\rangle_t/I_0$. Du point de vue de la source, les chemins parcourus d'un élément source à l'une des fentes $P_{1,2}$ sont notés $r_{i1,2}=(( X_{1,2}^2-\eta^2)^2+( Y_{1,2}^2-\xi^2)^2+d^2)^{\frac{1}{2}}$ (avec $\eta$ et $\xi$ étant des quantités angulaires). Pour une distance $d>> X_{1,2}$,  $Y_{1,2}$ , $\tau \rightarrow 0$ et on peut utiliser la même approximation utilisée pour l'Eq. \eqref{3.8}, ainsi on peut écrire :

\begin{gather}
\gamma_{12}(0)= \left\langle \vec{E}^*(\vec{p_1},t)\vec{E}(\vec{p_2},t) \right\rangle_t/I_0\\
\label{3.19}
\left|r_{i2}-r_{i1}\right|=\frac{(X^2_2+Y^2_2)-(X^2_1+Y^2_1)}{2d}-\frac{(X_2-X_1)\eta+(Y_2-Y_1) \xi }{d}\\
\label{3.20}
r_{i2}r_{i1}=d^2\\
\label{3.21}
\end{gather}

Le champ électrique produit par chaque élément de surface de la source $d\epsilon_i$ aux points $p_{1,2}$ est $\vec{E}_{i1,2}( \vec{r}_{i1,2},t)= \vec{E}_{0i1,2}( \vec{r}_{i1,2})e^{-i\omega (t-\frac{r_{i1,2}}{c})}$. La condition de cohérence temporelle étant ici assurée, i.e. $\left|r_{i2}-r_{i1}\right|\leq l_c$, l'Eq.\eqref{3.19} devient :

\begin{equation}
\gamma_{12}(0)= \iint_{E_s} \frac{I(E_s)}{r_{i2}r_{i1}} e^{-i2\pi(r_2-r_1)/\lambda}d\epsilon/I_0
\end{equation}

Des Eqs.\eqref{3.19}, \eqref{3.20} \& \eqref{3.21} et en posant $u=\frac{(X_2-X_1)}{\lambda d}$, $v=\frac{(Y_2-Y_1)}{\lambda d}$, les fréquences spatiales et $\Phi_{p_1,p_2}=\frac{2\pi}{\lambda}\frac{(X^2_2+Y^2_2)-(X^2_1+Y^2_1)}{2d}$ le déphasage, on trouve :

\begin{equation}
\gamma_{12}(0,u,v)=e^{-i\Phi_{p_1,p_2}} \frac{\iint I(\eta,\xi)e^{-i2\pi(u\eta+v\xi)} d\eta d\xi}{\iint I(\eta,\xi) d\eta d\xi}
\label{3.24}
\end{equation}

Il est ici tout à fait évident que pour une source très éloignée de l'objet diffractant, les chemins parcourus depuis la source peuvent être considérés comme étant les mêmes rendant le déphasage $\Phi_{p_1,p_2}=0$. De plus $\iint I(\eta,\xi)e^{-i2\pi(u\eta+v\xi)} d\eta d\xi$ n'est autre que la transformée de Fourier de l'intensité, que on va noter $\tilde{I(u,v)}$, sans oublier que $\gamma_{12}(0,u,v)$ n'est autre que la visibilité $V(0,u,v)$. Ainsi on peut enfin écrire :

\begin{equation}
V(0,u,v)=\frac{\tilde{I}(u,v)}{\tilde{I}(0,0)}
\label{3.25}
\end{equation}

On retrouve ainsi la formulation mathématique du \textbf{théorème de Van Cittert-Zernike}, qui stipule que \textbf{la visibilité complexe (ou degré de cohérence mutuelle) à comme valeur la transformée de Fourier normalisée de la distribution d'intensité de la source lumineuse.}\\

Ce théorème qui est issu à l'origine du travail du Néerlandais Pieter Hendrik van Cittert en 1934 sur la cohérence des rayonnements provenant des sources incohérentes lointaines, et qui a été repris, pour une reformulation plus simple, par son concitoyen Frits Zernike en 1938 , est la démonstration mathématique sans équivoque de l'intuition qu'avait eu Armand Hippolyte Louis Fizeau 70 ans plutôt (voir sa citation dans la figure introductive du chapitre \ref{chap:spec-interfero}).\\

\textbf{Rayon angulaire d'une source:}
Dans l'hypothèse d'une source considérée comme étant un disque uniforme de diamètre angulaire $\diameter$, on peut adopter les coordonnées polaires suivantes $\eta=\rho\cos\diameter_i$ et $\xi=\rho\sin\diameter_i$, où $\rho$ est la mesure angulaire sur la sphère céleste depuis le centre de la source ($0 \leq \rho \leq \diameter/2$). Dans ce cas la luminosité mesurée au niveau de l'objet diffractant ($p_1$ \& $p_2$) sera en fonction de l'intensité spécifique $I_0$ et du diamètre angulaire de la source $\diameter$ ; $L\propto I_0\pi\diameter^2/4$, sans oublier que les fréquences spatiales $u$ \& $v$ seront mesurées en fonction d'un angle de projection de la base interférométrique $PA$ qui peut être pris quelconque où bien considéré nul à cause de la symétrie de la source, avec $(u,v)=(f_s\cos PA, f_s\sin PA)$, où $f_s=\sqrt{u^2+v^2}=\frac{B}{\lambda}$. En prenant en compte tous ces éléments et à l'aide des propriétés de fonction de Bessel à l'ordre 0 \& 1; $\int_0^{2\pi} e^{ix\cos\diameter_i}d\diameter_i = 2\pi J_0(x)$ \& $\int_0^x x'J_0(x') dx'= xJ_1(x)$, l'Eq.\eqref{3.24} devient : 

\begin{equation}
V(B,\lambda,\diameter)=e^{-i\Phi_B} \frac{2J_1(\pi\diameter B/\lambda)}{\pi\diameter B/\lambda}
\label{3.26}
\end{equation}

Ceci est la visibilité complexe normalisée d'une source à symétrie de disque uniforme de diamètre $\mathbf{\diameter}$, mesurée avec une base interférométrique $B$ (ici $\left|p_1p_2\right|=B$) à une longueur d'onde $\lambda$. En pratique et pour un meilleur rapport signal à bruit (à cause de la turbulence atmosphérique) on utilise généralement le module de visibilité normalisé $V^2(B,\lambda,\diameter)$ qui vaut :

\begin{equation}
V^2(B,\lambda,\diameter)= \left(\frac{2J_1(\pi\diameter B/\lambda)}{\pi\diameter B/\lambda}\right)^2
\label{3.27}
\end{equation}

\begin{figure}[h!]
\centering
\includegraphics[height=0.5\hsize,draft=false]{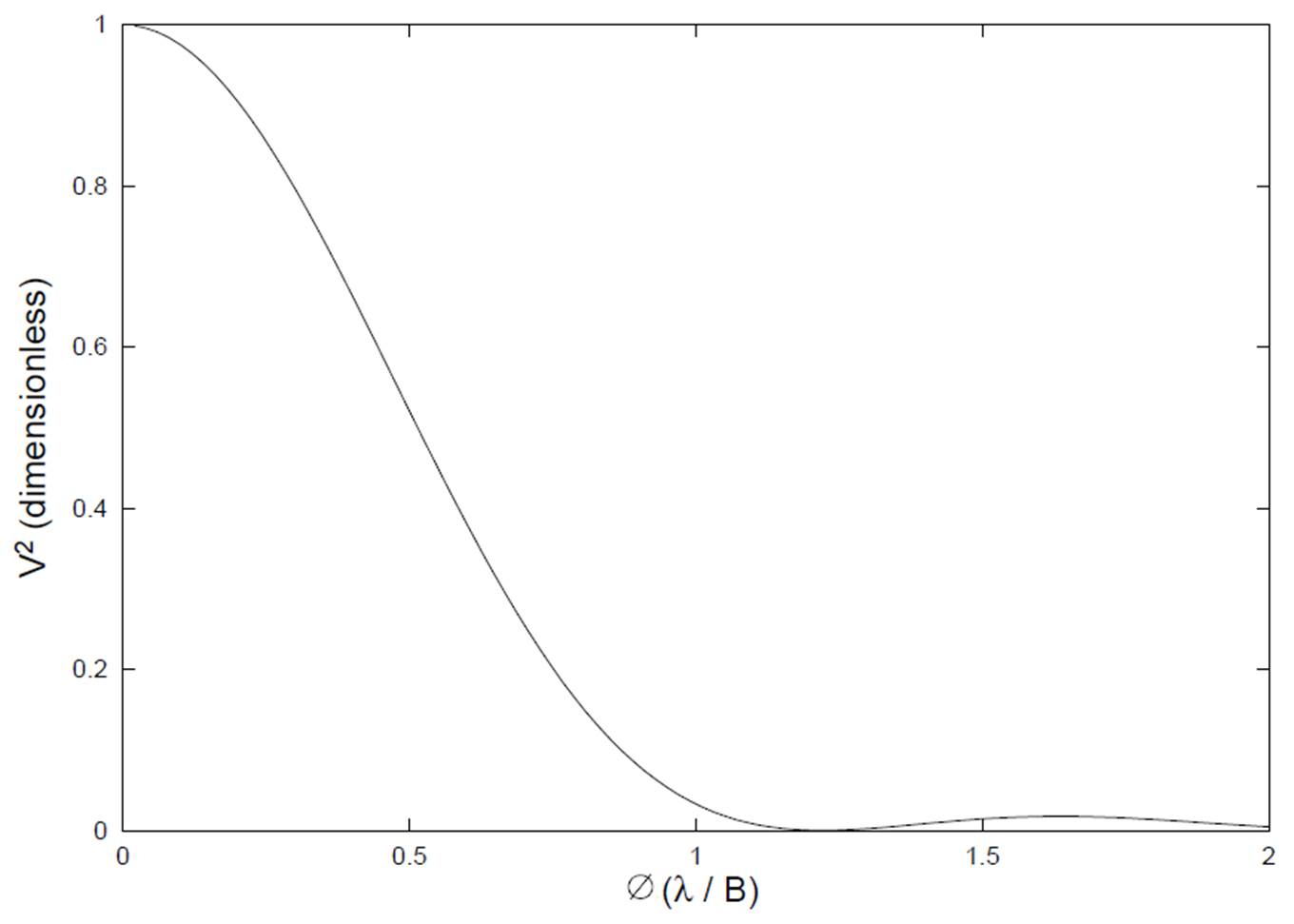}
\caption[Module de visibilité pour un disque de brillance uniforme.]{Module de visibilité normalisé en fonction de la résolution $\lambda/B$ pour un disque uniforme de diamètre $\diameter$ \citep{1999AAS...195.2301L}.}\label{Source_disk}
\end{figure}

Le tracé du module de visibilité $V^2$ en fonction du diamètre d'un disque uniforme $\diameter$ est représenté en fonction de la résolution (ou l'interfrange) $\lambda/ B$) dans la Fig.\ref{Source_disk} ci-dessus. On remarque bien que  pour $V^2=1$ la source n'est pas résolue, alors qu'elle l'est pour $V^2=0$. C'est à ce point-là précis où les franges disparaissent qu'on déduit le diamètre de la source $\diameter\approx 1.22 \lambda/B_0$ ($B_0$ la base qui annule les  franges d'interférences). C'est avec cette formule que Michelson \& Pease ont pu déterminer les diamètres angulaires des objets qu'ils avaient observés à partir de 1891, dans le visible. Sous l'hypothèse que leurs objets étaient des disques uniformes de diamètres finis, il faisait varier la séparation des miroirs (la distance $B$) jusqu'à ce qu'ils résolvent leurs objets et que les franges disparaissent, au-delà (de $B_{max}$ à $B_{min}$), les franges réapparaissent mais avec une moindre amplitude.\\

\textbf{Séparation angulaire de deux sources:}
La détermination de la séparation angulaire (notée ici $\Delta$) de deux sources non résolues est aussi possible en interférométrie avec les mêmes équations de bases et procédé du schéma (Fig. \ref{Equa_Interfero}). Tout d'abord une source non résolue peut être considérée comme étant ponctuelle, ce qui nous permet de la formuler via la distribution de Dirac $\delta$ (communément connue sous le nom de fonction de Dirac). En effet ceci peut être rapidement démontré via l'Eq.\eqref{3.24} qu'on peut réécrire comme suit : $\gamma_{12}(0,u,v)=\iint I'(\eta,\xi)e^{-i2\pi(u\eta+v\xi)} d\eta d\xi$, où $I'(\eta,\xi)$ est la distribution d'intensité normalisée à la source. La considération de celle-ci comme étant ponctuelle (i.e. $I'(\eta,\xi)= \delta(\eta,\xi)$) nous conduit à une visibilité complexe $\gamma_{12}(0,u,v)=1$, ce qui démontre effectivement qu'une source ponctuelle est une source non résolue.\\

En utilisant cette information, considérant maintenant deux sources non résolues au coordonnées $E_{s1}(\eta-\frac{\Delta}{2}, \xi=0)$ \& $E_{s2}(\eta+\frac{\Delta}{2}, \xi=0)$ à une distance $d$ de l'objet diffractant (ici la fréquence spatiale $v=0 \rightarrow f_s=u$), ce qui nous permet de traiter le degré de cohérence mutuelle complexe qu'en une seule dimension : $\gamma_{12}(0,u)=\int I'(\eta)e^{-i2\pi u\eta} d\eta$, avec $I'=\frac{1}{2}\left(\delta(\eta-\frac{\Delta}{2}) + \delta(\eta+\frac{\Delta}{2})  \right)$. De ce fait, et en utilisant la formule d'Euler, on a : $\gamma_{12}(0,u)=\frac{1}{2} \left( e^{-i\pi u\Delta}+e^{i\pi u\Delta} \right)=\cos(\pi u\Delta)$. Ce système optique (à deux sources ponctuelles) est résolu (au premier lobe de visibilité) pour $\gamma_{12}(0,u)=0 \Rightarrow \pi u\Delta = \frac{\pi}{2}$. Ainsi la séparation angulaire de deux sources non résolues est :
  
\begin{equation}
\Delta\approx \frac{\lambda }{2B_0}
\label{3.28}
\end{equation}

Sur notre écran on verra l'équivalent de deux interférogrammes (de source unique) qui se chevauche l'un l'autre, avec un interfrange  $\delta x'=\frac{\delta x}{2}$ (voir Eq. \eqref{3.9}), i.e. $\delta x'=\frac{F\lambda}{2B}$. Ainsi la séparation des maximas de chaque système de franges sur l'écran (à distance $F$ de l'objet diffractant) pour un système de source binaire est $\delta x'= F\Delta$. Ce qui nous amène à déduire que le pouvoir de résolution $\frac{1}{\Delta}$ est inversement proportionnelle à la longueur d'onde $\lambda$ et proportionnelle à la taille $B$  de l'objet diffractant. Là aussi, l'intuition de Fizeau est confirmée, à savoir que les franges disparaissent pour une taille d'objet assez étendue (ici $\Delta$).

%\subsubsection{L'Interférométrie Optique à Longue Base}
\subsection{L'Interférométrie Optique à Longue Base (OLBI)}
\begin{figure}[h!]
\centering
\includegraphics[height=0.5\hsize,draft=false]{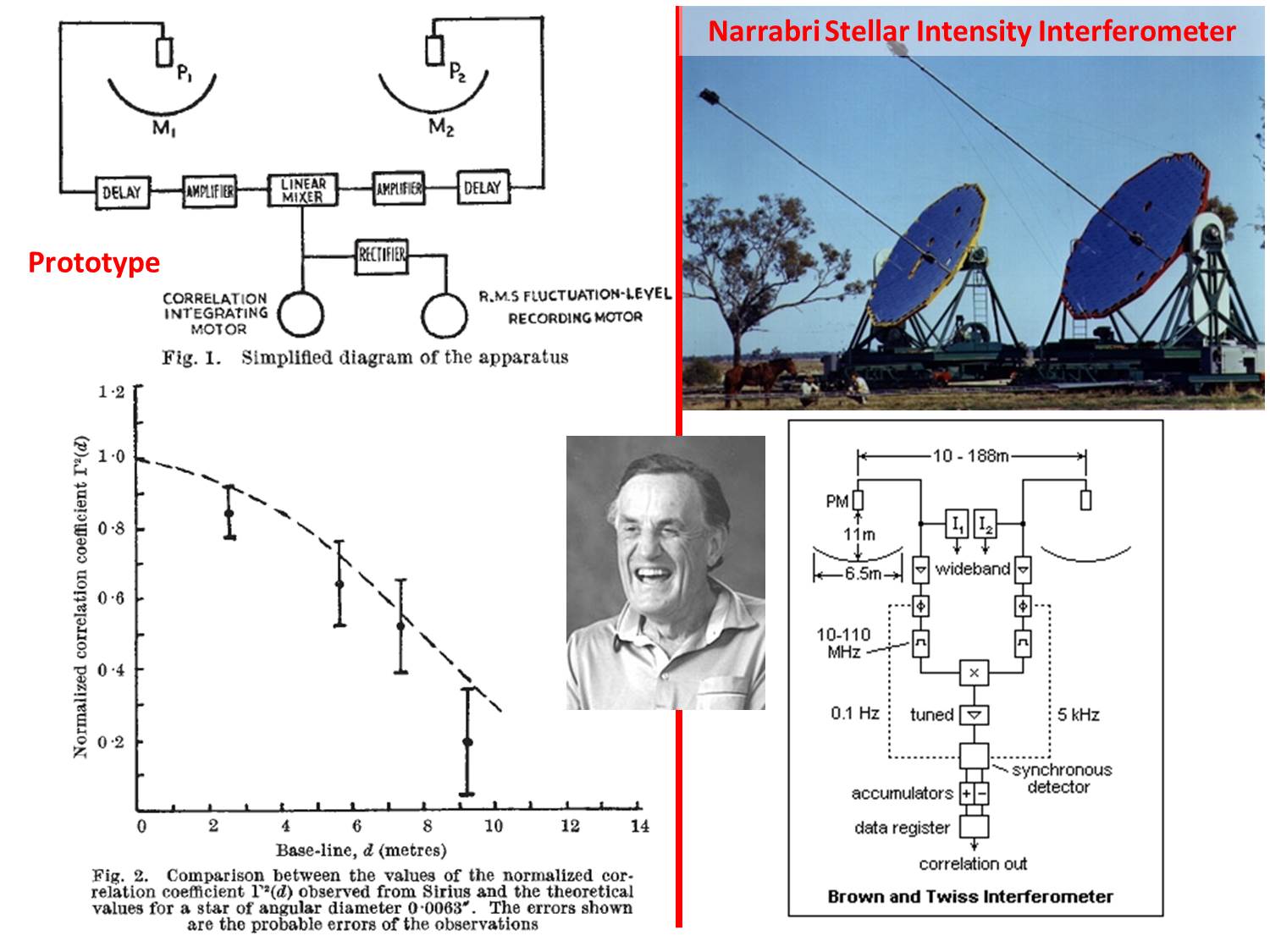}
\caption[L'interférométrie d'intensité à Longue Base de Hanbury Brown.]{L'Interférométrie Optique à Longue Base de Hanbury Brown.}\label{Hanbury_Brown}
\end{figure}

L'aventure de l'OLBI commença entre les années 1950 et 1960 avec Robert Hanbury Brown, qui avec l'aide du mathématicien Richard Twiss s'est rendu compte que l'une des techniques qu'il avait mise au point et utilisée en radioastronomie pouvait être adoptée pour mesurer optiquement les diamètres angulaires des étoiles. De 1955 et 1956 Hanbury Brown passa 60 nuits à essayer de faire des observations à Jodrell Bank près de Manchester avec un instrument baptisé "prototype"  (un interféromètre dit d'intensité qui mesure la corrélation entre les fluctuations des signaux électriques résultant des détecteurs photoélectriques installés sur chaque télescope, avec une base interférométrique pouvant aller jusqu'à 10 m -Fig.\ref{Hanbury_Brown}-) afin de vérifier la fonctionnalité de son concept. C'est au cours de cette période que Hanbury Brown parvint à mesurer le rayon angulaire de l'étoile la plus brillante de notre ciel nocturne ; Sirius ($\alpha$ Canis Majoris) avec un rayon estimé à 6.8 milli-seconde d'arc \citep{1956Natur.178.1046H}. Hanbury Brown et son équipe continuèrent leurs observations à l'aide d'un instrument plus sophistiqué à Narrabri en Australie et purent ainsi mesurer le diamètre angulaire de plusieurs étoiles chaudes (e.g. \citet{1968ARA&A...6...13B, 1974MNRAS.167..121H, 1974MNRAS.167..475H}). Cependant, l'utilisation de la technique de Hanbury Brown était limitée car elle n'offrait aucun accès direct au contraste des franges d'interférences issu de la lumière des deux télescopes (à l'instar des fentes de Young).

\begin{figure}[h!]
\centering
\includegraphics[height=0.5\hsize,draft=false]{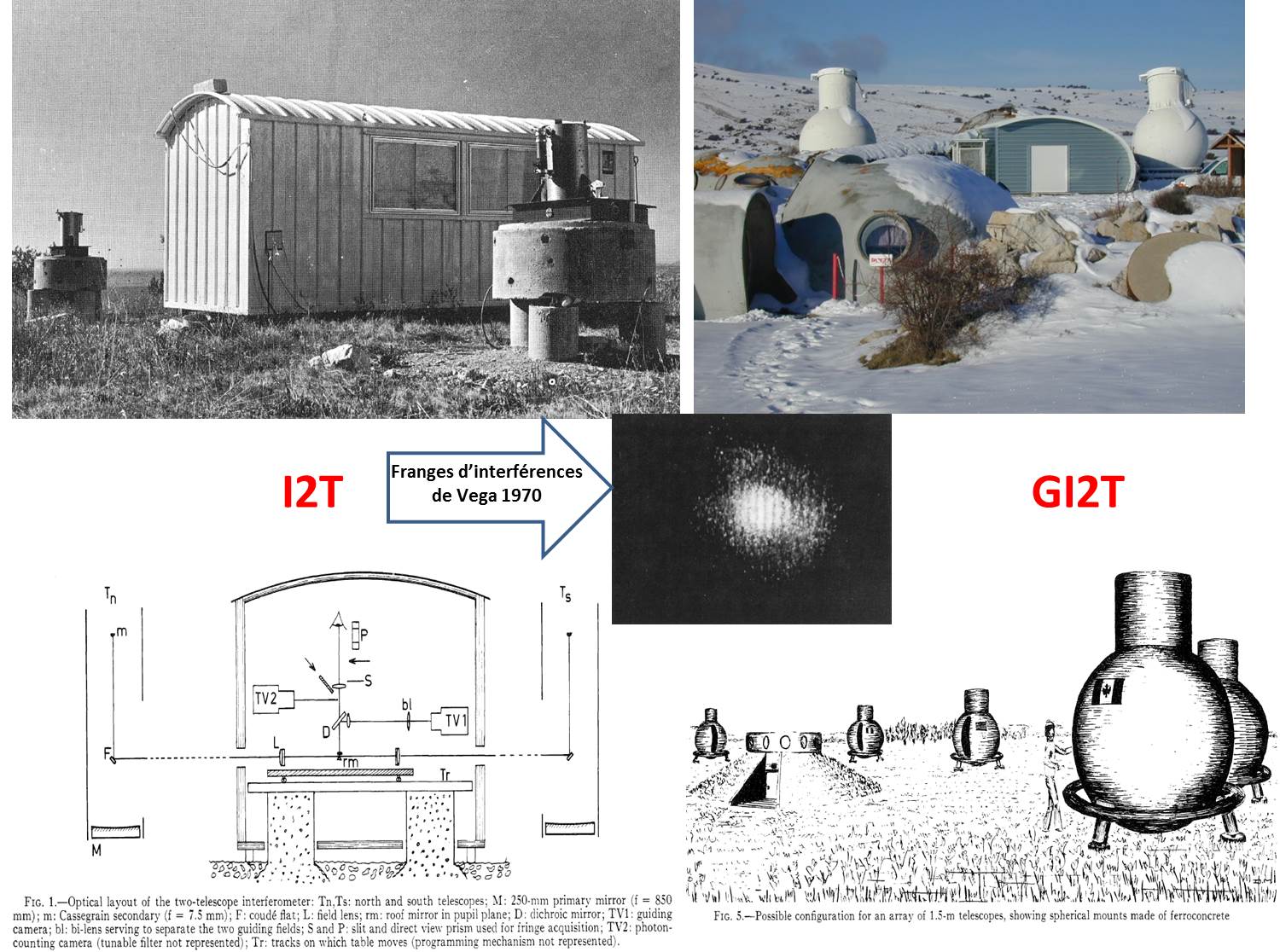}
\caption[L'interférométrie Optique à Longue Base d'Antoine Labeyrie.]{L'Interférométrie Optique à Longue Base d'Antoine Labeyrie.}\label{Antoine_Labeyrie}
\end{figure}

Ce n'est qu'à la suite des travaux de A. Labeyrie dans les années 1970, que l'OLBI (Optical long Baseline Interferometry) a connu un réel essor. En effet,  Labeyrie avec la technique de l'interférométrie des tavelures \citep{1970A&A.....6...85L} parvint, pour la première fois à combiner  la lumière de deux télescopes distincts, séparés de 12 m \citep{1975ApJ...196L..71L} et à observer les franges d'interférences de la cinquième étoile la plus brillante de notre voute céleste ; Vega -$\alpha$ Lyrae-  \citep{1975ApJ...196L..71L}, Fig.\ref{Antoine_Labeyrie}. Malgré le fait qu'à peu près à la même époque, le russe E.S. Kulagin avait aussi réussi à mesurer l'orbite du compagnon de Capella ($\alpha$ Aurigae de la constellation du cocher, faisant ainsi suite aux travaux de \citet{1922ApJ....56...40M}) à l'aide d'un interféromètre de 6 m à Pulkovo \citep{1970SvA....14..445K}, ce fut le travail de Labeyrie qui eut le plus d'impact sur l'avenir de l'interférométrie stellaire par la suite. Labeyrie ne s'arrêta pas là. Sur le plateau de Calern, avec son équipe, il initia d'abord le I2T (Interféromètre à 2 télescopes), puis le GI2T (Le Grand Interféromètre à 2 Télescopes) qu'il proposa à la conférence ESO de Genève en 1977 avec beaucoup d'ambition, et qu'il réussit à réaliser quelques années plus tard avec une table de recombinaison "REGAIN". Ce dernier a permis quelques belles \oe{}uvres scientifiques, telles que  la résolution de l'enveloppe de $\gamma$ Cas et la mise en évidence de sa rotation \citep{1989Natur.342..520M}. Labeyrie fut le premier aussi à proposer l'idée d'interféromètres spatiaux tel que "FLUTE" \citep{1980oits.conf.1020L} et TRIO \citep{1982vlbi.conf..477L}. Il proposa, depuis les années 2000 également des idées d'hypertélescopes spatiaux et terrestres dont l'un sur lequel il travaille en ce moment sur le site de la Moutière \citep{2013EAS....59....5L}.\\

\begin{figure}[h!]
\centering
\includegraphics[height=0.5\hsize,draft=false]{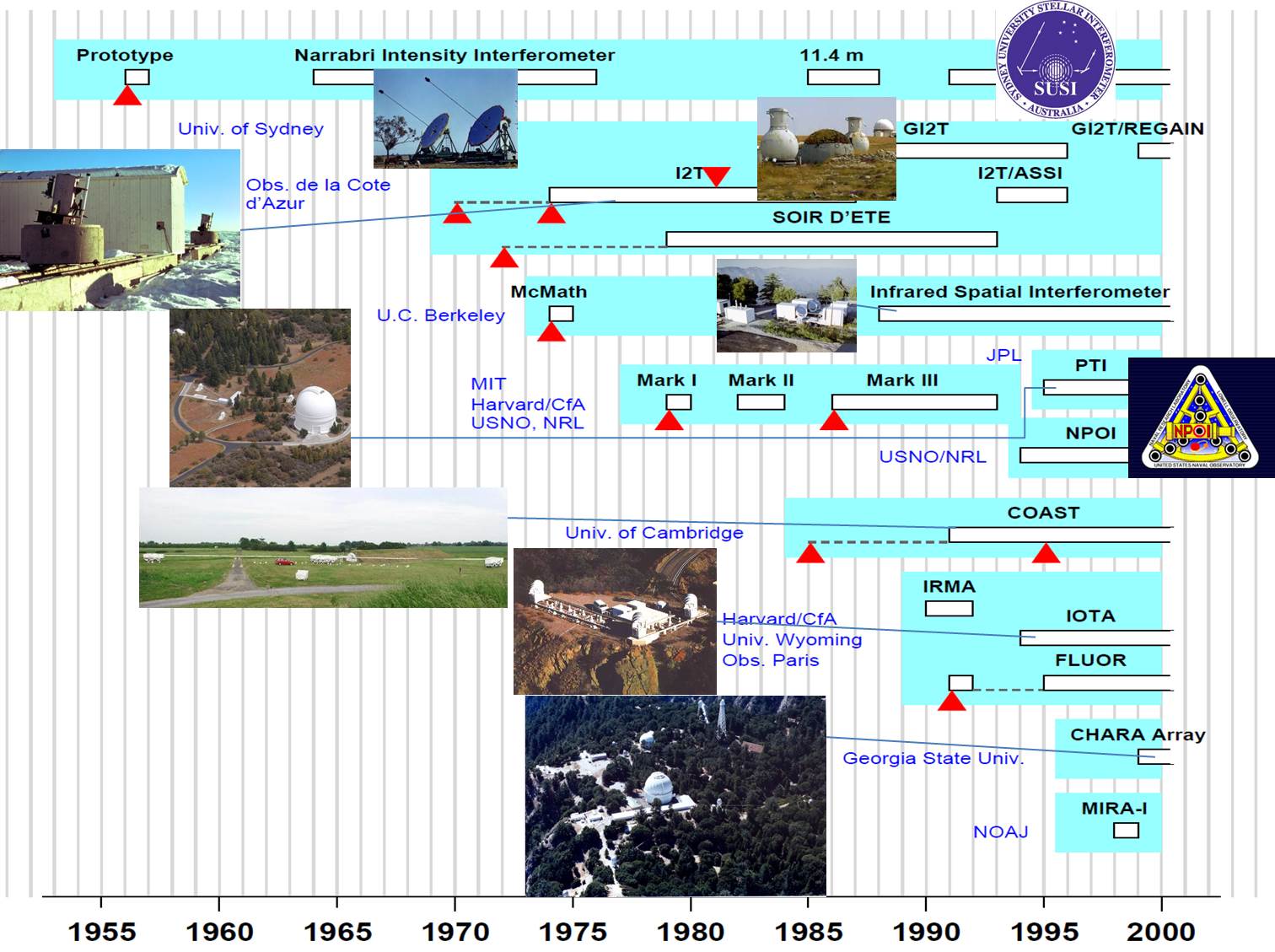}
\caption[L'interférométrie stellaire de 1950 à 2000.]{Les débuts de l'interférométrie stellaire de 1950 à 2000 (inspiré du diagramme de \citet{1999AAS...195.2301L}).}\label{Interfero_Moderne}
\end{figure}

\begin{figure}[h!]
\centering
\includegraphics[height=0.5\hsize,draft=false]{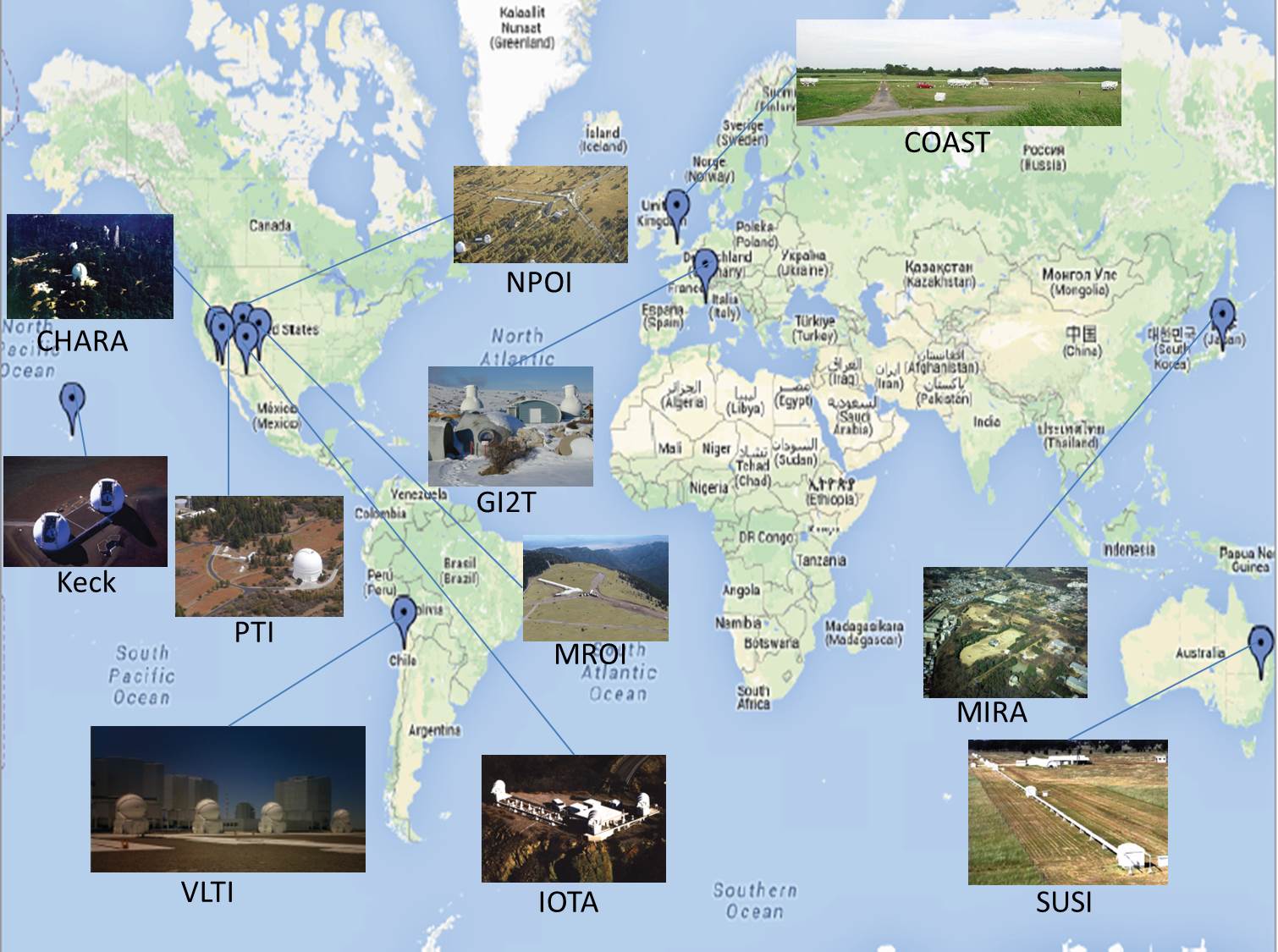}
\caption[Énumération et géolocalisation des interféromètres à longue base les plus importants du 20\up{iéme} siècle.]{Énumération et géolocalisation des interféromètres à longue base les plus importants du 20\up{iéme} siècle. De nos jours seulement 3 d'entre eux sont toujours opérationnels (CHARA, NPOI et le VLTI).}\label{OLBI_World}
\end{figure}

Depuis, le nombre d'interféromètres a rapidement augmenté, avant de décroitre dans les années 2000 faute de financements. Seul un nombre restreint restent encore opérationnels de nos jours, alors que les publications produites ne cessent de croître. La figure \ref{Interfero_Moderne} \& le tableau \ref{Lawsan_1999_2} résument assez bien le rapide développement de l'OLBI, des années 50 jusqu'aux années 2000, et la Fig.\ref{OLBI_World} énumère les interféromètres à longue base les plus importants encore opérationnels de nos jours.\\

\begin{table*}[htbp]
\centering
\caption[Principaux évènements de l'interférométrie à longue base stellaire de 1950 à 2000]{ Principaux évènements de l'interférométrie à longue base stellaire de 1950 à 2000 (inspiré de \citet{1999AAS...195.2301L}).}\label{Lawsan_1999_2}
\centering
\resizebox{\textwidth}{!}{\begin{tabular}{ccc}
\hline\hline
Année & Événements & Auteurs \& référence\\
\hline
1956 & Mesure du rayon angulaire de Sirius avec l'interféromètre d'intensité "prototype" & R. Hanbury Brown et RQ Twiss, Nature 177, 27 (1956)\\
1970 & Invention de l'interférométrie des tavelures & A. Labeyrie, Astron. Astrophys. 6, 85 (1970)\\
1972 & Franges hétérodyne à 10 microns & J. Gay et A. Journet, Nature Phys. Sci. 241, 32 (1973)\\
1974 & Franges hétérodyne à 10 microns avec des télescopes séparés & MA Johnson et al., Phys. Rev Lett. 33, 1617 (1974)\\
1974 & Détection directe de franges visibles avec des télescopes séparés et mesure de rayon angulaire de Vega & A. Labeyrie, Astrophys. J. 196, L71 (1975)\\
1979 & Première mesures de franges avec un interféromètre stellaire à suivi de phase stellaire & M. Shao et DH Staelin, Appl. Opt. 19, 1519 (1980)\\
1982 & Mesures de Frange à 2.2 microns & G.P. Di Benedetto et G. Conti, Astrophys. J. 268, 309 (1983)\\
1985 & Mesure de la clôture de phase aux longueurs d'onde optiques & JE Baldwin et al., Nature 320, 595 (1986)\\
1986 & Premier interféromètre entièrement automatisé pour l'astrométrie & M. Shao, MM Colavita et al., Astron. Astrophys. 193, 357 (1988)\\
1991 & Première utilisation des fibres de verre monomodes avec des télescopes séparés & V. Coudé du Foresto et ST Ridgway, ESO Proc. 39, 731 (1992)\\
1995 & Imagerie de synthèse optique avec des télescopes séparés & JE Baldwin et al., Astron. Astrophys. 306, L13 (1996)\\
\hline\hline
\end{tabular}}
\end{table*}

Néanmoins nous pouvons prédire quelques pistes de développement en énumérant les lignes majeures actuellement en cours ou en tests. Parmi ces pistes, nous pouvons citer l'interférométrie différentielle (qu'on aborde en détail dans la Sec.\ref{spectro_int}, qui a encore de beaux jours devant elle, avec le la mise en fonction prochaine de l'instrument MATISSE (Multi AperTure mid-Infrared SpectroScopic Experiment), en juin 2017 au VLTI, porté par l'OCA et qui pourra assurer une couverture spectrale comprise entre les bandes L \& N (en infra-rouge), pour l'étude des nébuleuses pouponnières donnant naissance aux étoiles et des étoiles jeunes entourées d'un fort environnement circumstellaire de type  Herbig Ae/Be propice à l'étude de la formation des planètes gazeuses (temps de formation 10 millions d'années environ) et des planètes rocheuses type terrestres (100 millions d'années de formation environ). Il y a aussi le projet GRAVITY (General Relativity Analysis via Vlt InTerferometrY), qui est en phase de test et qui est prévu pour combiner la lumière de 4 télescopes dans l'infra-rouge au VLTI, pour l'étude des noyaux actifs de galaxies, de disque et jets autour d'étoiles en formation ou au sein des micro-quasars, des trous noirs de masse intermédiaire au c\oe{}ur des amas globulaire et des planètes extrasolaires. Il y a aussi le LBT (Large Binocular Telescope) qui est constitué de deux télescopes de $8.4$ mètres de diamètre, fonctionnant à la fois en mode "Fizeau imaging" et en mode "Nulling", et qui pourrait nous réserver de beaux résultats scientifiques. Il existe aussi quelques projets audacieux et séduisants parmi lesquels on peut citer la combinaison de plusieurs télescopes très distancés les uns des autres via fibre optique (à condition de régler les différents problèmes liés au phénomène de dispersion) ;  Un projet d'une telle envergure a été suggéré, sous le nom de OHANA (Optical Hawaian Array for Nanoradian Astronomy)  pour relier, dans l'infra-rouge, les 7 télescopes de Mauna Kea, afin d'atteindre une résolution angulaire équivalente à un télescope monolithique de 800 m de diamètre. En 2006 un résultat encourageant à d'ailleurs été obtenu avec les deux télescopes du Keck (ayant des miroirs de 10 m de diamètre chacun et une séparation de 85 m l'un de l'autre) sur l'étoile 107 Herculis \citep{2006Sci...311..194P}. Autre idée audacieuse, les Hyper-télescopes, dont j'ai cité comme exemple les travaux tests actuels sur site de la Moutière, et qui consistent à utiliser les reliefs naturels de forme parabolique, pour installer plusieurs petits miroirs (tous pointés sur la même cible) et diriger la lumière collectée vers une gondole combinatrice qui se déplace au-dessus de la parabole via des câbles \citep{2013EAS....59....5L}. Sans oublier l'ambitieux projet d'interférométrie hétérodyne, qui consiste à la transposition de plusieurs signaux d'une même source en laboratoire (éventuellement préalablement acquis, de manière simultanée et indépendante n'importe où dans le monde). Bien que cette idée est fort prometteuse, elle rencontre cependant d'importants problèmes de bande passante au cours des acquisitions. Un tel projet avant-gardiste avait vu le jour au plateau de Calern, sous le nom de "SOIRDETE", dans les années 70 \citep{2014ipco.conf..181G}. Les 3 dernières idées de l'avenir de l'interférométrie (la fibre optique, l'hyper-télescope et l'interférométrie hétérodyne) sont des projets très couteux et qui nécessitent beaucoup d'investissement (humain et matériel) et d'investigation pour atteindre un jour une concrétisation optimale. Néanmoins, certains grands projets en cours de réalisation peuvent être détournés à des fins interférométriques et relativement à moindre coût, à l'instar de la proposition de \citet{2010RMxAC..38....1M}, qui consiste à implémenter plusieurs petits télescopes tout autour de l'E-ELT afin de combiner les lumières collectées interférométriquement, l'ESO bénéficiant déjà de l'expérience du VLTI.\\

%\clearpage

\subsection{Technique et caractéristiques de l'OLBI}\label{OLBI}

Dans cette sous-section nous allons aborder les caractéristiques et techniques mise en pratique en OLBI dans le but d'obtenir une mesure optimale. Et pour ce faire nous avons plusieurs éléments à prendre en compte, dont :\\

\textbf{Nombre de Bases interférométriques:}
L'interférométrie à deux télescopes distancés au maximum possible techniquement parlant nous apporte un gain considérable en termes de résolution mais aussi un manque non négligeable en termes de flux. Ainsi plus on a de télescopes, plus la couverture dans le plan de Fourier (qu'on explicite en détail, plus bas) est grande, meilleure sera l'information sur l'objet observé. 

Pour deux télescopes c'est simple on a qu'une seule base interférométrique, pour 3 télescopes c'est 3 et pour dénombrer le nombre de bases interférométriques $N_{base}$ qu'on peut obtenir, deux à deux d'un nombre $N_{tel}$ de télescopes sans répétition, on fait appel aux mathématiques via la loi binomiale $C^2_{N_{tel}}$, qu'on lit combinaison de 2 parmi $N_{tel}$ et qu'on formule comme suit :

\begin{equation}
N_{base}=C^2_{N_{tel}}=\frac{N_{tel}!}{2!( N_{tel}-2)}
\label{3.29}
\end{equation}

où $!$ désigne la factorielle d'un nombre entier positif $n$, avec $n!= \prod \limits_{1 \leq i\leq n} i$. Alors on peut écrire que le nombre de bases interférométriques possible est:

\begin{equation}
N_{base}=\frac{ N_{tel}(N_{tel}-1)}{2}
\label{3.30}
\end{equation}

Ainsi, pour 100 télescopes, par exemple, le nombre de bases interférométriques sera de 4950. Ce qui veut dire aussi un rapport de 4949 d'informations en plus comparé à une seule base interférométrique (2 télescopes), d'où l'enjeu du nombre de bases interférométriques.\\

\begin{figure}[h!]
\centering
\includegraphics[height=0.5\hsize,draft=false]{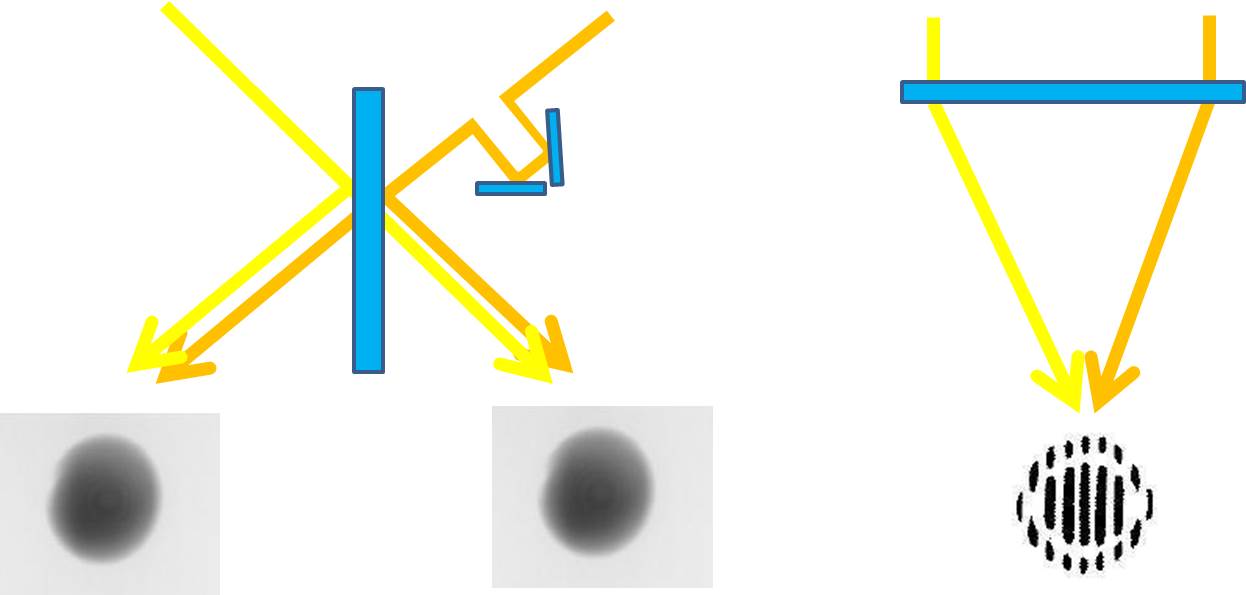}
\caption[Recombinaisons coaxiale et multiaxiale.]{Schéma représentant une recombinaison coaxiale à gauche et multiaxiale à droite.}\label{m-c_axiale}
\end{figure}

\textbf{Les différentes techniques de recombinaison cohérente:}
Une fois les faisceaux lumineux issus de deux télescopes collectés et égalisés, ils sont ensuite combinés à l'aide d'un instrument recombineur. Il existe différentes manières de recombinaisons possibles, dont je vais citer ici que les deux principales (voir Fig.\ref{m-c_axiale}):\\

\begin{itemize}
\item \textbf{La recombinaison coaxiale:} Les deux faisceaux lumineux de chaque base interférométrique sont combinés à l'instar de l'interféromètre de Michelson via une lame parallèle semi-réfléchissante. La cohérence temporelle est assurée ici à l'aide de miroirs piézoélectriques, produisant ainsi deux interférogrammes de type "teinte plate" dont l'enveloppe correspond à celle de la cohérence temporelle.

\item \textbf{La recombinaison multiaxiale:} Les faisceaux sont combinés à l'aide de lentilles ("convergentes", ou dites "coin d'air") pour obtenir un seul interférogramme de type tache d'Airy à franges. Ici et comme c'est le cas pour l'expérience de Young on joue plutôt sur la cohérence spatiale contrairement à la combinaison coaxiale qui elle joue sur la cohérence temporelle. Comme exemple, je cite AMBER, l'instrument avec lequel ont étés collectées toutes les données de mon sujet d'étude de thèse, qui utilise un combinateur multiaxial.\\
\end{itemize}

\textbf{La super-synthèse d'ouverture:}

L'OLBI implique la possibilité d'avoir des bases interférométriques qui peuvent atteindre jusqu'à plusieurs centaines de mètres. A cela s'ajoute la rotation de la terre qui fait également tourner les lignes de base à cause du suivi d'un objet pointé au cours d'une nuit d'observation. Ce qui accroit significativement la couverture des fréquences spatiales $u$ \& $v$, sous forme d'arcs d'ellipses, dont le référentiel est au centre du dispositif diffractant; c'est ce qu'on appelle \textbf{la couverture (u,v)} (ou bien le plan (u,v)).\\

\begin{figure}[h!]
\centering
\includegraphics[height=0.5\hsize,draft=false]{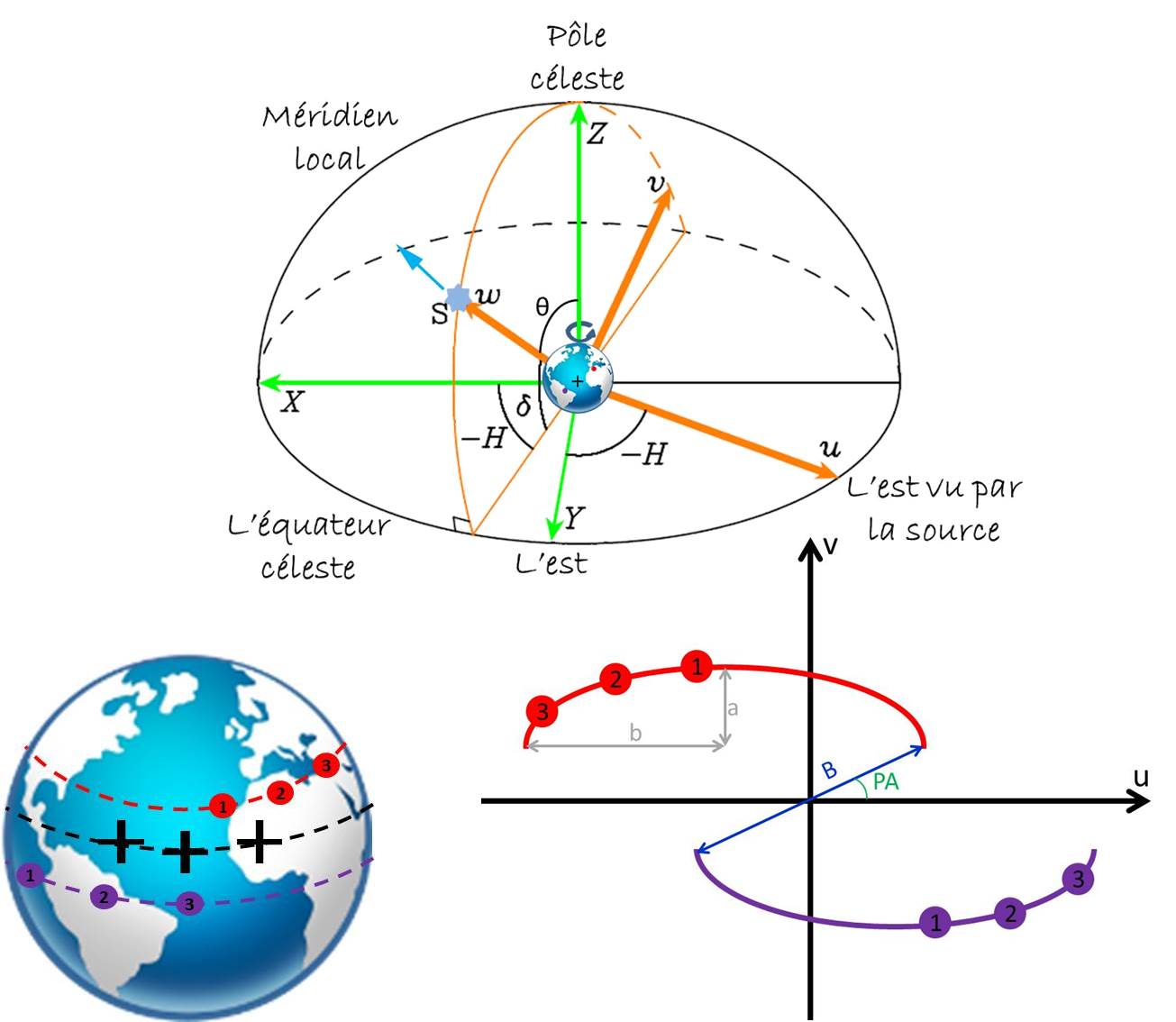}
\caption[La super-synthèse d'ouverture.]{Schéma décrivant la super-synthèse d'ouverture.}\label{Plan_uv}
\end{figure}

Les équations qui régissent la couverture (u,v) dépendent principalement de la longueur d'onde observée, de la longueur de la base $B$, des déclinaisons et angles horaires $(\delta_s,h_s)$ de la source et $(\delta_B,h_B)$ du centre de la base interférométrique (voir Fig.\ref{Plan_uv}). où l'angle horaire $h=TSL-AD$, avec $TSL$ le Temps Standard Local, et $AD$ l'Ascension Droite \citep{1974gegr.book..256F}.  

Selon la Fig.\ref{Plan_uv}, on peut écrire que:

\begin{equation}
\left( \begin{array}{c}u\\ v\\ w \end{array}\right)= \begin{pmatrix} \sin(h_s) & cos(h_s) & 0\\ -\sin(\delta_s)cos(h_s) & \sin(\delta_s)cos(h_s) & \cos(\delta_s)\\ \cos(\delta_s)cos(h_s) & -\cos(\delta_s)sin(h_s) & sin(\delta_s) \end{pmatrix}\frac{B}{\lambda} \left( \begin{array}{c}\cos(\delta_B)\cos(h_B)\\ \cos(\delta_B)\sin(h_B)\\ \sin(\delta_B) \end{array}\right)
\end{equation}

Les coordonnées $(B\cos(\delta_B)\cos(h_B),B\cos(\delta_B)\sin(h_B),B\sin(\delta_B))$ représentent ici les coordonnées $(X,Y,Z)$ des télescopes, et les coordonnées (0,0,0) représentent le milieu de la base interférométrique, avec $B=\sqrt{X^2+Y^2+Z^2}$.
Il faut bien noté que même si on a là 3 fréquences spatiales ; $u,v$ \& $w$, le plan (u,v) lui, qui nous permet d'avoir une idée sur la couverture d'observation dans l'espace de Fourier, est toujours considéré au niveau de la base interférométrique, i.e. à $w=0$ (voir Fig.\ref{Plan_uv}).
La période sidérale de rotation de la terre étant exactement de 23h56s U.T., sa vitesse angulaire est alors de $\omega_{terre}=7.27x10^{-5} rad/s$, ce qui entraine (pour une source résolue) une rapide variation de l'amplitude et de la phase de la visibilité, mesurée par la fréquence de frange $\nu_f=B\frac{d\cos(\theta)}{dh_s}= \omega_{terre}u\cos(\delta_s)$. De plus la rotation terrestre engendre une progression du module des fréquences spatiales sous forme d'arc d'ellipse dans l'espace de Fourier. Ce mouvement est décrit par la formule ci-dessous:\\

\begin{equation}
\frac{u^2}{a^2}+\frac{(v-v_0)^2}{b^2}=1,
\end{equation}

avec $a=\sqrt{X^2+Y^2}=B\cos(\delta_B)$, $b=a\sin(\delta_s)=B\cos(\delta_B)\sin(\delta_s)$ et $v_0=Z\cos(\delta_s)= B\sin(\delta_B )\cos(\delta_s)$. La Fig.\ref{Plan_uv} montre un exemple de l'élargissement du plan (u,v) pour deux télescopes imaginaires (rouge \& violet) évoluant sur 3 point chacun. Les coordonnées $u$ \& $v$ peuvent aussi être exprimées à l'aide de la longueur de la base $B$ et de l'angle de projection de la base $PA$, où $u=B\cos(PA)$ \& $v=B\sin(PA)$.\\

Cependant, pour augmenter la couverture en fréquences spatiales, mais surtout pour étudier des processus physiques qui induisent une dépendance chromatique de la carte d'intensité de l'objet observé, nous utilisons des observations interférométriques à plusieurs longueurs d'onde, obtenant de ce fait une grande quantité d'information avec une technique permettant de résoudre les astres à la fois spectralement, avec la spectroscopie, et spatialement, avec l'interférométrie. Cette technique combinée est appelée spectro-interférométrie. Mais avant d'aborder plus en profondeur cette technique, introduisons tout d'abord la spectroscopie.\\

\section{La Spectroscopie}\label{spectro}
La spectroscopie est sans doute le moyen qui a permis le plus d'avancées scientifiques en astrophysique à partir du 19\up{ième} et 20\up{ième} siècle.  Même si les débuts de cette science furent en physique avec la décomposition de la lumière blanche du Soleil par Isaac Newton à l'aide d'un prisme en 1665, toutes les autres découvertes influencèrent fortement le domaine de l'astrophysique. En effet,  c'est à Joseph von Fraunhofer en 1814 qu'on doit l'invention du spectroscope (auquel il a incorporé le réseau de diffraction inventé par David Rittenhouse en 1785). C'est d'ailleurs à l'aide de son invention que Fraunhofer a pu établir un catalogue du spectre solaire, dont certaines raies portent son nom, et d'y repérer de mystérieuses bandes sombres.\\

\begin{figure}[h!]
\centering
\includegraphics[height=0.5\hsize,draft=false]{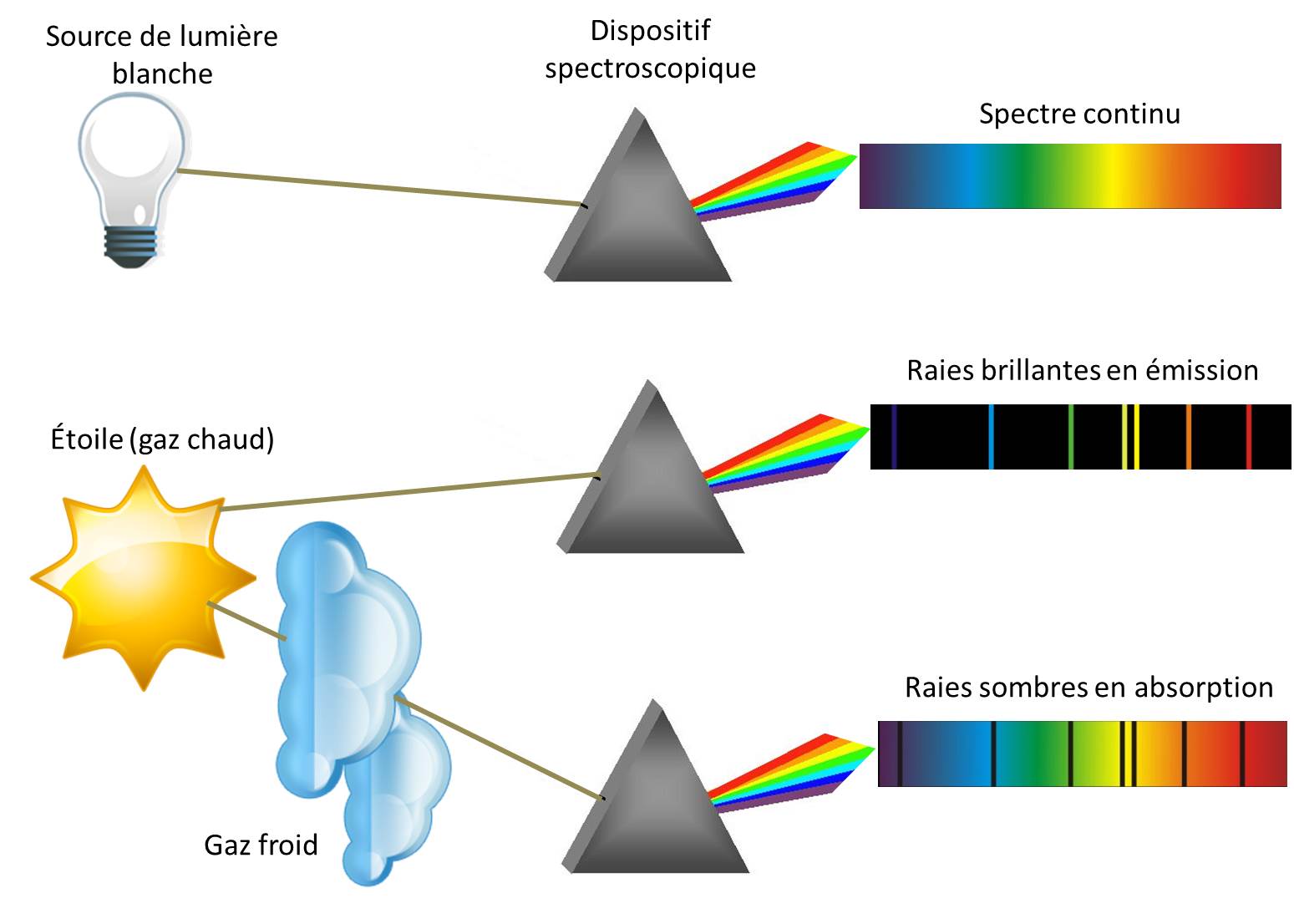}
\caption[Schéma descriptif des trois lois de la spectroscopie de Kirchhoff ]{Le Schéma descriptif des trois lois de la spectroscopie de Kirchhoff.}\label{Kirchhoff}
\end{figure}

S'appuyant sur les travaux de l'inventeur américain David Alter (1807-1881) qui avait suggéré l'idée que chaque élément chimique devait avoir sa propre signature d'émission spectrale spécifique (1854) et d'en apporter la preuve une année plus tard grâce à l'étude des propriétés optiques des gaz, et s'inspirant des études spectroscopiques de l'astronome suédois Anders Jonas \AA{}ngstr\"om (1814-1874) qui découvrit l'existence de l'hydrogène sur la photosphère solaire en 1862, Gustav Kirchhoff déduit en 1859, avec l'aide de Robert Bunsen, qu'un corps ne pouvait absorber que la quantité de radiations qu'il pouvait émettre. Il formula ainsi ses fameuses trois lois spectroscopiques qui décrivent chacune les trois différents spectres qu'on peut observer dans la nature (voir Fig.\ref{Kirchhoff}):\\

\begin{itemize}
\item	\textbf{Un spectre continu:} qui est produit par tout corps chaud lumineux,
\item	\textbf{Un spectre de raies d'émission:} qui est produit par tout gaz chaud (optiquement mince), nous renseigne sur la nature des atomes qui compose le gaz et sur leurs  niveaux d'énergie,
\item \textbf{Un spectre de raies d'absorption:} qui est produit par tout corps chaud incandescent entouré par un gaz relativement plus froid (c'est ainsi qu'ils purent conclure que le Soleil était essentiellement constitué d'un noyau chaud entouré d'un gaz relativement plus froid).\\
\end{itemize}

C'est le spectre continu qui inspira à Kirchhoff le concept d'un objet parfait dont le spectre électromagnétique ne dépend que de sa température; le concept du corps noir, qui a pu être modélisé par Planck en 1900, fortement influencé par les travaux de Kelvin, Stefan, Boltzmann, Paschen, Rayleigh et Wilhelm Wien. Ce modèle qui relie l'intensité spécifique $I_0$ à la température effective $T_{\rm eff}$ et la longueur d'onde $\lambda$ d'une source lumineuse est connu sous le nom de la loi de Planck, et est formulé comme suit:

\begin{equation}
I_0(\lambda,T_{\rm eff})=\frac{2hc^2}{\lambda^5}\frac{1}{e^{\frac{hc}{\lambda \sigma_{B} T_{\rm eff}}}-1},
\label{eq5}
\end{equation}

où $\sigma_{B}$ est la constante de Boltzmann, $c$ la vitesse de la lumière et $h$ la constante de Planck. Le pic d'intensité d'un corps noir (sa température maximale) est déterminée par la loi de Wein $\lambda_{max}=\frac{hc}{4.965\cdot\sigma_{SB} T_{\rm eff}}$. L'expérience du corps noir a pu être réalisé à l'aide d'un four chauffé à blanc, où se produisait un échange de température entre ses parois jusqu'à atteindre une température d'équilibre. A cette température correspondait une intensité lumineuse dont la loi de Rayleigh-Jeans (1900) prévoyait qu'elle serait proportionnelle à la température absolue et inversement proportionnelle au carré de la longueur d'onde, ce qui implique une valeur infinie pour les petites longueurs d'onde. Ce qui ne correspond pas du tout aux valeurs expérimentales dans le domaine des ultraviolets (d'où le nom de catastrophe ultraviolette). C'est cette ambigüité que Planck a su résoudre avec son équation.\\

La température d'une étoile est relié à sa luminosité $L$ via la loi de Stefan-Boltzmann; $L=\sigma_{SB} T_{\rm eff}^4S_*$, avec $\sigma_{SB} $ la constante de Stefan-Boltzmann et $S_*$ la surface de l'étoile. Il est important de distinguer à ce stade l'intensité spécifique $I_0$ qui représente l'intensité ponctuelle au centre de l'étoile (voir effet d'assombrissement centre-bord), le flux intégré ou spectre $F$ qui désigne la portion d'intensité observée sur un angle solide donné et la luminosité $L$ qui représente l'intensité totale émise par toute la surface de l'étoile. Enfin, en pratique ce qu'on observe sur une étoile est plutôt un corps gris qui est la combinaison d'un corps noir parfait dépendant uniquement de la température et de raies d'absorptions spécifiques à la composition chimique du gaz surfacique de ladite étoile.\\

Quelques années plus tard, en 1865, Kirchhoff \& Bunsen mirent en évidence le lien entre les spectres et les compositions chimiques des objets observés (faisant suite ainsi aux travaux de Foucault à partir de 1845), en observant sur certains spectres de flammes terrestres les mêmes bandes spectrales que celles décrites par Fraunhofer 51 ans auparavant, déterminant de ce fait la composition chimique du Soleil. Chose qui semblait encore impossible 30 ans plus tôt, tel qu'on pouvait lire dans le Cours de philosophie positive d'Auguste Comte en 1835.\\

Depuis et grâce à cette technique, on a pu déterminer l'abondance chimique  de toute étoile visible via  l'observation de sa photosphère, et d'ouvrir de ce fait le champ de la physique stellaire (astrophysique) et totalement la démarquer du domaine de l'astronomie. 
La spectroscopie a ainsi permis à Jules Janssen d'identifier l'atome d'hélium sur le spectre H$\alpha$ du limbe solaire, lors de l'éclipse totale solaire à Guntur en Inde le  18 Août 1868, avant sa détection terrestre, d'où le nom de cet élément en hommage au dieu grec du Soleil "Hélios", proposé par l'astronome britannique Sir J.N. Lockyer. En plus de l'hélium, les travaux de kirchhoff ont permis la découverte de bon nombre de nouveaux éléments chimique qui ont enrichit le  tableau périodique de Mendeleïev proposé en 1869.\\

De ce fait la spectroscopie permet aussi d'étudier l'influence de la métallicité sur l'activité stellaire, qui joue un grand rôle dans l'histoire évolutive des étoiles.  Bons nombre d'études ont étaient menées qui démontrent une importante corrélation entre l'abondance métallique, l'impact des zones radiatives et convectives d'une étoile et sa rotation \citep{2001A&A...373..555M, 2005A&A...443..581H, 2005A&A...429..581M}. Ainsi, plus la métallicité est faible, plus l'étoile est encline à être une chaude active à fort moment cinétique \citep{1999A&A...347..185M, 2006A&A...452..273M}. De plus la métallicité d'une étoile peut aussi impacter de manière significative la puissance de ses vents radiatifs au niveau de la photosphère (tel qu'abordé au chapitre \ref{Vent_radiatif}).\\ 

La spectroscopie nous permet aussi de classifier le type spectral des étoiles, de mesurer leur température effective de surface  (plus la photosphère de l'étoile est chaude, plus celle-ci émet un rayonnement spectrale riche en longueurs d'ondes courtes (bleu et violet)), de déduire la pression interne et aussi de prédire l'évolution des étoiles. En effet dès l'énoncé de la loi de Pogson (1856), qui nous permet de relier la luminosité d'une étoile à sa magnitude apparente via une source connue (le Soleil par exemple), il a été possible à Ejnar Hertzsprung et Henry Norris Russell en 1910 de classifier les étoiles selon leur type spectral et leur luminosité sous la forme d'un schéma qui porte le nom de diagramme de Hertzsprung-Russell.\\

Cette technique a permis  aussi de mettre en évidence l'influence du champ magnétique sur les étoiles, via l'effet Zeeman (1886), qui produit un dédoublement des raies sous l'effet du champ magnétique, telle la découverte de George Ellery Hale qui a démontré l'origine magnétique des tâches solaires avec le spectrohéliographe qu'il avait inventé en 1902, et qui réalisa ainsi les premiers magnétogrammes de l'humanité. Un effet analogue à cet effet, et qui crée des dédoublements des raies spectrales mais sous l'action d'un champ électrique est connu sous le nom d'effet Stark (1874-1957). C'est également à Johannes Stark qu'on doit la découverte de l'effet Doppler-Fizeau dans des faisceaux d'ions positifs, et pour lequel un prix Nobel lui a été décerné en 1919. Ces derniers trois effets (Zeeman, Doppler-Fizeau et à moindre mesure Stark) sont responsables d'importants changements morphologiques des raies spectrales et sont dominants dans les étoiles et leurs environments.\\

A ce propos l'effet Doppler (dont le nom complet est l'effet Doppler-Fizeau) est un effet très important dans l'observation du mouvement des étoiles et de leurs environnements proches, entre autre leur rotation, ainsi que la déduction des distances. Cet effet a été décrit la toute première fois par Christian Doppler en 1846 dans l'article Erratum sur la lumière colorée des étoiles doubles, cet effet a été vérifié pour les ondes acoustiques, en 1845, par le chercheur néerlandais Buys Ballot, et par Hippolyte Fizeau pour les ondes électromagnétiques en 1848.\\

Cet effet est aisément démontrable en imaginant deux référentiels alignés sur une droite; un placé sur une source émettrice d'un signal électromagnétique avec une fréquence $\nu_0=\frac{c}{\lambda_0}$ et se déplaçant avec une vitesse notée $v_{proj}$, et un référentiel observateur qui se déplace avec une vitesse $v_{obs}$. La distance parcourue, pour un battement, vu par le référentiel émetteur est $d_{em}=\frac{(c-v_{proj})}{\nu_0}$, mais du point de vue de l'observateur la distance parcourue durant ce laps de temps $T_{obs}= \frac{1}{\nu}=\frac{\lambda}{c}$  est $ d_{obs}=c T_{obs}=d_{em}+v_{rec}T_{obs} \Rightarrow d_{em}=\frac{c-v_{proj}}{\nu_0}=\frac{c-v_{obs}}{\nu}$. De cette dernière équation et en supposant que le référentiel observateur est fixe ($v_{obs}=0$), on peut formuler le décalage en longueur d'onde par effet Doppler-Fizeau comme suit:\\

\begin{equation}
\Delta\lambda=\frac{v_{proj}}{c}\lambda_0 
\label{3.34}
\end{equation}

De ce fait, un décalage vers le rouge (redshift) ou vers le bleu (blueshift) s'opère selon que la source s'éloigne ou s'approche de l'observateur (c'est ce phénomène qui a permis à Edwin Hubble, en 1929, d'énoncer sa fameuse loi et de déduire l'expansion de l'univers, renforçant de ce fait la théorie du Big bang, énoncée initialement par Georges Lemaître une année plus tôt sous le nom d'atome primitif). La photosphère des étoiles est constituée de gaz agité thermiquement, provoque un élargissement des profils de raies d'absorption de l'ordre de deux fois la vitesse thermique dudit gaz.\\

De même, l'effet Doppler s'applique aussi parfaitement à l'étude des étoiles (et de leur environnement proche) en rotation. En effet, là aussi un élargissement des raies photosphériques est observé de l'ordre de deux fois la vitesse de rotation observée. Dans le cas des rotateurs rapides qui sont souvent des étoiles actives chaudes, la vitesse de rotation est très grande devant la vitesse thermique. En supposant la vitesse de rotation équatoriale $v_{eq}$ et $i$ l'angle d'inclinaison entre l'axe de visée de l'observateur et l'axe de rotation de l'étoile, on observera pour chaque point à la surface photosphérique de l'étoile une vitesse de $\pm \vsini$ selon le sens de rotation. Importante donnée qui nous permettra via des modèles de déterminer la forme, l'inclinaison et la distribution latitudinale des températures et de l'intensité, sans quoi notre étoile pourrait se retrouver mal classée  dans le digramme HR, avec toutes les incohérences théoriques que cela pourrait entrainer. 
Enfin la mesure du $\vsini$ est sensible à l'assombrissement gravitationnel engendré par la rotation critique de certains rotateurs rapides, et la non considération de ce phénomène, accentué par un grand angle $i$, peut entrainer d'importants biais, tel que révélé par  \citet{2004MNRAS.350..189T}.\\

Pour de plus amples détails sur les profils de raies stellaires et leur élargissement naturel, thermique, collisionnel et rotationnel, voir le livre de David F. \citet{1988lsla.book.....G}, qui aborde le phénomène à la fois d'un point de vue microscopique et macroscopique.\\

La spectroscopie, comme on vient de le voir, est un outil puissant qui offre d'importantes informations sur les vitesses, cependant elle est limitée. Le spectre étant une intégration sur toute la surface de la source observée (le moment d'ordre 0 de la distribution de brillance), elle offre peu ou pas du tout d'informations spatiales (ou angulaire). Informations qui sont obtenues par l'interférométrie (voir plus haut). La combinaison de ces deux méthodes peut donc nous fournir des informations à la fois spatialement et dans le champ des vitesses. Cette technique qui est connue sous le nom de spectro-interférométrie ou d'interférométrie différentielle est expliquée ci-dessous.\\

\section{La Spectro-Interférométrie}\label{spectro_int}
L'aventure spectro-interférométrique a débuté avec \citet{1982AcOpt..29..361B}, qui a eu l'idée de comparer les tavelures (speckles) issues de longueurs d'ondes différentes d'un télescope monolithique, et s'est aperçu d'un déplacement photo-centrique entre chaque longueur d'onde. Ce dernier est aisément appréciable indépendamment  des variations du seeing (la largeur à mi-hauteur de la PSF liée la perturbation atmosphérique) et de la taille du speckle, contournant de ce fait le critère de Rayleigh en offrant la possibilité de mesurer tout déplacement plus petit que la taille de la tavelure elle-même. Beckers nomma cette nouvelle technique Differential speckle interferometry (DSI), ou interférométrie différentielle des tavelures en français.  Cette technique, qui fournit un nouveau paramètre astrophysique: le vecteur représentant la variation chromatique du photo-centre de l'objet comme une fonction de la longueur d'onde (qui est proportionnelle à la dépendance chromatique de la phase de la transformée de Fourier de la distribution de brillance de la source), a été étendue à une gamme de longueurs d'ondes plus importante et appliquée à l'interférométrie par \citet{1989dli..conf..249P} qui a établi les fondements de l'Interférométrie Différentielle (DI). Ce qui a permis pour la première fois de séparer des paramètres spatiaux et spectraux des objets formant la binaire Capella \citep{1992ASPC...32..477P} à l'OHP.\\

Depuis la DI n'a cessé de se développer et de connaitre  un grand essor à travers le monde, les deux plus grands interféromètres fonctionnant en mode différentiel construits à ce jour sont le Keck et le VLTI, dont les premières franges ont été observées en 2001. Le tableau ci-dessous (Tab.\ref{DI_resume}) récapitule la localisation ainsi que les caractéristiques des principaux interféromètres différentiels fonctionnels à travers le monde.\\

\begin{table*}[htbp]
\centering
\caption[Récapitulatif des interféromètres différentiels à travers le monde]{Récapitulatif des interféromètres en mode de fonctionnement différentiel à travers le monde, avec leur nom, lieu géographique, nombre de télescopes combinables $N_{tel}$, longueur de base maximale $B_{max}$, et domaine spectral $\lambda(\mu m)$ (voir Tab.\ref{UBVRIJHKLMNQ})}\label{DI_resume}
\centering
%\resizebox{\textwidth}{!}{
\begin{tabular}{||c|c|c|c|c||}
\hline\hline
\textbf{Nom} & \textbf{Lieu géographique} & $\mathbf{N_{tel}}$ & $\mathbf{B_{max}}$ & $\mathbf{\lambda(\mu m)}$\\
\hline
CHARA & Mont Wilson, USA & 6 & 330 & 0.45-2.5 (B-K)\\
\hline
ISI & Mont Wilson, USA & 3 & 70 & 8-13 (N)\\
\hline
KECK-I & Mauna Kea, USA & 2 & 85 & 2.2-10 (K-N)\\
\hline
NPOI & Anderson Mesa, USA & 6 & 435 & 0.45-0.85 (B-I)\\
\hline
PTI & Mt Palomar, USA & 2 & 110 & 1.5-2.4 (H-K)\\
\hline
SUSI & Narrabri, Australie & 2 & 640 & 0.4-0.9 (B-I)\\
\hline
VLTI & Cerro Paranal, Chili & 4 & 200 & 1.2-13 (J-N)\\
\hline\hline
\end{tabular}
%}
\end{table*}

%\begin{table*}[htbp]
%\centering
%\caption[Bandes spectrales UBVRIJHKLMNQ]{Bandes spectrales du système UBVRIJHKLMNQ. Sont donnés le nom de la bande et le domaine spectral $\lambda(\mu m)$}\label{UBVRIJHKLMNQ}
%\centering
%\resizebox{\textwidth}{!}{\begin{tabular}{|c|c|c|c|c|c|c|c|c|c|c|c|c|c|c|}
%\hline\hline
%Bande & U & B & V & $R_j$ & $I_j$ & $R_c$ & $I_c$ & J & H & K & L & M & N & Q\\
%\hline
%$\mathbf{\lambda(\mu m)}$ & 0.35 & 0.44 & 0.55 & 0.70 & 0.90 & 0.65 & 0.80 & 1.22 & 1.63 & 2.19 & 3.45 & 4.75 & 10.20 & 21.00\\
%\hline\hline
%\end{tabular}}
%\end{table*}

\subsection{Les mesurables en spectro-interférométrie}\label{mesurables_DI}
Les mesurables de la DI à l'instar de l'interférométrie classique sont déduits de la visibilité complexe via le théorème de Van Cittert-Zernike (voir Eq.\eqref{3.25}), où on aura simplement une dépendance supplémentaire sur la variable de la longueur d'onde $\lambda$. Dans ce cas, la visibilité complexe différentielle $V$ est:\\

\begin{equation}
V(u,v,\lambda)=\frac{\tilde{I}(u,v,\lambda)}{\tilde{I}(0,0,\lambda)}
\label{3.35}
\end{equation}

De cette équation, et de raisons pratiques liées au rapport signal à bruit (SNR) et autres biais instrumentaux, sont déduits les 3 principaux observables utilisés en interférométrie différentielle à savoir (e.g., entre une paire de télescopes notée $i$ \& $j$):\\

\begin{itemize}
\item \textbf{Le module de visibilité:} Formulé $V_{ij}^2(u_{ij},v_{ij},\lambda)= |V_{ij}(u_{ij},v_{ij},\lambda)|^2$, il est aisément mesurable et nous renseigne sur la résolution de l'objet (peu résolu $V^2 \rightarrow 1$, résolu $V^2 \rightarrow 0$), sur sa taille (voir Fig. \ref{Source_disk}) et sur le type d'objet (singulier ou binaire).

\item \textbf{La phase différentielle:} Formulée $\Phi_{ij}(u_{ij},v_{ij},\lambda)=arg\left(V_{ij}(u_{ij},v_{ij},\lambda)\right)-arg\left(V_{ij}(u_{ij},v_{ij},\lambda_0)\right)$, cette observable, qui est tout aussi aisément observable, est directement reliée au déplacement chromatique du photo-centre $\epsilon_{ij}(\lambda)=-\frac{\Phi_{ij}}{2\pi f_s}$ (1\up{ier} ordre du développement en série de MacLaurin de la phase, \citet{2001A&A...377..721J}, où $ f_s =B/\lambda$), ce qui nous permet d'avoir de précieuses informations sur la cinématique de l'objet.

\item \textbf{La clôture de phase:} Formulée $\Psi_{ij}(\lambda)= \sum \limits_{C^{N_{tel}}_2} \Phi_{ij}(u_{ij},v_{ij},\lambda)$, elle désigne la sommation totale des phases sur toutes les combinaisons possibles de tous les télescopes deux à deux. Cette observable est un peu plus couteuse en terme de perte d'information que les autres car le nombre de clôtures de phase mesurable est $N_{\psi}=\frac{(N_{tel}-1)(N_{tel}-2)}{2}$. Elle nous renseigne surtout sur la symétrie de l'objet, ainsi $\Psi \neq 0$ veut dire que l'objet est assymétrique, mais $\Psi=0$ ne permet pas d'affirmer le contraire.\\
\end{itemize}

De ce fait le nombre total d'observables est donc $N_{obs}=2N_{base}+N_{\Psi}$ (voir Eq.\eqref{3.30}). Cela dit, le degré de résolution de l'objet a un impact significatif sur les observables, ce qui crée une sorte de hiérarchie interferométrique, tel que \citet{1989dli..conf..249P} \& \citet{2003A&A...400..795L} l'ont démontré, via le moment de la distribution de flux au nième ordre (appelé facteur de super-résolution par \citet{1989dli..conf..249P}) et qui est noté :

\begin{equation}
M_n.u_1...u_n=\int{\int{I(\alpha)(\alpha-\alpha_0).\tiny{u}_1...(\alpha-\alpha_0).u_n}d\alpha^2},
\label{moment}
\end{equation}

où $ \tiny{u} = \tiny{B} / \lambda $ est la fréquence spatiale, en fonction de la base projeté $\tiny{B} $ et de la longueur d'onde $\lambda$, et $\alpha$ la position angulaire de l'objet observé dans le ciel. En supposant que l'objet est partiellement résolu, la visibilité $ V $ est alors proche de 1; sa taille est donc une fraction de $ B/\lambda $. Dans un tel cas, la plupart du flux se trouve dans une zone où $ | \tiny{u} \tiny{\alpha} |<< 1$, on peut donc écrire en utilisant le moment de la distribution du flux en première approximation pour 3 télescopes (par exemple) que le module de visibilité est:

\begin{equation}
| V | ^ 2 = 1-4 \pi ^ 2 (M _2.u .u),
\label{moment_V2}
\end{equation}

la phase différentielle est:
\begin{equation}
\phi = -2 \pi (M_1 u),
\label{moment_phi}
\end{equation}

et la clôture de phase est:
\begin{equation}
\Psi=\phi(u_{12})+\phi(u_{23})+\phi(u_{31})=\frac{4}{3}\pi^3(M_3 u_{12}.u_{23}.u_{31}). 
\label{moment_psi}
\end{equation}

Notons que $\phi$ est proportionnelle au moment du premier ordre de la distribution de flux, $ 1-| V | ^ 2 $ est proportionnel au moment du second ordre de la distribution de flux et $\Psi$ au 3\up{ème} ordre de la distribution de flux . Ainsi pour une source non résolue, on obtient tout d'abord des informations spatiales à partir de la phase différentielle, mais moins du module de visibilité et encore moins de la clôture de phase ($\Phi>|1-V^2|>\Psi$).\\

\begin{figure}[h!]
\centering
\includegraphics[height=0.7\hsize,draft=false]{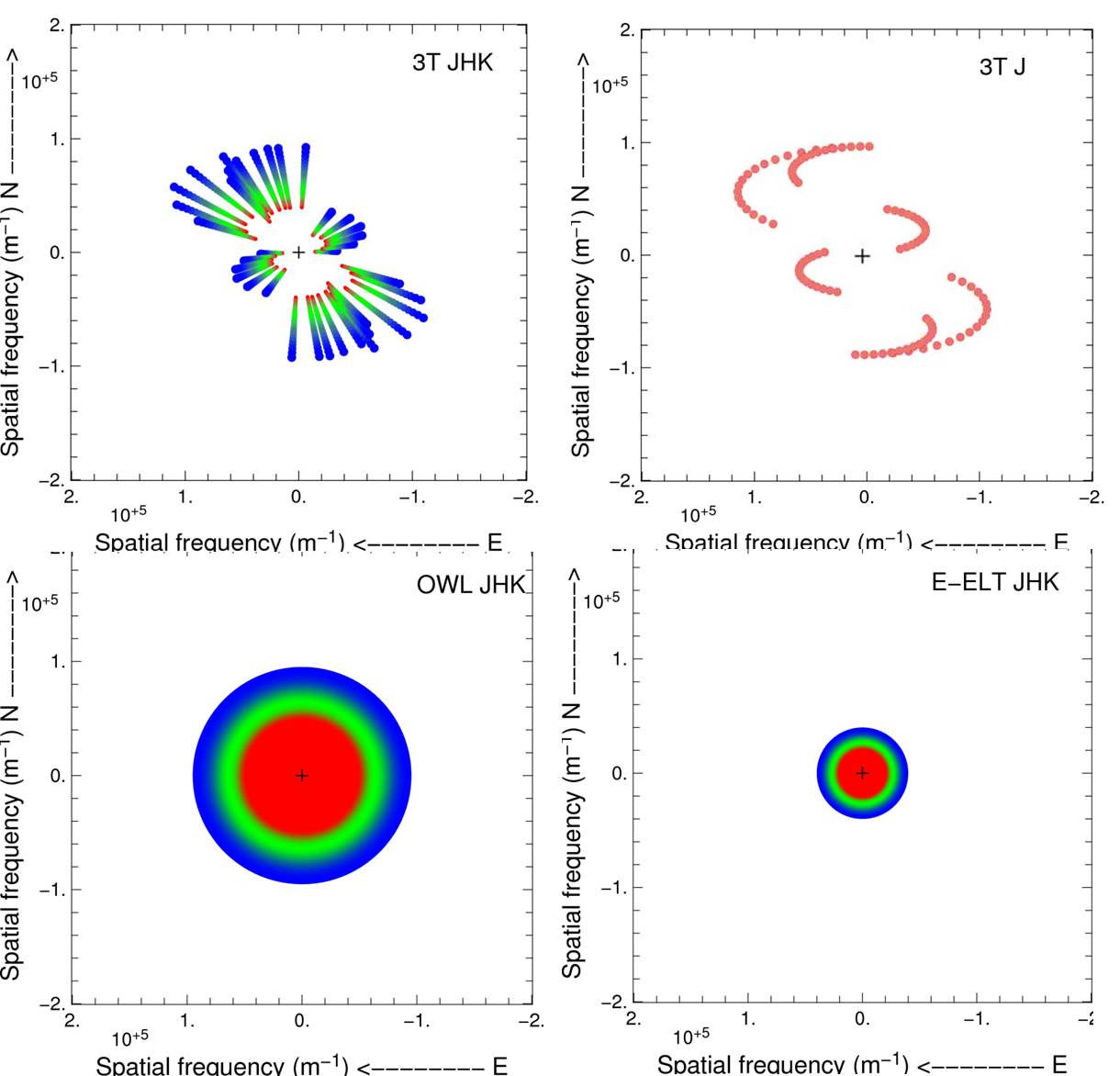}
\caption[Couverture (u,v) du plan spectrale]{Comparaison de différentes couvertures(u,v). En haut à droite: Une couverture(u,v) en bande spectrale $J$ à trois télescopes au VLTI. En haut à gauche: La même couverture mais en bandes spectrales $JHK$. En bas à droite: la couverture (u,v) $JHK$ du futur E-ELT (l'European Extremely Large Telescope), le télescope géant de $39.3$ $m$ de diamètre. En bas à gauche: la même couverture mais d'OWL (Overwhelmingly Large Telescope, appelé aussi OLT), le projet d'un télescope de $100$ $m$ de diamètre abandonné de l'ESO pour l'E-ELT pour des raisons budgétaires. Le code de couleur rouge, vert, bleu correspond respectivement au domaine des bandes spectrales J, H \& K dans le plan (u,v). (Source: cours de F. Millour, VLTI School 2010).}\label{flo_uv-covreage}
\end{figure}

Enfin, il faut noter que l'interférométrie différentielle à longues bases améliore grandement le champ de couverture (u,v) dans l'espace de Fourier, ce que nous permet d'avoir plus de flux et donc de plus amples informations sur l'objet observé (voir Fig.\ref{flo_uv-covreage}).\\

Ainsi, la DI nous permet d'avoir accès à plusieurs grandeurs une fois que notre objet est correctement résolu, entre autres : le spectre, le module de visibilité, la phase différentielle et la clôture de phase. Grandeurs qui nous renseigne sur le type (singulier ou binaire), grâce auxquels on peut déduire l'inclinaison de l'axe de rotation ainsi que la rotation différentielle de l'objet observé, la forme et l'aplatissement. Ainsi que des informations sur l'asymétrie de l'objet et la possibilité d'estimer dans certains cas l'image à haute résolution spatiale de l'objet sans passer par l'ajustement de modèles.\\

Ayant  évoqué dans le premier chapitre les objets de nos études et dans le second la méthode de mesure d'observation choisie (la DI infra-rouge entre autre). Dans la section suivante nous allons expliciter en détail, la déduction du diamètre angulaire d'un rotateur rapide (l'étoile Achernar) via la phase différentielle (notée ici $\phidiff$), mesurée par l'instrument AMBER au VLTI ; un instrument que nous allons aussi aborder sous tous ses aspects aussi bien caractéristique/fonctionnel que techniques (données, réductions et traitement). Mais pour plus amples détails, avec un large aperçu sur les méthodes, instrumentations, techniques et tous les résultats scientifiques utilisés et acquis en interférométrie, lire \citet{2010SerAJ.181....1J, 2011SerAJ.183....1J}.\\

\clearpage

\section{Application: Détermination du diamètre angulaire d'Achernar via la $\mathbf{\phidiff}$ d'AMBER}

La détermination du diamètre apparent des étoiles a depuis toujours intéressé les astronomes. Au moyen âge, le diamètre angulaire $\diameter$ des étoiles les plus imposantes du ciel nocturne était estimé à $2'$ (les limites de la résolution de l'\oe{}il humain). En 1632 Galilée fut le premier à imaginer un système simple mais ingénieux afin de mesurer le diamètre angulaire de Vega, où à l'aide d'une ficelle (d'épaisseur connue) rigide placée à la verticale et s'éloignant suffisamment de celle-ci, jusqu'à ce que le fil recouvre entièrement l'étoile, Galilée estima le diamètre angulaire de Véga à $5"$. Résultat bien évidement bien loin de la vraie valeur mesurée de nos jours avec nos moyens actuels, mais qui correspond assez bien à l'effet rajouté de la perturbation atmosphérique. Une méthode d'estimation théorique du diamètre angulaire fut proposée au 18\up{ième} siècle par Newton qui essaya de déduire à quelle distance de la Terre il faudrait placer le Soleil ($\diameter_{\odot}=30'$) pour que ce dernier ait une magnitude comparable à celle de Véga (i.e. magnitude 0). Ses calculs le menèrent à un diamètre angulaire de 0.2 milli-arc-sec (le type spectral de Véga étant différent de celui du Soleil, le calcul de Newton est alors partiellement biaisé, en réalité  Véga a un diamètre angulaire de 3 milli-arc-sec environ). Ce n'est qu'avec l'essor de l'interférométrie, à partir du 20\up{ième} que la détermination expérimentale du diamètre angulaire des étoiles fut possible, d'abord en utilisant la visibilité \citep{1921ApJ....53..249M}, puis via la phase différentielle $\phidiff$ qui permis en un premier temps à \citet{1986A&A...165L..13T} de déduire la taille angulaire de l'enveloppe de $\gamma$ Cas observée par le I2T $\diameter_{disk}=5.2$ mas, puis avec le GI2T de déduire via  un modèle dit de Poeckert, le diamètre angulaire de  la même étoile ($\diameter_*=1.7$ mas, \citet{1989Natur.342..520M}).\\

Dans la même optique, un travail de déduction des paramètres fondamentaux (diamètre angulaire équatorial $\diameq$ compris) du rotateur rapide Achernar fut entrepris, en ajustant les paramètres d'un modèle de rotateurs rapides aux mesures effectuées par l'instrument AMBER au VLTI lors d'une campagne d'observation, menée par mes directeurs de thèse fin 2009. J'ai rejoint l'équipe et ai commencé mon aventure interférométrique début 2010 où je devais principalement réduire les données et traiter les importants biais instrumentaux qui avaient entaché nos mesures et qui rendaient toute exploitation des données quasi-impossible, et que je détaille ci-dessous. Mais avant d'aller plus loin, familiarisons-nous d'abord avec notre étoile-cible (Achernar), notre grand système interférométrique (VLTI) et notre instrument de mesure (AMBER): \\

\subsection{Achernar}
Le mot Achernar dérive de l'arabe phonétique "Akhir al Nahr", qui signifie littéralement "fin de la rivière". C'est l'étoile principale de la constellation de l'Éridan, qui a été cataloguée par William Herschel en 1783. Achernar n'étant pas visible depuis l'hémisphère nord et donc depuis l'Europe, elle fut découverte dans le ciel australe par les navigateurs du 16\up{ième} siècle. L'Éridan \footnote{Fleuve mythologique grec qui accueillit le corps sans vie de Phaéton, fils d'Hélios (dieu du Soleil), foudroyé par Zeus pour avoir perdu le contrôle du char solaire et failli embraser le monde.} se terminait auparavant par Acamar ($\theta$ Eridani), une étoile visible à la limite de l'hémisphère nord depuis l'Europe. Dans la littérature, certains philosophes contemporains estiment qu'Achernar, Canopus et Fomalhaut sont les trois étoiles qui éclairent le ciel du purgatoire, tel que décrit par Dante Alighieri (1265-1321) dans son \oe{}uvre de "la Divine Comédie" (probablement influencé par le travail de l'astronome perse Al-Farghani (805-880)).\\

\begin{figure}[h!]
\centering
\includegraphics[height=0.7\hsize,draft=false]{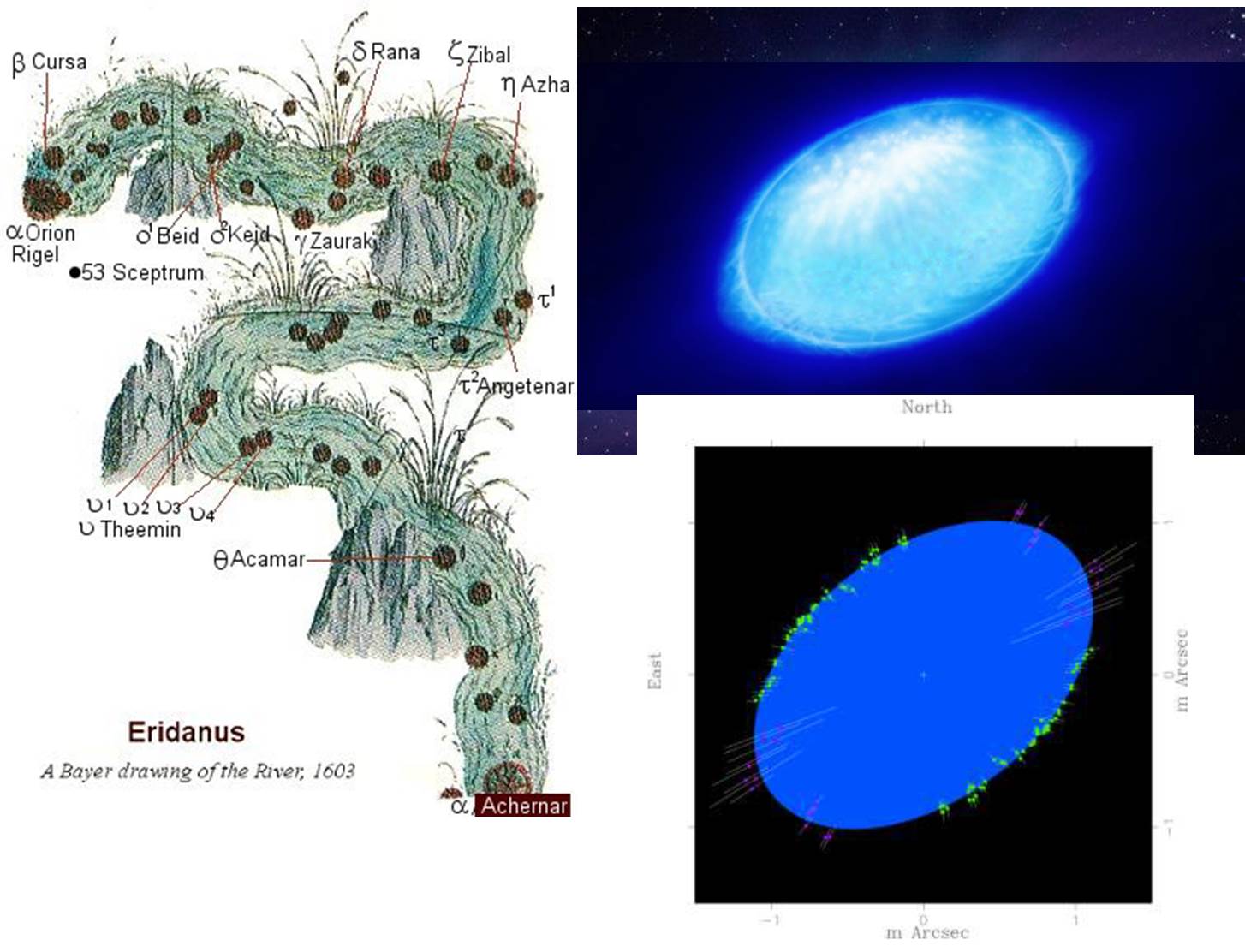}
\caption[Achernar et la constellation de l'Éridan]{A gauche: Une vue d'artiste de la constellation de l'Éridan (le fleuve mythique) et des étoiles qui le constitue. A droite en haut : une vue d'artiste d'Achernar. A droite en bas : Représentation de l'aplatissement d'Achernar  \citep{2003A&A...407L..47D}, une première à l'époque.}\label{Achernar}
\end{figure}

Achernar ($\alpha$ Eri, HR 472, HD 10144) est une étoile de masse $M = 6.1 \Msun$ \citep{1988BAICz..39..329H} éloignée de la terre d'une distance estimée entre $d = 44.1pc$ \citep{1997A&A...323L..49P} et $d = 42.7 pc$  \citep{2007A&A...474..653V}. De type spectral B6Vep, elle a une température effective apparente $\Tmean=15000$ $K$ \citep{2006A&A...446..643V}. Étoile tournant à une grande vitesse $\vsini = 225 \kms$ (estimée spectroscopiquement par \citet{1982ApJS...50...55S}), elle est fortement aplatie avec un rapport d'aplatissement généralement estimé à $1.41-1.56$ avec une précision de $3\%$ \citep{2003A&A...407L..47D}. Des études ultérieures \citep{2006A&A...453.1059K} rapportent une perte de masse polaire. Un petit disque résiduel a été observé par  \citet{2008ApJ...676L..41C}, et la présence d'un compagnon a été déterminée par \citet{2008A&A...484L..13K} avec un cycle de 7 ans environ, qui peut expliquer aussi la périodicité d'apparition/disparition du disque circumstellaire d'Achernar. Cependant, \citet{2008A&A...486..785K} a étudié la contribution de l'environnement circumstellaire (CSE) pour expliquer la possible importance du fort aplatissement apparent d'Achernar. Récemment \citep{2012A&A...545A.130D}, on a pu déterminer quatre paramètres fondamentaux d'Achernar, à savoir son rayon équatorial  $R_{eq}=11.6 \pm 0.6 \Rsun$, sa vitesse de rotation équatoriale $V_{eq}=298 \pm 9 \kms$, son inclinaison $i=101 \pm 5.2^\circ$ et l'angle de position du grand-axe $PA_{rot}=34.7 \pm 1.6^\circ$. Ces résultats ont été confirmés dans les barres d'erreur par \citet{2014A&A...569A..45H} \& \citet{2014A&A...569A..10D}.

\subsection{Le VLTI}
C'est sous l'égide de l'ESO (l'Observatoire Européen Austral) que fut décidé, en 1983, le lancement du projet du VLT (Very Large Telescope); la construction d'une série de télescopes de très grandes tailles, à fonctionnement individuel et indépendant, dotés d'instruments et d'imageurs photométriques, polarimétriques et spectroscopiques. Quatre ans plus tard, encouragé par les dernières avancées interférométriques (voir Fig.\ref{Interfero_Moderne}) et tous les résultats obtenus en matière de détermination du rayon angulaire de quelques étoiles et de leurs environnement proches, l'ESO décida d'inclure le mode interférométrique ce qui donna officiellement naissance au projet VLTI (Very large Telescopes Interferometer) en 1987.\\

Après une recherche, à travers le monde, du site adéquat pouvant accueillir ce projet dans des conditions météorologiques et atmosphériques optimales, l'ESO choisit en 1990 d'installer le VLT/VLTI au nord du Chili, aux portes du désert de l'Atacama, perché à 2635 m, sur le mont Paranal, au sommet d'une chaine montagneuse longeant, par l'ouest l'océan pacifique baigné par les courants froid descendant du "Gulf stream" (dit courant de Humboldt), ce qui limite considérablement l'évaporation océanique et de surcroit l'humidité. Inversement, l'air chaud ascendant du désert de l'Atacama, à l'est, aidé par le relief naturel du site empêche le peu d'humidité d'arriver au site d'observation, ce qui incite toute formation nuageuse à se former en contre-bas du flanc de montage sur la côte. Au-delà du désert de l'Atacama, la cordillère de Domeyko et surtout la cordillère des Andes protégeant le site de toute formation nuageuse coté est. Cette configuration naturelle fait que le ciel de Cerro Paranal est l'un des plus purs au monde (pluviométrie quasi-nulle, 350 à 360 nuits claires par an et perturbation atmosphérique très faible). Les travaux de construction furent entrepris une année plus tard et durèrent plus de 7 ans.\\

\begin{figure}[h!]
\centering
\includegraphics[height=0.9\hsize,draft=false]{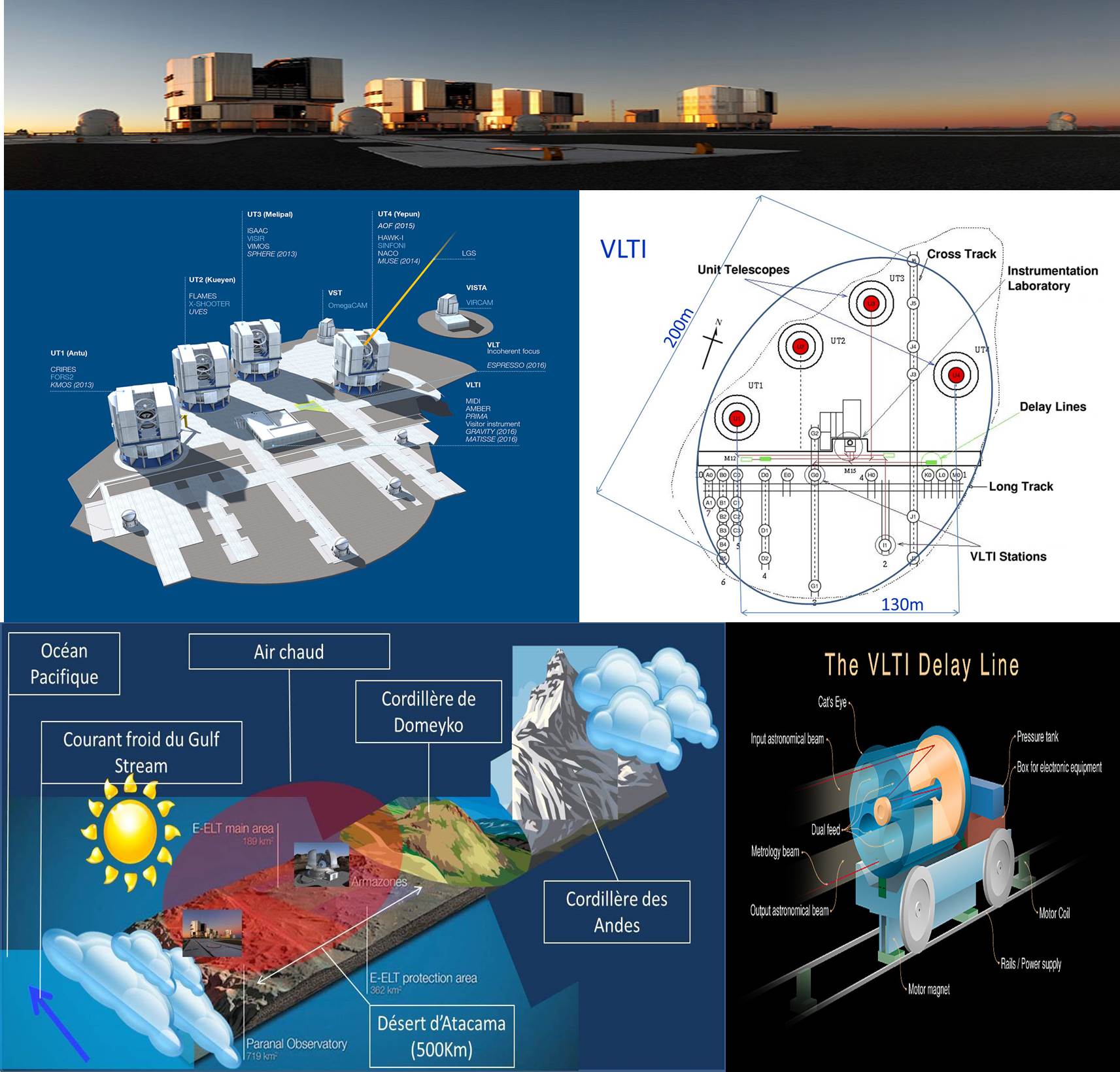}
\caption[Le VLTI sous tous ses angles]{\textbf{En haut:} Une vue panoramique sur l'ensemble des télescopes du VLTI (UT \& AT) -crédit ESO-. \textbf{Au milieu, à gauche:} Un plan détaillé 3D de l'ensemble des installations et instrument VLTI actuels et futurs -crédit ESO-. \textbf{Au milieu, à droite:} Un plan d'ensemble, vue du haut, du VLTI avec échelle des distances et toutes les positions possibles que peuvent adopter les AT (Source: ESO). \textbf{En bas, à gauche:} Un plan topologique qui explique le positionnement géographique unique au monde, avec des conditions météorologiques d'exception, du VLT sur le mont Paranal. \textbf{En bas, à droite:} Un schéma descriptif des chariots des lignes à retard, et se qui trouvent dans les galeries de tunnels souterrains qui relient l'ensemble des télescopes du VLTI entre eux -crédit ESO-.}\label{VLTI}
\end{figure}

L'orientation et l'emplacement des différents télescopes furent étudiés de sorte à optimiser les couvertures (u,v) des observations interférométriques. Deux types de télescopes furent installés. Les UT (Unit Telescopes), ont un miroir de 8.2 m chacun; ils sont fixes et au nombre de 4, et possède des noms Mapuches: Antu (Soleil) -UT1-, Kueyen (Lune) -UT2-, Melipal (croix du sud) -UT3- et Yepun (Venus) -UT4-. Ils sont dotés chacun d'un système puissant d'optique active pour palier à toutes déformations de leurs miroirs géants. Ils sont également pourvus de petites ouvertures qu'on ouvre quelques minutes avant le coucher du Soleil afin de veiller à l'équilibre thermique des dômes qui les abritent avant chaque nuit d'observation pour éviter toute perturbation atmosphérique à l'intérieur des dômes. Ils sont complétés par un deuxième type de télescopes, les AT (Auxiliary Telescope), mobiles et placés sur rails, ils ont des miroirs de 1.8 m de diamètre. Ces 4 télescopes ne sont prévus que pour fonctionner en mode interférométrique. Ils peuvent être placés selon un grand nombre de combinaisons de taille et d'orientation possible, près de 4000 en tout, ce qui permet une grande couverture de l'espace de Fourier, augmentant ainsi les capacités de reconstruction d'images du VLTI (voir Fig.\ref{VLTI}). Au début, le rôle des AT était régit par 2 sidérostats de 40 cm de diamètre, avant leur remplacement définitif en 2003. Les AT sont couverts par des dômes compacts et sont totalement autonomes.\\

Le sous-sol du VLTI abrite des tunnels contenant des lignes à retard qui servent à corriger la différence de chemin optique entre les télescopes. Les faisceaux lumineux sont ensuite dirigés vers les instruments de combinaison situés dans une salle appelée le "laboratoire focal". Actuellement le VLTI compte plusieurs instruments de recombinaison (AMBER, MIDI, PIONIER et PRIMA), mais à l'origine et lors de sa mise en marche en 2001, le VLTI ne comptait qu'un seul recombinateur test dans l'infra-rouge proche; VINCI (Vlt INterferometer Commissioning Instrument) qui ne pouvait combiner que la lumière de 2 télescopes, et qui était en fait une copie de l'instrument FLUOR de l'interféromètre IOTA. Les premières franges d'interférence furent acquises le 16 mars 2001 sur l'étoile $\alpha$ Hydrae. La plus exceptionnelle observation réalisée par VINCI a été celle qui a permis pour la première fois de mesurer l'aplatissement de l'étoile Achernar avec une soixantaine points de visibilité environ \citep{2003A&A...407L..47D}. Enfin, VINCI a permis de repérer puis de corriger un bon nombre de défauts et de régler un bon nombre de problème du VLTI, ouvrant  de ce fait la voie à une nouvelle génération de combinateurs plus puissants et plus performants.\\

\subsection{AMBER}
\subsubsection{Un peu d'histoire:}
Juste avant la mise en service du VLT entre 1996-1997, l'ESO constitua des groupes internationaux de réflexion (ESO Workshop on Science with the VLT Interferometer, 18-21 June 1996) afin de proposer et d'établir au "Laboratoire Focal" de nouveaux projets et instruments dédiés à l'interférométrie pour le VLTI, pour la combinaison d'au moins 3 Télescopes (avec prise en compte de la turbulence atmosphérique et correction via l'optique adaptative (dans le visible pour les AT, IR pour les UT), dans l'infra-rouge et dans le visible (bandes spectrale J,H,K et R). En 1997 a eu lieu, à Garching (Allemagne),  une réunion dédiée à l'instrumentation du VLTI  qui a débouché sur quatre décisions majeures:\\
\begin{itemize}
\item{} Le lancement des travaux de réalisation d'un recombinateur test dans l'infra-rouge proche, VINCI, en attendant la mise en place d'instruments plus performants. 
\item{} La discussion de la nécessité d'équiper les UT de système d'optique adaptative pour un usage en mode interférométrique et le lancement du projet MACAO (Multi-Application Curvature Adaptive Optics).
\item{}La mise en place d'un recombinateur à 2 télescopes dans l'infra-rouge moyen, en bande N (MIDI ; MID-infrared Interferometer), projet dirigé par le Max-Planck-Institut für Astronomie (MPIA) à Heidelberg. 
\item{}La mise en \oe{}uvre d'un combinateur à 3 télescopes (3T) dans l'infrarouge, bande J,H et K ; AMBER (Astronomical Multi-BEam combineR).\\
\end{itemize}

Ce n'est qu'en 1999 que le nom d'AMBER a été proposé et accepté et qu'un consortium de travail regroupant 5 instituts européens  (Observatoire de la Côte d'Azur, Laboratoire Universitaire de Nice, Laboratoire Astrophysique de l'Observatoire de Grenoble, Max-Planck Institut f\"ur Radioastronomie, et Observatorio Astrofisico di Arcetri) fut établi. Chapeauté par R. Petrov du Laboratoire de l'université de Nice de l'époque, il a été décidé de doter AMBER de 3 modes d'observations spectraux (3T\_JHK); Low Resolution (LR $\frac{\Delta\lambda}{\lambda}=35$),  Medium Resolution (MR $\frac{\Delta\lambda}{\lambda}=1500$), \& High Resolution (HR $\frac{\Delta\lambda}{\lambda}=12000$), et ce pour offrir un large choix d'observations et d'études des phénomènes locaux et globaux des objets ciblés. En 2000 a été établie la revue de conception préliminaire, où furent signés les accords entre l'ESO et les différents instituts en charge d'AMBER. En 2001 a été approuvée la revue de conception finale. En 2003 l'Europe accorda son acceptation préliminaire au projet. Et en 2004 a eu la 1\up{ière} livraison, l'assemblage et les tests à Paranal, avec acquisition des premières franges sur l'étoile Sirius via les sidérostats, avec succès.\\ 

AMBER se propose donc de mesurer à la fois, les spectres, les modules de visibilités différentielles, les phases différentielles et la clôture de phase (qui représente une avancée majeures par rapport à VINCI). Il a été conçu pour l'étude des noyaux galactiques actifs (AGN; Active Galactic Nuclei), des exo-planètes de types gazeuses géantes et chaudes (proches de leurs Soleil) et des étoiles jeunes (Young Stellar Object ou YSO). Dans les longueurs d'onde d'observation dans le proche infra-rouge, comprises entre $1-2.5$ $\mu m$, autour de la raie Brackett $\gamma$, AMBER permet une acquisition d'information simultanée de la photosphère et de l'environnement circumstellaire de la plupart des étoiles chaudes actives. Rajoutons à cela le fait qu'AMBER possède une des résolutions spatiales les plus fines parmis tous les instruments du VLTI (1.5 à 2 mas selon qu'on est en bande J ou K, pour une longueur de base interférométrique maximale, i.e. 200 m), où il est possible de résoudre une étoile de 5 $\mathrm{D}_{\odot}$ à $30$ $pc$ à 100\% et à 95\%  à une distance de 200 pc, ce qui permet d'étudier l'effet de l'aplatissement des étoiles et leurs assombrissements centre-bord et gravitationnel selon leurs distances et/ou tailles réelles, sans oublier l'aspect différentiel des phases qui permet l'étude cinématique des surfaces d'étoiles et de leur environnement. Tous ces éléments cités font d'AMBER un instrument d'exception pour l'étude des étoiles actives chaudes, des  rotateurs rapides et de leurs environnements proches, ce qui explique l'usage exclusif de cet instrument pour l'observation de l'ensemble des étoiles étudiées dans mon projet de thèse. Il est important de noter qu'AMBER n'est pas l'instrument le plus performant en ce qui concerne la calibration des modules de visibilité. Un manque qui peut être compensé par l'utilisation d'autres instruments complémentaires, tel que CHARA et ses 350m de base interférométrique par exemple.\\

\begin{sidewaysfigure}[b!]
\centering
\includegraphics[width=1.\hsize,height=0.2\hsize,draft=false,angle=0]{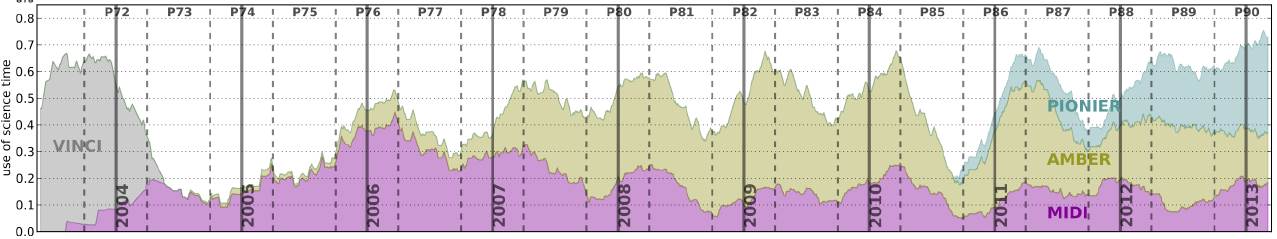}
\caption[Temps d'observation sur AMBER]{Temps d'observation sur les recombinateurs VINCI, MIDI, AMBER et PIONIER de 2003 à 2013 (J.P. Berger VLTI School 2013).}\label{stat_AMBER}
\end{sidewaysfigure}

Ce n'est qu'en 2005, en plus de quelques ajustements, que les premières retombées scientifiques furent réalisées (période d'observation P74) avec l'observation GTO (Guaranteed Time Observation) du disque de MWC297 \citep{2005prpl.conf.8395B}, de la Be $\kappa$ Canis Majoris \citep{2007A&A...464...73M}, la Wolf-Rayet (WR) $\gamma^2$ Velorum et la lumineuse variable bleue (LBV) $\eta$ Carinae \citep{2006SPIE.6268E..02M, 2007A&A...464..107M, 2007A&A...464...87W} en mode MR et HR avec les UT (le mode interférométrique AMBER avec les AT n'a été possible que depuis 2007). Depuis le nombre d'observations via AMBER n'a cessé de croitre jusqu'à nos jours (voir Fig.\ref{stat_AMBER}), en apportant de plus en plus de contributions, sur plusieurs types d'objets astrophysiques, aux méthodes d'ajustement de modèles (Model Fitting) et de reconstruction d'images (Image Reconstruction), y compris polychromatiques \citep{2011A&A...526A.107M}. Pour plus de détails sur la contribution scientifique d'AMBER lire \citet{2012POBeo..91...21P}. Dans les faits et pour résumer, AMBER est l'instrument qui a apporté le plus de contributions en matière de papiers scientifiques dans le domaine de l'interférométrie astronomique, où ce dernier détient, à ce jour, un tiers de toutes publications mondiales dans le domaine.\\

\clearpage

\subsubsection{Des télescopes au laboratoire focal:}
\begin{figure}[h!]
\centering
\includegraphics[height=0.7\hsize,draft=false]{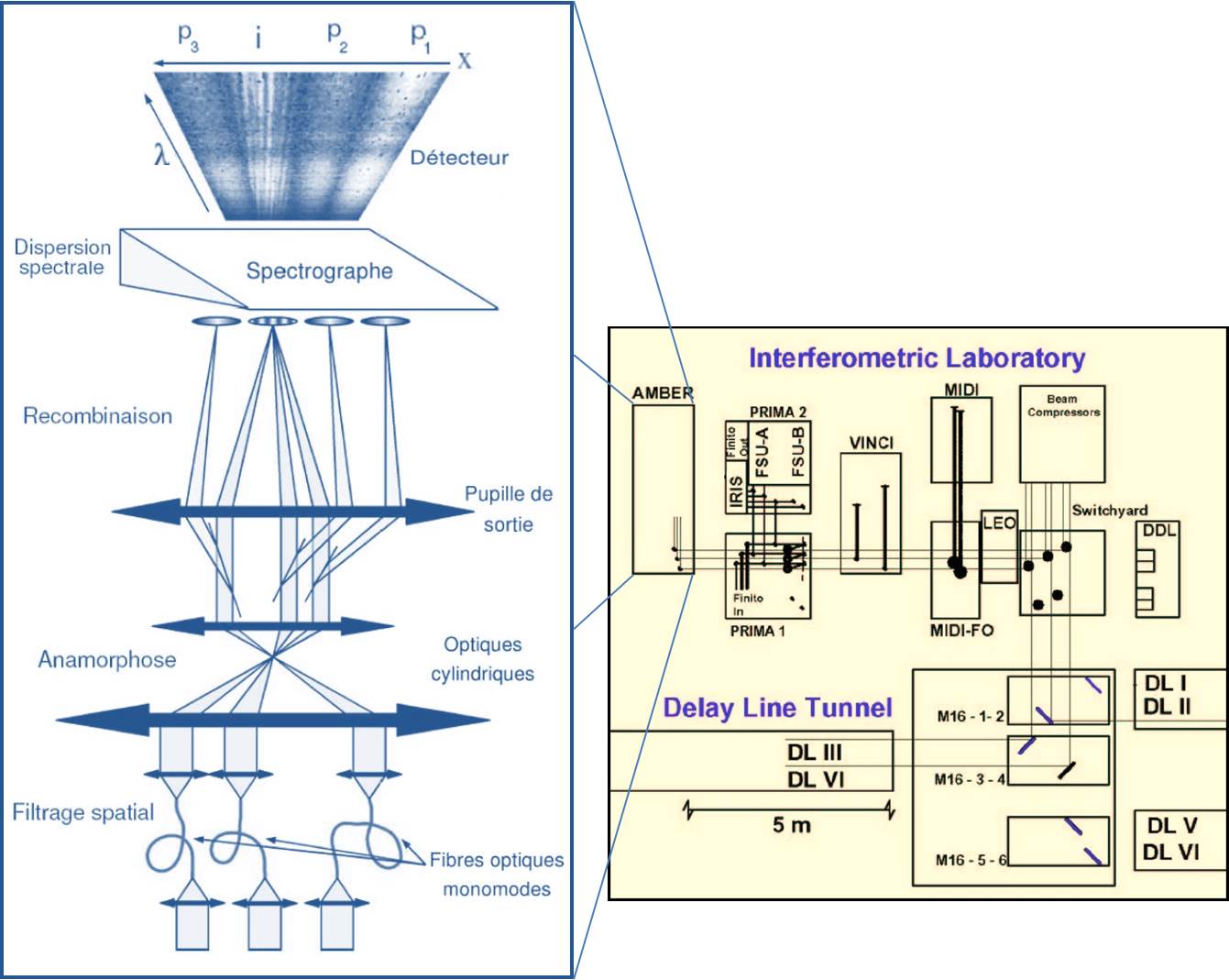}
\caption[Schéma descriptif d'AMBER]{Schéma descriptif d'AMBER;  Parcours des faisceaux lumineux depuis les tunnels sous-terrain, jusqu'au laboratoire focal VLTI. (Fig. inspirée de \citet{2006PhDT........46M} \& \citet{2004SPIE.5491..944G}).}\label{manip_AMBER}
\end{figure}

Le principe de fonctionnement d'AMBER, est décrit en détail ci-dessous et dans la Fig.\ref{manip_AMBER} : Une fois la lumière collectée par les télescopes  (avec correction en OA sur les UT par MACAO et par STRAP - System for Tip-tilt Removal with Avalanche Photodiodes- en tip-tilt pour les AT), les faisceaux empruntent les tunnels sous-terrain où ces derniers subissent une correction de l'OPD (Optical Path Difference) via les charriots de la Fig.\ref{VLTI} (en bas à droite). Via des miroirs, les faisceaux lumineux sont conduits à l'intérieur du laboratoire focal où ils doivent d'abord être conditionnés par un réducteur de faisceaux (beam compressor), avant d'arriver vers le recombinateur de notre choix (AMBER dans notre cas).\\

Pour une cohérence optimale des faisceaux, AMBER bénéficie de l'aide du capteur d'inclinaison IRIS (InfraRed Image Sensor)  qui peut mesurer et contrôler la dérive de l'image introduite à l'intérieur du VLTI entre le foyer coudé de chaque télescope (AT/UT) et le laboratoire focal, de 4 faisceaux simultanément, dans les trois bandes spectrales  J H \& K. Cette dérive est due aux effets de la dispersion atmosphérique latérale qui ne sont évidemment pas corrigés par STRAP/MACAO. Parallèlement, et depuis la période P80, un suiveur de franges peut être associé à AMBER: FINITO (Fringe-tracking Instrument of NIce and TOrino). Celui-ci mesure la variation de l'OPD de 3 faisceaux lumineux simultanément, induit par la turbulence atmosphérique, puis fournit les renseignements requis à la boucle de commande des lignes à retard pour compenser la perturbation, augmentant ainsi le temps d'exposition cohérent de quelques millisecondes à plusieurs secondes, ce qui rend possible la mesure de la clôture de phase et améliore significativement la précision et la sensibilité (ou SNR) de l'instrument auquel il est associé (AMBER/MIDI). Initialement les magnitudes limites AMBER atteintes, en bande K, avec les UT sont de 7, 4, et 1.5 respectivement pour les modes LR, MR, et HR, et de 5.1 et 1.6 en LR et MR avec les AT, mais avec FINITO il est possible d'atteindre une magnitude en H et K égale à 3, avec les AT, toute résolution spectrale confondue. Dans les fais et malgré tous ces grands et récents apports technologiques, le nombre d'étoiles chaudes actives abordables par AMBER (et MIDI) reste assez faible. Pour les Be classiques par exemple, seule une dizaine d'étoiles, ayant une magnitude K<3 sont observables (et jusqu'à 5 avec les AT's), avec AMBER en mode HR et avec le renfort de FINITO, depuis l'hémisphère Sud.\\

\subsubsection{Le fonctionnement d'AMBER:}
Une fois les faisceaux lumineux arrivés au recombinateur AMBER, ces derniers sont conduits par des fibres optiques monomodes correspondant au canal spectral d'observation choisi (J et/ou H et/ou K). Les fibres optiques permettent un filtrage spatial en ne gardant que la partie centrale de la tâche d'Airy, afin d'optimiser la cohérence interférométrique. Cela entraine par contre une variation de flux, qui peut être calibrée ultérieurement à l'aide de la mesure spectro-photométrique propre à chaque faisceau sur le détecteur, qui a été prévue à cet effet.  A la sortie des fibres optiques les faisceaux subissent une déformation anamorphique, à l'aide de deux lentilles cylindriques, et ce dans le but de réduire la taille des faisceaux, dans le sens d'enregistrement des longueurs d'ondes dans le détecteur, au minimum. Ce n'est qu'ensuite que s'opère la recombinaison grâce à des lames semi-réfléchissantes, avant d'être dispersés par un spectrographe (au choix, selon les trois modes LR, MR ou HR cités plus haut) et enfin enregistrés sur un détecteur de type Hawaï 512x512, selon l'ordre décrit par la Fig.\ref{manip_AMBER} ($P _3$, $i$ ,$P _2$, et $P _1$), où $i$ est le canal interférométrique et les $P_i$ ($i=1..3$) les trois canaux photométriques sur le détecteur, relatifs au faisceaux interférés. Il faut noter aussi que quelques pixels, avant le canal $P_1$ sont réservés à la mesure du courant noir (Dark current), à des fins de calibration photométrique rigoureuse, post-acquisition. Pour plus de détails sur l'aspect technique d'AMBER, lire \citet{2007A&A...464...13R} et/ou lire la documentation ESO d'AMBER (http://www.eso.org/sci/facilities/paranal/instruments/amber/doc.html).\\

\subsubsection{Observer avec AMBER:}
Comme nous venons de le voir, la perturbation atmosphérique reste le principal handicap de toute observation terrestre et nécessite un bon nombre de dispositifs correctifs tout le long de la chaine instrumentale d'acquisition. En effet celle-ci provoque en général 3 effets indésirables sur les mesures astronomiques/astrophysiques: \textbf{La scintillation} (la variation de l'intensité lumineuse qu'on peut observer à l'\oe{}il nu sur certaines étoiles) provoquée par l'étalement et/ou la concentration énergétique du front d'onde, \textbf{l'agitation} de l'image perçue au plan focal du télescope et provoquée par une fluctuation temporelle du front d'onde,  et enfin \textbf{l'étalement} de l'image provoqué par un déficit de cohérence spatiale à l'entrée de la pupille de l'instrument (L'observation en astrophysique, François \citet{Lebrun08}).\\

Les effets de ces nuisances ont des impacts significatifs sur l'interférométrie du point de vue d'une altération de la phase authentique des franges et d'une baisse du contraste \footnote{Des franges parfaitement contrastées ($C=V=1$, voir Eq.\eqref{3.10}) correspondent à une variation de l'intensité lumineuse jusqu'à son annulation (0 à 100\%). Et vis-versa des franges peu contrastées s'appliquent à une intensité lumineuse qui ne varie que très légèrement par rapport à l'intensité moyenne.}. Ces derniers, aléatoires et très variables dans le temps, se révèlent très compliqués à gérer pour les instruments recombinateurs de faisceaux, d'où l'utilisation, dans les limites d'un SNR tolérable, d'un temps de pose aussi réduit que possible. Ainsi, il faut jouer sur le temps d'intégration (NDIT - Number of individual Detector Integration Time-; de 1 à 100 secondes, selon l'agitation atmosphérique) des poses courtes (DIT -Detector Integration Time -; une moyenne de100 ms pour les AT et 50 ms pour les UT).\\

Ce qu'il faut surtout retenir ici c'est qu'en plus de l'étalement du flux de l'objet (qui est mesuré par la LSF - Line Spread Function- dans l'espace de Fourier et qui peut être inférieur au critère de Rayleigh, selon l'importance de l'agitation atmosphérique),  il y a aussi une perte de flux relativement importante qui est occasionnée par l'interférométrie (tout dépend de la configuration et de la longueur des bases), d'où la complexité technique et instrumentale pour corriger l'ensemble de ces effets durant l'acquisition des données de visibilités. Néanmoins les corrections, citées ci-haut, restent insuffisantes et la visibilité de notre objet source nécessite donc une correction par celle d'un objet non résolu (et/ou de diamètre angulaire connu), qu'on nomme "calibateur", et qui varie avec la fluctuation de l'atmosphère. Cette mesure connue sous le nom de "visibilité instrumentale" (qui n'est pas spécifique que pour AMBER) est l'entité par laquelle doit être divisée la visibilité mesurée pour déterminer la visibilité réelle et cette opération est effectuée lors de la réduction des données (voir ci-dessous). De préférence le calibrateur est choisi pour être stable, de même type spectral et aussi proche que possible de l'objet science, pour éviter d'avoir d'importantes différences de polarisation qui peuvent grandement influer sur la fonction de transfert \citep{2008SPIE.7013E..0FL}. La séquence d'observation optimale doit de préférence alterner (de 15 min à 1 h chacune) le calibrateur, l'objet science, calibrateur, science, ...etc. Enfin, notons aussi qu'un court échantillonnage du ciel est nécessaire pour effectuer, lors de la réduction des données (explicitées en détail ci-dessous), une correction relative à la réponse de la totalité des pixels de la caméra CCD vis-à-vis d'une lumière uniforme. Les nuits d'observation à Paranal pouvant être coupées en deux programmes, il faut aussi prendre en compte le temps, d'une demi-heure à peu près, nécessaire à la mise en route et configuration de l'ensemble des installations et instruments VLTI, et dans tous les cas d'un temps 15 à 20 min nécessaire au pointage, puis au calcul et enregistrement surtout d'importants paramètres primordials à la déduction des mesurables interférométriques (visibilités, phases, ...etc.) lors de la phase de réduction, via un procédé nommé P2VM (Pixel to Visibility Matrix), expliqué dans le sous chapitre ci-dessous.\\

Enfin, une observation sur AMBER, ou sur tout autre type d'instrument au VLT/VLTI, nécessite 3 étapes étalées sur au minimum une année: tout d'abord de préparation et de justifications scientifique et technique qu'il faut matérialiser par une proposition écrite (un proposal) via des formulaires d'applications téléchargés et déposés sur le compte du demandeur du site d'User Portal de l'ESO avant la clôture de la deadline des deux sessions annuelles réservées à chaque période d'observation. Une fois la proposal expertisée, quelques mois plus tard, via des experts et sous réserve d'acceptation, le responsable de projet et son équipe se retrouvent affectés d'un temps et d'un mode d'observation (qui peut être effectué à distance, comme le "service mode" par exemple). L'équipe se doit alors de minutieusement préparer ses observations à l'avance et veiller à choisir les bases et couverture (u,v) qui correspondent le plus aux types et natures des objets qu'elle souhaite étudier via le logiciel gratuit ASPRO2 (Astronomical Software to PRepare Observations) du JMMC (Jean-Marie Mariotti Center) ou avec VisCalc (Visibility Calculator) en libre utilisation sur le site de l'ESO. Le choix des calibrateurs est aussi très important et peut être préparé avec SearchCal du JMMC ou bien avec CalVin de l'ESO. La dernière étape consiste à résumer toutes les informations relatives à nos futures observations sur forme de fichier "OBs" (Observations Blocks) via le logiciel P2PP (Phase 2 Proposal Preparation). Chaque fichier doit contenir pour chaque type d'objet (science/calibrateur) et mode d'observation spectrale, des informations astronomiques qu'on peut aisément retrouver sur le site SIMBAD de l'université de Strasbourg et des informations techniques liées à l'exposition et le temps d'intégration souhaité.  Une fois tout cela accompli, il ne reste plus qu'à faire le voyage à Paranal ou bien observer à distance avec l'astronome de nuit via Skype (par exemple). Dans le second cas toutes les données météos et atmosphériques sont consultables en direct sur le lien du site ESO ambient condition database (http://archive.eso.org/asm/ambient-server?site=paranal). Enfin, il faut noter qu'en cas de mauvaises conditions météo, ce qui est très rare à Paranal, la nuit d'observation n'est ni remboursée ni échangée par l'ESO.%, il nous reste plus donc qu'à croiser les doigts et a espérer que la chance soit de notre côté.\\

\subsubsection{Réduction des données:}
Après chaque nuit d'observation réussie, les données brutes d'AMBER ("Rawdata") sont automatiquement  enregistrées dans les serveurs de l'ESO sous format FITS (Flexible Image Transport System). Durant toute l'année, elles sont à disposition de l'équipe chargée du projet où seul le PI (Principal Investigator) y a accès via son compte ESO pour téléchargement. Passé ce délai, les données deviennent publiques et toute personne ayant accès à internet, à travers le monde, peut librement les télécharger depuis le site de l'ESO Archive Query Form.\\

Avec les "Rawdata" en notre possession il nous est impossible de faire une quelconque étude scientifique sans d'abord leur faire subir toute une série de calibrations et de réductions. Un outil de réduction performant nommé "amdlib" (AMber Data LIBrary) a été mis au point, en langage C et interface graphique Yorick (http://yorick.sourceforge.net/), par le JMMC (à téléchargement libre : http://www.jmmc.fr/data\_processing\_amber.htm). Celui-ci a connu plusieurs versions et améliorations, et de nos jour la version 3.0.8 est utilisable sur des systèmes d'exploitations LINUX ou MAC sans installation préalable. Le concept de la réduction des données brutes est assez simple et est résumé ci-dessous sous forme de notes (savoir-faire acquis et conforté lors de ma participation aux deux dernières VLTI School, de 2010 et 2013):\\

\begin{itemize}
\item{}  Les premières calibrations d'AMBER s'opèrent via la "Bad Pixel Map" qui répertorie les pixels défectueux de la caméra et la "Flat Field map" qui recense la réaction des tous les pixels du CCD par rapport à un éclairage stable et uniforme (ces deux cartes sont périodiquement mises à jour et disponibles sur le site Quality Control and Data Processing de l'ESO -http://www.eso.org/observing/dfo/quality/AMBER/qc/qc1.html-). La calibration du rayonnement thermique ambiant "Dark Field"  et la calibration "Sky" (cité plus haut) sont automatiquement gérés indépendamment de l'exécution des commandes amdlib ; "amdlibLoadBadPixelMap" et "amdlibLoadFlatFieldMap". L'interférogramme calibré peut être visualisé via la commande "amdlibShowRawData".

\item{}  L'extraction des modules de visibilités et des phases différentielles des "Rawdata" avec un maximum de gain  en rapport signal-à-bruit (SNR) amdlib utilise une méthode appelée P2VM (Pixel to Visibility Matrix) qui est bien expliquée par \citet{2007A&A...464...29T}, et qui n'est qu'une version améliorée et optimisée de la méthode ABCD dont le principe est de relier la mesure de l'intensité de 4 points temporels différents (A, B, C \& D) sur une frange individuelle, durent une période de modulation d'intensité due à la perturbation atmosphérique (d'une pause à une autre), afin de déterminer le module de visibilité $v=\frac{\pi}{\sqrt{2}}\frac{\sqrt{(I_A-I_C)^2+(I_B-I_D)^2}}{I_{tot}}$ et la phase $\Phi=\arctan \left(\frac{I_A-I_C}{I_B-I_D}\right)$ \citep{1999PASP..111..111C}. La P2VM exige une série de mesures propres à AMBER réalisée grâce à une lumière artificielle instrumentale et cohérente qui éclaire toutes les trois voies d'entrée de l'instrument, faites d'abord sur chaque voie d'entrée, puis pour chaque configuration interférométrique, en phase puis en quadrature de phase (i.e. 10 mesures pour une recombinaison à 3 télescopes par exemple) afin de d'abord définir la "Visibility to Pixel Matrix", qui par inversion permet de déduire la P2VM et enfin de calculer les modules de visibilités, les phases et par conséquent les clôtures de phase (nos "OI data"). Il faut noter que chaque changement de mode spectral d'observation entraine systématiquement un recalcul et un enregistrement des paramètres de calibration nécessaires à la soustraction de la P2VM lors de la réduction des données. La commande qui permet de réaliser cela est "amdlibComputeAllP2vm" et celle qui permet sa visualisation est "amdlibShowP2vm".

\item{}  Une fois le calcul de la P2VM  effectué, il ne reste plus qu'à déduire les "OIdata"\footnote{Pour plus d'informations sur les OIDATA (ou OIFITS) voir la documentation: "A Data Exchange Standard for Optical (Visible/IR) Interferometry", Prepared by NPOI and COAST, First Version: 10 January 2001, Last Update: 7 April 2003 (ou bien \citet{2005PASP..117.1255P}), et qui est elle même inspirée du "Definition of the Flexible Image Transport
System", NOST, 1999.} pour chaque temps d'intégration de nos objets science et calibrateur avec la commande "amdlibComputeAllOiData". La visualisation des ces données optique interférométrique est permise individuellement via "amdlibShowOiData".

\item{}  Il nous est possible et même préférable (pour un gain de calcul et d'un espace mémoire ultérieur) de concaténer toutes nos données relatives à une série temporelle du même objet et de les assembler en une seule via "amdlibComputeAllOiData". 

\item{}  A ce stade, nous pouvons effectuer une sélection des séquences d'observation "frames" qui nous parait la plus appropriée, comme un pourcentage sur le SNR par exemple, avec la commande "amdlibPerformAllFrameSelection".

\item{}  Maintenant qu'on a la meilleure sélection possible des visibilités et phases, à la fois de l'objet science et du calibrateur, il ne reste plus qu'à effectuer la calibration, mais d'abord il faut prendre en compte les diamètres des calibrateurs (qu'on va aller chercher sur le site du CDS SIMBAD via la commande "amdlibSearchAllStarDiameters" (de la routine "amdlibCalibrate").

\item{} Dernière étape avant d'effectuer la calibration finale, et non des moindres: le calcul de la fonction de transfert. En effet, dans la pratique le flux cohérent mesuré sur l'interférogramme est proportionnel au produit degré complexe de cohérence mutuelle (Eq. \eqref{3.24}, mesurée dans des conditions idéales) par une seconde composante rattachée aux spécificités de l'instrument et liée aux conditions atmosphériques qui varie d'un temps de pose à un autre et qui est connue sous le nom de la fonction de transfert atmosphérique et instrumentale. amdlib et avec toutes les données instrumentales et atmosphériques collectées, durant l'observation, permet de calculer la fonction de transfert via "amdlibComputeAllTransferFunction" et de visualiser cette dernière avec "amdlibShowTransferFunctionVsTime".

\item{}  La calibration est maintenant à notre portée, il ne faut que soustraire la fonction de transfert de nos visibilités et phases des étoiles sciences et de calibration (tout trou dans la fonction de transfert, causé par un échantillonnage irrégulier, peut être comblé par une méthode d'ajustement au choix), et corriger le diamètre apparent des visibilités des calibrateurs via un modèle de disque uniforme. La calibration est effectuée grâce à la commande "amdlibCalibrateOiData". Mathématiquement, la visibilité complexe d'un objet peut être formulée comme suit : $V_{obj}=|V_{obj}|e^{i\Phi_{obj}}$. De ce fait la calibration d'une visibilité d'un objet science $V_{sci}$  par  celle d'un calibrateur $V_{cal}$ est $\frac{V_{sci}}{V_{cal}}=\frac{|V_{sci}|}{|V_{cal}|}e^{i(\Phi_{sci}-\Phi_{obj})}$. Autrement dit, la calibration des modules de visibilités se divise et celles des phases se soustrait.\\
\end{itemize}

A la fin de la réduction nous devrons obtenir des données optiques interférométriques propres et optimisées sous forme de fichiers OIFITS (Optical Interferometers Flexible Image Transport System), comprenant les visibilités, les phases différentielles et la clôture de phase calculées durant la réduction comme suit : $\Psi=\Phi_{12}+\Phi_{23}-\Phi_{13}$. Malheureusement, le spectre, qui n'est pas consideré comme une donnée OIFITS standard, n'est pas calibré par amdlib, et il doit être calibré via des codes et méthodes propres à chacun.\\

\clearpage

\subsection{Observation AMBER d'Achernar et biais instrumentaux résiduels:}\label{sec_3.5.4}
\begin{figure}[h!]
\centering
\includegraphics[height=0.4\hsize,draft=false]{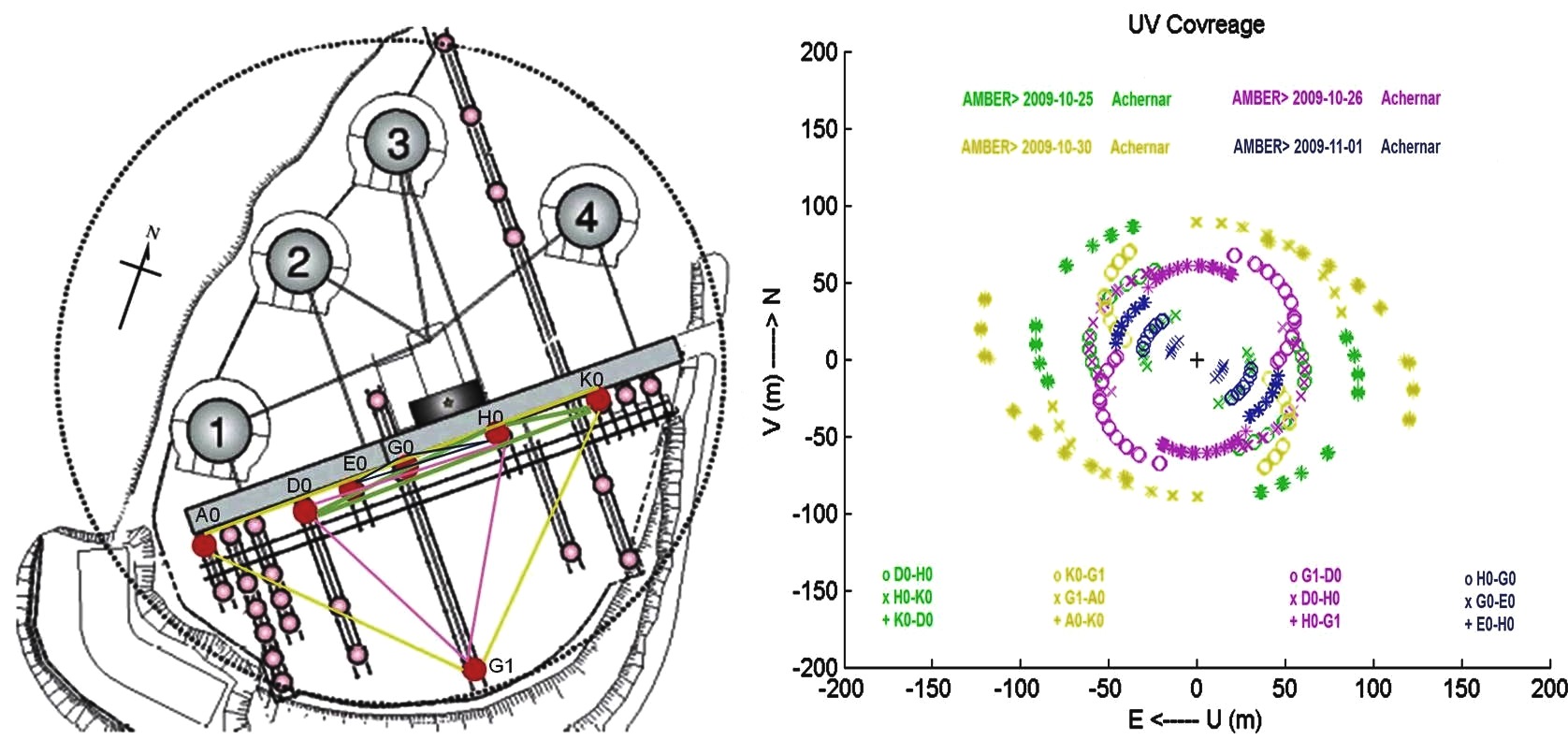}
\caption[Couvertures (u,v) VLTI/AMBER d'Achernar adoptées fin 2009]{Couvertures (u,v) VLTI/AMBER d'Achernar adoptées lors des quatre nuits d'observation fin 2009.}\label{Ach_uv_cov}
\end{figure}

Une campagne de mesure haute résolution (HR) autour de la raie Br$\gamma$ (en bande K; $2.14-2.19\mu m$) sur Achernar via AMBER (appuyé par FINITO) a été menée par mes directeurs de thèse (et leurs collaborateurs) en mode GTO à Paranal, et avec plusieurs configurations d'AT -voir le plan (u,v) sur la Fig.\ref{Ach_uv_cov}-, pendant 4 nuits durant la période 84 (ID 084.D-0456) fin 2009 ( pour le 25, 26 \& 30 Octobre et le 01 Novembre 2009, plus exactement).\\

Une première réduction standard utlisant amdlib des données collectées se révéla insuffisante à cause de données OIFITS finies hautement biaisées et scientifiquement inexploitables. La première mission qui m'a été confiée au début de ma thèse à Nice, en janvier 2010, était donc d'identifier la ou les source(s) de ces biais puis de les corriger via des traitements appropriés post-réductions. A l'aide d'avis experts d'AMBER au sein du laboratoire J.L.Lagrange (tel que Florentin Millour), deux sources de biais, non traités par une réduction standard, ont été mises au jour:\\

\begin{figure}[h!]
\centering
\includegraphics[height=0.6\hsize,draft=false]{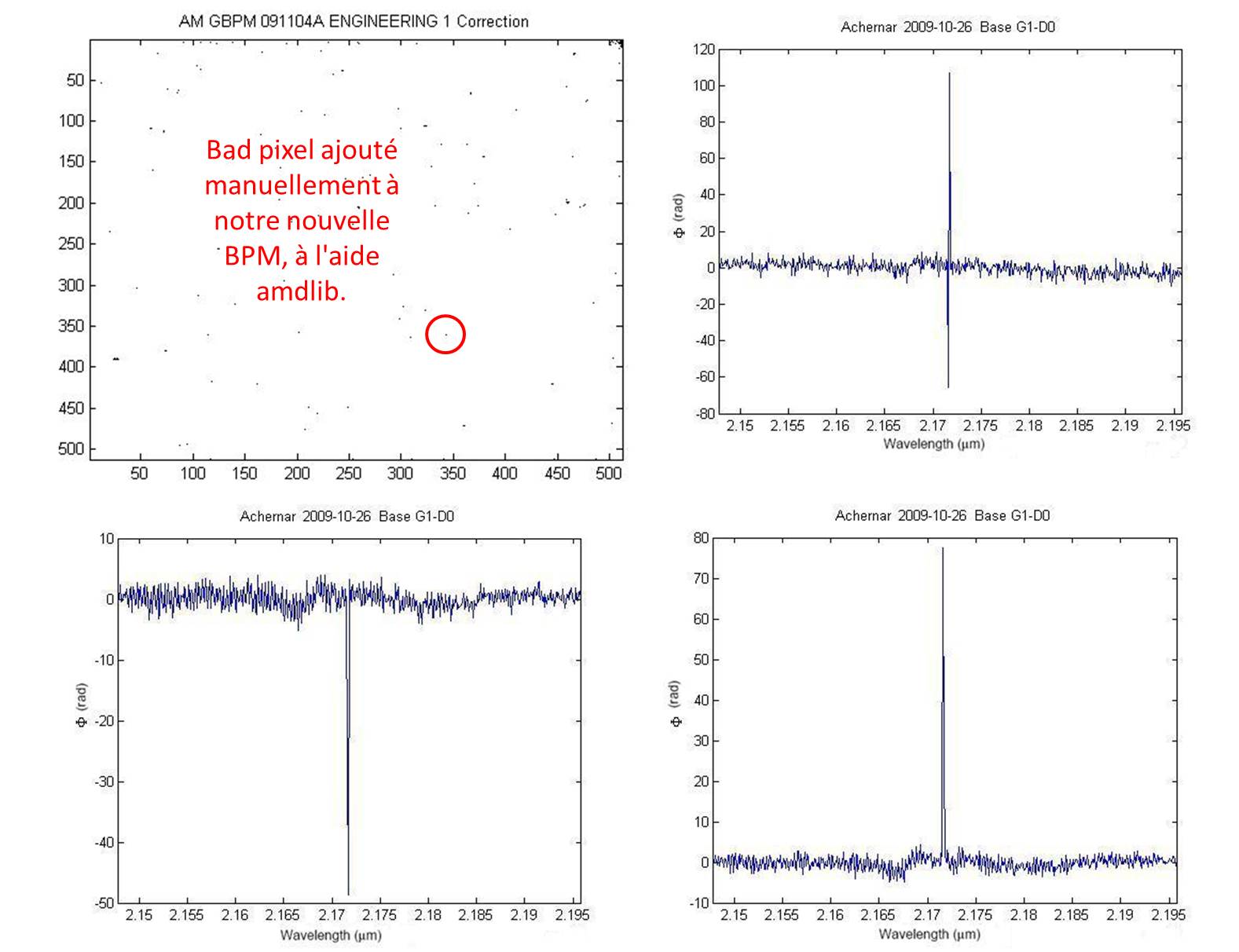}
\caption[Biais du "very bad pixel" et ses conséquences sur nos données Achernar]{Biais du very bad pixel et ses conséquences nos données Achernar.}\label{biais_Ach_1}
\end{figure}

\begin{figure}[h!]
\centering
\includegraphics[height=0.6\hsize,draft=false]{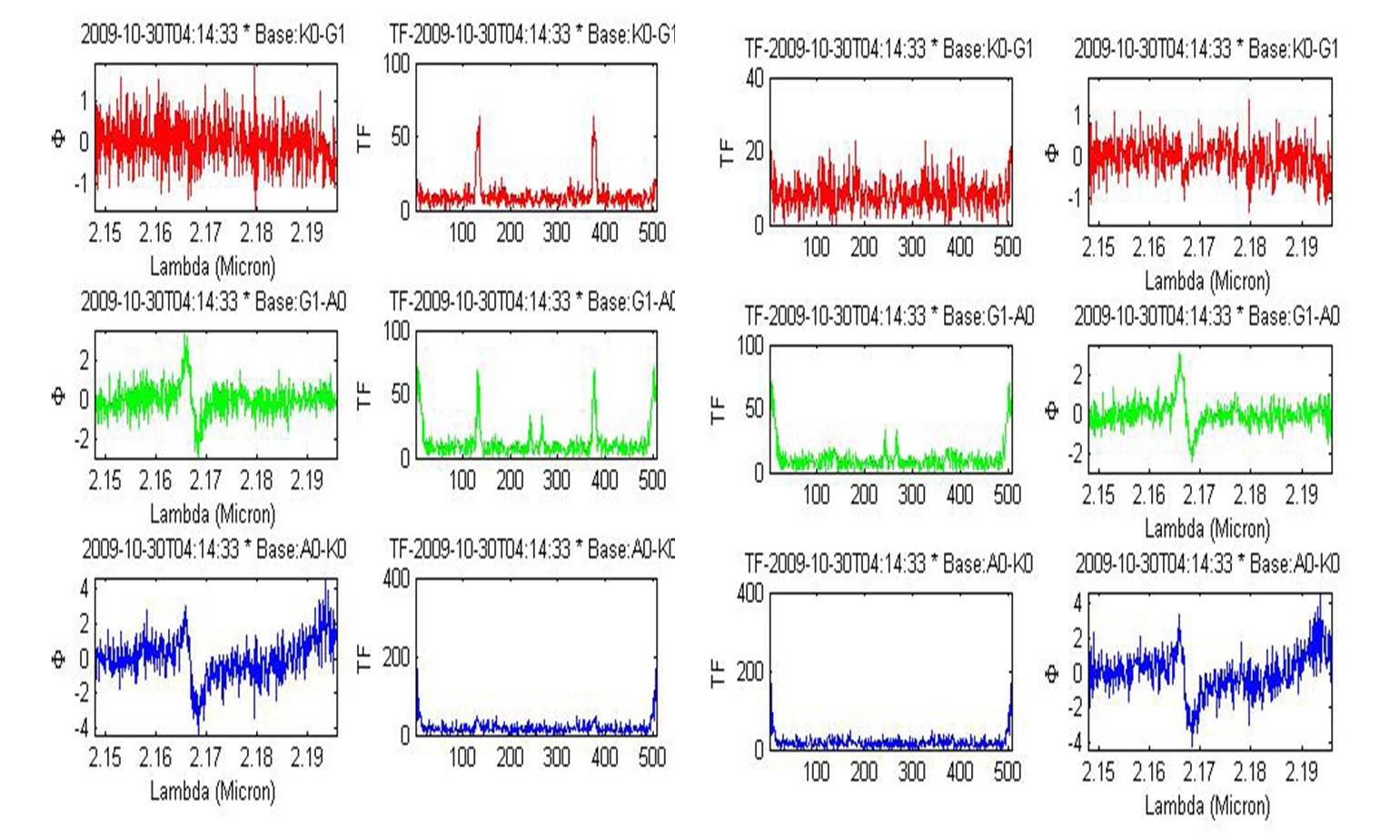}
\includegraphics[height=0.6\hsize,draft=false]{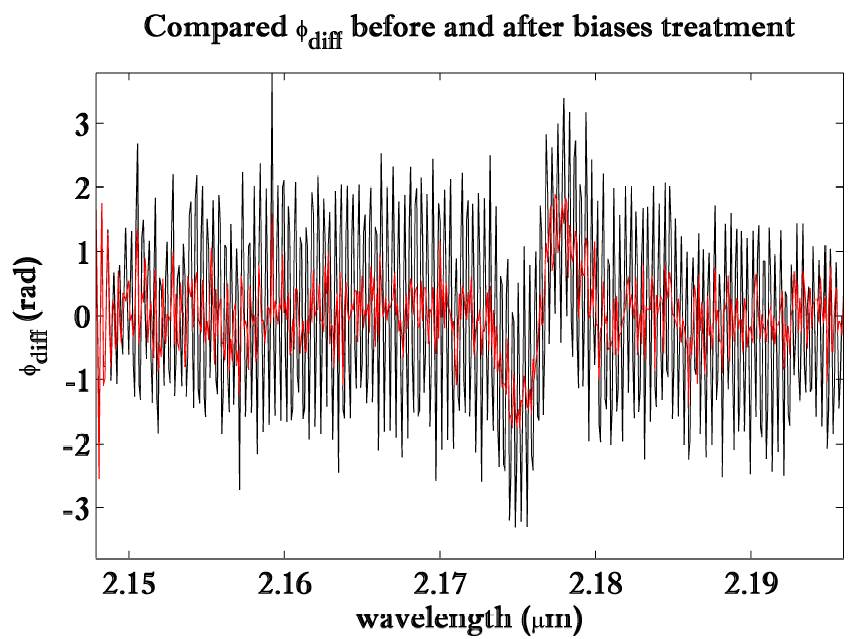}
\caption[Biais hautes fréquences et son traitement sur nos données Achernar]{ Biais hautes fréquences et son traitement sur nos données Achernar.}\label{biais_Ach_2}
\end{figure}

\begin{enumerate}
\item Un mauvais pixel non répertorié dans la dernière "badpixelmap" (BPM du 04/11/2009) mise en ligne sur le site de l'ESO à l'époque et qui affecte grandement les données réduites (voir Fig.\ref{biais_Ach_1}). Affectueusement surnommé "the very bad pixel" par notre équipe, il fut identifié en repérant d'intenses pics (points aberrants ou spikes) inhabituels mais récurents, à la même longueur d'onde (même abscisse), sur toutes nos données (visibilités et phases).

\item Une lame séparatrice dichroïque défaillante de 2008 à 2009. Celle-ci est un composant du laboratoire focal du VLTI (non d'AMBER) et qui induit des "chaussettes" (biais de modulations à hautes fréquences) par effet Fabry-Pérot (voir Fig.\ref{biais_Ach_2}). Cette lame dichroïque a été changée en Janvier 2010, et depuis la qualité des données collectés s'est grandement améliorée.\\
\end{enumerate}

\begin{figure}[h!]
\centering
\includegraphics[height=0.6\hsize,draft=false]{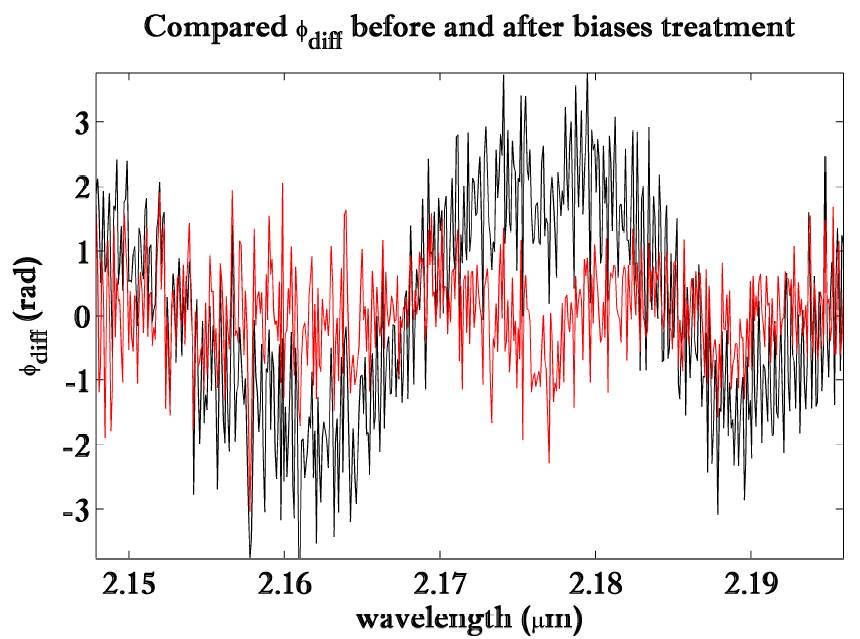}
\caption[Biais basses fréquences et son traitement sur nos données Achernar]{Biais basses fréquences et son traitement sur nos données Achernar.}\label{biais_Ach_3}
\end{figure}

Pour obtenir des données OIFITS plus robustes, il fallait régler le problème occasionné par le biais $N^\circ 1$ qui concerne le "very bad pixel". Ce dernier fut rajouté manuellement à notre "badpixelmap" (le fichier BPM\_04/11/2009) via une routine d'une ancienne version d'amdlib2.0 (qui n'existe plus sur les versions plus récentes). Cela étant fait, et ayant repéré les autres sources de biais et leurs causes, il ne me restait plus qu'à les traiter via des algorithmes adéquats qu'il fallait réaliser, puis mettre à l'épreuve et enfin valider sur d'autres données AMBER en notre possession -Altair, $\delta$ Aquilae \& Fomlahaut-, qui datent de 2008 et montrent le même biais dû à la lame dichroïque. Sur celles-ci, j'ai constaté un 3\up{ième} biais à basses fréquences (uniquement sur les spectres et les phases différentielles $\phidiff$), qui se présente sous forme d'ondulations systématiques persistantes en fonction de la longueur d'onde (voir Fig.\ref{biais_Ach_3}) et qui ont été imputées à un effet Pérot-Fabry aussi, provoqué par un polariseur à deux prismes rapprochés avec un trou d'air au milieu, appelé prisme de Glan-Taylor, composant défectueux d'AMBER à l'époque, mais ce problème a été réglé depuis. Il est à noter que l'effet de l'ondulation ne disparait pas après la calibration, parce que les phases (science/calibrateur) se soustraient et non se devisent comme c'est le cas pour le spectre ou pour le module de visibilité (voir précédente section). Idem pour les "chaussettes" qui sont encore présentes après la calibration car celles-ci ne sont pas toujours à la même longueur d'onde pour l'étoile science et le calibrateur.\\

Ainsi, le biais de haute fréquence dû à la lame dichroïque VLTI a été corrigé (sur tous nos mesurables) dans l'espace de Fourier où le pic lié à cet effet (également appelé chaussettes) était aisément identifiable. Celui-ci a été remplacé par un bruit gaussien de même amplitude que le continuum avant d'opérer une transformation de Fourier (TF) inverse pour retrouver un signal filtré de ce biais (Fig.\ref{biais_Ach_2}). Le biais basses fréquences occasionné par le prisme Glan-Taylor d'AMBER a quant à lui été traité par la soustraction ajustée par une sinusoïde sur le signal (Fig.\ref{biais_Ach_3}). De ce fait, l'ensemble des traitements opérés sur nos données est énuméré ci-dessous, mesurable par mesurable:\\

\textbf{Pour les modules de visibilité ($V^2$):}
Deux traitements furent nécessaires: un traitement des "spikes" résiduels qui ont échappés aux logiciels de réduction d'amdlib et qui sont bien en dehors de l'écart type du signal et qui ont été remplacés par la valeur moyenne de celui-ci (Achernar étant peu résolue, le $V^2$ reste quasi-constant). Le second traitement a concerné le biais haute fréquence causé par la lame dichroïque.\\

\textbf{Pour la phase différentielle ($\phidiff$):}
Même traitement que pour $V^2$ auquel s'ajoute le traitement du biais haute fréquence (décrit ci-haut) et un traitement de pente ou de courbure de la $\phidiff$, provoqué des "spikes" résiduels lors du calcul de la P2VM d'amdlib (lors de la normalisation longueur d'onde par longueur d'onde, tout dépend si le "spike" est sur l'un des bord de l'intervalle des longueurs d'onde -pente- ou au centre -courbure-). Dans ce dernier une simple soustraction d'un ajustement par un polynôme d'ordre 1 ou 2 s'avère suffisante.\\

\textbf{Les phases de clôture ($\Psi$):}
Le même traitement que pour $V^2$ était suffisant.\\

Il est important de noter à ce stade que le spectre n'est pas considéré comme étant une donnée OIFITS et n'est par conséquent pas calibré par amdlib. Tout calibration/traitement du spectre doit se faire via les propres moyens du manipulateur de données. La manipulation des données OIFITS est possible via divers outils, par exemple la librairie OIFITSlib\footnote{http://www.mrao.cam.ac.uk/research/optical-interferometry/oifits/} en langage IDL (Interactive Data Language) de John Monnier avec laquelle j'ai travaillé au début de ma thèse et que j'ai adapté afin d'inclure le spectre en tant que donnée OIFITS à part entière, avant que je m'oriente vers une librairie OIFITS-MATLAB\footnote{http://www.antonyschutz.com/Software.html} (MATrix LABoratory) développée par Antony Schutz au sein du laboratoire J.L.Lagrange, que j'ai adapté selon mes besoins aussi, et grâce à laquelle j'ai pu développer une routine MATLAB de lecture des données OIFITS "ShowOidata" AMBER (à l'instar de celle IDL d'Anthony Meilland et Yorick sur amdlib de Florentin Milllour) -voir Annexe-. Ainsi, le traitement des spectres est résumé ci-dessous:\\

\begin{figure}[h!]
\centering
\includegraphics[height=0.7\hsize,draft=false]{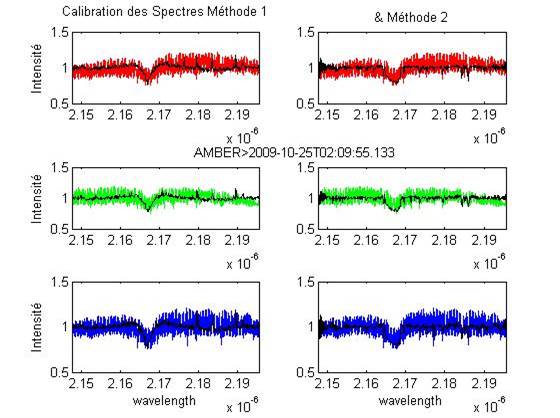}
\caption[Calibration spectrale des données d'Achernar]{Calibration spectrale des données d'Achernar.}\label{calib_spec_Ach}
\end{figure}

\textbf{Pour les spectres (Flux):}
Idem que pour la $\phidiff$, où il faut traiter les biais basses et hautes fréquences et les "spikes" résiduels. A cela doit s'ajouter la calibration spectrale que j'ai entrepris d'abord par une méthode proposée par \citet{2006PhDT........46M} et inspiré par \citet{1996ApJS..107..281H}, dont le principe est de calibrer le spectre science par celui du calibreur à l'exception de la raie Br$\gamma$ qui est calibrée par un profil dit de Voigt (convolution d'une Gaussienne et d'une Lorentzienne) -méthode 1-. Finalement, j'ai opté pour une autre méthode qui convient mieux au spectre des bandes K en haute résolution, où j'effectue une opération d'auto calibration/normalisation par un ajustement polynomial d'ordre 4 en HR (3 en MR et de 2 en LR) sur notre spectre à l'exception de la raie Br$\gamma$ qui est calibrée par une reconstruction en ondelette Daubechies de celle-ci, avec un moment d'ordre 4 (qui reproduit assez bien la forme générale de la raie non calibrée) -méthode 2- (voir Fig.\ref{calib_spec_Ach}).\\

\begin{figure*}[h!]
\centering
\includegraphics[height=0.7\hsize,draft=false]{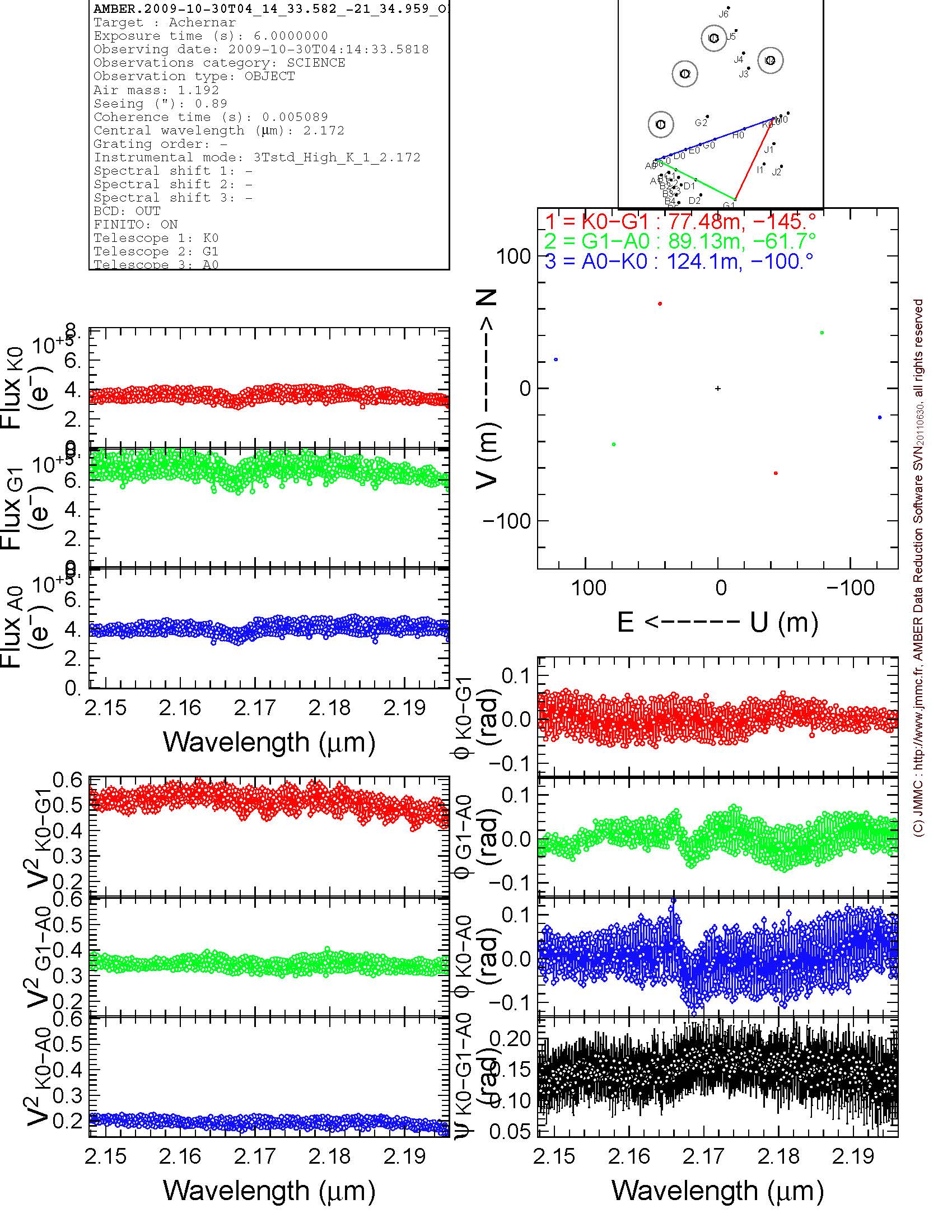}
\includegraphics[height=0.7\hsize,draft=false]{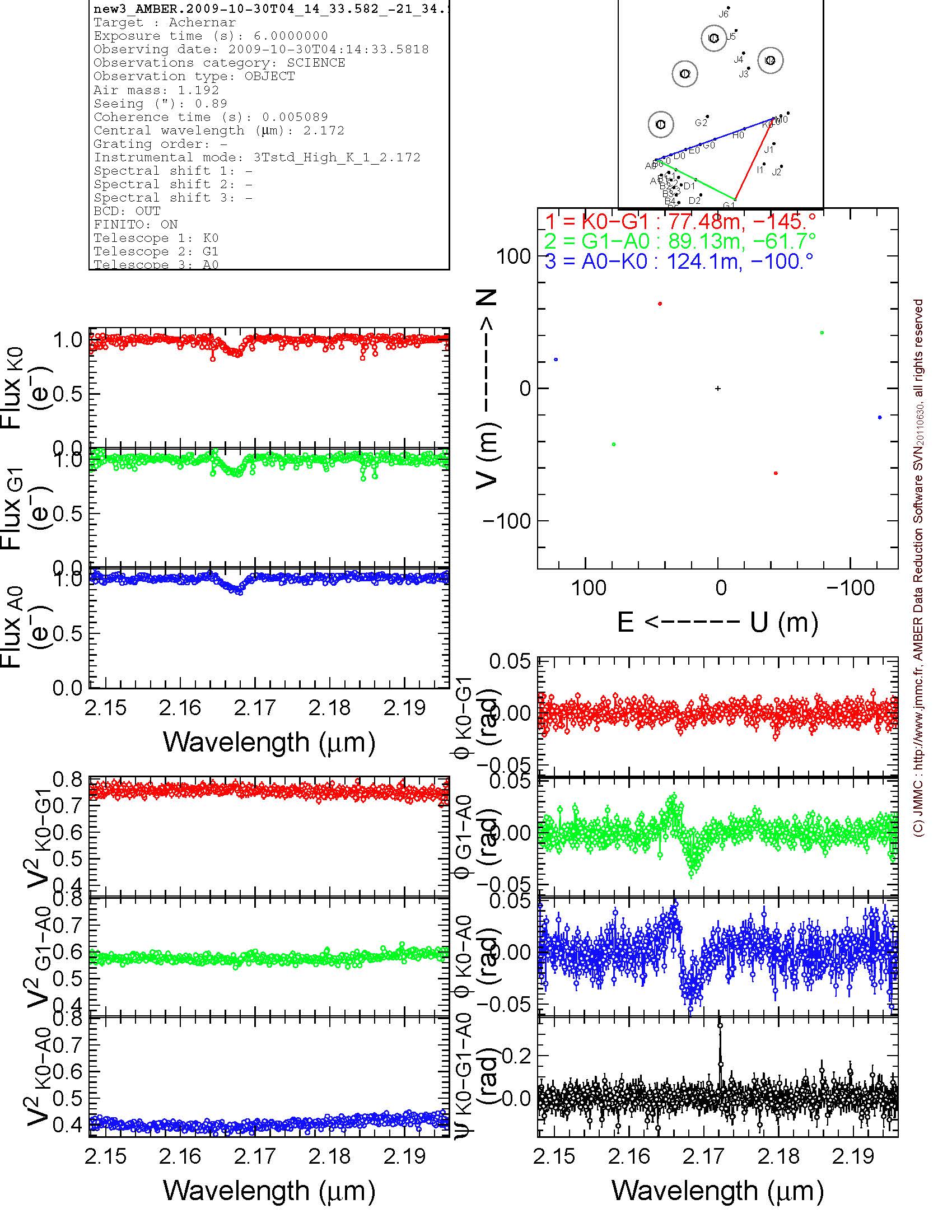}
\caption[Comparaison de l'ensemble des données OIFITS d'Achernar avant et après traitement]{Comparaison de l'ensemble des données OIFITS d'Achernar avant (en haut) et après traitement (en bas).}\label{compar_trait_Achernar}
\end{figure*}

L'ensemble des traitements une fois appliqué sur toutes nos observables (spectre, module de visibilité, phase différentielle et clôture de phase) sont présentés dans la Fig.\ref{compar_trait_Achernar} où on constate la qualité des données OIFITS avant et après traitement.\\

Une dernière et importante correction des données AMBER concerne les longueurs d'onde qui ne correspondent pas tout à fait à la spectroscopie observée. En effet et à cause d'une complication mécanique sur le moteur de positionnement du réseau du spectromètre AMBER, les observables sont légèrement décalées en longueurs d'onde, d'un léger décalage qu'il faut bien sûr prendre en compte et corriger à l'aide d'une référence à portée de main et peu couteuse, la raie Br$\gamma$ (seule raie dans la bande K) dont on connait précisément l'emplacement en longueur d'onde $\lambda_{Br_\gamma}= 2.165$ $\mu m$. Même si le décalage est minime (de l'ordre de 0.1\%) il est très important de le corriger en décalant nos mesurables interférométriques, sur les bonnes longueurs d'onde, pour que le centre de la raie spectrale Br$\gamma$ coïncide avec la valeur théorique.\\

\begin{figure}[ht]
\centering
\includegraphics[width=0.4\hsize,draft=false]{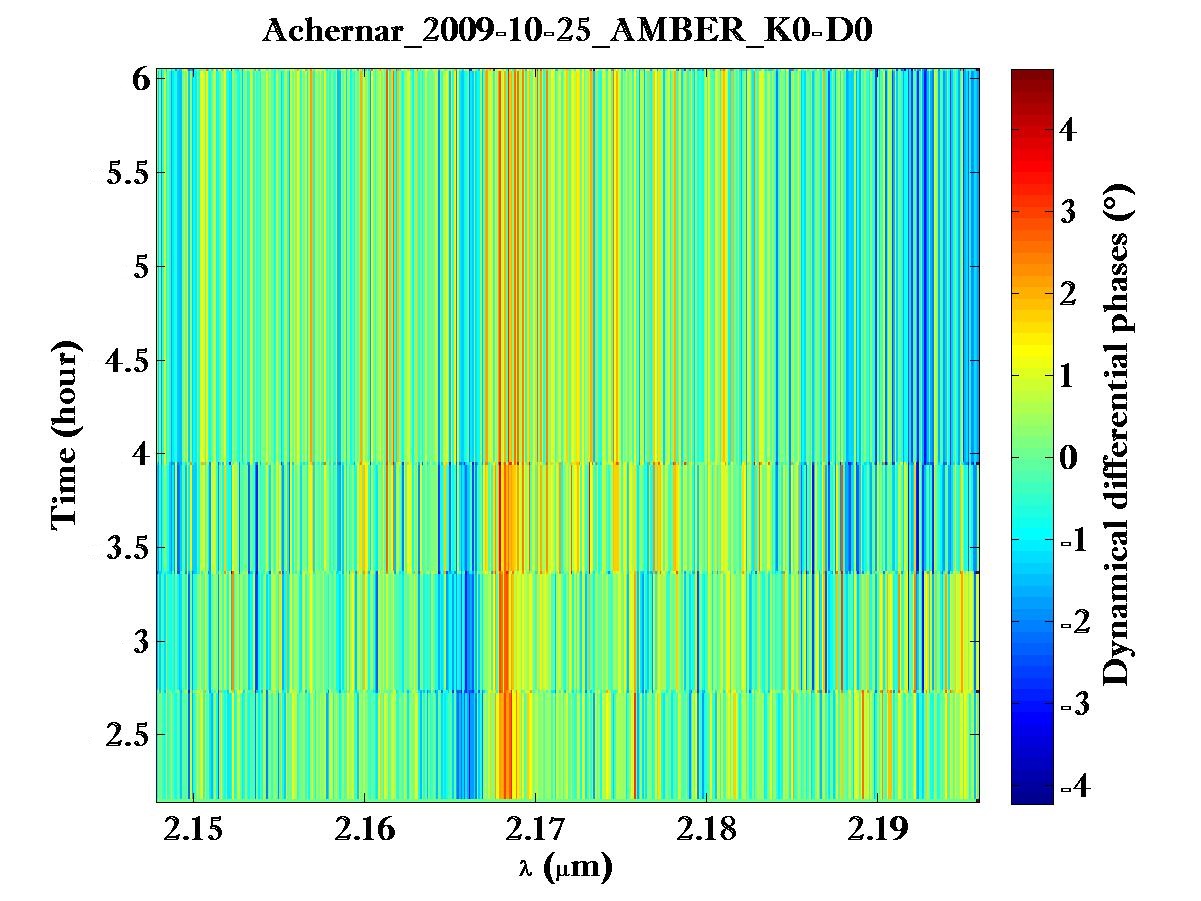}
\includegraphics[width=0.4\hsize,draft=false]{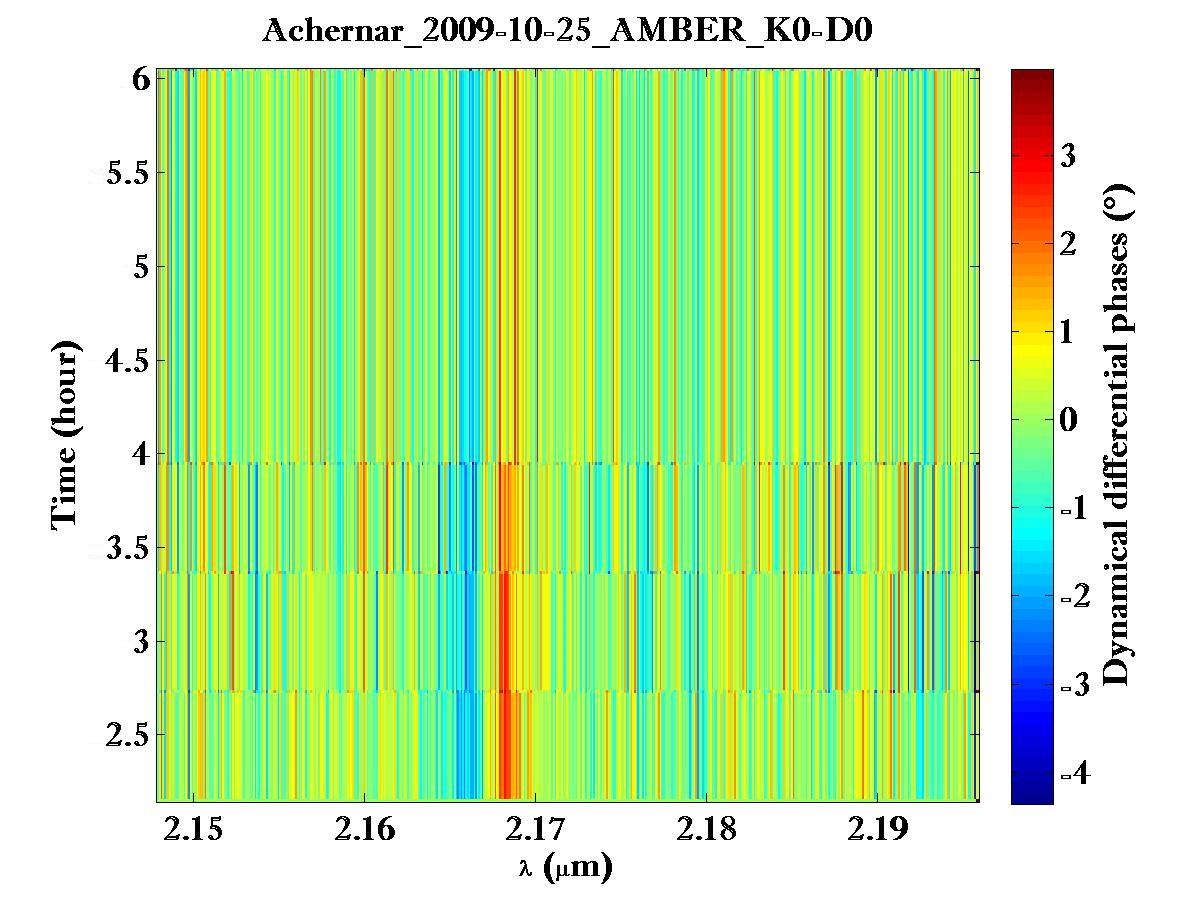}
\caption[Carte des $\phidiff$ dynamique d'Achernar: Observation sur AMBER/VLTI 25-10-2009]{À gauche: la carte des phases différentielles dynamique d'Achernar avant traitement des biais pour la nuit d'observation du 25-10-2009 sur le triplet D0-H0-K0. À droite: Après traitement, où l'effet de la rotation autour de la raie Br$\gamma$ est plus évident.}\label{Ach_dyn_diff_phi1}
\end{figure}

\begin{figure}[ht]
\centering
\includegraphics[width=0.4\hsize,draft=false]{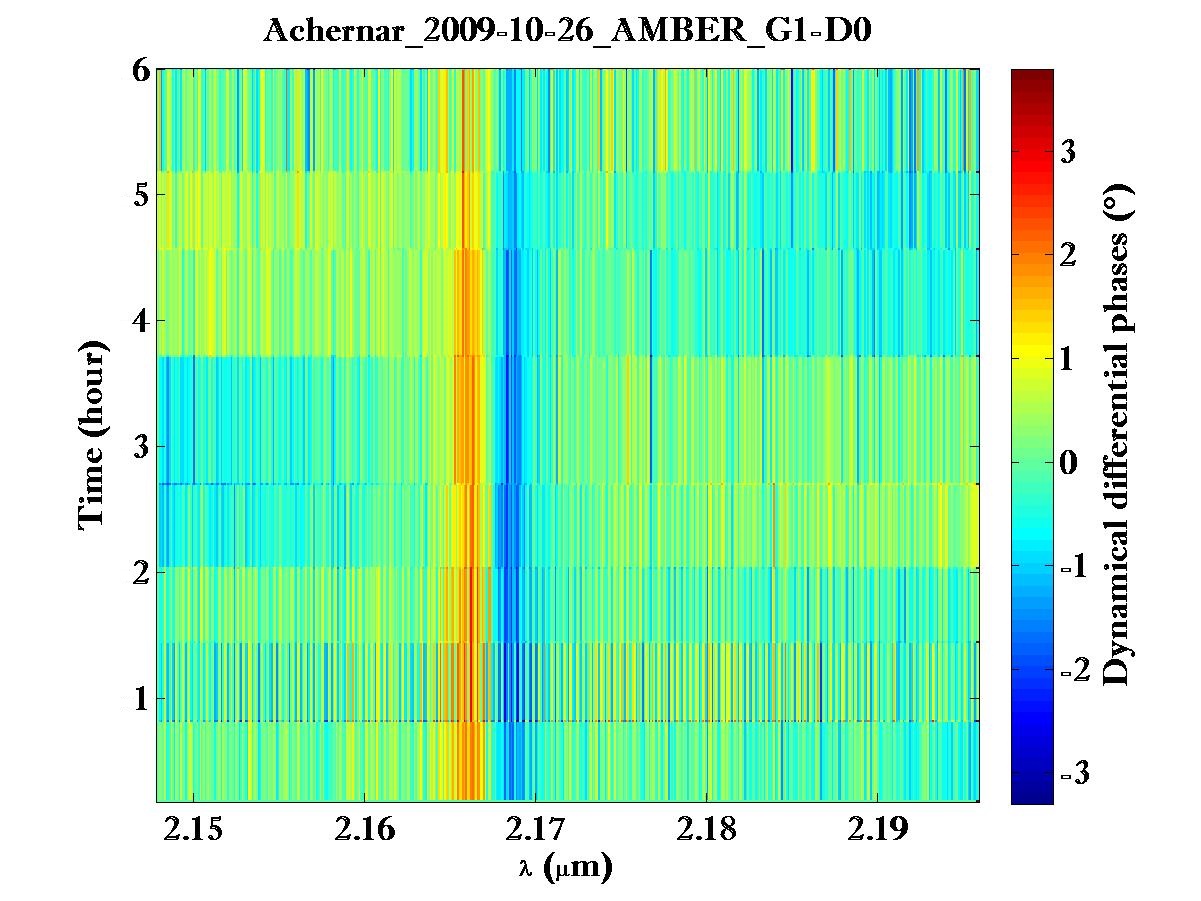}
\includegraphics[width=0.4\hsize,draft=false]{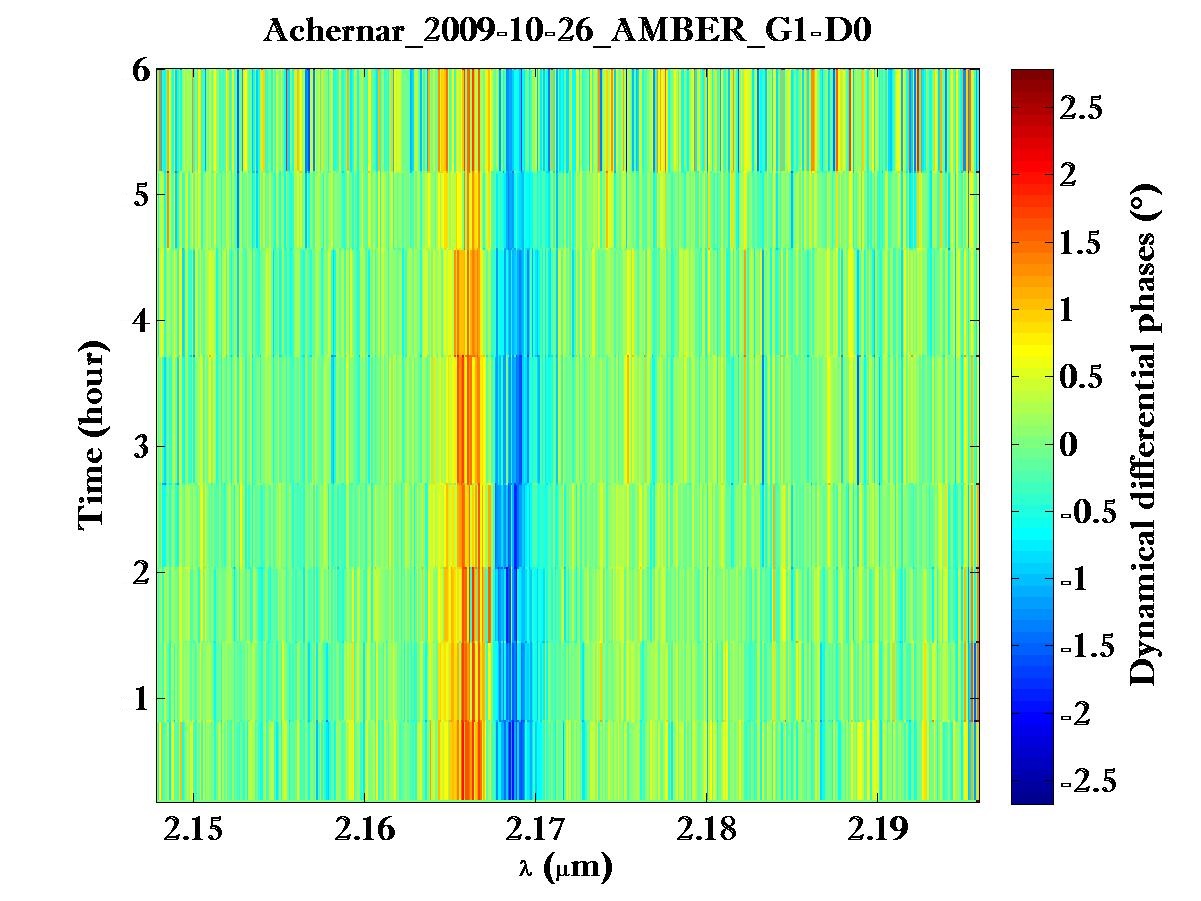}
\caption[Carte des $\phidiff$ dynamique d'Achernar: Observation sur AMBER/VLTI 26-10-2009]{À gauche: la carte des phases différentielles dynamique d'Achernar avant traitement des biais pour la nuit d'observation du 26-10-2009 sur le triplet D0-H0-G1. À droite: Après traitement, où l'effet de la rotation autour de la raie Br$\gamma$ est plus évident.}\label{Ach_dyn_diff_phi2}
\end{figure}

\begin{figure}[ht]
\centering
\includegraphics[width=0.4\hsize,draft=false]{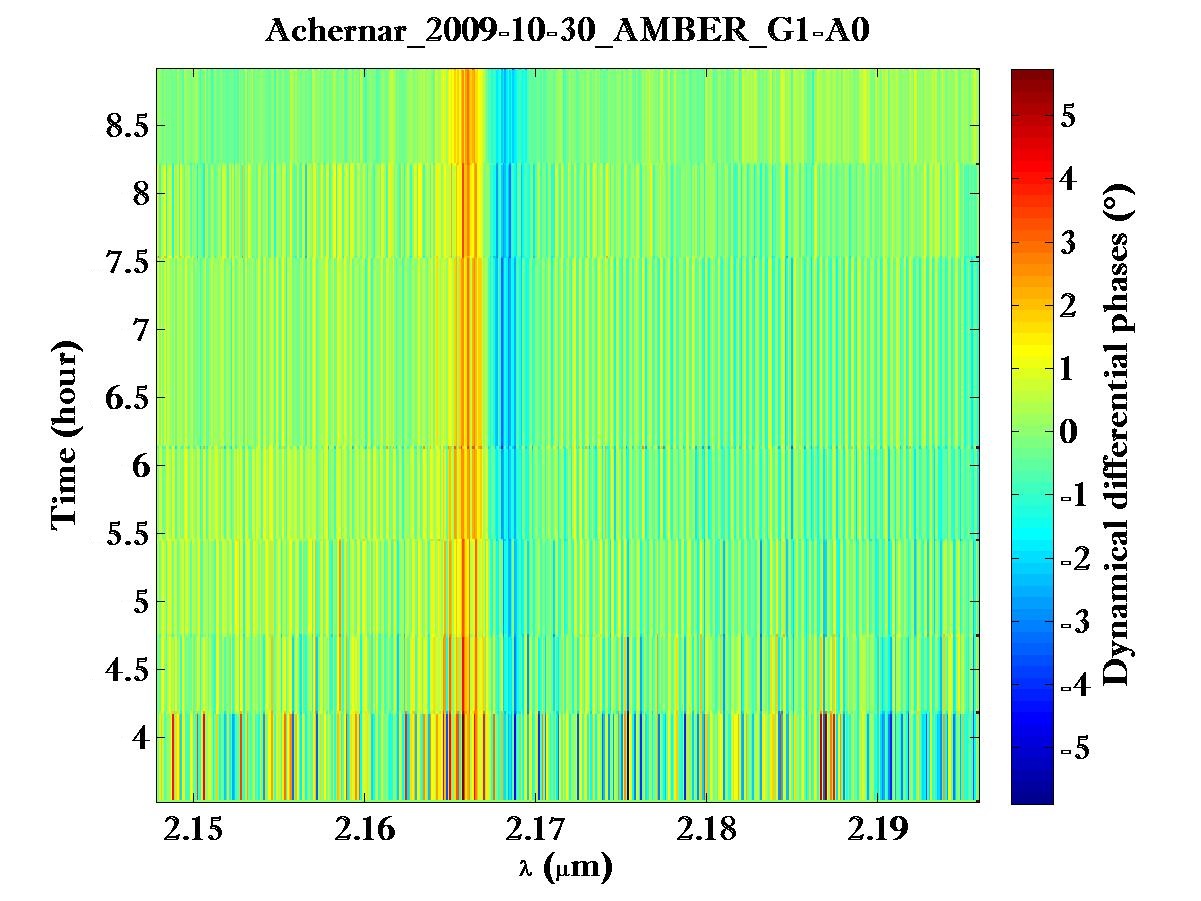}
\includegraphics[width=0.4\hsize,draft=false]{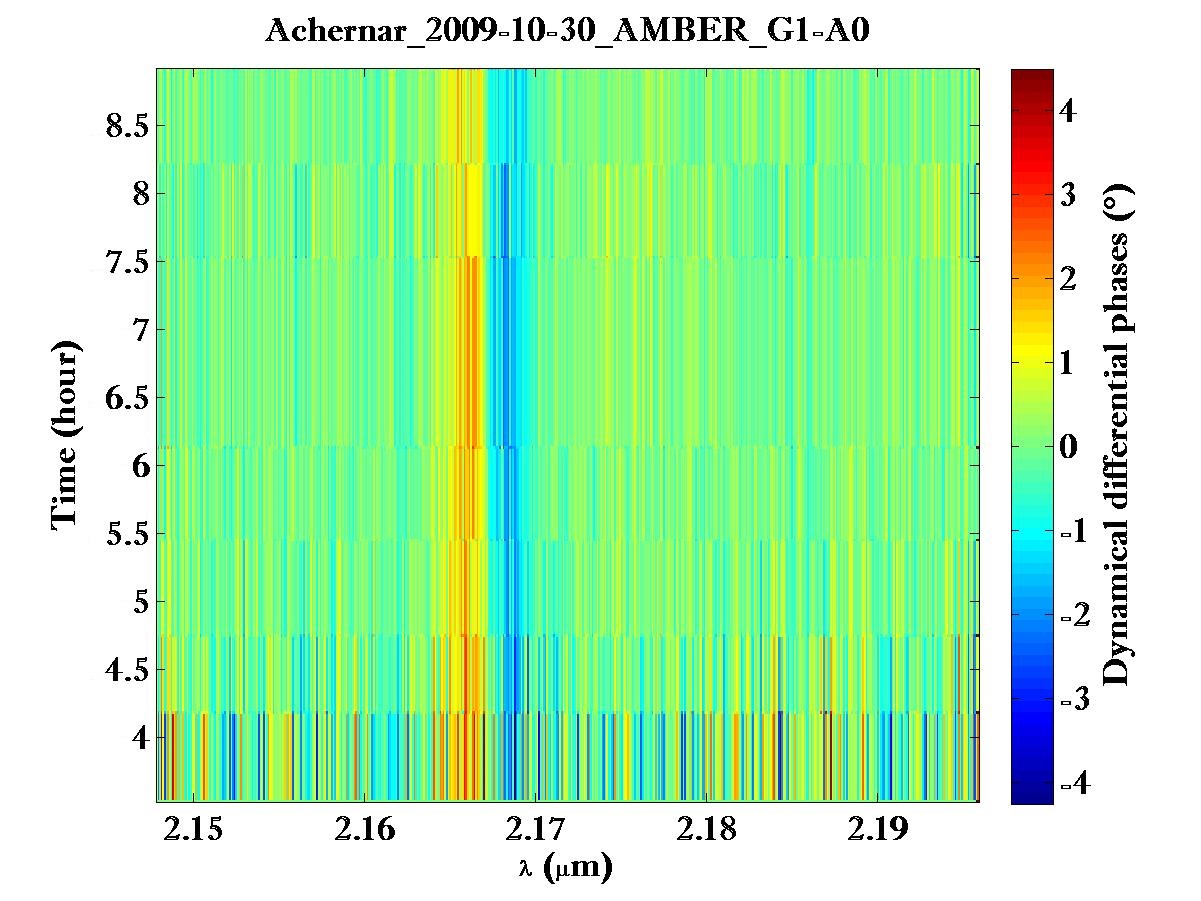}
\caption[Carte des $\phidiff$ dynamique d'Achernar: Observation sur AMBER/VLTI 30-10-2009]{À gauche: la carte des phases différentielles dynamique d'Achernar avant traitement des biais pour la nuit d'observation du 30-10-2009 sur le triplet A0-K0-G1. À droite: Après traitement, où l'effet de la rotation autour de la raie Br$\gamma$ est plus évident.}\label{Ach_dyn_diff_phi3}
\end{figure}

\begin{figure}[ht]
\centering
\includegraphics[width=0.4\hsize,draft=false]{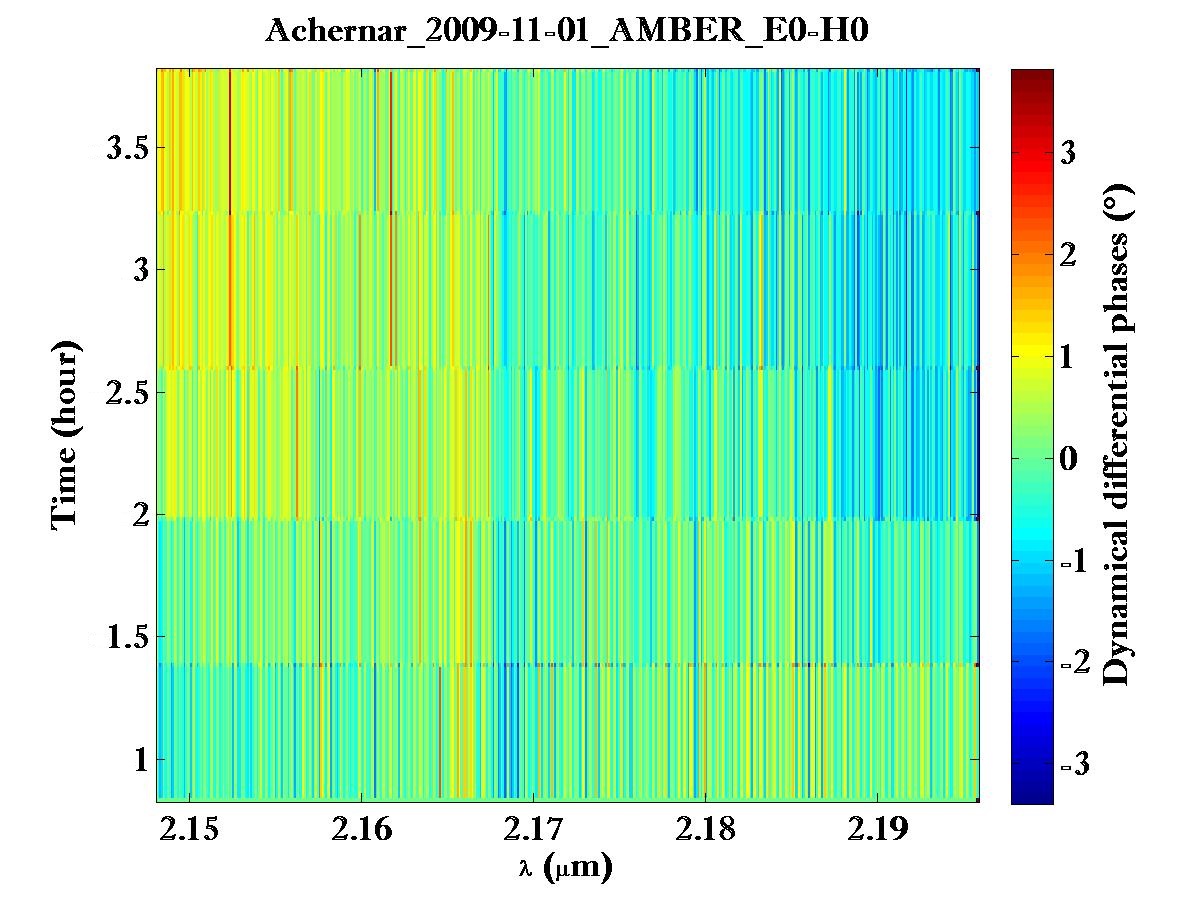}
\includegraphics[width=0.4\hsize,draft=false]{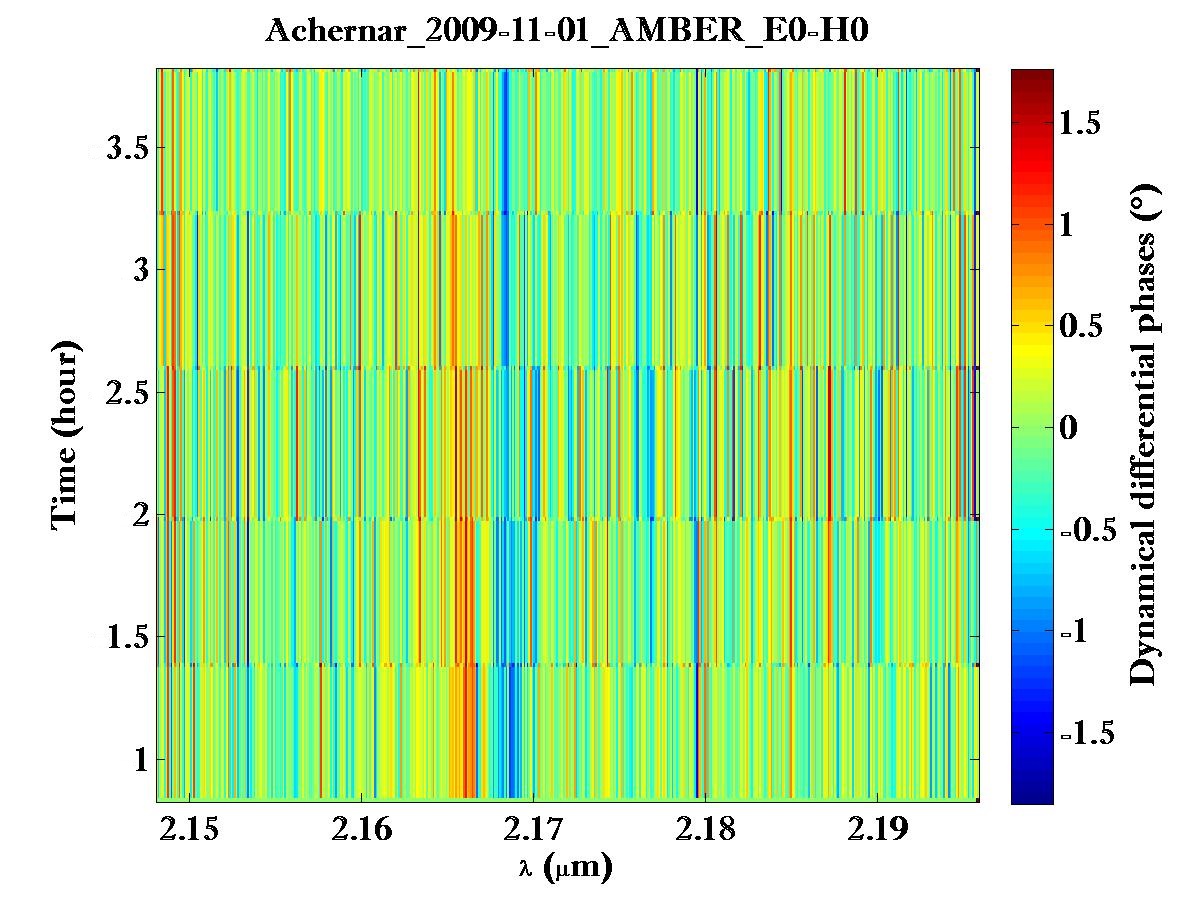}
\caption[Carte des $\phidiff$ dynamique d'Achernar: Observation sur AMBER/VLTI 01-11-2009]{À gauche: la carte des phases différentielles dynamique d'Achernar avant traitement des biais pour la nuit d'observation du 01-11-2009 sur le triplet H0-G0-E0. À droite: Après traitement, où l'effet de la rotation autour de la raie Br$\gamma$ est plus évident.}\label{Ach_dyn_diff_phi4}
\end{figure}

Afin de mieux visualiser l'effet du traitement sur la $\phidiff$ d'Achernar (notre principal atout pour l'étude de celle-ci), nous avons mis au point le concept d'une carte des phases différentielles dynamique, qui nous permet d'effectuer une comparaison entre l'ensemble des données $\phidiff$ (avant et après le traitement), pour chaque base interférométrique de chaque nuit observée, ainsi qu'on peut le voir sur les quatre figures (Figs \ref{Ach_dyn_diff_phi1}, \ref{Ach_dyn_diff_phi2}, \ref{Ach_dyn_diff_phi3} \& \ref{Ach_dyn_diff_phi4}).\\

\begin{figure}[ht]
\centering
\includegraphics[width=0.4\hsize,draft=false]{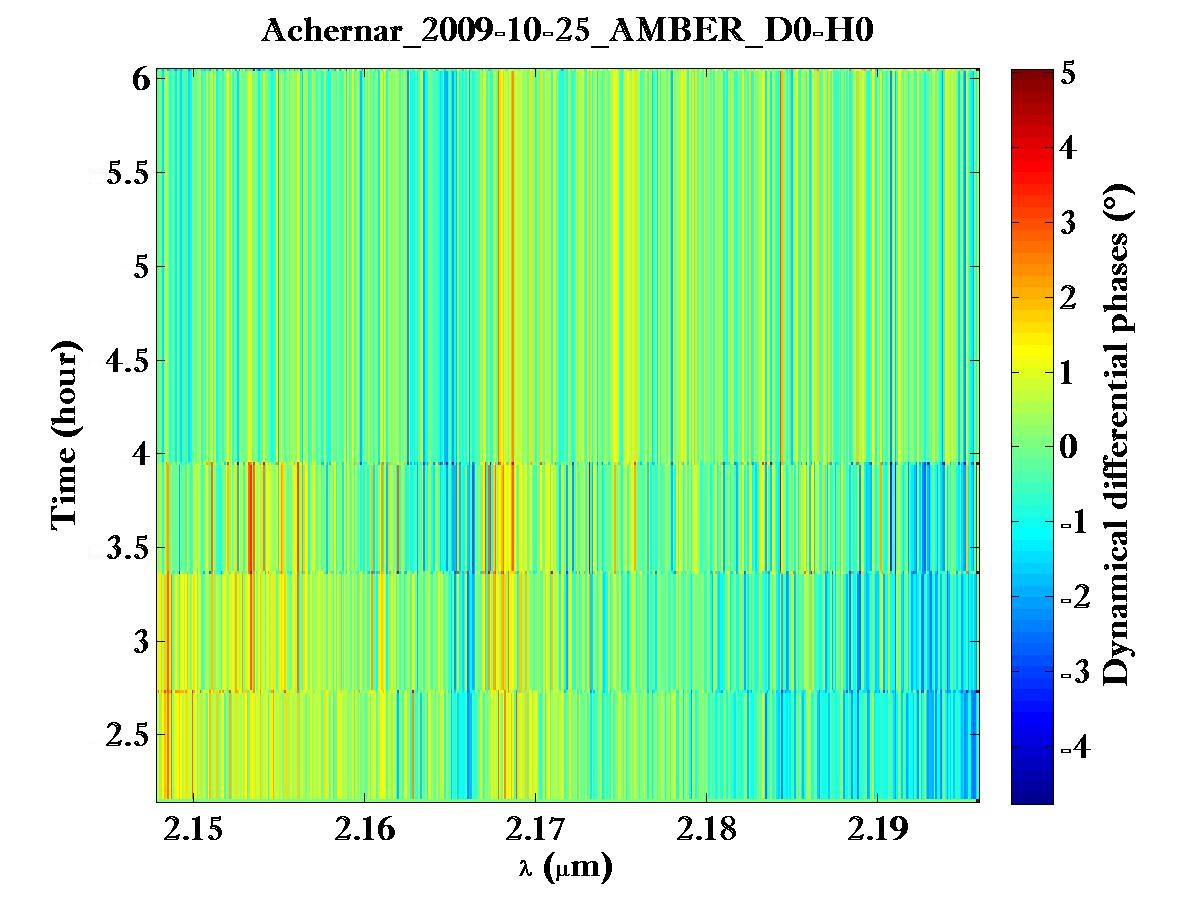}
\includegraphics[width=0.4\hsize,draft=false]{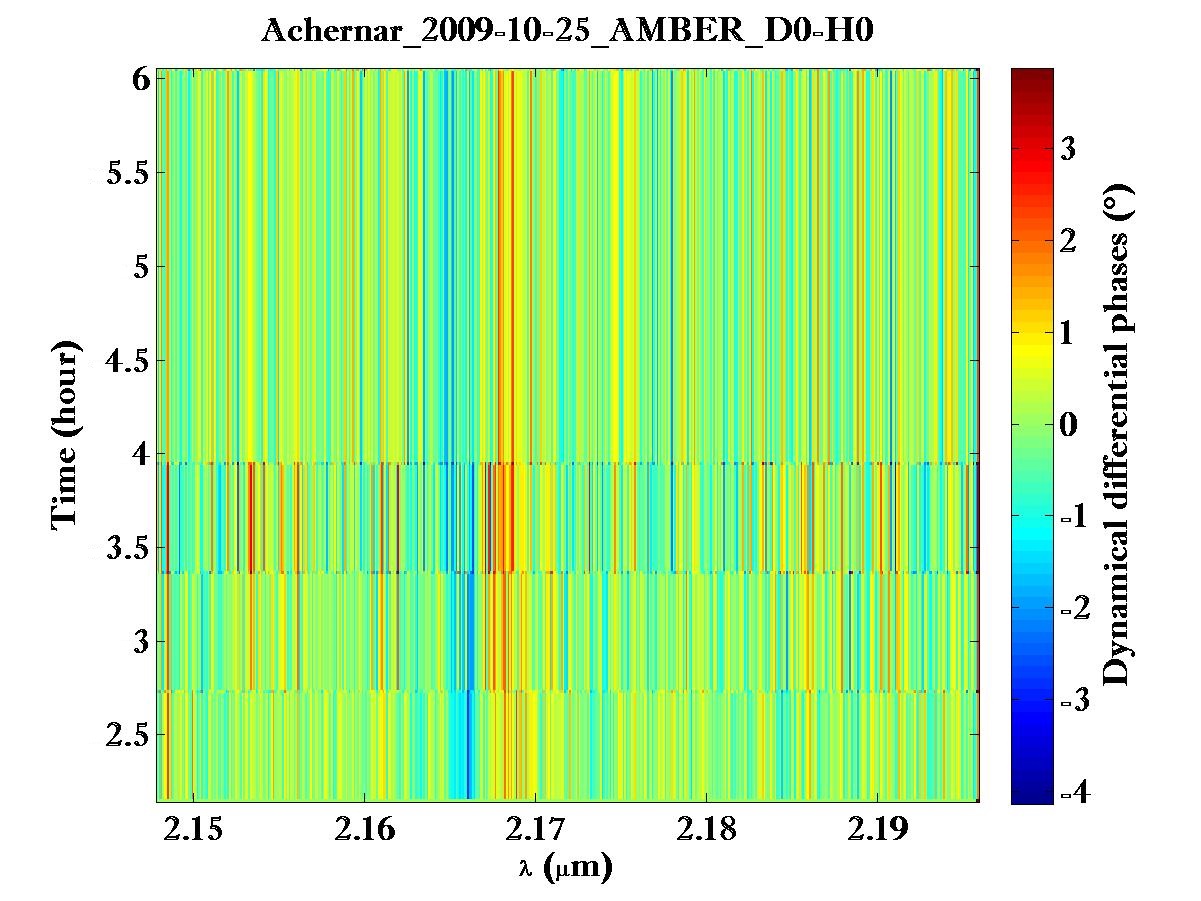}
\end{figure}
\begin{figure}[ht]
\centering
\includegraphics[width=0.4\hsize,draft=false]{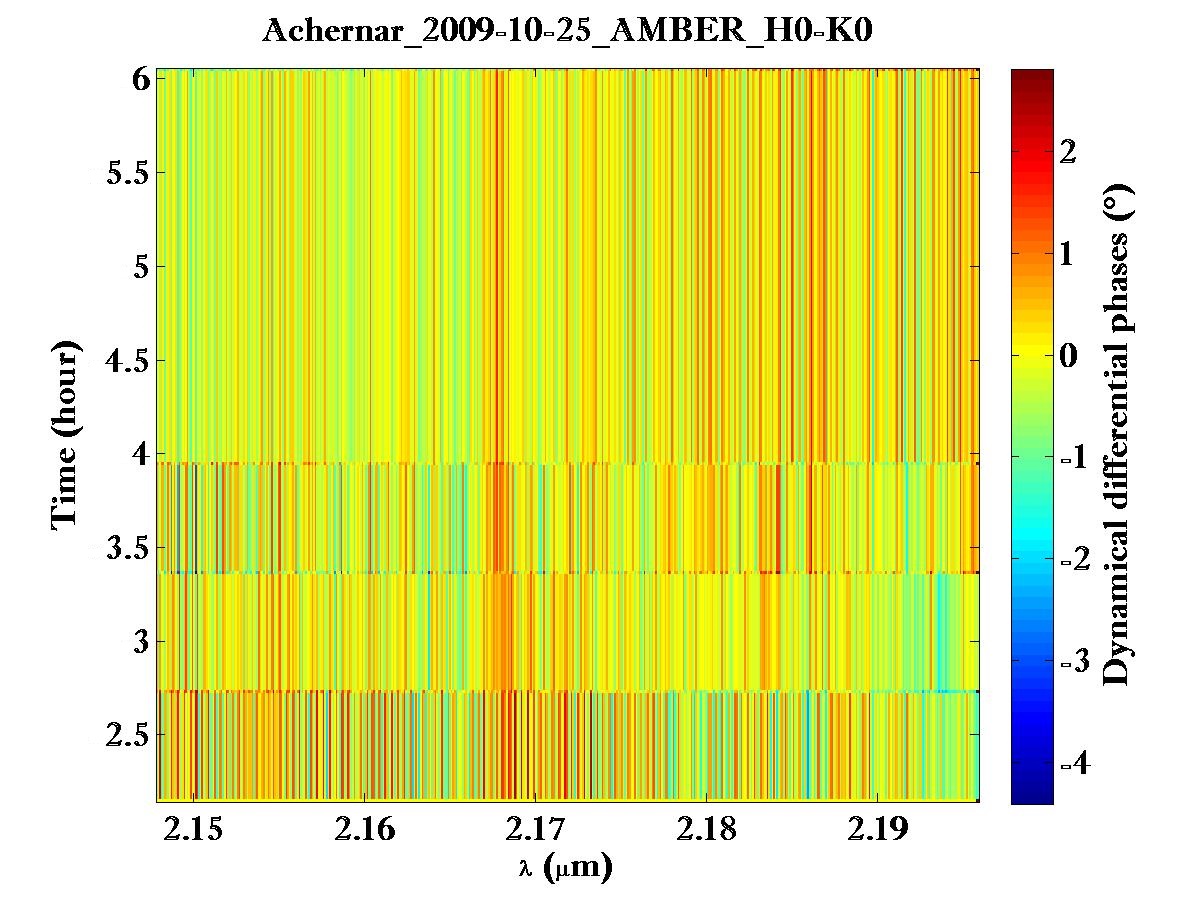}
\includegraphics[width=0.4\hsize,draft=false]{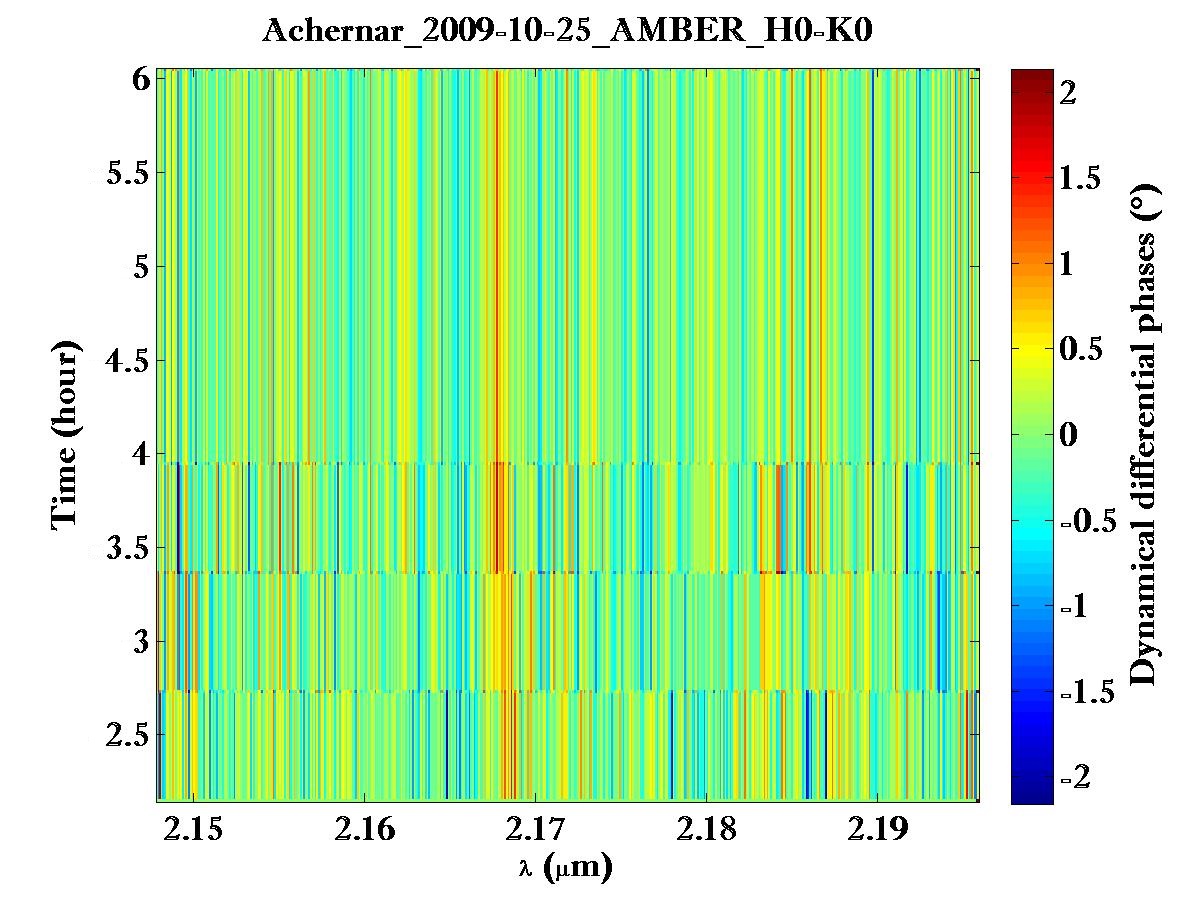}
\end{figure}

\clearpage

\begin{figure}[ht]
\centering
\includegraphics[width=0.4\hsize,draft=false]{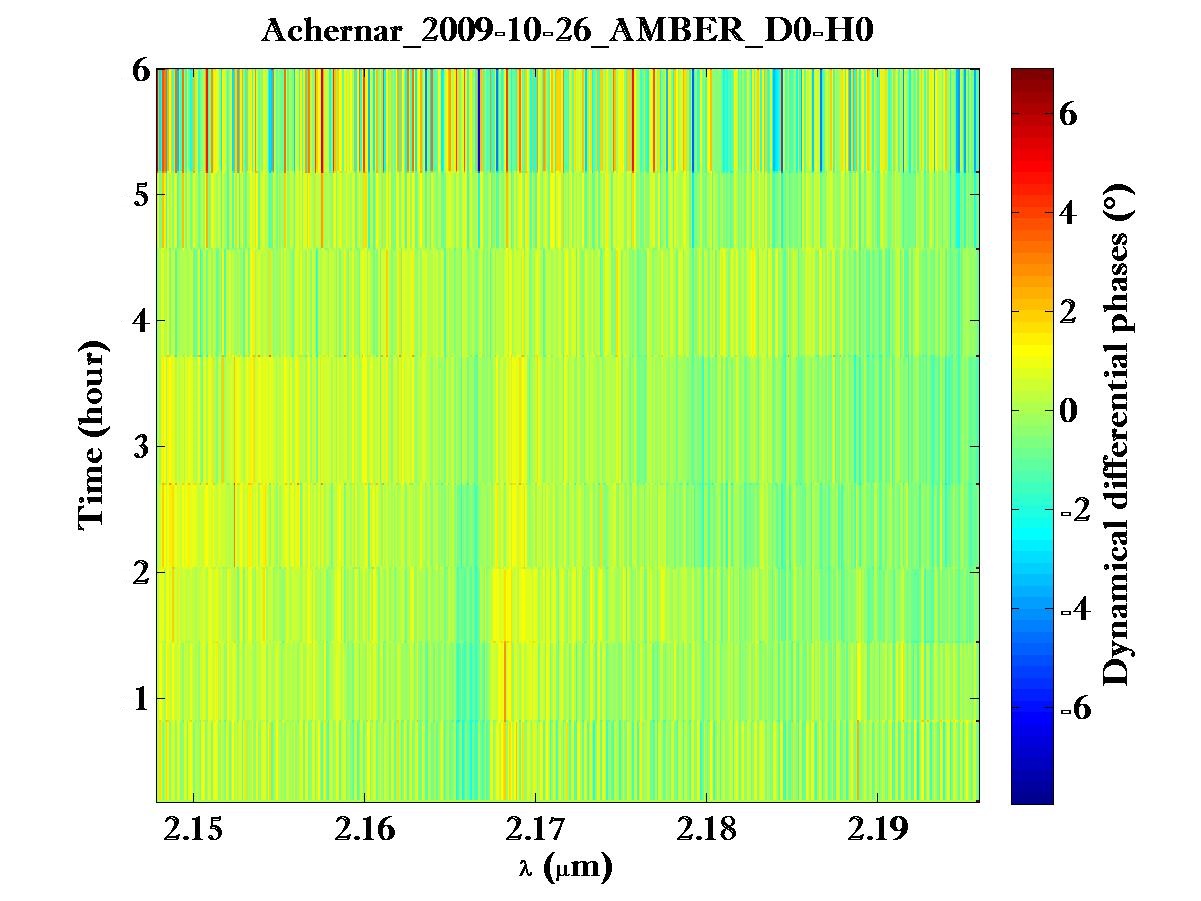}
\includegraphics[width=0.4\hsize,draft=false]{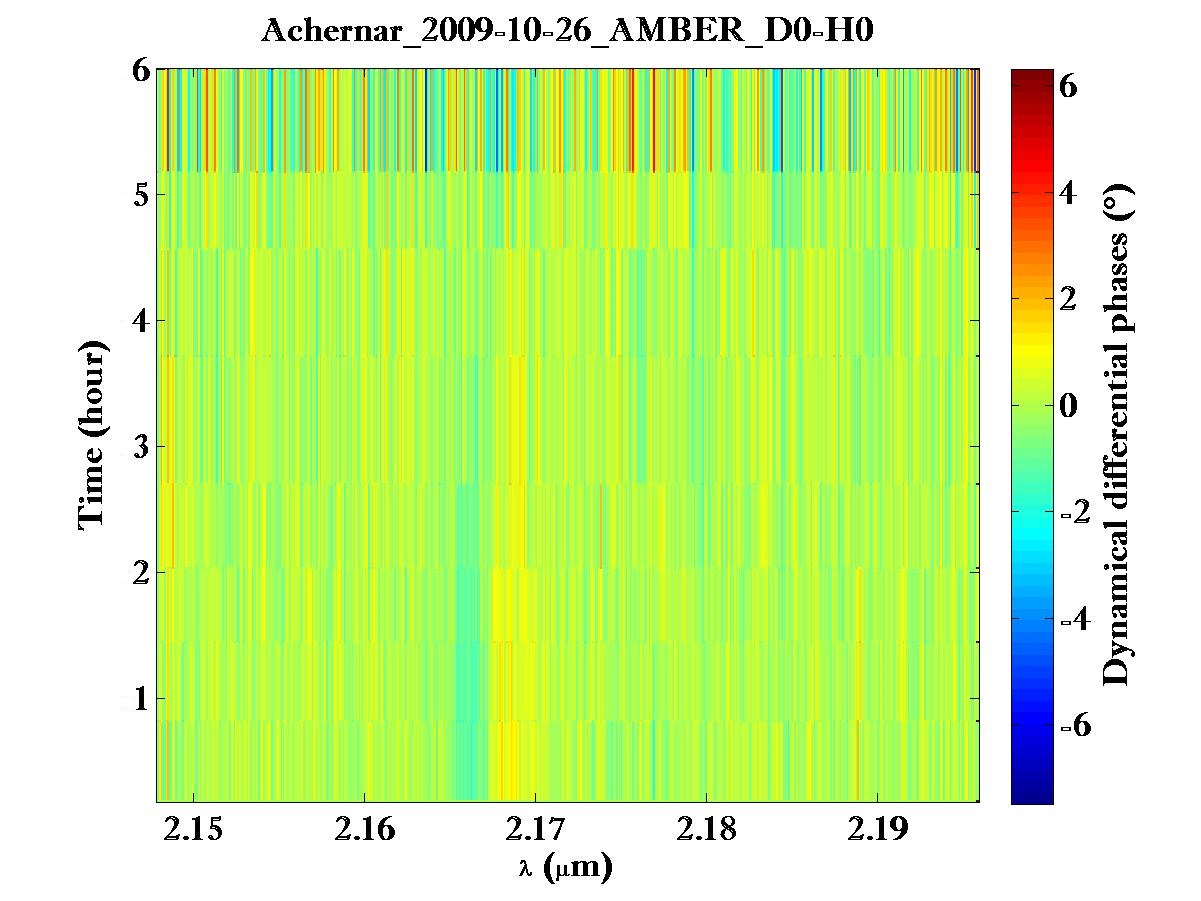}
\end{figure}
\begin{figure}[ht]
\centering
\includegraphics[width=0.4\hsize,draft=false]{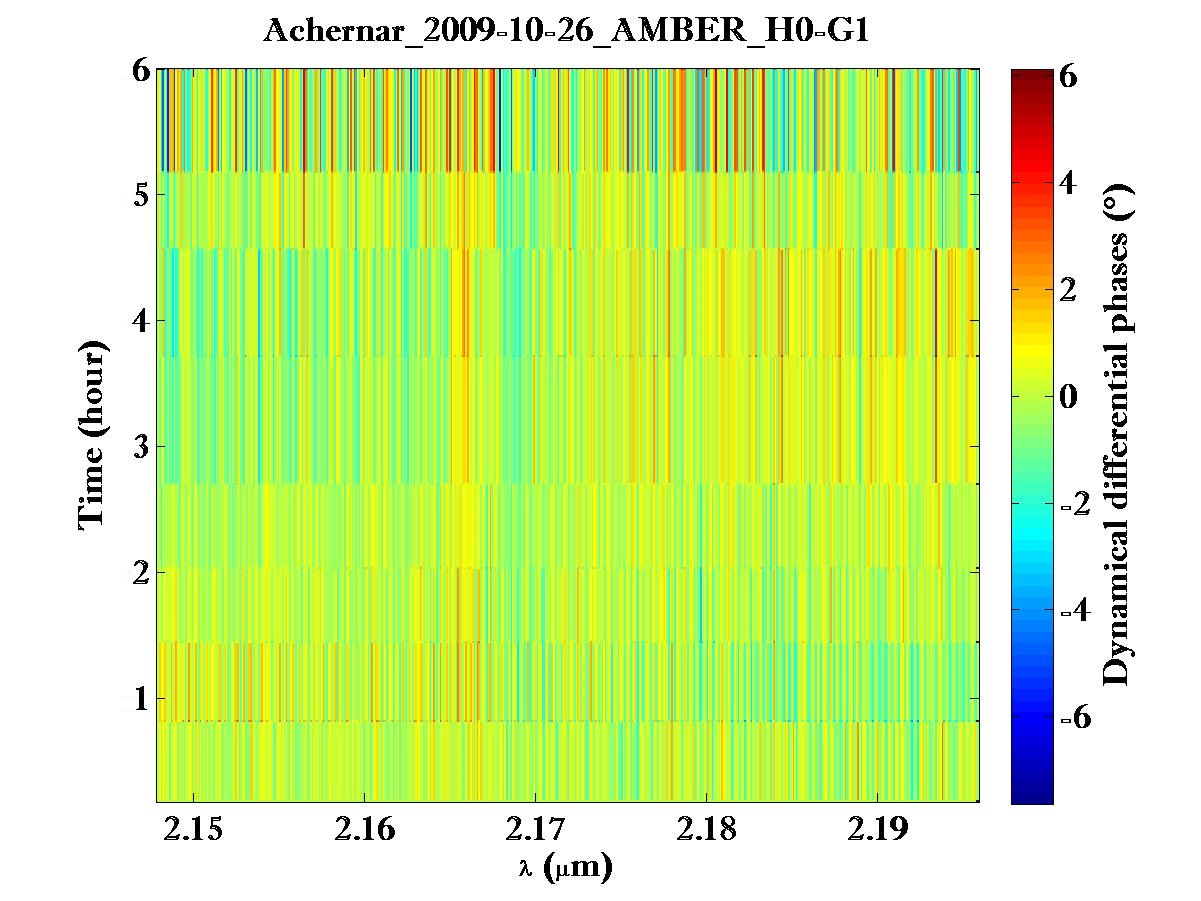}
\includegraphics[width=0.4\hsize,draft=false]{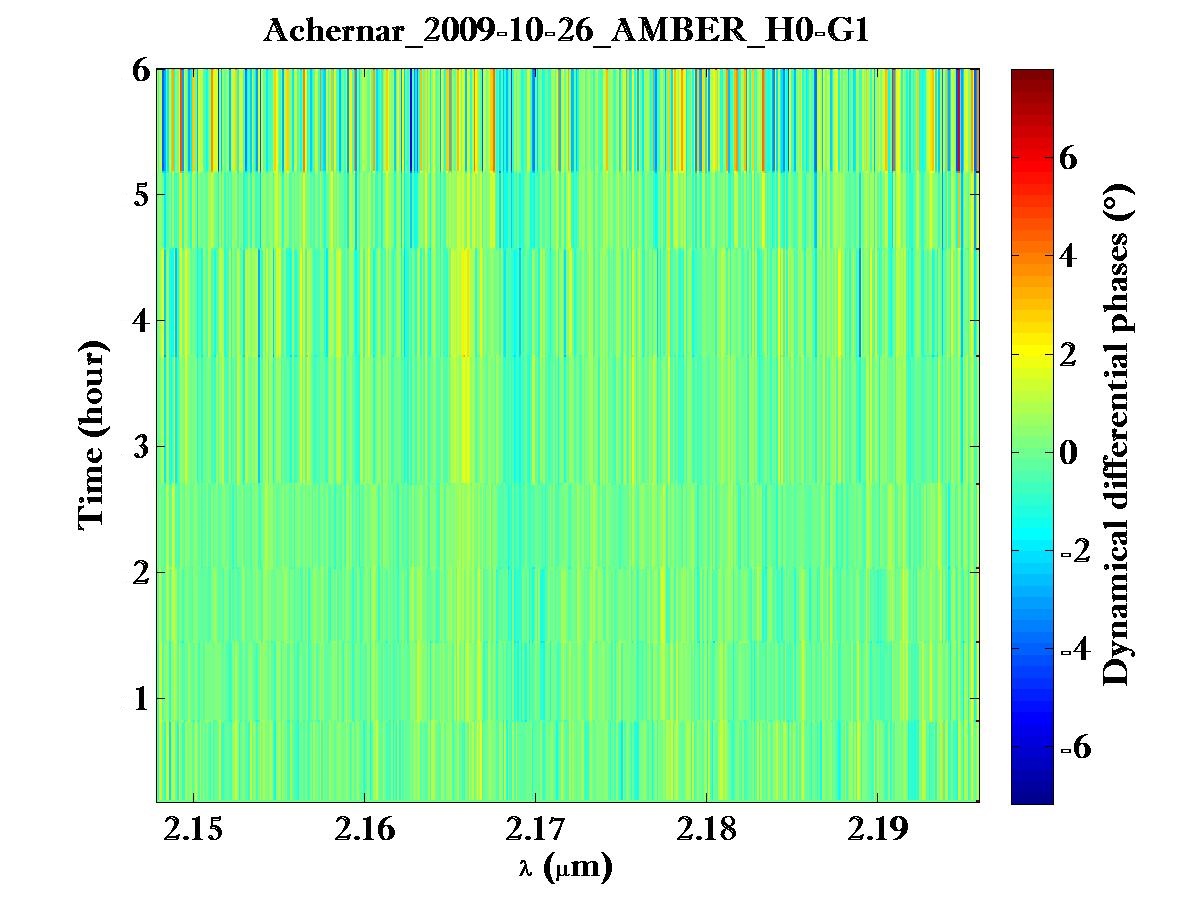}
\end{figure}

\clearpage

\begin{figure}[ht]
\centering
\includegraphics[width=0.4\hsize,draft=false]{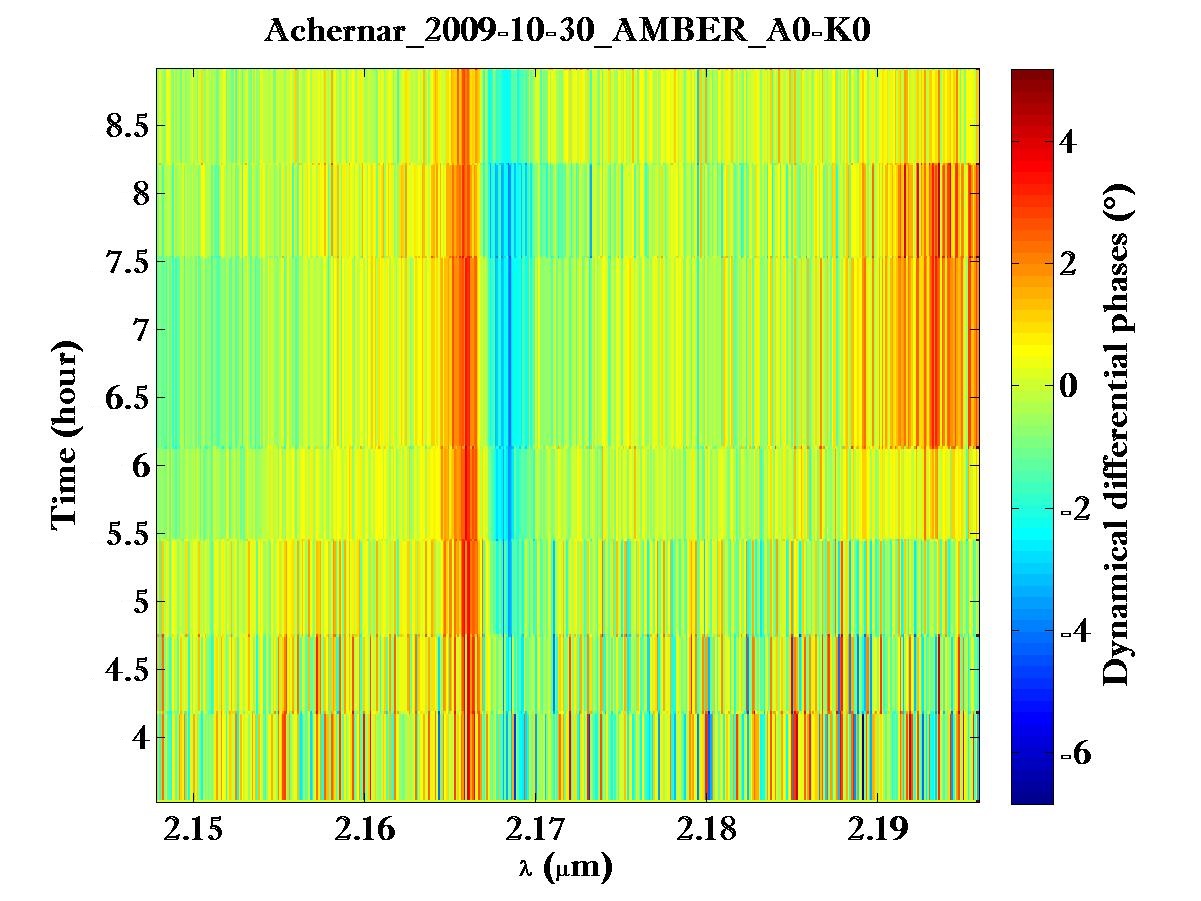}
\includegraphics[width=0.4\hsize,draft=false]{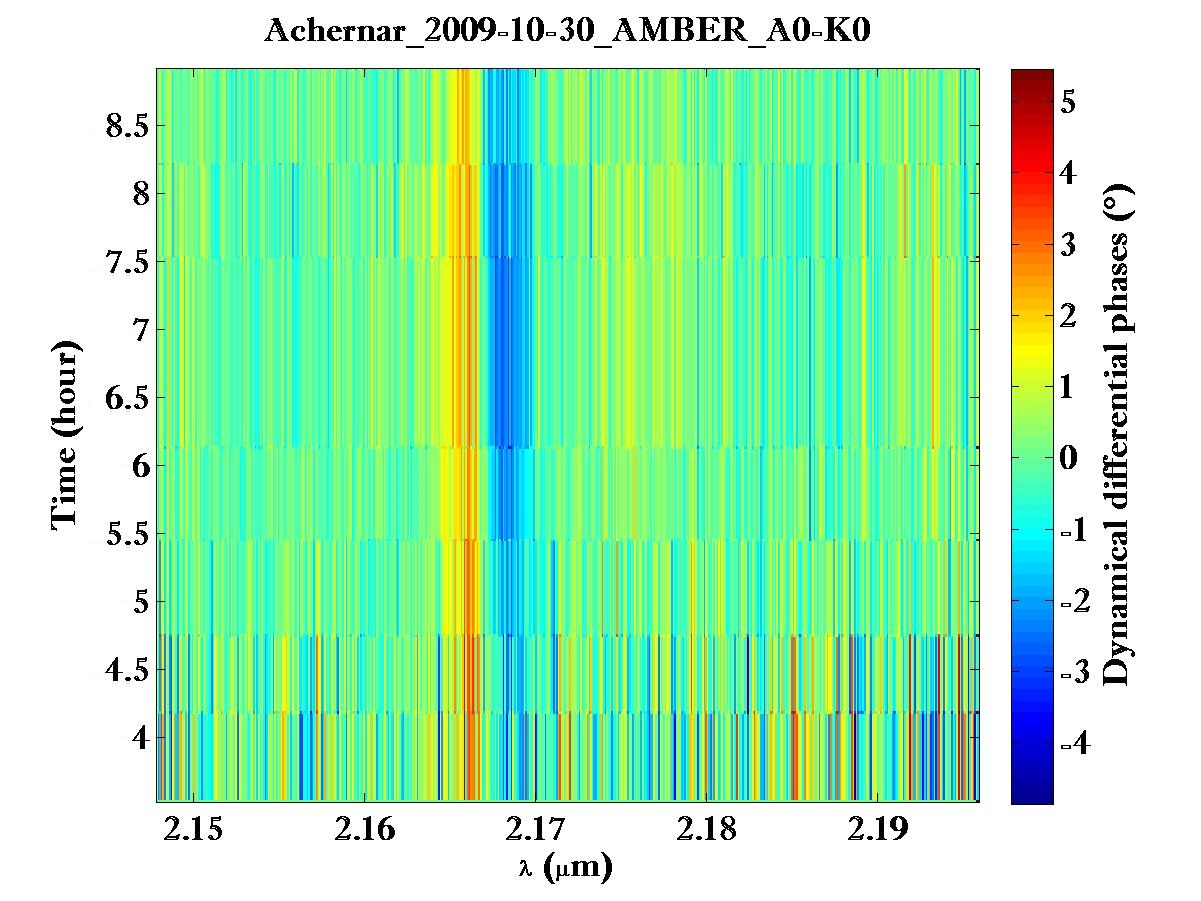}
\end{figure}
\begin{figure}[ht]
\centering
\includegraphics[width=0.4\hsize,draft=false]{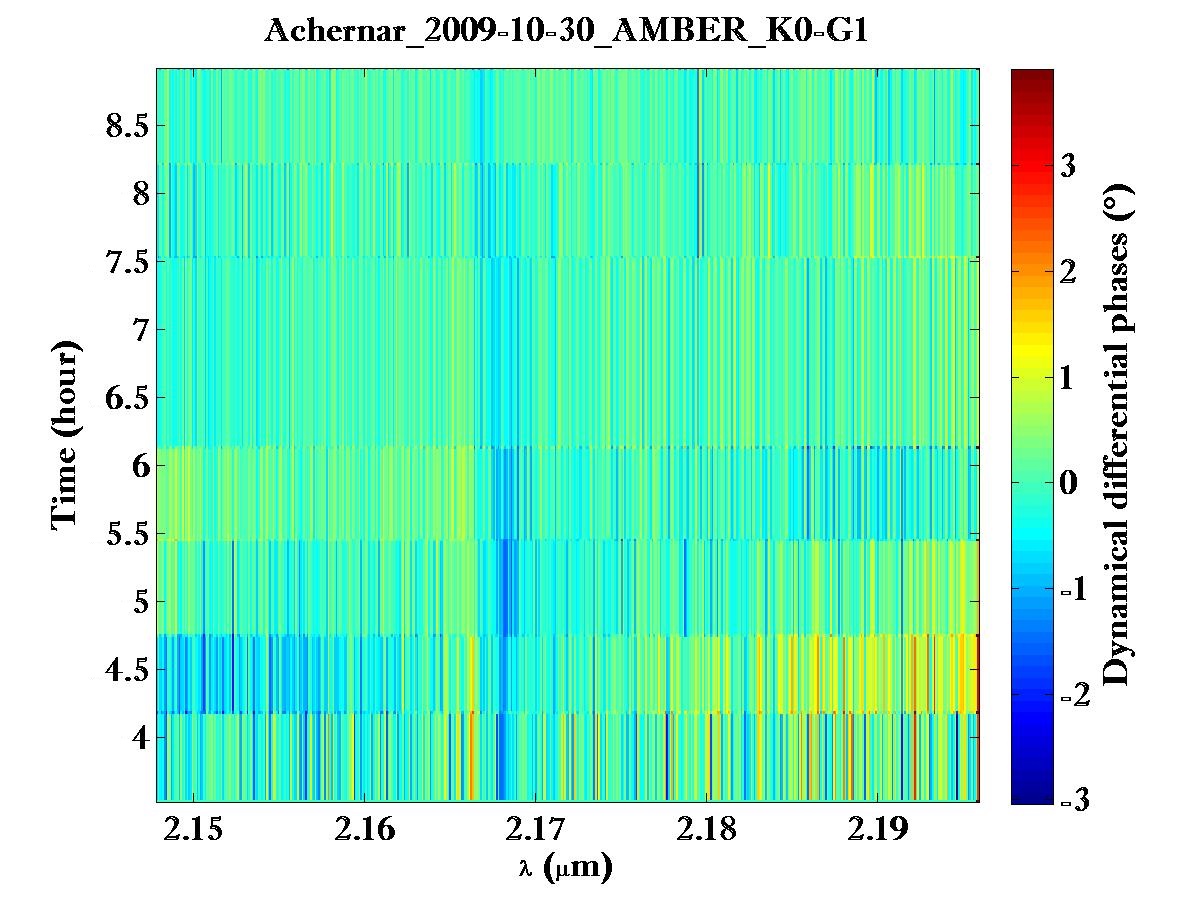}
\includegraphics[width=0.4\hsize,draft=false]{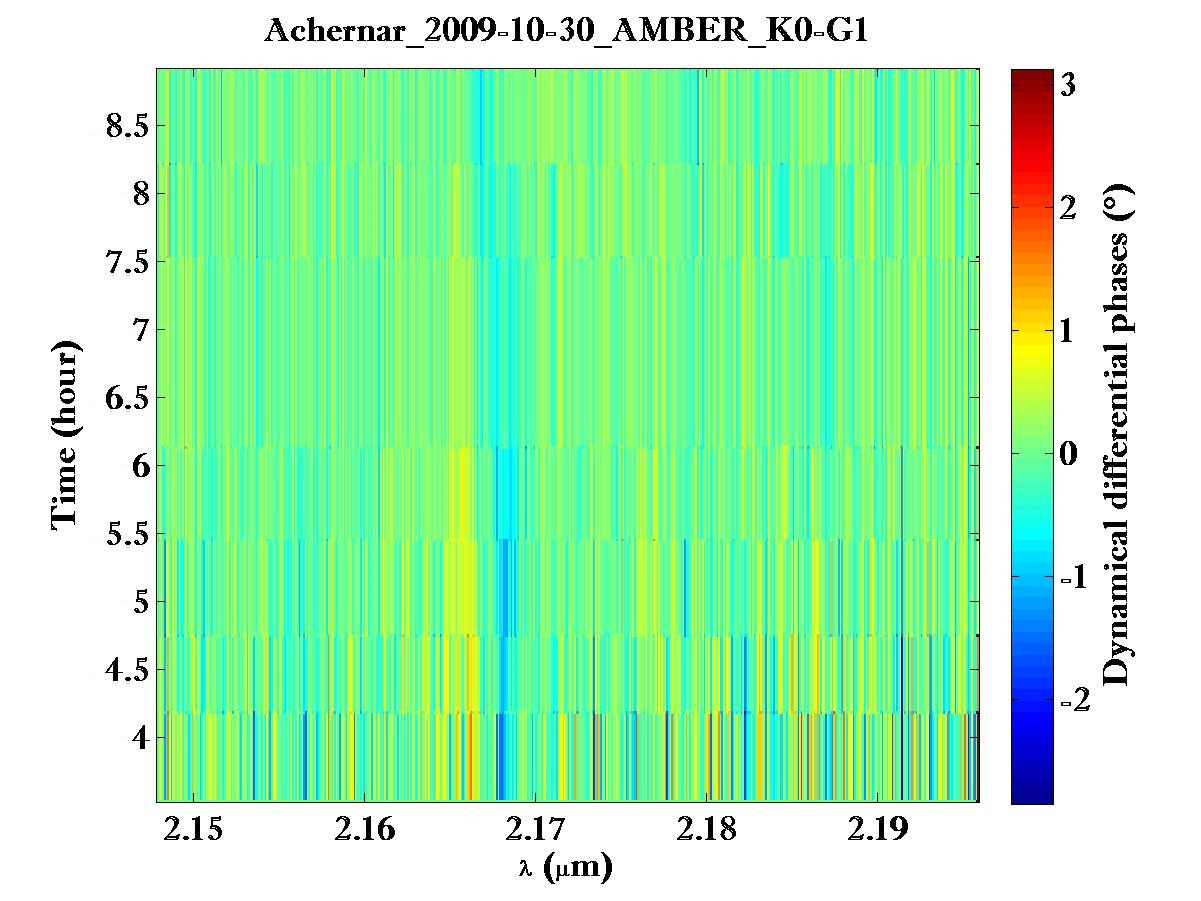}
\end{figure}

\clearpage

\begin{figure}[ht]
\centering
\includegraphics[width=0.4\hsize,draft=false]{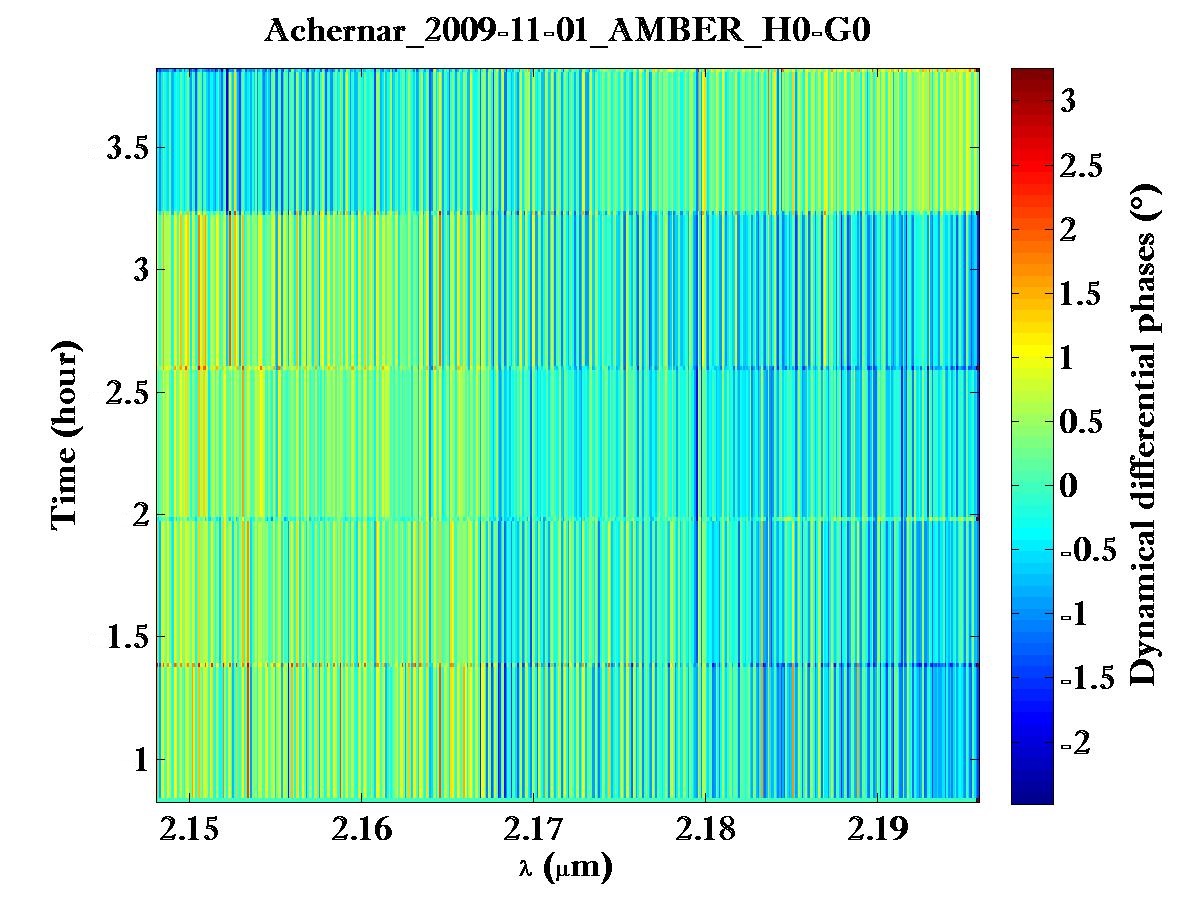}
\includegraphics[width=0.4\hsize,draft=false]{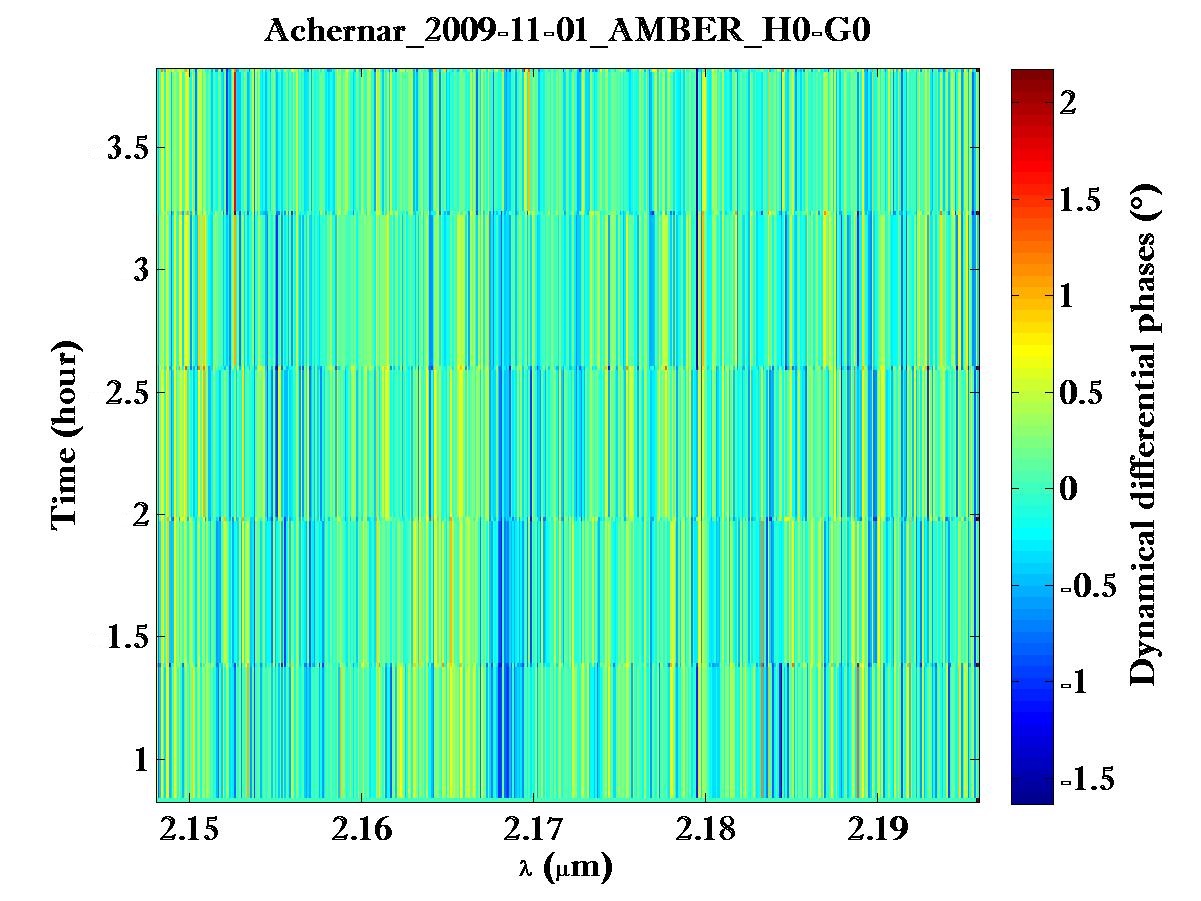}
\end{figure}
\begin{figure}[ht]
\centering
\includegraphics[width=0.4\hsize,draft=false]{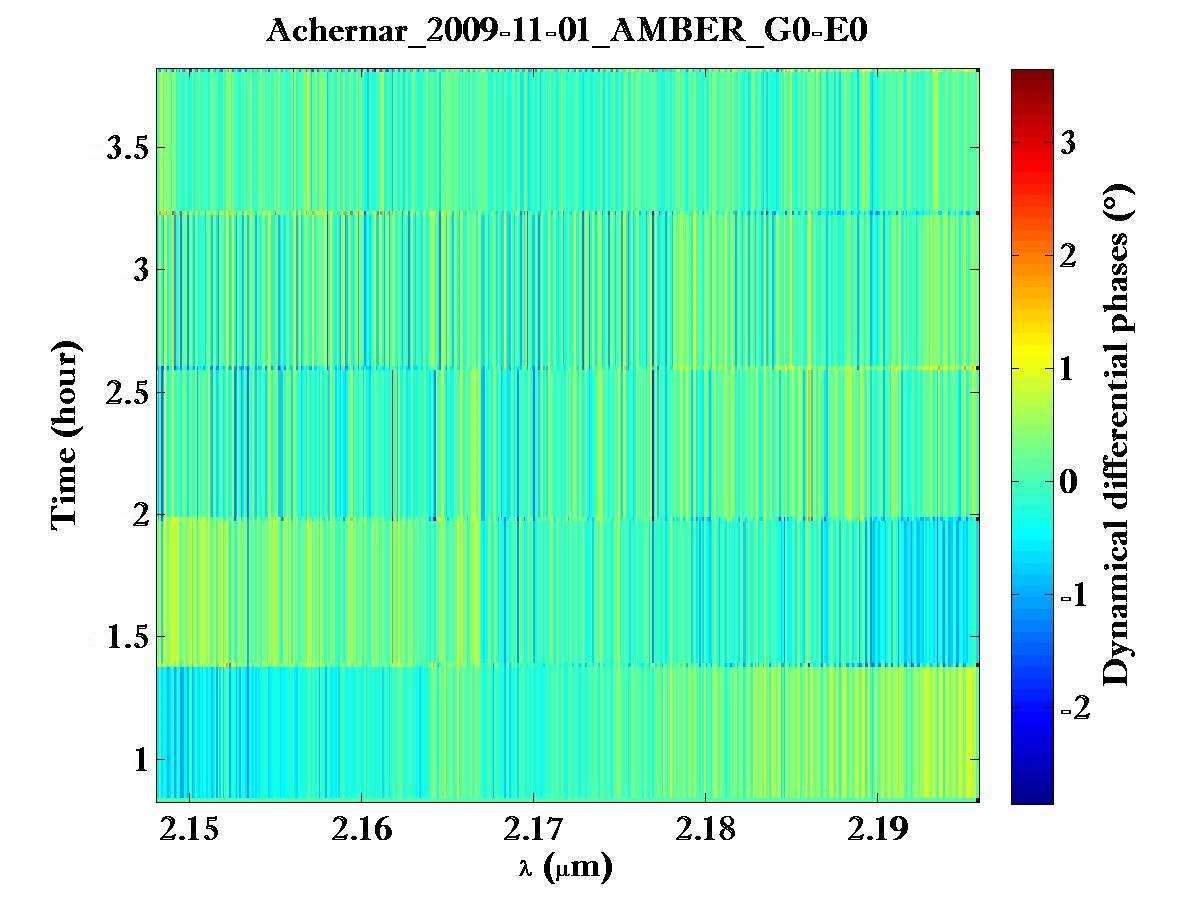}
\includegraphics[width=0.4\hsize,draft=false]{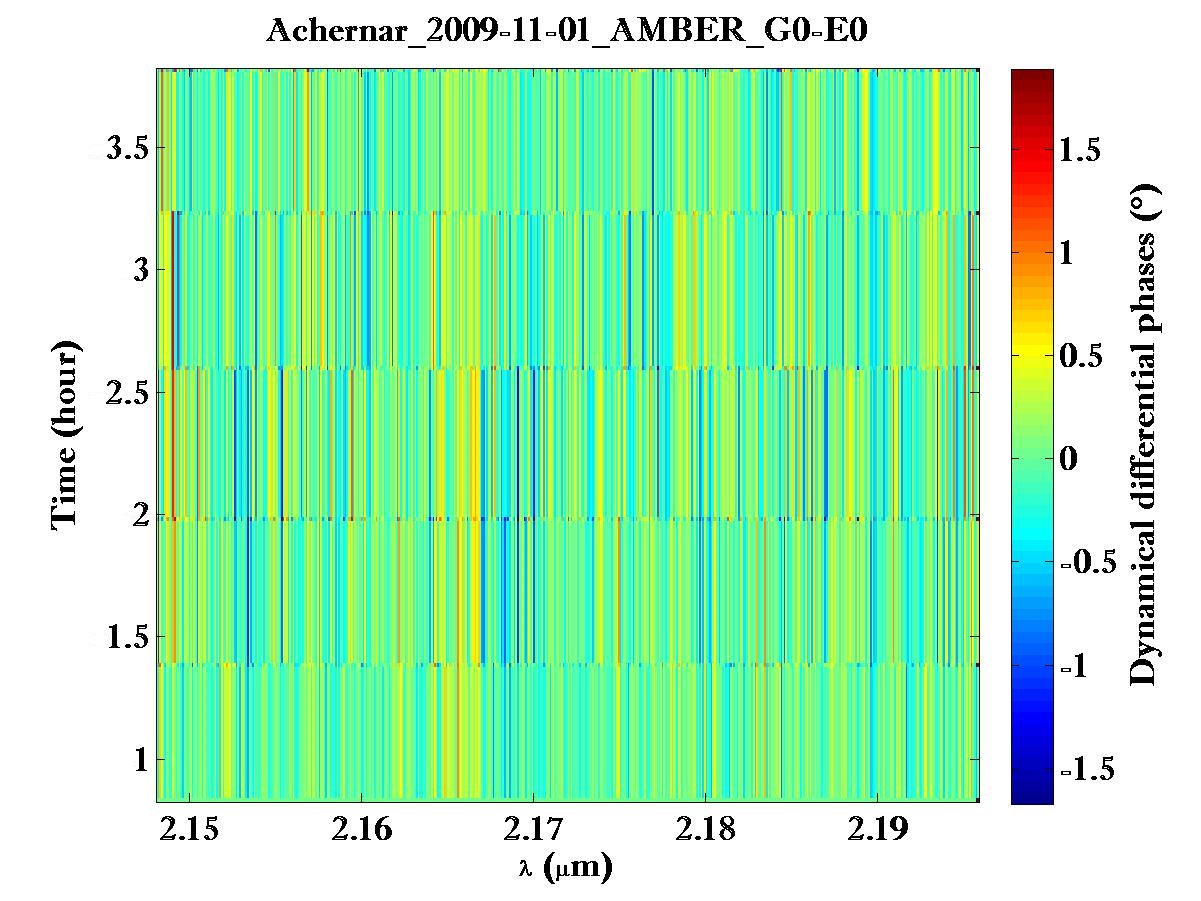}
\end{figure}

\clearpage

\begin{figure}[h!]
\centering
\includegraphics[height=0.6\hsize,draft=false]{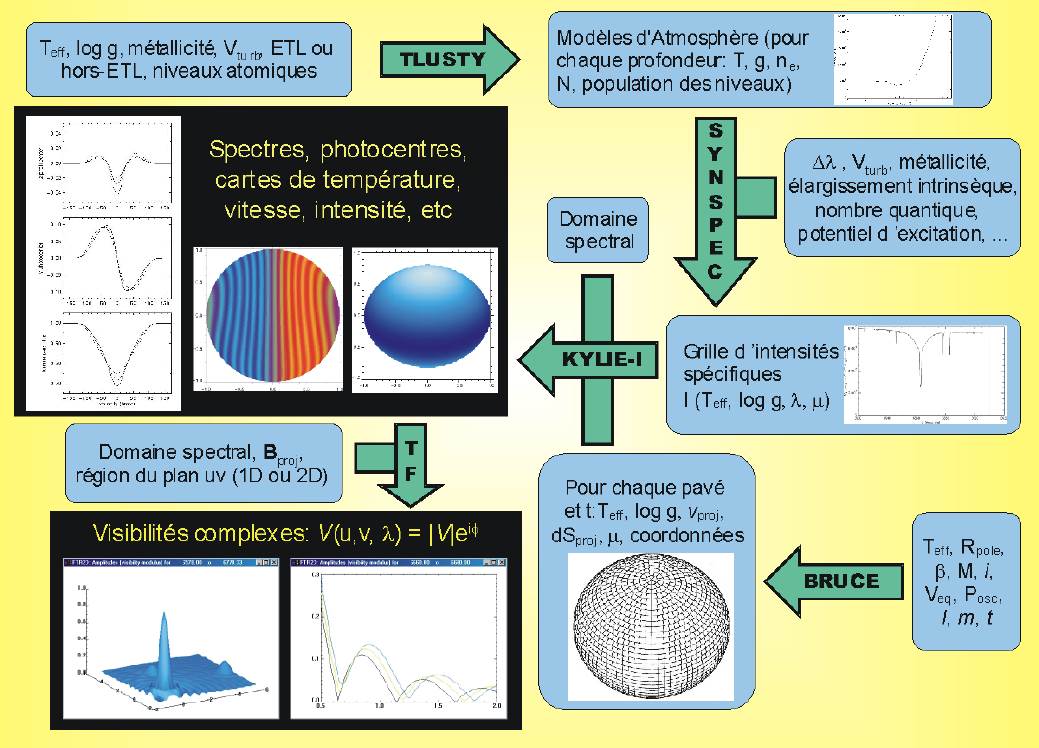}
\caption[Diagramme synoptique de CHARRON]{Diagramme synoptique du modèle physique CHARRON pour l'étude des surfaces stellaires sous l'angle de l'interférométrie optique à longue base \citep{2003A&A...407L..47D}.}\label{Syno_Charron}
\end{figure}

Une fois tout le long et périlleux processus, de demande d'observations, d'observations, de réduction et de traitement des données avec tout le savoir-faire et la compréhension scientifique, instrumentale et technique qui va avec, il nous est maintenant possible de nous occuper de l'étude et de l'interprétation scientifique de nos mesures. Dans le cas des données Achernar 2009 (mesurées, réduites puis traitées) d'AMBER, elles ont été modélisées à l'aide du code CHARRON (Code for High Angular Resolution of Rotating Objects in Nature ; \citet{2012A&A...545A.130D}) développé par Domiciano de Souza en 2003 pour l'étude des rotateurs rapides. Cet outil robuste et puissant élaboré en langage IDL (Interactive Data Language), modélise à la fois les profils de raie via Tlusty/Synspec \citep{1995ApJ...439..875H}, les cartes de vitesse via un modèle inspiré du code de BRUCE \citep{1997PhDT........24T} et d'intensités avec prise en compte des assombrissements centre-bord et gravitationnel via un modèle inspiré aussi du code de KYLIE \citep{1997MNRAS.284..839T, 1997PhDT........24T} pour en extraire les mesurables interférométriques et en déduire leurs paramètres fondamentaux par ajustement $\chi^2$ (voir le diagramme synoptique de ce modèle dans la Fig.\ref{Syno_Charron}). Pour Achernar, seules les $\phidiff$ ont étés ajustées en raison d'un manque d'information à extraire des autres mesurables, à cause justement de la trop faible résolution d'AMBER sur cette étoile (notre couverture (u,v) est en dessous du premier lobe de visibilité), ainsi que décrit dans la sous-section "Les mesurables en interférométrie différentielle" (sous-section \ref{mesurables_DI}; $\phidiff> |1-V^2|>\Psi$). Ce travail auquel j'ai participé, ayant fait l'objet d'une publication dans A\&A, m'a permis de jeter les bases de mon modèle SCIROCCO présenté dans le chapitre \ref{chap:scirocco}.\\

\includepdf[pages=1-10]{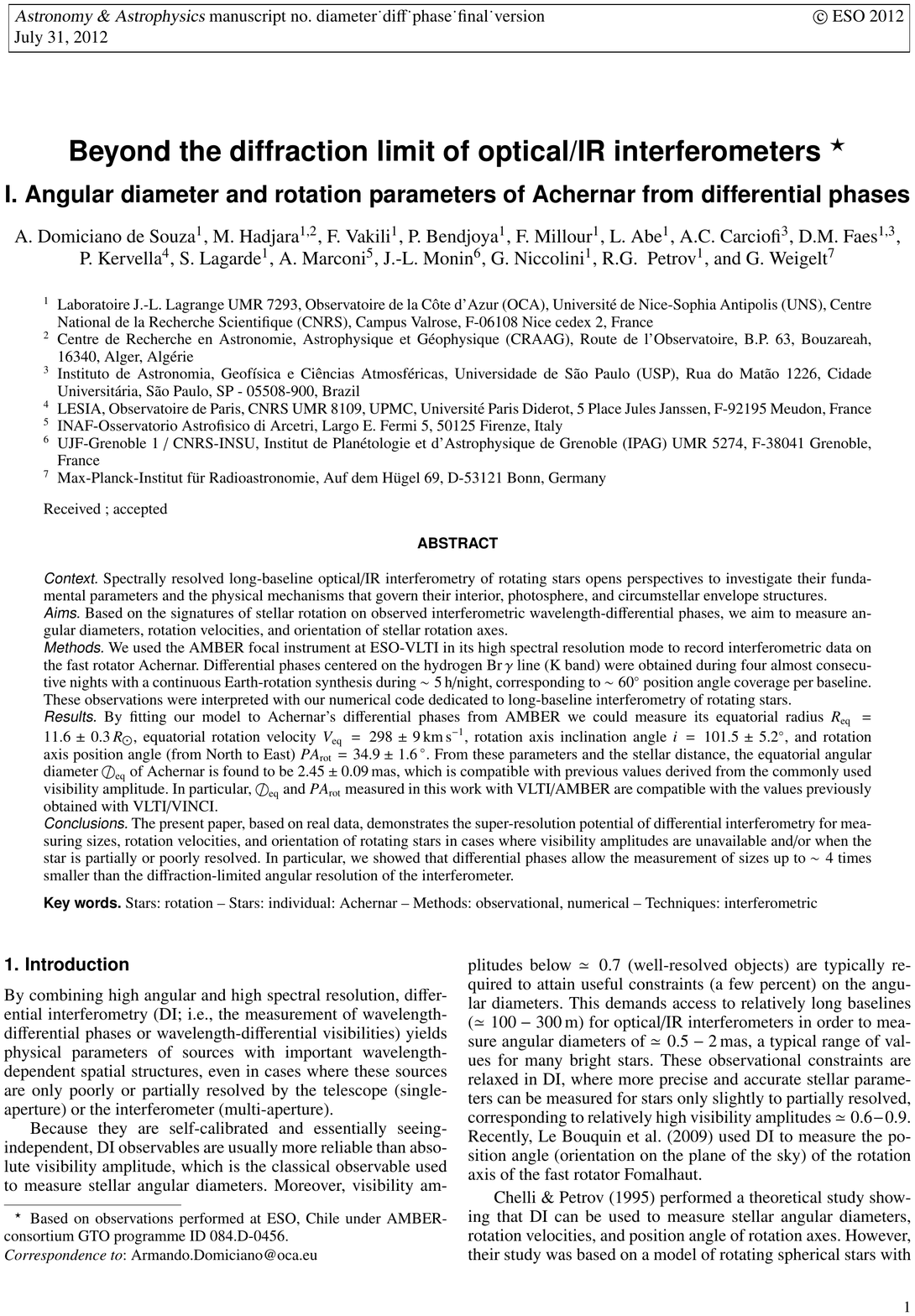}
\newpage
\chapter{SCIROCCO : Un Code pour l'analyse des données Spectro-Interférométrique}
\begin{figure}[h!]
\centering
 \includegraphics[height=0.5\hsize,draft=false]{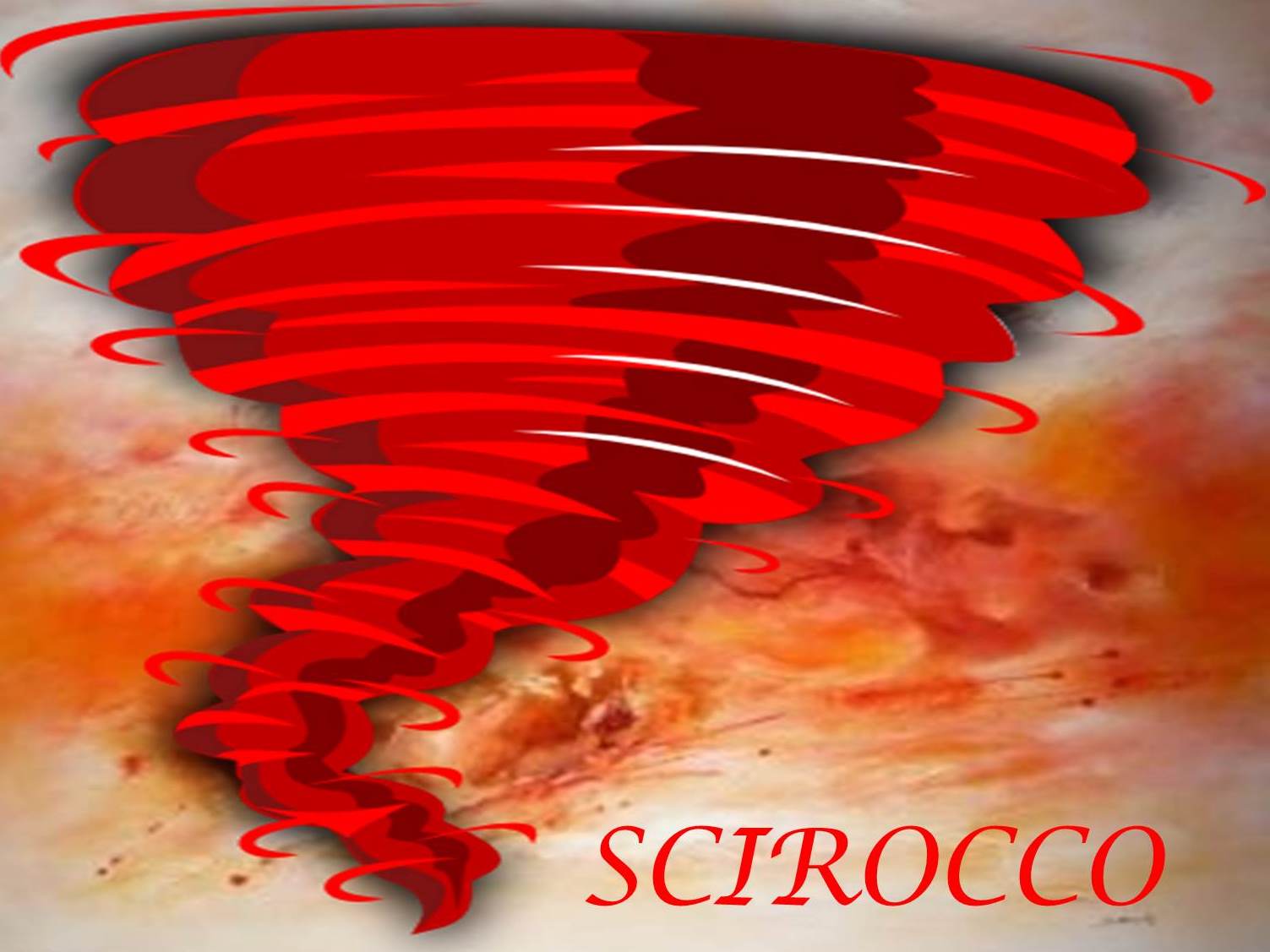}
\end{figure}
\label{chap:scirocco}
\minitoc

%%%%%%%%%%%%%%%%%%%%%%%%%%%%%%%%%%%%%%%%
\def\vsini{v_\mathrm{eq} \sin i} 
\def\kms{\mathrm{km.s}^{-1}}
\def\phidiff{\phi_\mathrm{diff}}
\def\Rsun{\mathrm{R}_{\odot}}
\def\Lsun{\mathrm{L}_{\odot}}
\def\Msun{\mathrm{M}_{\odot}}
\def\Tmean{\overline{T}_\mathrm{eff}}
\def\diameq{\diameter_\mathrm{eq}}
\def\chir{\chi_\mathrm{r}}
\def\chimin{\chi_\mathrm{min}}
\def\chirmin{\chi_\mathrm{min,r}}
%%%%%%%%%%%%%%%%%%%%%%%%%%%%%%%%%%%%%%%%

Juste après la soumission de notre article \citet{2012A&A...545A.130D} (\hyperlink{page.85}{voir p.85}), j'entrepris le développement d'un modèle numérique de rotateurs stellaires rapides, destiné aux observations spectro-interférométriques, avec une approche analytique faisant appel à une modélisation numérique aussi simple que possible. En effet les modèles sur les rotateurs et/ou leurs environnement proche sont relativement peu nombreux dans la communauté astrophysique, et peu destinés à l'interférométrie optique à longue pose. Parmi ceux-ci je cite CHARRON (Code for High Angular Resolution of Rotating Objects in Nature ; \citet{2012sf2a.conf..321D}),  décrit à la fin du chapitre \ref{chap:spec-interfero}, et ESTER (Evolution STellaire En Rotation ; \citet{2013ascl.soft05001R, 2013sf2a.conf..101R}). Ce dernier est un code d'évolution stellaire à deux dimensions développé en C++/ python qui permet de modéliser les paramètres suivants depuis les couches intérieures jusqu'en surface: la pression, la densité, les températures, la vitesse angulaire (rotation différentielle) et la circulation méridionale de l'ensemble d'un volume d'étoiles de masse excédant deux fois celle du soleil. Il existe aussi des codes de simulation de disque circumstellaire parmi lesquels on compte ; SIMECA (SIMulation pour Etoiles Chaudes Actives ; \citet{2008EAS....28..135S}) qui permet sous l'hypothèse de la physique du transfert de rayonnement en équilibre thermodynamique local (ELT) de modéliser:  les distributions de densité, de température, et les champs de vitesse radiale et azimutale des enveloppes circumstellaires à usages photométrique, spectroscopique et interférométrique. Ou encore le cas du modèle Be disk,  proposé par les canadiens \citet{2009ApJ...699.1973S}, plus récent que SIMECA, il est aussi dans le même état d'esprit, et comme son nom l'indique est destiné à l'étude des CSE (CircumStellar Environment) des BE. On peut également citer le modèle d'étude de déplacement du photo-centre d'enveloppe d'étoiles Be en équilibre thermodynamique local (ETL), et en interférométrie IR, proposé par \citet{2012ApJ...744...19K}.\\

Sinon on compte aussi l'existence de codes plus poussés, combinant à la fois la modélisation d'étoiles en rotation avec leurs environnements circumstellaires. C'est le cas d'HDUST développé par A.Carciofi et J.Bjorkman \citep{2006ApJ...639.1081C} qui est un puissant code de transfert radiatif Monte Carlo, qui permet d'obtenir des spectres et des cartes d'intensité en lumière naturelle et/ou polarisée pour des CSE d'étoiles massives incluant du gaz et de la poussière. Néanmoins ce genre de modèle nécessites un temps et une  puissance de calcul assez élevées.\\

Ainsi, ma contribution dans ce domaine de recherche a été de créer un code analytique simple, mais efficace, destiné à interpréter les observations spectro-interférométriques des étoiles en rotation rapide (travail principal sur lequel est axé mon code) avec quelques essais d'intégration d'effets supplémentaires, tels que les pulsations non radiales (PNR) et/ou les disques circumstellaires autour de ces étoiles. Mon modèle s'appelle SCIROCCO, qui est l'acronyme de :\\

\textbf{SCIROCCO:} \underline{S}imulation \underline{C}ode of \underline{I}nterferometric-observations for \underline{RO}tators and \underline{C}ir\underline{C}umstellar \underline{O}bjects. Il est basé sur un modèle polychromatique semi-analytique pour les rotateurs rapides prenant en compte, le rayon angulaire, la vitesse de rotation à la surface de l'étoile (avec et sans rotation différentielle) incluant l'inclinaison l'axe de rotation intrinsèque par rapport à la ligne de visée ( le $V_{\rm eq}\sin i$ ), et l'aplatissement apparent de cette étoile.\\

Ce code (écrit en langage Matlab) peut prendre en compte plusieurs profils de raie dit synthétiques (gaussien, lorentzien, profil de Voigt) aussi bien que ceux issus des modèles d'atmosphère stellaire (Kurucz-Tlusty/Synspec, Phoenix...etc.). Il applique sur des cartes d'iso-vitesses projetées et obtenues par effet Doppler. Le tout est combiné à des cartes d'intensité projetées sur le ciel incluant (ou non) l'effet de l'assombrissement centre-bord, et/ou l'assombrissement gravitationnel (l'effet von Zeipel) avec des paramètres de température et de longueurs d'onde adéquats. L'intensité au continuum est tout simplement tirée de l'équation du corps noir (fonction de Planck, pour une longueur d'onde et une température effective moyenne données). Cette approche nous permet d'extraire les informations interférométriques recherchées: visibilité, spectre, photo-centres, phases, et clôtures de phase ... etc., et ce pour des bases interférométriques bien déterminées. Le but final de cette démarche est de comparer les observables simulées aux observations effectuées sur un spectro-interféromètre à longue base, et de trouver les paramètres qui concordent le mieux: températures effectives, vitesse de rotation, inclinaison, rayon angulaire, aplatissement, coefficient de rotation différentielle, ... etc. Ceci est l'objet de la section suivante.\\

\section{Forme et vitesses de surface d'un rotateur}

Tel qu'illustré dans la Fig.\ref{roche_model}, la forme d'une étoile évolue avec celle de sa vitesse de rotation. L'observation de la rotation stellaire est détaillée dans le chapitre \ref{chap:rota}, mais la mesure de la photosphère aplatie des étoiles en rotation rapide, bien que prédite théoriquement dès les années 20, n'a été obtenue qu'à partir des années 2000, grâce à l'interférométrie qui permet d'atteindre des résolutions spatiales de l'ordre de la milli-arc-seconde nécessaire à la mesure directe de cet aplatissement.\\

Ainsi, la détermination du degré d'aplatissement d'une étoile et de sa carte de vitesse de rotation surfacique respective sont deux caractéristiques fondamentales stellaires nécessaires à l'élaboration de tout modèle (destiné à l'étude des rotateurs rapides). A titre comparatif ; le Soleil, seule étoile assez proche de la terre pour nous permettre d'étudier sa forme et sa surface à l'\oe{}il nu, nous révèle une forme très proche de la sphère parfaite d'un rayon moyen de $959".28 \pm 0".15$ \citep{2004ApJ...613.1241K}, et une valeur de $959".78\pm0.19"$ à une longueur d'onde de $535.7$ $nm$ mesurée par Picard Sol au plateau de calern caussols entre 2011 et 2013 \citep{2014A&A...569A..60M}, avec une variation entre les deux rayons polaire et équatorial, de seulement $9.0 \pm 1.8 mas$ \citep{2003SoPh..217...39R}. Ceci traduit un coefficient d'aplatissement $\left(\frac{R_{eq}-R_{pol}}{R_{eq}}\right)$ de l'ordre de $0.005\%$, alors que l'aplatissement peut atteindre pour certains rotateurs rapides (notamment des Be) $20\%$ à $30\%$. La comparaison des vitesses de rotation équatoriale/polaire\footnote{Les vitesses de rotation équatoriale et polaire varient selon le type spectral et la classe de luminosité des étoiles. Etant donné que la force centrifuge agit différemment selon la latitude stellaire, elle est bien plus prononcée à l'équateur. Ce qui se répercute sur la vitesse angulaire de ce corps (non solide) en rotation, selon la latitude relativement à l'axe de rotation. C'est ce qu'on appelle \textbf{la rotation différentielle}, et qui est également présent au Soleil.} est toute aussi impressionnante, car avec une vitesse de rotation moyenne équatoriale de $2$ $\kms$; i.e. une période de rotation de $25$ jours à l'équateur et de $34$ j à $85^\circ$ de latitude près des pôles \citep{1990ApJ...351..309S}, le Soleil tourne donc 100 à 150 fois moins vite que certains rotateurs tournant à une vitesse proche de leurs vitesse critique.\\

\begin{figure*}[ht!]
\centering
 \includegraphics[height=0.3\hsize,draft=false]{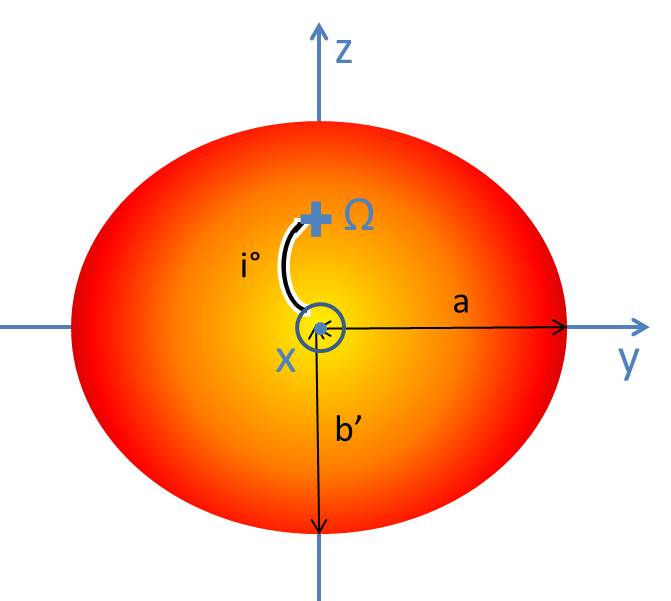}
 \includegraphics[height=0.3\hsize,draft=false]{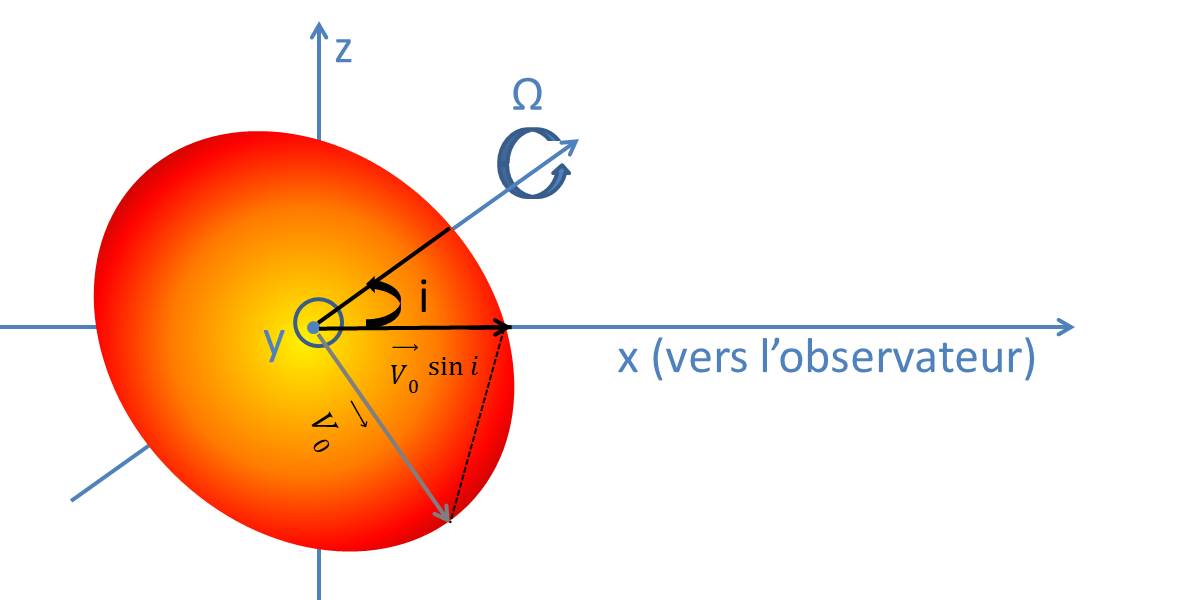}
 \includegraphics[height=0.1\hsize,draft=false]{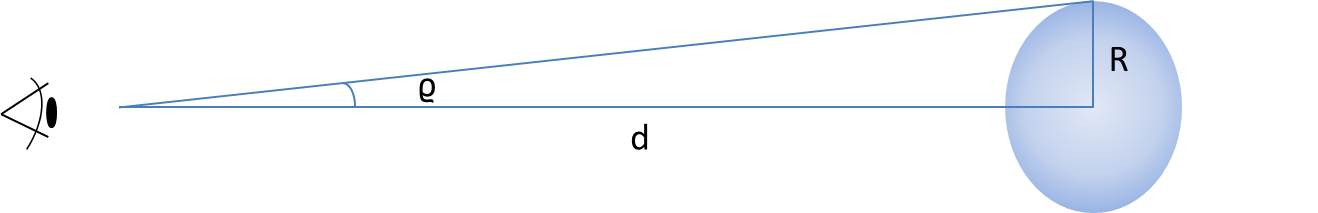}
 \caption[Système de référence adopté pour une étoile en rotation]{\textsl{\textbf{En haut à gauche:}} Système de référence adopté pour une étoile en rotation (étoile aplatie avec un demi grand axe \textit{a} et un demi petit axe \textit{b}, ici le demi petit axe apparent est $b '= a.b / (a + (b-a) \cos (i))$\footnotemark;  respectant les équations de l'ellipso\"{i}de de révolution). Notons que pour $i=90^\circ$ $\rightarrow$ $a=R_{eq}$ et $b'=b=R_{pol}$. La croix indique le point où l'axe de rotation intercepte la surface de l'étoile. Cet axe forme un angle \textit{i} avec la direction de l'observateur (axe x) et sa projection sur le ciel est parallèle à l'axe z. Notons ici que l'angle d'orientation choisi de l'étoile est nul, autrement l'étoile serait inclinée. \textsl{\textbf{En haut à droite:}} Le paramètre $\vsini$ tel que perçu par un observateur sur terre selon l'axe $x$, où pour $i=90^\circ$ la vitesse de rotation perçue sera $V_0=v_{eq}$ qui n'est autre que la vitesse de rotation équatoriale.  \textsl{\textbf{En bas:}} Le rayon angulaire $\rho=\frac{R}{D}$ tel que perçu par l'observateur à une distance $d$ pour l'étoile ayant un rayon $R$. Ici aussi et pour une orientation dite "dans le plan équatorial" ("edge on") ($i=90^\circ$) le rayon angulaire apparent sera celui du rayon équatorial $\rho=\frac{\diameq}{2}$.}\label{reference}
\end{figure*}

Avant d'aborder les principes et équations régissant la forme d'un rotateur rapide et sa carte des vitesses de surface, je vais tout d'abord adopter un système de notations et de référence appropriés, schématisé dans la Fig. \ref{reference}.\\

\footnotetext{Cette formule a été déduite du fait que le demi petit axe apparent $b'$ est inversement proportionnelle à l'angle $i$ (ex. $b'=\frac{1}{f(\cos i)}$). En effet, en supposant que cette fonction est linéaire, et connaissant les conditions limites (à $i=0^\circ$ $\rightarrow$  $\cos i=1$ $\rightarrow$  $b'=a=R_{eq}$ \& à $i=90^\circ$ $\rightarrow$  $\cos i=0$ $\rightarrow$  $b'=b=R_{pol}$) on déduit bien que $\frac{1}{b'}=\frac{b-a}{ab}\cos i+\frac{1}{b}$.}

\subsection{Formalisme théorique de la rotation stellaire}
\begin{figure}[ht!]
\centering
 \includegraphics[height=0.5\hsize,draft=false]{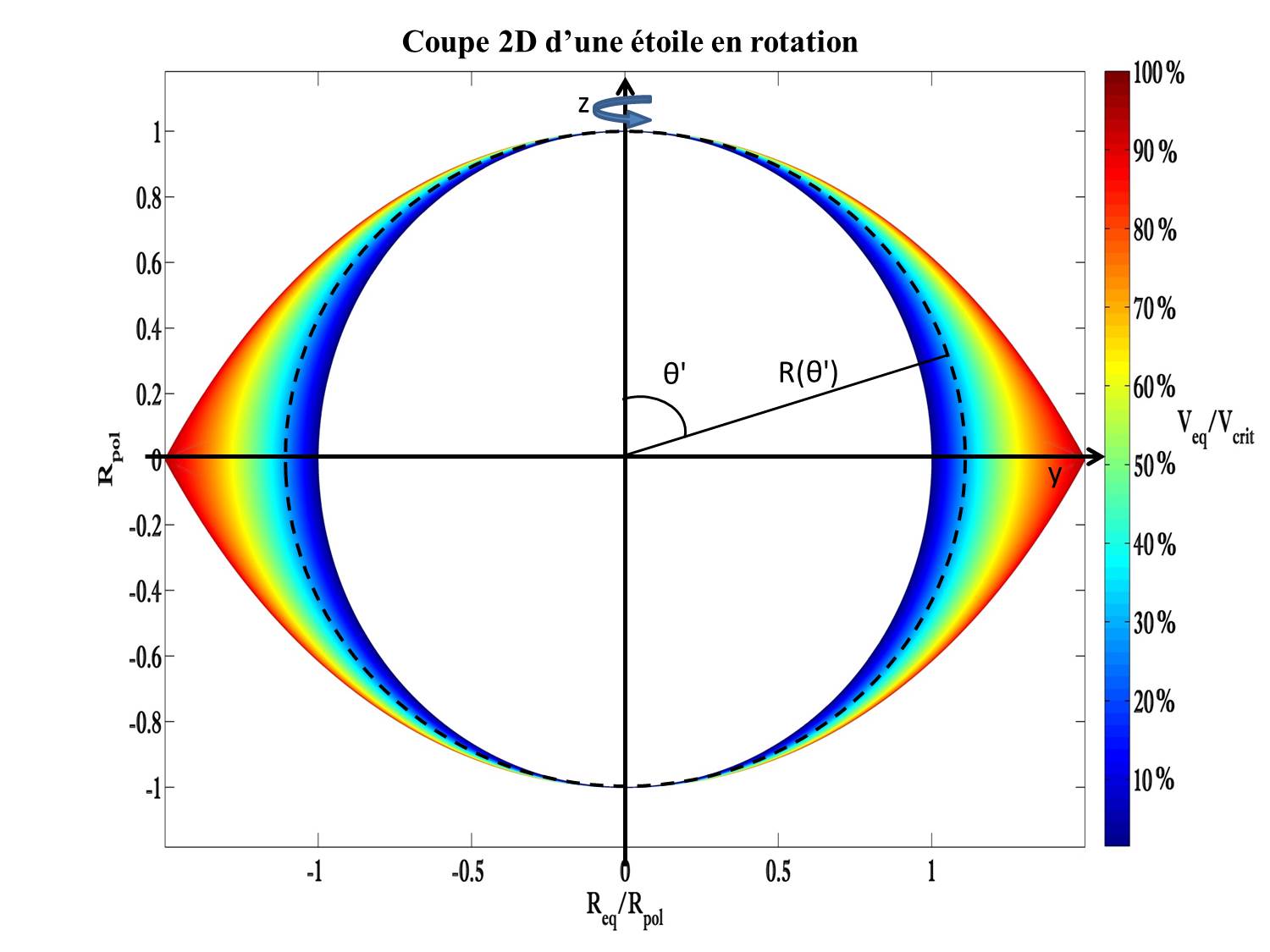}
\caption[Référentiel 2D adopté pour une étoile en rotation]{Référentiel 2D adopté pour une étoile en rotation uniforme. Ici $(R(\theta'),\theta')$ respectivement le rayon et la co-latitude sont les coordonnées polaires associées aux coordonnées cartésiennes $(y,z)$, pour une étoile tournant à une vitesse angulaire équatoriale $\Omega_{eq}=\frac{v_{eq}}{R_{eq}}$.}\label{2D_rot_star}
\end{figure}

Considérons une coupe 2D aux coordonnées polaires (rayon, co-latitude)=$(R(\theta'),\theta')$ (où $0\leq\theta'\leq\pi$) d'une étoile "par l'équateur" ("edge on") (à inclinaison $i=\frac{\pi}{2}$), de rayon équatorial $R_{eq}=R(\theta'=\frac{\pi}{2})$ et de masse $M$ en rotation uniforme avec une vitesse angulaire équatoriale $\Omega_{eq}=\frac{v_{eq}}{R_{eq}}$. L'étoile est parfaitement sphérique pour une vitesse de rotation nulle $R(\theta')=R_{eq}=R_{pol}$ (voir Fig.\ref{2D_rot_star}).\\

Ainsi, la force s'exerçant sur un élément à la surface d'une telle étoile serait à la fois gravitationnelle et centrifuge, i.e.:\\

\begin{equation}
\vec{F}(R(\theta'),\theta')=m\Omega^2 R(\theta')\sin^2\theta'\vec{U_\theta}-\frac{GmM}{R^2(\theta')}\vec{U_r}
\label{4.1}
\end{equation}

Où $G$ est la constante gravitationnelle de Newton et $m$ la masse ponctuelle de l'élément aux coordonnées $(R(\theta'),\theta')$. Dorénavant et pour plus de commodité $ R(\theta')$ sera simplement noté $R$. Notons qu'ici la vitesse de rotation aux pôles est nulle, d'où la dépendance en $\sin^2\theta'$ de la force centrifuge. En s'intéressant à l'\textit{équipotentielle de surface gravito-rotationnelle stellaire} $\Xi$ (où $\vec{F}=-m\vec{\nabla}\Xi(R)$), on peut reprendre les équations théoriques de \citet{1963ApJ...138.1134C}, de sorte que l'Eq.\eqref{4.1} devient:\\

\begin{equation}
\Xi(R,\theta')=\frac{GM}{R}+\frac{\Omega^2}{2} R^2\sin^2\theta'= \frac{GM}{R_{pol}}
\label{4.2}
\end{equation}

En effet, comme le potentiel gravito-rotationnel est constant à la surface du rotateur (tel que démontré par \citet{1926ics..book.....E}), cette constante ne peut être que le potentiel gravitationnel polaire, le potentiel rotationnel étant nul à cette co-latitude et avec l'hypothèse de $R_{pol}$ constant. Plus rigoureusement, un développement du module de la gravité de surface $|\vec{g}|=|-\vec{\nabla}\Xi(R,\theta')|$ en coordonnées polaires conduit à:\\

\begin{equation}
|\vec{g}|=\left[ \left(\frac{GM}{R^2}-\Omega R\sin^2\theta'\right)^2+\Omega R^2\sin^2\theta'\cos^2\theta\right]^\frac{1}{2}
\label{4.3}
\end{equation}

Ainsi, au niveau de l'équateur ($\theta'=\frac{\pi}{2}$) et pour une vitesse critique $v_{eq,crit}$, l'étoile perd son équilibre et se disloque car $|\vec{g}|=0$. L'Eq.\eqref{4.3} se ramène alors à la vitesse critique de libération $v_{crit}=\sqrt{\frac{GM}{R_{eq, crit}}}$ (l'Eq.\eqref{v_crit}), avec $R_{eq,crit}$ est le rayon critique. L'insertion de l'Eq.\eqref{v_crit}) dans l'Eq.\eqref{4.2}, toujours pour $\theta'=\frac{\pi}{2}$, nous donne:\\

\begin{equation}
\frac{R_{eq,crit}}{R_{pol}}=\frac{3}{2}
\label{4.4}
\end{equation}

Avec $\Omega_{eq,crit}=\frac{v_{eq,crit}}{R_{eq,crit}}$, et toujours au niveau de l'équateur, pour $\theta'=\frac{\pi}{2}$, on déduit de l'Eq.\eqref{4.2} le degré de sphéricité $D=\frac{R_{pol}}{R_{eq}}$:\\

\begin{equation}
D=\frac{R_{pol}}{R_{eq}}=1-\frac{v^2_{eq}R_{pol}}{2GM}=\left(1+\frac{v^2_{eq}R_{eq}}{2GM}\right)^{-1}
\label{4.5}
\end{equation}

Dans le cas d'une vitesse nulle les deux rayons polaire et équatorial restent identiques et dans ce cas $D=1$. Autrement l'étoile est aplatie et $D<1$. Des deux dernières équations (\eqref{4.5} \& \eqref{4.6}), avec $D_{crit}=\frac{R_{pol}}{R_{eq,crit}}$ et $\Omega_{crit}=\frac{v_{eq,crit}}{R_{eq,crit}}$, on déduit que:\\

\begin{equation}
\left(\frac{\Omega}{\Omega_{crit}}\right)^2=2\frac{1-D}{D}\left(\frac{3}{2}D\right)^3
\label{4.6}
\end{equation}

En définissant maintenant $r(\theta')=\frac{R(\theta')}{R_{eq}}$, l'Eq.\eqref{4.2} peut être réécrite à l'aide de $r(\theta')$ et de $D$ comme suit \citep{2002A&A...393..345D}:\\

\begin{equation}
r(\theta')^3+\left(\frac{1}{1-D}\frac{1}{\sin^2\theta'}\right)(1-r(\theta'))=0
\label{4.7}
\end{equation}

\begin{figure}[ht!]
\centering
\includegraphics[height=0.5\hsize,width=0.6\hsize,draft=false]{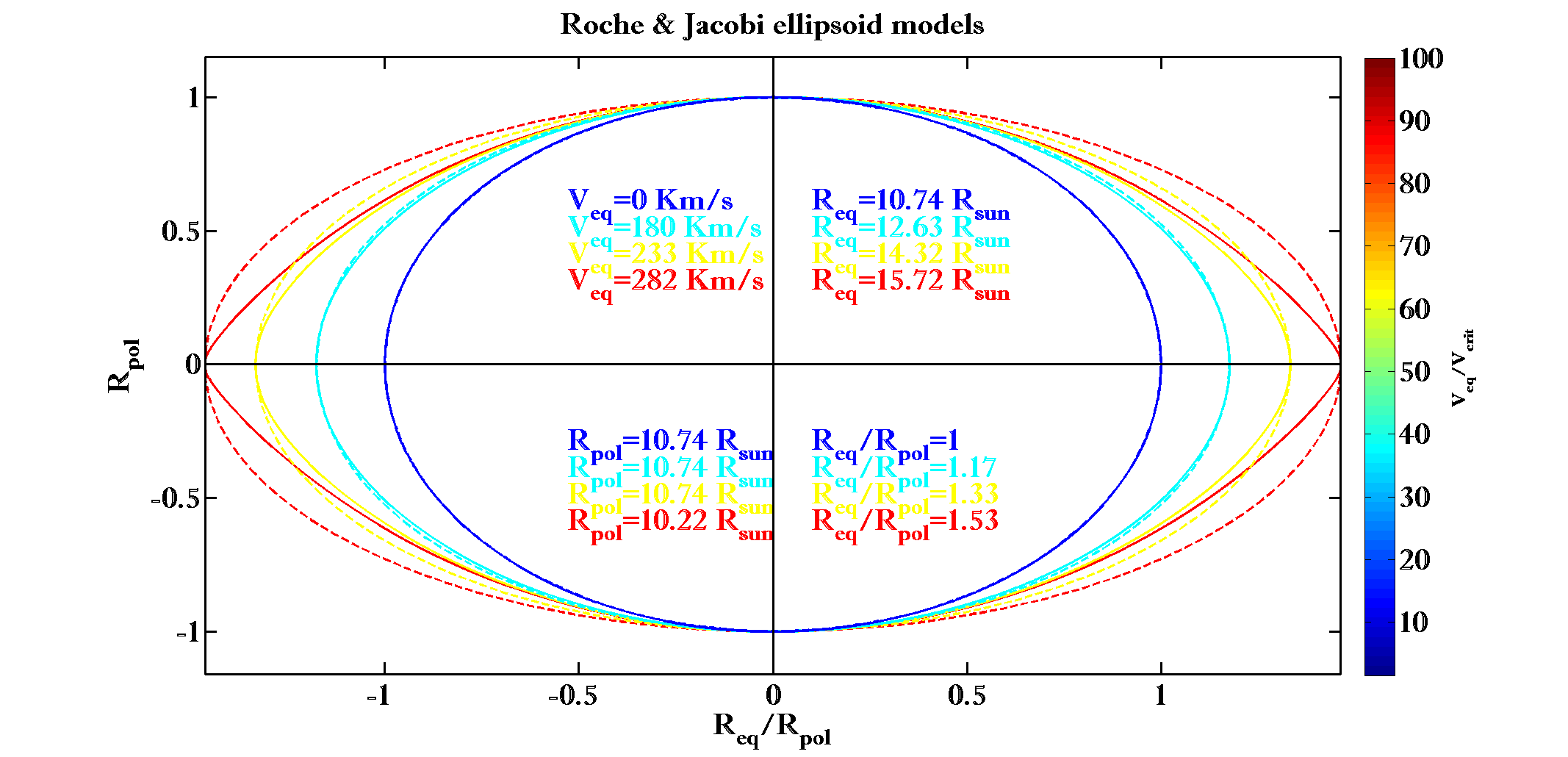}
\caption[Modèles de Roche et l'ellipsoïde de Jacobi]{Comparaison des formes d'aplatissement issues du modèle de Roche (ligne continue) et du modèle ellipsoïde de Jacobi (ligne discontinue), pour une étoile ayant les propriétés suivantes : $M=6.1\Msun$, $Rp=10.74\Rsun$, $d=50pc$ et une vitesse équatoriale $v_{eq}=0$ à $v_{eq,crit}$ $\kms$. Sur la figure sont notées les valeurs de $v_{eq}$, $R_{pol}$, $R_{eq}$ et du rapport des deux, i.e. $\frac{1}{D}$, selon un code de couleur du bleu au rouge en pourcentage du rapport $\frac{v_{eq}}{v_{eq,crit}}$.}\label{comp_roch-ellip}
\end{figure}

La solution de l'équation cubique (\ref{4.7}) a été proposée par Kopal's (1987):\\

\begin{equation}
r(\theta')=D\,_2F_1\left(\frac{1}{3},\frac{2}{3};\frac{3}{2};\gamma^2\right)=D\frac{\sin\left(\frac{1}{3}\arcsin(\gamma)\right)}{\frac{1}{3}\gamma},
\label{4.8}
\end{equation}

avec $_2F_1\left(\frac{1}{3},\frac{2}{3};\frac{3}{2};\gamma^2\right)$ étant la série hypergéométrique $_2F_1$ à argument $\gamma^2=\left(\frac{\Omega}{\Omega_{crit}}\sin\theta'\right)^2$ (voir Eq.\eqref{4.6}). L'Eq.\eqref{4.8} correspond à la forme projetée dans le plan équatorial ("edge on") d'une étoile aplatie par la rotation selon le modèle dit de Roche tel que représenté dans les Figs.\ref{roche_model} \& \ref{2D_rot_star}. Des modèles comme CHARRON et ESTER utilisent rigoureusement le modèle de Roche pour produire la forme aplatie des rotateurs stellaires étudiés. Cela-dit certains codes, pour des raisons de simplicité numérique, utilisent le raccourci de la forme d'ellipsoïde dite de Jacobi, où le degré de sphéricité $D$ (l'une des deux formules de l'Eq.{4.5}) est considéré comme étant le rapport entre les deux demi grand et petit axes de l'ellipsoïde de révolution, facilitant de ce fait les différentes manipulations de projections relatives aux angles d'inclinaison $i$ et d'orientation -ou projection de l'axe de rotation - $PA_{rot}$. Cette méthode (d'ellipsoïde) est d'ailleurs utilisée par Hdust et je l'ai adopté pour SCIROCCO. La Fig.\ref{comp_roch-ellip} représente la différence entre les deux modèles, de Roche et d'ellipsoïde de Jacobi pour une étoile en rotation à différentes vitesses. On ne remarque pas de grandes différences en dessous de $\frac{v_{eq}}{v_{eq,crit}}=80\%$. Au-delà la différence de surface projetée entre les deux modèles est entre 2 à 5\% en configuration vue par l'équateur ("edge on"). Sinon mis à part cette légère différence en terme de flux, cela n'a aucun impact sur les paramètres fondamentaux déduits lors des ajustements avec les données d'observations (voir les résultats très proches obtenus sur Achernar par SCIROCCO et par CHARRON, \citet{2014A&A...569A..45H}), car pour les deux configurations, les rayons polaire et équatorial restent strictement les mêmes.\\

\clearpage

\subsection{Vitesses surfaciques stellaires}
\begin{figure}[ht!]
\centering
\includegraphics[height=0.4\hsize,width=0.4\hsize,draft=false]{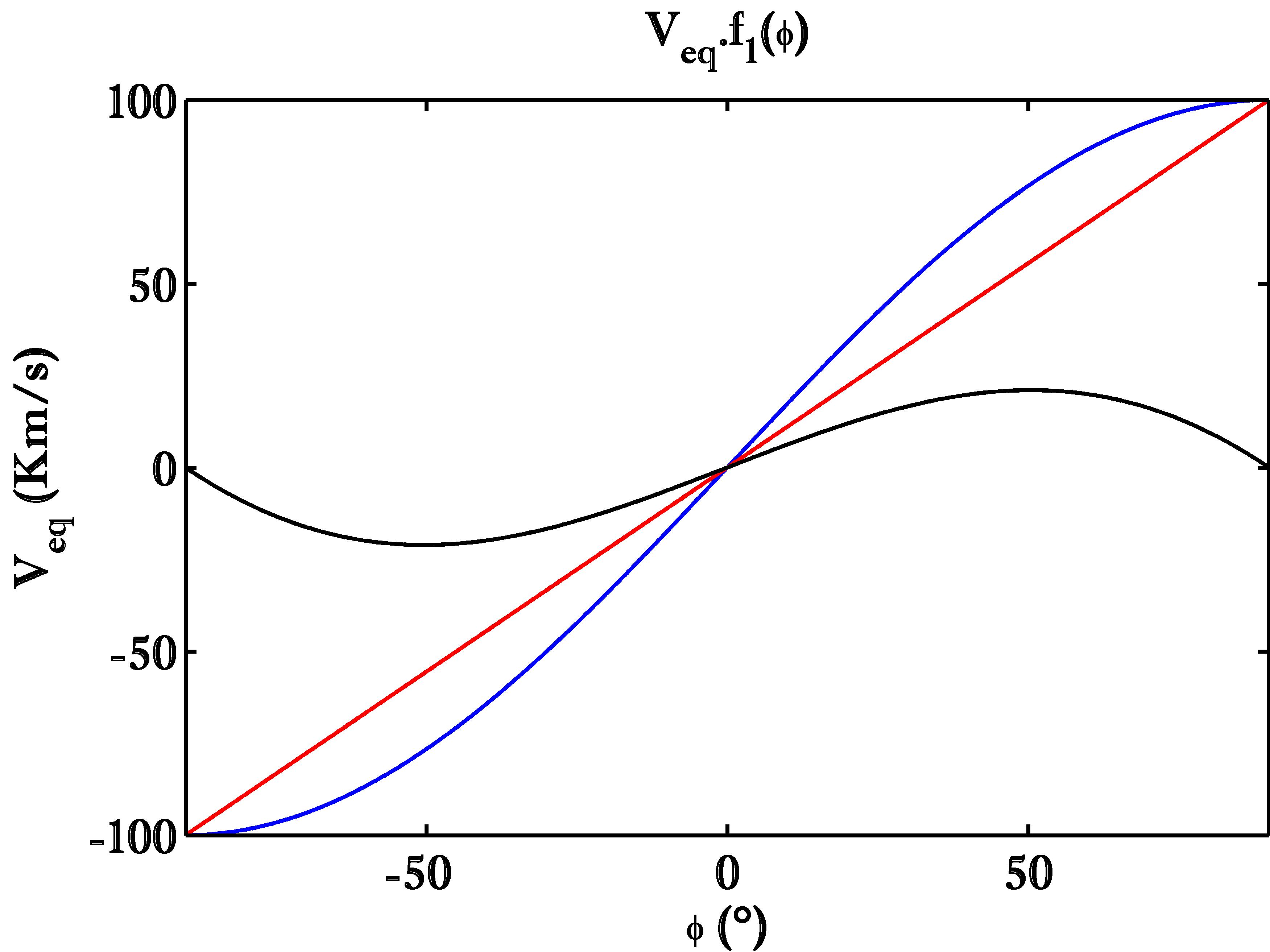}
\caption[Comparaison d'un profil de vitesse $v_{eq}.f_1(\phi)$ linéaire et sinusoïdale]{Comparaison d'un profil de vitesse $v_{eq}.f_1(\phi)$ linéaire (en rouge) et sinusoïdal (en bleu), ainsi que la différence entre les deux (en noir), pour une étoile en rotation avec une vitesse $v_{eq}=100$ $\kms$.}\label{veq_lin-sin0}
\end{figure}

\begin{figure}[ht!]
\centering
\includegraphics[height=0.4\hsize,width=0.5\hsize,draft=false]{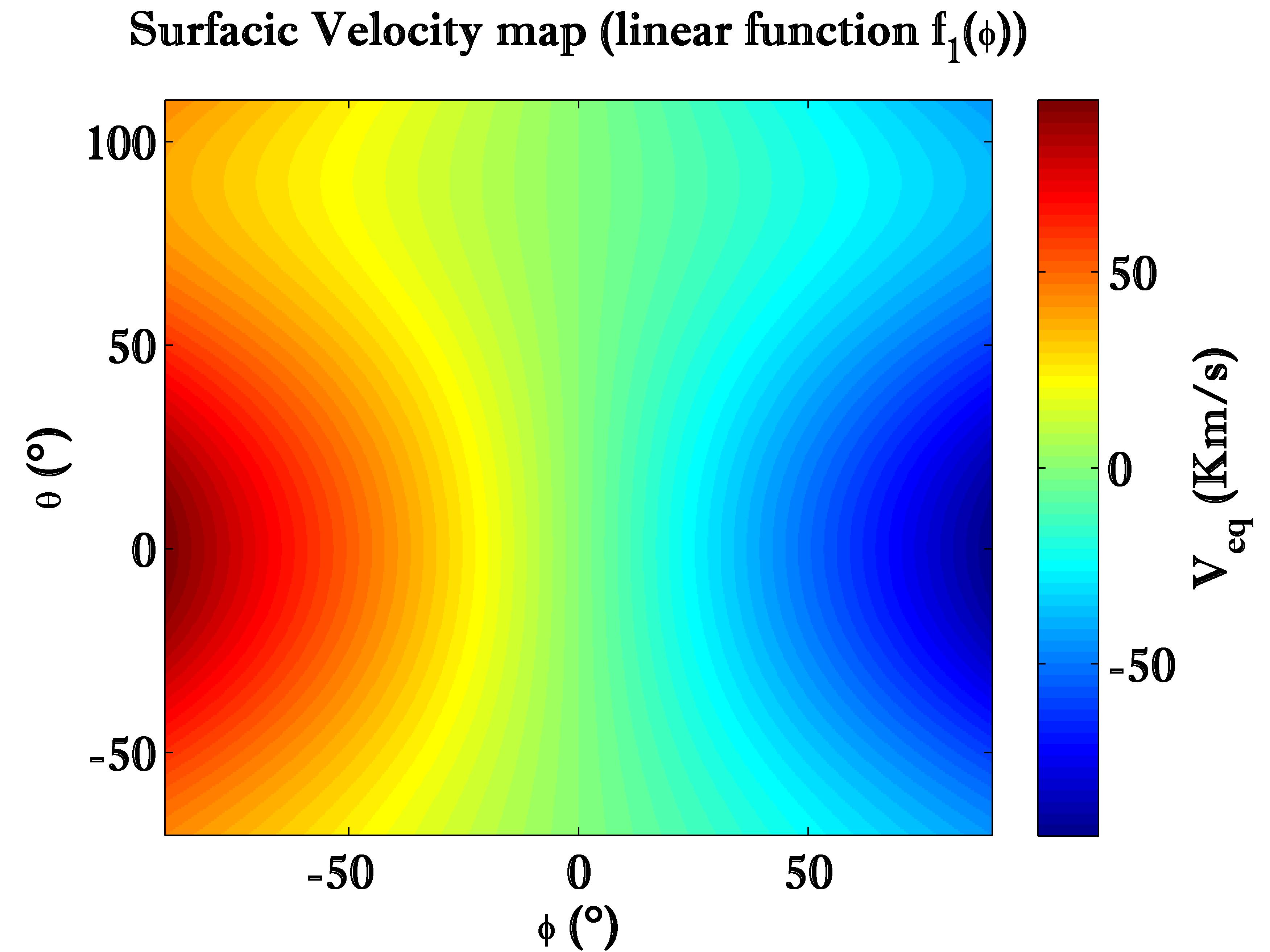}
\includegraphics[height=0.4\hsize,width=0.5\hsize,draft=false]{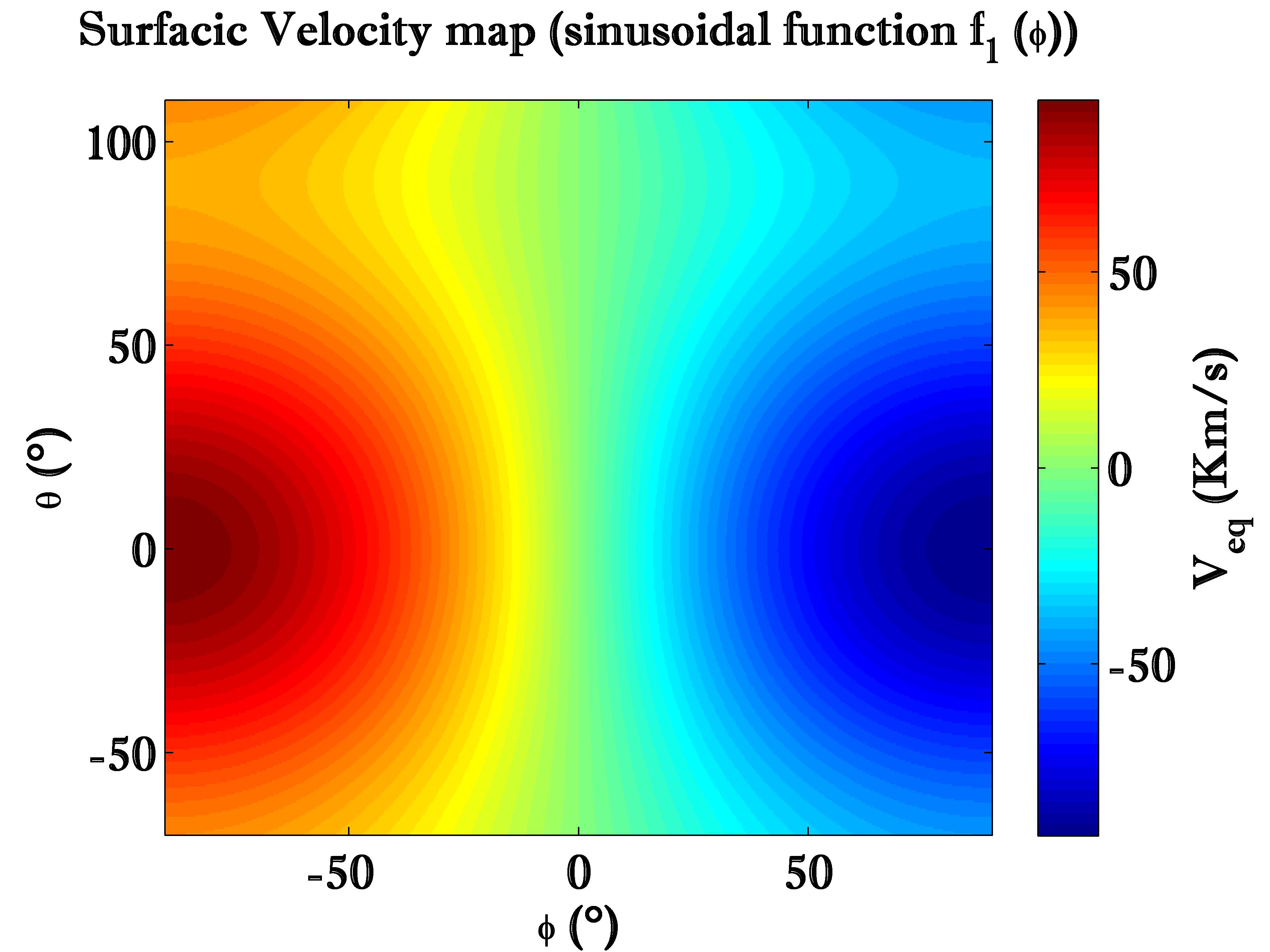}
\includegraphics[height=0.4\hsize,width=0.5\hsize,draft=false]{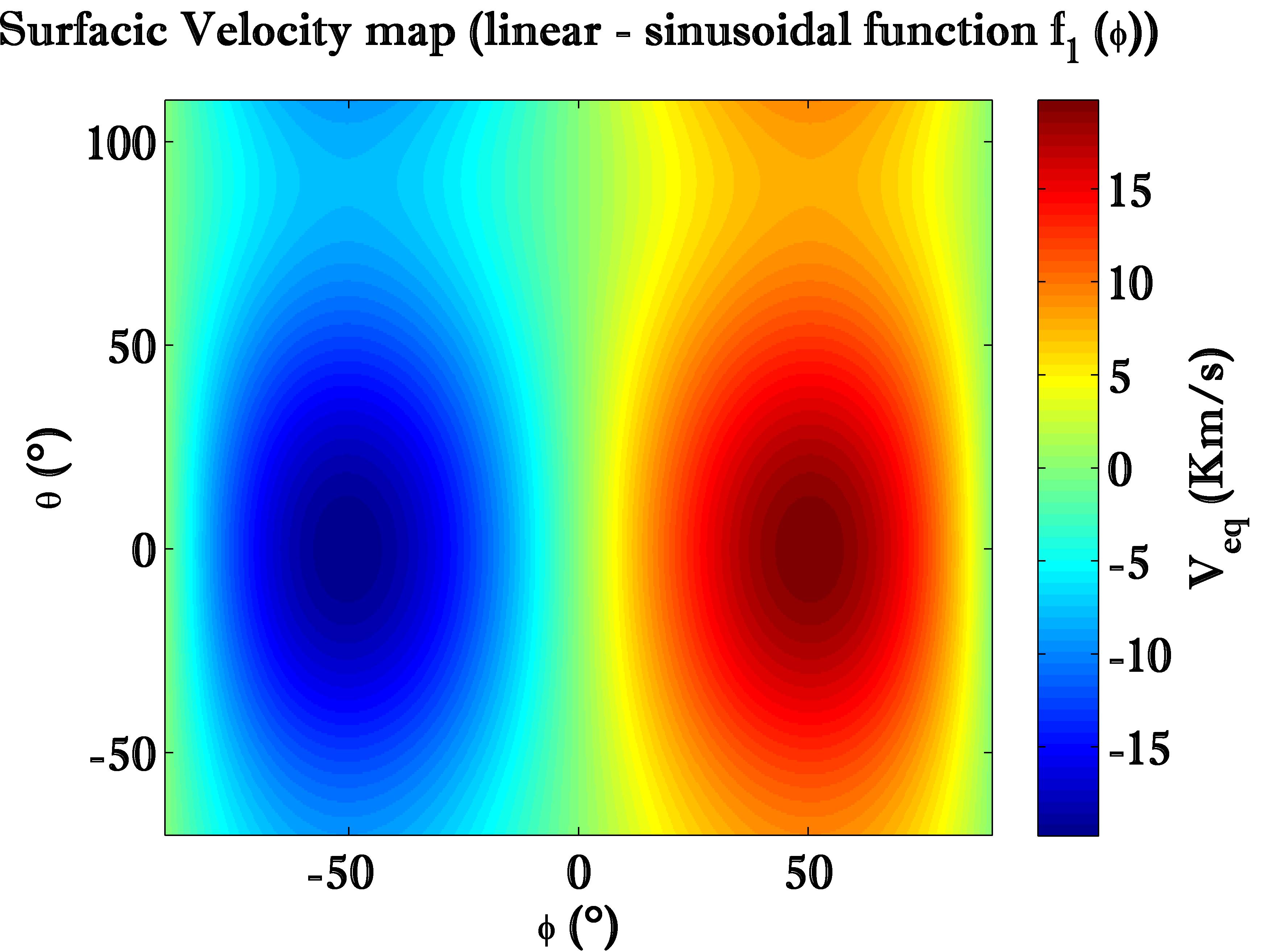}
\caption[Cartes 2D des vitesses surfaciques stellaires 1]{Exemple d'une étoile en rotation avec une vitesse $v_{eq}=100$ $\kms$ et un angle d'inclinaison $i=70^\circ$.}\label{veq_lin-sin1}
\end{figure}

La vitesse de rotation angulaire $\Omega$ d'une étoile peut se décomposer en deux composantes; une composante à dépendance longitudinale $F_1(\phi)$ et une autre co-latitudinale $F_2(\theta)$, où $(\theta,\phi)$ représentent respectivement la latitude et la longitude à la surface stellaire, comme suit:

\begin{equation}
\Omega(\theta,\phi)=\frac{v_{eq}}{R_*}F_1(\phi)F_2(\theta),
\label{4.9}
\end{equation}

où $v_{eq}$ est la vitesse de rotation équatoriale dans l'hypothèse d'une étoile parfaitement sphérique de rayon $R_*$. La dépendance en longitude ($-\frac{\pi}{2}\leq\phi\leq\frac{\pi}{2}$) peut être considérée de manière classique comme étant linéaire où
$F_1(\phi)=\frac{2}{\pi}*\phi$, ou bien d'une manière plus évoluée avec $F_1(\phi)=\sin(\phi)$ (voir Fig.\ref{veq_lin-sin0} \& \ref{veq_lin-sin1}).\\

La dépendance en latitude $\left( -\frac{\pi}{2}\leq\theta\leq\frac{\pi}{2}\right)$ peut être formulée comme suit : $F_2(\theta)=\frac{\Omega(\theta)}{\Omega_{eq}}$ et comprend essentiellement l'effet de la rotation différentielle, que j'ai introduite plus haut. En effet, ce phénomène a été mis en évidence par Carrington en 1863 sur le Soleil, avec la mesure de la différence de vitesse de rotation en fonction de la latitude, via l'observation des déplacements des taches solaires. Ces observations ont révélé que la différence relative de la vitesse angulaire est égale à une certaine fonction $\left(\frac{\Omega_{eq}-\Omega(\theta)}{\Omega_{eq}}=\alpha f(\theta)\right)$. La version la plus triviale connue de $f(\theta)$ est $\sin^2\theta$, ce qui veut dire que la quantité $\frac{\Omega_{eq}-\Omega(\theta)}{\Omega_{eq}f(\theta)}$ est une constante. Celle-ci est appelée coefficient de la rotation différentielle, noté ici $\alpha$. Pour le Soleil la fonction $f(\theta)$ est aussi connue avec un terme additif proportionnel à $\sin^4\theta$ (Snodgrass 1984). $\alpha$ est donc positif lorsque la vitesse équatoriale est plus élevée que celle d'une latitude proche des pôles (aux pôles les vitesses sont nulles), comme il peut être aussi théoriquement négatif (ce qui n'a jamais été encore observé à ce jour). La rotation différentielle est un phénomène important et qui a un effet non négligeable sur les mécanismes physiques stellaires. Ainsi et pour le Soleil par exemple, qui a une vitesse angulaire équatorial $\Omega_{eq}=14^\circ37$ par jour pour un $\alpha=0.22$, la rotation différentielle contribue d'une manière significative au maintien de l'effet dynamo en organisant les lignes du champ magnétique \citep{1999ApJ...523..827M}. La rotation différentielle (pour des $\vsini > 15$ $\kms$) jouerait également un rôle important sur les abondances de certains éléments chimiques tel que le Lithium (Li) qui peut être détruit sous l'effet d'un bon brassage causé par cet effet \citep{2002A&A...393L..77R, 2003A&A...398..647R}. La rotation différentielle reste néanmoins très difficile à observer, hormis pour le Soleil, à cause des limites de la résolution spatiale de nos instruments actuels y compris en interférométrie. \citet{1997MNRAS.291....1D} et \citet{2002MNRAS.329L..23C} l'ont néanmoins mesuré sur l'étoile AB Doradus de type K0V en utilisant la technique d'imagerie Doppler. Indirectement, cela peut se déduire également pour le paramètre de la rotation différentielle sur les émissions périodiques de certaines raies chromosphériques H et K du Ca II \citep{1996ApJ...466..384D}, ou bien grâce à une étude minutieuse du spectre stellaire à très haute résolution dans l'espace de Fourier, méthode proposée par Carroll 1933a, 1933b et qui a été reprise par la suite par bons nombres d'auteurs ; e.g. \citet{1977ApJ...211..198G, 1982ApJ...258..201G}, \citet{1981ApJ...248..274B}, \citet{2001A&A...376L..13R}, \citet{2002A&A...384..155R, 2002A&A...393L..77R, 2003A&A...398..647R}, et \citet{2004A&A...418..781D} avec une étude théorique pour la détermination de la rotation différentielle et de l'inclinaison des étoiles en rotation en Interférométrie Différentielle optique à longue base.\\

De l'Eq.\ref{4.9} on peut donc déduire la carte 2D stellaire des vitesses surfaciques aux coordonnées sphériques $v_s(\theta,\phi)$, centrée sur la latitude d'inclinaison $i$ comme suit:\\

\begin{equation}
v_s(\theta,\phi)=v_{eq}(\frac{2}{\pi}\phi)(1-\alpha\sin^2(\theta+\frac{\pi}{2}-i))
\label{4.10}
\end{equation}

Enfin, la détermination de la carte 2D des vitesses projetées et ses dimensions angulaires (selon le schéma de la Fig\ref{reference}) n'est que le passage en coordonnées cartésiennes $(y,z)$\footnote{La coordonnée $x$ elle disparait étant dans le sens de la projection (de l'observateur).} de la sphère aplatie, avec le degré d'aplatissement apparent $D'=\frac{b'}{a}$ dû à l'inclinaison $i$, le tout pondéré par $\sin i$. De ce fait la carte 2D stellaire des vitesses surfaciques  $v_{proj}(y,z)$, observée depuis la Terre est simplement:\\

\begin{equation}
v_{proj}(y,z)=v_s(y,z)\sin i
\label{4.11}
\end{equation}

\begin{figure}[ht!]
\centering
\includegraphics[height=0.4\hsize,width=0.5\hsize,draft=false]{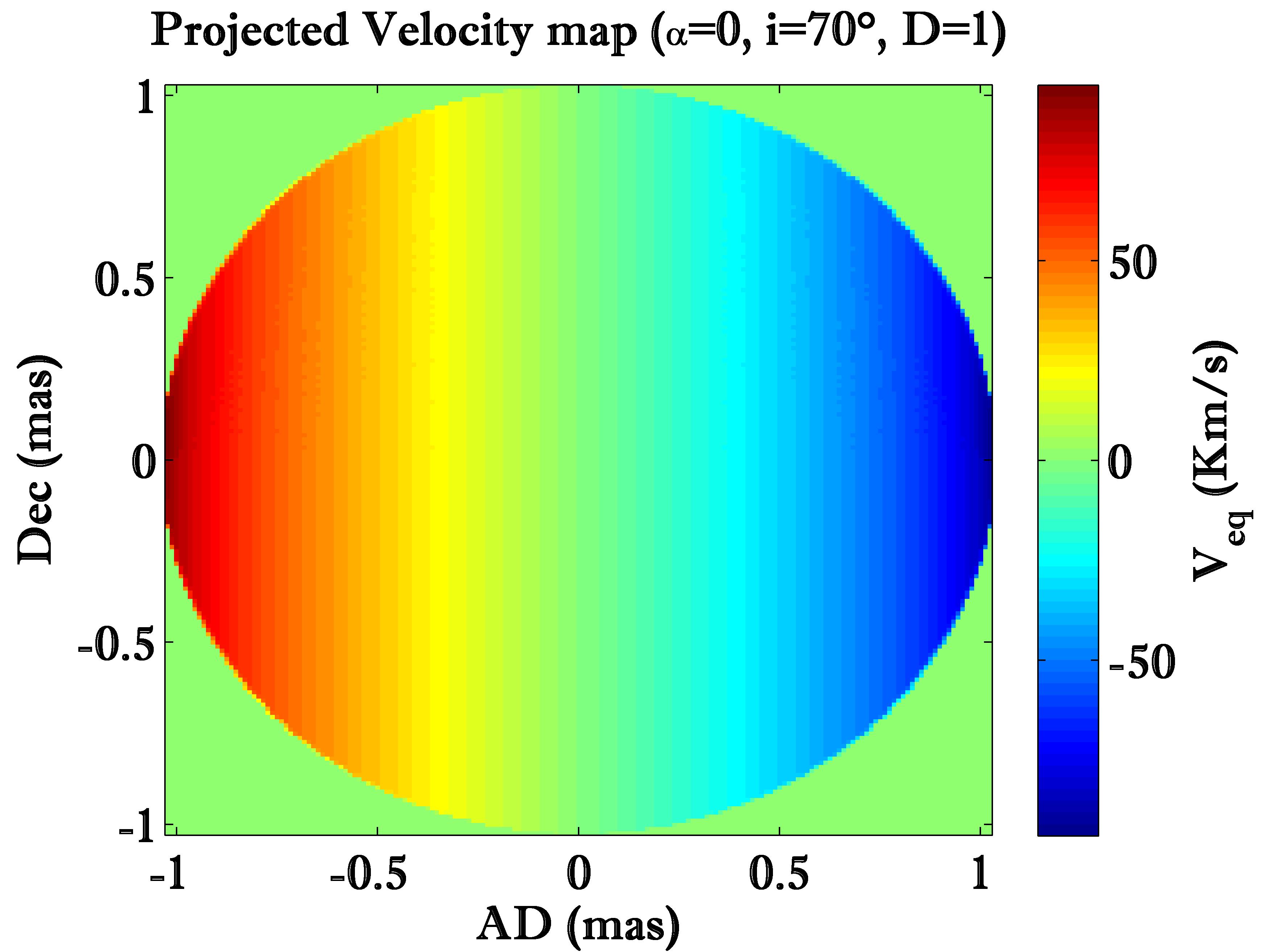}
\includegraphics[height=0.4\hsize,width=0.5\hsize,draft=false]{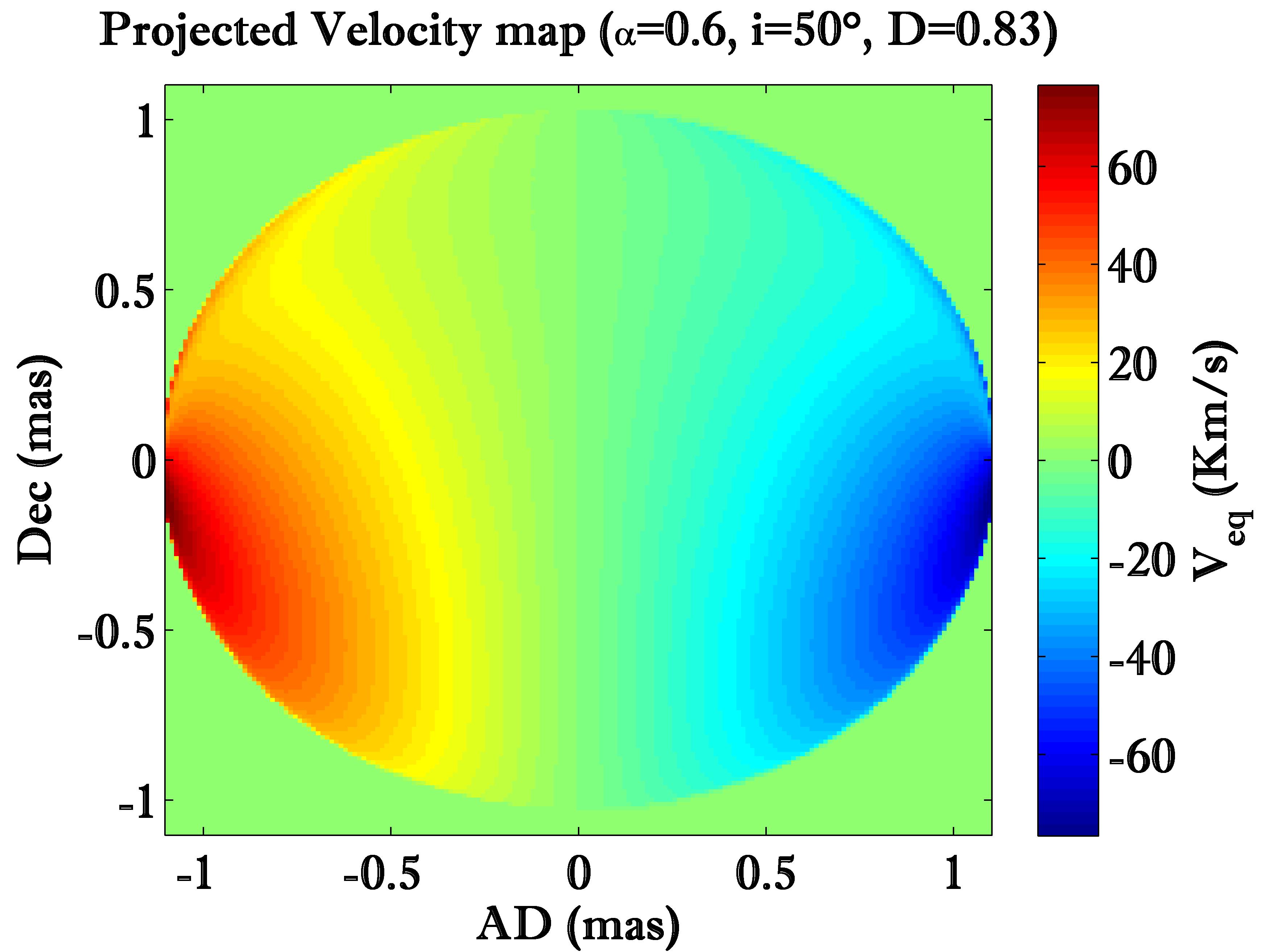}
\includegraphics[height=0.4\hsize,width=0.5\hsize,draft=false]{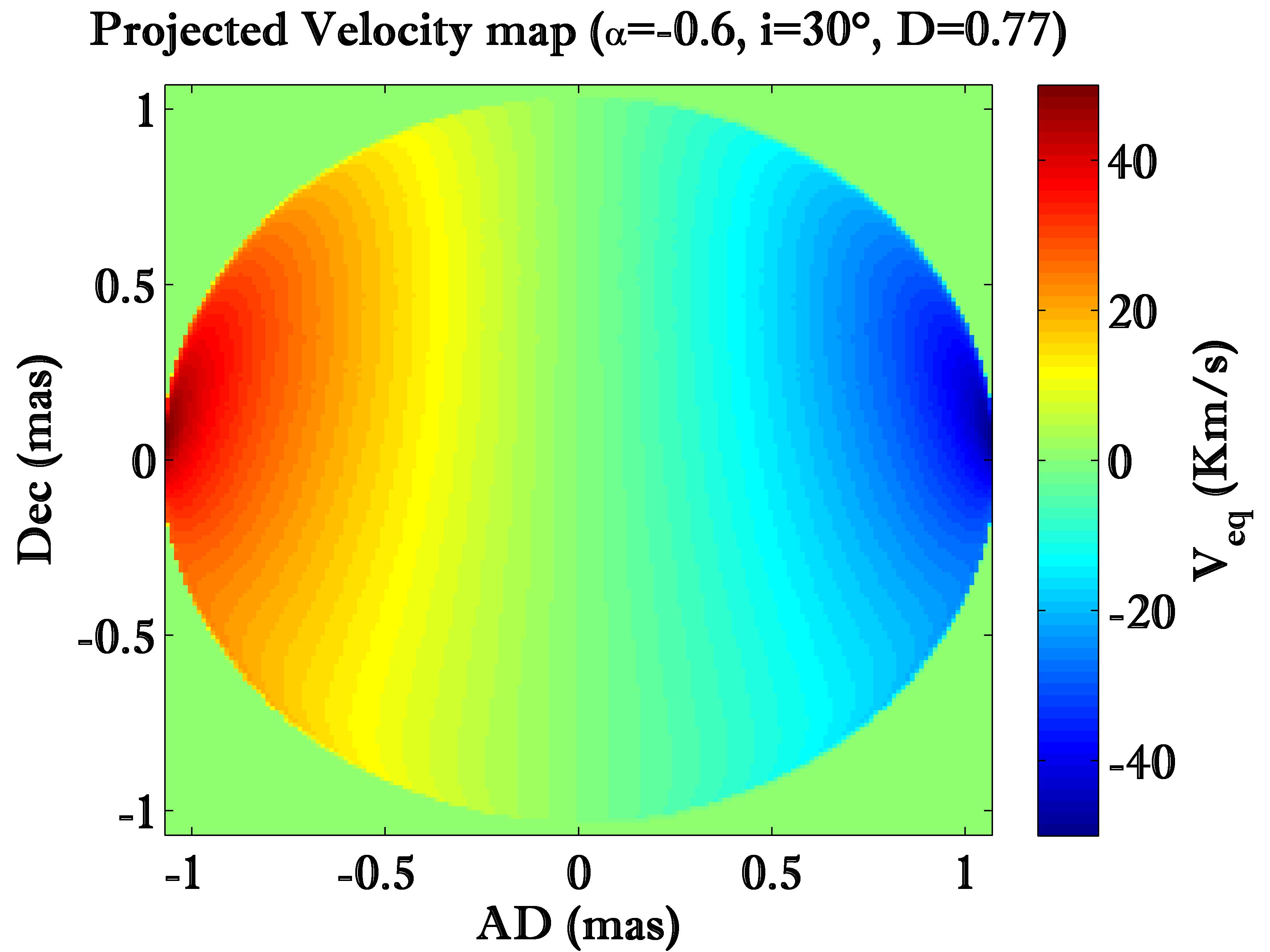}
\includegraphics[height=0.4\hsize,width=0.5\hsize,draft=false]{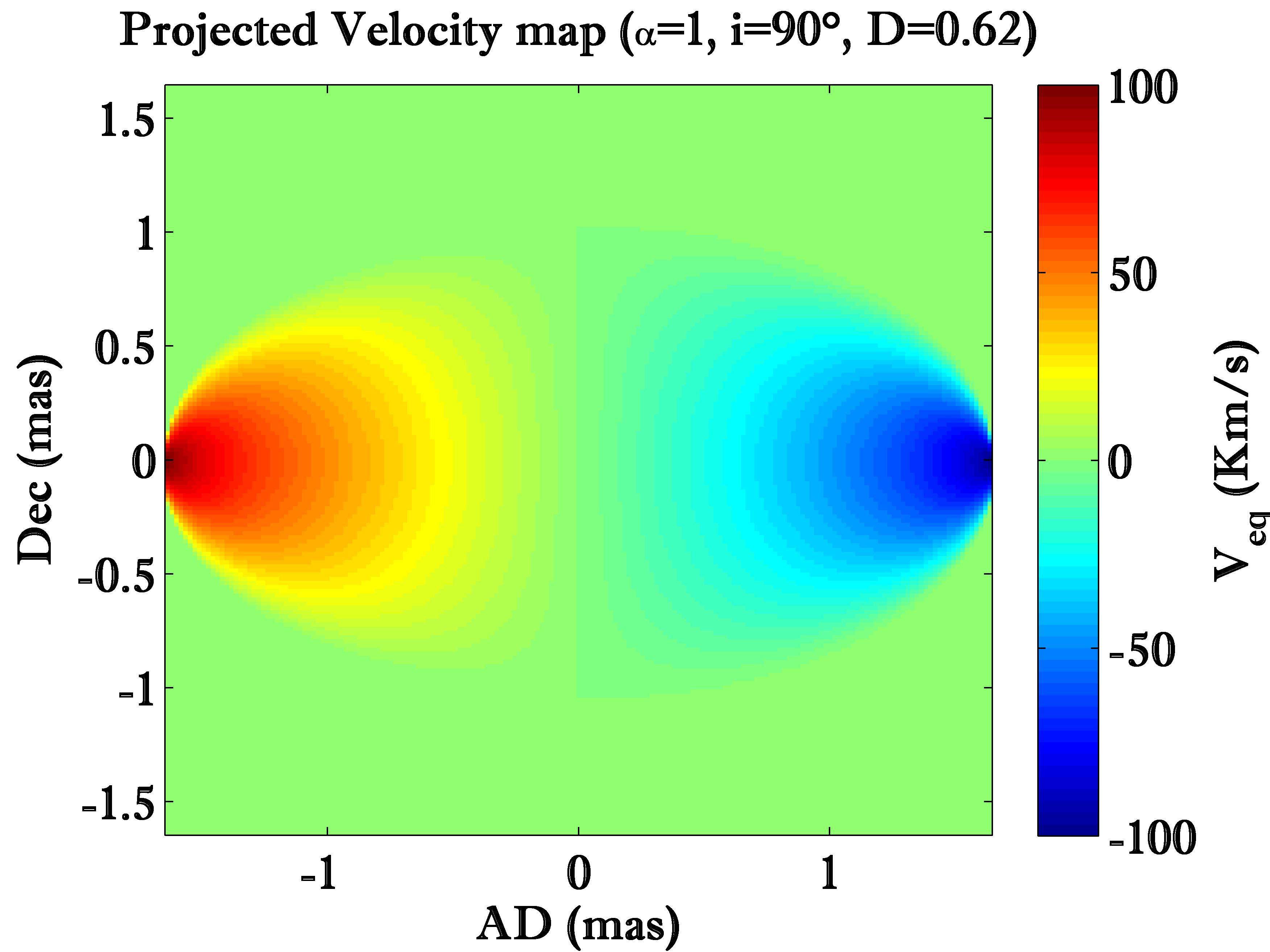}
\caption[Cartes 2D des vitesses surfaciques stellaires 2]{Exemple d'une étoile en rotation avec une vitesse $v_{eq}=100$ $\kms$ et un angle d'inclinaison $i=70^\circ$.}\label{veq_lin-sin2}
\end{figure}

La Fig.\ref{veq_lin-sin2} représente différentes configurations de carte de vitesses projetées d'une étoile théorique de rayon $R=11$ $\Rsun$ qui se trouve à une distance $d=50$ $pc$, pour différents jeux de vitesse équatoriale $v_{eq}$, d'inclinaison $i$, d'angle d'orientation $PA_{rot}$, de degré de sphéricité $D$ et de coefficient de rotation différentielle $\alpha$.\\

\clearpage

Le choix de type de la fonction $f_1(\phi)$ n'a pas d'impact réél sur les résultats d'utilisation des paramètres fondamentaux lors des ajustements avec les données observées par des tests que j'ai effectués. Par conséquent le choix de celle-ci fut celui de l'équation linéaire (le choix classique). Pour la fonction $f_2(\theta)$, j'ai fixé à $\alpha=0$ le coefficient de la rotation différentielle, c'est à dire, une rotation rigide. Ceci est justifié par le fait que les données observées n'ont pas une résolution spatiale suffisante pour révéler l'effet de la rotation différentielle \citep{2004A&A...418..781D}.\\

\section{Carte d'intensité d'un rotateur stellaire dans le continuum}
La manière la plus rigoureuse de synthétiser des cartes d'intensité stellaires consiste à se servir des équations du transfert radiatif (ETR; assez bien décrites dans \citet{2003A&A...407L..47D}). Une approximation suffisante pour la qualité temps et volume de calcul, consiste à utiliser la loi du corps noir de Planck (Eq.\eqref{eq5} du Chapitre \ref{chap:spec-interfero}), qui relie l'intensité spécifique $I_0$ à la température effective $\Tmean$ d'une source observée à une longueur d'onde $\lambda$. La photosphère d'une étoile observée est impactée par le phénomène de profondeur optique liée à la géométrie 3D de sa photosphère, correspondant à l'assombrissement centre-bord (limb darkening). De plus, si l'étoile a un moment cinétique assez élevé ($\frac{v_{eq}}{v_{eq,crit}}>80\%$) un second assombrissement, lié à la différence d'échelle de hauteur pôles-équateur causée par l'aplatissement, s'ajoute à l'assombrissement centre-bord. Cet effet est connu sous le nom d'assombrissement gravitationnel. Les principes et équations de bases de ces deux types d'assombrissements sont décrits par la suite\\

\subsection{Assombrissement gravitationnel}
La forte rotation d'une étoile engendre en plus de la déformation de celle-ci (étudiée ci-haut), des flux polaires qui entrainent des températures plus élevées aux pôles qu'à l'équateur, ce qui a pour effet visible direct une brillance polaire plus évidente qu'à l'équateur. Cet effet a été théoriquement étudié par \citet{1924MNRAS..84..665V} via une loi qui porte son nom. Dans l'hypothèse d'une rotation rigide, il a déterminé que le flux radiatif (voir la section n$^\circ$ \ref{spectro} du Chapitre \ref{chap:spec-interfero}) d'une étoile en rotation uniforme est proportionnel à la gravité effective locale de celle-ci. De ce fait on peut relier les températures effectives de surface d'une étoile $T_{eff}(\theta')$ au module de sa gravité de surface ($g_{eff}(\theta')=|\vec{g}(\theta')|$). $T_{eff}(\theta')$ \& $g_{eff}(\theta')$ varient tous les deux d'une latitude à une autre, sur une portion de la co-latitude $\theta'$ pôle-équateur-pôle (où $\theta'=\frac{\pi}{2}+\theta$, $\theta$ étant la latitude), selon la vitesse de rotation de l'étoile. Von Zeipel a montré que dans ces conditions, $T_{eff}(\theta')$ est strictement proportionnel à une loi de puissance en $g_{eff}(\theta')$:\\

\begin{equation}
\frac{T_{eff}(\theta')}{g_{eff}(\theta')^\beta }=Const
\label{4.12}
\end{equation}

$\beta$ est appelé coefficient d'assombrissement gravitationnel, et détermine la distribution de la température (intensité et gravité) surfacique, de l'équateur aux pôles, qui est propre à chaque rotateur et dont on va débattre la détermination plus bas. Le module de gravité de surface effective déduit précédemment dans (Eq.\ref{4.3}) peut être réécrit à l'aide de $D$, $\frac{\Omega}{\Omega_{crit}}$ et $r(\theta')$, comme suit \citep{2002A&A...393..345D}:\\

\begin{equation}
g_{eff}(\theta')=g_{pol}D^2\left(\frac{2}{3D}\right)^3\left\{\left[(r(\theta') \left(\frac{\Omega}{\Omega_{crit}}\right)^2\sin\theta'\cos\theta'\right]^2\\
+\left[\frac{1}{(r^2(\theta')}\left(\frac{3}{2}D^3\right)-\left(\frac{\Omega}{\Omega_{crit}}\sin\theta'\right)^2\right]^2\right\}^\frac{1}{2},
\label{4.13}
\end{equation}

où $g_{pol}=\frac{GM}{R_{pol}^2}$ désigne la gravité de surface au niveau des pôles. La constante de l'Eq.\ref{4.12} peut donc prendre la forme de $Const=\frac{T_{pol}}{g_{pol}^{\beta}}$, ce qui me permet de simplifier la variation des températures co-latidunales d'une surface stellaire par:\\

\begin{equation}
T_{eff}(\theta',\phi)= T_{pol}\left(\frac{g_{eff}(\theta',\phi)}{g_p}\right)^\beta= T_{pol}\left(g_n(\theta',\phi)\right)^\beta
\label{4.14}
\end{equation}

$g_n$ est la gravité de surface normalisée. Sachant que les gravités et températures de surface ne varient pas en longitude $\phi$ pour chaque latitude, je peux obtenir la carte d'intensité du continuum $I_0(\lambda,\theta,\phi)$ assombrie par effet gravitationnel selon la vitesse de rotation $v_{eq}$ de l'étoile, en utilisant la loi de Planck (Eq.\eqref{eq5}) via $T_{eff}(\theta,\phi)$:\\

\begin{equation}
I_0(\lambda,\theta,\phi)=\frac{2hc^2}{\lambda^5}\frac{1}{e^{\frac{hc}{\lambda \sigma_{SB} T_{\rm eff}(\theta,\phi)}}-1}
\label{4.15}
\end{equation}

L'intensité, la température et la gravité de surface sont par conséquent plus élevées là où la force centrifuge est plus faible (voir nulle), i.e. c'est aux pôles, où l'on constate l'intensité, la température et la gravité de surface les plus elevées. A l'inverse, c'est à l'équateur que ces dernières sont les plus faibles. L'écart de ces valeurs entre les pôles et l'équateur est d'autant plus important que la vitesse de rotation s'approche de la vitesse critique. Autrement dit, et pour une étoile sans rotation, les valeurs $I_0$,$T_{eff}$ et $g$ seront partout les mêmes sur la surface de l'étoile. Concernant la carte d'intensité 2D, en coordonnées sphériques, pour passer de $\theta'$ à $\theta$ il suffit juste de décaler notre carte de $\frac{\pi}{2}$. Pour passer aux cordonnées cartésiennes, j'ai adopté la projection orthographique. La projection centrée autour d'une longitude 0 et d'une latitude $\frac{\pi}{2}-i$, pour une étoile inclinée d'un angle $i$ et de degré de sphéricité apparent $D'$ est définie par :\\

\begin{gather}
 y=\frac{R_{pol}}{D'} \cos\theta\sin\phi \notag  \\
 \label{4.16}
 z=R_{pol}\left[\cos\left(\frac{\pi}{2}-i\right)\sin\theta-\sin\left(\frac{\pi}{2}-i\right)\cos\theta\cos\phi\right]
\end{gather}

L'aplatissement apparent, en fonction de l'inclinaison $i$ et le diamètre angulaire apparent se déroule ainsi de la même manière que pour les cartes d'iso-vitesses explicitées plus haut. Le dernier point important concerne la détermination des températures aux pôles $T_{pol}$ et à l'équateur $T_{eq}$ à partir de la température effective moyenne $\Tmean$. Pour cela on a recours à la luminosité $L$ (voir la section 3.3 du Chapitre \ref {chap:spec-interfero} précèdent), qu'on peut réécrire en utilisant l'Eq.\eqref{4.14} comme suit:\\

\begin{equation}
L=\sigma_B\Tmean^4S_*=\sigma \int T_{eff}^4(\theta',\phi)dS=\sigma \int  \left(\frac{T_{pol}}{g_{pol}^\beta} \right)^4 g_{eff}^{4\beta}(\theta',\phi)dS,
\label{4.17}
\end{equation}

où $S_*$ est la surface de l'étoile. Sachant que $\frac{T_{pol}}{g_{pol}^\beta}=Const$, on peut donc définir une nouvelle constante $C=\sigma\left(\frac{T_{pol}}{g_{pol}^\beta}\right)^4=\sigma\left(\frac{T_{eff}(\theta',\phi)}{g_{eff}(\theta',\phi)^\beta}\right)^4$, et qui peut se réécrire à partir de l'Eq.\eqref{4.17} comme:\\

\begin{equation}
C=\frac{\sigma\Tmean^4S_*}{\int g_{eff}(\theta',\phi)^{4\beta} dS}
\label{4.18}
\end{equation}

Cette expression de C aisément calculable grâce aux paramètres connus qu'elle contient, nous permet de déterminer les températures aux pôles $T_{pol}$ et à l'équateur $T_{eq}$ à partir du module de gravité effective de surface $g_{eff}$:\\

\begin{equation}
T_{eff}(\theta',\phi)=\left(\frac{C}{\sigma}\right)^{0.25} g_{eff}(\theta',\phi)^\beta 
\label{4.19}
\end{equation}

\textbf{Détermination de $\mathbf{\beta}$}\\

\begin{figure}[ht!]
\centering
\includegraphics[height=0.4\hsize,draft=false]{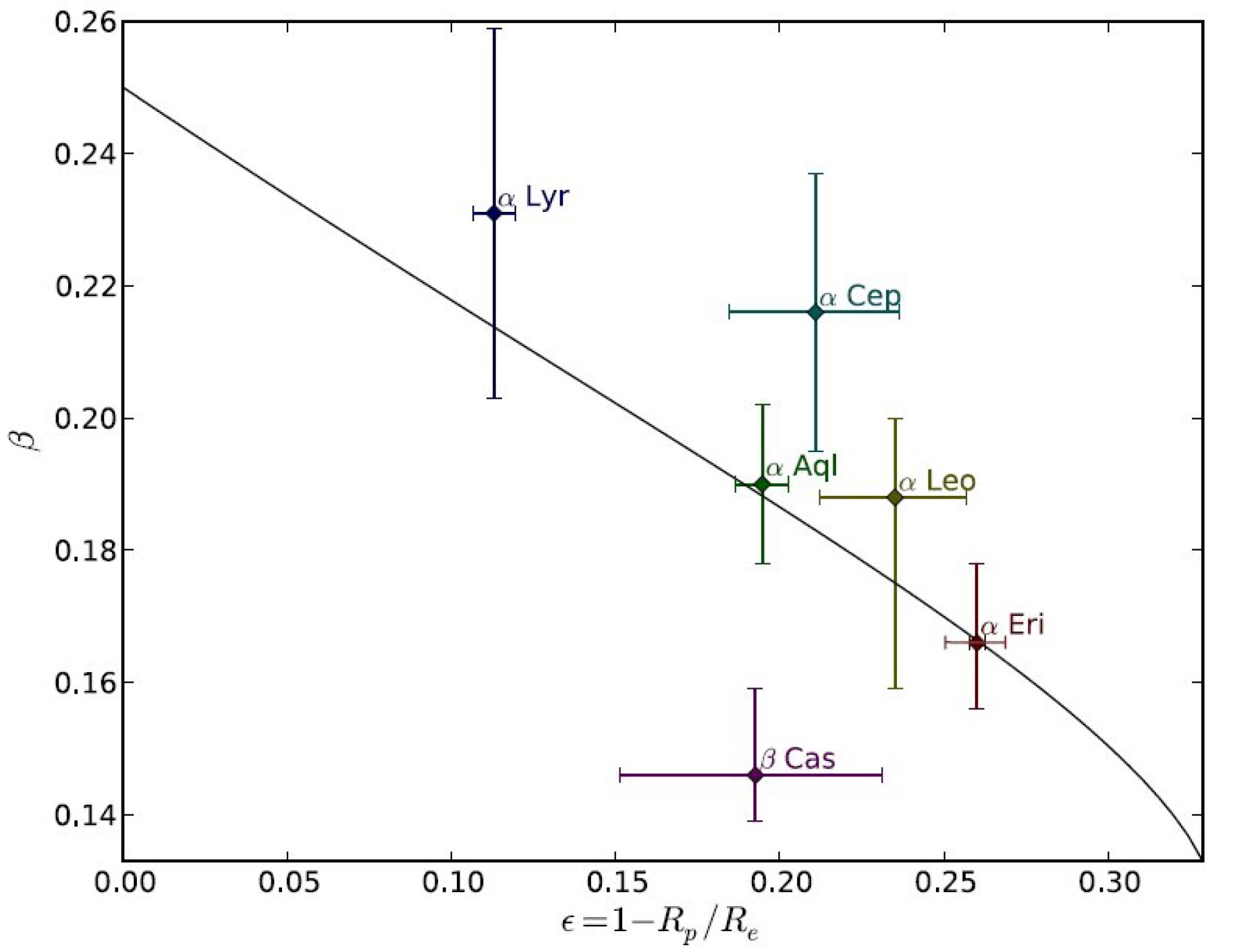}
\caption[Détermination de $\beta$]{Variation du coefficient d'assombrissement gravitationnel $\beta$ en fonction de $\in=1-\frac{R_{pol}}{R_{eq}}$ \citep{2014A&A...569A..10D}).}\label{beta_Rieutord}
\end{figure}

De la loi de Stefan-Boltzmann, qui est considérée en première approximation de la distribution surfacique du flux radiatif avec l'hypothèse de lois conservatrices de rotation (force centrifuge obtenue à partir d'un potentiel), découle naturellement que $T_{eff}(\theta)\propto(g_{eff}^{o.25})$. En effet, la luminosité étant proportionnelle à la gravité de surface (voir la démonstration rigoureuse dans \citet{2011A&A...533A..43E}), la valeur approximative de $\beta$ qui est de l'ordre de $0.25$. Pour des étoiles à couches convectives, \citet{1967ZA.....65...89L} a démontré que $\beta=0.08$, et \citet{2011ApJ...732...68C} recommande l'adoption d'un $\beta=0.19$ pour la modélisation d'étoiles radiatives en rotation. Cela dit, \citet{2011A&A...533A..43E} proposent une solution pour la détermination d'une valeur $\beta$ adaptée à chaque rotateur selon sa vitesse de rotation et son aplatissement $\left(\in=1-\frac{R_{pol}}{R_{eq}}=1-D\right)$ dans leur modèle ESTER (voir Fig.\ref{beta_Rieutord}, ci-dessous).\\

En effet, ces derniers, en incluant la force radiative ($F_r\propto\nabla T$) dans leur modélisation en plus des forces gravito-rotationnelles (que j'ai utilisées plus haut), ont pu déterminer la luminosité d'une étoile en rotation, aux pôles et à l'équateur $L_{pol}=g_{pol}e^{2\omega^2\frac{D^3}{3}}$ \& $L_{eq}=g_{eq}(1-\omega^2)^\frac{-2}{3}$, où $\omega=\frac{\Omega}{\Omega_{crit}}$ (la racine carré de l'Eq.\eqref{4.6}). De ce fait et via $\omega$ on peut reformuler l'Eq.\eqref{4.5} comme $D=\left(1+\frac{\omega2}{2}\right)^{-1}$ et déduire que $\frac{g_{eq}}{g_{pol}}=D^2(1-\omega^2)$. Le rapport des deux luminosités est donc $\frac{L_{eq}}{L_{pol}}=\left(\frac{T_{eq}}{T_{pol}}\right)^4=\frac{(1-\omega^2)^{\frac{1}{3}}}{(1+\frac{\omega^2}{2})^2}e^{-2\omega^2\frac{D^3}{3}}$. De ces deux dernières équations et via le logarithme népérien de l'Eq.\eqref{4.14}. On peut enfin rigoureusement formuler l'expression de $\beta$, dans le cas d'étoiles radiatives en rotation:\\

\begin{equation}
\beta=\frac{1}{4}-\frac{1}{6}\frac{\ln(1-\omega^2)+\omega^2D^3}{\ln(1-\omega^2)-2\ln(1+\frac{\omega^2}{2})}
\label{4.20}
\end{equation}

Le développement limité au premier ordre de $\beta$ , pour des petites valeurs de $\omega$ (qui est d'ailleurs toujours $<1$) réduit l'expression de $\beta$ à $\beta=\frac{1}{4}-\frac{1}{6}\omega^2$, ou bien en fonction de $\in=1-D=1-\frac{R_{pol}}{R_{eq}}$, après approximation ($\omega<<1$ $\Rightarrow$ $\omega^2\simeq2(1-D)=2\in$), i.e. $\beta=\frac{1}{4}-\frac{1}{3}\in$. Bien que la variation du coefficient assombrissement gravitationnel n'influe quasiment pas sur l'ajustement de nos données, qui n'ont pas encore une résolution spatiale suffisante pour cela, tel que démontré et discuté dans le Chapitre des résultats (Chap.\ref{chap:appli}), c'est cette dernière valeur que j'ai décidé d'adopter de manière automatique dans SCIROCCO.\\

\begin{figure}[ht!]
\centering
\includegraphics[height=0.4\hsize,width=0.4\hsize,draft=false]{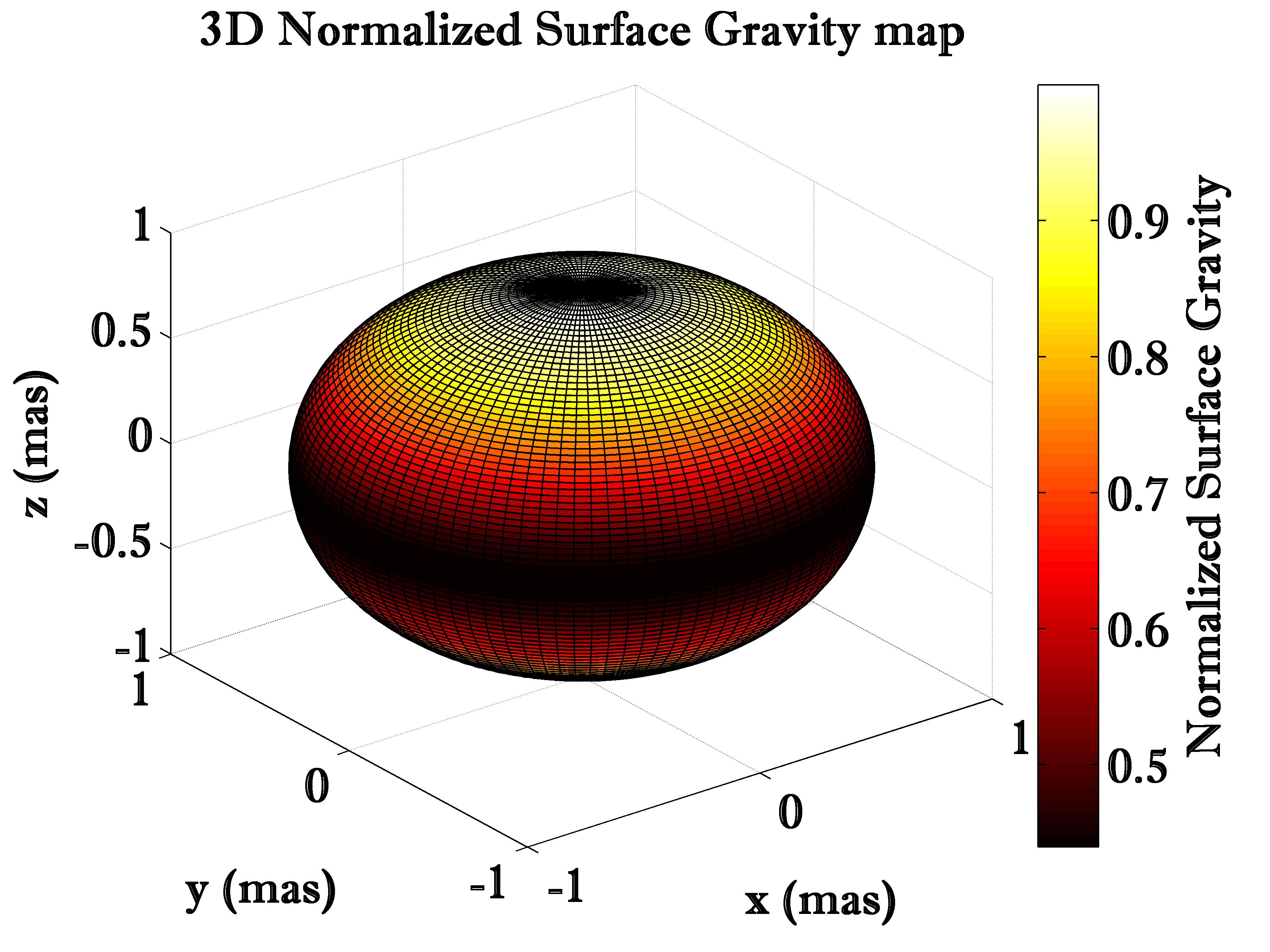}
\includegraphics[height=0.4\hsize,width=0.4\hsize,draft=false]{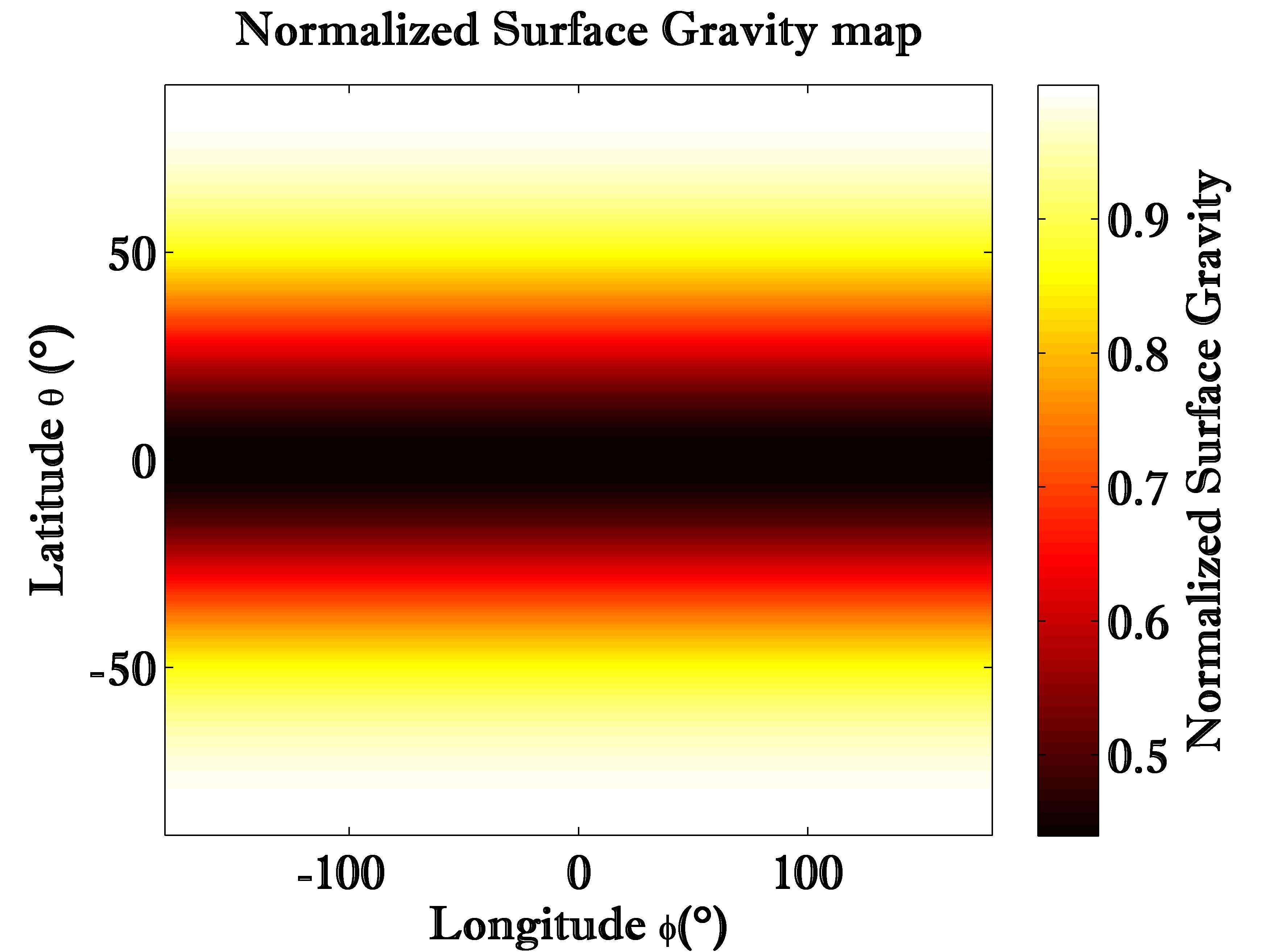}
\includegraphics[height=0.4\hsize,width=0.4\hsize,draft=false]{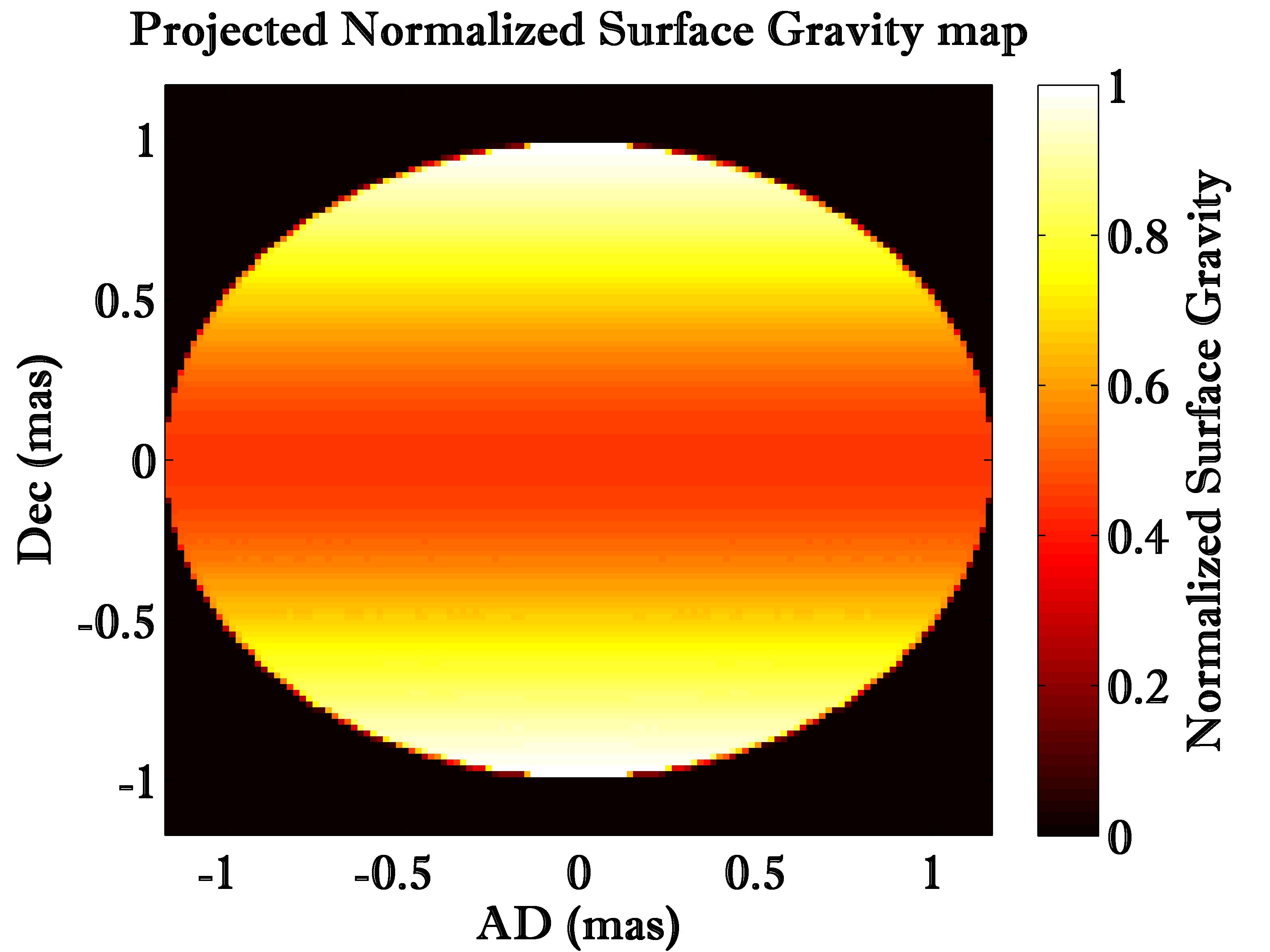}
\includegraphics[height=0.4\hsize,width=0.4\hsize,draft=false]{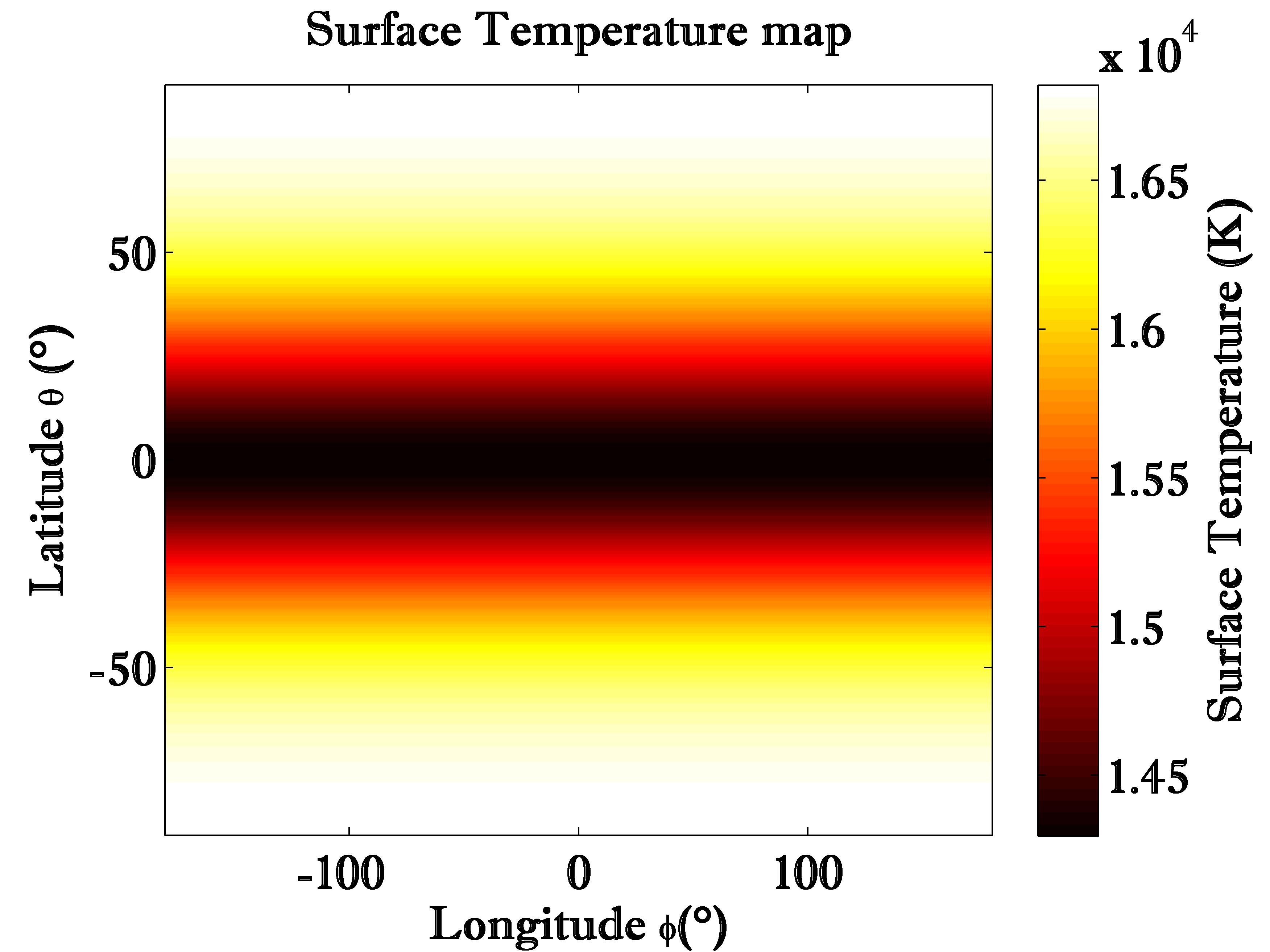}
\includegraphics[height=0.4\hsize,width=0.4\hsize,draft=false]{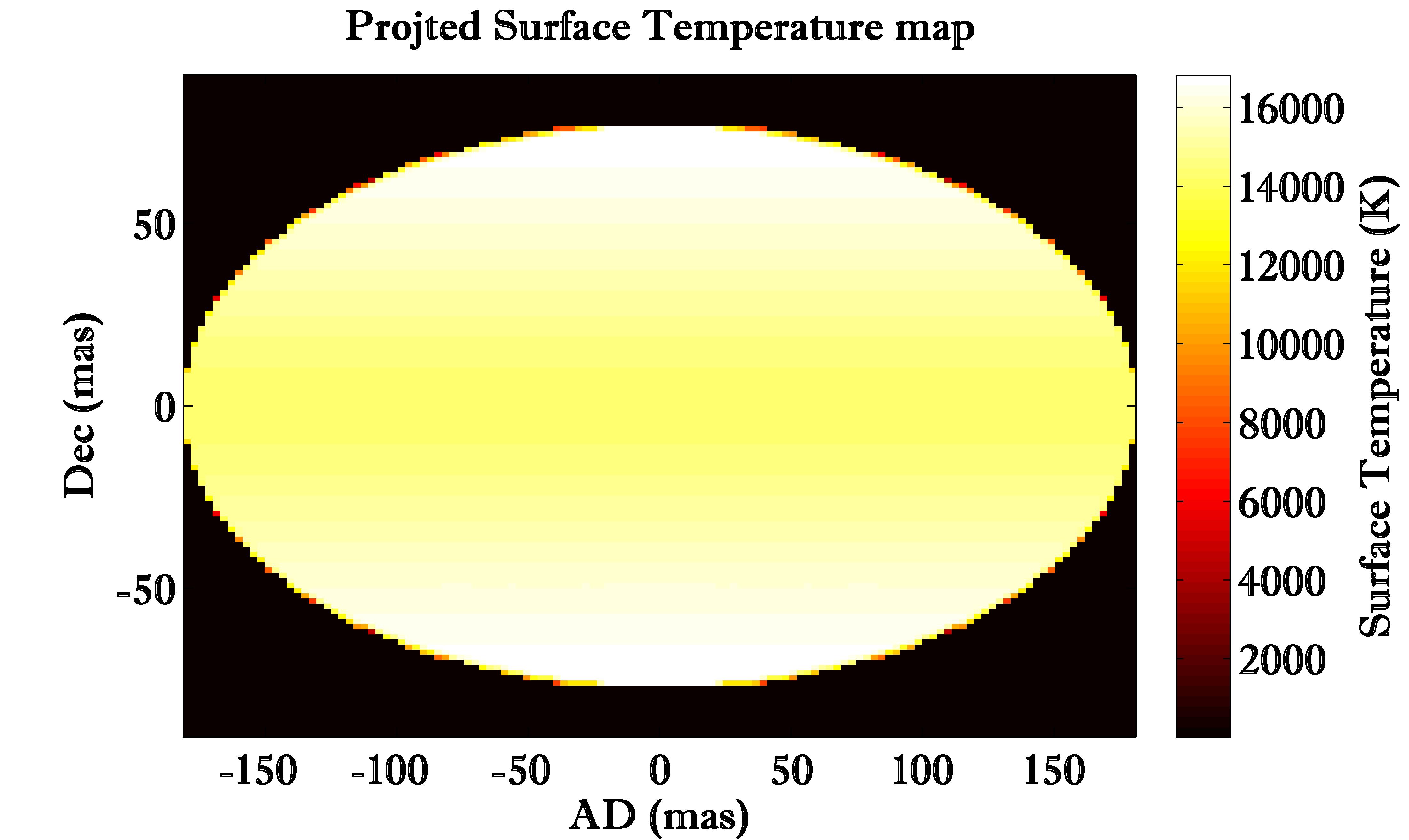}
\includegraphics[height=0.4\hsize,width=0.4\hsize,draft=false]{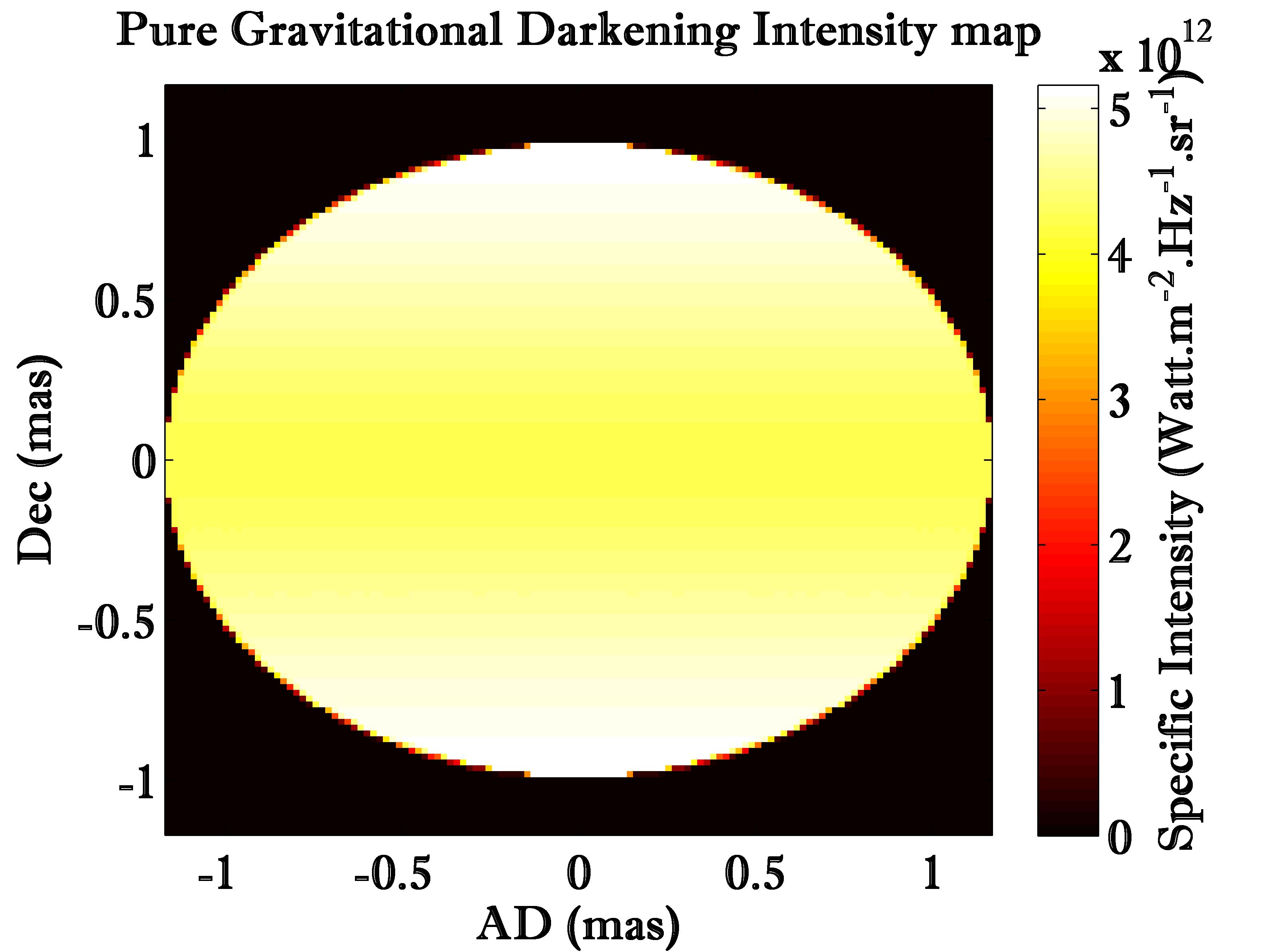}
\caption[Cartes 2D d'intensité assombrie gravitationnellement]{Exemple de cartes 2D/3D de température et d'intensité assombrie gravitationnellement. La variation de la gravité de surface, du pôle à l'équateur, est ici de $[\log G_{eq}, \log G_{pol}]=[3.14, 3.29] cm/s2$.}\label{int_2D}
\end{figure}

La Fig.\ref{int_2D} montre un exemple de carte des températures, une carte des gravités de surface, une carte des intensités et leurs projections respectives, à longueur d'onde Br$\gamma$ ($\lambda=\lambda_0=2.166$ $\mu m$), pour un étoile de $6.1$ $\Msun$, de rayon équatorial de $11$ $\Rsun$, à une distance $d=50$ $pc$, tournant à une vitesse $v_{eq}=200$ $\kms$ et "edge on" ($i=90^\circ$) avec une température effective moyenne $\Tmean=15000$ $K$. De l'Eq.{4.5} est déduit le degré de sphéricité $D=0.84$ ($\frac{R_{eq}}{R_{pol}}=1.19$), le rapport $\frac{v_{eq}}{v_{eq,crit}}=69\%$ ($\omega=0.87$) et $\log L/L_\odot=3.66$, et $[T_{eq},T_{pol}]=[14330,16820]K$ (à partir de l'Eq.\eqref{4.19}) et ici $\beta=0.2$ (déterminé par la formule simplifiée de l'Eq.\eqref{4.20}).\\

\clearpage

\subsection{Assombrissement centre-bord}
La morphologie d'une étoile, qu'elle soit sphérique ou ellipsoidale, subit l'effet de la profondeur optique centre-bord et la brillance de la photosphère visible. En effet un photon issu du centre du disque de l'étoile observée traverse moins d'atmosphère qu'un photon provenant du bord. Ce phénomène, connu comme l'assombrissement centre-bord a été observé pour la toute première fois sur une photographie du Soleil en avril 1845 par les physiciens français Louis Fizeau (1819-1896) et Lion Foucault (1819-1868). Depuis, un bon nombre de modèles décrivent la variation d'intensité sur un disque stellaire à travers le diagramme HR. Citons : la loi linéaire \citep{1921MNRAS..81..361M}, puis quadrique \citep{1977A&A....61..809M, 1985A&AS...60..471W, 1990A&A...230..412C}, ou bien en racine carrée \citep{1992A&A...259..227D}, ou encore algorithmique \citep{1970AJ.....75..175K}. Pour ma part j'ai opté pour une loi non-linéaire déduite du modèle ETL d'atmosphère stellaire proposé par \citet{2000A&A...363.1081C}, qui se rapproche le plus des données réelles, appliquées une carte 2D d'intensité normalisée $I_{nLD}$ purement "assombrissement centre-bord":\\

\begin{equation}
I_{nLD}(\lambda,\theta,\phi)=\frac{I_{\rm c}({\lambda,\theta,\phi})}{I_{\rm 0}(\lambda,T_{\rm 
eff})}=1-\sum_{k=1}^{4}a_{k}(\lambda)(1-\mu(\theta,\phi)^{\frac{k}{2}}),
\label{4.21}
\end{equation}

où $\mu$ est le cosinus de l'angle $A_{\theta,\phi}$ entre la normale à la surface au point considéré et de la direction d'observation et $a_k$ \& $k$, respectivement, le coefficient et l'ordre polynômial du modèle "assombrissement centre-bord". Les coefficients $a_k$ dit de Claret ont été rigoureusement calculés et tabulés pour de multiples configurations sur plusieurs paramètres du modèle ETL d'atmosphère stellaire ATLAS9\footnote{http://kurucz.harvard.edu/programs.html} de Robert Kurucz \citep{1970SAOSR.309.....K}, à savoir : la vitesse initiale de turbulence de l'atmosphère stellaire $VT$ (en $\kms$) (et que je prends par défaut $VT_\odot=2$ $\kms$), la gravité de surface $\log g$ (en $cm/s^2$), la température effective (en $K$), la métallicité $\log [metal/H]$ (en $\log[ Sun unit ]$) (qui est par défaut égale à la métallicité solaire, i.e. $=0$), et enfin la bande spectrale de l'ultraviolet "u" à l'infrarouge "K",  en passant par le visible "V"  [u v b y U B V R I J H K] (voir Tab.\ref{UBVRIJHKLMNQ}). Ces quatre paramètres sont tabulés et accessibles depuis le site internet de VizieR\footnote{http://vizier.u-strasbg.fr/viz-bin/VizieR-3?-source=J/A\%2bA/363/1081/atlas} pour chaque configuration.\\

\begin{table*}[htbp]
\centering
\caption[Bandes spectrales UBVRIJHKLMNQ]{Bandes spectrales du système photosphérique UBVRIJHKLMNQ données en fonction du nom de la bande et du domaine spectral $\lambda(\mu m)$}\label{UBVRIJHKLMNQ}
\centering
\resizebox{\textwidth}{!}{\begin{tabular}{|c|c|c|c|c|c|c|c|c|c|c|c|c|c|c|}
\hline\hline
Bande & U & B & V & $R_j$ & $I_j$ & $R_c$ & $I_c$ & J & H & K & L & M & N & Q\\
\hline
$\mathbf{\lambda(\mu m)}$ & 0.35 & 0.44 & 0.55 & 0.70 & 0.90 & 0.65 & 0.80 & 1.22 & 1.63 & 2.19 & 3.45 & 4.75 & 10.20 & 21.00\\
\hline\hline
\end{tabular}}
\end{table*}

\begin{figure}[ht!]
\centering
\includegraphics[height=0.4\hsize,width=0.4\hsize,draft=false]{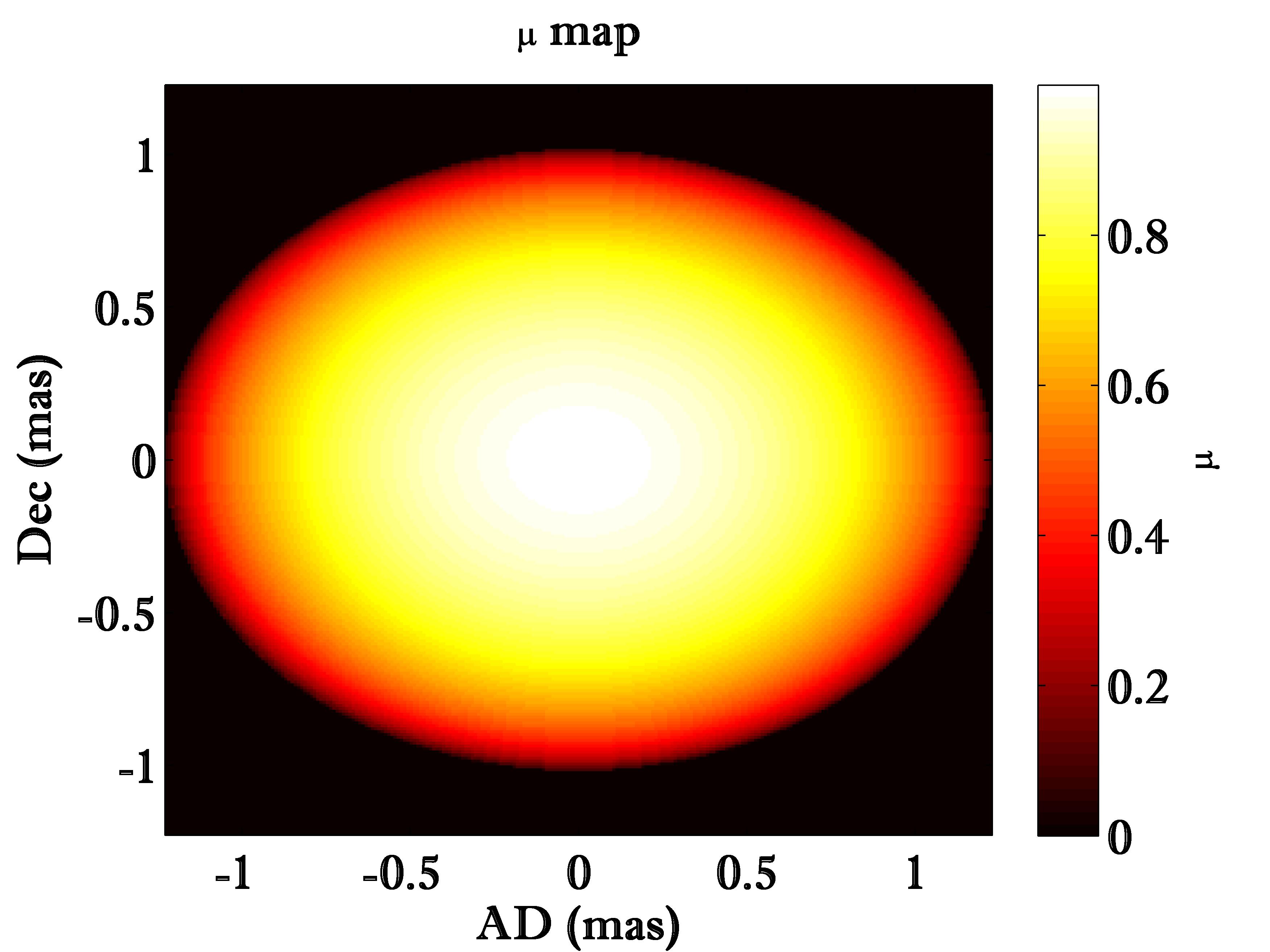}
\includegraphics[height=0.4\hsize,width=0.4\hsize,draft=false]{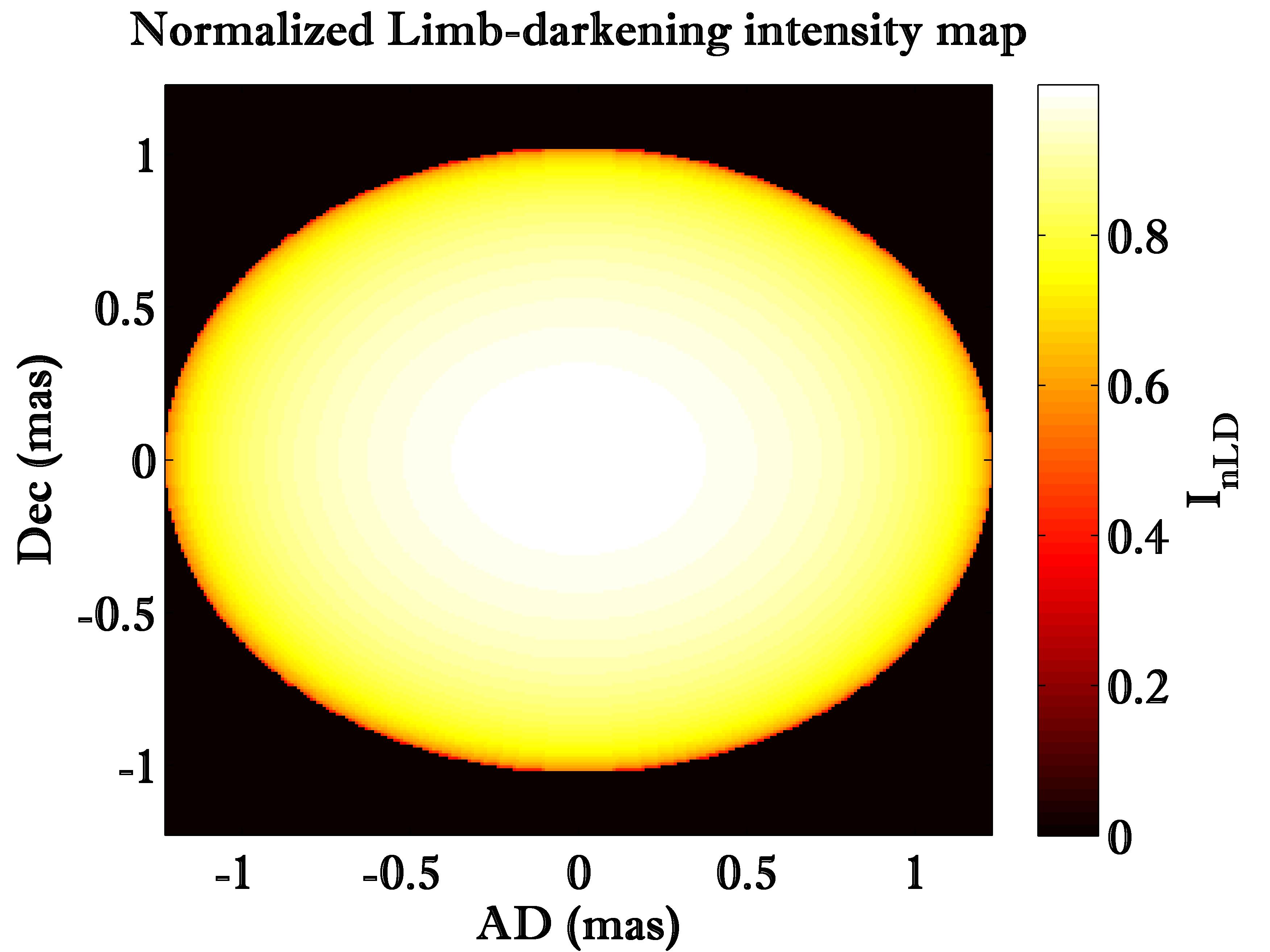}
\includegraphics[height=0.4\hsize,width=0.4\hsize,draft=false]{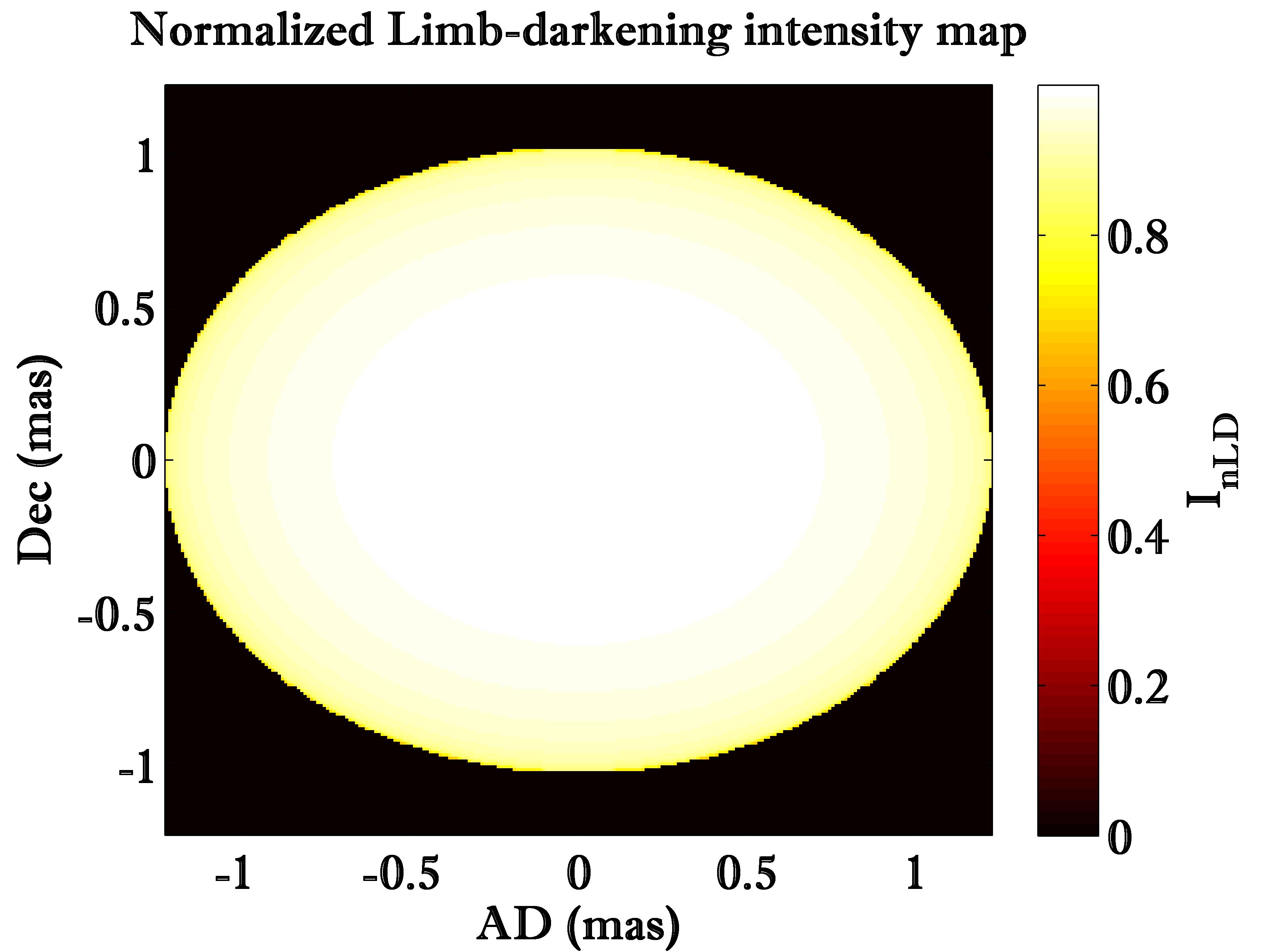}
\includegraphics[height=0.4\hsize,width=0.4\hsize,draft=false]{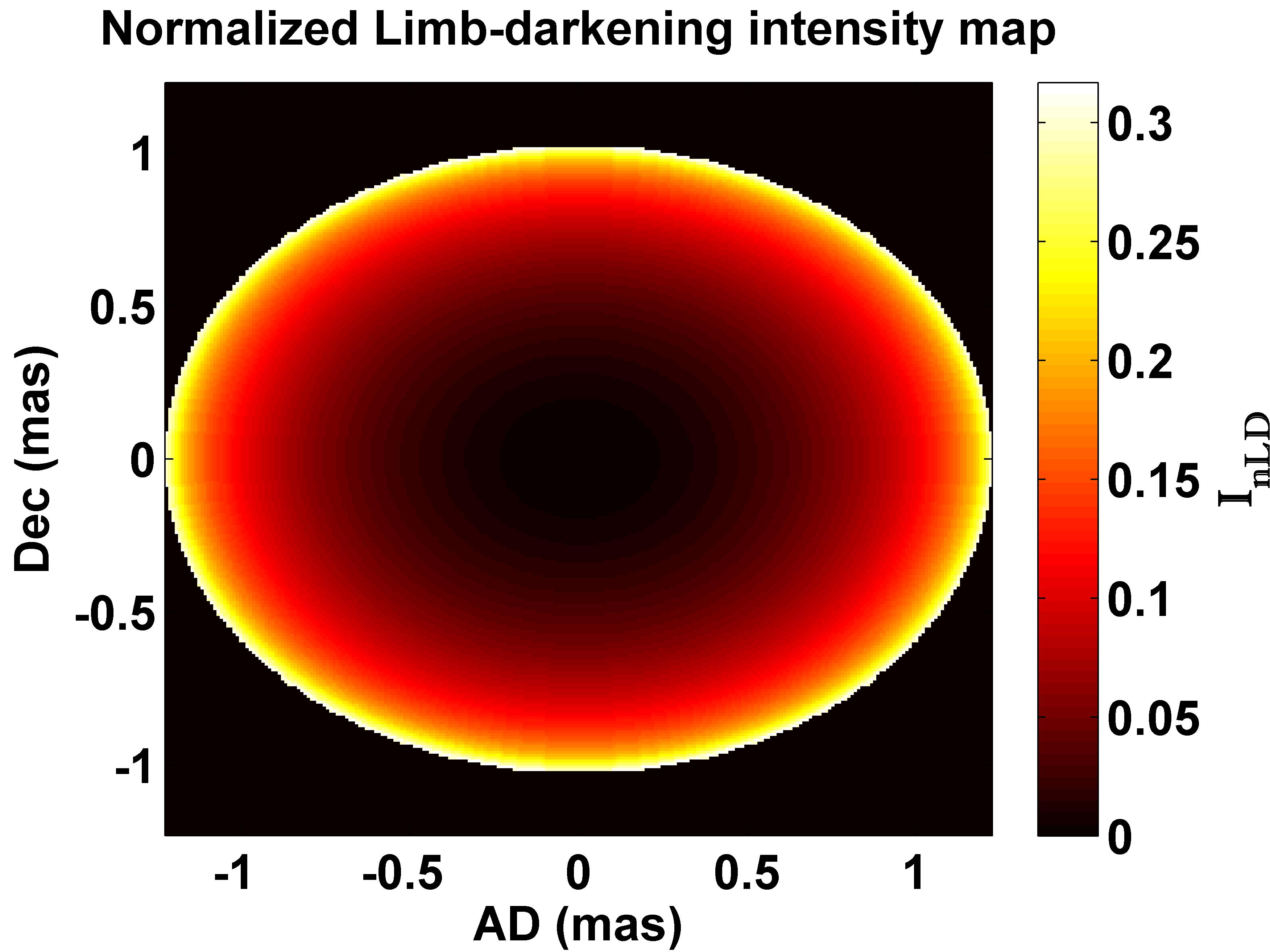}
\caption[Cartes 2D d'intensité assombrie centre-bord]{Exemple de cartes 2D d'intensité assombrie centre-bord ($\lambda=\lambda_0=2.166$ $\mu m$). \textbf{En haut:} \textit{à gauche:} La carte 2D de $\mu$ à la surface stellaire. \textit{A droite:} La carte 2D d'intensité en bande spectrale ultraviolet "u".  \textbf{En bas:} \textit{à gauche:} La carte 2D d'intensité en bande spectrale infrarouge "K". Et \textit{à droite:} La carte 2D de la différence d'intensité en "u" et "K".}\label{InLD_diff}
\end{figure}

Les cartes 2D d'intensité normalisées "assombrissement centre-bord" ("Limb Darkening") $I_{nLD}(\theta,\phi)$ sont numériquement calculées pour un disque en fonction de $\mu=\cos A_{\theta,\phi}$, aplati en forme d'ellipse au rapport des demi-axes qui est égal au degré de sphéricité apparent $D'$, spécifique à l'aplatissement réel et à l'angle d'inclinaison $i$, et ayant le même rayon angulaire équatorial $R_{eq}$ selon les équations \eqref{4.5}. Un exemple de cartes d'intensité 2D $I_{nLD}(\theta,\phi)$ pour deux bandes spectrales différentes est illustré dans la Fig.\ref{InLD_diff}.\\

\begin{figure}[ht!]
\centering
\includegraphics[height=0.4\hsize,width=0.4\hsize,draft=false]{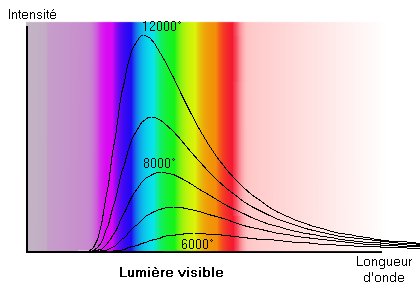}
\caption[Tracé corps noir]{Tracé du corps noir, où on remarque bien que la variation de la température propre à chaque intensité est plus importante en UV (bande "U") qu'en IR (bande "K").}\label{blackbody}
\end{figure}

On remarque que l'effet de l'assombrissement centre-bord est mieux visible dans l'ultraviolet qu'en infrarouge (effet de la profondeur optique) ce qui met en évidence la variation de la température propre à chaque intensité du disque stellaire et qui est plus importante en UV (bande "U") qu'en IR (bande "K"). Voir aussi la Fig.\ref{blackbody} du corps noir.\\

Enfin, et pour plus de rigueur, j'ai entrepris la modélisation de cartes 2D $I_{nLD}$ cohérente avec la variation de température et de gravité de surface propre à chaque latitude $\theta$, suivant les Eqs.\eqref{4.13} \& \eqref{4.14}, à l'aide des coefficients $a_k$ et en jouant sur les deux paramètres de gravité de surface $\log g$ et température effective ($T_{eff}$), tout en respectant l'effet de l'inclinaison. L'intensité au continuum regroupant les deux assombrissements devient donc:\\  

\begin{equation}
I_{\rm c}({\lambda,\theta,\phi})= I_{\rm 0}(\lambda,T_{\rm eff}(\theta,\phi))I_{nLD}(\lambda,\theta,\phi)
\label{4.22}
\end{equation}

\begin{figure}[ht!]
\centering
\includegraphics[height=1.2\hsize,width=1.2\hsize,draft=false]{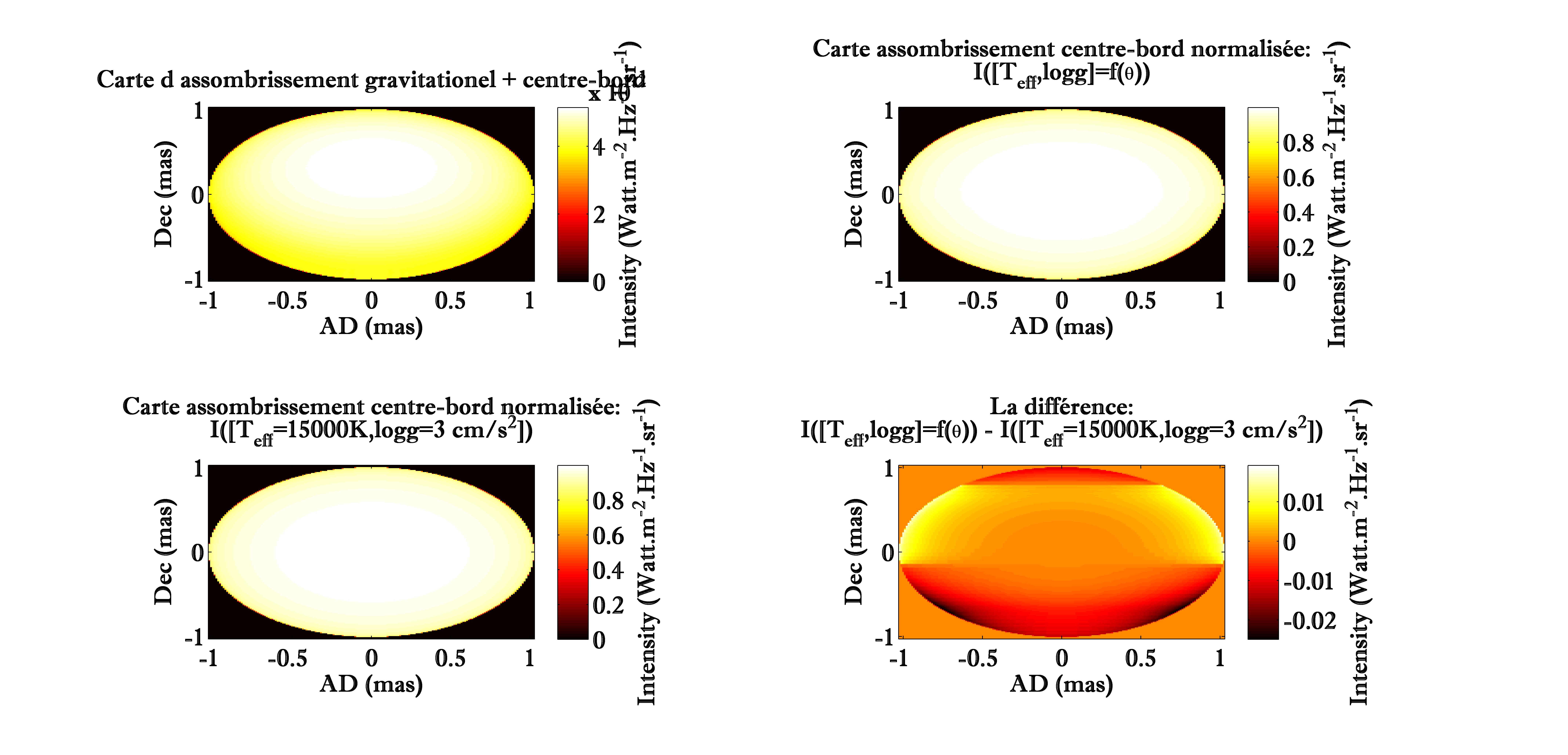}
\caption[Carte d'intensité 2D au continuum \& Carte $I_{nLD}=f(\theta)$]{Carte d'intensité 2D au continuum \& Carte $I_nLD=f(\theta)$ pour $i=30^\circ$}\label{Icontinu1}
\end{figure}

\begin{figure}[ht!]
\centering
\includegraphics[height=1.2\hsize,width=1.2\hsize,draft=false]{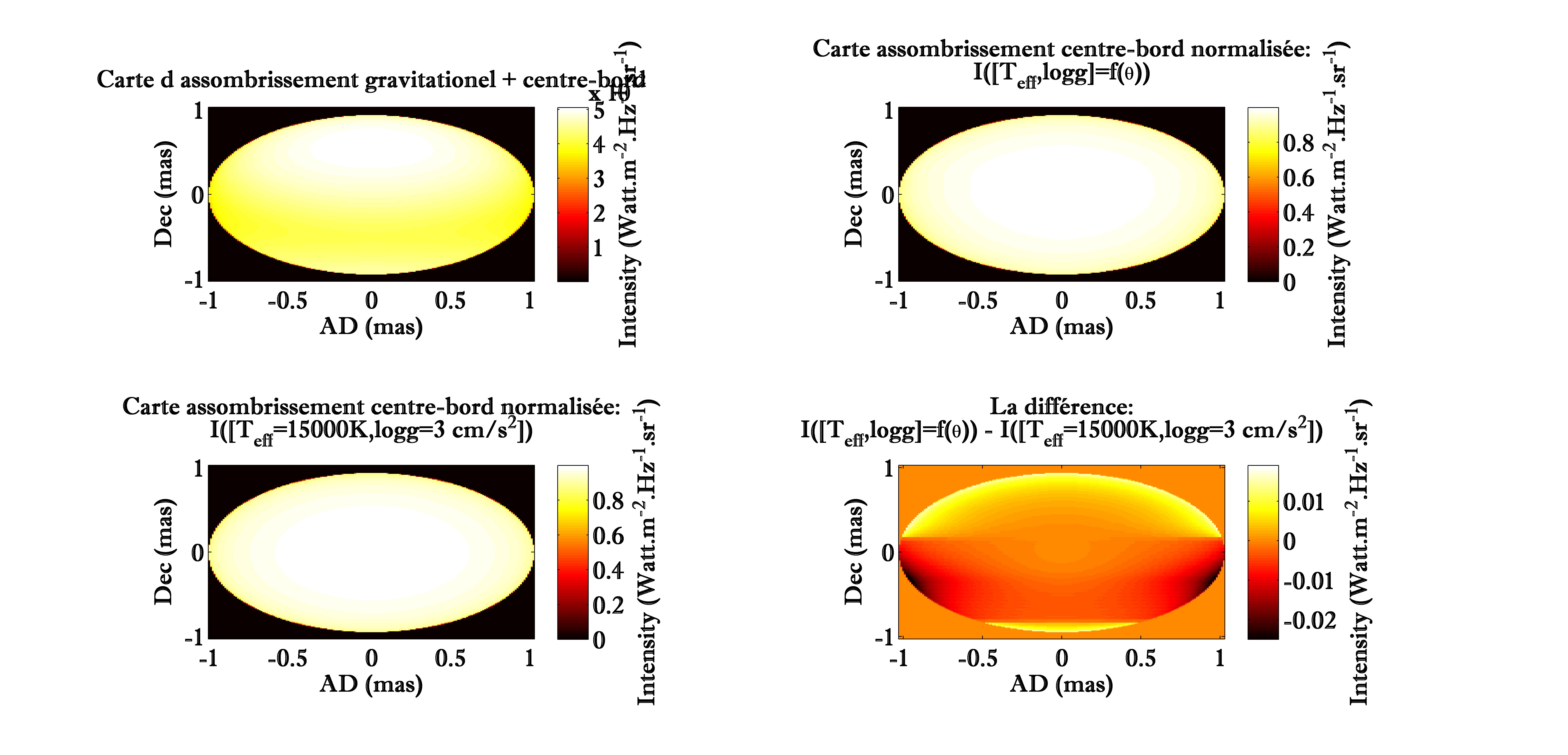}
\caption[Carte d'intensité 2D au continuum \& Carte $I_{nLD}=f(\theta)$]{Carte d'intensité 2D au continuum \& Carte $I_nLD=f(\theta)$ pour $i=60^\circ$}\label{Icontinu2}
\end{figure}

\begin{figure}[ht!]
\centering
\includegraphics[height=1.2\hsize,width=1.2\hsize,draft=false]{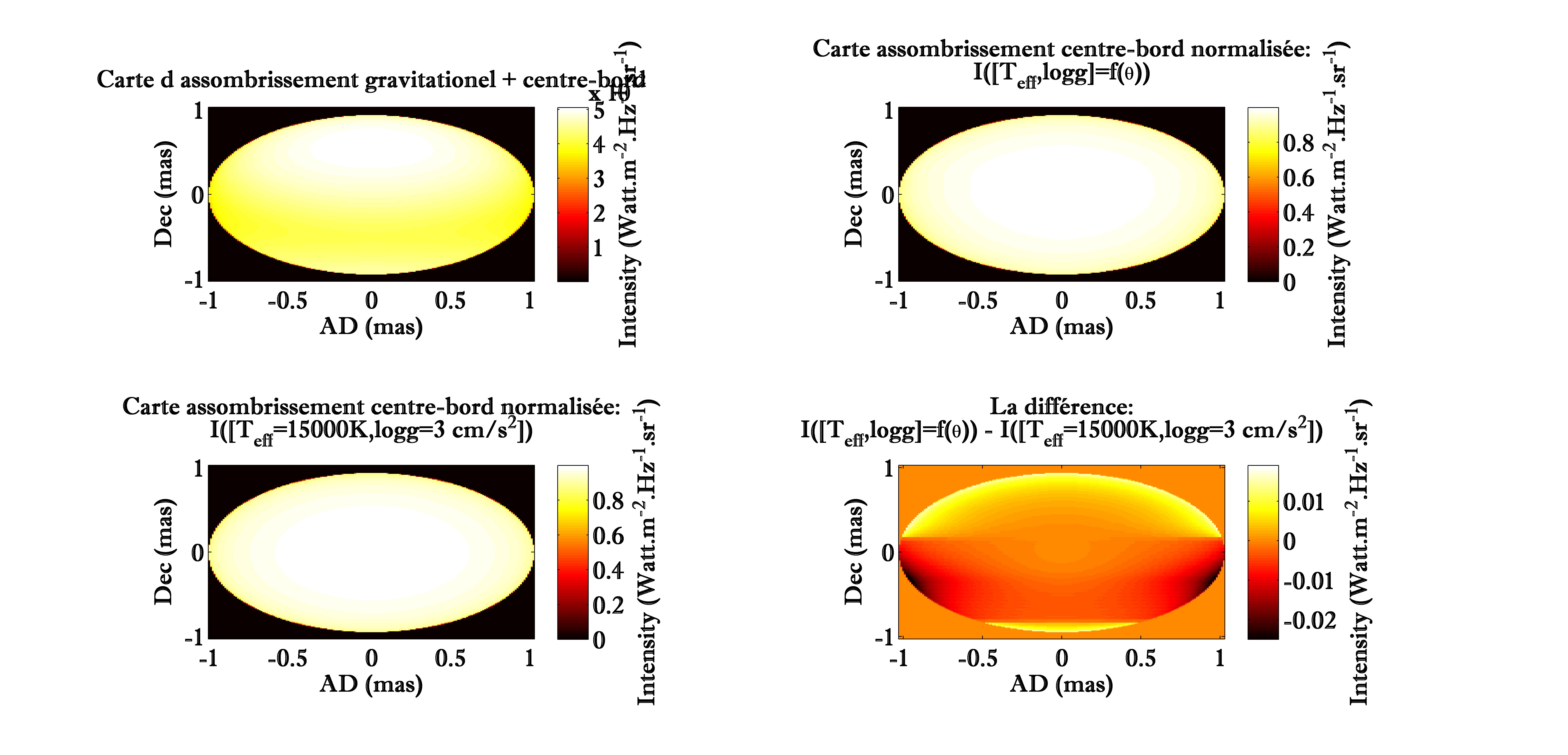}
\caption[Carte d'intensité 2D au continuum \& Carte $I_{nLD}=f(\theta)$]{Carte d'intensité 2D au continuum \& Carte $I_nLD=f(\theta)$ pour $i=90^\circ$}\label{Icontinu3}
\end{figure}

Les Figs.\ref{Icontinu1}-\ref{Icontinu3}, avec 3 jeux de 4 figures, illustrent une étoile de mêmes paramètres que celle de la Fig.\ref{int_2D}, et pour des inclinaisons respectives de $i= 30^\circ$, $60^\circ$ \& $90^\circ$, pour des cartes d'assombrissement gravitationnel + centre-bord, i.e. des cartes d'intensités au continuum $I_c$ pour bien voir l'effet des latitudes en terme de la variation de $[T_{eff}, \log g]$, des cartes d'assombrissement centre-bord normalisées variables $[T_{eff}, \log g]=[11000...19000K, 3...3.5 Cm/s2]$, des cartes d'assombrissement centre-bord normalisées fixes $[T_{eff}, \log g]_{mean}=[15000K, 3 Cm/s2]$, et la différence entre les deux cartes $[T_{eff}, \log g]$ fixes et variables.\\

Les effets d'assombrissement tels qu'abordés dans le chapitre \ref{chap:rota} peuvent avoir un impact significatif sur l'interprétation astrophysique des mesures obtenues. En effet, on peut observer pendant longtemps une étoile et croire qu'elle est parfaitement ronde avec une faible $v_{eq}$ et interpréter son assombrissement comme étant un pur assombrissement centre-bord, alors qu'il s'agit en fait d'un rotateur rapide observé selon l'un de ses pôles (pole-on, $i=0^\circ$ ou $i=180^\circ$) avec l'assombrissement gravitationnel qui lui est lié. Ce fut le cas pour Vega, qui est une étoile référence souvent prise pour l'échelle des magnitudes \citep{2006Natur.440..896P}. Il est bien été établi aujourd'hui que l'assombrissement gravitationnel a un impact majeur sur la mesure du $\vsini$ \citep{2004IAUS..215...23F, 2004MNRAS.350..189T}. Il a été démontré aussi, que les deux types d'assombrissement ont un effet sur la forme des raies (\citet{1963ApJ...138.1134C, 1965ApJ...142..265C} et \citet{1966ApJ...146..152C}).\\

\clearpage

\section{Spectroscopie d'un rotateur}\label{sec_4.3}

\begin{figure}[ht]
\centering
\includegraphics[width=0.5\hsize,height=0.5\hsize,draft=false]{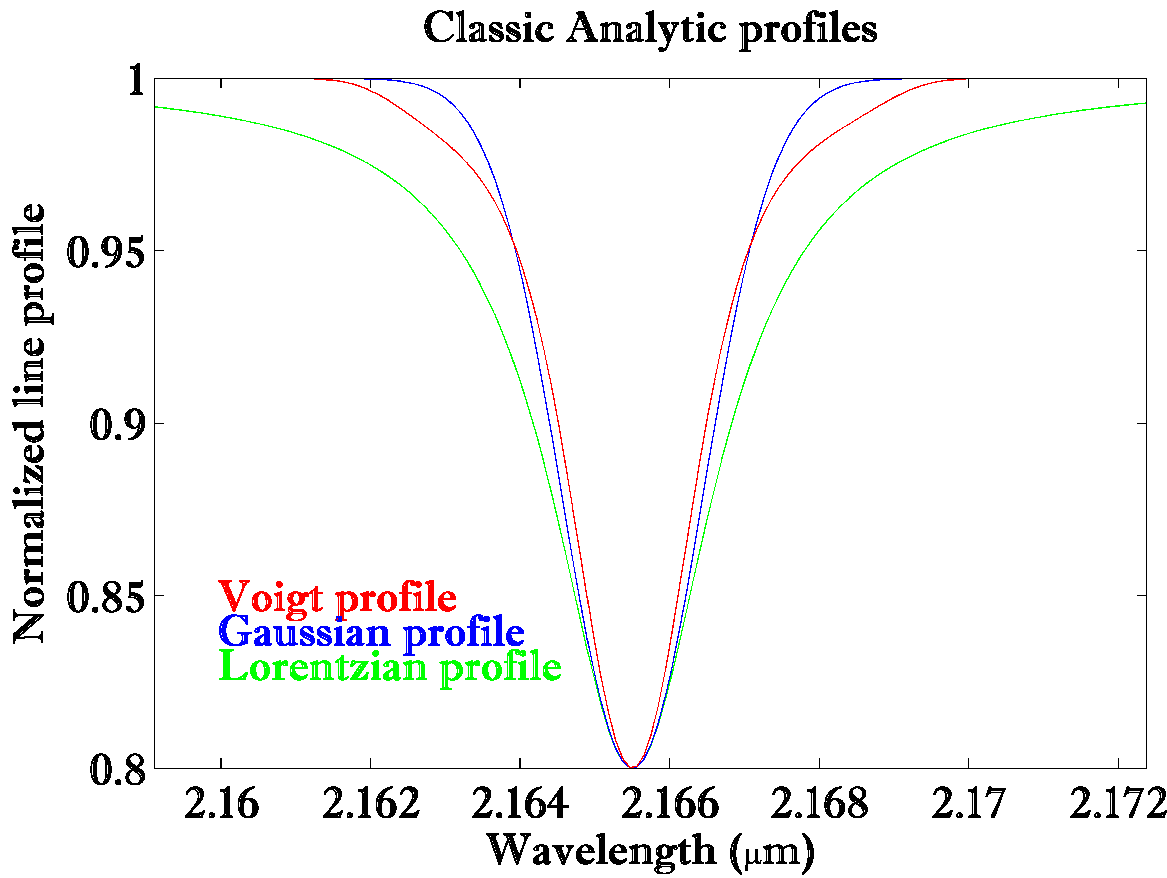}
\caption[Profils de raie analytiques]{Exemples de profils de raies analytiques classiques (Gaussien, Lorentzien, et profil de Voigt) normalisés, simulant la raie Br$\gamma$ entre $\lambda= 2.15$ $\mu m$ \& $2.18$ $\mu m$.}\label{fig_profil1}
\end{figure}

Un autre aspect des rotateurs stellaires dont il faut tenir compte dans toute modélisation, est celui de leurs profil de raies spectrales élargies par effet Doppler-Fizeau, dû à la rotation formulée par l'Eq.\eqref{3.34}. SCIROCCO a été bâti pour être léger et rapide afin de modéliser des phénomènes complexes à partir d'approximations basiques, tout en restant physiquement correct. Dans cet esprit j'ai incorporé dans mon code des profils de raie analytiques rapides d'exécution. J'ai examiné dans un premier temps des profils simples (Gaussienne, Lorentzienne et Voigt) définis respectivement par:\\ 

\begin{equation}
\left\{ \begin{array}{l}
H_{\rm Gauss}(\lambda)=1-H_{0}\left[-\pi H_{0}^{2}\frac{(\lambda-\lambda_{0})^{2}}{W^2}\right]\\
H_{\rm Lorentz}(\lambda)=1-\left[\frac{H_{0}}{1+(\frac{\lambda-\lambda_{0}}{W/2})^{2}}\right],\\
H_{\rm Voigt}(\lambda)=(H_{Gauss} \ast H_{Lorentz})(\lambda)
\end{array} \right.
\label{eq1}
\end{equation}

où $H_0$ est l'amplitude de profil de raie, $W$ sa largeur à mi-hauteur avec $\ast$ désignant le signe de l'opération de convolution (Fig.\ref{fig_profil1}).\\

La difficulté ici était la détermination de $H_0$ et $W$. Une solution consistait à recourir aux profils de raies de synthèse, issus de robustes modèles d'atmosphères stellaires que j'ai ajusté numériquement par un profil de Voigt. Les codes de simulation d'atmosphères stellaires existent, parmi lesquels on peut citer les modèles ETL : ATLAS9 \citep{1970SAOSR.309.....K} et MARCS -Model Atmospheres in Radiative and Convective Scheme- \citep{1975A&A....42..407G}, ainsi que le modèle PHOENIX ressuscité des cendres d'un ancien code appelé SNIRIS- \citep{2010ascl.soft10056B} qui peut être utilisé en ETL ou non-ETL, ou encore le code non-ETL Tlusty \citep{1995ApJ...439..875H}. Des codes produisant des spectres à partir de ces derniers modèles ont été développés, parmi lesquels Synspec, qui ouvre la possibilité de calculer les spectres synthétiques à partir de Kurucz ou de Tlusty. Une version IDL plus simple et plus directe, nommée "Synsplot", permet de rapidement déduire les spectres synthétiques stellaires (intensité spécifique ou le flux intégré) à partir de grilles tabulées et pré-calculées Kurucz et Tlusty, avec une simple ligne de commande sur laquelle il faut définir les paramètres importants dont; la température $T_{eff}$ effective, la gravité de surface effective $\log g$, l'intervalle des longueurs d'onde $[\lambda_{debut},\lambda_{fin}]$, la résolution $\delta\lambda$, l'angle $\mu$, et le type de grille de modèles d'atmosphères à prendre en compte (Kurucz ou Tlusty)...etc. J'ai finalement opté pour le logiciel Synspec utilisant le modèle ETL Kurucz, qui permet de modéliser ouvertement le spectre des étoiles chaudes et actives. Ce logiciel puissant, une fois les paramètres d'entrée établis, produit plusieurs fichiers de sortie pour chaque configuration, parmi lesquels on a un fichier de type ".7" qui désigne le flux intégré et un fichier du type ".10" qui concerne l'intensité spécifique. C'est ce dernier, que j'ai utilisé par la suite comme profil de raie pour $\mu=0$.\\

\begin{figure}[ht]
\centering
\includegraphics[width=0.4\hsize,height=0.4\hsize,draft=false]{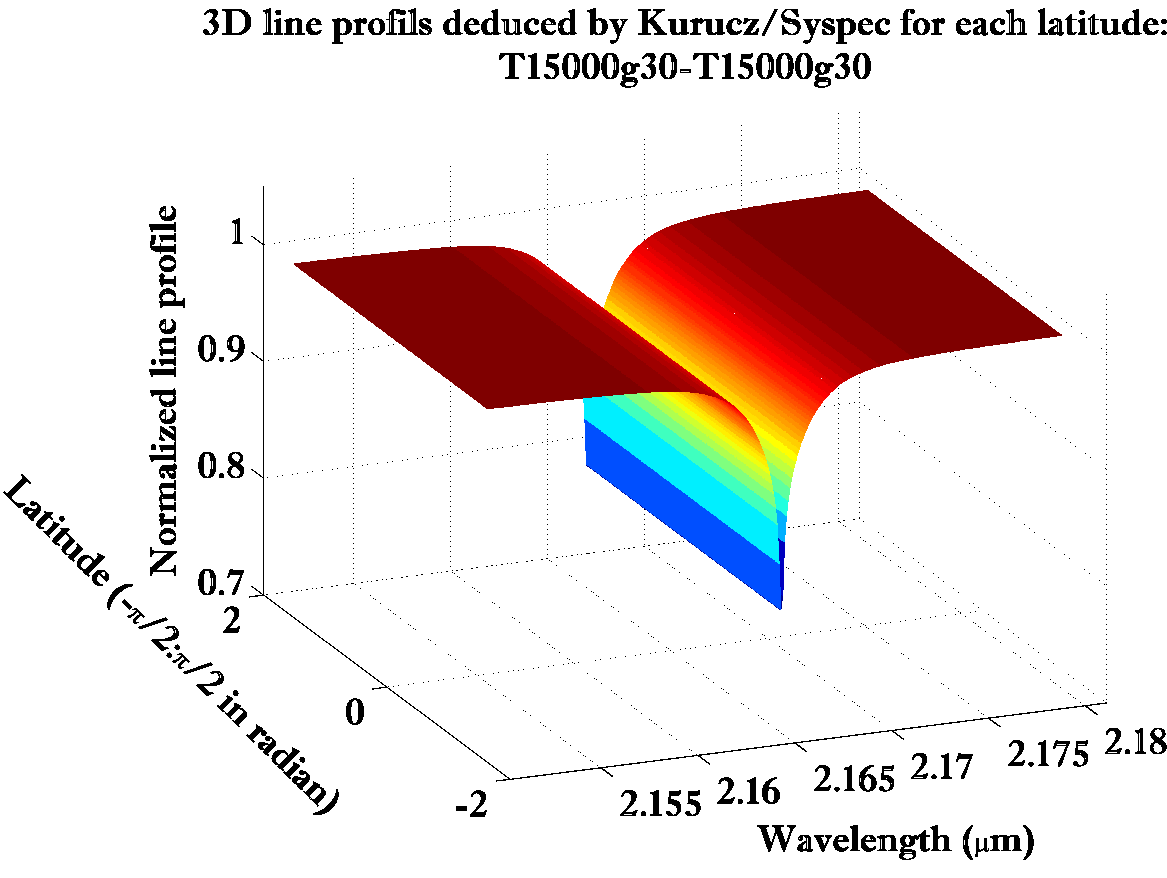}
\includegraphics[width=0.4\hsize,height=0.4\hsize,draft=false]{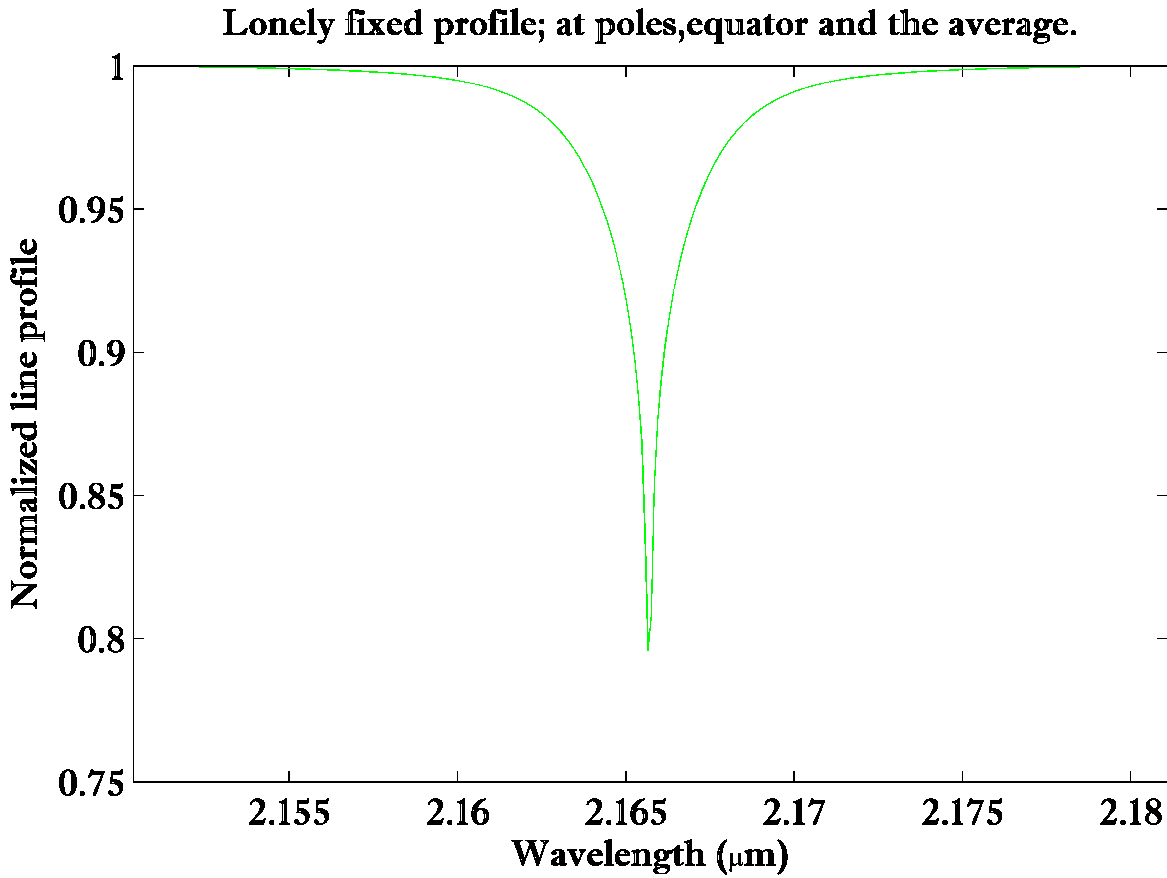}
\caption[Profils de raie Synplot]{Exemple de profil de raie produit par  Synplot à partir du modèle Kurucz/Synspec, pour $[T_{eff},\log g_{eff}]_{mean}=[15000$ $K,30$ $cm/s^2]$ normalisés autour de la raie Br$\gamma$ entre $\lambda= 2.15$ $\mu m$ \& $2.18$ $\mu m$.}\label{fig_profil2}
\end{figure}

\begin{figure}[ht]
\centering
\includegraphics[width=0.4\hsize,height=0.4\hsize,draft=false]{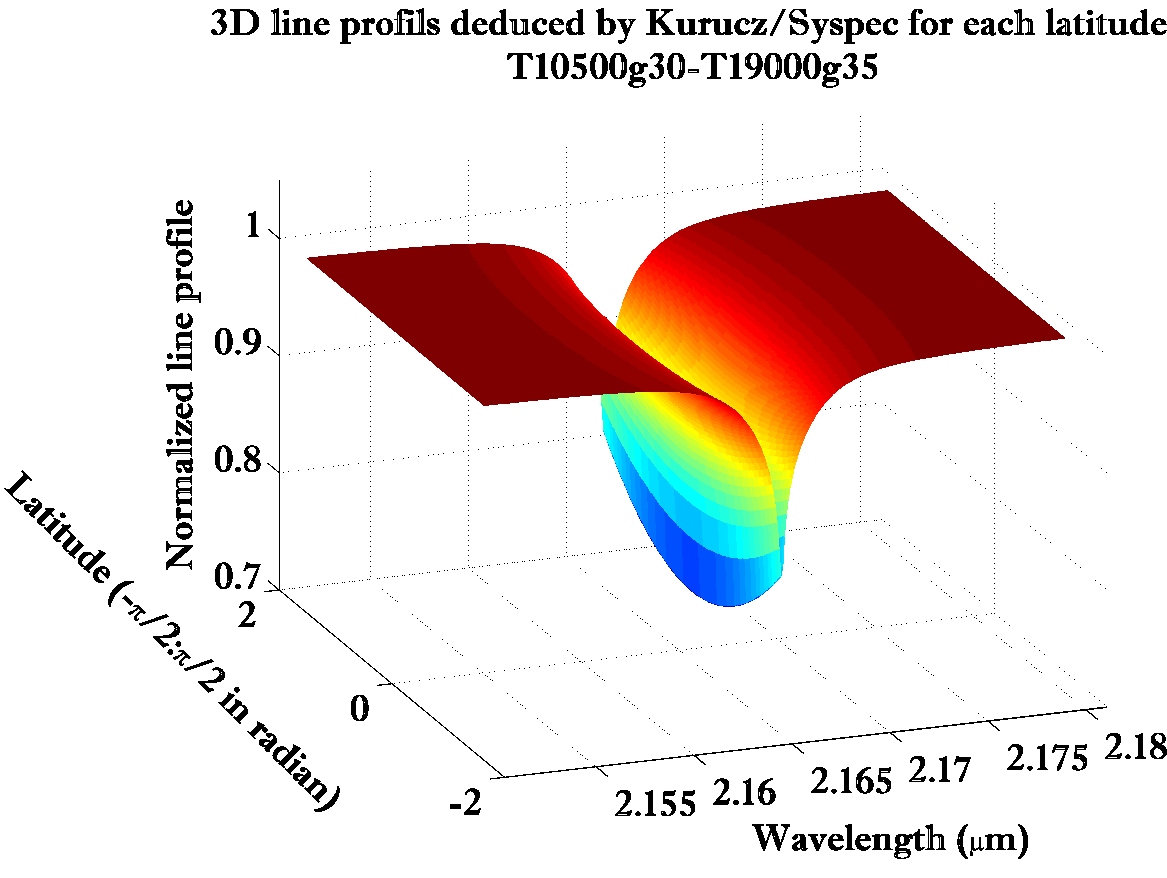}
\includegraphics[width=0.4\hsize,height=0.4\hsize,draft=false]{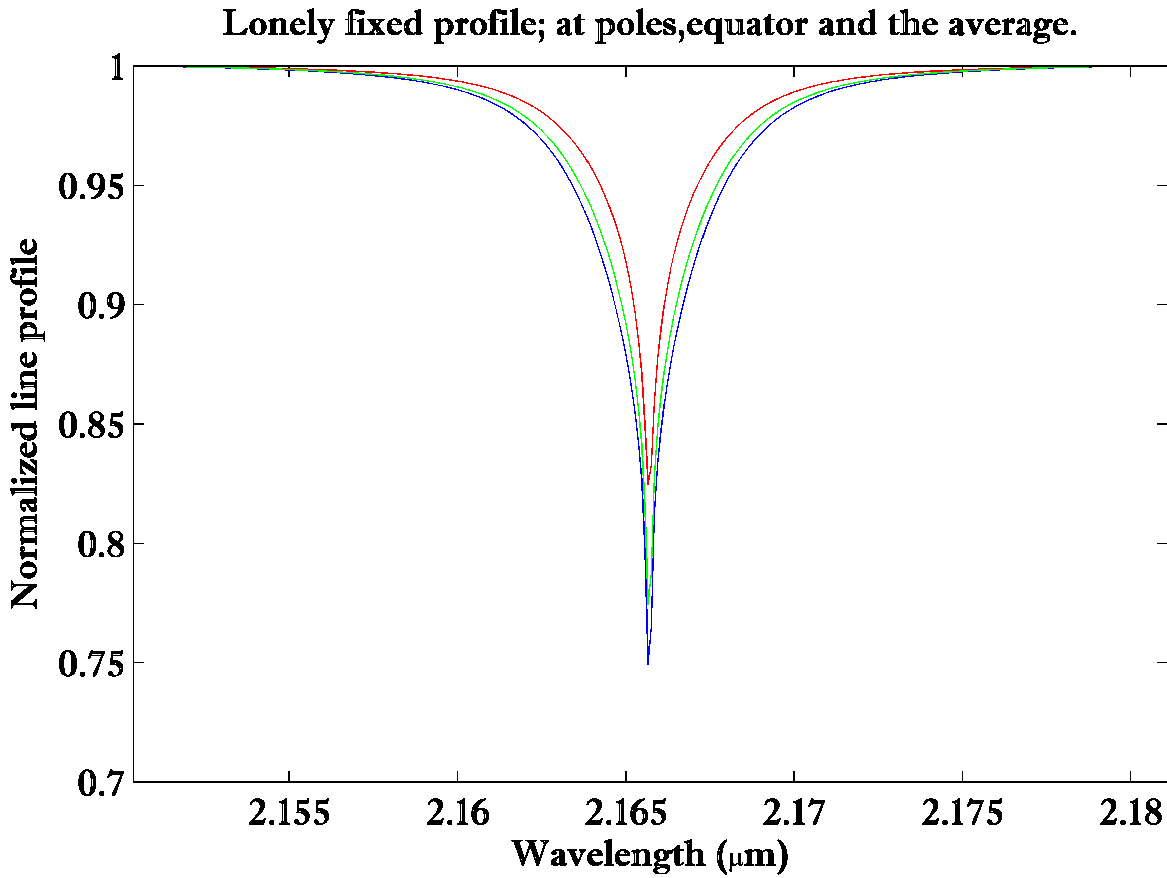}
\caption[Profils de raie 3D Synplot]{Exemple de profils de raie Synplot (Kurucz/Synspec) pour $[T_{eff},\log g_{eff}]=[10500...19000$ $K,30...35$ $cm/s^2]$ (équateur/pôles) normalisés autour de la raie Br$\gamma$ $\lambda= 2.15$ $\mu m$ \& $2.18$ $\mu m$.}\label{fig_profil3}
\end{figure}

Ensuite, on a la possibilité d'obtenir un profil de raie pour une configuration de $\Tmean$ et $\log g_{eff}$ moyennée (voir Fig.\ref{fig_profil2}), ou obtenir une configuration de profil de raie adaptée à chaque latitude, via un petit code IDL que j'ai élaboré et qui me permet d'obtenir un profil de raie 3D qui varie en fonction de la latitude $\theta$ pour chaque couple de valeurs $[T_{eff}(\theta),\log g_{eff}(\theta)]$. Les trous ici en latitude sont simplement comblés par un simple ajustement linéaire (Fig.\ref{fig_profil3}). Notons, que l'effet centre-bord n'est pas pris en compte dans le profil de raie, car celui-ci étant traité lors de la simulation de la carte d'intensité, il est néanmoins pris en compte dans la simulation globale de l'étoile.\\

\begin{figure}[ht]
\centering
\includegraphics[width=0.4\hsize,height=0.4\hsize,draft=false]{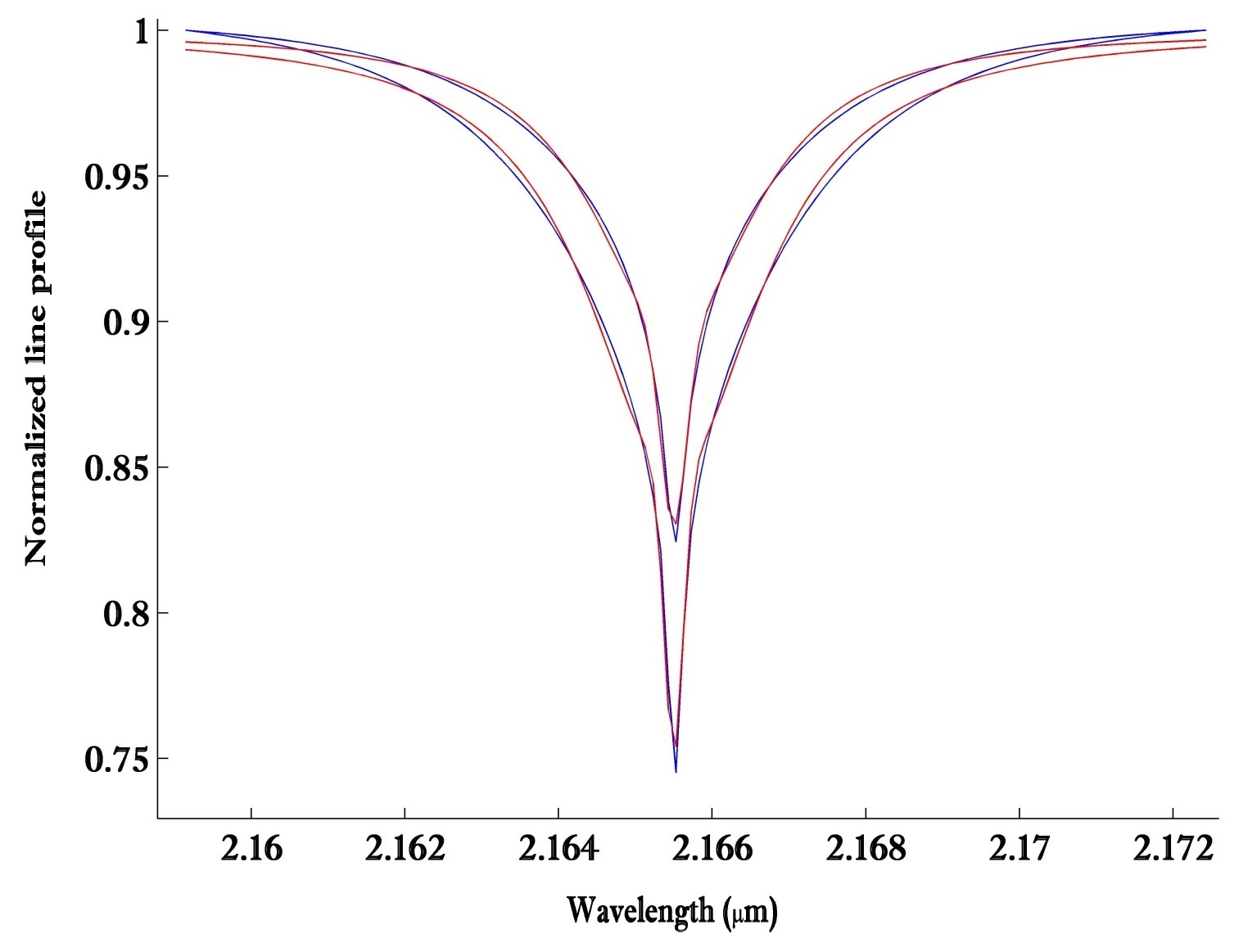}
\caption[Comparaison des profils Synplot/Synplot analytique]{Profils de raie normalisés sur l'étendue de la longueurs d'onde entre $\lambda= 2.15\mu m$ \& $2.18 \mu m$, pour la raie Br$\gamma$ ; Synplot en bleu, et ajustés par la fonction pseudo-Voigt à 5 paramètres en rouge. Pour un profil de raie $[T_{eff},\log g_{eff}]=[19000$ $K,3.5$ $cm/s^2]$ (aux pôles) en haut, et un autre $[T_{eff},\log g_{eff}]=[10500$ $K,3.0$ $cm/s^2]$ (à l'équateur) en bas.}\label{fig_profil7}
\end{figure}

\begin{figure}[ht]
\centering
\includegraphics[width=0.4\hsize,height=0.4\hsize,draft=false]{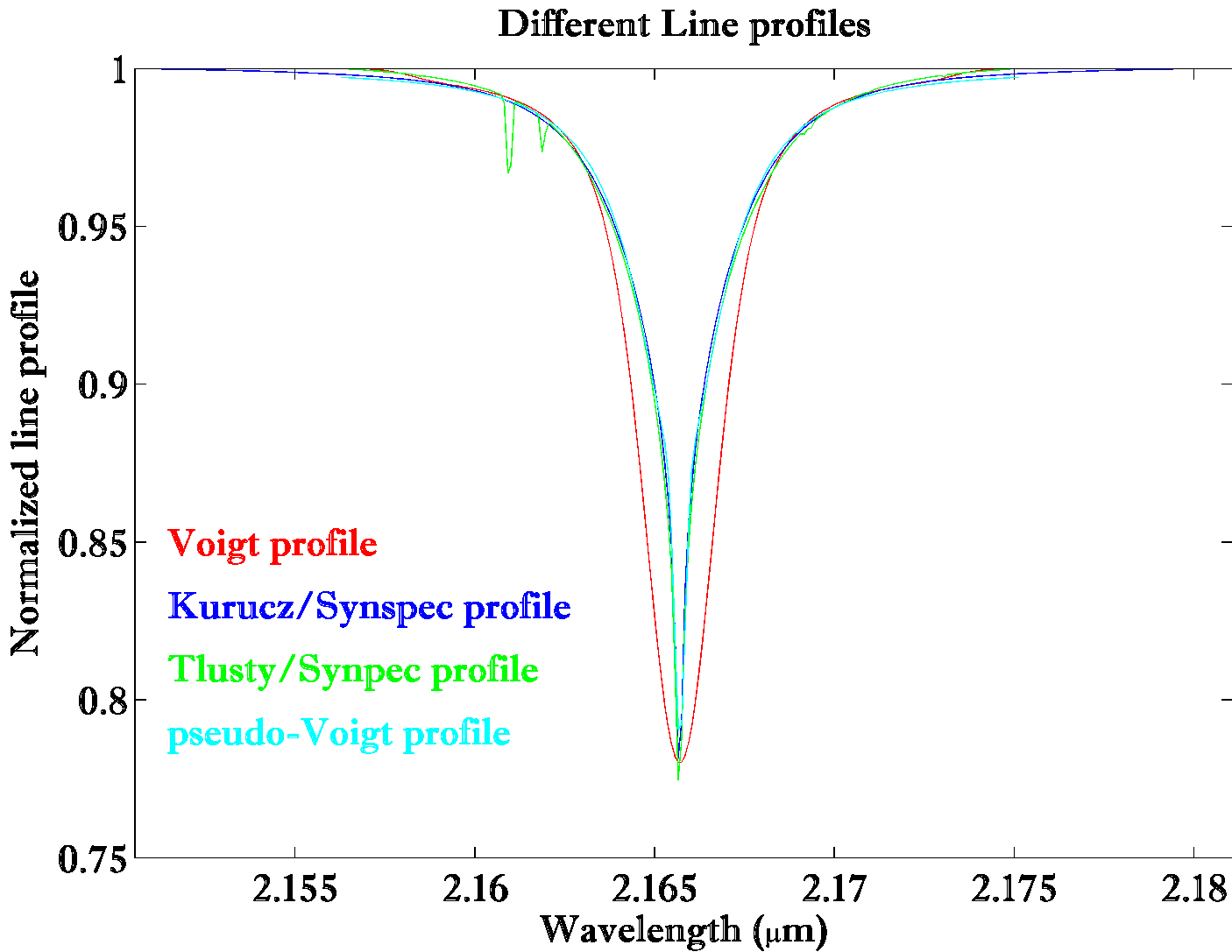}
\caption[Comparaison de tous les profils de raie]{Profils de raie normalisés sur l'étendue de la longueurs d'onde entre $\lambda= 2.15$ $\mu m$ \& $2.18$ $\mu m$, pour la raie Br$\gamma$ que SCIROCCO peut utiliser; pour $[T_{eff},\log g_{eff}]_{mean}=[15000$ $K,30$ $cm/s^2]$ ; profil de Voigt (en rouge), Kurucz/Synspec (en bleu foncé), Tlusty/synspec (en vert) et pseudo-Voigt (en bleu clair).}\label{fig_profil6}
\end{figure}

Pour une version rapide de mon code j'ai d'abord essayé de déduire une formule analytique des spectres synthétiques basée sur le modèle Kurucz/Synplot. Ainsi et pour un ajustement optimal, j'ai eu recours à la fonction de pseudo-Voigt $H_{pv}=\eta H_{Lorentz}+(1-\eta) H_{Gauss}$ à 5 paramètres ; les amplitudes et largeurs à mi-hauteur de la Gaussienne et de la Lorentzienne et un coefficient de pondération $\eta$. Malgré de nombreux essais, je ne suis jamais parvenu à un ajustement parfait entre un profil de raie Synplot et son ajustement pseudo-Voigt (Fig.\ref{fig_profil7}). Cette petite différence principalement dans les ailes entre les deux profils entraine une différence non négligeable, dans les résultats de certains paramètres fondamentaux obtenus lors des ajustements modèle/données observées, à savoir le rayon angulaire équatorial et la vitesse de rotation équatoriale des étoiles. J'ai pu estimer à $10\%$ la différence des résultats obtenus sur ces deux paramètres fortement corrélés, avec les deux profils de raie différents (voir le chapitre \ref{chap:appli} où une étude détaillée sur l'impact de chaque paramètre stellaire sur la phase différentielle $\phidiff$  a été menée). Le profil de raie reste un paramètre sensible qui démontre son importance pour la modélisation correcte de rotateurs stellaires rapides, déstinée aux observations interferométriques à longue base. La Fig.\ref{fig_profil6} regroupe l'ensemble des profils de raie utilisés dans mon étude comparative.\\

Au final, j'ai opté pour l'utilisation rigoureuse du profil de raie issu du modèle d'atmosphère Kurucz/Synspec basé sur Synplot.\\% La Fig.{} représente une carte de profil de raie $H_{synplot}$ 3D pour plusieurs longueurs d'onde en bande K, et pour une étoile de .\\

\clearpage

\section{Cartes d'intensité d'un rotateur rapide suivant le décalage Doppler-Fizeau sur le profil de la raie d'absorption photosphérique}

Une fois que j'ai simulé la carte d'iso-vitesses pour une étoile donnée, sa carte d'intensités au continum (assombries par les deux effets gravitationnel et centre-bord) et son profil de raie, il ne me reste plus qu'à déduire les cartes d'intensités monochromatiques $I({\lambda,\theta,\phi})$, conformément à l'équation suivante :

\begin{equation}
I({\lambda,\theta,\phi}) = I_{\rm c}({\lambda,\theta,\phi}) 
H\left({\lambda+\lambda_0\frac{{V_{\rm proj}({\theta,\phi})}}{c}}, 
\theta,\phi \right)
\label{eq2.23}
\end{equation}

\begin{figure}[ht]
\centering
\includegraphics[width=0.5\hsize,draft=false]{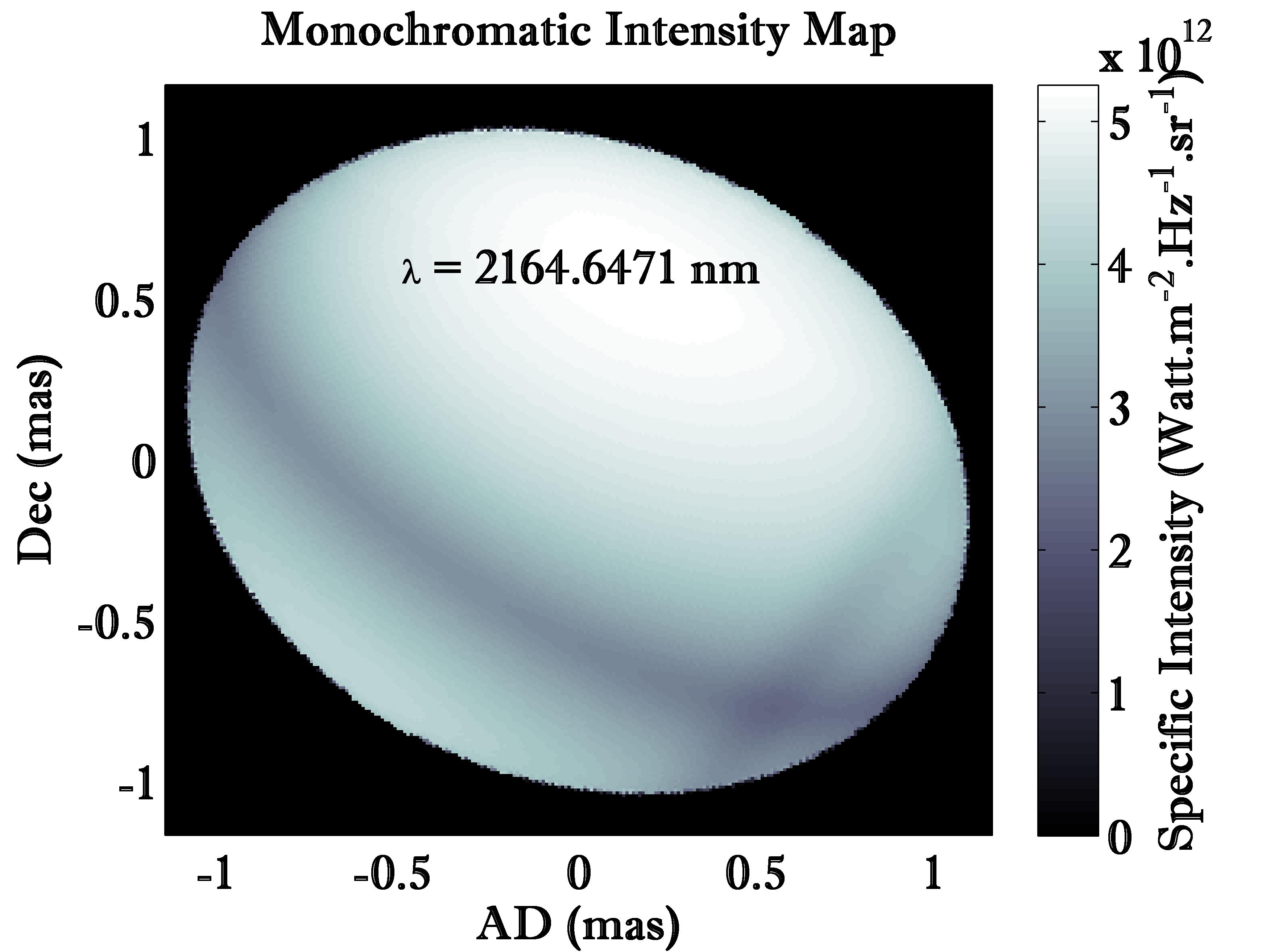}
\includegraphics[width=0.5\hsize,draft=false]{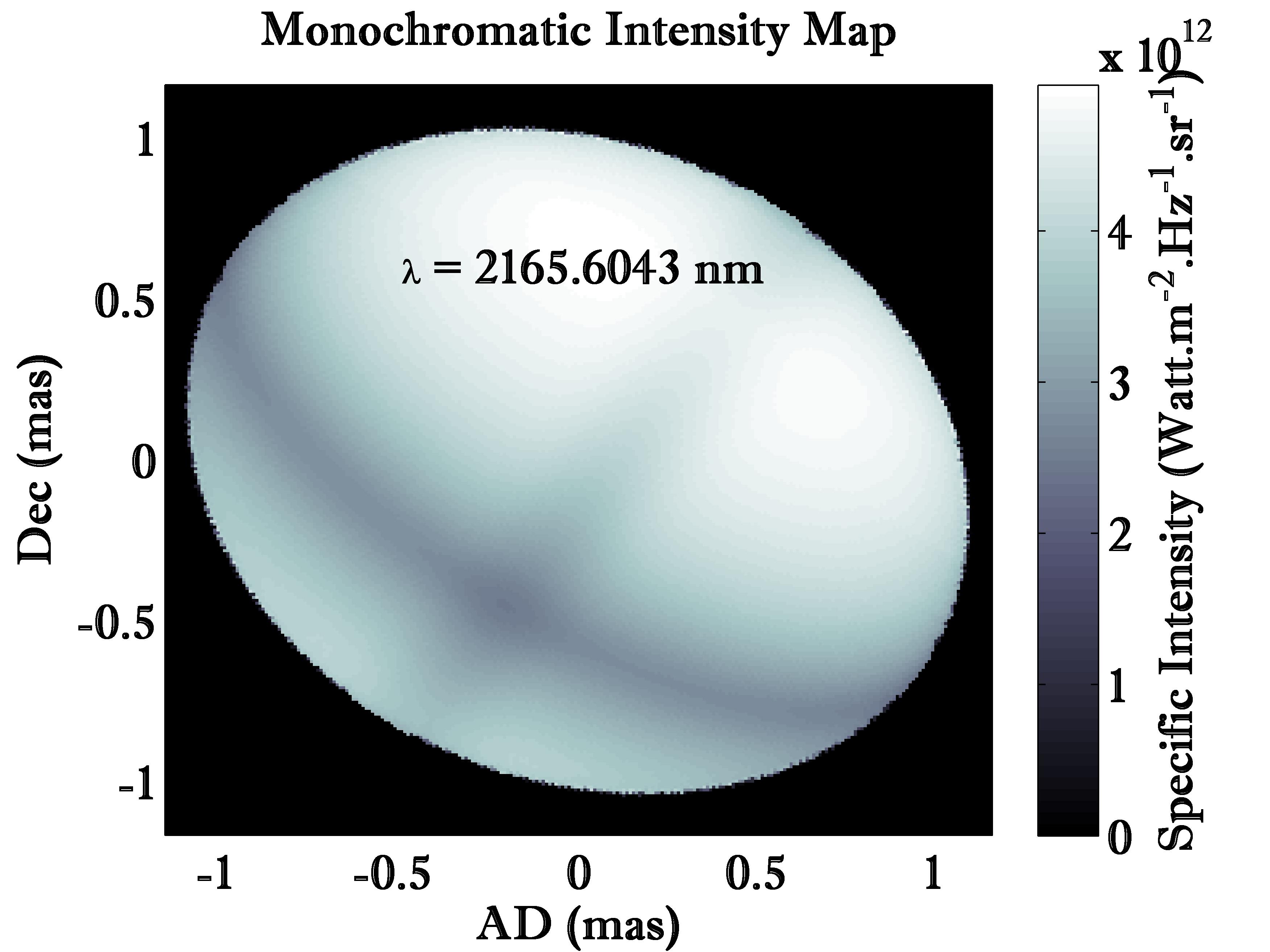}
\includegraphics[width=0.5\hsize,draft=false]{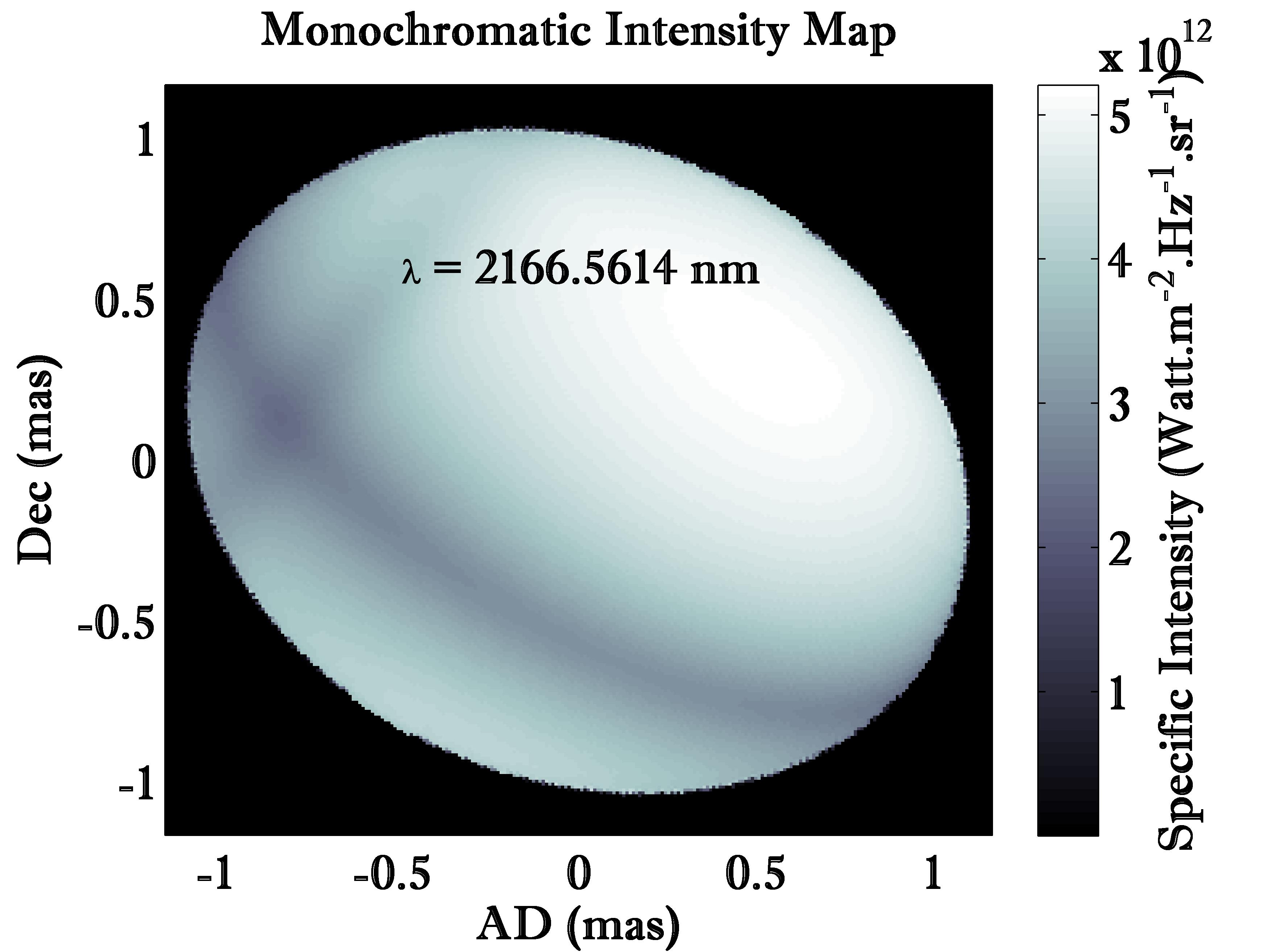}
\caption[Cartes d'intensités monochromatiques]{Cartes d'intensité monochromatique d'Achernar vues à différentes longueurs d'onde $\lambda$ autour de la raie Br$\gamma$.}\label{mono_Int}
\end{figure}

Enfin et concernant l'angle $PA_{rot}$ , défini comme étant l'angle de la projection de l'axe de rotation de l'étoile sur le plan référentiel de l'observateur (du Nord vers l'Est), il suffit juste d'incliner mes cartes selon cet angle. La Fig.\ref{mono_Int} représente la carte d'intensité monochromatique d'une étoile ayant la même configuration qu'Achernar (Voir l'article I ; \citet{2012A&A...545A.130D}), pour 3 longueurs d'onde $\lambda$ autour de la raie Br$\gamma$, où on voit le déplacement de la raie dans le sens de rotation de l'étoile. Notons ici, qu'il aurait été possible d'utiliser de manière plus physique les intensités spécifiques des profils de raie pour l'établissement des cartes d'intensités mais cela aurait été très couteux en temps de calcul, en plus du manque de flexibilité car faisant appel à un outil externe au code SIROCCO (Tlusty/Synspec).\\

\clearpage

\section{Les observables interférométriques}

Les observables qu'on peut directement obtenir des cartes d'intensité monochromatique sont, le flux intégré (le spectre) $F$:\\

\begin{equation}
F(\lambda)=\sum_y\sum_z I(\lambda,y,z)
\label{F}
\end{equation}

et les déplacements de photo-centres $P_y$ \& $P_z$ selon $y$ et $z$:\\

\begin{equation}
P_{y,z}(\lambda)=\frac{\sum_y\sum_z y,zI(\lambda,y,z)}{\sum_y\sum_z I(\lambda,y,z)}
\label{Pyz}
\end{equation}

\begin{figure}[ht]
\centering
\includegraphics[height=0.5\hsize,draft=false]{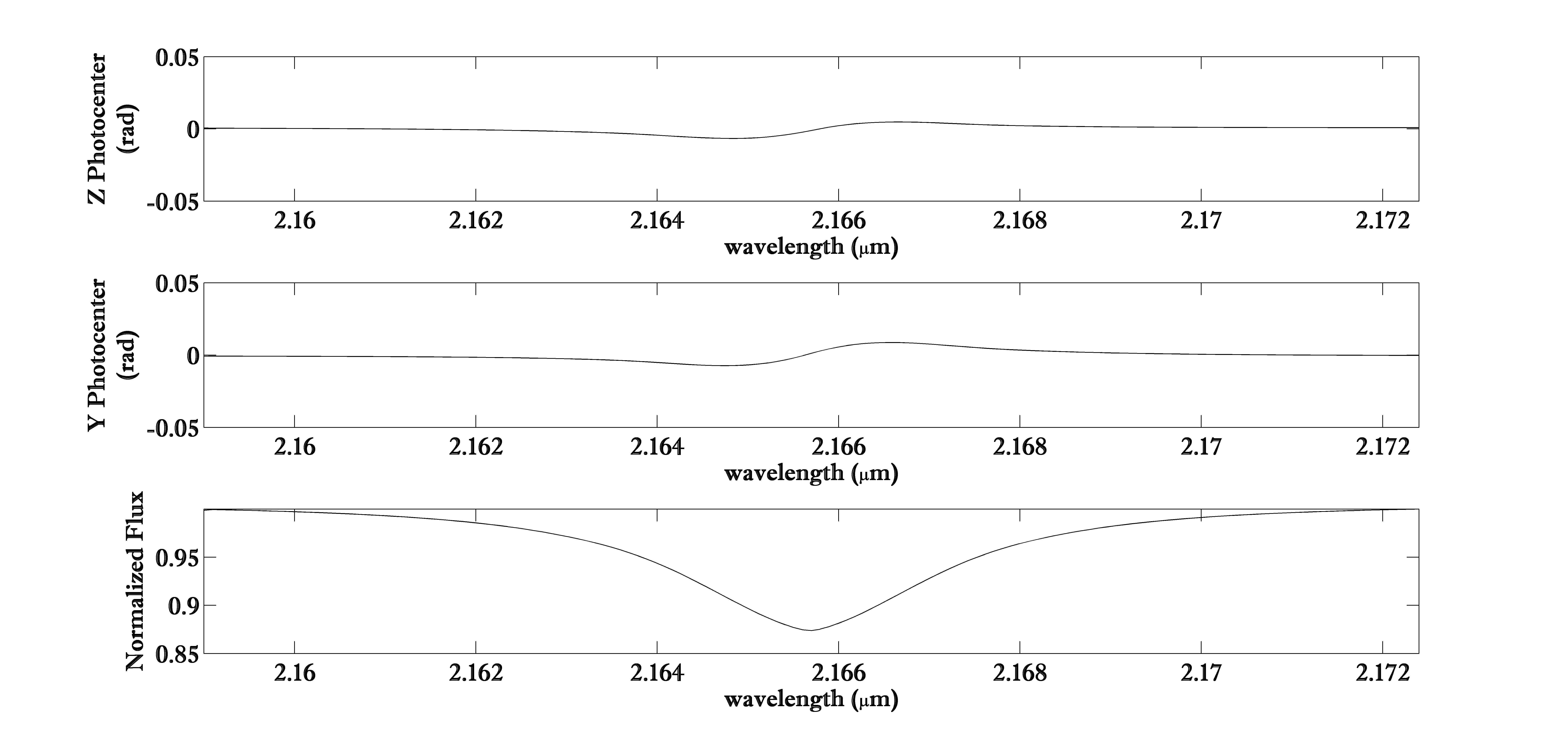}
\caption[Flux et déplacements photo-centriques d'Achernar]{Flux et déplacements photo-centriques d'Achernar en fonction de la longueur d'onde $\lambda$ autour de la raie Br$\gamma$.}\label{intefero-meus_Ach1}
\end{figure}

Un exemple de ces trois quantités mesurées pour l'étoile Achernar (aux paramètres cités dans notre papier I), est représenté dans la Fig.\ref{intefero-meus_Ach1}.\\

D'une manière indirecte dans l'espace de Fourier, on peut également déduire le module de visibilité $V^2(\lambda,u,v)$, la phase différentielle $\phidiff(\lambda,u,v)$ et la clôture de phase $\Psi(\lambda,u,v)$ (voir la section \ref{mesurables_DI}). Pour une précision optimale des valeurs de ces mesurables en tout point de fréquence spatiale (u,v), j'ai utilisé la transformation discrète de Fourier (DFT), qui s'écrit comme:\\
\begin{equation}
V(\lambda,u,v)=\sum\sum_{y,z}I(\lambda,y,z)e^{-2\pi(uy + vz)};
\label{DFT}
\end{equation}

\begin{figure}[ht]
\centering
\includegraphics[height=0.5\hsize,width=1.1\hsize,draft=false]{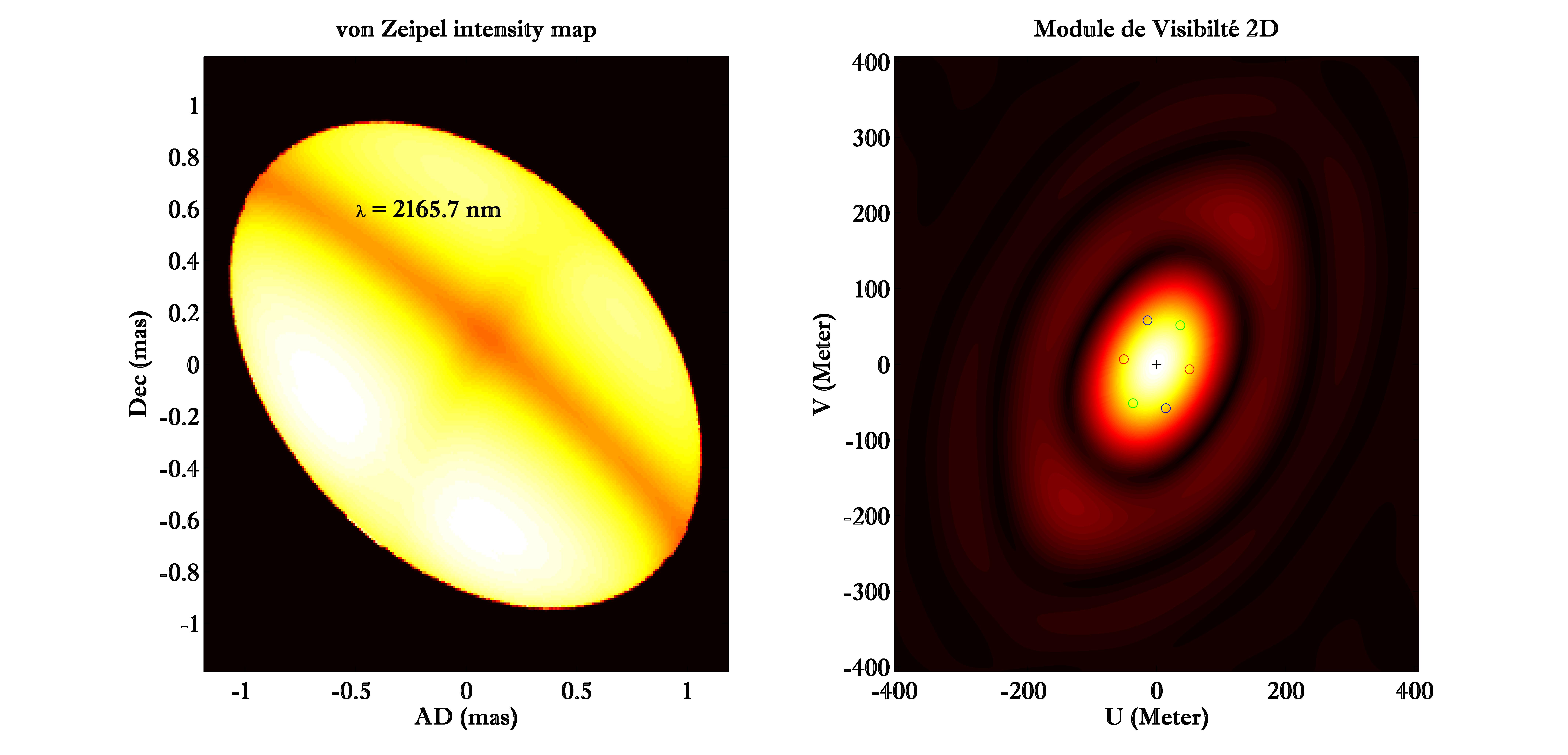}
\includegraphics[height=0.4\hsize,draft=false]{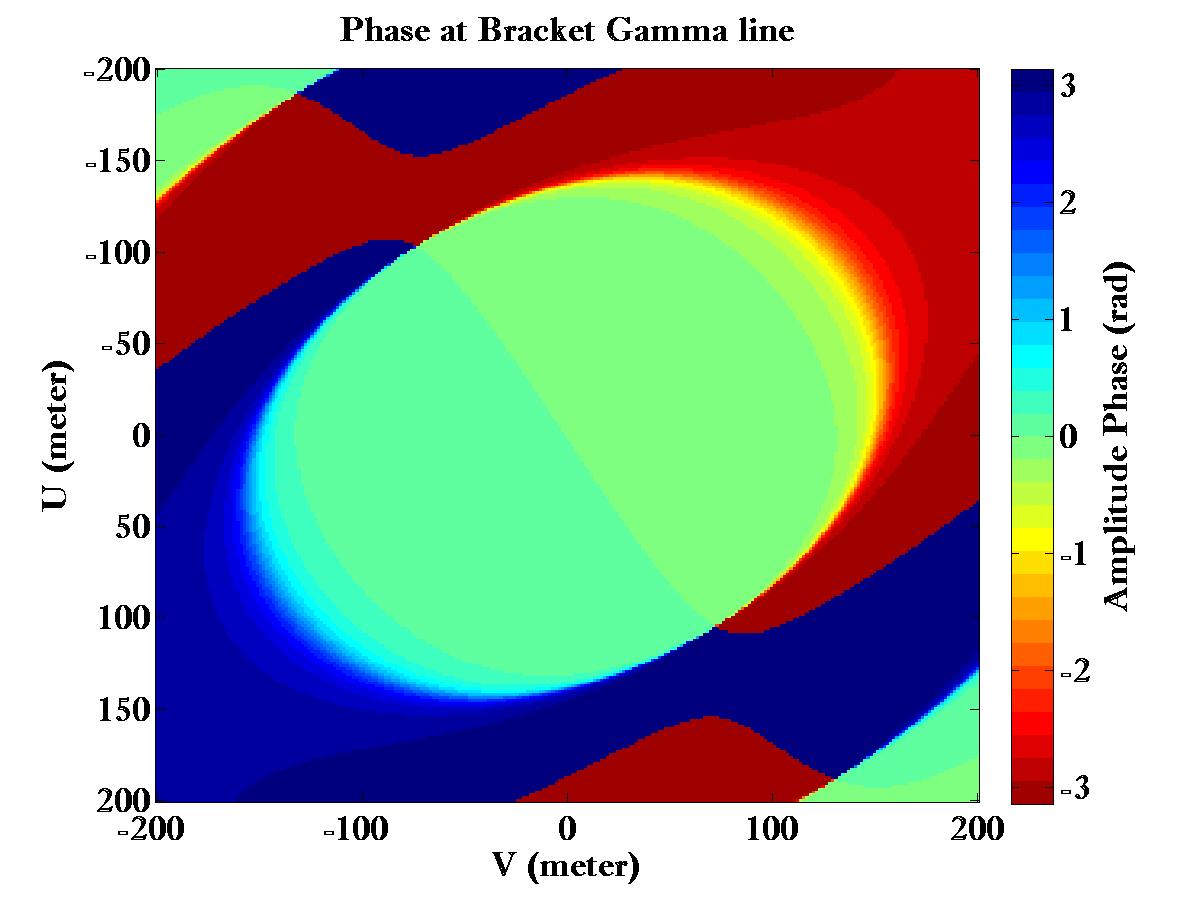}
\includegraphics[height=0.5\hsize,draft=false]{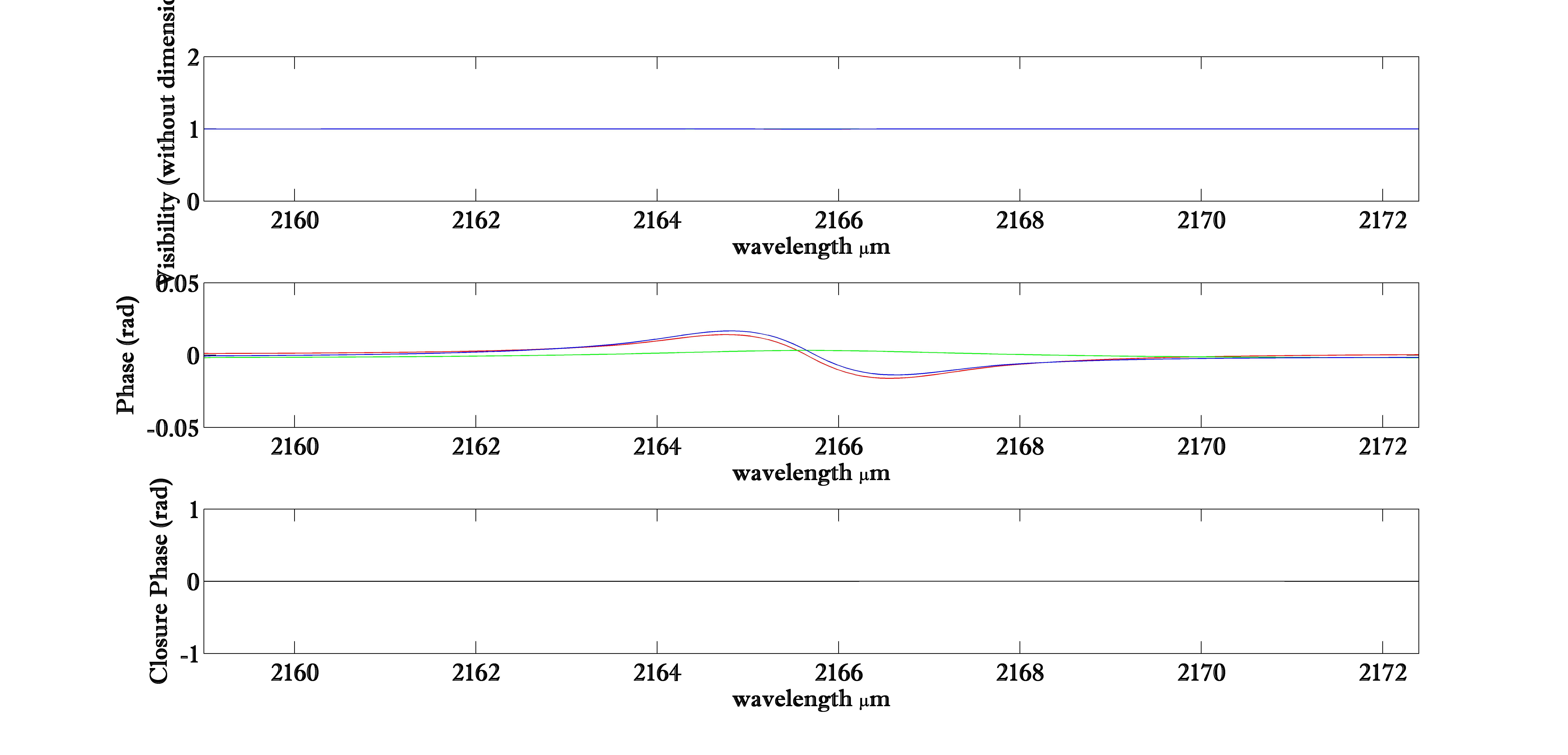}
\caption[Mesures interférométriques d'Achernar]{\textbf{En haut:} Représentation de la carte d'intensité d'Achernar et les cartes 2D du module de visibilité et des phases differentielles, le long de la raie Br$\gamma$. \textbf{En bas:} Les mesures interférométriques modélisées d'Achernar en fonction de la longueur d'onde $\lambda$ le long de la raie Br$\gamma$, à savoir, $V^2$, $\phidiff$ et $\Psi$ pour 3 coordonnées de bases différentes.}\label{intefero-meus_Ach2}
\end{figure}

Un exemple de ces mesures modélisées, pour Achernar (toujours aux paramètres cités dans notre article I), est représenté dans la Fig.\ref{intefero-meus_Ach2}.\\

\clearpage

\begin{figure}[hb]
\centering
\includegraphics[height=1.1\hsize,draft=false,angle=270]{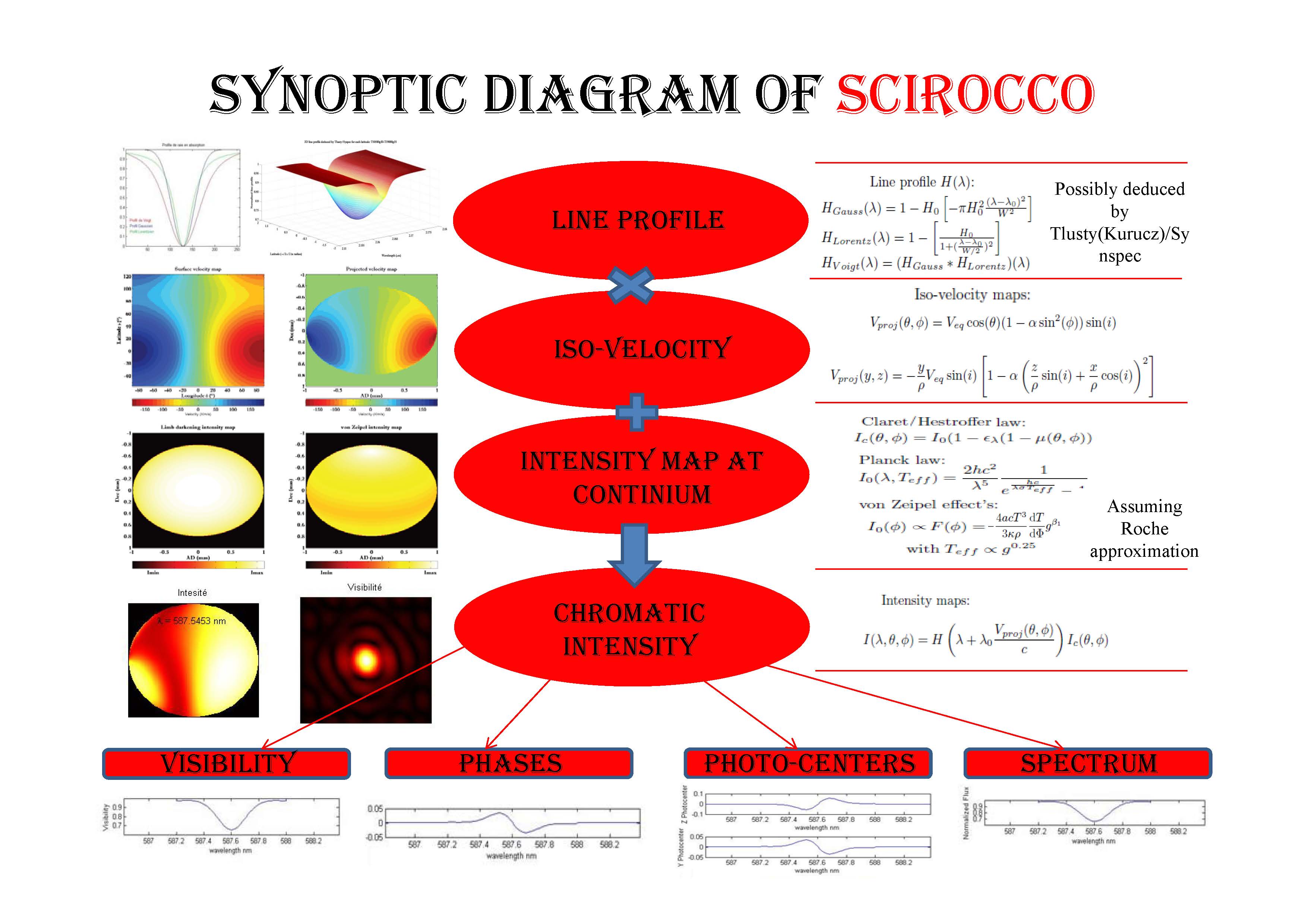}
\caption[Diagramme synoptique de SCIROCCO]{Diagramme synoptique de SCIROCCO.}\label{Syno_scirocco}
\end{figure}

La Fig.\ref{Syno_scirocco} représente de manière synoptique toutes les étapes et composantes de mon modèle SCIROCCO. Le résultat de ce travail, entrepris dès le début de 2012 a fait l'objet d'un poster présenté lors du colloque SF2A en Juin 2012 à Nice. Il est publié dans la contribution qui suit (juste après la Fig.\ref{Syno_scirocco}):\\

\clearpage

\includepdf[pages=1-6]{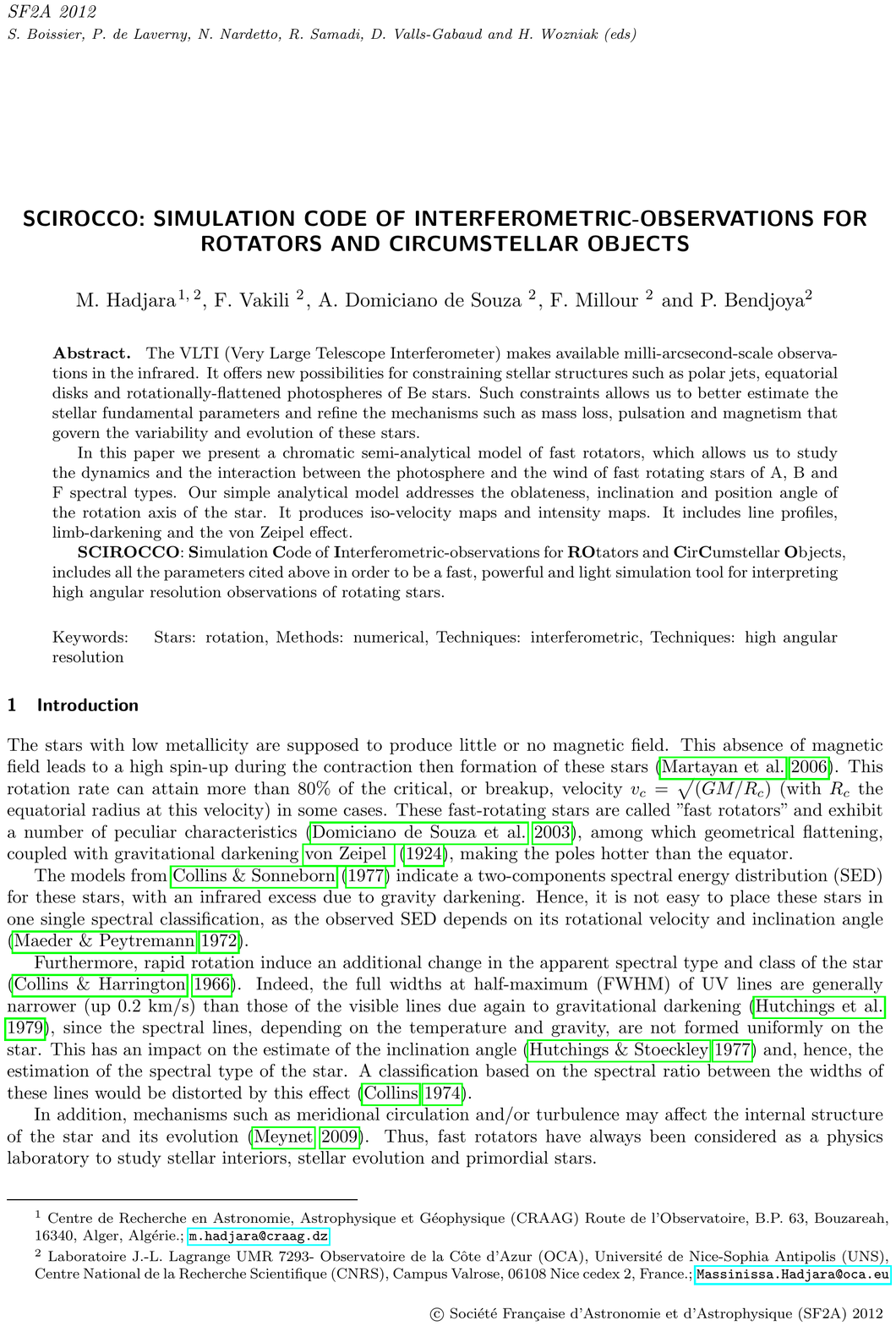}
\chapter{Application de SCIROCCO à la rotation stellaire}
\label{chap:appli}
\minitoc

%%%%%%%%%%%%%%%%%%%%%%%%%%%%%%%%%%%%%%%%
\def\vsini{v_\mathrm{eq} \sin i} 
\def\kms{\mathrm{km.s}^{-1}}
\def\phidiff{\phi_\mathrm{diff}}
\def\Rsun{\mathrm{R}_{\odot}}
\def\Lsun{\mathrm{L}_{\odot}}
\def\Msun{\mathrm{M}_{\odot}}
\def\Tmean{\overline{T}_\mathrm{eff}}
\def\diameq{\diameter_\mathrm{eq}}
\def\chir{\chi_\mathrm{r}}
\def\chimin{\chi_\mathrm{min}}
\def\chirmin{\chi_\mathrm{min,r}}
%%%%%%%%%%%%%%%%%%%%%%%%%%%%%%%%%%%%%%%%

Je présente dans ce chapitre l'étude menée sur l'impact de certains paramètres stellaires sur la phase différentielle $\phidiff$ simulée par SCIROCCO. Je décris ensuite les résultats basés sur l'utilisation de SCIROCCO, appliqué aux données interférométriques de 4 rotateurs rapides, c.à.d. Achernar, Altair, $\delta$ Aquilae et Fomalhaut.\\

\section{Impact de certains paramètres stellaires sur la $\phidiff$}\label{sec_5.1}

\begin{figure*}[ht]
\centering
\includegraphics[width=0.49\hsize,height=0.6\hsize,draft=false]{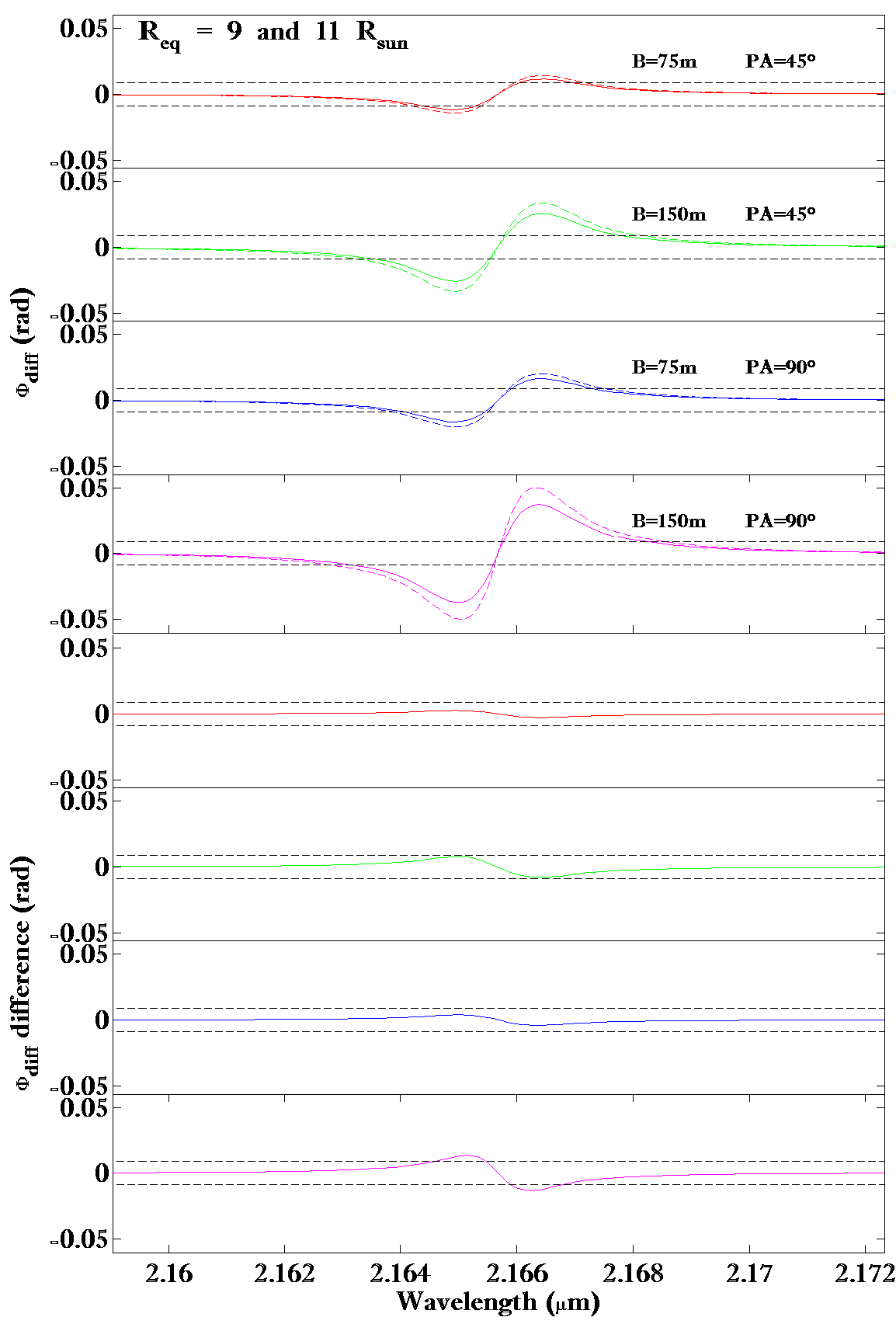}
\includegraphics[width=0.49\hsize,height=0.6\hsize,draft=false]{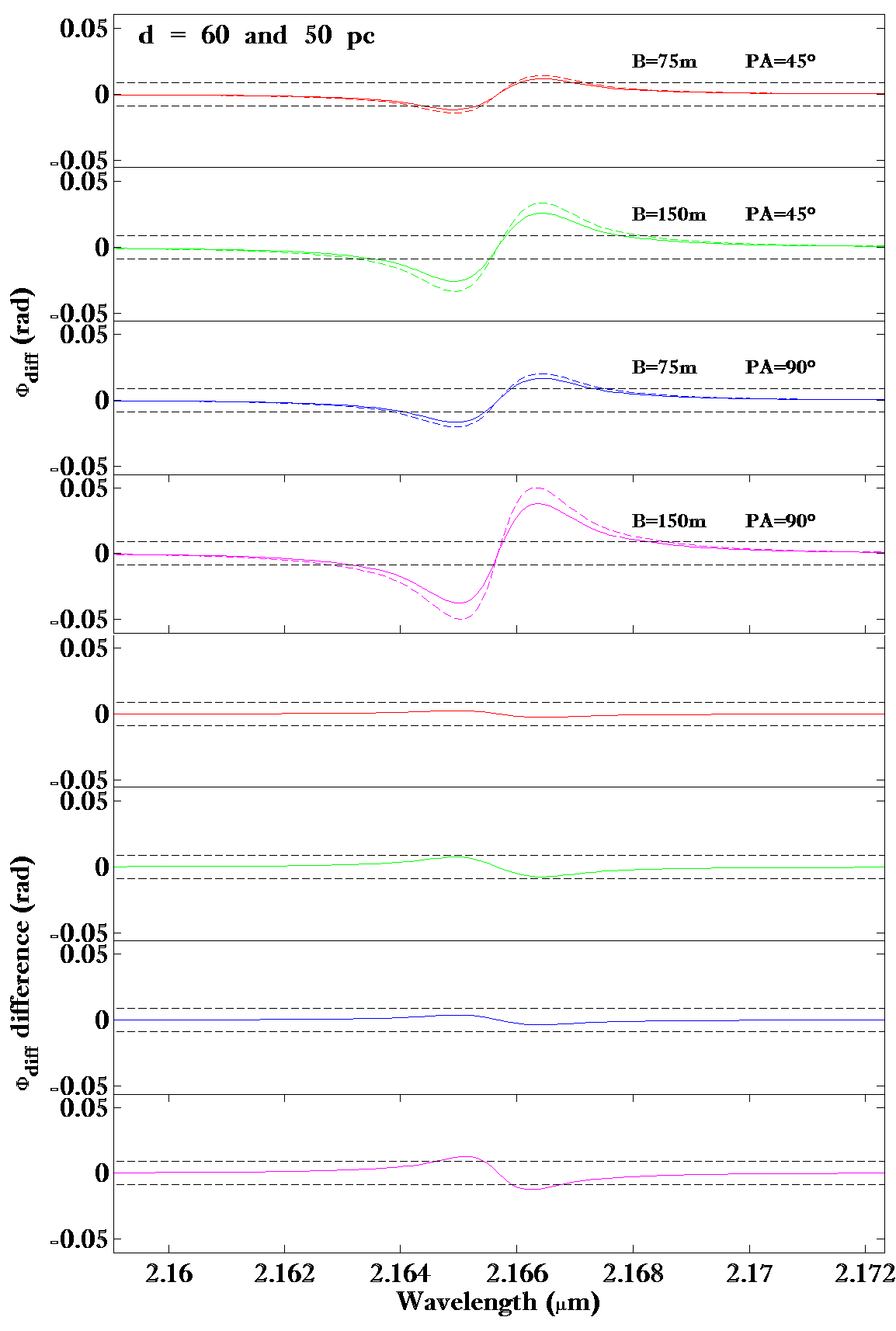}
\caption[Dépendance de phases différentielles simulées $\phidiff$ avec $R_{eq}$ et $d$]{\textbf{En haut}: La dépendance des phases différentielles $\phidiff$ simulées en $R_ {eq}$ et $d$ (ligne continue: modèle testé; ligne en pointillés: modèle de référence tel que décrit ci-dessus). Les paramètres du modèle testé sont identiques à ceux du modèle de référence, sauf que $R_{eq}=9\Rsun$ (à gauche) et $d=60pc$ (à droite). Les valeurs du modèle testé et celui de référence diffèrent de $18\%$ pour $R_{eq}$; ils sont indiqués dans la partie supérieure gauche des figures. Tous les modèles de $\phidiff$ ont été calculés autour de la raie $Br_\gamma$ avec une résolution spectrale de $12000$ pour quatre lignes de base projetées ($B_{proj}=75$ $m$ \& $150$ $m$ et $PA= 45^\circ$ et $90^\circ$), qui sont des valeurs typiques obtenues avec le VLTI/AMBER. Ces longueurs d'onde et lignes de base se traduisent par des amplitudes de visibilité entre $0,6$ et $0,9$ pour les modèles stellaires étudiés, ce qui correspond à une étoile partiellement résolue. Les lignes  horizontales en pointillés représentent la barre d'erreur typique de AMBER pour la $\phidiff$ ($\sigma_\phi=0.5^\circ=-0,0087$ $rad$). La signature de la $\phidiff$ d'une étoile en rotation apparaît comme une courbe en forme de S (ou de $\tild$), visible à l'intérieur de la raie $Br_\gamma$, avec une amplitude supérieure à $\sigma_\phi$ à toutes les lignes bases choisies. \textbf{En bas:} La différence entre les $\phidiff$ testées et le modèle de référence (courbe solide moins celle en pointillés de la partie supérieure). Le continuum de l'ensemble des courbes est lui égal à zéro.} \label{fig_model1}
\end{figure*}

\begin{figure*}[ht]
\centering
\includegraphics[width=0.49\hsize,height=0.6\hsize,draft=false]{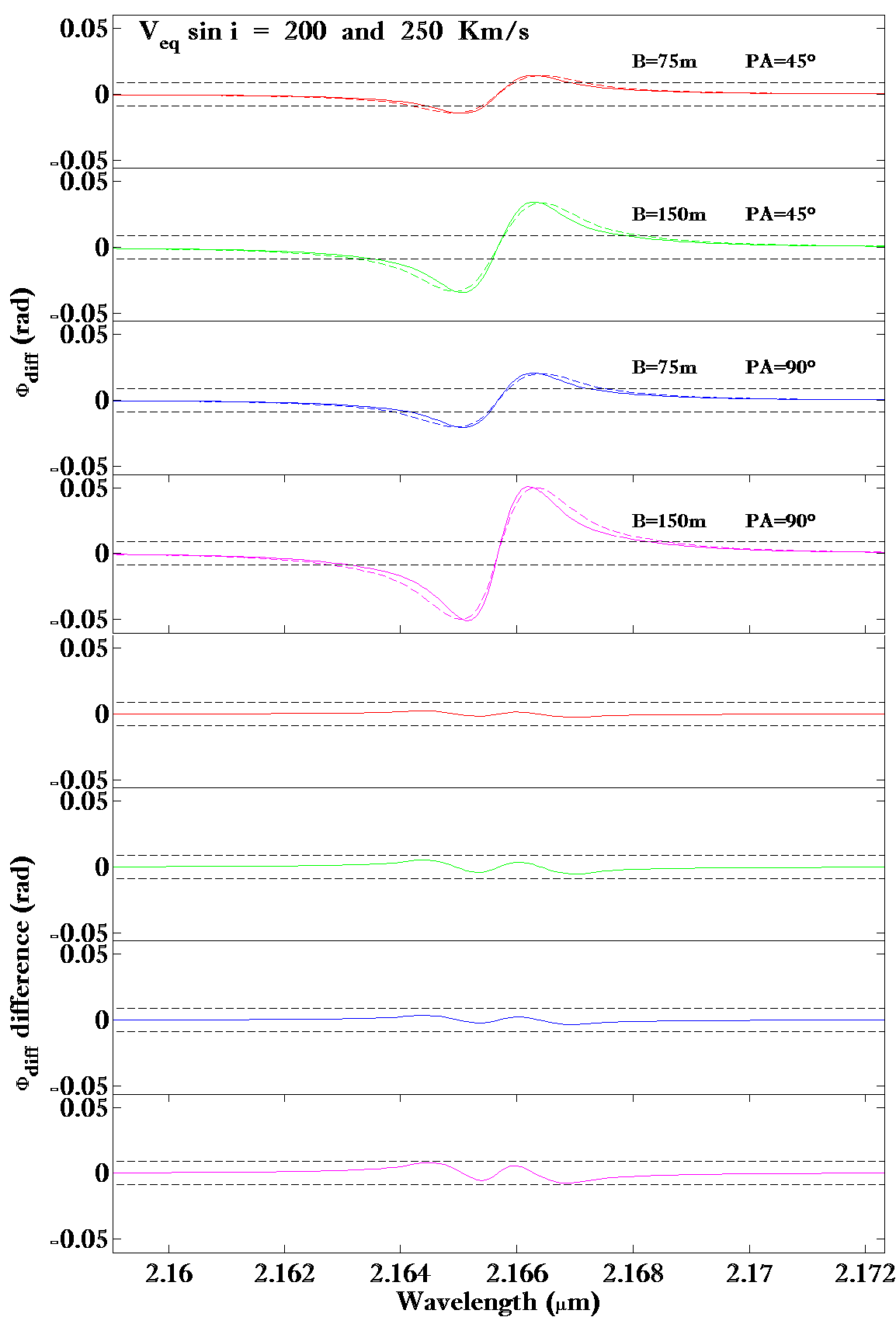}
\includegraphics[width=0.49\hsize,height=0.6\hsize,draft=false]{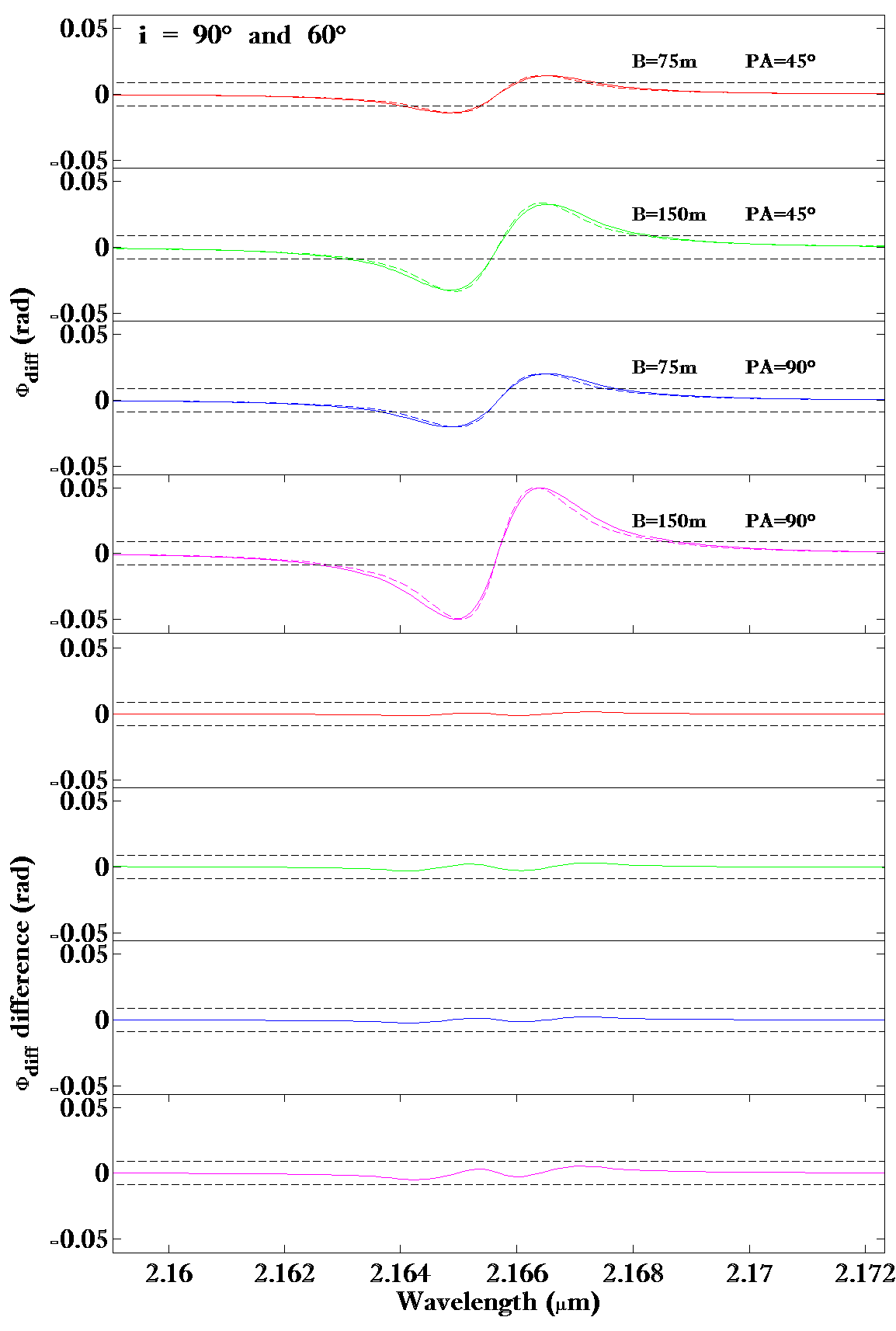}
\caption[Dépendance de phases différentielles simulées $\phidiff$ avec $\vsini$ et $i$]{Similaire à la Fig.\ref{fig_model1} mais pour la dépendance de la $\phidiff$ en $\vsini$ et $i$. La partie de gauche montre que $\phidiff$ en $Br_\gamma$ est sensible à une variation de $20\%$ du $\vsini (=200$ $\kms$ pour le modèle testé). La partie de droite montre quant à elle que $\phidiff$ dépend aussi de l'angle $i$ (évolution de $60^\circ$ à $90^\circ$), mais dans une moindre mesure par rapport à $R_{eq}$, $d$, et $\vsini$.} \label{fig_model2}
\end{figure*}

\begin{figure*}[ht]
\centering
\includegraphics[width=0.49\hsize,height=0.6\hsize,draft=false]{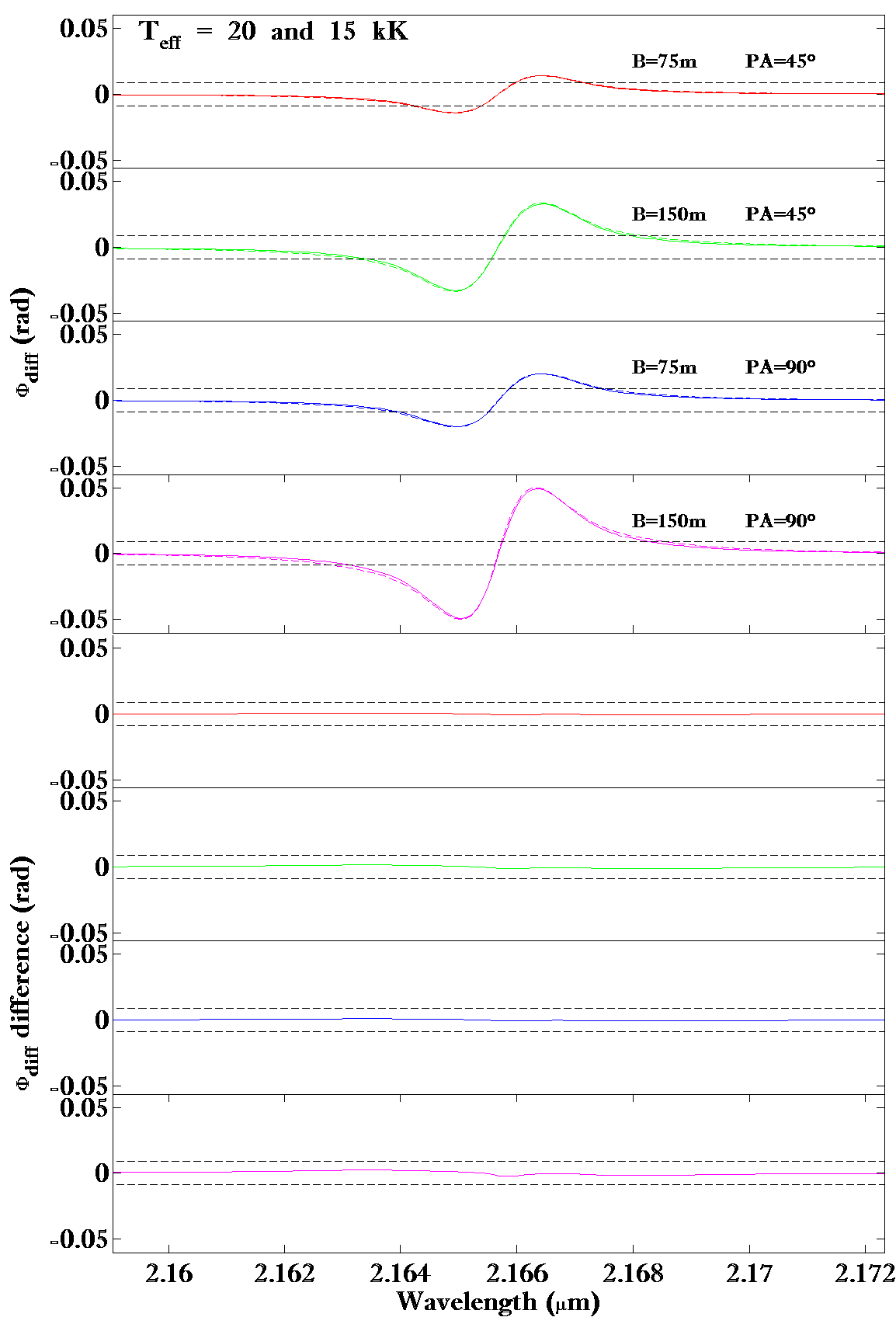}
\includegraphics[width=0.49\hsize,height=0.6\hsize,draft=false]{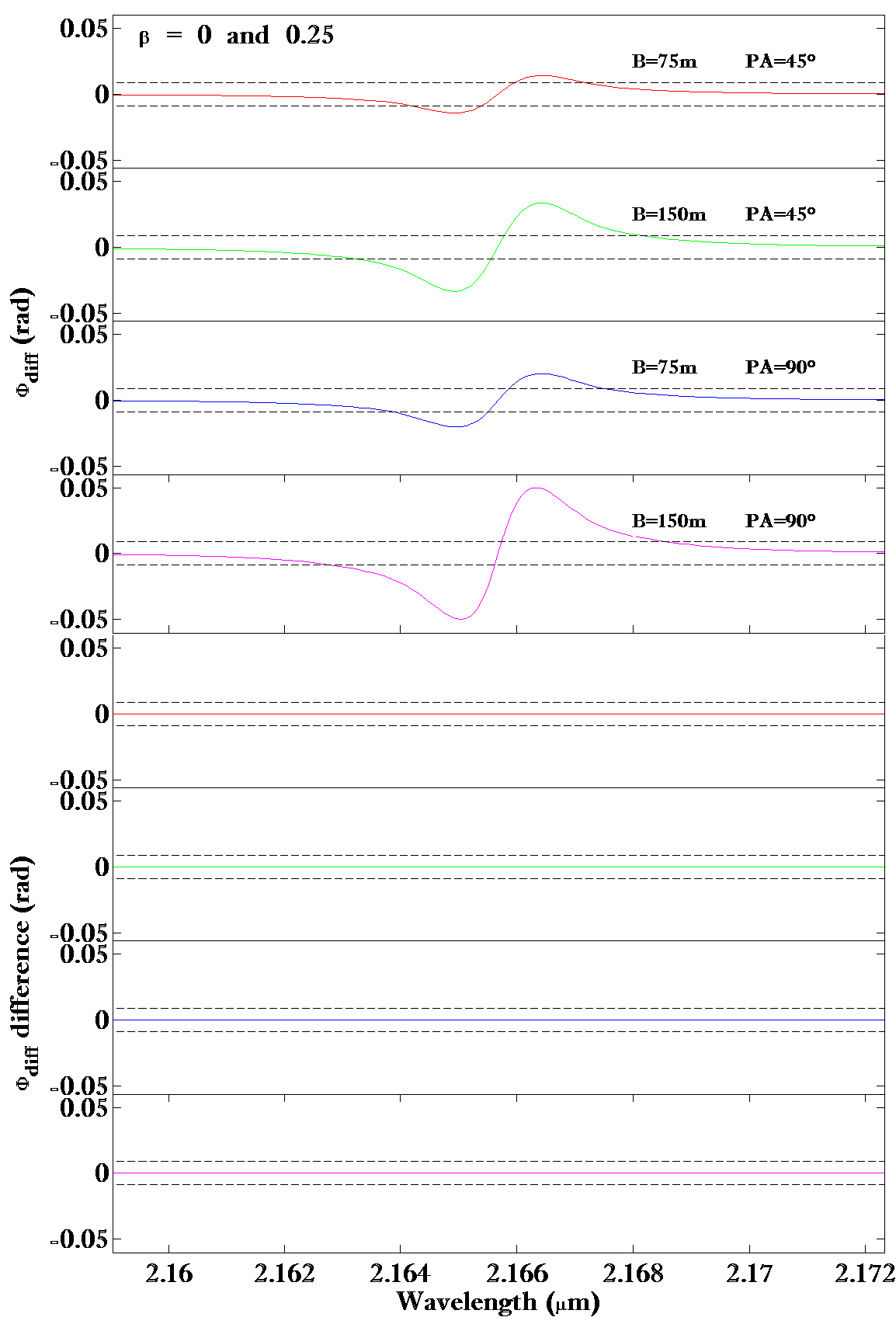}
\caption[Dépendance de phases différentielles simulées $\phidiff$ avec $\Tmean$ et $\beta$]{Similaire à la Fig.\ref{fig_model1} mais pour la dépendance de la $\phidiff $ en $\Tmean$ et $\beta$. $\Tmean$ passe de $15000$ $K$ à $20000$ $K$, et $\beta$ varie de $0.25$ (valeur de von Zeipel) à $0.0$ (pas d'assombrissement gravitationnel). Les simulations SCIROCCO de $\phidiff$ autour de la raie $Br_\gamma$ sont peu sensible à ces paramètres et voire pas du tout à $\beta$.} \label{fig_model3}
\end{figure*}

\begin{figure*}[ht]
\centering
\includegraphics[width=0.49\hsize,height=0.6\hsize,draft=false]{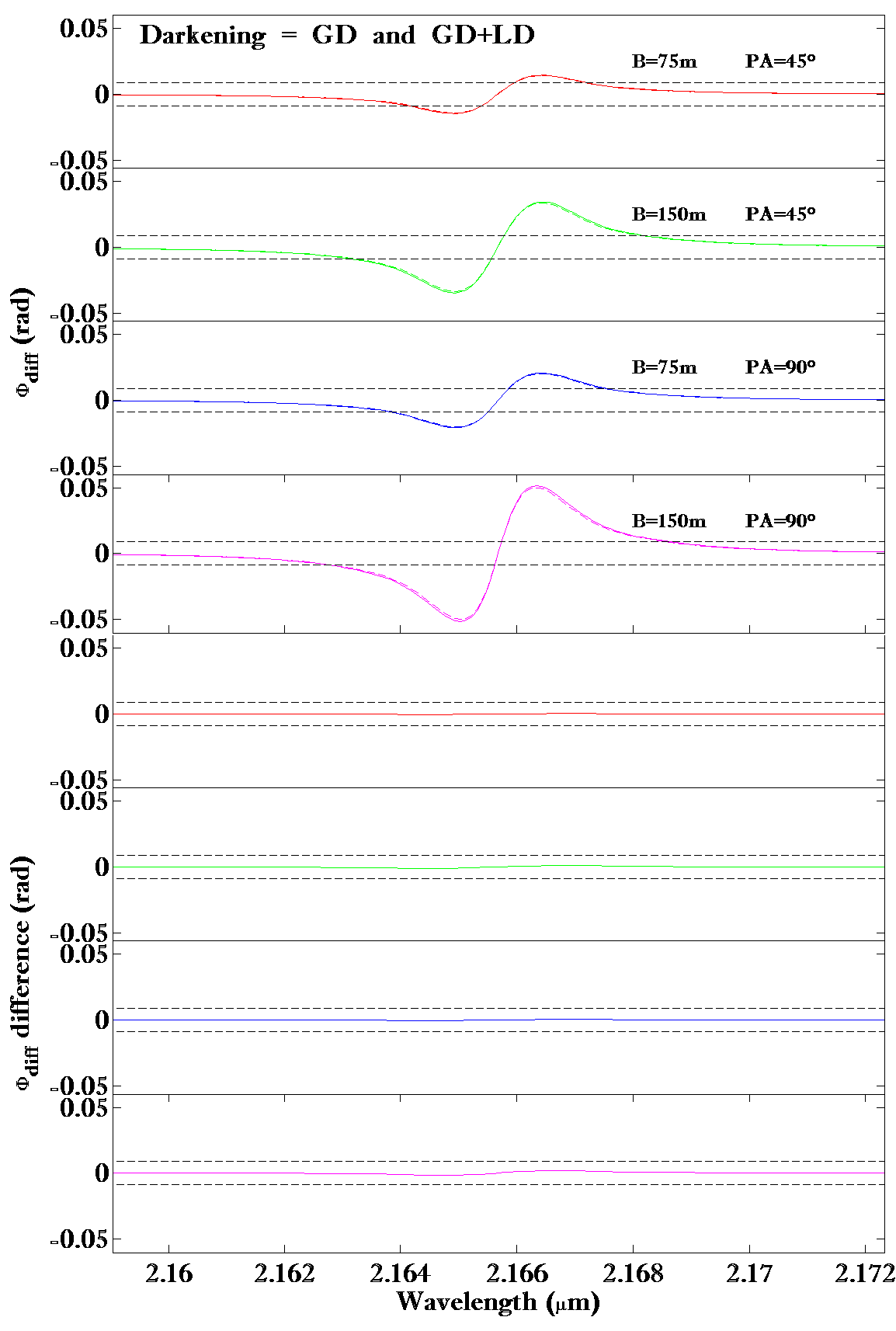}
\includegraphics[width=0.49\hsize,height=0.6\hsize,draft=false]{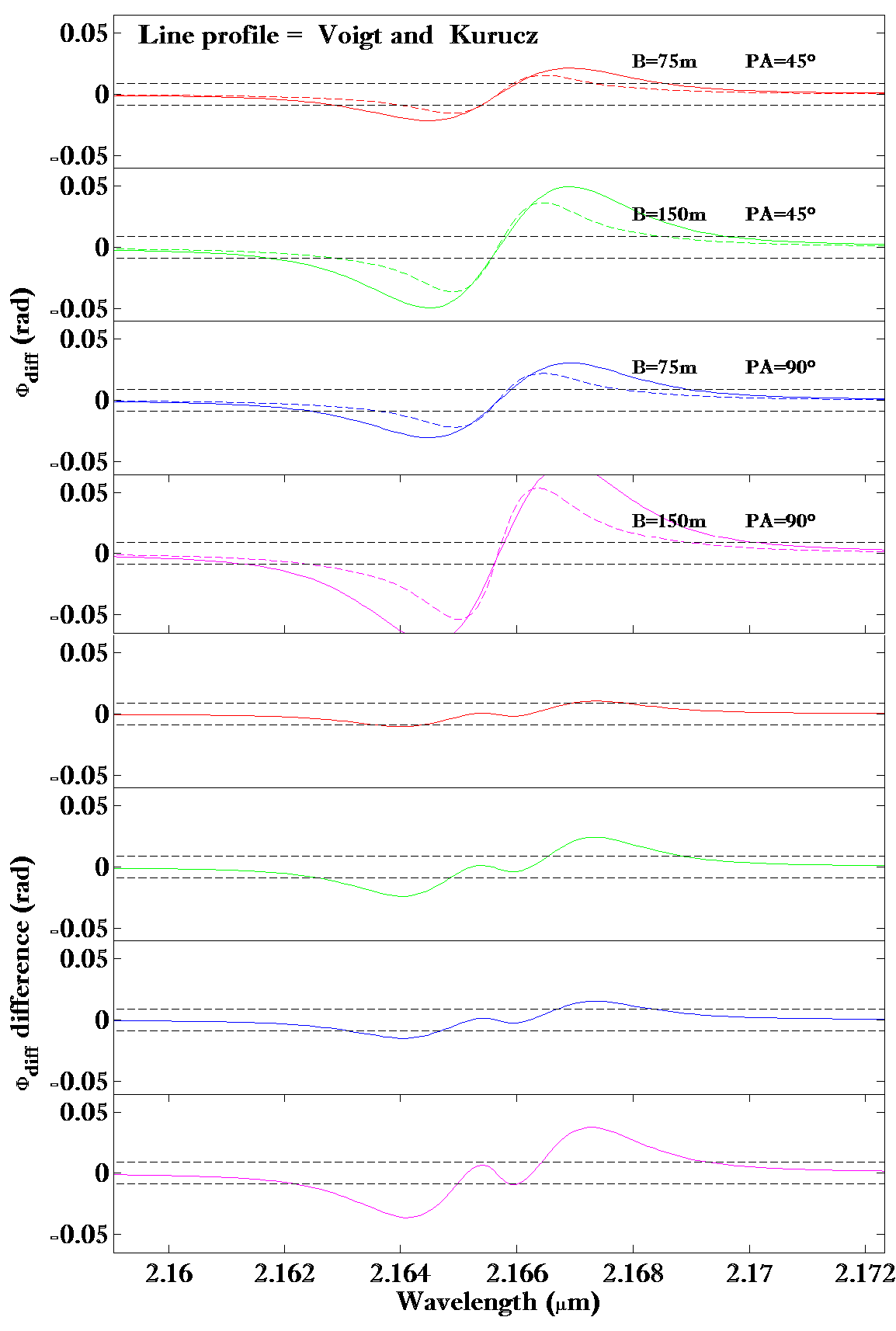}
\caption[Dépendance de phases différentielles simulées $\phidiff$ avec l'assombrissement]{Similaire à la Fig.\ref{fig_model1} mais pour la dépendance de $\phidiff$ à ``l'effet d'assombrissement'' (trait continu: l'effet de l'assombrissement gravitationel uniquement  -pas d'effet d'assombrissement centre-bord-; lignes en pointillés: le modèle de référence tel que décrit plus haut -assombrissement gravitationnel et assombrissement centre-bord-), et la dépendance de $\phidiff$ au ``type de profil de raie'' (trait continu: profil de raie analytique de Voigt, lignes en pointillés: profil de raie obtenu avec Kurucz/Synspec). La $\phidiff $ à $Br_\gamma$ n'est pas fortement sensible à l'effet d'assombrissement mais est fortement sensible au profil de raie utilisé.} \label{fig_model4}
\end{figure*}

\begin{figure*}[ht]
\centering
\includegraphics[width=0.49\hsize,height=0.6\hsize,draft=false]{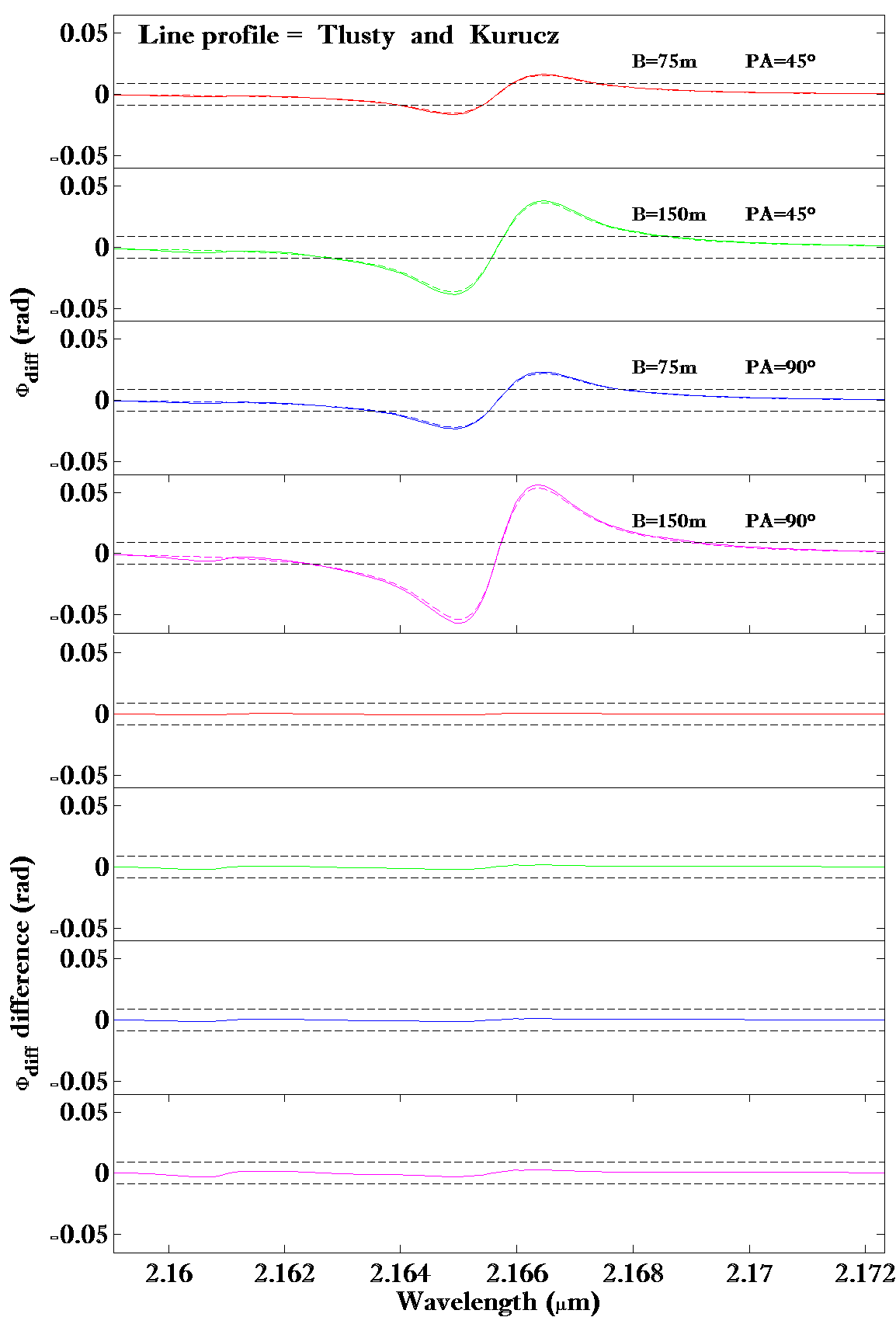}
\includegraphics[width=0.49\hsize,height=0.6\hsize,draft=false]{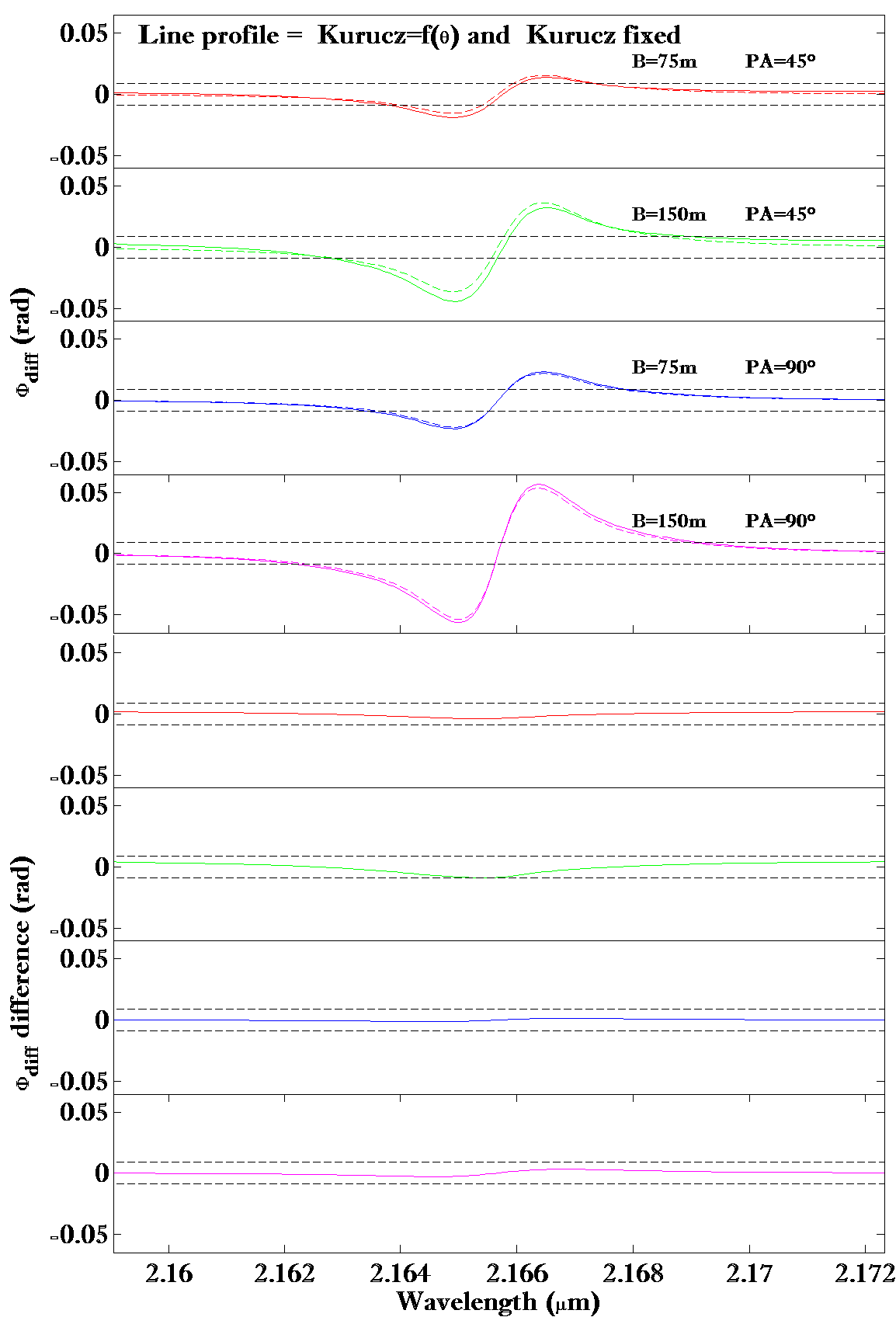}
\caption[Dépendance de phases différentielles simulées $\phidiff$ avec le profil de raie]{Similaire à la Fig.\ref{fig_model1} mais pour la dépendance de $\phidiff$ au ``type de profil de raie'' (trait continu: profil de raie obtenu avec Tlusty/Synspec, ligne en pointillés: profil de raie obtenu avec Kurucz/Synspec) et ``le type de profil de raie'' en fonction de la latitude (ligne continue: le profil de raie obtenu avec un profil de raie Krucz/Synspec  qui varie en fonction de la latitude, lignes en pointillés: ligne profil obtenu avec un profil de raie fixe Kurucz/Synspec -en considérant la moyenne de $[T_ {eff},\log g]$ de l'étoile). La $\phidiff$ en $Br_\gamma$ n'est pas fortement sensible à l'effet de l'assombrissement mais est fortement sensible au type de profil de raie utilisé. La $\phidiff$ en $Br_\gamma$ n'est pas fortement sensible aux profils de raie utilisés, qu'ils soient fixes ou variables en fonction de la latitude $\theta$.}\label{fig_model5}
\end{figure*}

Les paramètres étudiés ici sont:
\begin{itemize}
\item[-] Le rayon équatorial $R_{eq}$ et la distance $d$ (Fig.\ref{fig_model1}),
\item[-]Le $\vsini$ et l'angle d'inclinaison $i$ (Fig.\ref{fig_model2}), 
\item[-] La température effective effective $\Tmean$ et le coefficient d'assombrissement gravitationnel $\beta$ (Fig.\ref{fig_model3}),
\item[-] L'impact des deux assombrissements "Gravity Darkening" (GD) et/ou "Limb Darkening" (LD) (Fig.\ref{fig_model4} de gauche). 
\item[-] Et enfin l'important impact du profil du raie ; Voigt, Krurucz, Tlusty, qu'il soit fixe ou bien variable en fonction de la latitude $\theta$ (Fig.\ref{fig_model4} de droite \& Fig.\ref{fig_model5}). 
\end{itemize}

La simulation est menée sur une étoile hypothétique proche d'Achernar, aux mêmes valeurs que celles prises dans \citet{2012A&A...545A.130D} et \citet{2014A&A...569A..45H}, c.à.d pour:

\begin{itemize}
\item[-] $R_{\rm eq} = 11$ $\Rsun$,
\item[-] $d = 50$ $pc$,
\item[-] $\vsini = 250$ $\kms$,
\item[-] $i = 60^\circ$,
\item[-] $M = 6.1$ $\Msun$,
\item[-] $\Tmean = 15000$ $K$,
\item[-] $PA_{\rm rot} = 0^\circ$,
\item[-] $\beta = 0.25$, 
\item[-] Pas de  rotation différentielle, 
\item[-] Avec l'effet des deux assombrissements "gravity darkening" \& "limb darkening", 
\item[-] Et un profil de raie Kurucz/Synspec. 
\end{itemize}

Pour 4 configurations interférométriques à deux longueurs de base $B_{proj}=75$ $m$ \& $150$ $m$ et deux angles de projection $PA=45^\circ$ \& $90^\circ$, l'influence de chaque paramètre étudié sur la $\phidiff$ à la raie Br$\gamma$ est clairement et succinctement discutée dans l'entête de chaque figure ci-dessous.\\

\clearpage

Il est important de noter que l'étude de l'influence du profil de raie sur les mesures interférométriques en général et sur la $\phidiff$ autour de $Br_\gamma$ en particulier, m'ont permis d'une part de démontrer l'impact important de ce paramètre, et aussi l'impossibilité d'utiliser un simple profil analytique dans ce genre de simulation (tel que c'était envisagé au départ pour SCIROCO). En effet, la confrontation de mon code avec les observations via une méthode d'ajustement particulière (explicitée dans la section-ci-après) à 4 paramètres libres, à savoir $R_{eq}$, $v_{eq}$, $i$ et $PA_{rot}$, a révélé une forte dépendance des paramètres dynamiques $R_{eq}$ et $v_{eq}$ (qui sont d'ailleurs fortement corrélés et liés -voir Eq.\eqref{4.5}-). Ces derniers sont fortement sensibles au type de profil de raie (où j'ai d'ailleurs relevé une différence d'environ 10\% entre un profil de raie de Voigt analytique et un autre issu de Kurucz/Synspec -voir Fig.\ref{fig_profil7}-), alors que les paramètres géométriques liés à l'orientation ($i$ et $PA_{rot}$) eux restaient insensibles à ce type de changement.\\

\clearpage

\section{Détermination des paramètres fondamentaux stellaires via SCIROCCO}
Tout le travail de réduction/traitement des données interférométriques AMBER, présenté dans l'application du Chap.\ref{chap:spec-interfero}, ainsi que l'ensemble de simulations SCIROCCO (voir Chap.\ref{chap:scirocco} pour les 4 rotateurs rapides Achernar, Altair, $\delta$ Aquilae \& Fomalhaut, ont fini par aboutir, récemment (juin 2014), à un papier A\&A, dont je suis le premier auteur, et que je joins ci-après.  La confrontation de mon modèle avec les observations interférométriques des 4 étoiles, citées ci-haut, a été réalisée grâce à la méthode d'ajustement de minimisation du $\chi^2$ en utilisant un code Matlab à accès libre (à partir de site d'échange de Matlab ; Matlab central file exchange) que j'ai modifié/adapté selon mes besoins: le code nommé " The Generalized Nonlinear Nonanalytic Chi-Square Fitting code, developped by N. Brahms" a été développé par N. Brahms (de Université de Berkeley, Californie USA). Ce code qui effectue l'ajustement via la minimisation du $\chi^2$ avec estimation de l'incertitude sur des erreurs de mesures connues, peut utiliser plusieurs librairies Matlab, comme l'algorithme de convergence rapide de Levenberg-Marquardt (LM), tout comme il utilise aussi la méthode Monte-Carlo sur l'ensemble des données afin de s'assurer du calcul optimal sur les incertitudes. Ainsi j'ai utilisé ce code afin d'ajuster les phases différentielles simulées aux observations et à contraindre les paramètres libres ($R_{eq}$, $v_{eq}$, $i$, et $PA_{rot}$) et leurs incertitudes respectives. La formule du $\chi^2$ que j'utilise peut être écrite comme suit:\\

\begin{equation}
\chi^2_{\phidiff}(R_{eq},v_{eq},i,PA_{rot})=\sum_i\frac{[\phi_\mathrm{diff,i}-\phi_\mathrm{diff,SCIROCCO}(u_i,v_i)]^2}{\sigma_i^2}
\label{xi2}
\end{equation}

Où $\sigma_i$ est l'incertitude (ou variance) réciproque à chaque $\phi_\mathrm{diff,i}$ observée. En pratique, et pour traiter toutes les données simultanément, je mets toutes les $N_{obs}$ $\phidiff$ (toute base confondue) à la suite, les unes derrières les autres pour un ajustement optimal. Ne travaillant que sur 4 paramètres libres ($R_{eq}$, $v_{eq}$, $i$, et $PA_{rot}$), le degré de liberté (Degree Of Freedom) est donc $D.O.F=4$, ce qui me permet d'enfin exprimer $\chi^2_{red}$ ainsi:\\

\begin{equation}
\chi^2_{red}=\frac{\chi^2_{\phidiff}}{N_{obs}-D.O.F}
\label{xi2red}
\end{equation}

\begin{figure}[ht]
\centering
\includegraphics[width=0.8\hsize,draft=false]{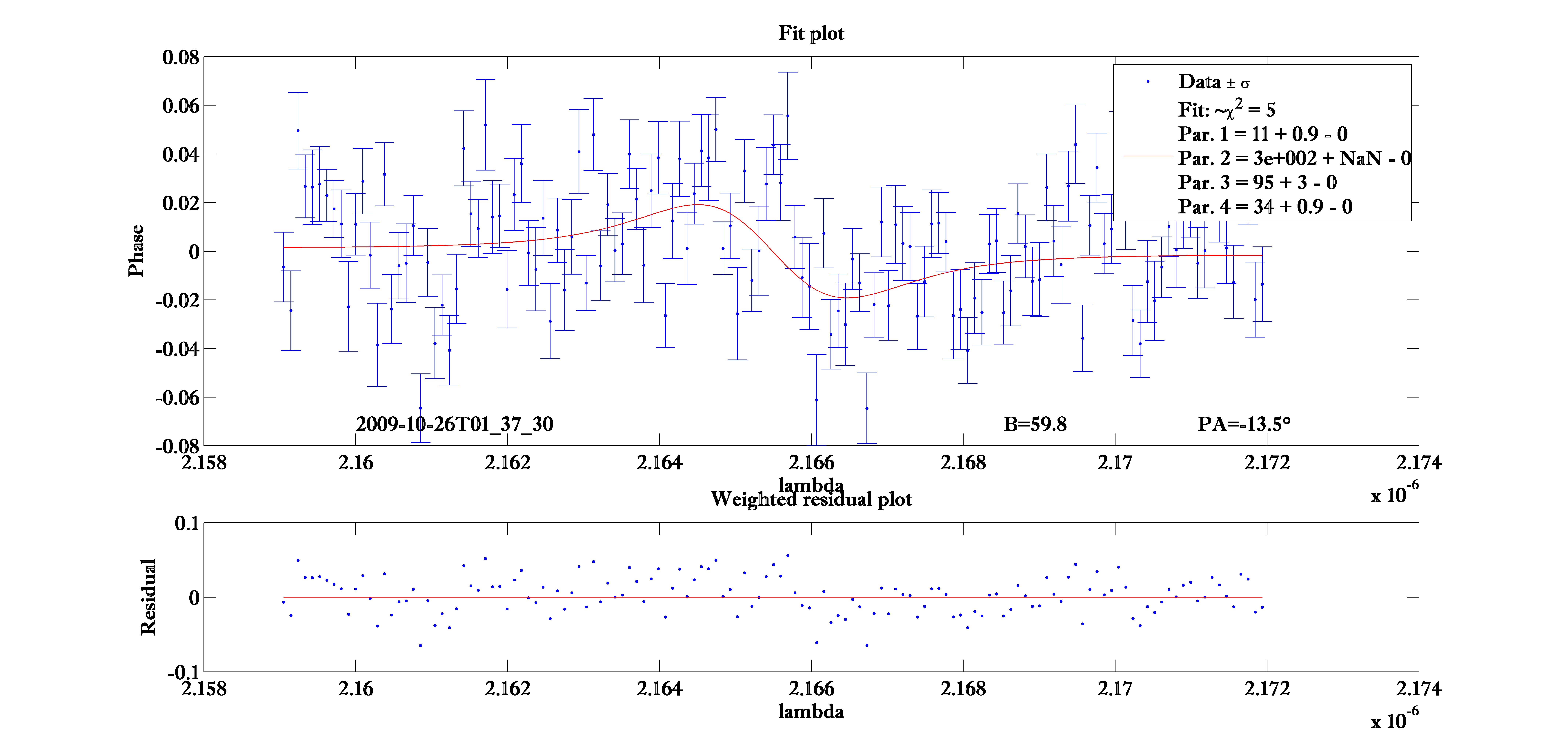}
\caption[Exemple d'ajustement $\chi^2$ sur une seule $\phidiff$]{Exemple d'ajustement $\chi^2$ sur une seule $\phidiff$ représentée en barres d'erreur en bleu pour la $\phidiff$ observée, et en ligne rouge continue pour celle modélisée. En dessous la représentation des résidus (la différence entre la $\phidiff$ simulée et observée, longueur d'onde par longueur d'onde).}\label{exm_xi2}
\end{figure}

Ci-dessous, la Fig.\ref{exm_xi2} montre un petit exemple d'ajustement $\chi^2$ sur une seule $\phidiff$ via le code que j'utilise.\\

\clearpage

\begin{figure}[ht]
\centering
\includegraphics[width=0.6\hsize,draft=false]{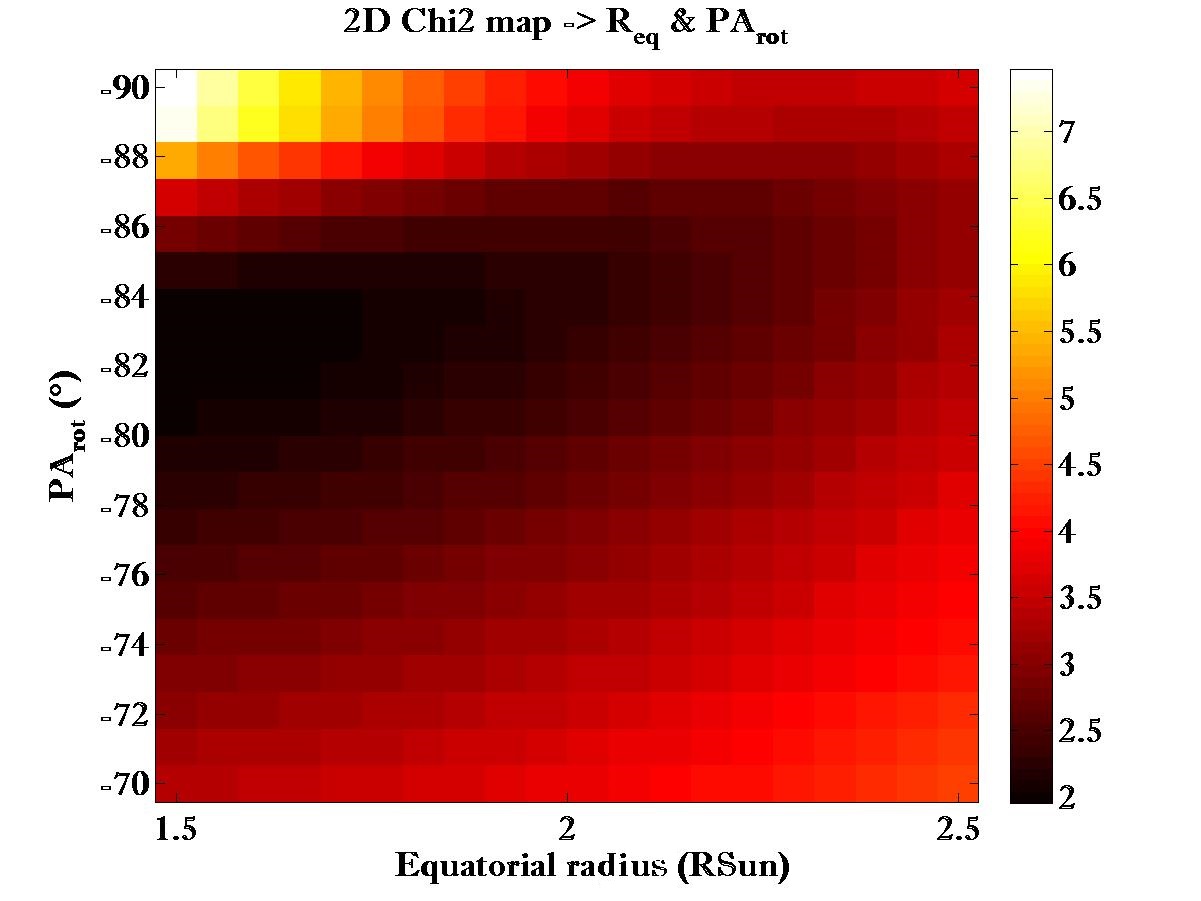}
\caption[Exemple de carte 2D $\chi^2$]{Exemple de carte 2D de $\chi^2$ ajusté sur des $\phidiff$ d'Achernar, avec un $v_{eq}$ et un $i$ fixés respectivement à $282$ $\kms$ et  $57^\circ$ et une résolution de 128x128 pix, pour une variation de $PA_{rot}$ et $ R_{eq}$ sur 20 
points chacun. On observe clairement le minimum du $\chi^2$ ici (la partie sombre en haut à gauche).}\label{exm_xi2_2D}
\end{figure}

Tout ce travail sur le $\chi^2$ me permet aussi de tracer des cartes 2D à deux paramètres libres, afin de mieux voir la dépendance des paramètres variables entre eux. Ci-dessous, exemple (Fig.\ref{exm_xi2_2D}) de carte $\chi^2$ à paramètres $v_{eq}$ \& $i$ fixes et à paramètres $PA_{rot}$ \& $R_{eq}$ libres, pour des simulations de $\phidiff$ sur  l'étoile Achernar.\\

Il faut rappeler que j'ai décidé d'utiliser uniquement la $\phidiff$ dans mes ajustements, car et comme j'ai l'ai expliqué à la fin du Chap.\ref{chap:spec-interfero}, cette mesure interférométrique reste celle qui détient le plus d'informations sur l'étoile, particulièrement quand celle-ci est peu résolue. Ainsi, ci-après est présentée ma principale contribution dans le cadre de mon travail de thèse, ce qui est aussi mon premier papier A\&A en tant que premier auteur.\\

\includepdf[pages=1-14]{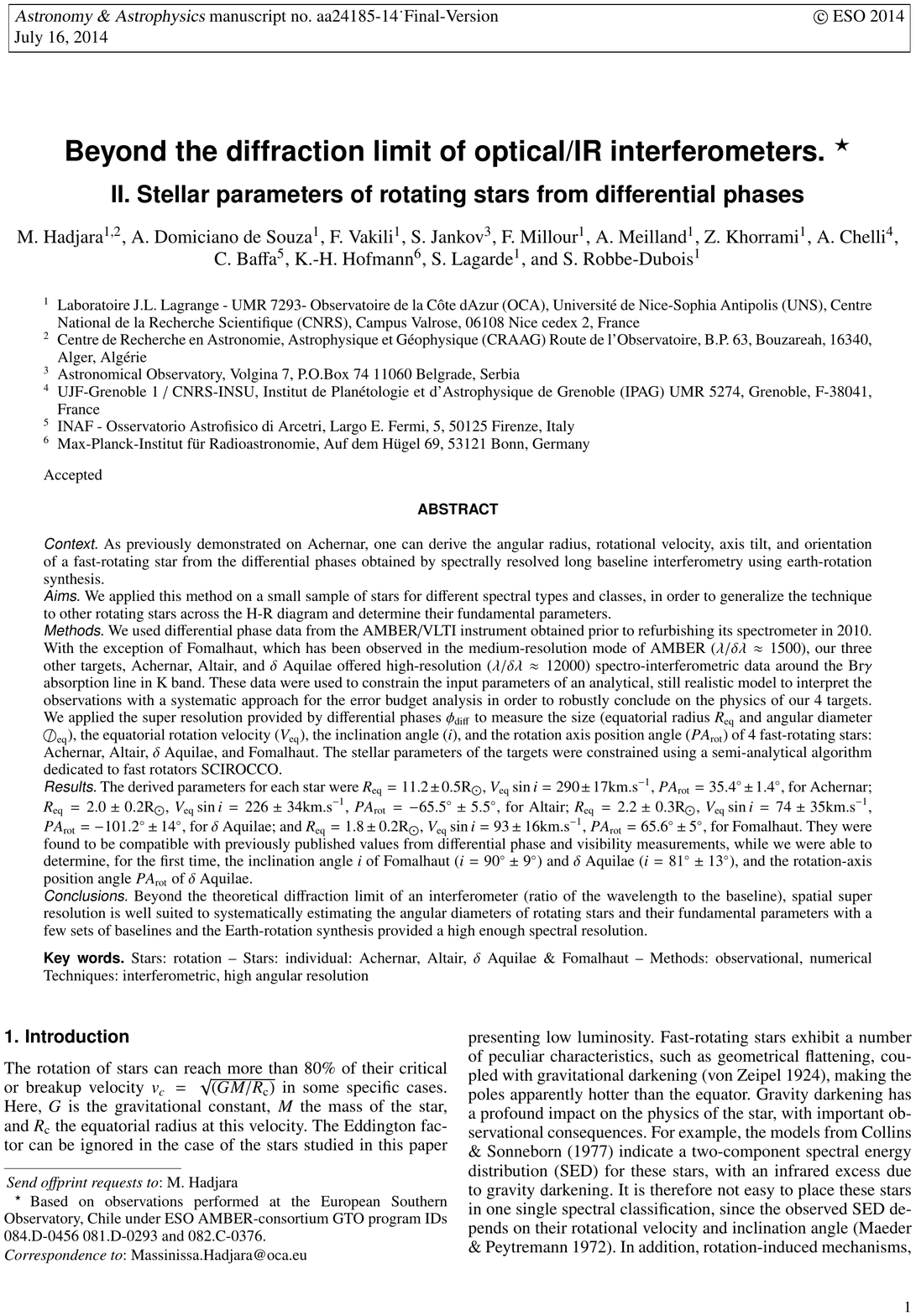}

\begin{figure}[ht!]
\centering
\includegraphics[width=0.6\hsize,draft=false]{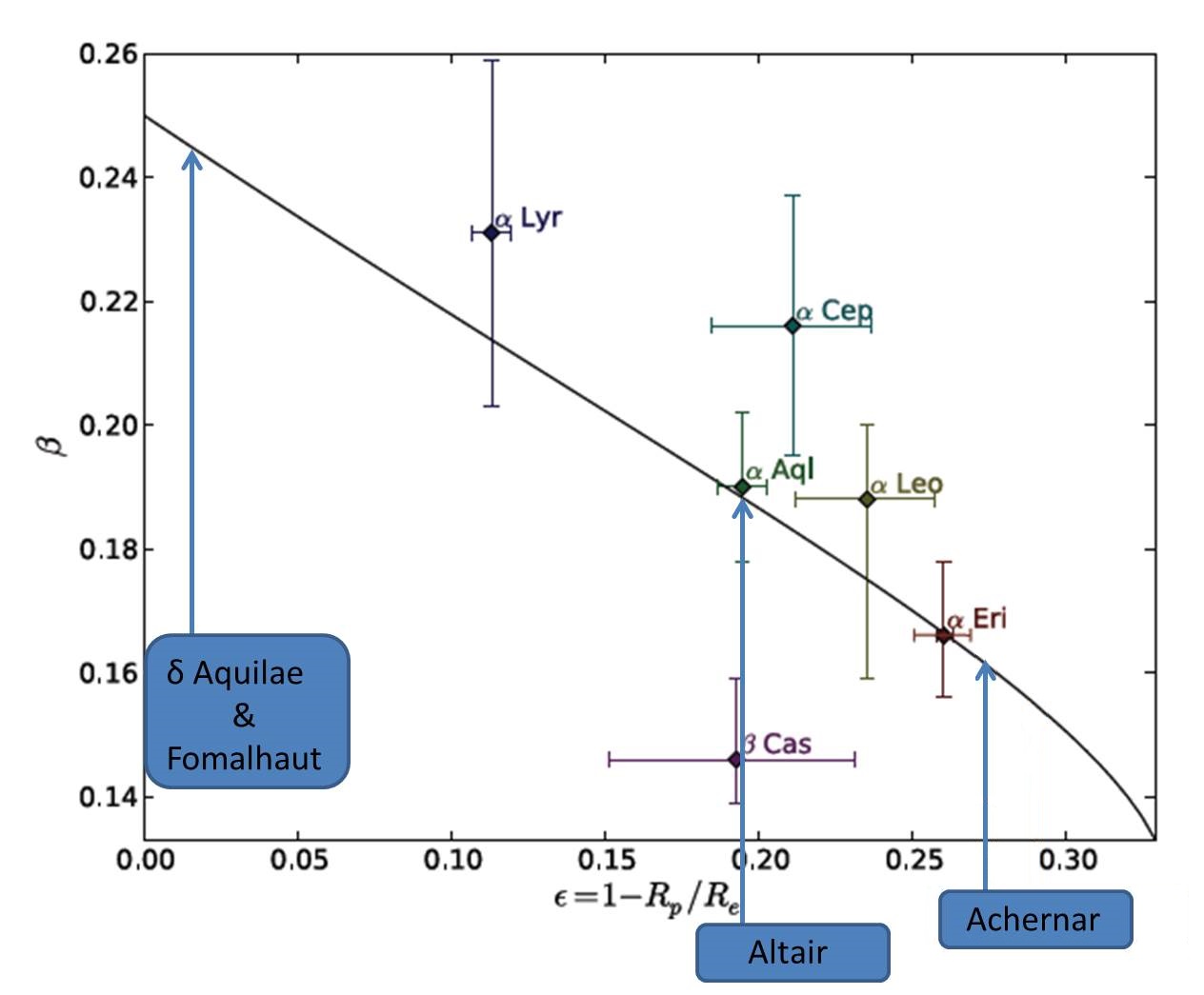}
\caption[Nos 4 étoiles sur le diagramme de Rieutord]{Représentation de l'emplacement de nos 4 étoiles étudiées sur la Fig\ref{beta_Rieutord}.}\label{comp_pap}
\end{figure}

Comme complément à cet article je tiens à rajouter que d'après la méthode de Rieutord \citep{2011A&A...533A..43E}, qui lie $\beta$ à la valeur de $1-\frac{R_{pol}}{R_{eq}}$ (i.e. à son aplatissement -voir Chap.\ref{chap:scirocco}-), les valeurs de $\beta$ peuvent êtres déduites pour chacune des 4 étoiles étudiées, ainsi que leur représentation dans la Fig.\ref{comp_pap}.\\

\begin{itemize}
\item[] Pour Achernar: $\frac{R_{eq}}{R_{pol}}=1.42$, $1-\frac{R_{pol}}{R_{eq}}=0.2957$ $\rightarrow$ $\beta=0.16$.
\item[] Pour Altair: $\frac{R_{eq}}{R_{pol}}=1.22$, $1-\frac{R_{pol}}{R_{eq}}=0.1803$ $\rightarrow$ $\beta=0.19$.
\item[] Pour $\delta$ Aquilae: $\frac{R_{eq}}{R_{pol}}=1.22$, $1-\frac{R_{pol}}{R_{eq}}=0.0196$ $\rightarrow$ $\beta=0.24$.
\item[] Pour Fomalhaut: $\frac{R_{eq}}{R_{pol}}=1.22$, $1-\frac{R_{pol}}{R_{eq}}=0.0196$ $\rightarrow$ $\beta=0.24$.
\end{itemize}

Tout ce qui a été présenté jusqu'à présent ne concernant que la modélisation et l'étude des rotateurs rapides. Dans le chapitre suivant (Chap.\ref{chap:scriocco-pot}), je présente des études primaires SCIROCCO impliquant d'autres phénomènes astrophysiques tels que : les pulsations non radiales (PNR), les disques circumstellaires, et les taches stellaires (ou transites d'exoplanète).
\chapter{SCIROCCO : un code à usages multiples}
%\begin{figure}[h!]
%\centering
% \includegraphics[height=0.5\hsize,draft=false]{Chapitre4/SCIROCCO}
%\end{figure}
%\label{Sec_4.6.1}
\label{chap:scriocco-pot}
\minitoc

%%%%%%%%%%%%%%%%%%%%%%%%%%%%%%%%%%%%%%%%
\def\vsini{v_\mathrm{eq} \sin i} 
\def\kms{\mathrm{km.s}^{-1}}
\def\phidiff{\phi_\mathrm{diff}}
\def\Rsun{\mathrm{R}_{\odot}}
\def\Lsun{\mathrm{L}_{\odot}}
\def\Msun{\mathrm{M}_{\odot}}
\def\Tmean{\overline{T}_\mathrm{eff}}
\def\diameq{\diameter_\mathrm{eq}}
\def\chir{\chi_\mathrm{r}}
\def\chimin{\chi_\mathrm{min}}
\def\chirmin{\chi_\mathrm{min,r}}
%%%%%%%%%%%%%%%%%%%%%%%%%%%%%%%%%%%%%%%%

%\section{SCIROCCO : un code à usages multiples}\label{Sec_4.6.1}

Mon modèle, qui ne portait pas encore de nom début 2012, a été conçu à la base pour l'étude des rotateurs rapides. Le hasard a voulu, lors d'un échange avec R. Petrov et S. Jankov, me faisant part de leur intention de soumettre une demande GTO d'observation VLTI/AMBER consacrée à l'étude des pulsations non radiales (PNR) de l'étoile Be $\eta$ Cen, de voir s'il y avait possibilité de détecter ces oscillations par interférométrie via AMBER. Ces collègues m'ont suggéré d'intégrer à mon code l'aspect des PNR pour pouvoir l'inclure dans la demande d'observation qui devait être soumise quelques jours plus tard. Trouvant l'idée séduisante et beaucoup plus par curiosité scientifique, j'entrepris d'inclure le phénomène des PNR dans mes modélisations. Les résultats de mon étude ainsi que mon nom furent inclus dans leur proposition d'observation, qui avait reçu un avis favorable de la part de l'ESO sur AMBER. Ce qui m'a valu une année plus tard la position de principal observateur, avec un beau voyage à Paranal d'initiation à l'observation sur le fantastique instrument que représente le VLTI dans les Andes Chiliennes. La PNR ayant été bien débattue dans le chapitre \ref{chap:rota}, je vais résumer en quelques lignes, dans cette section, tout le travail qui a été entrepris avec SCIROCCO.\\

\section{Simulation des Pulsations Non Radiales (PNR)}

La simulation du phénomène des PNR ainsi que leur possible détection en interférométrie était un point auquel les chercheurs se sont intéressés assez rapidement dès le début des années 90 \citep{1991ESOC...36...77V, 2001A&A...377..721J}. Pour ma part je me suis contenté de rajouter l'effet de la pulsation sur mes cartes d'iso-vitesses radiales et ses conséquences sur la carte d'intensité dans le continuum. Ainsi, pour la carte des iso-vitesses je n'ai eu qu'à rajouter la carte des vitesses $v_{pnr}(\theta,\phi)$ décrites par les harmoniques sphériques via la fonction associée de Legendre. Ce travail a fait l'objet d'un article dans une présentation EAS (European Astronomical Society) lors de ma participation à l'école de reconstruction d'images - applications astrophysiques, organisée par l'OCA du 18 juin au 22 juin 2012 à Fréjus (France). La version du papier est jointe ci-dessous:\\

\clearpage

\includepdf[pages=1-12]{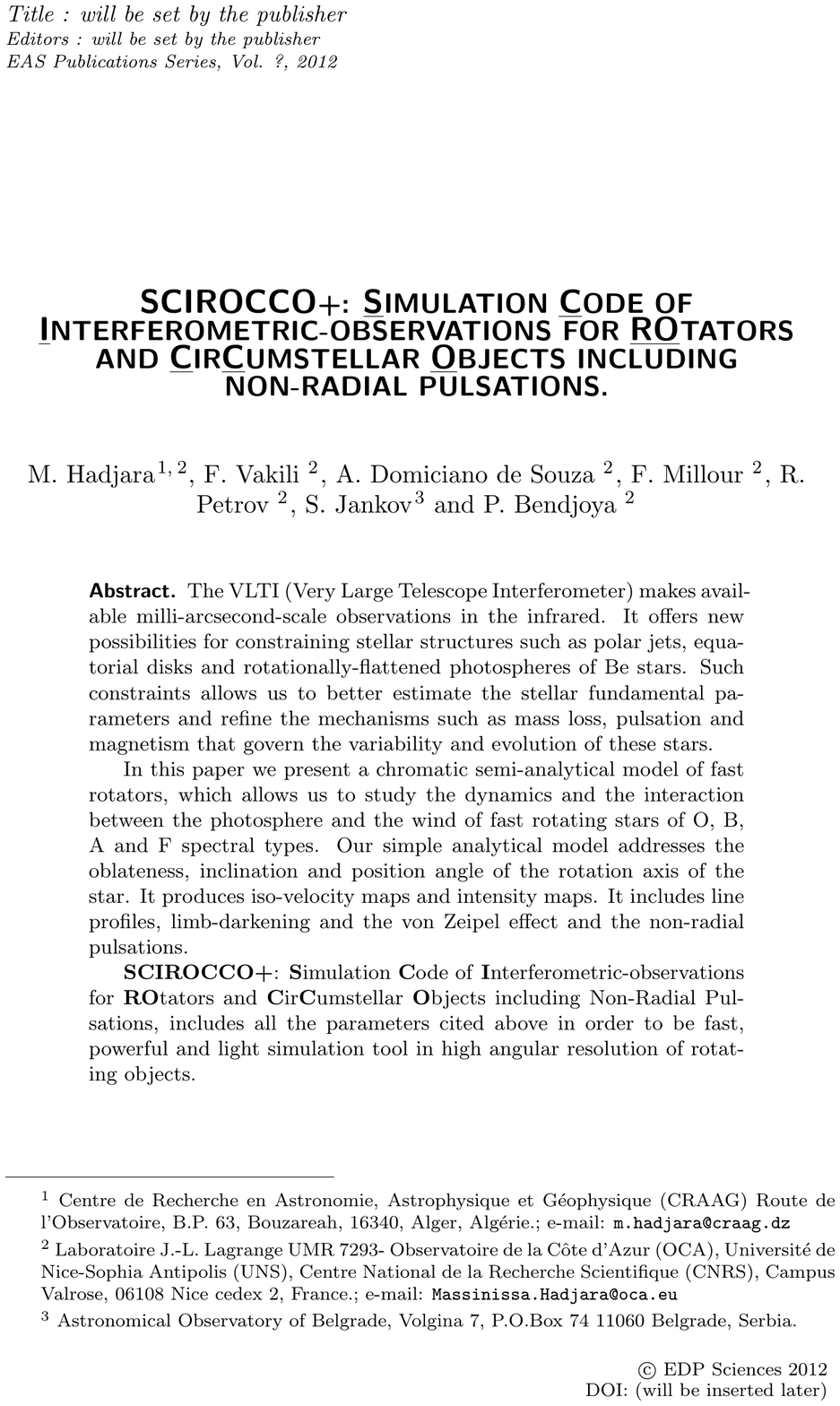}

Suite à ce travail, j'ai entrepris d'étudier la question de l'impact des pulsations non-radiales sur la carte d'intensité et j'ai proposé une solution simple: des modèles théoriques spectroscopiques non-ETL complets des pulsations non radiales en rotation des étoiles de type spectral précoce ("early-type"), développés par \citet{1997MNRAS.284..839T}, ou par \citet{1999A&A...351..582H} sur l'étude des amplitudes des oscillations d'étoiles de la séquence principale, excitées stochastiquement. Les résultats de tels modèles, confortés et validés par les observations, ont été tabulés. Ainsi pour les étoiles pulsantes de type spectral B par exemple en retrouve des périodicités $>0.2$ jour, pour des vitesses $v_{nrp}$ comprises entre $5$ $\kms$ \& $30$ $\kms$, ce qui correspond à des variations de luminosité relative $\delta L=\frac{L-L_0}{L_0}$ de $1\%$ à $6\%$. Me servant de ce type de données, je me suis contenté de déduire la variation relative de la température à la surface de l'étoile pulsante, suivant la loi de Stefan-Boltzmann, i.e. $\delta T=\frac{T-\Tmean}{\Tmean}=(\delta L +1)^{\frac{1}{4}}-1$. Ainsi, et toujours pour les étoiles pulsantes de type B, j'ai estimé la variation de la température $\delta T_{max}$ à $0.25\%$ de $\Tmean$, ce qui correspond à une variation de $\delta T=0.05\%$ de $\Tmean$ pour chaque $v_{nrp}=1$ $\kms$. De ce fait je déduis de manière assez simple et rapide la carte de distribution des températures à partir de la carte iso-vitesses due au phénomène PNR $T(\theta,\phi)=\Tmean+T_{nrp}(\theta,\phi)$, avec:\\

\begin{equation}
T_{nrp}(\theta,\phi)= \delta T\frac{v_{nrp}(\theta,\phi)}{v_{puls}}
\label{Tnrp}
\end{equation}

\begin{figure}[ht]
\centering
\includegraphics[width=0.45\hsize,height=0.48\hsize,draft=false]{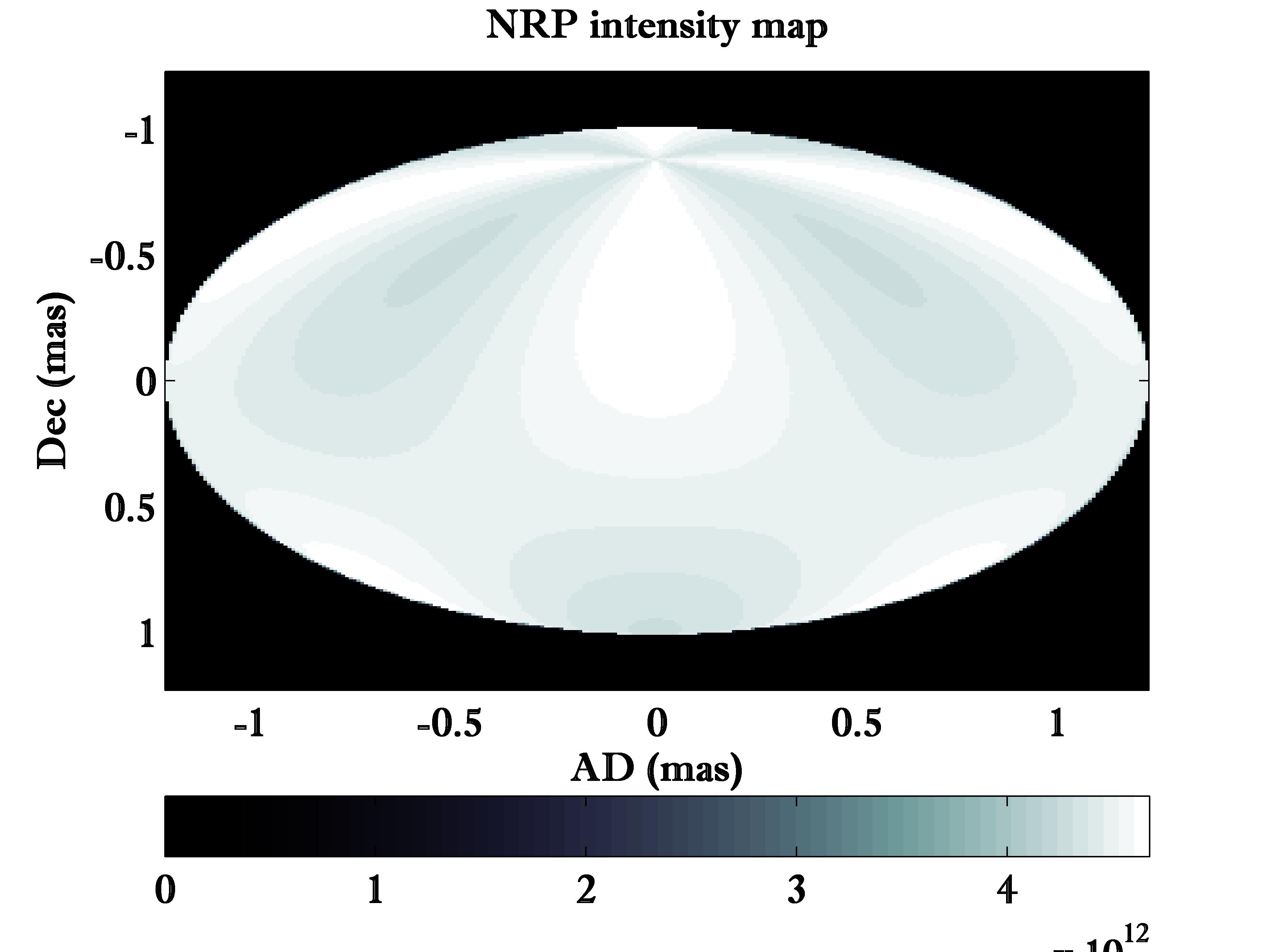}
\includegraphics[width=0.45\hsize,height=0.48\hsize,draft=false]{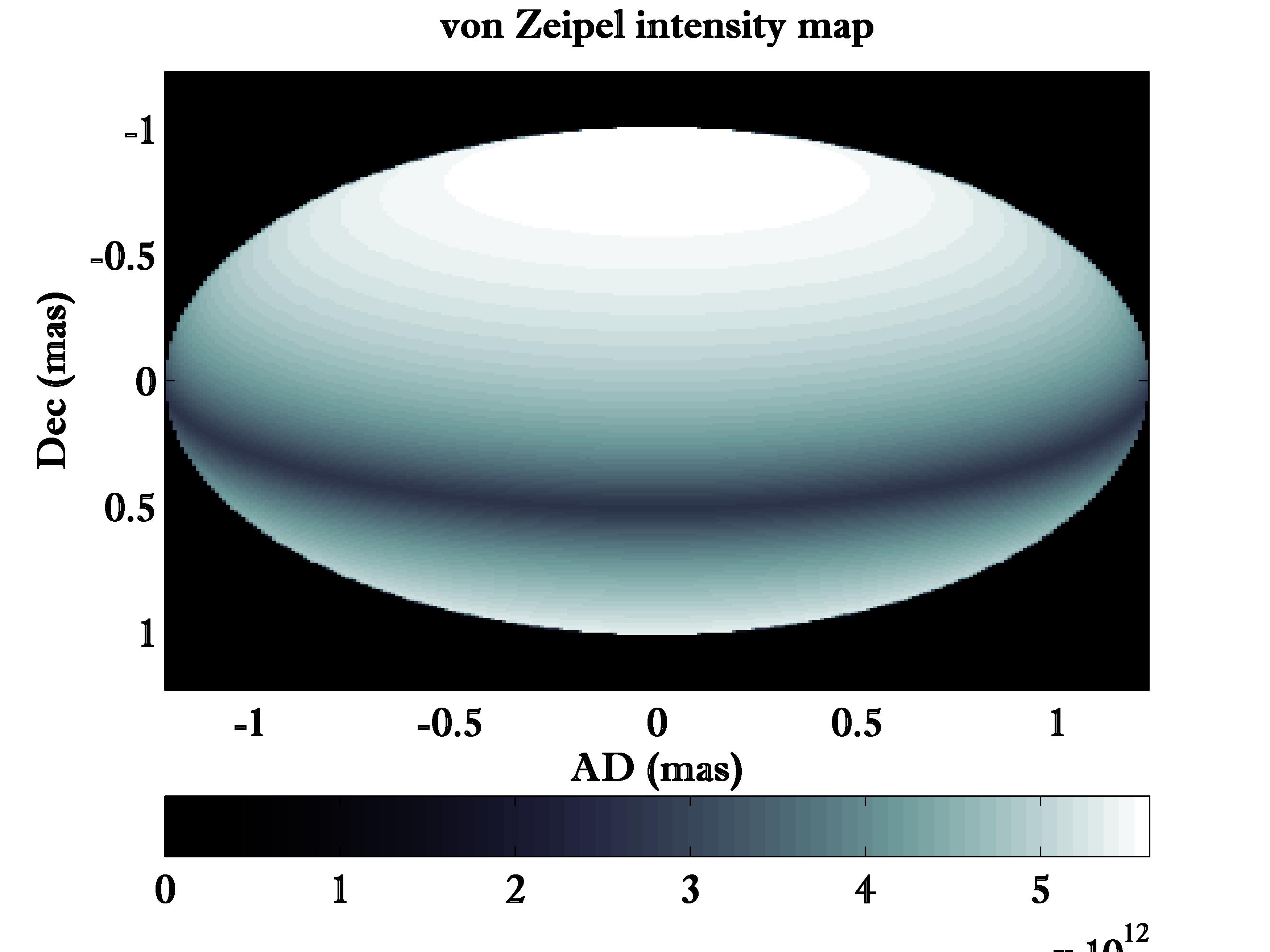}
\includegraphics[width=0.45\hsize,height=0.48\hsize,draft=false]{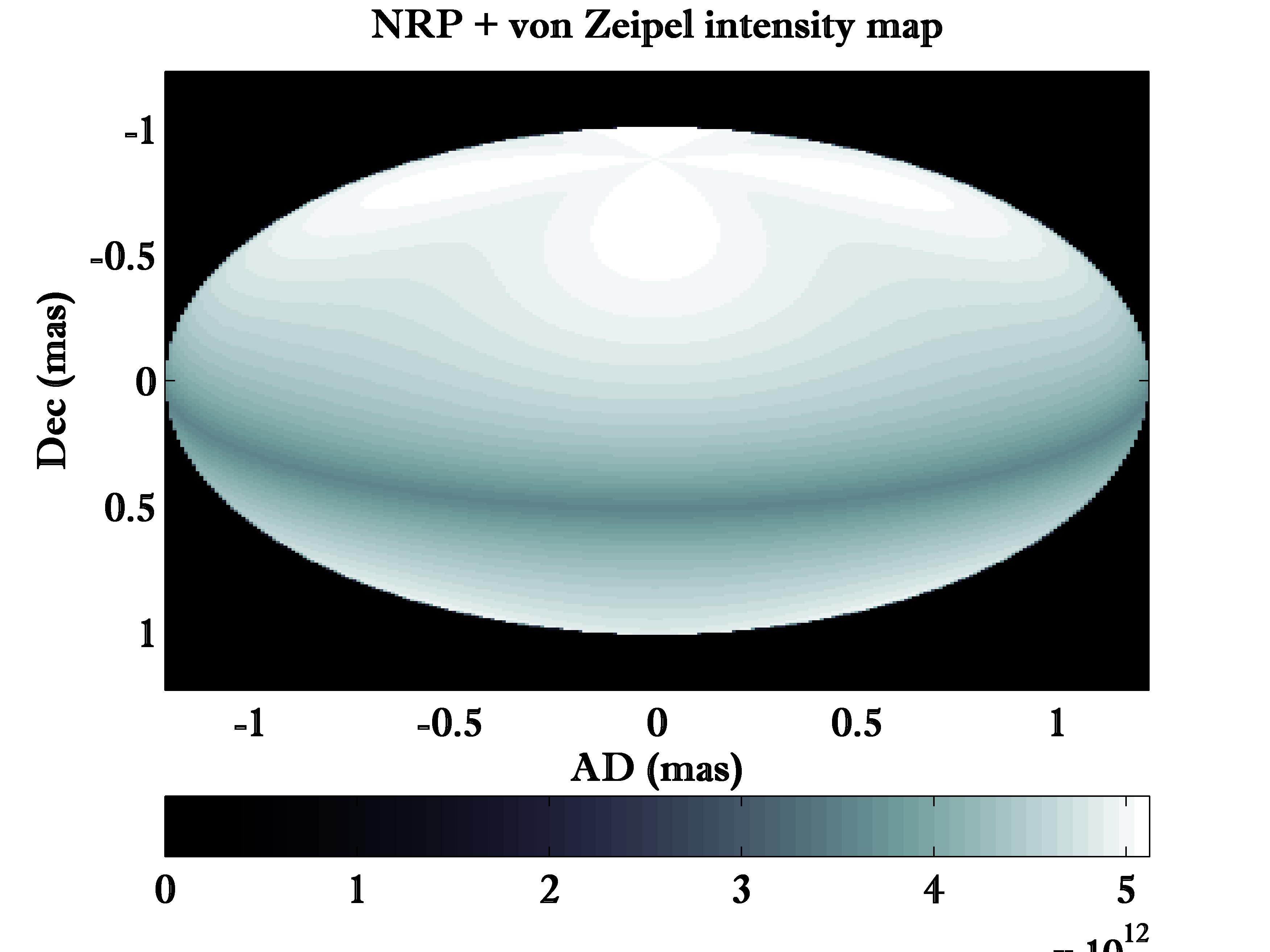}
\caption[Carte d'intensité avec pulsation non-radiale (PNR)]{\textbf{En haut, à gauche:} Une carte d'intensité due à l'effet pur de la pulsation non radiale $(Watt.m^{-2}.Hz^{-1}.Sr^{-1})$. \textbf{En haut, à droite:} Une carte d'intensité purement causée par l'effet de l'assombrissement gravitationnel. \textbf{Et en bas:} une carte d'intensité regroupant les deux effets NRP et gravity darkening, pour une simulation de l'étoile $\eta$ Cen (avec $PA_{rot}=0^\circ$).}\label{Int_puls}
\end{figure}

Je calcule enfin, à partir de la loi du corps noir de Planck (Eq.\eqref{eq5}), la carte d'intensité $I_0(\lambda,\theta,\phi)$ impactée par l'effet des pulsations non radiales. La Fig.\ref{Int_puls} représente les cartes d'intensité (PNR, gravity darkening et/ou les deux) pour le même cas de figure que l'étoile étudiée dans l'article "SCIROCCO+" \citep{2013EAS....59..131H}, traitant de la simulation de l'étoile $\eta$ Cen (avec $PA_{rot}=0^\circ$), et $\delta T=1\%$ par $1$ $\kms$. Cela correspond plus aux céphéides et aux RR Lyrae, cependant juste pour accentuer l'effet sur la carte d'intensité pour valeur de démonstration.\\

\clearpage

\begin{figure}[ht]
\centering
\includegraphics[width=0.6\hsize,draft=false]{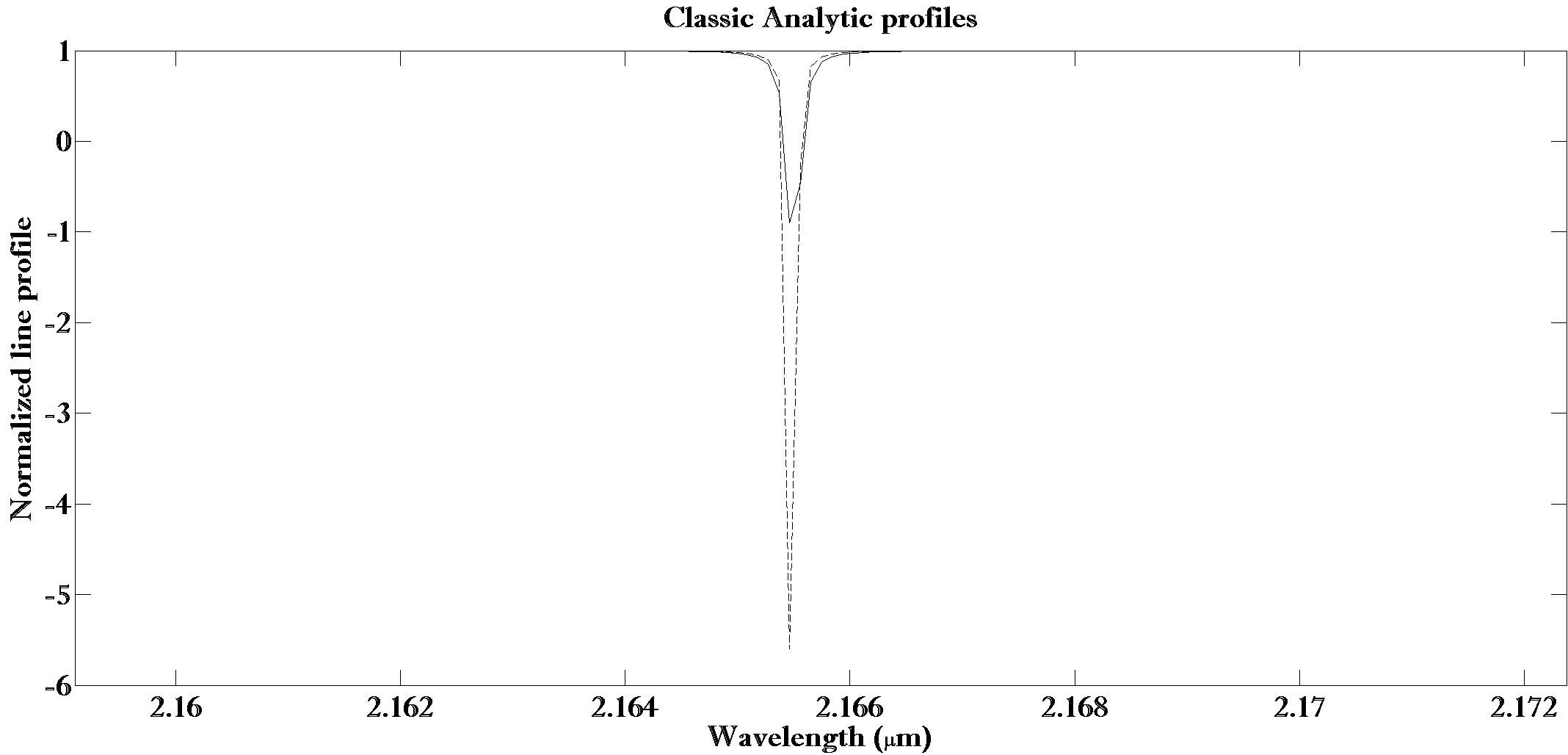}
\includegraphics[width=0.45\hsize,draft=false]{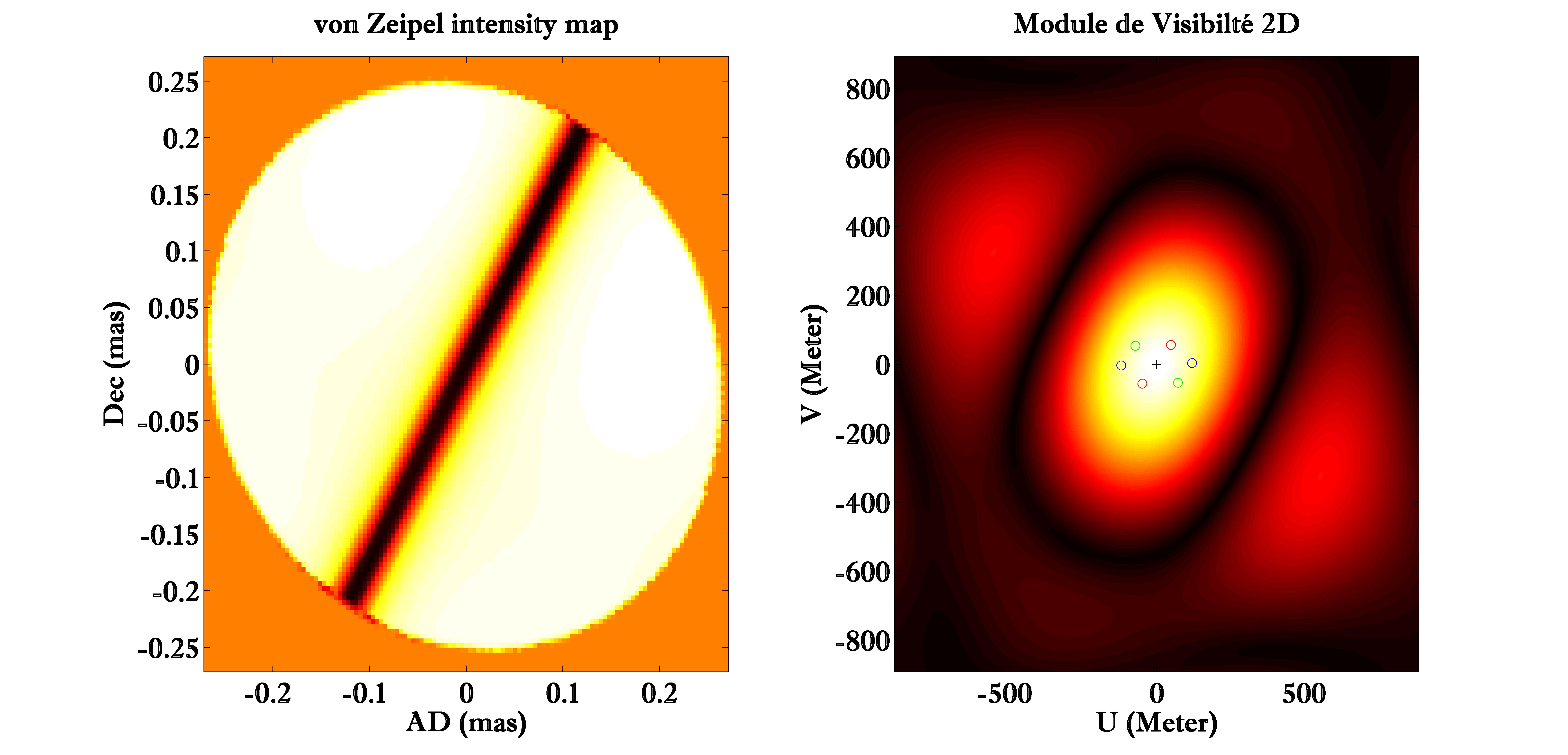}
\includegraphics[width=0.45\hsize,draft=false]{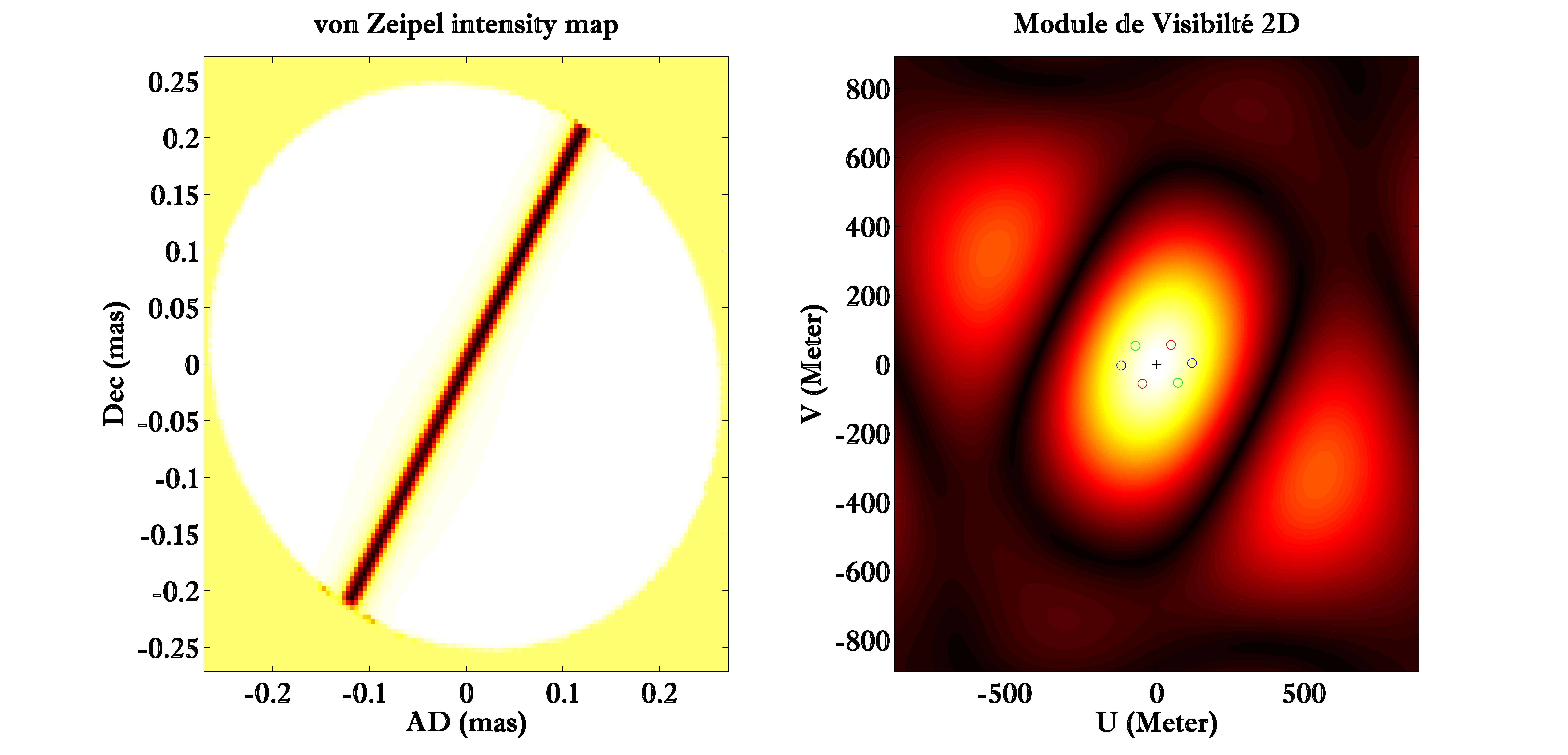}
\includegraphics[width=0.8\hsize,draft=false]{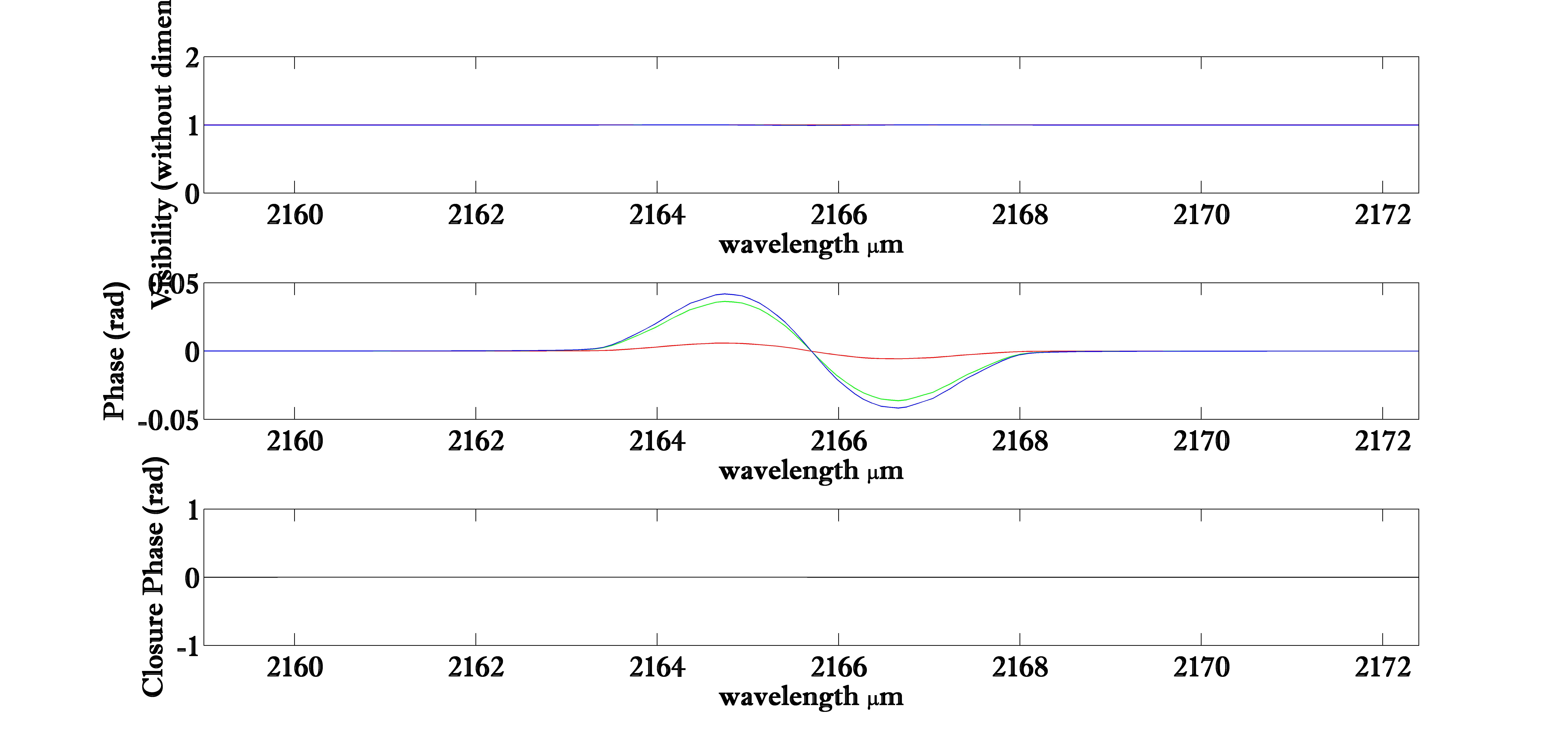}
\includegraphics[width=0.8\hsize,draft=false]{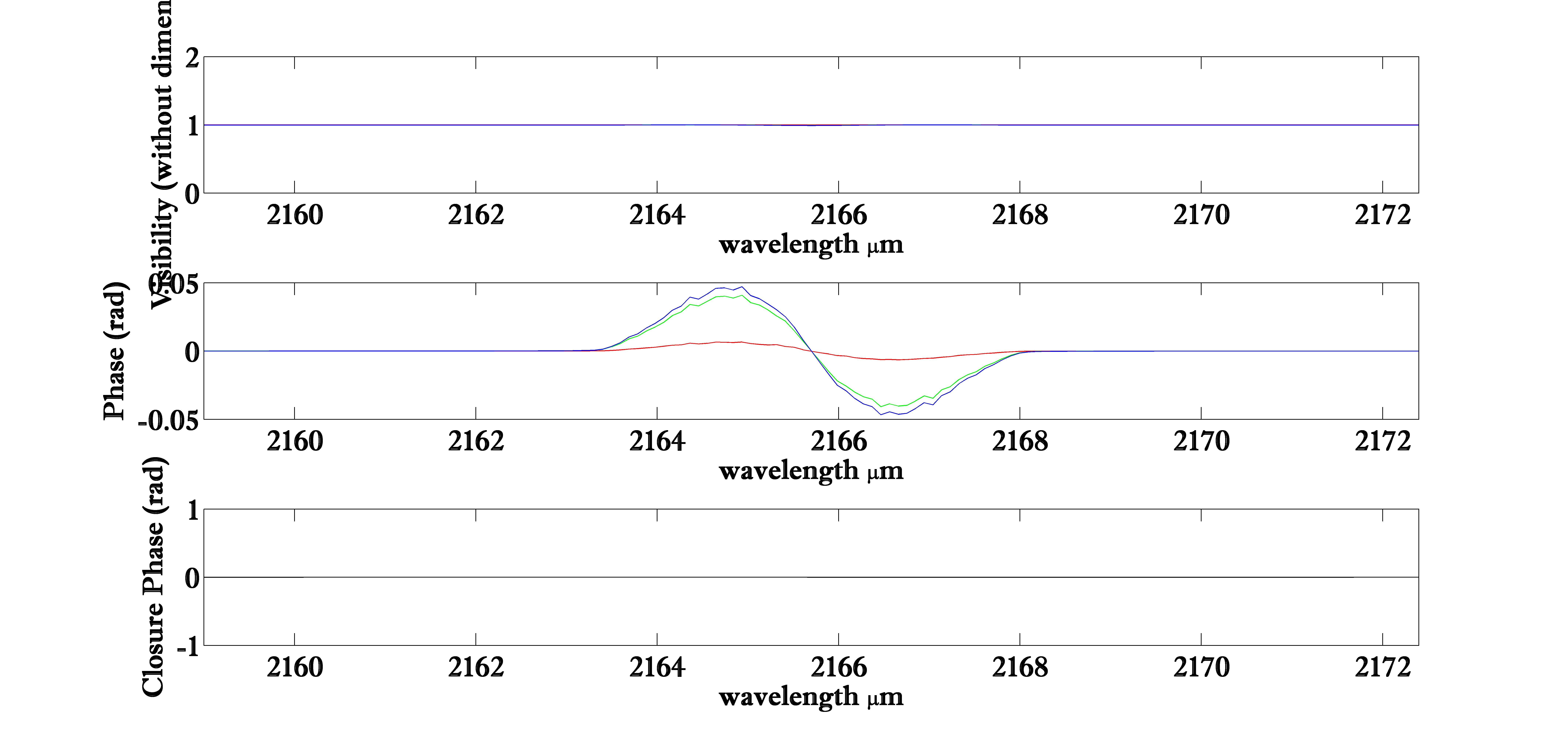}
\caption[Impact du profil de raie sur la pulsation non-radiale (PNR)]{\textbf{En haut:} Deux profils de Voigt, l'un fin (en pointillés) et l'autre un peu plus large (moins profond, en ligne continue). \textbf{Deuxième rangée:} La représentation de la carte d'intensité monochromatique 2D et de sa carte de module de visibilité, à la longueur d'onde Br$\gamma$, à gauche pour le large profil de raie et à droite pour le plus fin. \textbf{Troisième rangée:} Les visibilités, $\phidiff$ et clôture de phase des points (u,v) représentés sur la carte $V^2$ du large profil de raie. Aucun impact des PNR sur nos mesures interférométriques. \textbf{Quatrième et dernière rangée:} Les visibilités, $\phidiff$ et clôture de phase des points (u,v) représentés sur la carte $V^2$ du fin profil de raie. L'impact des PNR est clairement mis en évidence sur les $\phidiff$.}\label{raie_npr}
\end{figure}

Un autre effet important abordé dans ce travail incluant les PNR est celui du profil de raie. En effet, et tel que démontré par \citet{1996IAUS..181...247S}, le déplacement du photo-centre (ou la $\phidiff$ entre autres), et tous les observables interférométriques en général, sont très sensibles au profil de raie. Plus la raie est fine (largeur à mi-hauteur plus petite) plus les observables interférométriques sont sensibles aux effets de la pulsation non-radiale. J'ai tenté de simuler cet effet avec SCIROCCO et les résultats sont conformes à ce qu'avait prédit \citet{1996IAUS..181...247S}. La Fig.\ref{raie_npr} démontre d'ailleurs clairement que le profil de raie à plus petite largeur à mi-hauteur (le plus fin) est celui qui est le plus sensible aux PNR sur les $\phidiff$. Notons aussi une rotation rigide (le coefficient de rotation différentielle $\alpha=0$) pour une masse $M=9.5$ $\Msun$, $d=95$ $pc$ et un $PA_{rot}=30^\circ$ d'$\eta$ Cen, en plus des paramètres déjà considerés lors de mon étude théorique sur cette étoile dans les proceedings \citet{2013EAS....59..131H}. De ce fait le rayon angulaire équatorial d'$\eta$ Cen $\diameq=0.27$ $mas$ pour un degré de sphéricité $D=0.85$, un coefficient d'assombrissement gravitationnel $\beta=0.2$, une vitesse équatoriale $v_{eq}=67\%v_{crit}$ et des températures pôles/équateur $[T_{eq},T_{pol}]=[13251, 16749]K$.\\

La raie Brackett $\gamma$, étant la raie majoritairement observée en IR, est considérée comme étant une raie large, et de ce fait il n'est pas possible de directement observer l'effet des PNR en IR avec cette raie sur des mesures individuelles. Par contre, une analyse temporelle du photo-centre ou de la $\phidiff$ (eg. \citet{2001A&A...377..721J}) permet cela. Ainsi, la campagne de mesures prospectives HR d'une demi-nuit sur $\eta$ Cen 2013 ne nous a pas permis de directement constater les effets des pulsations non radiales mais nous a néanmoins révélé la présence d'un disque circumstellaire éventuel autour de l'étoile (voir l'échantillon de données juste ci-dessous). Ceci m'a poussé à intégrer les simulations des disques en plus des étoiles à rotation et de l'effet des PNR à SCIROCCO, contribuant ainsi activement à une nouvelle demande d'observation 2014 pour $\eta$ Cen. Notre nouvelle demande d'observation AMBER/VLTI étant aussi acceptée, j'ai pu mener des observations à distance (par Skype) avec Paranal depuis l'OCA dans le courant de mars 2014. Un court échantillon de mon travail sur la simulation d'étoile \& disque en rotation est résumé dans la section ci-dessous (juste après l'échantillon de données $\eta$ Cen).\\

\clearpage

\section{Simulation étoile \& disque}

\begin{figure}[ht!]
\centering
\includegraphics[width=0.8\hsize,draft=false]{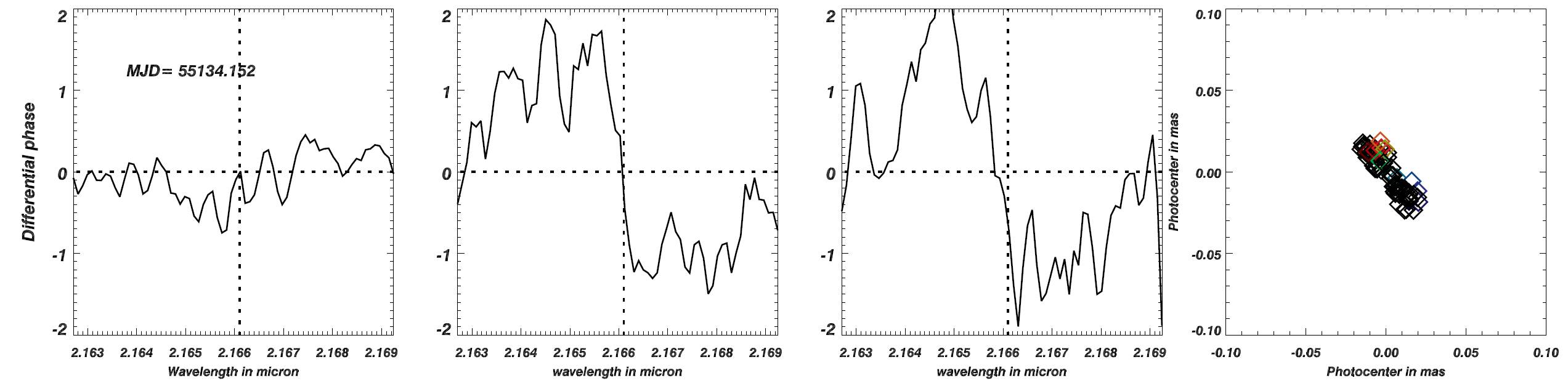}
\includegraphics[width=0.8\hsize,draft=false]{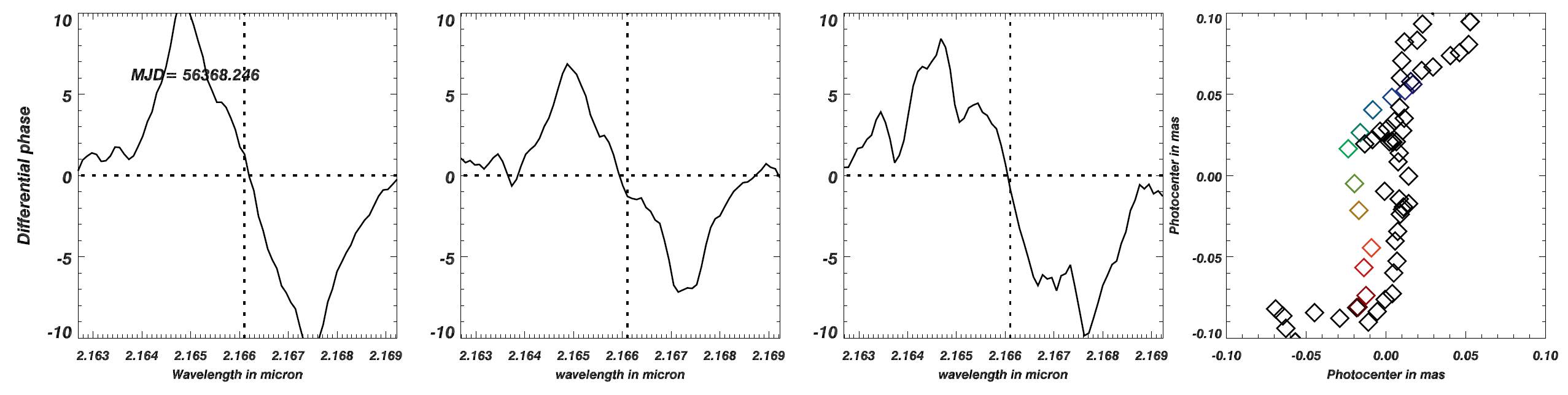}
\caption[Déplacement photo-centre d'Achernar et d'$\eta$ Cen]{En haut: Les phases différentielles à gauche et le déplacement du photo-centre leurs correspondant à droite pour Achernar. On remarque bien qu'ici il s'agit bien d'une étoile en rotation avec un angle $PA_{rot}\propto 30^\circ$. En bas : Les phases différentielles à gauche et le déplacement photo-centre leurs correspondant à droite pour $\eta$ Cen. On remarque ici un déplacement de photo-centre assez large, qu'on a interprété comme étant celui du disque, avec couplage de deux effets; rotation plus expansion.}\label{dep_phot}
\end{figure}

Après avoir constaté des formes en doubles $\tild$ sur les $\phidiff$, des "M" sur les spectres et des "W" sur les modules de visibilités, autour de la raie Br$\gamma$ (\hyperlink{page.180}{voir p.180}), on a fortement suspecté la présence d'un disque autour de l'étoile $\eta$ Cen, tel que rapporté par \citet{2012ApJ...744...19K}. Une autre possibilité de tels constatations sur les mesures interférométriques est l'effet de la sur-résolution, tel que soulevé par \citet{2012A&A...538A.110M}. Une étude comparative\footnote{à l'aide d'un code qui m'a généreusement était proposé et fourni par Olivier Chesneau, et à qui je tiens à rendre un solennel hommage via ce manuscrit.} du déplacement du photo-centre sur les deux étoiles; Achernar et $\eta$ Cen (voir Fig.\ref{dep_phot}), nous a poussé à fortement soupçonner l'existence d'un disque en rotation expensive autour d'$\eta$ Cen.\\

Pour simuler les disques circumstellaires via SCIROCCO, j'ai eu recours aux équations qu'on trouve dans littérature, à savoir la vitesse rotationnelle projetée du disque, telle que formulée par \citet{1996A&A...311..945S}, en coordonnées polaires $(r,\theta_{pol})$:\\

\begin{equation}
v_{rot,disk}(r,\theta_{pol})=v_0\sin\theta_{pol}\left(\frac{R_*}{r}\right)^\kappa,
\label{vdisk1}
\end{equation}

où $R_*$ est le rayon équatorial de l'étoile, $v_0$ n'est autre que la vitesse de rotation équatoriale de l'étoile (lorsque $r=R_*$), et $\kappa$ étant un paramètre définissant le type de rotation du disque ; $\kappa=0$ pour une rotation dite constante, $\kappa=0.25$ pour un modèle de disque proposé par Ara\'ujo et al. 1994, $\kappa=0.5$ pour une rotation Képlérienne, $\kappa=1$ pour une rotation à moment angulaire conservé et $\kappa=-1$ pour une rotation rigide. J'ai aussi pris en compte une composante de vitesse expansive du disque, et qui est directement lié aux vents radiatifs engendrés par l'étoile (tel qu'abordé dans le Chap.\ref{chap:rota}). La formulation d'une telle vitesse a était proposé par \citet{1975ApJ...195..157C}, qui l'ont formulée comme suit en coordonnées polaires:\\

\begin{equation}
v_{exp,disk}(r,\theta_{pol})=v_{term}\cos\theta_{pol}(1-r)^{0.5},
\label{vdisk2}
\end{equation}

\begin{figure}[ht]
\centering
\includegraphics[width=0.45\hsize,height=0.4\hsize,draft=false]{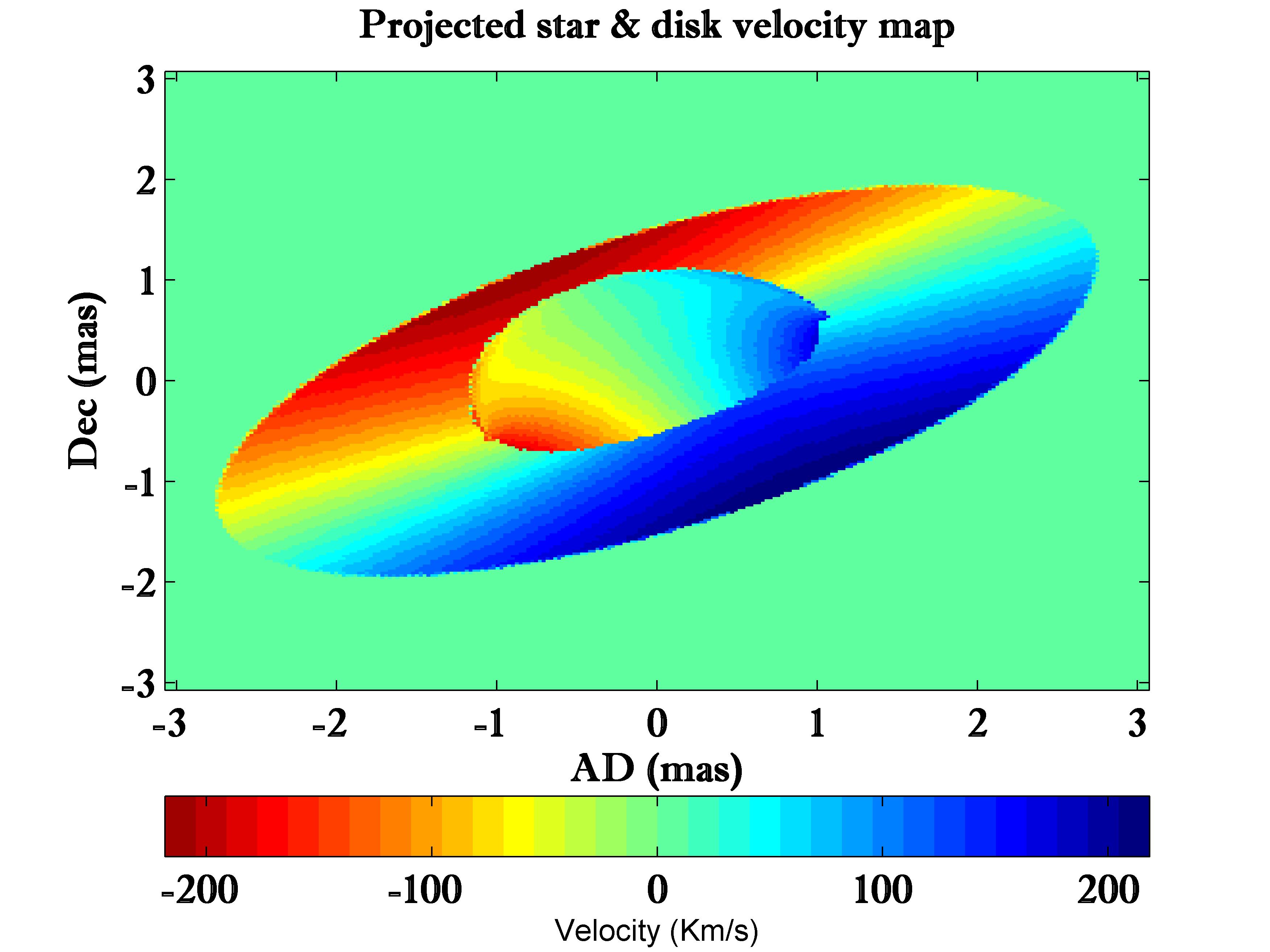}
\includegraphics[width=0.45\hsize,height=0.4\hsize,draft=false]{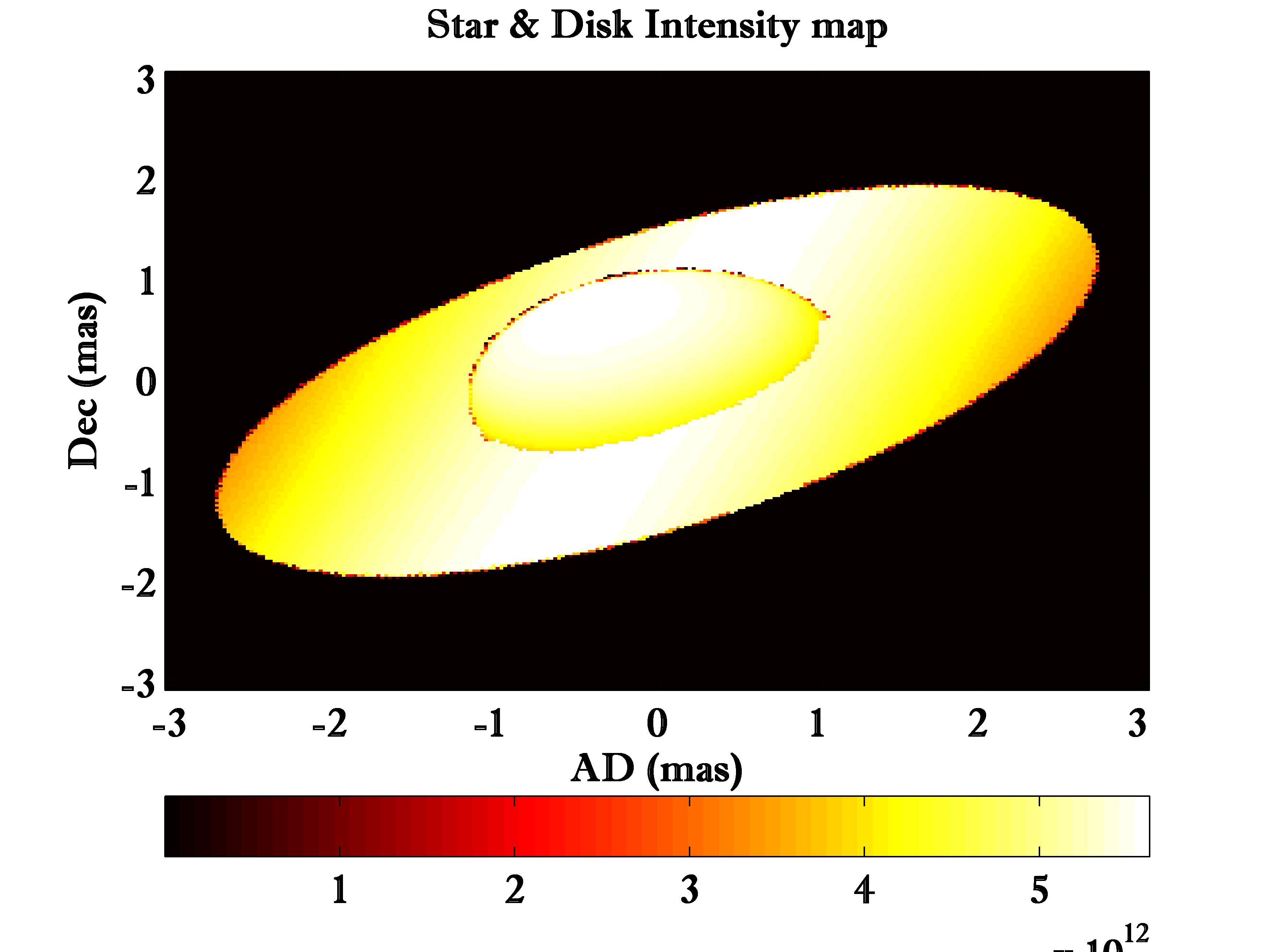}
\caption[Simulation d'étoile \& disque]{La carte des iso-vitesses 2D d'une étoile fictive et son disque à gauche, et à droite la carte d'intensité 2D dans le continuum.}\label{stardisk1a}
\end{figure}

\begin{figure}[ht]
\centering
\includegraphics[width=0.6\hsize,height=0.3\hsize,draft=false]{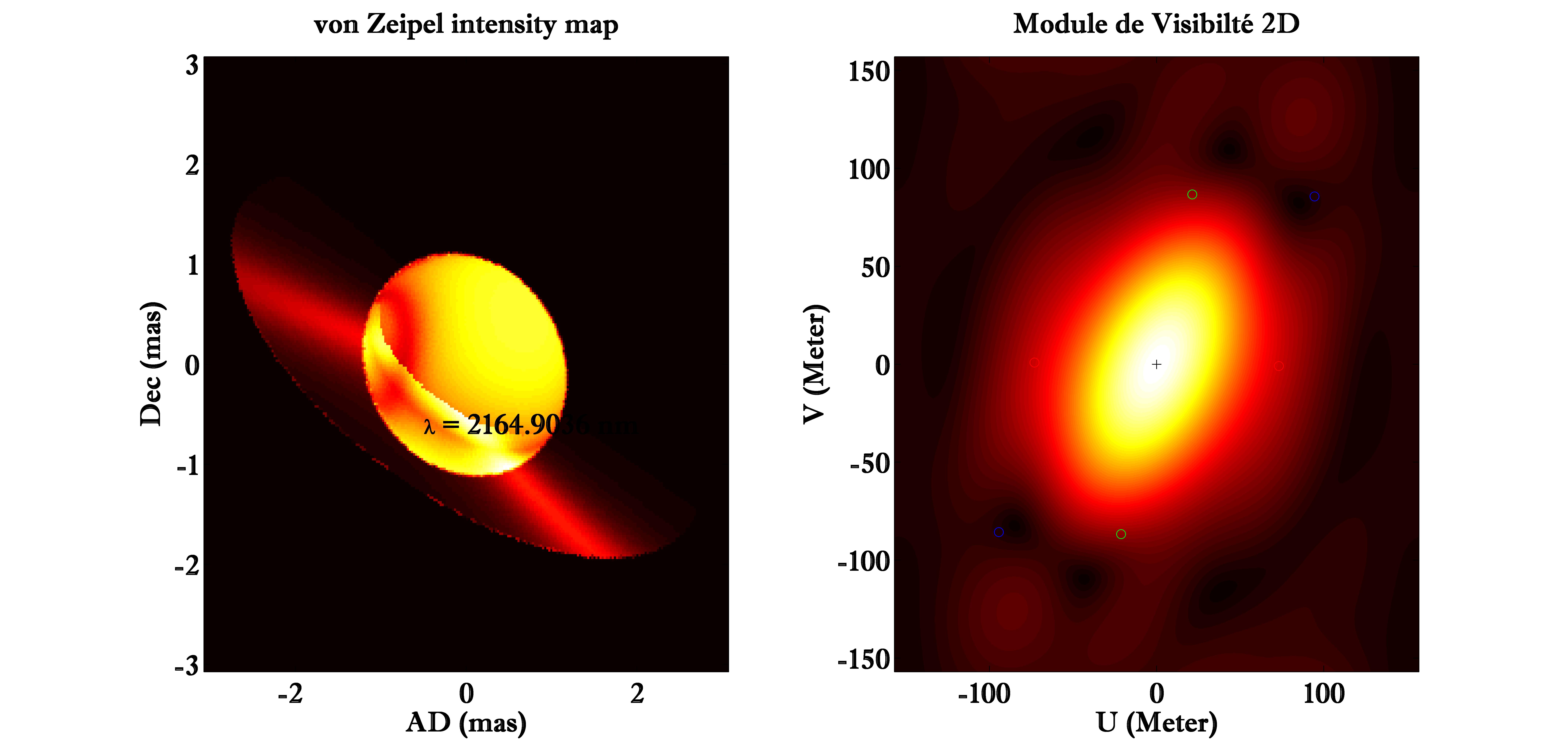}
\includegraphics[width=0.6\hsize,height=0.3\hsize,draft=false]{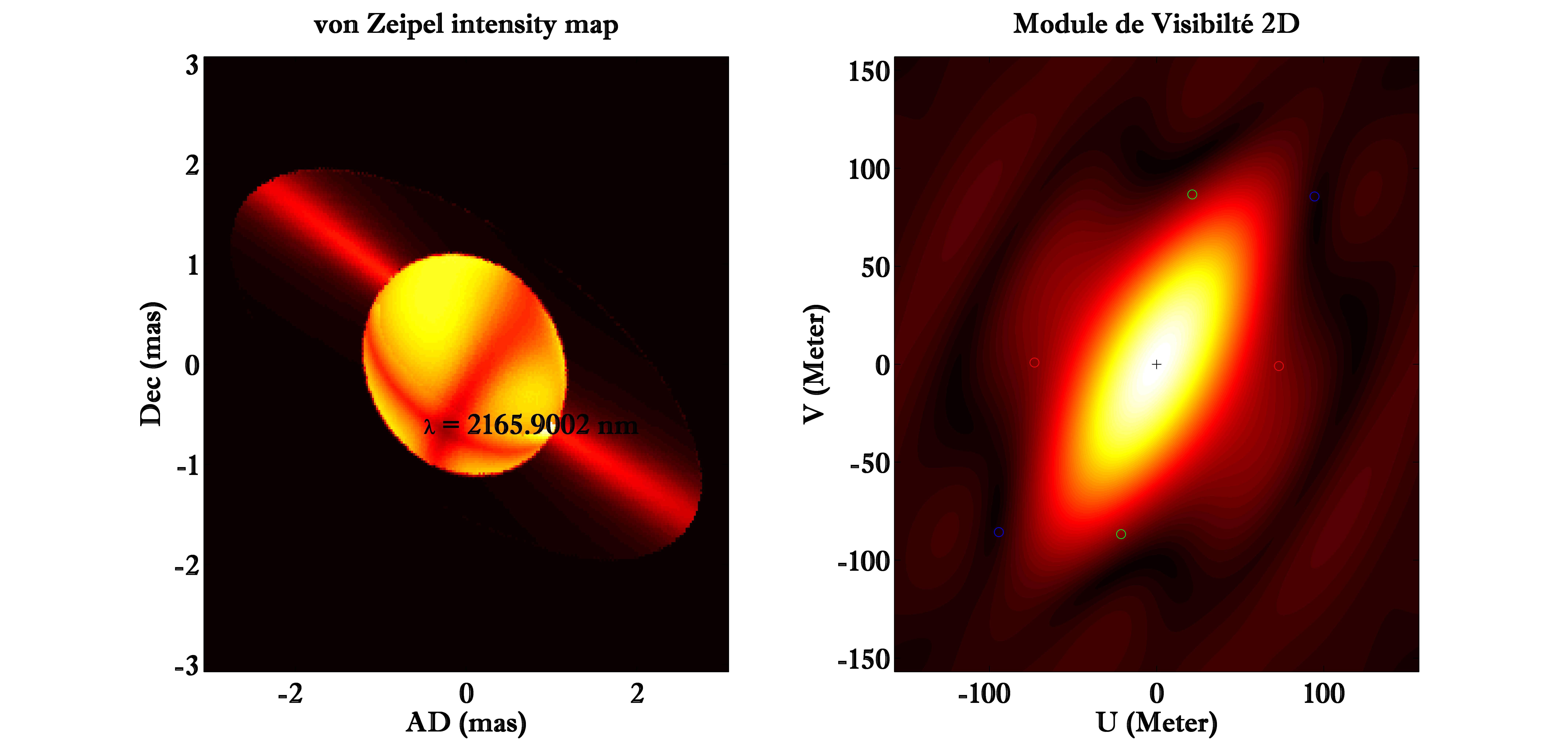}
\includegraphics[width=0.6\hsize,height=0.3\hsize,draft=false]{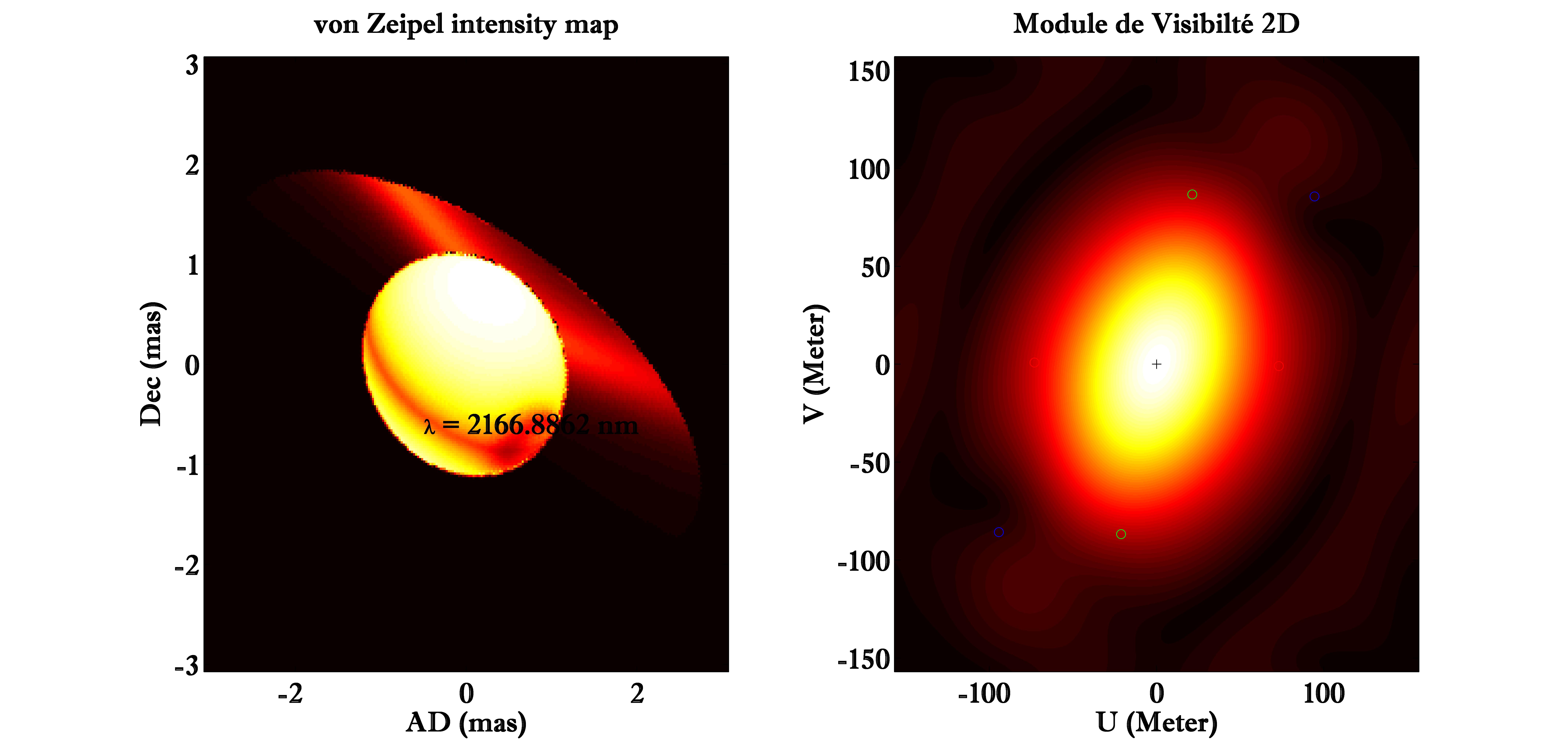}
\caption[Déplacement de la raie sur une étoile + disque]{Les représentations des cartes d'intensités monochromatique et leurs cartes de module de visibilité 2D correspondantes pour 3  longueurs d'onde différentes en IR autour de la raie Br$\gamma$.}\label{stardisk1b}
\end{figure}

\begin{figure}[ht]
\centering
\includegraphics[width=0.8\hsize,draft=false]{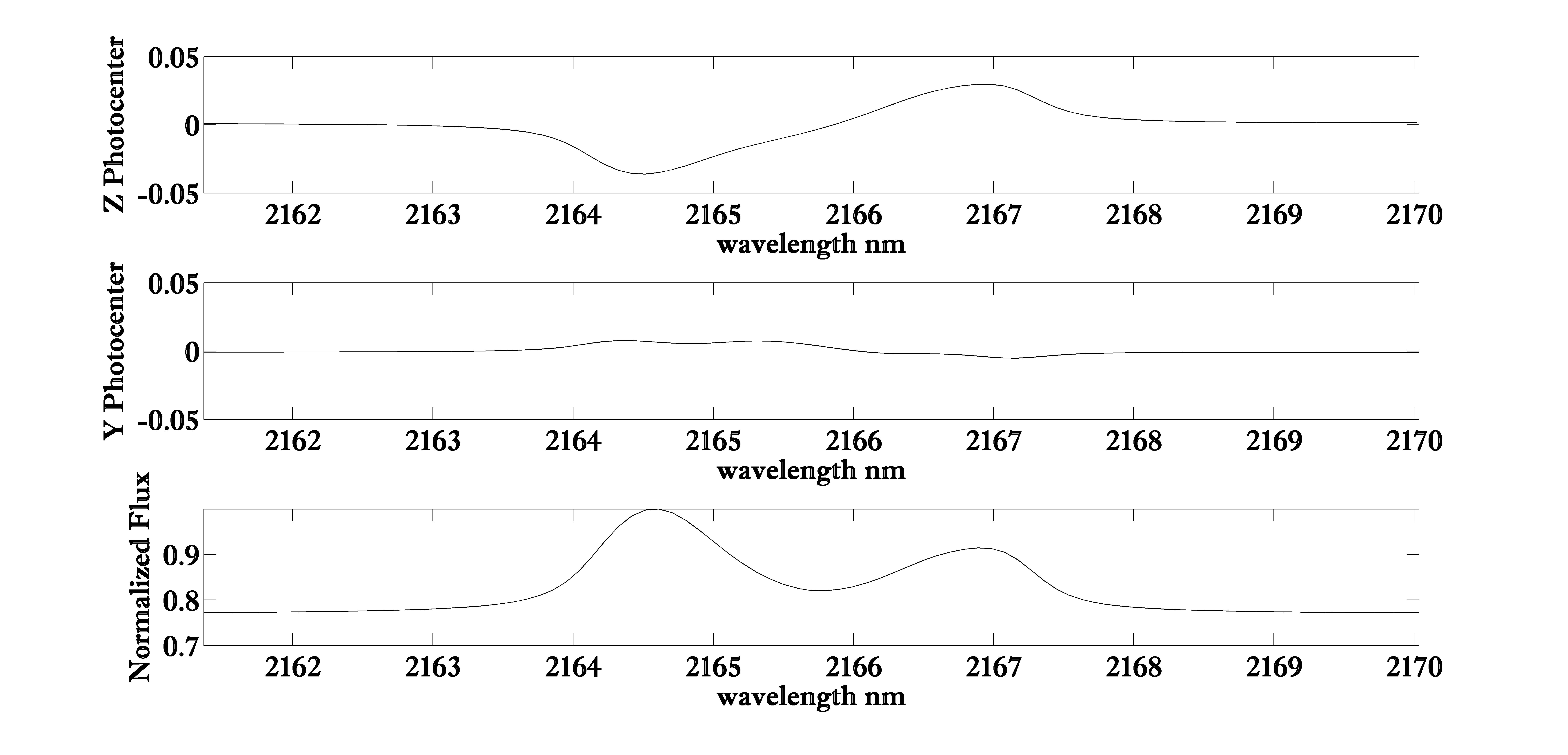}
\includegraphics[width=0.8\hsize,draft=false]{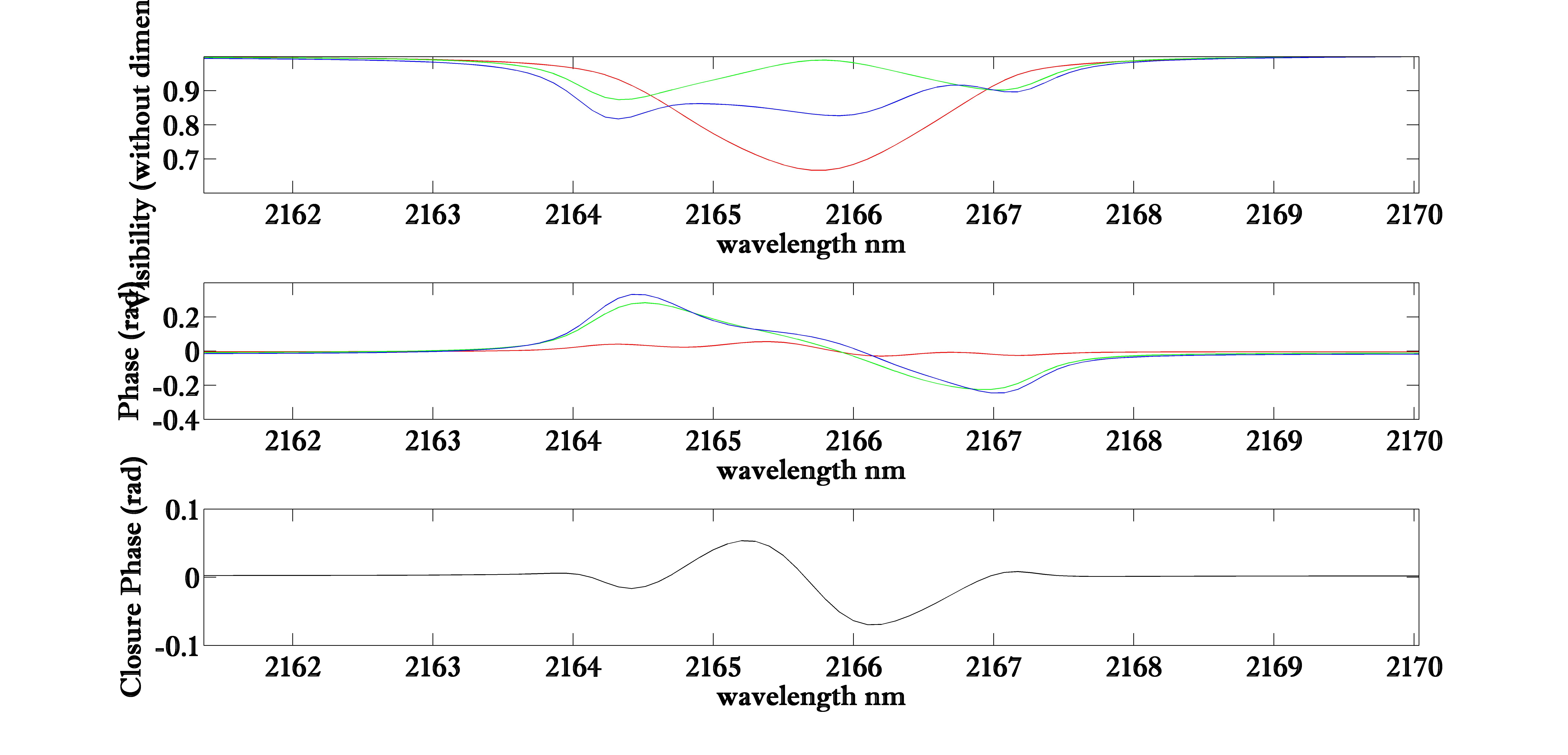}
\caption[Observables interférométriques d'une étoile \& disque]{\textbf{En haut:} Déplacements de photo-centres selon les axes $Y$ et $Z$ ainsi que le spectre de l'ensemble étoile+disque représenté dans la Fig.\ref{stardisk1a}. \textbf{En bas:} Les modules de visibilités $V^2$, les $\phidiff$ et la phase de clôture $\Psi$ du couple étoile+disque en fonction des coordonnées du plan de Fourier $(u,v)$, avec les bases interférométriques $[73.10m,90.73^\circ]$ en rouge $[89.29m,13.84^\circ]$ en vert et $[127.5m,47.75^\circ]$ en bleu, représentés dans les cartes de module de visibilité 2D de la Fig.\ref{stardisk1b}.}\label{stardisk2}
\end{figure}

$v_{term}$ étant la vitesse asymptotique terminale du vent radiatif. Ainsi, la carte des iso-vitesses totales du disque circumstellaire est : $v_{disk}(r,\theta_{pol})=v_{rot,disk}(r,\theta_{pol})+v_{exp,disk}(r,\theta_{pol})$. Le code que j'ai développé prend aussi en considération l'intensité du disque comme étant une gaussienne 2D, à amplitude proportionnelle à celle de l'étoile, tout en prenant en compte l'effet de l'anisotropie du disque \citep{1994MNRAS.268..845O}, l'opacité du disque, son inclinaison via l'angle $i$, ainsi que l'angle $PA_{rot}$. Concernant, le profil de raie du disque j'ai opté pour une méthode simple qui consiste à déduire la largeur équivalente du profil du raie (en émission du disque, voir Lois de Kirchhoff du Chap.\ref{chap:spec-interfero}) à partir de ceux de l'étoile. Sinon on peut aussi recourir à des profils de raie plus rigoureux via le code Tlusty/Synspec dont les dernières versions peuvent aussi générer les spectres synthétiques des disques. Dans l'exemple qui suit (voir Fig.\ref{stardisk1a}, \ref{stardisk1b}\& \ref{stardisk2} ci-dessous) j'ai essayé de simuler une étoile fictive avec son disque circumstellaire, dont voici les principales caractéristiques : une masse $M=6.1$ $\Msun$, un rayon équatorial $R_{eq}=11.6$ $\Rsun$, à une distance $d=44.1$ $pc$, tournant à une vitesse équatoriale $v_{eq}=v_{rot,disk}=300$ $\kms$ et une vitesse asymptotique terminale $v_{term}=150$ $\kms$, le disque, au rayon $2.5$ fois le $R_{eq}$ de l'étoile, est considéré comme étant opaque avec un mouvement képlérien $\kappa=0.5$ avec une distribution d'intensité isotropique.\\

On observe bien les mêmes caractéristiques que celles observées sur l'échantillon d'$\eta$ Cen (\hyperlink{page.180}{voir p.180}). A savoir des doubles $\tild$ sur les $\phidiff$, des "M" sur les spectres et des "W" sur les modules de visibilités. Bien que les modélisations SCIROCCO sur l'étoile+disque ne soient pas encore très poussées, le travail fictif présenté ici, démontre assez bien toute la possibilité future de la réalisation d'une telle étude sur $\eta$ Cen, ainsi que tout le potentiel de SCIROCCO en la matière.\\

\clearpage

\includepdf[]{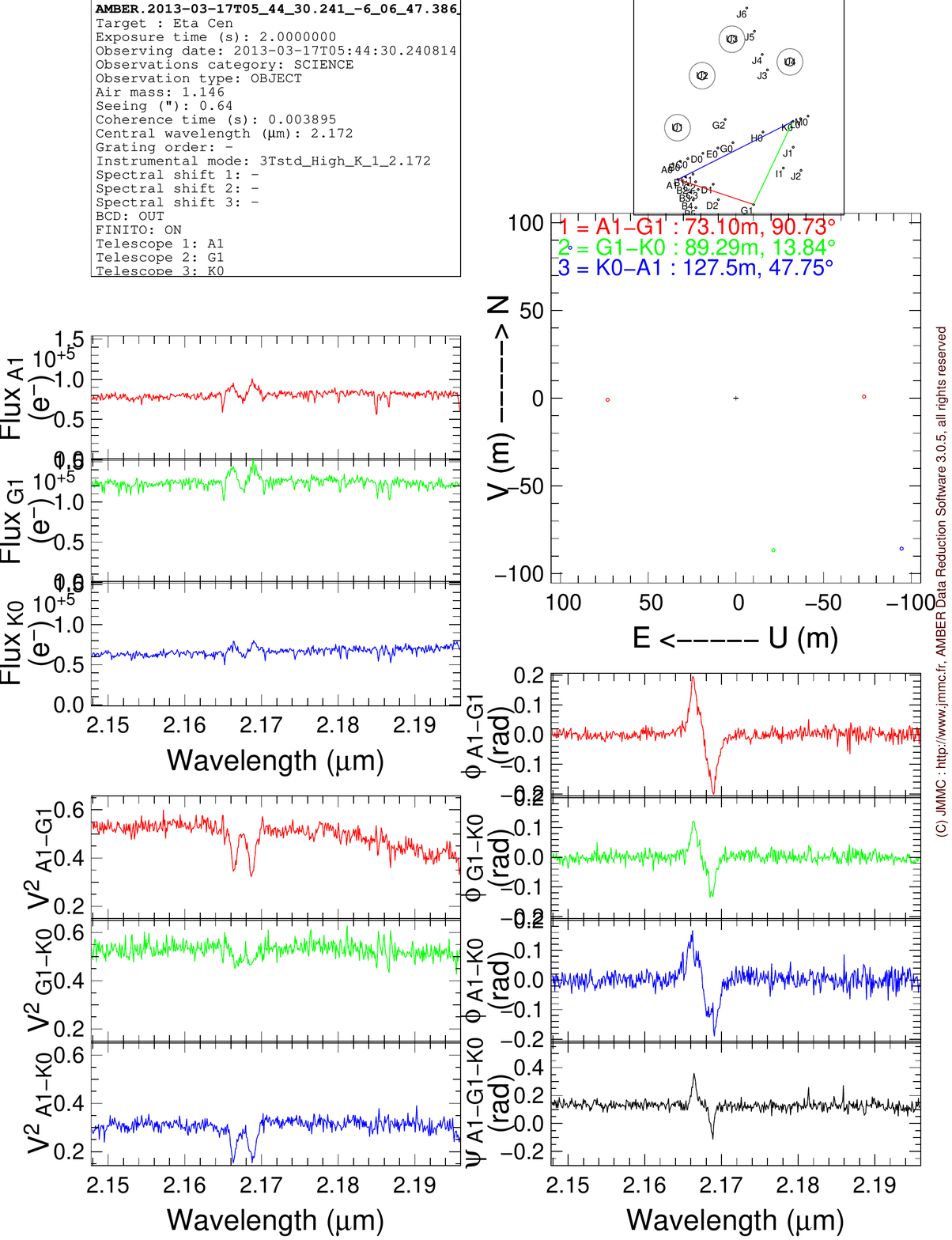}

\section{Taches stellaires \& exoplanètes}
\begin{figure}[ht]
\centering
\includegraphics[width=1.6\hsize,draft=false, angle=270]{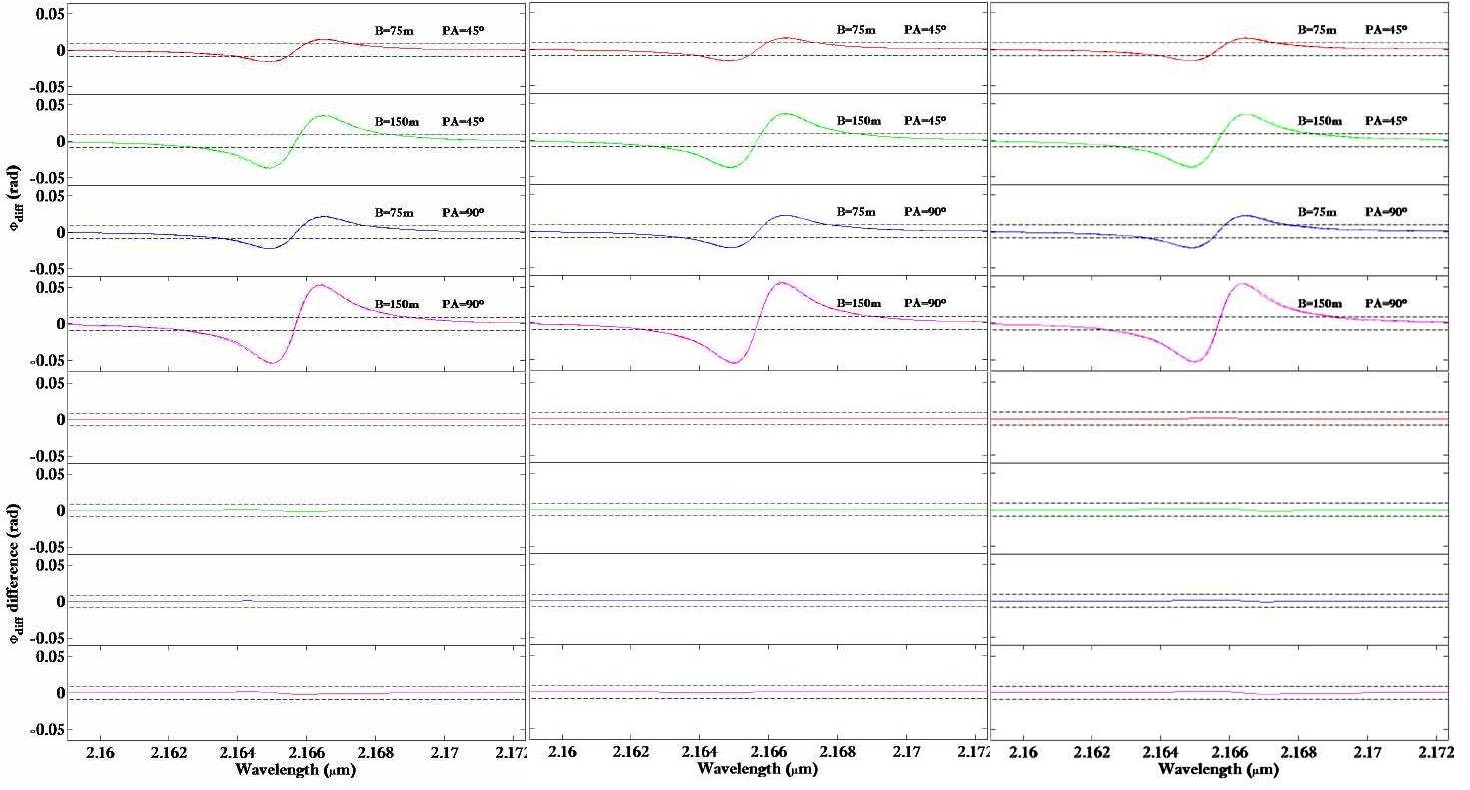}
\caption{Les phases différentielles résultantes. L'effet mesurable est très faible.}\label{exorot}
\end{figure}

J'ai tenté de modéliser très grossièrement, sur une idée suggérée par Farrokh Vakili, l'effet provoqué par une exoplanète transitant devant la photosphère de son étoile parent (ou une tache stellaire à la surface de l'étoile en rotation) sur certains observables interférométriques tels que le spectre et la phase différentielle $\phidiff$. Dans l'exemple ci-dessous, je montre le résultat d'une telle simulation. L'étoile théorique est proche d'Achernar (avec les mêmes paramètres et bases interférométriques que celui de \citet{2014A&A...569A..45H}) avec une planète ayant un dixième de son diamètre, à 3 positions différentes. Je représente pour chaque cas, dans la Fig.\ref{exorot}; une carte d'intensité, le flux normalisé (sans exoplanète, et les différences selon les 3 configurations), et  la phase différentielle (avec planète en continu, sans en discontinue, et la différence pour les 3 configurations de la planète, et les coordonnées du plan $(u,v)$).\\

\begin{figure}[ht]
\centering
\includegraphics[width=0.32\hsize,height=0.27\hsize,draft=false]{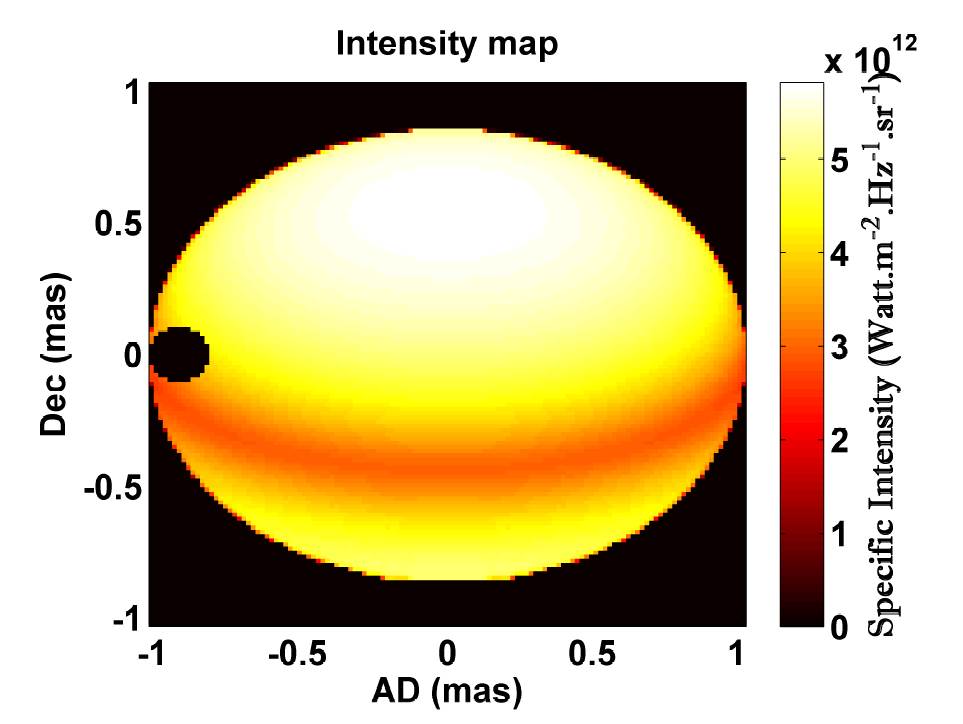}
\includegraphics[width=0.32\hsize,height=0.27\hsize,draft=false]{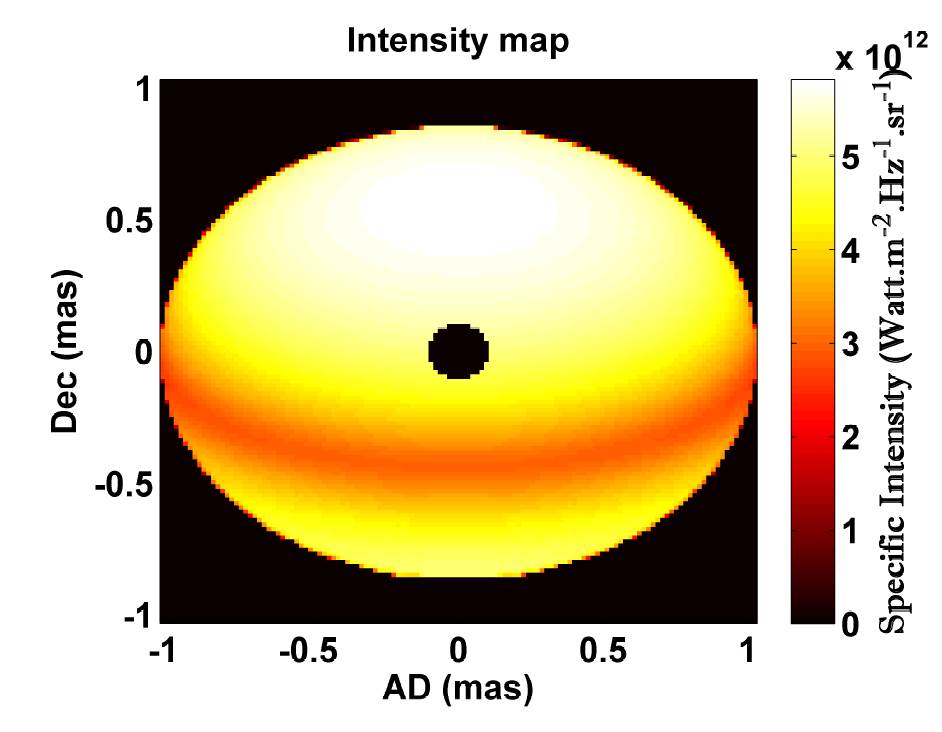}
\includegraphics[width=0.32\hsize,height=0.27\hsize,draft=false]{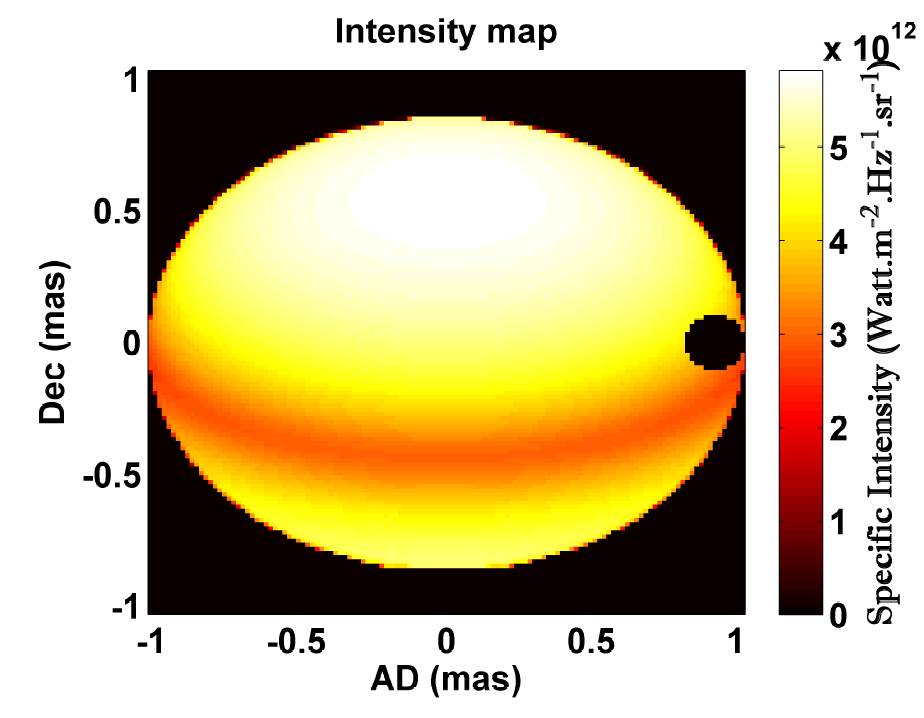}
\caption[Simulation du transit d'une planète devant la photosphère d'une étoile en rotation]{Simulation du transit d'une planète devant la photosphère d'une étoile en rotation.}
\end{figure}

\begin{figure}[ht]
\centering
\includegraphics[width=0.56\hsize,height=0.55\hsize,draft=false]{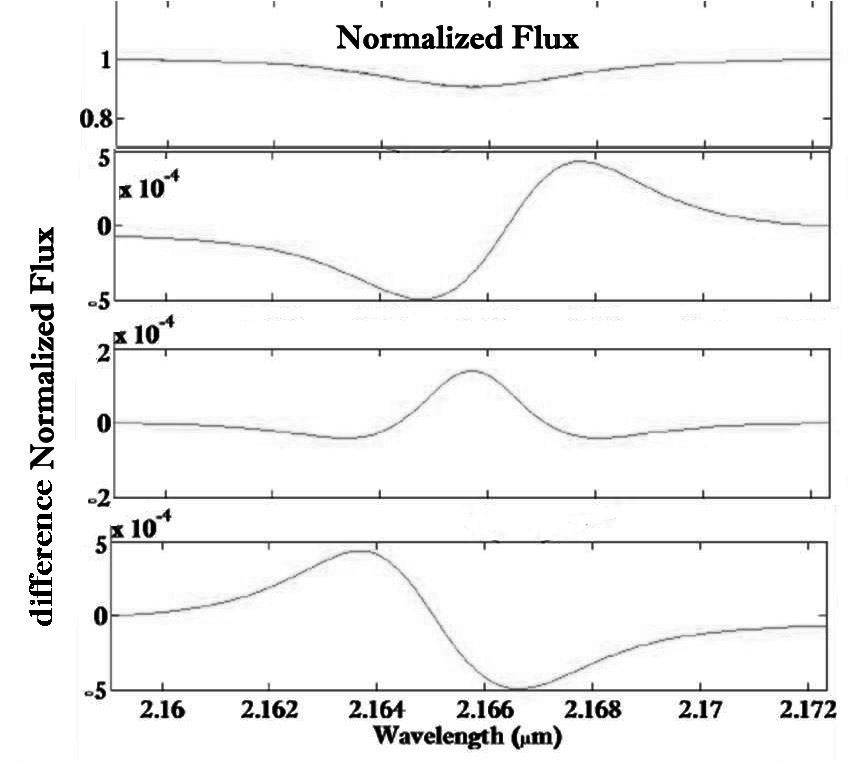}
\caption{Le flux normalisés sans la planète (la courbe du haut) et avec les 3 étapes de transit, représentées ci-haut, de celle-ci (les courbes du bas).}
\end{figure}

\clearpage

On remarque bien ici que l'impact du passage d'une exoplanète (ou d'une tache stellaire) sur le spectre est faible et sur les $\phidiff$ (ou bien sur le déplacement du photo-centre). Il faut soit avoir des bases kilométriques, ou bien que la planète (ou tache) soit très grande, car l'effet de tels phénomènes liées à l'intensité est en dessous des deux lignes parallèles en pointillé (des derniers tracés $\phidiff$ de la Fig.\ref{exorot}) représentant l'incertitude instrumentale détectable de nos jours (entre autre AMBER, avec ici $\sigma_\phi=\pm0.5^\circ$). En effet, \citet{1988ESOC...29..235P} avait mis en évidence que les rapports signal à bruit en DI et en spectroscopie $\frac{SNR(DI)}{SNR(spectro)}$ pour les effets fondamentaux était de l'ordre de $\frac{2R_*}{\frac{\lambda}{2B}}$, à condition que l'étoile soit résolue à minima (i.e. $\frac{2R_*}{\frac{\lambda}{2B}}<1$). Alors que dans notre cas Achernar est partiellement résolue. Néanmoins ces petites variations peuvent être accentuées via une analyse temporelle qui peut se débarrasser des effets systématiques (effets instrumentaux liés à l'OPD chromatique par exemple), en soustrayant la moyenne du signal par exemple, ce qui élimine les effets systématiques \citep{1988ESOC...29..235P}, ou bien via l'analyse de Fourier qui est une combinaison optimale des différentes mesures, pour laquelle le SNR évolue comme la racine carré du nombre de mesures temporelles (méthode utilisée par \citet{2001A&A...377..721J} pour le traitement et mise en évidence des PNR par exemple).\\

Bien qu'un travail plus détaillé et rigoureux reste à faire, ce dernier chapitre démontre que SCIROCCO a le potentiel d'être un code à usage multiple en interférométrie. Pour conclure ce chapitre, je résume dans un poster tout ce qui a été décrit dans le chapitre présent, et que j'ai eu l'opportunité de présenter, à la VLTI School Barcelonnette, organisée par A. Chiavassa en septembre 2013.\\

\includepdf[]{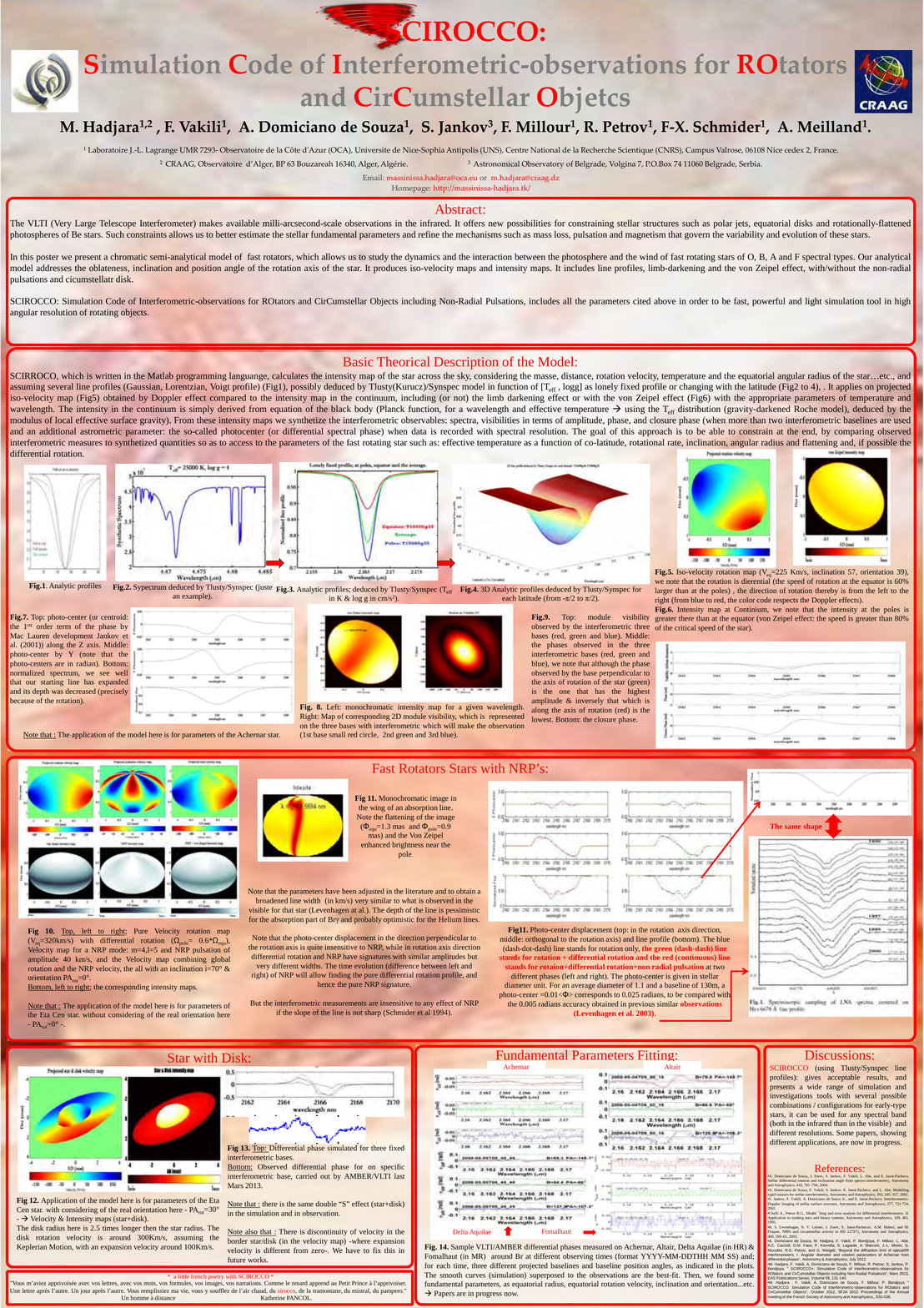}

Ainsi et après avoir introduit mes cibles et objets d'études dans le Chap.\ref{chap:rota}, parlé du contexte spectro-interférométrique historique et mondial, dans lequel s'insère mon travail (dans le Chap.\ref{chap:spec-interfero}), puis montré mes principales contributions dans le domaine (Chap.\ref{chap:scirocco} et Chap.\ref{chap:appli}) et enfin présenté tout le potentiel de SCIROCCO dans le Chap.\ref{chap:scriocco-pot}, il ne me reste plus qu'à conclure par un chapitre épilogue qui regroupe une discussion, les conclusions ainsi que des perspectives (Chap.\ref{chap:conclu}, ci-dessous), suivi par une annexe regroupant quelques uns des travaux secondaires que j'ai pu réaliser au cours de ma formation doctorale (Annexe.\ref{chap:annexe1}).\\
\chapter{Conclusions et Perspectives}
\label{chap:conclu}

\section{Conclusions}

Le sujet de recherche de ma thèse de doctorat choisi initialement avec mon directeur, était à caractère instrumental destiné à l'imagerie à très haut contraste en vue de la détection de planètes extrasolaire. Il s'agissait du concept DIFFRACT (DIFFerential RemApped Coronagraphic Telescope) que je présente en Annexe.\ref{chap:annexe1}. Etant chercheur permanent à l'Observatoire d'Alger, et ne pouvant être qu'épisodiquement et à courte durée à l'Observatoire de la Côte d'Azur, il ne m'était pas possible de m'impliquer de manière continue dans un travail de montage et de tests optiques complexes en laboratoire étant donné aussi qu'il n'était pas possible de faire ce type de recherche instrumentale en Algérie non plus.\\

L'existence de données observationnelles obtenues par l'équipe niçoise (A. Domiciano, F. Vakili et P. Bendjoya) fin 2009 avec l'instrument AMBER du VLTI sur l'étoile toupie "Achernar" offrait une opportunité immédiate et réelle pour un travail d'analyse et d'interprétation astrophysique. Naturellement, l'équipe m'a proposé de m'y investir ce que j'ai fait malgré la qualité dégradée des données en raison d'artéfacts instrumentaux qui rendaient l'extraction de l'information astrophysique très difficile.
Cela a radicalement changé mon plan de thèse car, si les corrections des artéfacts en question représentaient un défi en soi (beaucoup des données souffrant de ce problème par d'autres équipes avaient été laissées de côté), il  fallait me reconvertir pour découvrir le  monde fascinant, néanmoins pointu sur le plan technique, de l'interférométrie. Ce travail étant réalisable à distance, j'ai pu le mener à bien depuis Alger. Quelques mois plus tard, j'ai eu la possibilité de participer à la VLTI School, du 17 au 28 avril 2010 à Porquerolles organisée par Olivier Chesneau. Cela m'a grandement servi à parfaire mes connaissances en interférométrie optique à longue base et d'acquérir les outils de réduction et d'analyse comme les "OIFITS AMBER" pour Achernar grâce au logiciel "amdlib". Ma première année de thèse a donc été consacrée à me familiariser avec les principes de l'interférométrie et sa pratique; ce que j'ai présentée dans le Chap.\ref{chap:spec-interfero}.\\

Après avoir localisé la source et l'origine des biais présents sur les données AMBER d'Achernar de 2009, j'ai développé une méthode de traitement adéquate pour interpréter les mesures (voir Sec.\ref{sec_3.5.4}). La deuxième année de thèse m'a permis de valider cette méthode de traitement sur des données AMBER antérieures, également fortement biaisées, et accessibles dans les archives publiques du VLTI à ESO. Elles concernaient les étoiles, Altair, $\delta$ Aquilae et Fomalhaut (elles sont décrites en détail dans mon article A\&A, \citet{2014A&A...569A..45H}). Entre-temps j'ai commencé à m'intéresser à la physique des cibles stellaires à l'origine de ces données. Je me suis investi dans l'aspect de leur modélisation orientée vers l'interférométrie, et ce faisant acquérir des connaissances sur la rotation stellaire ainsi que sur l'intérêt scientifique de son étude (voir le Chap.\ref{chap:rota}). Ainsi, au cours de la fin de l'année 2011 je me suis investi dans la modélisation de cartes d'intensité 2D de l'hémisphère visible des étoiles théoriques, tout comme j'ai participé à la rédaction de l'article \citet{2012A&A...545A.130D}, où j'ai plus particulièrement rédigé la section sur l'acquisition, réduction et analyse de données, lors de ma brève présence à l'OCA en 2011.\\

Plus tard grâce à une aide financière du programme européen Fizeau d'Opticon FP7, j'ai pu passer 4 mois à l'OCA en 2012. Cela me permis de me concentrer sur le développement des bases de mon modèle SCIROCCO, qui était initialement dédié aux  simulations de rotateurs rapides (voir Chap.\ref{chap:scirocco}). Grâce à ce travail j'ai participé aux rencontres SF2A de juin 2012 à Nice \citep{2012sf2a.conf..533H}. A cette période je suis rentré en contact avec R. Petrov et S. Jankov pour participer à la demande d'observation ESO VLTI/AMBER sur le rotateur pulsant $\eta$ Cen, où j'ai fait une étude théorique sur l'effet des pulsations non-radiales (PNR) intégrée dans SCIROCCO (Chap.\ref{chap:scriocco-pot}). Le résultat de ce travail a été présenté à l'école d'été "Reconstruction d'images ; applications astrophysiques", organisée fin juin 2012 à Fréjus \citep{2013EAS....59..131H}. L'année 2012 fut donc riche en termes de contribution scientifique où mon travail sur DIFFRACT en collaboration avec Fatmé Allouche a abouti à un article SPIE \citep{2012SPIE.8446E..7EA} ainsi que mon travail sur les données d'Achernar avec AMBER en 2009 sous la direction de mes responsables de thèse \citep{2012A&A...545A.130D}.\\

L'année 2013 fut enrichissante: j'ai eu alors l'opportunité d'acquérir le savoir-faire de comment mener des observations directement sur place à l'Observatoire de Paranal/Chili avec VLTI/AMBER sur $\eta$ Cen. J'ai mené tout au long de cette période un grand nombre d'ajustements et de corrections de SCIROCCO, qui m'ont permis de déterminer l'impossibilité d'utiliser des profils de raie analytiques (Gaussienne, Lorentzienne ou profil de Voigt) pour l'extraction des paramètres fondamentaux stellaires, où seuls les profils de raie issus des codes de simulation d'atmosphère stellaire (Kurucz-Tlusty/Synspec par exemple) rendaient l'interprétation réaliste. Cela m'a permis de manipuler ce type de logiciel, Synplot entre autres, tel que discuté dans la Sec.\ref{sec_4.3}. Ma participation à la VLTI School 2013 a renforcé mes connaissances dans le maniement de certains logiciels dédiés à l'interférométrie, et en particulier ceux concernant la reconstruction d'images, à savoir MIRA et BSMEM, ce qui m'a permis de participer à la fin de l'année 2013 à l'équipe conduite par A. A. Domiciano de Souza qui travaillait sur la reconstruction d'image d'Achernar observé fin 2012 sur l'instrument PIONIER/VLTI (voir les détails de l'Annexe.\ref{chap:annexe1}). Ce travail a d'ailleurs abouti à un article A\&A ; \citet{2014A&A...569A..10D}, dont je suis co-auteur pour ma modeste contribution.\\

L'année 2014 s'est déroulée intégralement à l'OCA, où j'ai entamé et acquis le savoir-faire de réalisation de cartes d'assombrissement centre-bord et leurs  variation en fonction de la latitude $\theta$, tout comme l'assombrissement gravitationnel, (voir Chap.\ref{chap:scirocco}). J'ai aussi étudié et déterminé l'effet  des différents paramètres stellaires sur la phase interférométrique différentielle $\phidiff$ en fonction de la longueur d'onde que j'ai simulée grâce à SCIROCCO (voir Sect.\ref{sec_5.1}). J'ai également mené une seconde série de télé-observations sur $\eta$ Cen avec le VLTI/AMBER en mars 2014, qui m'ont permis d'acquérir de l'expérience des demandes d'observation ESO effectuées en 2012 \& 2013 et en même temps d'obtenir une bonne série de mesures sur des étoiles intéressantes telles que Regulus et $\eta$ CMa, que je souhaite traiter et interpréter dès que possible.\\

\section{Perspectives}

Dans le chapitre \ref{chap:scriocco-pot}, j'ai montré que SCIROCCO est un code qui possède un fort potentiel pour un usage multiple en interférométrie. Certes, toutes les études menées impliquant les PNR, un disque circumstellaire ou incluant des taches stellaires (ou exoplanètes en transit) restent primaires et nécessitent de plus amples investigations et expertises. Cela dit tout le travail de modélisation SCIROCCO sur les rotateurs rapides n'a été encore vérifié que sur les $\phidiff$ issus de l'instrument AMBER/VLTI avec un maximum de 3 bases interférométriques dans le domaine de l'IR. Il serait donc très intéressant de confronter mon modèle, toujours concernant les rotateurs stellaires, avec d'autres observables (visibilités, phase de clôture), sur d'autres instruments à plus de 3 bases interférométriques tel que PIONIER qui en possède 6, ou sur d'autres domaines spectraux, avec l'instrument CHARA par exemple, dans le domaine du visible. Il serait aussi intéressant d'essayer de nouvelles méthodes d'ajustement non conventionnelles (ex. la méthode proposée par \citet{1995A&AS..109..389C, 1995A&AS..109..401C}).\\ 

Enfin, l'avènement dans un futur proche, d'une seconde  génération d'instruments interférométriques tels que MATISSE et GRAVITY au VLTI, où les limites instrumentales sont poussées à l'extrême, pourrait aussi apporter son lot de découvertes, notamment sur la dynamique photosphérique, les environnements circumstellaire et proto-planétaire. SCIROCCO pourrait très bien être adapté pour l'étude de telles cibles en s'appuyant sur ces nouvelles limites instrumentales.\\

En effet, MATISSE (Multi AperTure mid-Infrared SpectroScopic Experiment), et dont la mise en service est prévue pour 2017, pourra combiner les faisceaux de 4 télescopes (y compris les UT) simultanément dans l'infrarouge moyen (bande L, M \& N). Développé pour l'étude des disques protoplanétaires et des AGNs, il est peu adapté à l'étude des surfaces d'étoiles, à cause d'une faible résolution spatiale, mise à part quelques étoiles très proches, géantes ou supergéantes.  Par contre il peut parfaitement étudier des environnements de rotateurs rapides, et par conséquent l'influence de la rotation sur l'éjection de matière de certains rotateurs critiques ou quasi-critiques (les étoiles Be classiques par exemple).\\

Quant à GRAVITY (General Relativity Analysis via Vlt InTerferometrY), dont la mise en service est prévue pour 2016,  pourra combiner les faisceaux de 4 télescopes UT (et 6 pour les AT) simultanément dans l'infrarouge (bande K) avec une résolution spectrale maximale de 4000. Principalement développé pour faire de l'astrométrie (mesure de phase) à très grande précision sur le centre galactique, il sera parfaitement adapté à l'étude des surfaces stellaires des rotateurs rapides (tout comme AMBER) par la méthode présentée dans cette thèse. Cependant, GRAVITY diffèrera d'AMBER sur 3 points majeurs, à savoir ; le nombre de combinaisons de télescopes, où avec  4 télescopes, GRAVITY offrira une meilleure couverture (u,v). La grande précision sur les phases de GRAVITY donnera accès à des étoiles plus petites que ce qui est actuellement réalisable avec AMBER. Enfin, GRAVITY ayant une plus faible résolution spectrale, il ne permettra pas l'étude aussi fine de vitesses de rotation élevées comme celles observées avec AMBER.\\

Notons aussi qu'au-delà des instruments cités ci-dessus, un travail de prospective en interférométrie visible (domaine le plus adaptés à l'étude des surfaces d'étoiles, car ayant une meilleure résolution spatiale et donnant accès à de nombreuses raies photosphériques) a été initié dans la communauté  afin de développer une nouvelle génération d'instruments visibles pour CHARA et le VLTI (prototype FRIEND).\\

\appendix

\chapter{Autres Travaux}
\label{chap:annexe1}

Dans cette annexe  je présente brièvement d'autres travaux scientifiques parallèles que j'ai dû mener ; à savoir un code Matlab de lecture des données OIFITS, une étude sur un instrument de détection directe d'Exoplanète (voir papier SPIE où mon nom y est en second auteur ; \citet{2012SPIE.8446E..7EA}), et un travail de reconstruction d'image sur Achernar en nième auteur dans un papier A\&A \citet{2014A&A...569A..10D}).

\section{OIFITS sur Matlab}

Ci-dessous est présentée un exemple de figure qui permet de visualiser simultanément toutes les données OIFITS (spectres, modules de visibilités, phases différentielles et la phase de clôture), ainsi que d'importantes informations annexes, telles que : le plan de couverture $(u,v)$, l'instrument, les bases interférométriques (télescopes), l'objet observé, la date et heure d'observation, le temps d'exposition, ...etc. Cette figure résulte d'un code Matlab "ShowOidata" que j'ai développé à la base pour la lecture des données OIFITS AMBER, et qui s'appuie sur une librairie OIFITS-MATLAB (MATrix LABoratory) -que j'ai adaptée à mes besoins- développée au sein du laboratoire J.L.Lagrange par Antony Schutz.

\includepdf[]{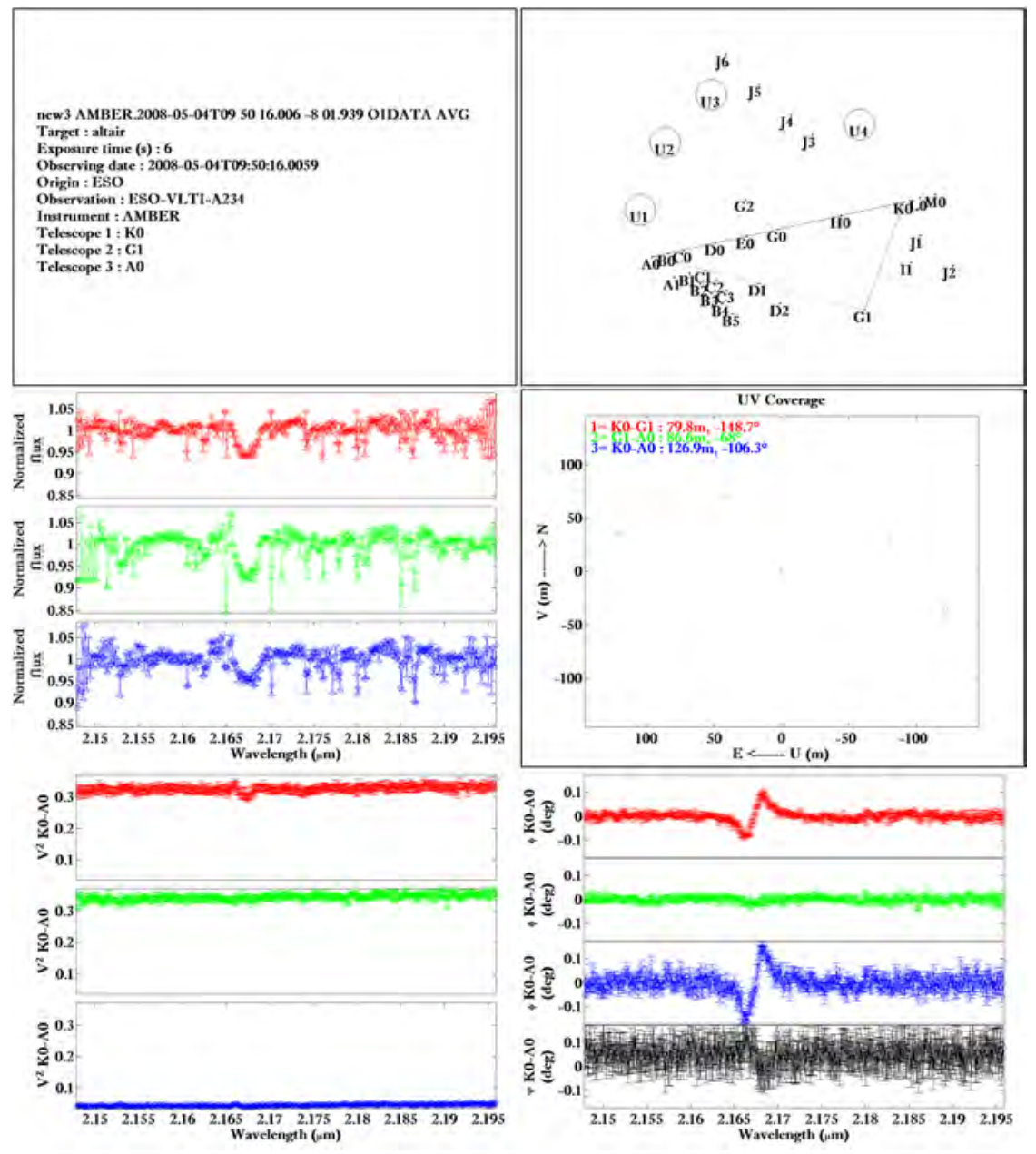}

\section{Reconstruction d'image}

Dans le courant du mois d'octobre 2013, j'étais invité à participer au travail de reconstruction d'image sur des données PIONIER d'Achernar, avec Gaetan Dalla Vedova, Florentin Millour et Armando Domciano de Souza. Mon travail a d'abord été l'installation, compréhension et exécution des logiciels MIRA\footnote{Multi-aperture Image Reconstruction Algorithm est un algorithme de reconstruction d'image à partir des données fournies par des interféromètres optiques. Ecrit en Yorick par Eric Thiébaut, du Centre de Recherche Astrophysique de Lyon \citep{2008SPIE.7013E..1IT}. MIRA procède par réduction directe d'une vraisemblance pénalisée. Cette pénalité est la somme de deux termes: un terme de vraisemblance (le $\chi^2$) qui impose l'accord du modèle avec les données, en plus d'un terme de régularisation pour tenir compte des appropries, qui sont tenus de lever nombreuses dégénérescences en raison de la faible densité d'échantillonnage des fréquences spatiales.} \& BSMEM \footnote{BiSpectrum Maximum Entropy Method : est aussi un logiciel pour la reconstruction d'image à partir des données d'interférométrie optique. Il a d'abord été écrit en Fortran et complété par David Buscher en 1992, à l'université de Cambridge, pour démontrer la reconstruction directe maximale d'entropie à partir de données de d'ouverture de synthèse optique \citep{1994IAUS..158...91B}. BSMEM applique une approche entièrement bayésienne pour le problème inverse de trouver l'image la plus probable et met en \oe{}uvre un algorithme de descente de gradient pour maximiser la probabilité a posteriori d'une image, en utilisant l'entropie de l'image reconstruite comme probabilité priori.}. Je me suis ensuite concentré exclusivement sur le logiciel BSMEM sans apporter aucune modification, tandis que mon collègue Gaetan lui s'est chargé d'adapter le logiciel MIRA aux besoins de la reconstruction. Ci-dessous, l'article qui contient la reconstruction d'image à partir des données PIONIER d'Achernar et juste après une figure de reconstruction d'image obtenue par BSMEM. Les résultats obtenus par les deux méthodes ont été jugé équivalents.

\includepdf[]{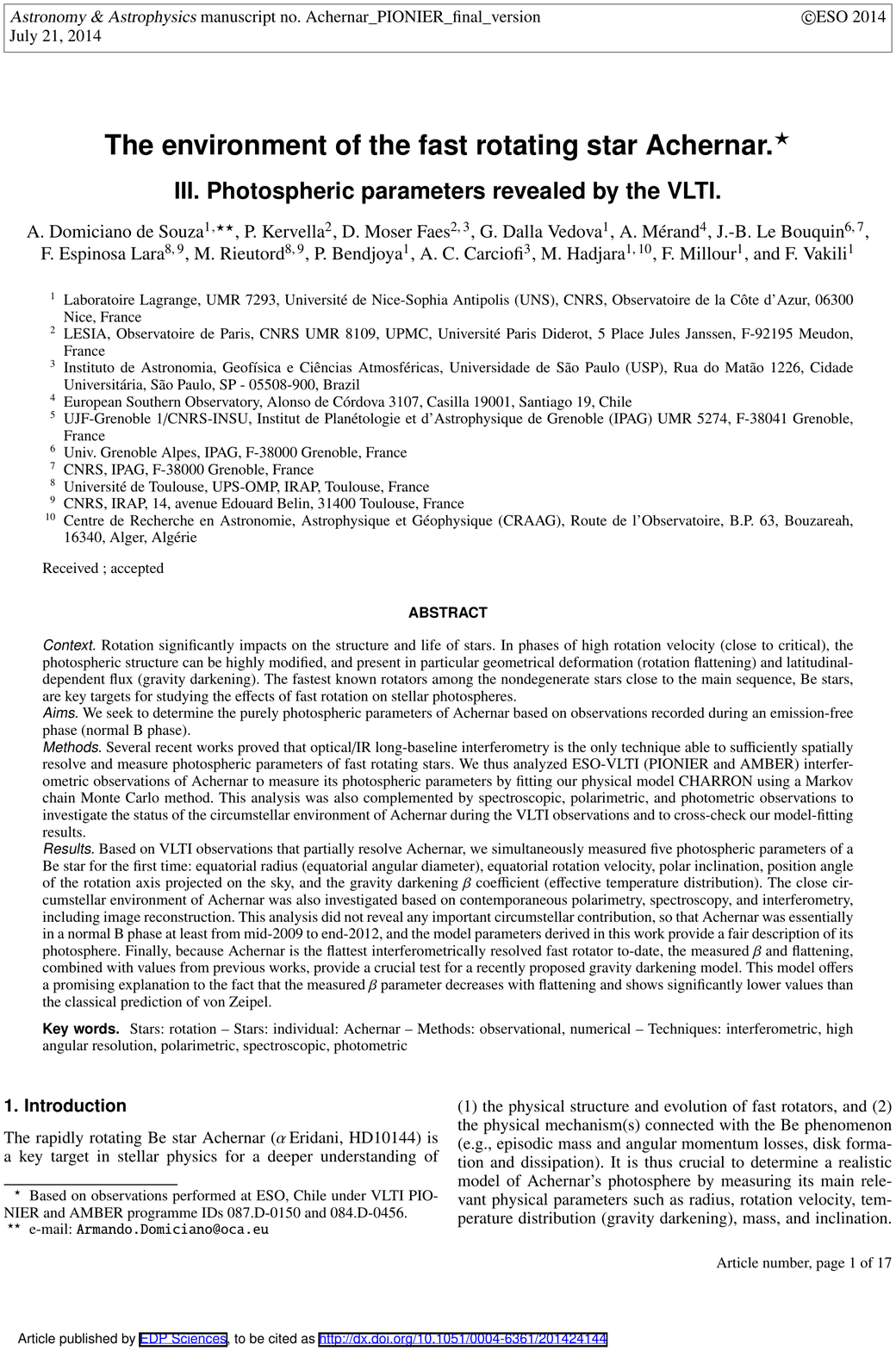}

\begin{figure}[ht]
\centering
\includegraphics[width=0.6\hsize,draft=false]{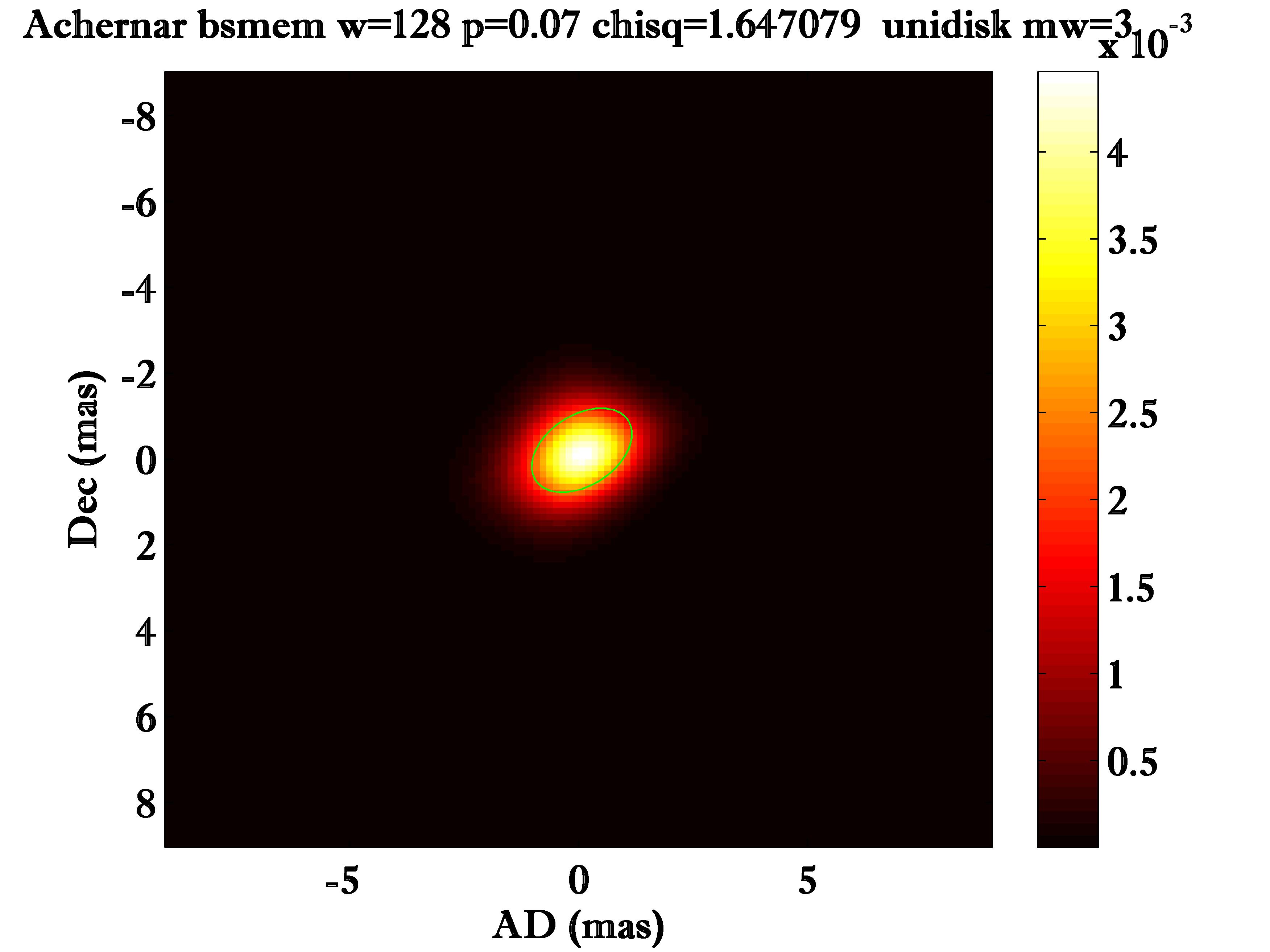}
\caption[Reconstruction d'image des données PIONIER Achernar par BSMEM.]{Reconstruction d'image des données PIONIER Achernar par BSMEM pour une image de 128x128pixels, pour une résolution 0.07mas, avec un approprié d'un disque uniforme de 3 mas de diamètre. A titre indicatif j'ai y rajouté un contour d'ellipse ayant les paramètres d'un rayon de grand axe de $1.5 mas$, un rapport d'aplatissement grand axe /petit axe de $1.53$, et un angle $PA_{rot}=36.9^\circ$.}\label{Achernar_bsmem}
\end{figure}

\clearpage

\section{DIFFRACT}

Ma contribution au projet DiffRACT: DIFFerential Remapped Aperture CoronagraphicTelescope, qui comprenait des simulations en optique de Fourier, se résume dans l'article ci-dessous :

\includepdf[pages=1-6]{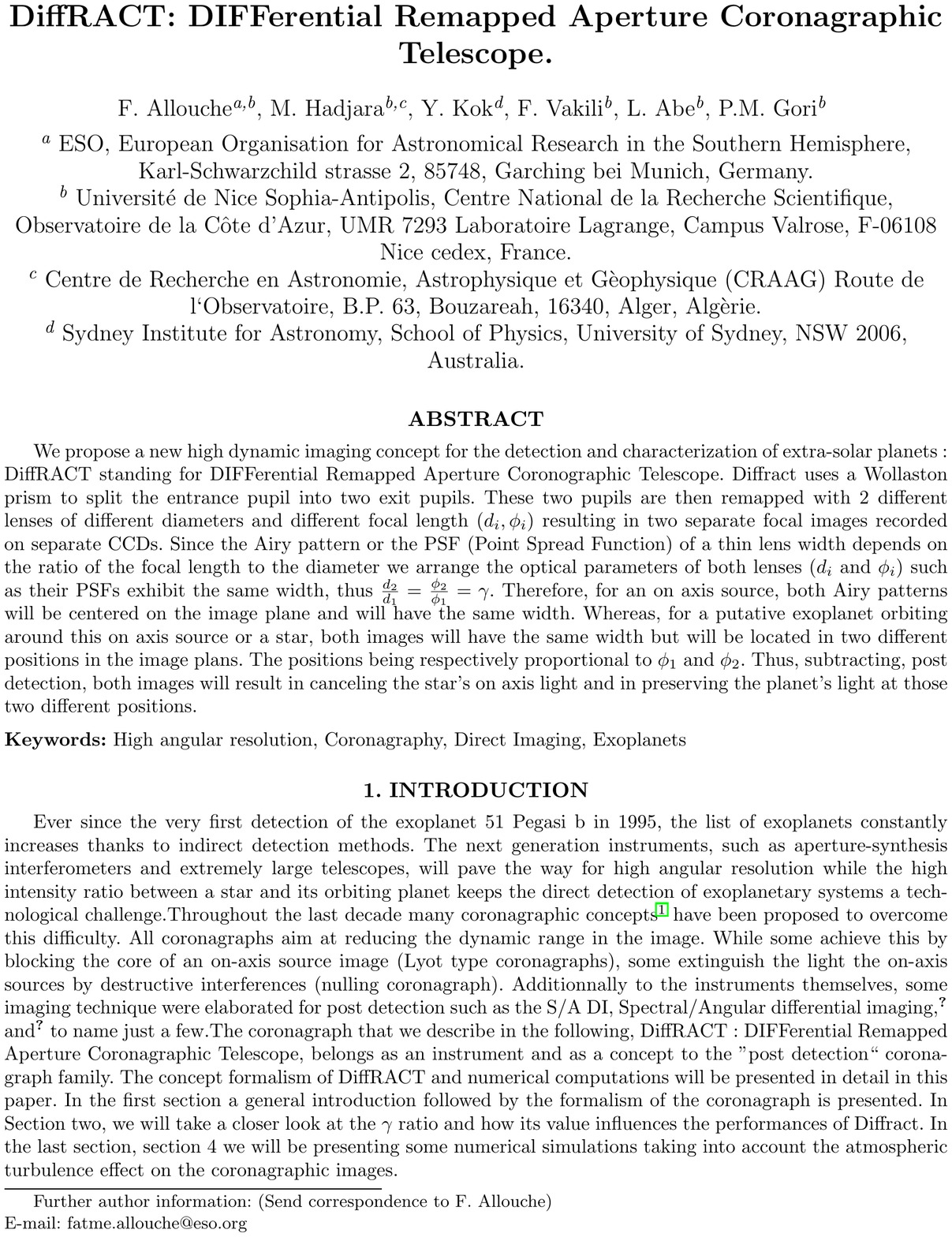}
\includepdf[]{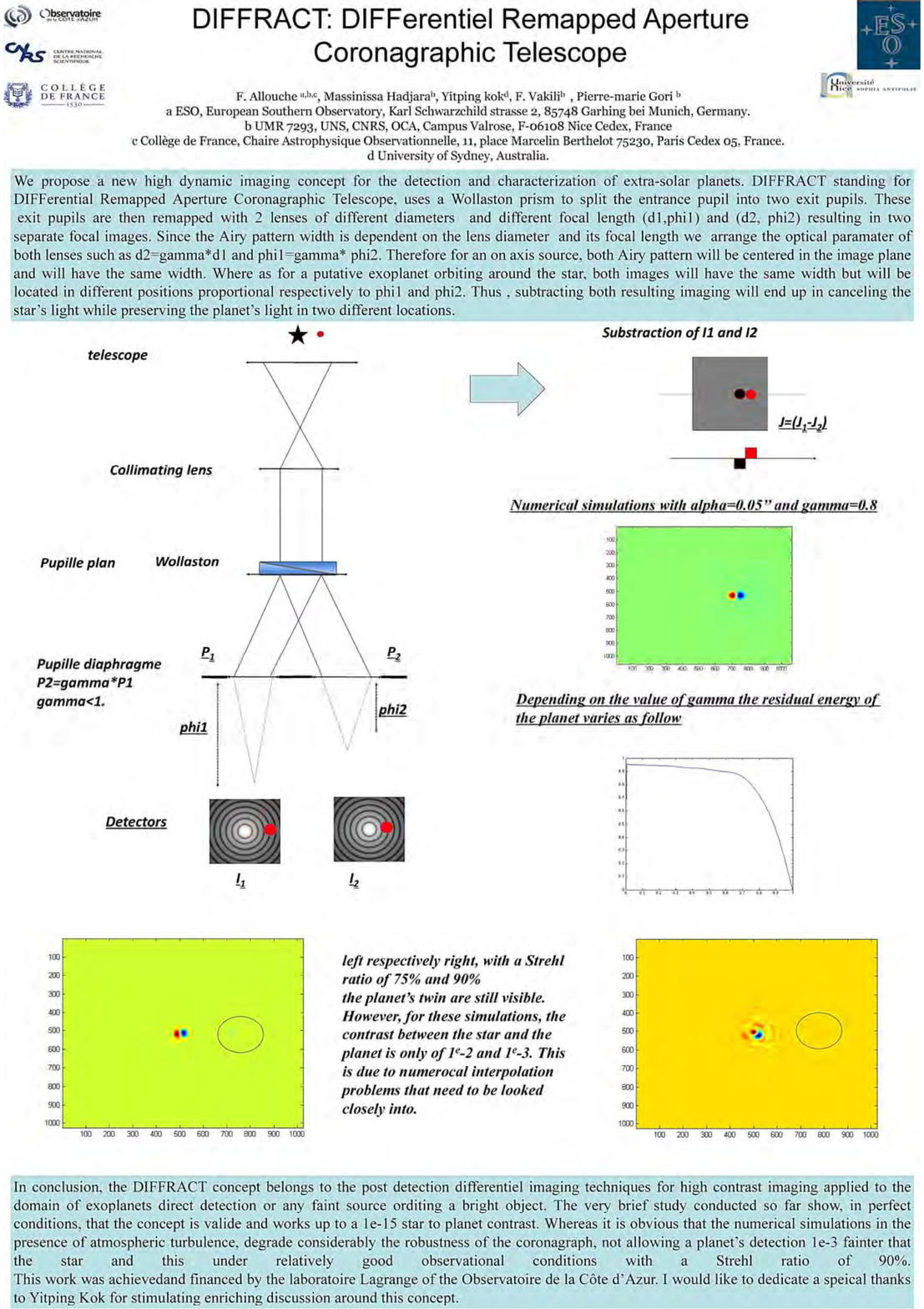}

\section{Contexte de mon travail de thèse}

Mon travail pour le doctorat s'est déroulée à l'observatoire d'Alger avec quelques missions régulières annuelles d'un mois à trois à l'observatoire de la Côte d'Azur, à l'exception de l'année 2013-2014 qui s'est déroulée en quasi-totalité à l'OCA. Avec une aide financière de quelques centaines d'euros de chacune de mes instituts de cotutelle, conditions que j'avais adaptées pour ma thèse . J'ai appris ainsi, en plus de travail scientifique entamé en toute autonomie, à gérer les différentes demandes de financement, quasi-trimestrielle. Ma formation doctorale m'a permis d'acquérir donc une solide expérience dans l'élaboration de projets scientifiques et de demandes de subvention, comme :

\begin{itemize}
\item[-] Une demande de projet CRAAG (observatoire d'Alger) sur le domaine de l'interférométrie stellaire et dont son CS a approuvé la mise en marche.

\item[-] Deux demandes de financement Opticon Fizeau en 2012 et 2014 (toutes deux acceptées).

\item[-] Un projet CMEP, puis un PICS franco-Algérien que j'avais géré de A à Z, dans l'espoir d'obtenir plus amples soutien financier pour mes projets scientifique. Ces projets ont tous les deux été approuvés par le CNRS côté français, mais sont restés hélas sans suite coté Algérien.

\item[-] Et enfin une mission d'observation au VLTI Paranal/Chili financée en grande partie par mon projet CRAAG, avec une notable aide financière de l'équipe MPO/OCA et de l'ESO.
\end{itemize}

J'estime que toutes ces expériences m'ont permis de développer des aptitudes de gestion de projet, y compris dans des conditions relativement complexes.
\bibliographystyle{aa}
\bibliography{Biblio_Massi14}

\cleardoublepage
\begin{vcenterpage}
\noindent\rule[2pt]{\textwidth}{0.5pt}
\\
{\large\textbf{Résumé :}}
La résolution angulaire d'un télescope optique s'améliore comme l'inverse du diamètre de sa pupille. Au sol, la turbulence atmosphérique réduit cependant ce pouvoir. La construction de télescopes optiques de très grand diamètre au-delà de 50 m constitue un défi technologique difficilement envisageable aujourd'hui. Afin de remédier à cette limite et à la perturbation atmosphérique, une alternative consiste à combiner de manière cohérente la lumière collectée par 2 télescopes (ou plus) séparés de plusieurs dizaines de mètres. L'interférométrie optique à grand nombre de télescopes pose toutefois des obstacles techniques qui ne peuvent être  surmontés que depuis peu, apportant néanmoins des résultats uniques pour les astrophysiciens théoriciens.
\\
Avec la construction du VLTI (Very Large Telescope Interferometer, de l'observatoire européen ESO pour l'hémisphère sud) il est désormais possible d'effectuer des observations avec des résolutions de l'ordre des milli-arc-secondes, en particulier dans l'infrarouge avec l'instrument AMBER (Astronomical Multiple BEam Recombiner). Ces nouvelles possibilités nous permettent de mieux contraindre les structures stellaires telles que les jets polaires, les disques équatoriaux et les photosphères aplaties des étoiles en rotation. Ainsi l'estimation des paramètres fondamentaux stellaires permet d'explorer en détail les mécanismes de la perte de masse, la pulsation et le magnétisme qui régissent la variabilité et l'évolution stellaire.
\\
Cette thèse présente les résultats d'observations d'étoiles en rotation rapide menées sur le spectro-interféromètre AMBER du VLTI dans ses modes haute et moyenne résolutions spectrales. Les mesures effectuées sont les visibilités estimées sur trois bases simultanées, les phases différentielles en fonction de la longueur d'onde et des phases de clôtures avec, pour certaines nuits une bonne couverture du plan $(u,v)$. Les données utilisées sont issues de plusieurs campagnes d'observation, y compris d'archives. Ces dernières étaient fortement dégradées par les défauts optiques d'AMBER, et affectés par des bruits classiques d'interférométrie optique à longue base en IR: défauts du détecteur, bruit de lecture, instabilités du suiveur de franges, ...etc. Leur analyse a nécessité la mise au point d'outils numériques de réduction spécifiques pour atteindre les précisions nécessaires à l'interprétation de mesures interférométriques. Pour interpréter ces mesures j'ai développé un modèle semi-analytique chromatique d'étoile en rotation rapide qui m'a permis d'estimer, à partir des phases différentielles; le degré d'aplatissement, le rayon équatorial, la vitesse de rotation, l'angle d'inclinaison, l'angle position de l'axe de rotation de l'étoile sur le ciel, la distribution de la température effective locale et de la gravité à la surface de l'étoile dans le cadre du théorème de von Zeipel. Les résultats concernant 4 étoiles massives de types spectraux B, A et F m'ont permis de les caractériser pour les mécanismes évoqués ci-dessus et d'ouvrir ainsi la perspective d'études plus systématiques d'objets similaires en étendant ultérieurement ces études à la relation photosphère-enveloppe circumstellaire.\\
{\large\textbf{Mots clés :}}
Étoiles: rotation rapide, Méthodes: observationnelle, Techniques numérique: interférométrique, haute résolution angulaire.
\\
\noindent\rule[2pt]{\textwidth}{0.5pt}
\end{vcenterpage}

\begin{vcenterpage}
\noindent\rule[2pt]{\textwidth}{0.5pt}
\begin{center}
{\large\textbf{Long baseline spectro-interferometric observing and modeling of stars and their close environment\\}}
\end{center}
{\large\textbf{Abstract:}}
The angular resolution of an optical telescope improves as the inverse of the pupil diameter. On the earth, the atmospheric turbulence reduced however that power. The construction of optical telescopes with very large diameter beyond 50 m is a difficult technological challenge to imagine today. To address this limitation and to the atmospheric disturbance, an alternative is to combine coherently the light collected by two telescopes (or more) separate by several tens of meters.  Optical interferometry of many telescopes poses however technical obstacles that can be overcome only recently, nevertheless providing unique results for theorists astrophysicists.
\\
With the construction of the VLTI (Very Large Telescope Interferometer of the European Observatory ESO for the southern hemisphere) it is now possible to make observations with resolutions of the order of milli-arc-seconds, especially in IR with AMBER instrument (Astronomical Multi Beam Recombine). These new capabilities allow us to better constrain the stellar structures such as polar jets, equatorial disks and flattened photospheres of rotating stars. Thus the estimation of stellar fundamental parameters allows to explore in detail the mechanisms of mass loss, pulsation and magnetism governing the variability and the evolution of the stars.
\\
This thesis presents the results of fast rotating stars observations carried out on the AMBER spectro-interferometer VLTI in its high \& medium spectral resolutions modes during my PhD thesis. The measurements are the visibilities, estimated on three simultaneous bases, the differential phases depending on the wavelength and the closure phases, with good coverage of (u,v) plane for some nights. The data used are from several observation campaigns, including archives. Those were highly degraded by the optical defects of AMBER, and assigned by standard optical interferometry long base IR noises: defects of the detector, reading noise, fringes follower instabilities ... etc. Their analysis required the development of specific digital reduction tools to reach the necessary precision for the interferometric measurements interpretation. In order to interpret those measures I developed a chromatic semi-analytical model of rapidly rotating star that allowed me to estimate, from the differential phases; the degree of flattening, the equatorial radius, the rotation velocity, the angle of inclination, the position angle of the star rotation axis in the sky, the local distribution of the effective temperature and the surface gravity of the star within the von Zeipel theorem. The results for four massive stars of spectral types B, A and F have allowed me to characterize the mechanisms discussed above and thus open some prospect for more systematic studies of similar objects, with extending later these studies to the relationship photosphere - circumstellar envelope.\\
{\large\textbf{Keywords:}}
Stars: fast rotation, Methods: observational, numerical Techniques: interferometric, high angular resolution.
\\
\noindent\rule[2pt]{\textwidth}{0.5pt}
\end{vcenterpage}

\end{document}